# Wide-Field InfraRed Survey Telescope-Astrophysics Focused Telescope Assets

# WFIRST-AFTA

## Final Report

by the

Science Definition Team (SDT) and WFIRST Project

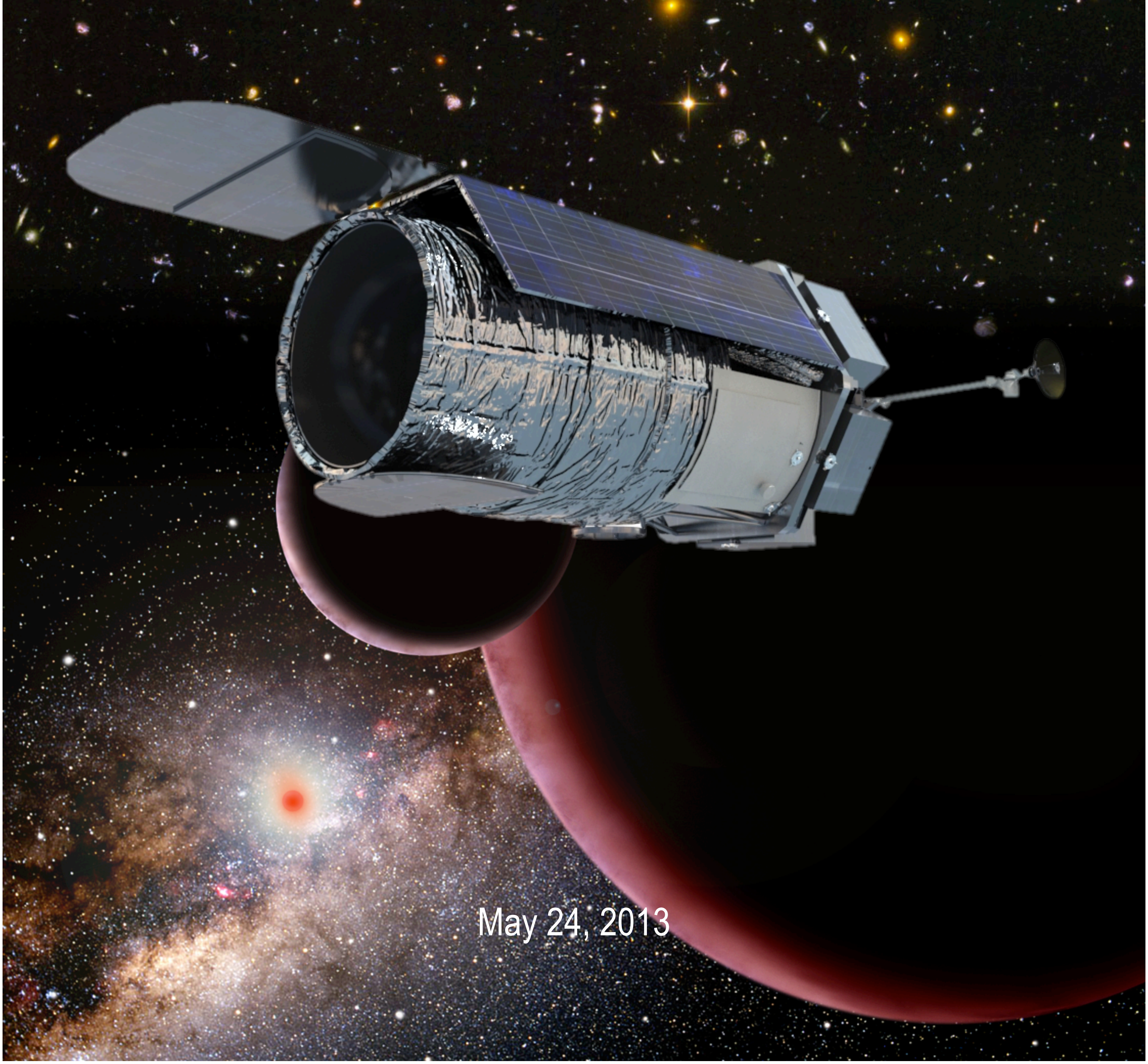

May 24, 2013

# **W**ide-**F**ield **I**nfra**R**ed **S**urvey **T**elescope-**A**strophysics **F**ocused **T**elescope **A**ssets

# **WFIRST-AFTA**

# Final Report

## Science Definition Team

D. Spergel[1], N. Gehrels[2]
J. Breckinridge[3], M. Donahue[4], A. Dressler[5], B. S. Gaudi[6], T. Greene[7], O. Guyon[8], C. Hirata[3]
J. Kalirai[9], N. J. Kasdin[1], W. Moos[10], S. Perlmutter[11], M. Postman[9], B. Rauscher[2], J. Rhodes[12], Y. Wang[13]
D. Weinberg[6], J. Centrella[14], W. Traub[12]

Consultants
C. Baltay[15], J. Colbert[16], D. Bennett[17], A. Kiessling[12], B. Macintosh[18], J. Merten[12], M. Mortonson[6], M. Penny[6]
E. Rozo[19], D. Savransky[18], K. Stapelfeldt[2], Y. Zu[6]

Study Team
C. Baker[2], E. Cheng[20], D. Content[2], J. Dooley[12], M. Foote[12], R. Goullioud[12], K. Grady[2], C. Jackson[21], J. Kruk[2]
M. Levine[12], M. Melton[2], C. Peddie[2], J. Ruffa[2], S. Shaklan[12]

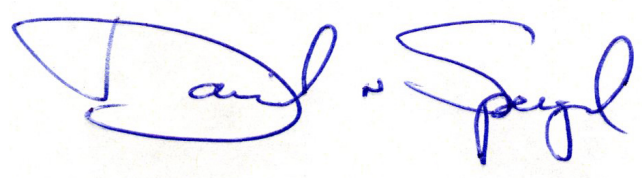

David Spergel, SDT-Co-Chair Date

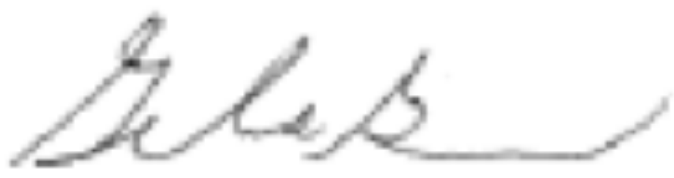

Neil Gehrels, SDT Co-Chair Date

1 Princeton University
2 NASA/Goddard Space Flight Center
3 California Institute of Technology
4 Michigan State University
5 Carnegie Institution for Science
6 Ohio State University
7 NASA/Ames Research Center
8 University of Arizona
9 Space Telescope Science Institute
10 Johns Hopkins University
11 University of California Berkeley/Lawrence Berkeley National Laboratory
12 Jet Propulsion Laboratory/California Institute of Technology
13 University of Oklahoma
14 NASA Headquarters
15 Yale University
16 Infrared Processing and Analysis Center/California Institute of Technology
17 University of Notre Dame
18 Lawrence Livermore National Laboratory
19 Stanford Linear Accelerator Laboratory
20 Conceptual Analytics
21 Stinger Ghaffarian Technologies

## EXECUTIVE SUMMARY

**A 2.4-meter space telescope equipped with a very wide-field infrared camera would revolutionize astrophysics. The transfer of this telescope asset to NASA is a boon to the US scientific community that should be utilized to produce transformative science.**

The National Academy of Sciences decadal survey for astronomy, "New Worlds/New Horizons" ranked WFIRST as the top priority for space-based astronomy for the coming decade, recognizing the importance of the science it would produce in areas ranging from the study of dark energy to the search for exoplanets. The DRM1 design from the previous Science Definition Team (SDT) confirmed that an unobstructed, 1.3-m aperture WFIRST would be among the highest 'science-per-dollar' missions ever built.

**If used for the WFIRST mission, the 2.4-meter telescope would be significantly more capable than the smaller versions of WFIRST studied in previous SDTs.** As we show here in this Report, by dividing the time on WFIRST-2.4 between dark energy, microlensing, and guest observer observations, gains of 50-100% or more are expected for all. Furthermore, the 2.4-m telescope makes possible new science at the limit of, and beyond, what was achievable with earlier WFIRST designs. WFIRST-2.4 is not only capable of wide-field near-IR observations that are key to many of the frontier areas of astrophysics, but will do so at the resolution of HST.

The 2.4-m aperture of WFIRST-2.4 collects almost 3 times as much light as the unobstructed 1.3-m SDT-DRM1 and offers a factor of 1.9 improvement in spatial resolution (point spread function effective area–- PSF), which itself provides another factor-of-two improvement in accomplishing many science programs. Compared to SDT-DRM1, the design presented in our report has a ~30% smaller field-of-view (FOV) that moderates these gains, but the higher spatial resolution takes the WFIRST-2.4 program to the "next level" and sets it apart from DRM1 and — for the dark energy program specifically — the Euclid mission. In particular, having the greater speed of the larger aperture and the sharper PSF allows WFIRST-2.4 to reach a factor-of-two deeper per unit time over an unprecedentedly large field for a large space telescope, ~90 times bigger than the HST–ACS FOV, and ~200 times bigger than the IR channel of WFC3 that has been a tremendously successful scientific tool.

To fully take advantage of this larger telescope, the wide-field instrument uses higher performance infrared detectors that provide 4 times as many pixels in about the same size package as the current generation of detectors. The WFIRST-2.4 concept uses eighteen of these detectors in a single focal plane to cover 0.28 $deg^2$ of sky in a single image at a pixel scale of 0.11 arcsec/pix. In the mission presented in this report, WFIRST-2.4 is deployed in a 28.5 degree inclined geosynchronous orbit and will operate for a minimum of 5 years (and for 6 years if the optional coronagraph is added). The geosynchronous orbit allows continuous downlink to the ground, enabling a much higher science data rate. WFIRST-2.4 is designed to be robotically serviceable, should a future robotic servicing capability be deployed in geosynchronous orbit.

**Box 1**

*WFIRST-2.4 will have an imaging field more than 200 times larger than the Hubble Space Telescope. The Hubble Deep Field — a picture containing thousands of faint, distant — pushed the study of galaxy evolution to new limits. A WFIRST-2.4 "Degree Deep Field" will — in a single picture — will include millions of galaxies. Displayed at the resolution of a computer screen, a Degree Deep Field would cover a wall 20-feet long and 10 feet high. A room whose four walls display 4 different deep field images will convey as never before the vastness of the universe of galaxies. Electronic display walls would allow visitors to touch the surface to magnify the image. NASA could make 'wallpaper' of these and other WFIRST-2.4 images for distribution to schools, museums, and the public. As illustrated below with the small orange box, only 10% of the Hubble Deep Field (whose full size is shown by the open box) contains a wealth of detail about the evolution of galaxies across cosmic time.*

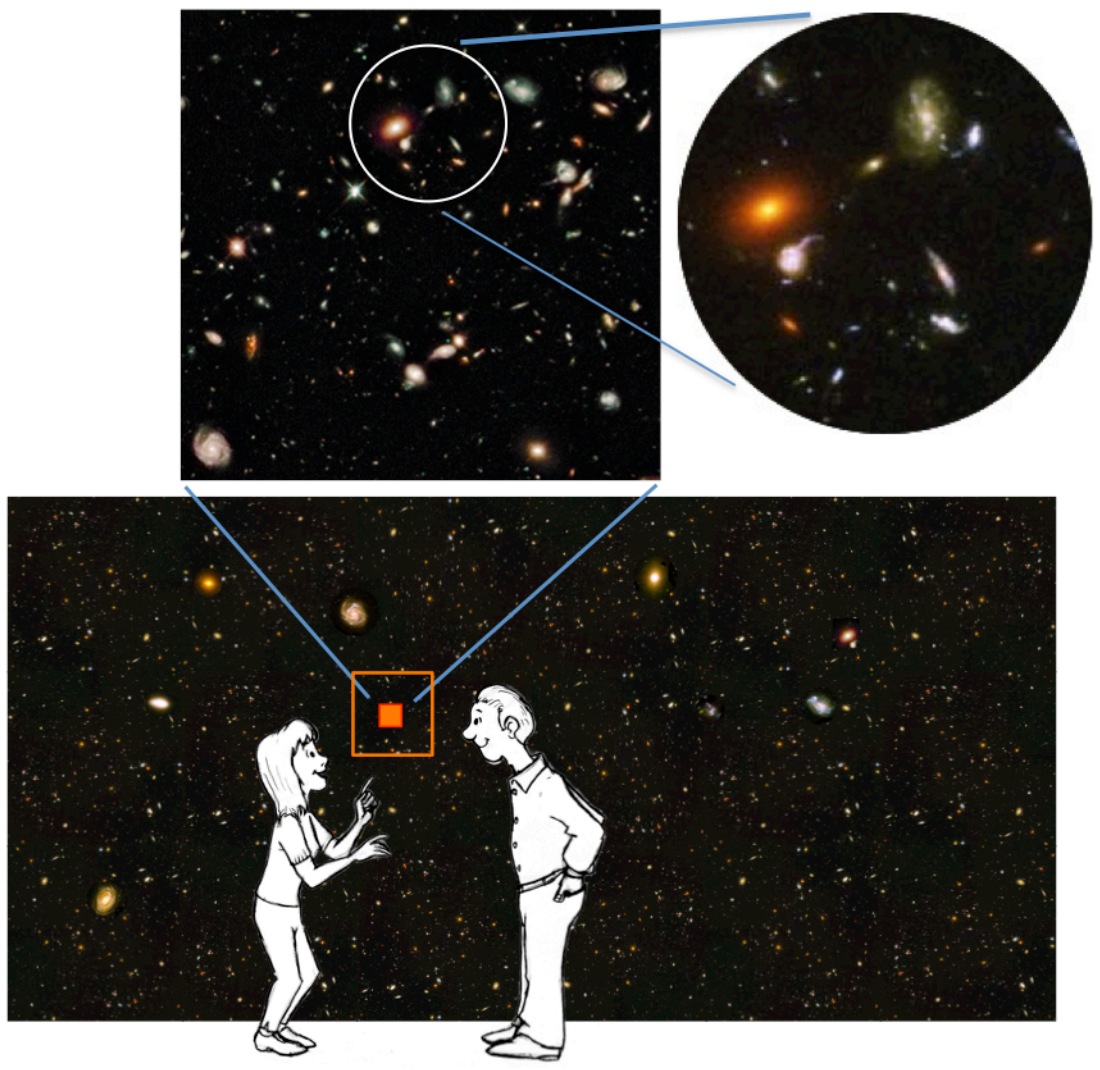

**WFIRST-2.4 will measure the expansion history of the universe and growth of structure to better than 1% in narrow redshift bins using several independent methods with aggregate precision of 0.2% or better on multiple cosmological observables. Observations could reveal the nature of dark energy, the driver of cosmic acceleration, which is one of the great mysteries of modern science**. In 2011, the Nobel Prize committee awarded its physics prize to three American scientists who used observations of supernovae to provide the first convincing evidence for cosmic acceleration. The use of a 2.4 meter telescope will enhance our ability to use the three basic techniques that were the focus of previous WFIRST mission studies (BAO, supernova, and gravitational lensing) and will also significantly improve the use of two other important techniques: galaxy cluster counts and redshift space distortions. Specific strengths of WFIRST-2.4 for dark energy studies include:

- In addition to using lines of hydrogen atoms in galaxies to trace the large-scale distribution of matter at redshifts between 1 and 2, WFIRST-2.4's larger collecting area will enable it to also use lines of ionized oxygen in more distant galaxies. This added capability enables measurements of large-scale structure out to a redshift of 2.95, just two billion years after the big bang.
- WFIRST-2.4's added sensitivity will enable the detection of a 60% higher density of galaxies in 10% less time at z=1.5 in its spectroscopic survey vs. DRM1. With this higher density of galaxies, the redshift space distortion measurements (measuring the effects of cosmic acceleration on structure growth) are also improved.
- WFIRST-2.4 will use an Integral Field Unit (IFU) to take spectra of ~2700 distant supernovae discovered in its synoptic imaging survey. This enables more precise distance measurements and reduces uncertainties associated with cosmic evolution.
- WFIRST-2.4 will make deeper observations of galaxy lensing. In the high latitude survey, WFIRST will measure the shapes of 70 galaxies per square arcminute over 2000 square degrees. These deeper observations will complement the wider and shallower observations by the Large Synoptic Space Telescope (LSST) and the European-led Euclid telescope and will enable precision measurements of non-Gaussian features in the lensing maps, thus, significantly improving the dark energy measurements.
- These deeper lensing observations will also enable accurate determination of the masses of clusters. Cluster abundances could then be used as another complementary method to determine the effects of dark energy on the growth rate of large structure.

**WFIRST-2.4 will perform a microlensing survey that will greatly advance our understanding of the demographics of extrasolar planets by complementing Kepler's census of close-in planets.** The growing realization that our Galaxy is filled not only with billions of stars, but also with billions of planets has been the most important astronomical discovery of the new millennium. Far more stable than the Hubble Space Telescope or any ground-based telescope, WFIRST-2.4 would be able to make important contributions to our understanding of the diversity of planets in the universe and towards discovering habitable planets beyond our solar system with its microlensing survey:

- WFIRST-2.4 will systematically survey the cold, outer regions of planetary systems throughout the Galaxy, detecting ~3000 total bound exoplanets in the range of 0.1-10,000 Earth masses, including **~1000** "Super-Earths" (roughly 10 times the mass of Earth), **~300** Earth-mass planets, and **~40** Mars-mass planets. This will enable the measurement of the mass function of cold exoplanets to better than ~10% per decade in mass for masses >0.3 $M_{Earth}$, and an estimate of the frequency of Mars-mass embryos accurate to ~15%.
- When combined with Kepler, WFIRST-2.4 will provide statistical constraints on the characteristic density of rocky planets in the outer habitable zones of sunlike stars.
- WFIRST-2.4 can detect analogs to all of the planets in our Solar System more massive than Mercury, allowing us to properly place our Solar System in the context of other planetary systems for the first time, and providing crucial constraints on the frequency and habitability of potentially habitable worlds.
- WFIRST-2.4 will detect large numbers of orphan planets, planets ejected from their solar system, even those with masses smaller than the Earth. These planets cannot be detected any other way and their abundance is an important test of planet formation theories.
- A microlensing exoplanet survey with WFIRST-2.4 is substantially superior to other WFIRST designs,

resulting in 25% more planet detections, significantly better sensitivity to small planets, and enabling detailed characterization of the majority of the detected planetary systems.

**With the spatial resolution of HST's powerful WFC3/IR camera and more than 200 times its field of view, WFIRST-2.4 will impact a broad range of astrophysics.** As part of this study, we engaged the broader astronomical community in identifying potential observational programs and have mapped the capabilities of the WFIRST-2.4 telescope into the 20 key science questions and 5 discovery areas identified by the decadal survey (see Table ES-1). We anticipate that the telescope will make important contributions towards at least 20 of these 25 areas. Key examples include:

- WFIRST-2.4 will make an accurate three-dimensional map of the distribution of dark matter across 2000 square degrees of sky. This map will reveal new insights into the relation between baryonic structures and dark matter. When combined with WFIRST-2.4 tracing galaxies across cosmic time through its high-resolution images, this dark matter map could be a Rosetta stone for our understanding of the process of galaxy formation and evolution.
- WFIRST-2.4 will be able to make "Degree-Deep-Fields" almost 100 times larger than the famed Hubble deep fields. These samples of millions of galaxies over billions of years will make fundamental contributions to our understanding of galaxy evolution. Especially because these WFIRST-2.4 observations will be made further into the near-IR, these studies of galaxy evolution will be more powerful with respect to the early universe than the Hubble Deep Fields. Moreover, there are many vital research programs where the idea is not just to collect a large sample but to image particular structures that extend over a degree or more of sky, areas that cannot be covered with HST or JWST
- WFIRST-2.4 will determine the positions and motions of more than 200 million stars in our Galaxy to unprecedented precision. With its large collecting area, it will be able to study stars up to 500 times fainter than the soon-to-be-launched European GAIA telescope, a dedicated astrometric mission. Its larger aperture will enable WFIRST-2.4 to achieve a given astrometric (positional) precision for a faint star nine times faster than a 1.3-meter unobscured version of WFIRST. These measurements should be able to trace the dark matter distribution not only in our Galaxy, but also in our poorly understood dwarf companions, an important laboratory for determining the mass and interaction properties of dark matter.
- WFIRST-2.4's wide field and superb imaging will make it the ideal telescope to find counterparts to the gravitational wave merger events that will likely be detected in large numbers by the upgraded LIGO experiment in Washington and Louisiana. More generally, WFIRST-2.4 will provide a newer, deeper view of the transient universe.
- WFIRST-2.4 will enable detailed studies of the properties of stars not only in our own Milky Way, but also over the full extent of all neighboring galaxies. These comparative studies will greatly inform our understanding of both stellar evolution and galaxy formation.

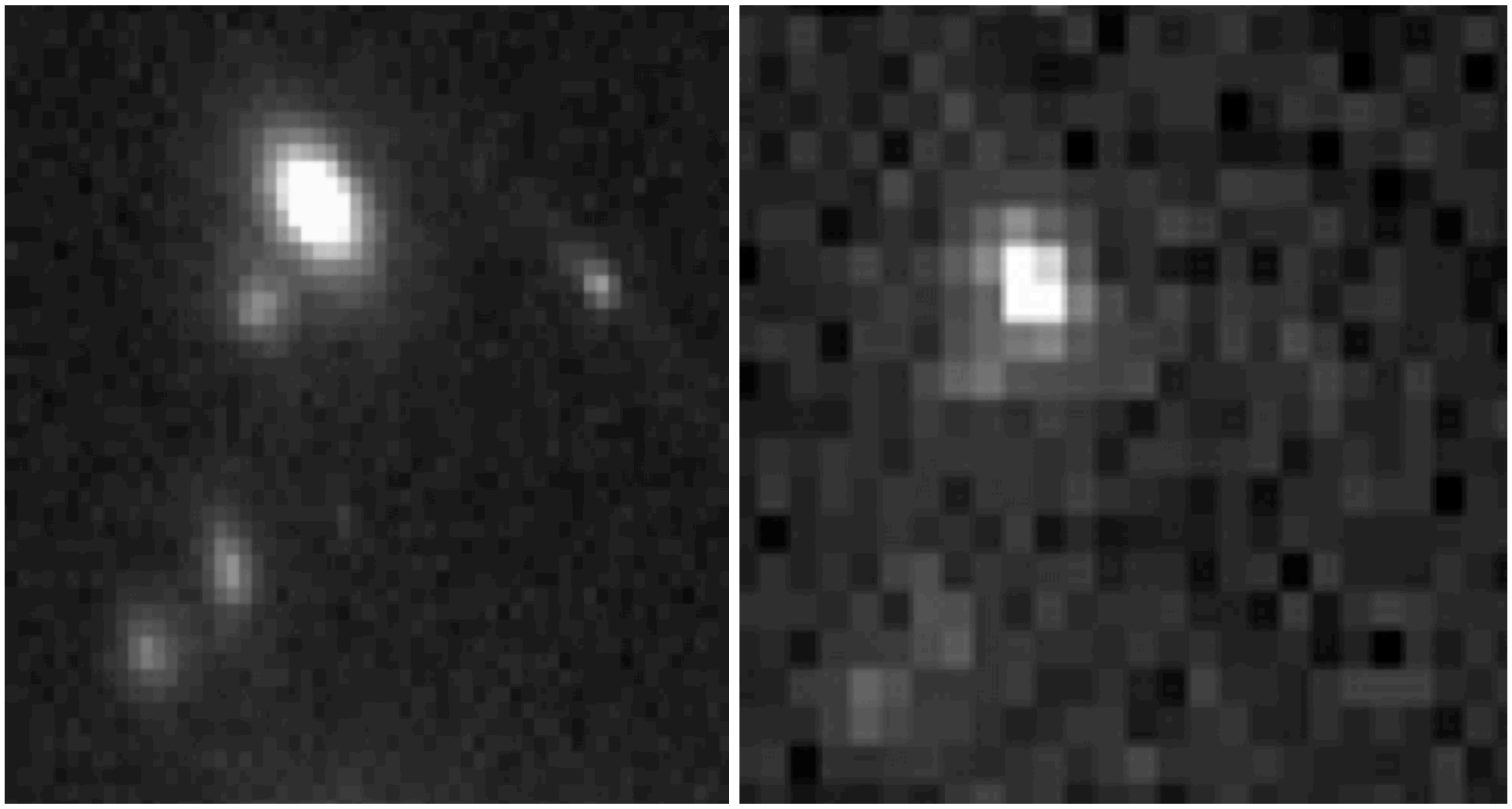

*Figure ES-1: This figure shows a simulated J band image of a few of the galaxies in a CLASH cluster observed with the WFIRST-2.4 high latitude survey (left) and with the Euclid wide field survey (right). At infrared wavelengths, the WFIRST-2.4 survey will be 15 times more sensitive and will have 8 times finer pixels by area. Euclid's images are best at optical wavelengths, where nearby galaxies emit most of their light. Light from more distant galaxies is redshifted into the infrared wavelengths where WFIRST observes.*

These examples are just a handful of the programs that the astronomy community identified as potential programs in a series of ~50 white papers collected in Appendix A. Historical experience with HST and other great observatories suggests that this list does not include many of the most important results that will be achieved by WFIRST-2.4, as many of the visionaries who will use the telescope for novel applications are still in high school!

**A coronagraph on WFIRST-2.4 would be an exciting extension in its capability that would not only characterize giant planets around the nearest stars, but also be an important step towards detecting habitable exoEarths:**

- The coronagraph would be able to detect new planets around many of the nearest stars. The combination of ground-based searches and Kepler observations suggest that there are roughly four planets per star. These planets are a mix of hydrogen-envelope gas giants, rocky planets, and icy Neptune and Saturn like planets. With the baseline contrast level, WFIRST-2.4 will be able to detect a substantial number of gas giants and icy planets. If an enhanced coronagraph proves possible, WFIRST-2.4 might even be capable of detecting rocky, Super-Earth-like, planets around a small number of nearby stars.
- WFIRST-2.4 would be able to directly image and take spectra of many of the nearest planets discovered by the indirect method of radial velocity searches. This will be an important next step in characterizing the properties of these diverse planets.
- WFIRST-2.4 would be able to image dust and debris disks, measuring large-scale structure and revealing signposts to planets.
- The development of a coronagraph for WFIRST-2.4 directly responds to the NWNH first medium scale priority to mature technology development for exoplanet imaging, providing an important stepping-stone towards building a future telescope capable of searching nearby planets for signs of life.
- The likely additional cost of including a coronagraph (and the associated observing time) on WFIRST-2.4 would be approximately the same cost as an Explorer mission, but taking advantage of the existing 2.4-meter telescope would give it enormous discovery power.

**If launched early in the JWST mission lifetime, WFIRST-2.4 would significantly enhance the science return from the James Webb Space Telescope (JWST).** By surveying thousands of square degrees of the sky, WFIRST-2.4 should be able to detect galaxies forming in the first few hundred million years after the big bang and could potentially detect supernova explosions occurring in the early universe. JWST's powerful spectroscopic capabilities will enable the detailed study of these rare objects. WFIRST-2.4 will discover myriads of strongly lensed quasars and galaxies that will be exciting objects for JWST follow-up observations. This powerful combination of WFIRST-2.4 wide field survey finding rare, but important, objects and JWST performing detailed follow-up study should also yield important results in galactic and extragalactic astronomy. Several of the white papers in Appendix B describe some of the science enabled by coincident operation of JWST and WFIRST-2.4.

**The existence of a high quality optical telescope assembly (OTA) reduces both the cost and risk associated with this mission.** While far more capable than a smaller version of WFIRST, our internal cost estimates suggest that the mission costs are very similar. The cost savings associated with using an existing OTA balances the additional cost of building a spacecraft for a large telescope. The OTA, one of the long-lead items of any astronomical mission, is complete and most of its associated risks are retired.

The key technology for the WFIRST-2.4 concept is the development of H4RG detectors for the camera, which is on the critical path for the mission. A modest investment of $5 M in fiscal year 2014 could significantly reduce the mission risk and produce a significant cost savings over the lifetime of the mission.

**Based on its initial study, the science definition team concludes that NASA could use the 2.4-meter telescope asset to build a more capable version of the WFIRST mission at comparable cost to the previously studied versions of the telescope and with lower technical and schedule risk.**

**Table ES-1: WFIRST-2.4 capabilities mapped into the 20 key science questions and 5 discovery areas identified in the New Worlds New Horizons decadal survey report.**

Italics denote improvements relative to DRM1
BOLD improvement denotes unique WFIRST-2.4 capability
Red denotes capability that requires joint JWST + WFIRST-2.4 operations

**DISCOVERY SCIENCE**

| | Key Observation | Improvement over DRM1 | Section |
|---|---|---|---|
| *Identification and characterization of nearby habitable exoplanets* | Characterize tens of Jupiter-like planets around nearby stars.<br><br>Potential to detect Earth-like planets around nearest stars | **Coronagraph** | 2.5.2<br>A-6, A-8 |
| *Gravitational wave astronomy* | Detect optical counterparts | *Ability to detect fainter sources* | A-52 |
| *Time-domain astronomy* | Repeated observations | *3x more sensitive, well matched to LSST* | A-48 |
| *Astrometry* | Measure star positions and motions | *Achieve same level of accuracy 9x faster* | 2.3.3<br>A-6, A-17, A-18<br>A-19, A-22, A-23<br>A-24, A-25, A-26 |
| *The epoch of reionization* | Detect early galaxies for follow-up by JWST, ALMA, and next generation ground-based telescopes | ***~10x increase in JWST targets*** | **2.3.1**<br>A-40, A-44, A-45<br>A-46, **B-4** |

**ORIGINS**

| | Key Observation | Improvement over DRM1 | Section |
|---|---|---|---|
| *What were the first objects to light up the universe, and when did they do it?* | Detect early galaxies and quasars for follow-up by JWST, ALMA, and next generation ground-based telescopes | ***~10x increase in high z JWST target galaxies***<br><br>***Very high-z supernova*** | **2.3.1**<br>A-43, A-45, A-46<br>**B-5** |
| *How do cosmic structures form and evolve?* | Trace evolution of galaxy properties | *1.9x sharper galaxy images* | A-31, A-32, A-39<br>A-47, **B-13** |
| *What are the connections between dark and luminous matter?* | High resolution 2000 sq. deg map of dark matter distribution and still higher resolution maps in selected fields<br><br>Dark Matter distribution in dwarfs to rich clusters | *Double the number density of lensed galaxies per unit area.*<br><br>*Capable of observing 200-300 lensed galaxies/arcmin$^2$*<br><br>*Astrometry of stars in nearby dwarfs* | A-25, A-26, A-33<br>A-35, A-36, A-37<br>A-38, A-50 |
| *What is the fossil record of galaxy assembly from the first stars to the present?* | Map the motions and properties of stars in the Milky Way + its neighbors<br><br>Find faint dwarfs | *3x increase in photometric sensitivity + 9x increase in astrometric speed*<br><br>*JWST follow-up* | A-21, A-22, A-25 A-26, A-27, A-28 A-29, A-30, **B-19** |
| *How do stars form?* | Survey stellar populations across wide range of luminosities, ages and environments | **IFU spectroscopy**<br><br>*3x more sensitive + 1.9x sharper galaxy images* | A-11, A-12, A-13 A-14, A-15, A-16 A-47, **B-8, B-11** |
| *How do circumstellar disks evolve and form planetary systems?* | Image debris disks | **Coronagraph** | 2.5.2 |
| *How did the universe begin?* | Measure the shape of the galaxy power spectrum at high precision; test for signatures of non-Gaussianity and stochastic bias | *Higher space density of galaxy tracers; higher space density of lensed galaxies* | 2.2 |

**UNDERSTANDING THE COSMIC ORDER**

| | Key Observation | Improvement over DRM1 | Section |
|---|---|---|---|
| *How do baryons cycle in and out of galaxies, and what do they do while they are there?* | Discover the most extreme star forming galaxies and quasars | | 2.3.4 |
| *What are the flows of matter and energy in the circumgalactic medium?* | | | |
| *What controls the mass-energy-chemical cycles within galaxies?* | Study effects of black holes on environment | **IFU Spectroscopy** | A-34 |
| *How do black holes grow, radiate, and influence their surroundings?* | Identify and characterize quasars and AGNs, black hole hosts<br><br>Use strong lensing to probe black hole disk structure | *Excellent match to LSST sensitivity*<br><br>*1.9x sharper images* | A-41, A-43, A-48 |
| *How do rotation and magnetic fields affect stars?* | | | |
| *How do the lives of massive stars end?* | Microlensing census of black holes in the Milky Way | | A-18 |
| *What are the progenitors of Type Ia supernovae and how do they explode?* | Study supernova Ia across cosmic time<br><br>Detect SN progenitors in nearby galaxies | **IFU Spectroscopy** | **B-7** |
| *How diverse are planetary systems?* | Detect 3000 cold exoplanets and complete the census of exoplanetary systems throughout the Galaxy. | *60% increase in the number of Earth size and smaller planets detected by microlensing, improved characterization of the planetary systems* | 2.5.1, 2.5.2.3 A-6, A-7, A-8 **B-15, B-17** |
| | Detects free-floating planets<br><br>**Joint lensing studies with JWST**<br><br>Images of exozodiacal disks around nearby stars | <br><br>**IFU**<br><br>**Coronagraph** | |
| *Do habitable worlds exist around other stars, and can we identify the telltale signs of life on an exoplanet?* | Develop precursor coronagraph for TPF<br><br>Characterize number of planets beyond snow line to understand origins of water | **Coronagraph**<br><br>*60% increase in the number of Earth size and smaller planets detected by microlensing* | **2.5.2**<br><br>2.5.1 |

**FRONTIERS OF KNOWLEDGE**

| | Key Observation | Improvement over DRM1 | Section |
|---|---|---|---|
| *Why is the universe accelerating?* | Use SN as standard candles<br><br>Use BAO to measure distance as a function of redshift<br><br>Use lensing to trace the evolution of dark matter<br><br>Use rich clusters to measure the growth rate of structure | *~2x improvement in SN distance measurements and significantly improved control of systematics*<br><br>*60% higher density of galaxies for the redshift survey*<br><br>*~2x increase in source density*<br><br>*Capable of observing 200-300 lensed galaxies/arcmin$^2$* | 2.2 |
| *What is dark matter?* | Characterize dark matter subhalos around the Milky Way<br><br>Characterize dark matter in clusters<br><br>Strong lenses | *~9x increase in astrometry speed*<br><br>*~1.9x sharper galaxy images*<br><br>***~JWST follow-up of strong lenses*** | A-22, A-24, A-25<br>2.3.2, A-38<br>**B-5** |
| *What are the properties of neutrinos?* | Measure neutrino effects on growth rate of structure and shape of galaxy power spectrum | *~2-3x increase in lensed galaxies per unit area*<br><br>*~2x increase in number density of spectroscopic galaxies* | |
| *What controls the mass, radius, and spin of compact stellar remnants?* | | | |

# 1 INTRODUCTION

This report contains the findings of a NASA-appointed Science Definition Team (SDT) to study the Astrophysics Focused Telescope Assets (AFTA) implementation of the Wide-Field Infra-Red Survey Telescope (WFIRST) mission. It follows the study of the previous SDT, chaired by J. Green and P. Schechter, of other implementations as described in their 2012 report.[1] In October 2012, the Director of the Astrophysics Division of NASA's Science Mission Directorate charged the WFIRST-AFTA SDT to work with the WFIRST project office to produce a design reference mission (DRM) for WFIRST, herein referred to as WFIRST-2.4, using an existing 2.4 meter telescope which is being made available to NASA. The existing telescope significantly reduces the development risk of the WFIRST-2.4 mission. This document fulfills that charge.

The NRC's 2010 decadal survey of astronomy and astrophysics, "New Worlds, New Horizons" (henceforth NWNH) gave WFIRST the highest priority for a large space mission. The NWNH science goals for WFIRST are quite broad. They include: tiered infrared sky surveys of unprecedented scope and depth; a census of exoplanets using microlensing; measurements of the history of cosmic acceleration using three distinct but interlocking methods (weak lensing, baryon acoustic oscillations, supernova standard candles); and a guest observer/investigator program. What brought these very different science goals together was the realization, across the astronomical community, that recent advances in infrared detector technology have, for the first time, made it possible to launch a wide field infrared telescope with a very large number of diffraction limited "effective pixels" in the focal plane. The telescope prescribed for WFIRST in the NWNH report had a primary mirror diameter of 1.5 m.

In addition to the science and payload envisioned for WFIRST by NWNH, NASA has charged the current SDT with considering an optional addition of a coronagraph instrument to the payload for direct imaging of exoplanets. The rationale for the addition is that the increase in the primary mirror size from 1.5 m to 2.4 m results in a finer angular resolution (smaller point spread function) that enables meaningful coronagraphy.

In October 2011 ESA selected Euclid as one of two Cosmic Vision medium class missions. Euclid is also a wide field telescope (1.2 m primary mirror), but the majority of its detectors (36 of 52) work at optical rather than infrared wavelengths. While its optical pixels properly sample diffraction-limited images, its infrared pixels are eight times coarser in area and do not take advantage of the angular resolution that is gained by going into space. The Euclid prime-mission science is more limited than WFIRST with emphasis on dark energy surveys using the weak lensing and baryon acoustic oscillations. In January 2012 the NRC's *Committee on the Assessment of a Plan for US Participation in Euclid* recommended a modest contribution to Euclid, saying "This investment should be made in the context of a strong U.S. commitment to move forward with the full implementation of WFIRST in order to fully realize the decadal science priorities of the NWNH report."

Also in October 2011, the Nobel Prize in Physics was awarded to Saul Perlmutter (a member of the SDT), Adam Riess and Brian Schmidt "for the discovery of the accelerating expansion of the Universe through observations of distant supernovae." The mechanism of that acceleration is unknown. WFIRST-2.4 will measure the expansion history of the universe and growth of structure to better than 1% in narrow redshift bins using several independent methods, constraints that will greatly narrow the range of possible mechanisms.

For exoplanets, we are in the midst of a revolution in detections and understanding. Five different techniques have resulted in nearly a thousand firm detections and an even larger number of candidates to be confirmed. The microlensing technique has detected free-floating planets with an indication that there are as many unbound planets as there are orbiting stars. When combined with the results from Kepler, WFIRST-2.4 will produce the first statistically complete census of exoplanets. With the optional coronagraph it would make the first large sample of direct images of exoplanets.

Consistent with its charge, the SDT herein presents the WFIRST-2.4 DRM. Similar to Design Reference Mission 1 (DRM1) of the previous SDT study, this 5-year mission includes a wide field instrument with a single focal plane array for imaging and spectroscopy. The spectroscopy is accomplished by rotating a grism into the field for slitless spectroscopy. Changes from DRM1 are as follows:

- On-axis telescope
- 2.4-m vs. 1.3-m primary mirror
- Slitless spectroscopy with a grism instead of prism
- Integral Field Spectrograph for supernova slit spectroscopy instead of a slitless prism

- Optional coronagraph instrument for direct imaging of exoplanets (along with an additional year of operations)
- Usage of HgCdTe IR detectors with 4k x 4k, 10μm pixel (H4RG) instead of the 2k x 2k, 18μm pixel H2RGs
- Higher operating temperature
- 28.5° inclined geosynchronous orbit

While most of these changes improve the telescope science performance, the higher operating temperature (due to the "as-is" use of the telescope) leads to the loss of sensitivity at 2.0-2.4 microns. For most science problems, this loss is outweighed by the gain in sensitivity shortwards of 2.0 microns. However, for Milky Way studies, this will reduce the ability to explore highly extincted star forming regions and to detect ultracool brown dwarf stars. The on-axis telescope design leads to a larger point spread function than an off-axis design. Thus, the improvement in PSF effective area is a factor of 1.9 rather than the ratio of the areas of the primaries.

The orbit baseline for the WFIRST-2.4 DRM is a geosynchronous orbit with a 28.5 degree inclination, as compared to Sun-Earth L2 for DRM1, which allows continuous downlink to the ground, enabling a much higher science data rate.

The previous SDT also had a DRM2 version that had a smaller primary mirror (1.1 m), fewer detectors and fewer years of design lifetime than DRM1. The performance relative to DRM1 was improved by baselining H4RG detectors with more pixels. The design rationale for DRM2 was to reduce overlap with other observatories and to reduce cost. A comparison of DRM1, DRM2 and WFIRST-2.4 is given in Table 1-1

The choice of the more-advanced H4RG detectors for WFIRST-2.4 is based on the rapid state of advancement in detector manufacturing. Ground instruments are being built with 4k x 4k detectors and we have judged that H4RGs will be the state-of-the-art at the time of the WFIRST-2.4 development. A vigorous program is underway to space qualify the devices in time for WFIRST.

The detector layout for WFIRST-2.4 is shown in Figure 1-1. The size of the detectors is shown as a footprint on the sky to illustrate the wide field of view. The field of view is a factor of ~200 larger than the instruments on HST (WFC3/IR) and JWST (NIRCam), enabling deep sky surveys of unprecedented size.

This report gives the science return from WFIRST-2.4 and a comparison to other observatories in Section 2 and a detailed description of the mission in Section 3.

To present definite forecasts for the scientific performance of WFIRST-2.4, we have adopted a notional observing program for a 5-year prime mission, which is summarized in Table 1-2 and detailed in subsequent sections. The key elements of this program are: a high-cadence (15 min) survey of 2.81 $deg^2$ in the Galactic bulge to discover exoplanets through microlensing (1.2 years in six 72-day campaigns), an imaging and spectroscopic survey of 2000 $deg^2$ at high Galactic and ecliptic latitude (1.9 years), a 3-tiered imaging and spectroscopic survey for supernovae (0.5 years of observing spread over 2 years with a 5-day cadence), and 1.4 years devoted to a competed Guest Observer program. The actual allocation of observing time will be decided closer to the launch of WFIRST-2.4, based on scientific and technical developments between now and then. If the optional coronagraph is adopted, we assume that the prime mission will be extended to 6 years, enabling a robust coronagraph science program (interleaved in time with other programs) without reducing the reach of other programs. At the end of its prime mission, WFIRST-2.4 will remain a facility of exceptional scientific power, and we include some discussions below of what could be accomplished in an extended mission.

| WFIRST Version | CATE Date | Primary Mirror Dia. (m) | Pixel Scale (as/pix) | Active FOV ($deg^2$) | Science Detectors | Notes |
|---|---|---|---|---|---|---|
| SDT #1: Interim DRM | 2011 | 1.3 | 0.18 | 0.29 | 36 H2RG-18 | 1 – 4x7 Imaging FPA |
| | | | 0.45 | 0.26/ea | | 2 – 2x2 Spec FPAs |
| SDT #1: DRM1 | N/A | 1.3 | 0.18 | 0.375 | 36 H2RG-18 | Imaging & Spec in single FPA with GRS and SN prisms in a filter wheel |
| SDT #1: DRM2 | 2012 | 1.1 | 0.18 | 0.585 | 14 H4RG-10 | Imaging & Spec in single FPA with GRS and SN prisms in a filter wheel |
| SDT #2: WFIRST-2.4 | 2013 | 2.4 | 0.11 | 0.281 | 18-H4RG-10 | Imaging & Spec in single FPA with GRS grism in a wheel |
| | | | 0.11 | 9.45 $as^2$ | 1 H2RG-18 | IFU for SN spectra |
| | | | | | | Optional coronagraph for exoplanet imaging |

**Table 1-1: Comparison to past WFIRST Design Reference Missions.**

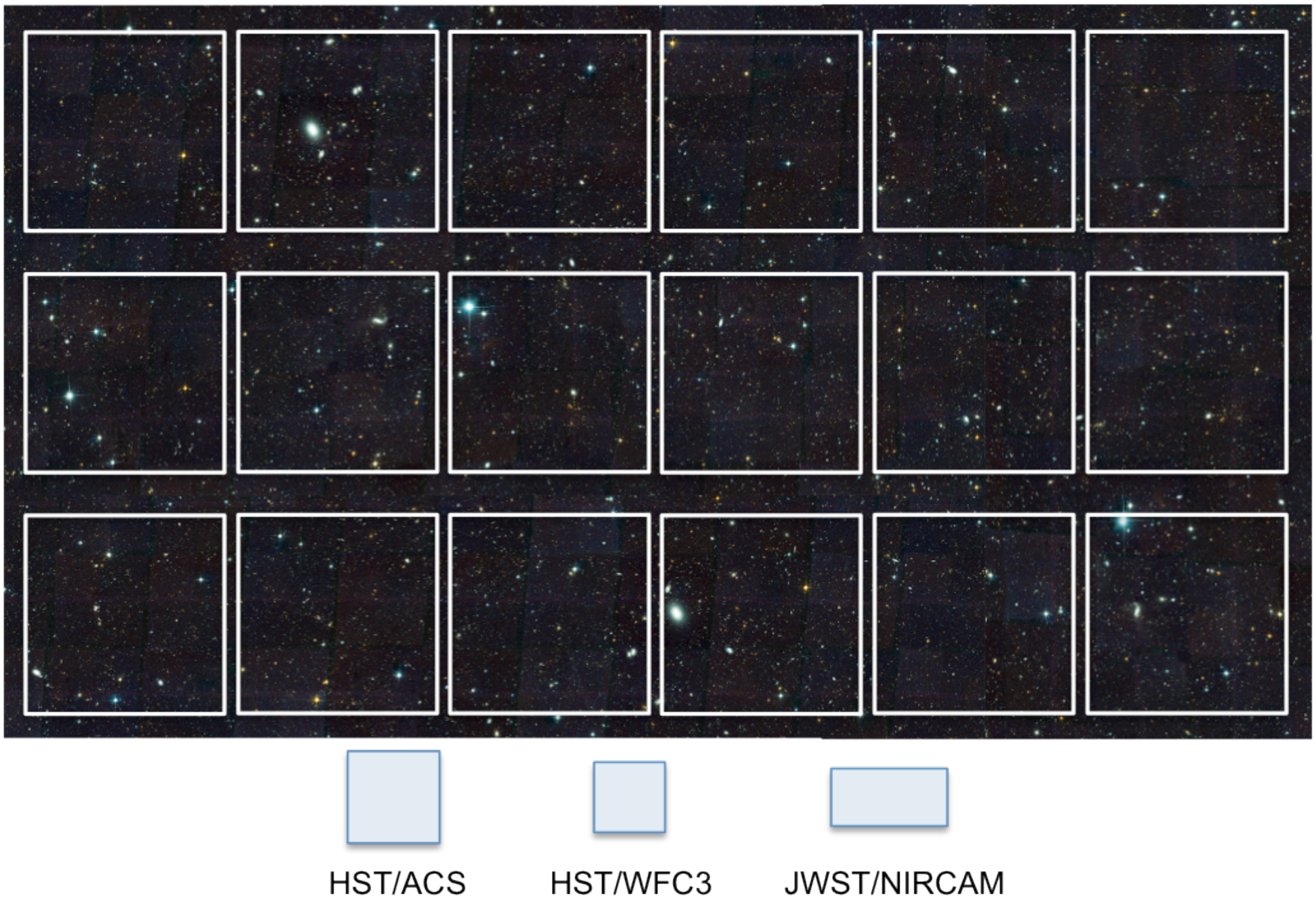

**Figure 1-1: Field of view comparison, to scale, of the WFIRST-2.4 wide field instrument with wide field instruments on the Hubble and James Webb Space Telescopes. Each square is a 4k x 4k HgCdTe sensor array. The field of view extent is about 0.79 x 0.43 degrees. The pixels are mapped to 0.11 arcseconds on the sky.**

**WFIRST-2.4 Design Reference Mission Capabilities**

| | | | | | | |
|---|---|---|---|---|---|---|
| Imaging Capability | 0.281 deg$^2$ | | 0.11 arcsec/pix | | 0.6 – 2.0 $\mu$m | |
| Filters | Z087 | Y106 | J129 | H158 | F184 | W149 |
| Wavelength ($\mu$m) | 0.760-0.977 | 0.927-1.192 | 1.131-1.454 | 1.380-1.774 | 1.683-2.000 | 0.927-2.000 |
| PSF EE50 (arcsec) | 0.11 | 0.12 | 0.12 | 0.14 | 0.14 | 0.13 |
| Spectroscopic Capability | Grism (0.281 deg$^2$) | | | IFU (3.00 x 3.15 arcsec) | | |
| | 1.35 – 1.95 $\mu$m, R = 550-800 | | | 0.6 – 2.0 $\mu$m, R = ~100 | | |

**Baseline Survey Characteristics**

| Survey | Bandpass | Area (deg$^2$) | Depth | Duration | Cadence |
|---|---|---|---|---|---|
| Exoplanet Microlensing | Z, W | 2.81 | n/a | 6 x 72 days | W: 15 min<br>Z: 12 hrs |
| HLS Imaging | Y, J, H, F184 | 2000 | Y = 26.7, J = 26.9<br>H = 26.7, F184 = 26.2 | 1.3 years | n/a |
| HLS Spectroscopy | 1.35 – 1.95 $\mu$m | 2000 | 0.5x10$^{-16}$ erg/s/cm$^2$ @ 1.65 $\mu$m | 0.6 years | n/a |
| SN Survey | | | | 0.5 years (in a 2-yr interval) | 5 days |
| Wide | Y, J | 27.44 | Y = 27.1, J = 27.5 | | |
| Medium | J, H | 8.96 | J = 27.6, H = 28.1 | | |
| Deep | J, H | 5.04 | J = 29.3, H = 29.4 | | |
| IFU Spec | 7 exposures with S/N=3/pix, 1 near peak with S/N=10/pix, 1 post-SN reference with S/N=6/pix<br>Parallel imaging during deep tier IFU spectroscopy: Z, Y, J, H ~29.5, F184 ~29.0 | | | | |

**Guest Observer Capabilities**

| 1.4 years of the 5 year prime mission | | | | | | |
|---|---|---|---|---|---|---|
| | Z087 | Y106 | J129 | H158 | F184 | W149 |
| Imaging depth in 1000 seconds ($m_{AB}$) | 27.15 | 27.13 | 27.14 | 27.12 | 26.15 | 27.67 |
| $t_{exp}$ for $\sigma_{read} = \sigma_{sky}$ (secs) | 200 | 190 | 180 | 180 | 240 | 90 |
| Grism depth in 1000 sec | S/N=10 per R=~600 element at AB=20.4 (1.45 $\mu$m) or 20.5 (1.75 $\mu$m)<br>$t_{exp}$ for $\sigma_{read} = \sigma_{sky}$: 170 secs | | | | | |
| IFU depth in 1000 sec | S/N=10 per R~100 element at AB=24.2 (1.5 $\mu$m) | | | | | |
| Slew and settle time | chip gap step: 13 sec, full field step: 61 sec, 10 deg step: 178 sec | | | | | |

**Optional Coronagraph Capabilities**

| 1 year in addition to the 5-year primary mission, interspersed, for a 6-year total mission | |
|---|---|
| Field of view | Annular region around star, with 0.2 to 2.0 arcsec inner and outer radii |
| Sensitivity | Able to detect gas-giant planets and bright debris disks at the 1 ppb brightness level |
| Wavelength range | 400 to 1000 nm |
| Image mode | Images of full annular region with sequential 10% bandpass filters |
| Spectroscopy mode | Spectra of full annular region with spectral resolution of 70 |
| Polarization mode | Imaging in 10% filters with full Stokes polarization |
| Stretch goals | 0.1 arcsec inner annulus radius, and super-Earth planets |

**Table 1-2: WFIRST-2.4 design reference mission observing program. The quoted magnitude/flux limits are for point sources, 5$\sigma$ for imaging, 7$\sigma$ for HLS spectroscopy.**

# 2 THE WFIRST-2.4 SCIENCE PROGRAM

## 2.1 Science Overview

The 2010 Decadal Survey, *New Worlds, New Horizons*, sought to advance two of the highest priority astrophysics programs — the quest to understand the acceleration of the universe, and the search for other worlds — in a way that would also substantially benefit the research of the larger community. Without a budget commensurate with such ambitious goals, the Survey found a novel solution that joined these science programs on the framework of a modest-sized telescope with two unique attributes: (1) the high sensitivity of near-IR observations achievable from space, (2) wide-field of view. Together, these would provide science capability orders-of-magnitude beyond previous missions. In addition to the two core programs, there would be large surveys and a competitive guest observer program that would benefit a wide segment of the astronomical community.

By choosing a 1.5-m telescope, the minimum possible aperture to achieve mission goals, WFIRST — the Wide-Field Infrared Space Telescope, promised all this within a budget that could be provided during the 2011-2020 decade, although resources proved to be even more constrained than expected. A consequence of the small aperture, however, was that that no part of the program --- the dark energy (extragalactic survey) program, the microlensing planet search, or the guest observer program, could be supported at the desired level, as described by the Science Prioritization Panels of the Survey and the proposals made by the community. Although the Science Definition Team's DRM1 design confirmed that an unobstructed, 1.3-m aperture WFIRST would be among the highest 'science-per-dollar' missions ever built, the science had already been "descoped" relative to community ambitions.

The unexpected availability of the 2.4-m telescope has the potential to restore the robustness and balance of these science programs, and more. As we show here in this Report, by dividing the time on WFIRST-2.4 between dark energy, microlensing, and guest observer observations, gains of 50-100% or more are expected for all. Furthermore, the 2.4-m makes possible new science at the limit of, and beyond, what was achievable with earlier WFIRST designs. This 2.4-m version of WFIRST combines the resolution of the Hubble Space Telescope with the wide-field near-IR field-of-view that will revolutionize several different frontier areas of astrophysics.

The 2.4-m aperture of WFIRST-2.4 collects almost 3 times as much light as the unobstructed 1.3-m SDT-DRM1 and offers a factor of 1.9 improvement in spatial resolution (point spread function area at H band), which itself provides another factor-of-two improvement in accomplishing many science programs. Compared to SDT-DRM1, the design presented in our report has a ~25% smaller field-of-view (FOV) that moderates these gains, but the higher spatial resolution takes the WFIRST-2.4 program to the "next level" and sets it apart from DRM1 and — for the dark energy program specifically — the Euclid mission. In particular, having the greater speed of the larger aperture and the sharper PSF allows WFIRST-2.4 to reach a factor-of-two deeper per unit time over an unprecedentedly large field for a large space telescope, ~90 times bigger than the HST–ACS FOV, and ~200 times bigger than the IR channel of WFC3 that has been a tremendously successful scientific tool.

The following sections explain how this capability will be exploited for the core programs of dark energy and microlensing planet finding. The surveys, while remarkably deep by today's standards, are shallow compared to WFIRST-2.4's full capability. This is done in order to cover large areas of the sky for cosmological studies, and for repeating thousands of times a small number of fields to identify distant supernovae as a dark energy tool, and to search for microlensing events due to stars and their planets that intercept the lines of sight to hundreds of thousands of stars in the Galactic bulge. In this introduction to the science sections we provide some examples of making much deeper exposures with the WFIRST-2.4 wide-field camera, to accomplish high-priority science objectives not possible with any other mission, and to provide EPO material bound to amaze and inspire.

Most of these types of programs will be pursued in the Guest Observer program, the part of WFIRST-2.4 that will offer broad support to many fields of astrophysics in the tradition of HST, no doubt with the same astonishing results of new, creative, field-changing science. In an extended mission, the Guest Observer program would likely become the dominant part of the WFIRST mission. HST has demonstrated clearly that the combination of a powerful facility and peer-reviewed proposals has the greatest impact in advancing the extraordinarily broad field of astrophysics research.

Because of the giant FOV and high sensitivity, even the "deep field" science — accomplished by Hubble as an observatory project — will be within reach of Guest Observer proposals. The pioneering Hubble

Deep Fields revolutionized our view of the young universe. WFIRST-2.4 will produce "Degree-Deep-Fields" almost 100 times larger (See Figure 1-1 and Box 1 in the Executive Summary). Their samples of millions of galaxies at redshifts 1<z<3 will make fundamental contributions to understanding galaxy evolution over this critical epoch of galaxy growth, and provide mural-sized views of the deep universe that will be gems for public engagement.

The deep fields will be invaluable because of their huge samples of distant galaxies, like the dark energy surveys described below, but much deeper. Especially because these WFIRST-2.4 observations will be made further into the near-IR, these studies of galaxy evolution will be more powerful with respect to the early universe than the Hubble Deep Fields. Moreover, there are many vital research programs where the idea is not just to collect a large sample — in these cases the `subject' extends over a degree or more over the sky, areas that cannot be covered with HST or JWST.

For example, the wide and deep capability of WFIRST-2.4 will have remarkable application to studies of the nearby universe. Deep pictures of the Andromeda galaxy taken with ground-based telescopes have revealed a faint network of structure surrounding our nearest-neighbor that records how it was assembled. What ground-based resolution can do for Andromeda, WFIRST-2.4 will do for hundreds of our Galaxy's other, more distant neighbors (Box 2). Color-magnitude diagrams using far-red and infrared bands can constrain ages and metallicities for millions of stars and trace assembly histories for the main body of each galaxy as well as its faint stellar halo. With such observations, a context for the development of our own Galaxy will be within reach.

Regions of star formation, for example, the Taurus association, stretch across many degrees of the sky. The near-IR sensitivity of WFIRST-2.4 will make possible a full census of stars in these systems to a depth far beyond what has been possible with ground-based and space-based telescopes (Box 3). Panchromatic, wide-field imaging will provide our clearest picture to date of the star formation process and the shape of the initial mass function, and the dependence of these on the global environment. This will include brown dwarfs and even planets down to a few Jupiter masses that have been ejected from their parent systems — such information will be crucially important in understanding the evolution of our own solar system. High precision color-magnitude diagrams for Galactic globular clusters will exploit sharp features in the infrared color-magnitude relation of stars to determine much more accurate ages, metallicities, reddenings, and distances than have been possible with visible data alone. Selected swaths of the Milky Way's disk will map star formation in our Galaxy over the last 6 billion years of cosmic history, a time when the `cosmic' rate of star formation has been rapidly falling.

**Box 2**

*Our closest neighbor galaxy is Andromeda, a giant spiral galaxy much like our own Galaxy. Studies comparing the two have shown many similarities, but also some notable differences, in when generations of stars were born and how they enriched the abundances of chemical elements such as carbon, nitrogen, oxygen, calcium, iron — elements that are essential for building both planets and life. As this picture taken with a ground-based telescope shows, the Andromeda galaxy has a halo of faint stars with streams and arcs of higher density — these are the remnants of small galaxies that have been captured by Andromeda over the last few billion years. Studying these features in detail will reveal how the Andromeda galaxy was assembled.*

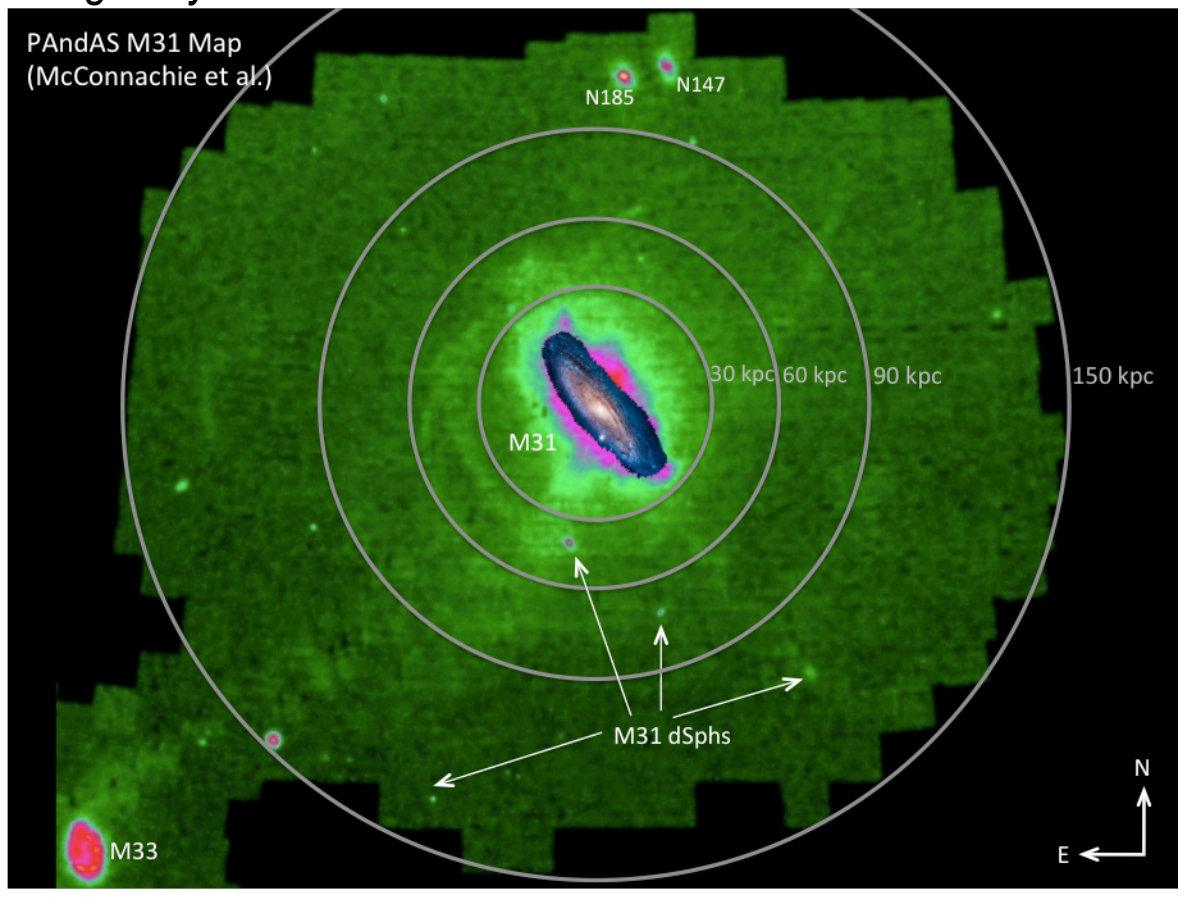

*What ground-based telescopes have done for Andromeda, the wide field of WFIRST-2.4 can do for as many as a hundred nearby galaxies, producing spectacular images of other galaxies with detail now only available for the nearest few. The contribution to our understanding of the formation and evolution of galaxies will be enormous.*

The potential for breakthroughs in the history of stellar populations — for our own Galaxy and those nearby — is also promising. The combination of wide field, sharp PSF, and superior sensitivity is precisely the combination needed to make substantial progress in this foundational research that impacts much of astrophysics. We will show that WFIRST-2.4 is an enabling facility in this area, something important that was beyond the original conception of the WFIRST mission.

Looking further out into deep space, and further back in time, the principal epoch of galaxy growth will

**Box 3**

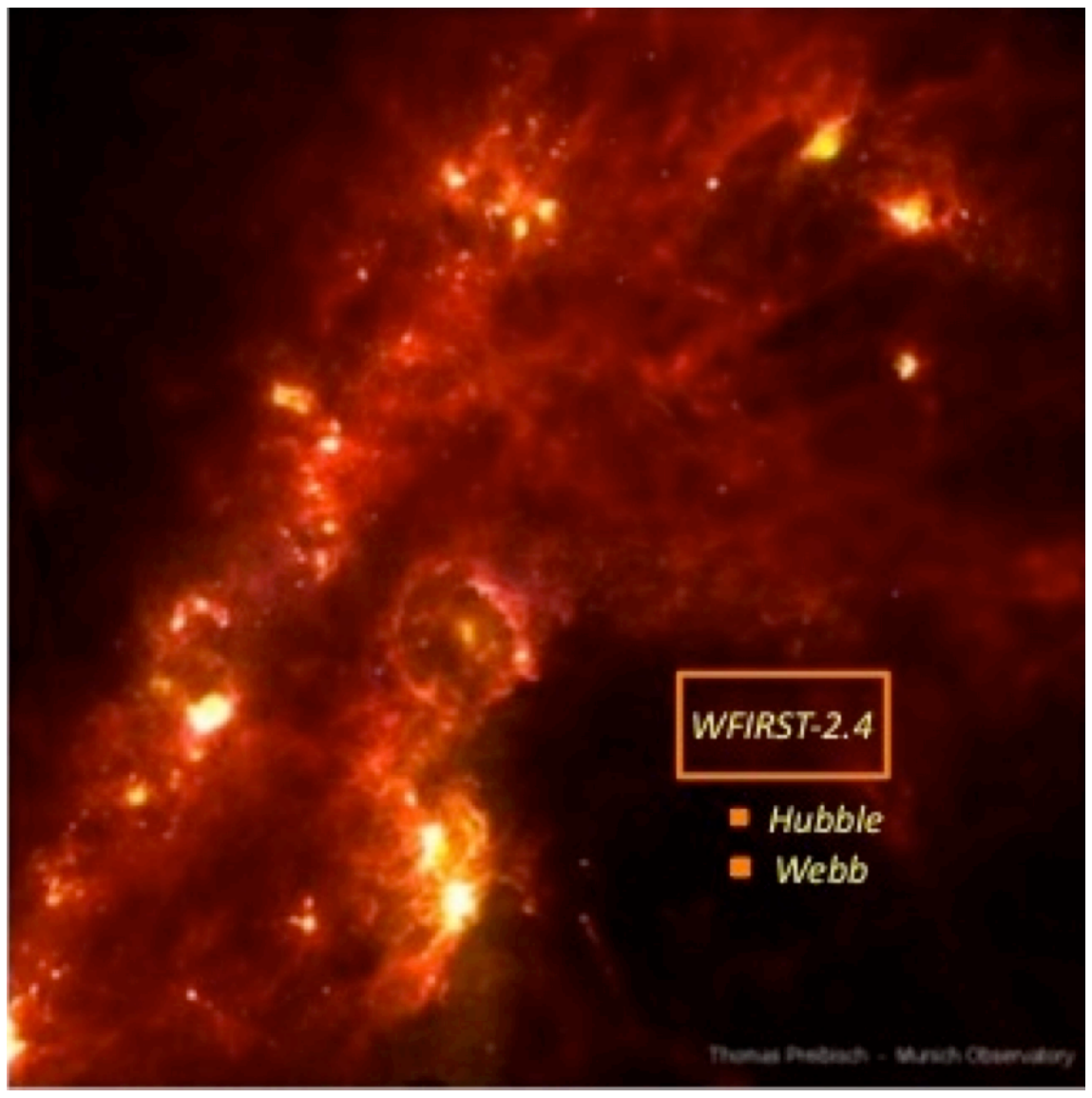

*The Taurus Star Forming Cloud (above) is the birthplace of thousands of new stars. The much bigger field of WFIRST-2.4 compared to HST or JWST means that its deep, panoramic view will encompass all these newcomers, from the youngest, smallest stars, buried in dust, to free-floating Jupiter-sized planets, flung out of their 'solar systems'. Finding these ejected planets is key to understanding the evolution of our own Solar System. Deep imaging of the richest star clusters in the Milky Way, like the 'globular cluster' shown below (left), and the rich star fields of our Galaxy (right) will be a treasure trove for understanding the history of our home galaxy back to ancient times.*

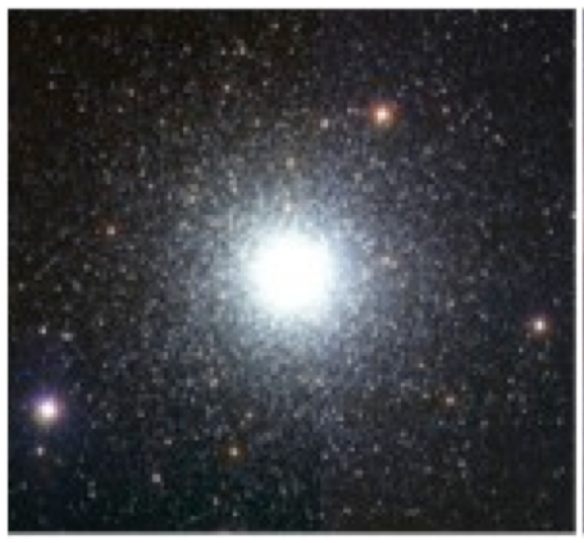

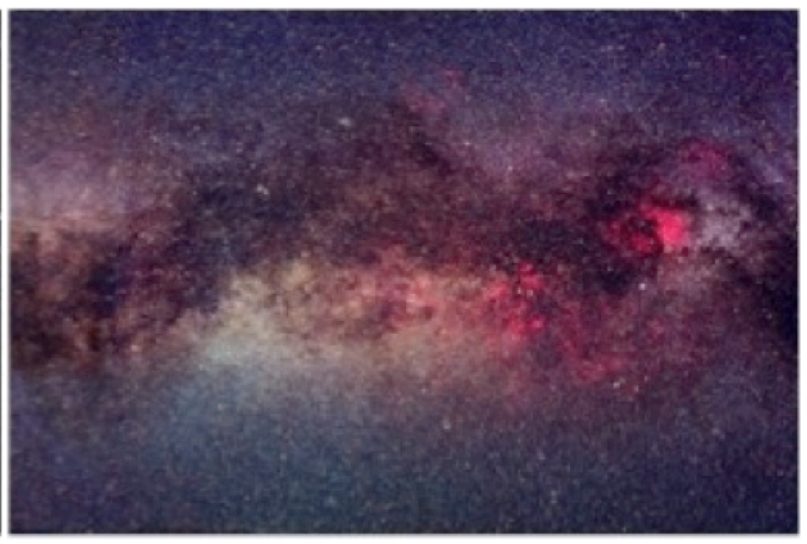

be well probed by the wide-and-deep imaging capability of WFIRST-2.4, complementing visible-light ground-based surveys made with the Large Synoptic Survey Telescope (LSST), and the small-field, but extraordinarily sensitive spectroscopic capability of JWST. A particular scientific focus of high priority is the linking of galaxy evolution with large-scale structure (Box 4). For example, elliptical galaxies, with their prevalently old populations, prefer the denser regions of the universe. Ground-based telescopes can cover areas large enough to map large-scale structure on the required scale of tens of megaparsecs (~1 degree at z=1 — a few WFIRST-2.4 fields), but such images lack the resolution for accurate morphological classification, and the depth to probe to small galaxies. HST or JWST imaging over such a scale is, again, impractical. WFIRST-2.4 can do it all — cover thousands of square degrees to map the large-scale structure, and obtain deep, high precision photometry to identify and characterize the full mass-range of these young versions of older galaxies, investigating their structure and star-formation history in the first half of cosmic time.

Even observing the birth of large-scale-structure during "cosmic dawn" is possible with WFIRST-2.4, using deep photometry to select Lyman-break galaxies from $6 < z < 10$, tracing the rapid growth of denser environments and connecting this to the emergence of Lyman-alpha-emitters and Lyman-break systems in the birth phase of today's galaxies.

These are examples of the exciting science that WFIRST-2.4 will make possible, for the first time, but they are truly the tip of the iceberg. As optimistic as the creators of HST were, they never imagined how prolific it would be and how much it would change the field of astrophysics. In the following sections we will describe in detail some areas of science where the impact of WFIRST-2.4 can be accurately quantified now. This includes the dark energy program with its three-pronged campaign of baryon acoustic oscillations, weak lensing of galaxies, and Type Ia supernovae to measure the evolution of space-time and the possible departures from gravitation as described by the theory of general relativity. This program will produce a vast extragalactic survey — a map for JWST to follow — that will inform a wide range of galaxy studies, including quasars, star formation histories, and rich clusters as telescopes for observing the earliest galaxies.

We describe the gains to be had with WFIRST-2.4 for the second core program, the microlensing search for thousands of planets down to below Mars mass, in orbit around F, G, K, and M stars and "beyond the snow line" — the essential and ideal complement to the Kepler mission. WFIRST-2.4 offers substantial advantages over previous WFIRST designs in determining absolute, not just relative, masses of these planets. We also in-

**Box 4**

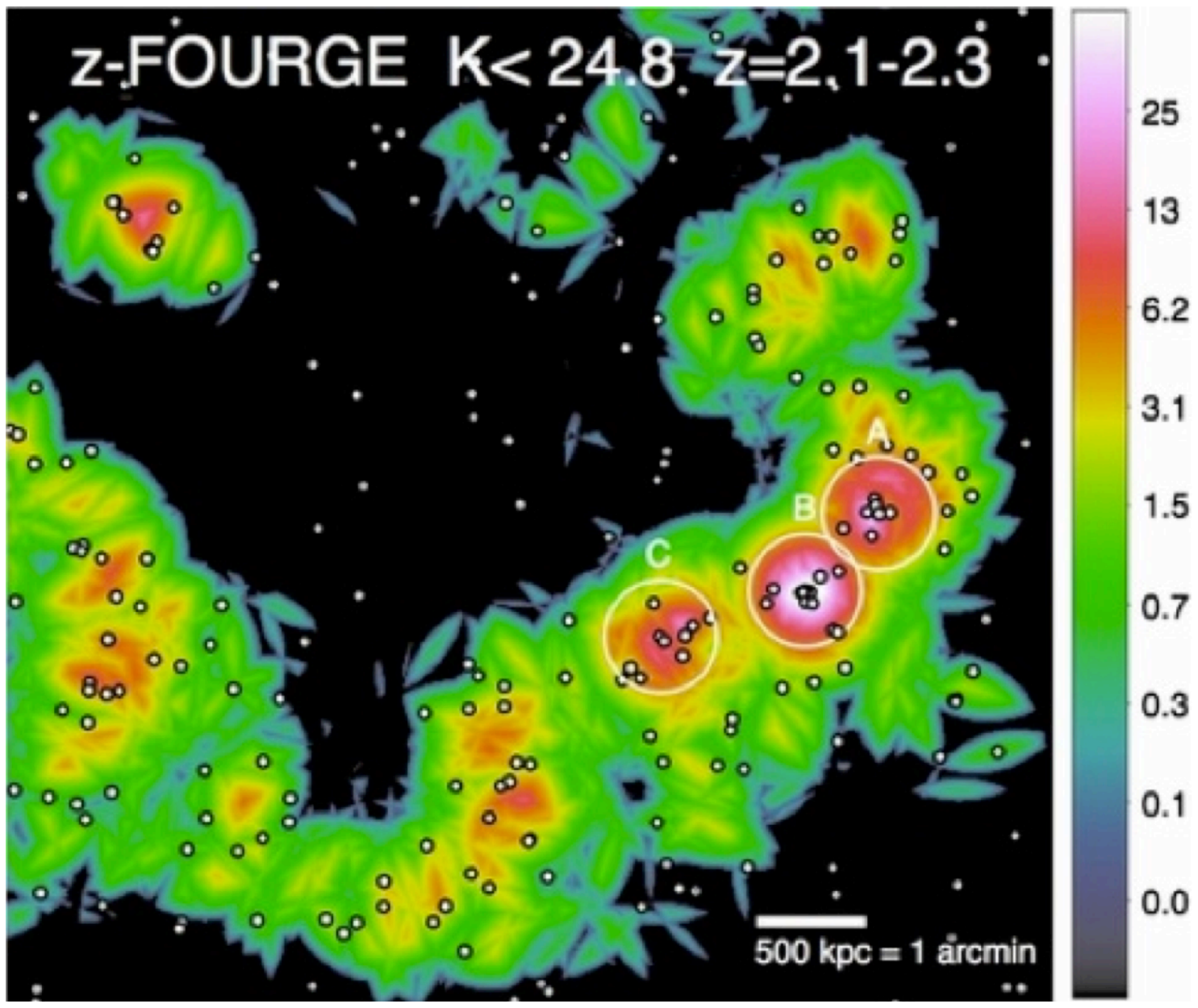

*The 'teen years' for galaxies began about 3 billion years after the Big Bang and lasted about 3 billion years more. WFIRST-2.4 can tie the evolution of galaxies to the growth of 'large-scale structure', as the universe evolves from primal smoothness to a tapestry ranging from vast voids with few galaxies to clusters packed with galaxies. The picture above[2] shows a dense region coalescing when the universe was 3 billion years old, the time the most massive galaxies (called 'ellipticals') went through their growth spurt. A single WFIRST-2.4 field would cover an area 16 times larger, providing a sample of thousands of early galaxies over a wide range of environments. The bottom picture[3] shows a theoretical model of even earlier large-scale structure, when the universe was only 1 billion years old — about the time the seeds for such galaxies were sown. WFIRST-2.4 can trace the growth of elliptical galaxies, and the slower-developing 'spiral' galaxies, in this changing environment. This picture covers an area of a single WFIRST-2.4 field.*

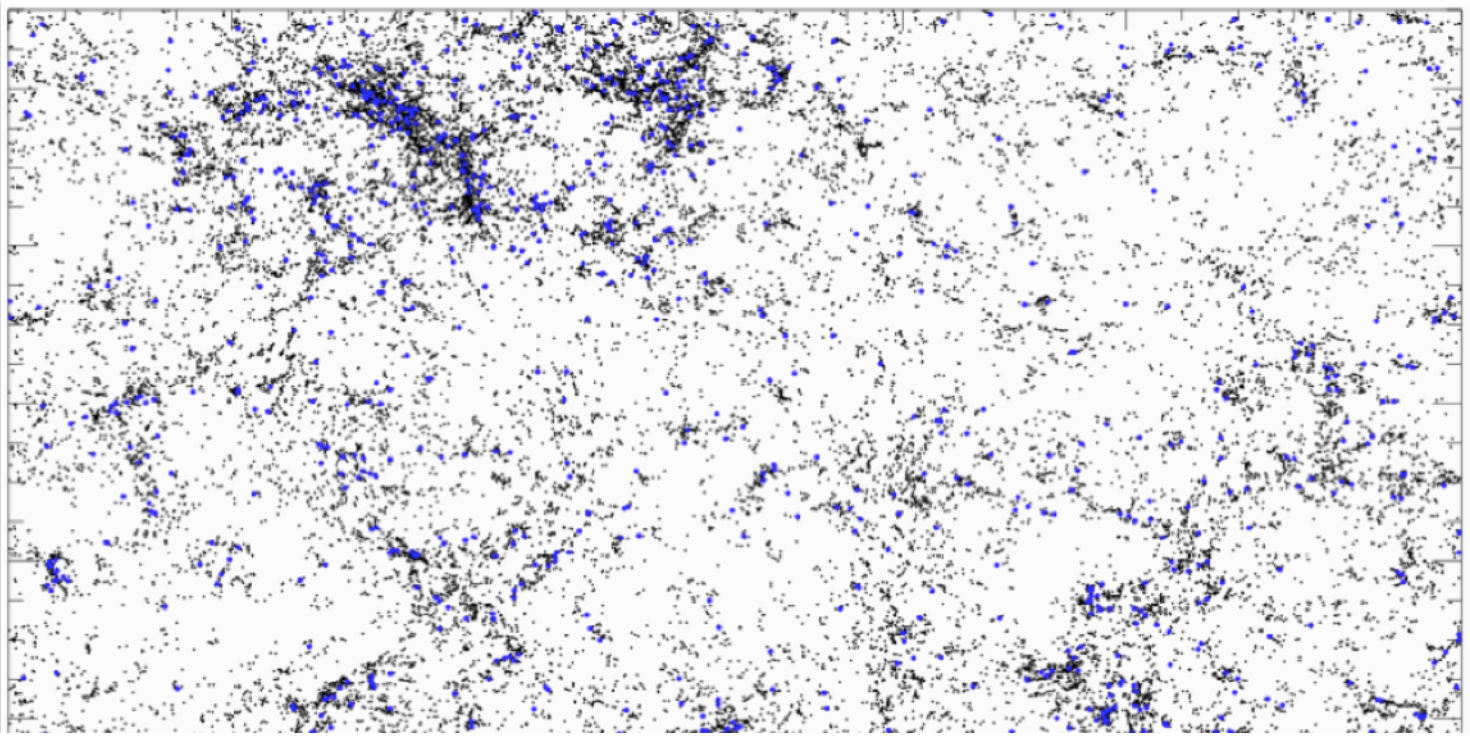

vestigate the exciting possibility of pushing ahead one of the Decadal Survey's highest priorities: not just finding, but *characterizing* planets, something that can be pursued by adding an extremely high-contrast coronagraph to WFIRST-2.4. A game changing feature of this large aperture telescope is that it can provide the first platform to test this new and rapidly improving technology and take the first scientific steps in describing the properties of extrasolar planets, something that the last two Decadals have identified as one of, if not *the*, most important fields in astrophysics in 2020 and beyond. Directly imaging planets around nearby stars is a step not possible with other versions of WFIRST, but it is one that resonates with the spirit of the mission as described in *NWNH.*

The relation and synergy with other major facilities in the 2020 and beyond time frame are explored. This and the science programs motivate the final section on the science requirements we have chosen for this first design reference mission of WFIRST-2.4. These science requirements are the foundation for the Observatory plan we present in Section 3.

### 2.2 Dark Energy & Fundamental Cosmology

The accelerating expansion of the universe is the most surprising cosmological discovery in many decades, with profound consequences for our understanding of fundamental physics and the mechanisms that govern the evolution of the cosmos. The two top-level questions of the field are:

1. Is cosmic acceleration caused by a new energy component or by the breakdown of General Relativity (GR) on cosmological scales?
2. If the cause is a new energy component, is its energy density constant in space and time, or has it evolved over the history of the universe?

A constant energy component, a.k.a. a "cosmological constant," could arise from the gravitational effects of the quantum vacuum. An evolving energy component would imply a new type of dynamical field. Gravitational explanations could come from changing the action in Einstein's GR equation, or from still more radical modifications such as extra spatial dimensions. Observationally, the route to addressing these questions is to measure the histories of cosmic expansion and growth of structure with the greatest achievable precision over the widest accessible redshift range.

As defined by NWNH, one of WFIRST's primary mission goals is to "settle fundamental questions about the nature of dark energy, the discovery of which was one of the greatest achievements of U.S. telescopes in recent years." (Following common practice, we will use "dark energy" as a generic term that is intended to encompass modified gravity explanations of cosmic acceleration as well as new energy components.) It will do so using three distinct surveys that enable complementary measurements of the expansion history and structure growth. In each case, the larger collecting area and higher angular resolution of WFIRST-2.4 afford significant advantages relative to the DRM1 design. The WFIRST-2.4 dark energy program is summarized below and described at greater length in Appendix C. Further background on the measurement and forecast methods can be found in the Green et al. report[1], in papers by Wang et al.[4,5,6] on cosmological constraints from galaxy redshift surveys, and in the comprehensive review article of Weinberg et al.[7]

Figure 2-1 presents an overview of the WFIRST-2.4 dark energy program. With the observing strategy

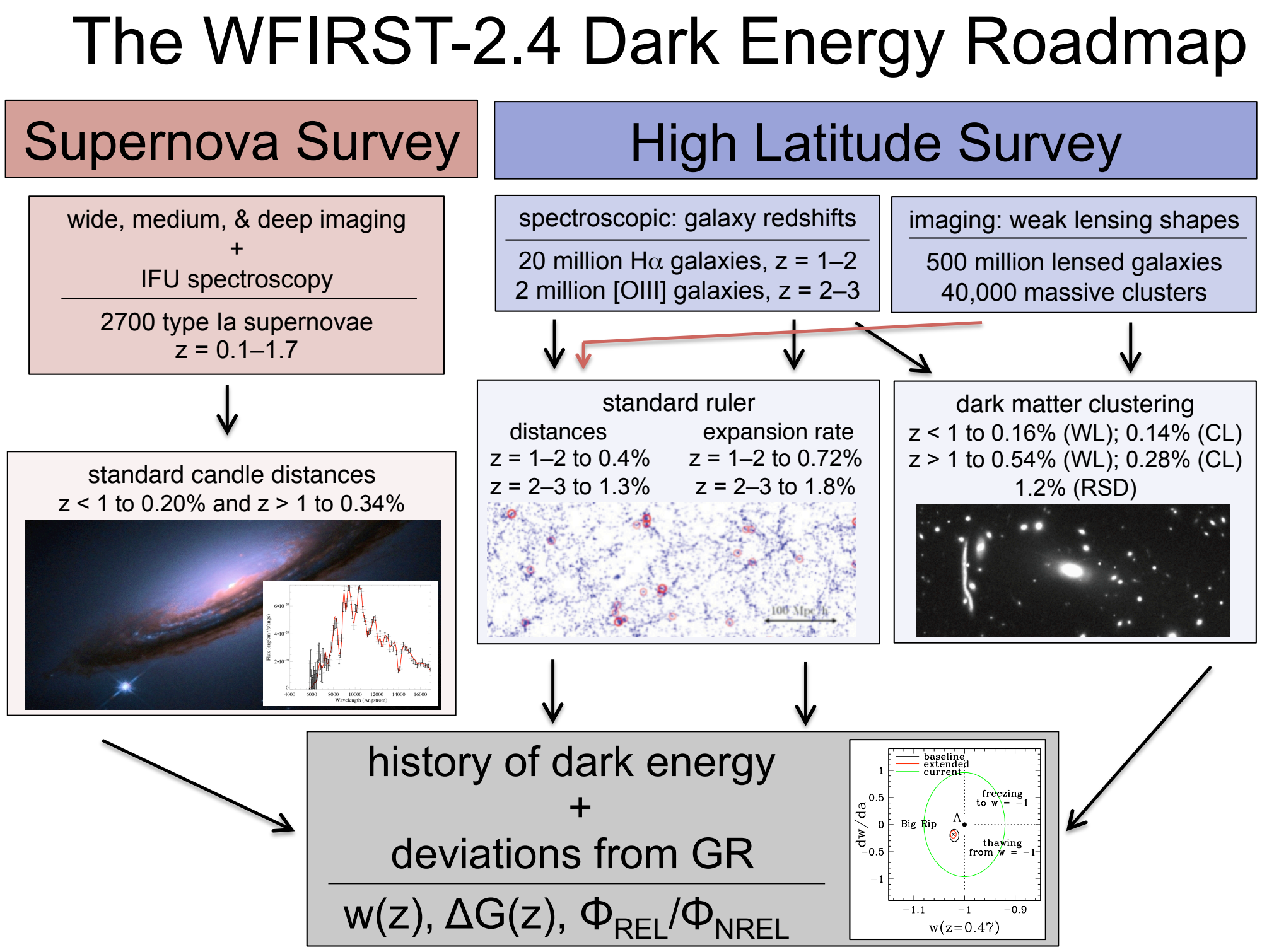

**Figure 2-1: A high-level view of the WFIRST-2.4 dark energy program. The supernova (SN) survey will measure the cosmic expansion history through precise spectrophotometric measurements of more than 2700 supernovae out to redshift z = 1.7. The high-latitude survey (HLS) will measure redshifts of more than 20 million emission-line galaxies and shapes (in multiple filters) of more than 500 million galaxies. The former allow measurements of "standard ruler" distances through characteristic scales imprinted in the galaxy clustering pattern, while the latter allow measurements of matter clustering through the "cosmic shear" produced by weak gravitational lensing and through the abundance of galaxy clusters with masses calibrated by weak lensing. As indicated by crossing arrows, weak lensing measurements also constrain distances, while the galaxy redshift survey provides an alternative measure of structure growth through the distortion of redshift-space clustering induced by galaxy motions. Boxes in the middle layer list the forecast aggregate precision of these measurements in different ranges of redshift. These high-precision measurements of multiple cosmological observables spanning most of the history of the universe lead to stringent tests of theories for the origin of cosmic acceleration, through constraints on the dark energy equation-of-state parameter w(z), on deviations ΔG(z) from the growth of structure predicted by General Relativity, or on deviations between the gravitational potentials that govern relativistic particles (and thus weak lensing) and non-relativistic tracers (and thus galaxy motions).**

and error model described in the following subsections and Appendix C, the supernova survey will measure "standard candle" distances with an aggregate precision of 0.20% at z < 1 (error-weighted z = 0.50) and 0.34% at z > 1 (z = 1.32). Galaxy clustering provides "standard rulers" for distance measurement in the form of the baryon acoustic oscillation (BAO) feature and the turnover scale of the galaxy power spectrum. The galaxy redshift survey (GRS) enables measurements of the angular diameter distance $D_A(z)$ and the expansion rate H(z) using H$\alpha$ emission line galaxies at 1 < z < 2 and [OIII] emission line galaxies at 2 < z < 3. The aggregate precision of these measurements ranges from 0.40% to 1.8% (see Figure 2-1). The imaging survey will enable measurements of dark matter clustering via cosmic shear and via the abundance of galaxy clusters with mean mass profiles calibrated by weak lensing; we expect 40,000 $M \geq 10^{14}M_{sun}$ clusters in the 2000 $deg^2$ area of the high-latitude survey. These data constrain the amplitude of matter fluctuations at redshifts 0 < z < 2, and they provide additional leverage on the distance-redshift relation. Treating the fluctuation amplitude $\sigma_m(z)$ as a single-parameter change, we forecast aggregate precision of 0.14% (z < 1) and 0.28% (z>1) from clusters and 0.16% (z<1) and 0.54% (z>1) from cosmic shear. In the GRS, the distortion of structure in redshift space induced by galaxy peculiar velocities provides an entirely independent approach to measuring the growth of structure, with forecast aggregate precision of 1.2% at z = 1 – 2 and 3.2% at z = 2-3.

These high-precision measurements over a wide range of redshifts in turn lead to powerful constraints on theories of cosmic acceleration. If the cause of acceleration is a new energy component, then the key physical characteristic is the history w(z) of the equation-of-state parameter $w = P/\varepsilon$, the ratio of pressure to energy density. A cosmological constant has w = -1 at all times, while dynamical dark energy models have $w \neq -1$ and an evolutionary history that depends on the underlying physics of the dark energy field. If the cause of acceleration is a breakdown of GR on cosmological scales, then it may be detected in a deviation between the measured growth history G(z) and the growth predicted by GR given the measured expansion history. Alternatively, some modified gravity theories predict a mismatch between the gravitational potential inferred from weak lensing (in Figure 2-1, cosmic shear and clusters) and the gravitational potential that affects motions of non-relativistic tracers, which governs redshift-space distortions (RSD) in the galaxy redshift survey.

This flow diagram necessarily simplifies some key points, most notably (a) that the SN distance scale is calibrated in the local Hubble flow while the BAO distance scale is calibrated in absolute ($H_0$-independent) units, so they provide complementary information even when measured at the same redshift, and (b) that cosmic shear, cluster abundances, and RSD probe the expansion history as well as the growth history. The combination of WFIRST-2.4 dark energy probes is far more powerful than any one probe would be in isolation, allowing both cross-checks for unrecognized systematics and rich diagnostics for the origin of cosmic acceleration. We discuss the anticipated constraints on theoretical models below, in §2.2.3, after first summarizing the plans for the surveys themselves.

### 2.2.1 *The Supernova Survey*

The WFIRST-2.4 supernova (SN) survey has two components: multi-band imaging to discover supernovae and measure light curve shapes, and Integral Field Unit (IFU) spectroscopy to confirm the Type Ia classification, measure redshifts and spectral diagnostics, and perform the synthetic photometry used to measure luminosity distance. Our forecast for WFIRST-2.4 indicates that it will discover and monitor more than 2700 spectroscopically confirmed Type Ia supernovae out to redshift z = 1.7 in six months of observing time, spread over a 2-year interval. Space-based near-IR observations are the only feasible way to observe large samples of supernovae at z > 1, and at every redshift the use of near-IR wavelengths and a stable, space-based imaging platform with sharp PSF reduces the critical systematic uncertainties associated with dust extinction and photometric calibration. Relative to DRM1, the larger collecting area and sharper PSF of WFIRST-2.4 enables a survey with more supernovae, a more uniform redshift distribution, and the reduced systematics afforded by IFU spectroscopy.

Supernova observations take place with a five-day cadence, with each interval of observations taking a total of 30 hours of imaging and spectroscopy. There are three tiers to the survey: a shallow survey over 27.44 $deg^2$ for SNe at z < 0.4, a medium survey over 8.96 $deg^2$ for SNe at z < 0.8, and a deep survey over 5.04 $deg^2$ for SNe out to z = 1.7. This design is based on optimizing the dark energy Figure of Merit from SNe at fixed observing time, given our assumptions about statistical and systematic errors described below.

The survey uses the wide-field imager to discover supernovae in two filter bands: Y and J for the shallow, low redshift tier and J and H for the medium and deep tiers. However, the "photometric" measurements used for the cosmological analysis are derived from observations with the IFU spectrometer. Even though IFU observations are one object at a time, they prove more efficient than slitless spectroscopy over the 0.28 $deg^2$ field of view because the exposure time can be chosen for each supernova individually instead of being driven by the faintest object in the field, and because IFU observations have dramatically lower sky noise per pixel, allowing much better isolation of the faint SN signal.[8] Even more important, IFU spectrophotometry reduces systematic uncertainties by (a) removing the need for K-corrections, as the same rest-frame wavelength range can be chosen for SNe at different redshifts, (b) allowing better separation of SN and background galaxy light as a function of wavelength, (c) providing spectral diagnostics that can help distinguish intrinsic color variations from the effects of dust extinction[9] and match high and low-z SNe with similar properties to suppress evolutionary effects in the mix of supernovae. Photometric calibration is a limiting systematic uncertainty in current ground-based SN surveys (e.g., Sullivan et al.[10]) and a critical consideration for a space-based survey. An IFU system should achieve a much more accurate level of calibration for a supernova survey than broad-band filter photometry (or synthetic photometry from slitless spectra), because each pixel on the detector is illuminated by light of a fixed wavelength for both the source and backgrounds and because the total number of pixels to be calibrated is small, making it feasible to scan a standard star along the length of each slice in the IFU.

For each supernova, we obtain seven IFU spectra on the light curve with a roughly 5-day rest-frame cadence, from -10 rest-frame days before peak to +25 rest-frame days past peak, with S/N = 3 per pixel (S/N = 15 per synthetic filter band). In addition, we obtain one deep spectrum when the supernova is near its peak, with S/N = 10 per pixel, to provide one highly accurate photometric data point, identify supernova subtypes, and perhaps measure spectral diagnostics that can reduce statistical or systematic errors. Finally, we obtain one reference spectrum after the supernova has faded, for galaxy subtraction, with S/N = 6 per pixel. On average, in each set of observations roughly six hours is spent on imaging discovery, 11 hours on spectra for light curves, and 13 hours on the deep spectra and reference spectra. Forecasts for the number of SNe, statistical errors, and systematic errors as a function of redshift can be found in Appendix C.1.

### 2.2.2 *The High Latitude Survey*

The WFIRST-2.4 high latitude survey (HLS) will image 2000 $deg^2$ in the four near-IR bands listed in Table 2-1, which we will refer to as Y, J, H, and F184. The HLS will support an enormous range of science, as discussed in this report. From the point of view of dark energy, the most important contribution of the HLS imaging survey is weak lensing (WL) measurements. At fixed total observing time, the statistical power of a weak lensing survey generally increases with larger survey area (and shallower images), until the exposures become short enough that detector read noise dominates over sky noise. The HLS follows a dithering strategy (see Appendix C.2) so that images in J, H, and F184 are fully sampled for shape measurements, even when some exposures of a galaxy are lost to chip gaps, cosmic rays, or detector defects. We have chosen the exposure time of 184 secs for the HLS so that read noise and sky noise make roughly equal contributions in individual exposures. The survey area then follows from the total observing time devoted to the HLS imaging survey, which for the WFIRST-2.4 DRM is 1.3 years. Characteristics of the imaging survey are summarized in Table 2-1.

Figure 2-2 shows the imaging depth achieved by the WFIRST-2.4 HLS, in comparison to LSST (after 10 years of operation) and Euclid. In an AB-magnitude sense, WFIRST-2.4 imaging is well matched to the i-band depth of LSST. The AB magnitude limits for LSST in g and r are fainter; however, a typical z>1 LSST weak lensing source galaxy has r-J or i-H color of about 1.2, so even here the WFIRST-2.4 imaging depth remains well matched, and of course the angular resolution is much higher. The Euclid IR imaging is ~2.5 magnitudes shallower than WFIRST, and it is severely undersampled, so it cannot be used for galaxy shape measurements. The Euclid optical imaging depth is comparable to LSST, but in a single very wide filter rather than the ~5 optical bands necessary for photometric redshift determination.

The imaging component of the WFIRST-2.4 HLS will make weak lensing shape measurements of nearly 500 million galaxies over an area of 2000 $deg^2$. Weak lensing directly probes the clustering of matter between the observer and the lensed source galaxies, so it measures the growth of structure without uncertainties associated with galaxy bias. The weak lensing signal also depends on distances to the sources and the lensing matter distribution, so it provides expansion history constraints that are competitive with those from other methods. However, away from massive clusters the typical distortion of source galaxy shapes is only about 1%. Measuring the lensing signal with high precision, in the face of intrinsic ellipticity variations that are ~ 0.4 rms, requires enormous galaxy samples *and* exquisite control of systematic errors. Space-based measurements offer potentially enormous advantages for weak lensing because of high angular resolution and stability of the observing platform, allowing accurate characterization of the instrumental point-spread function (PSF). The WFIRST-2.4 HLS has been designed with control of systematics as a paramount consideration. The large aperture of WFIRST-2.4 yields a high surface density of lensed source galaxies, ~ 65 arcmin$^{-2}$ in the HLS (see Table 2-1) and potentially 200-300 arcmin$^{-2}$ in longer, targeted observations, much higher than any other ground-based or space-based facilities equipped for large area surveys. The expected source densities for LSST and Euclid (and WFIRST-2.4) depend on assumptions about which galaxies can be used for shape measurements; for the same criteria adopted here (S/N > 18, $r_{gal}/r_{PSF}$ > 0.8, $\sigma_e$ < 0.2 per component) we find $n_{eff}$ = 15 arcmin$^{-2}$ and 35 arcmin$^{-2}$ for LSST and Euclid, respectively.[11]

| | Band (μm) | Exp Time (sec) | Time Required (Days/1000 deg²) | Point Source Depth | Extended Source Depth | PSF EE50 (arcsec) | Weak Lensing $n_{eff}$ (galaxies/arcmin²) |
|---|---|---|---|---|---|---|---|
| Y | 0.927-1.192 | 5 x 184 | 50 | 26.8 | 25.6 | 0.12 | n/a |
| J | 1.131-1.454 | 6 x 184 | 59 | 26.9 | 25.7 | 0.12 | 54 |
| H | 1.380-1.774 | 5 x 184 | 50 | 26.8 | 25.7 | 0.14 | 61 |
| F184 | 1.683-2.000 | 5 x 184 | 50 | 26.2 | 25.2 | 0.14 | 44 |
| Grism | 1.350-1.950 | 6 x 362 | 118 | $4.6\times10^{-17}$ | $1.0\times10^{-16}$ | 0.18 | n/a |

**Table 2-1: Characteristics of the HLS. The dither strategy (Appendix C.2) has eight passes at each location in the 2000 deg² survey area (nine in J). Here we list exposure numbers and depths (5σ for point sources and exponential sources with $r_{eff}$ = 0.3 arcsec) at ≥90% fill factor, accounting for chip gaps and cosmic rays. The "union" lens sample with a good shape measurement in at least one band has $n_{eff}$ = 68 arcmin$^{-2}$, while summing J and H bands yields a deeper catalog with $n_{eff}$ ~ 75 arcmin$^{-2}$.**

The abundance of rich galaxy clusters as a function of mass and redshift offers an alternative route to measuring the growth of structure. The key uncertainty in this approach is accurate calibration of the cluster mass scale --- the average virial mass of clusters at redshift z as a function of a mass-correlated observable such as galaxy richness or X-ray luminosity --- which must be known to sub-percent accuracy to exploit the statistical potential of cluster surveys. The HLS imaging survey is an ideal tool for carrying out this calibration through measurements of the average weak lensing profiles of large cluster samples. The clusters themselves can be identified in WFIRST-2.4 imaging, together with optical imaging from LSST, or from X-ray surveys (the eROSITA mission in particular) or radio surveys that utilize the Sunyaev-Zel'dovich (SZ) effect.

In addition to statistical measurements, the HLS imaging survey can produce maps of the projected or 3-dimensional dark matter distribution. The dark matter maps from the 2 deg² COSMOS survey[12] have been among the most popular cosmological results from HST, yielding tests of theoretical models, an accessible illustration for public outreach, and even inspiration for art works. Figure 2-3 illustrates the high fidelity of mass maps that can be made with the high source density reached by WFIRST-2.4.

Over the same 2000 deg² area as the HLS imaging survey, the WFIRST-2.4 DRM incorporates a grism slitless spectroscopic survey (also called the galaxy redshift survey) to map the distribution of emission line galaxies. The primary target is Hα (0.6563 μm) at 1.05 < z < 2, but it is also possible to extend to higher redshift, 1.7 < z < 2.9, using galaxies with strong [OIII] emission (0.5007 μm). The large aperture of WFIRST-2.4 allows a survey significantly deeper than that planned for DRM1 (though in a different redshift range), and much deeper than the Euclid galaxy redshift survey, thus providing much more complete sampling of structure in the high-redshift universe. Figure 2-4 shows the limiting flux of the HLS spectroscopic survey for a point source and for an extended source with $r_{eff}$ = 0.3 arcsec and an exponential profile. Our forecasts, based on recent estimates for the Hα and [OIII] galaxy luminosity functions by J. Colbert et al. (in preparation), predict ~ 20 million Hα galaxies, and ~ 2 million [OIII] galaxies at z > 2 (see Table 2-2).

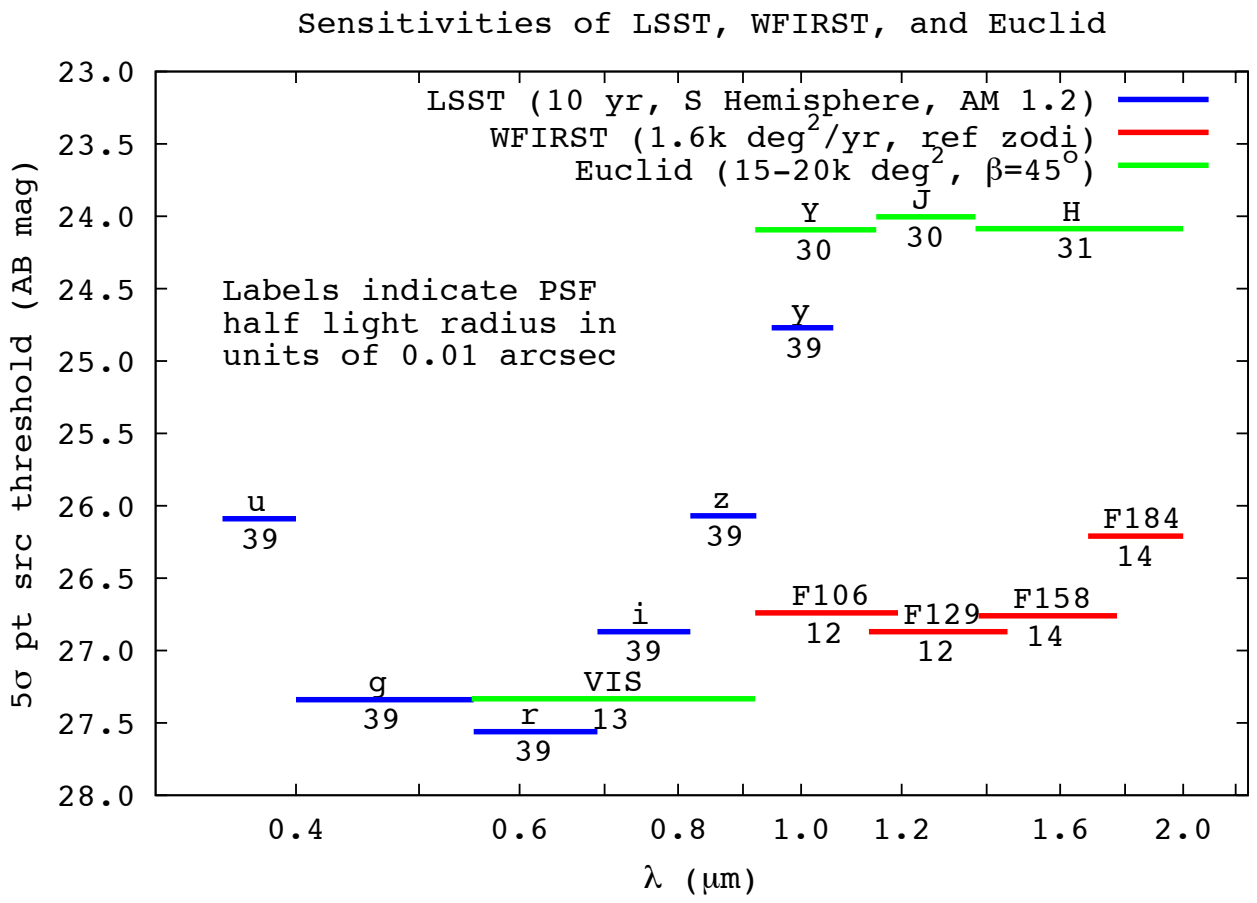

**Figure 2-2: Depth in AB magnitudes of the WFIRST-2.4 high-latitude survey (red), Euclid (green), and LSST (blue) imaging surveys. Labels below each bar indicate the size of the PSF (specifically, the EE50 radius) in units of 0.01 arcsec. The near-IR depth of the WFIRST-2.4 is well matched to the optical depth of LSST (10-year co-add).**

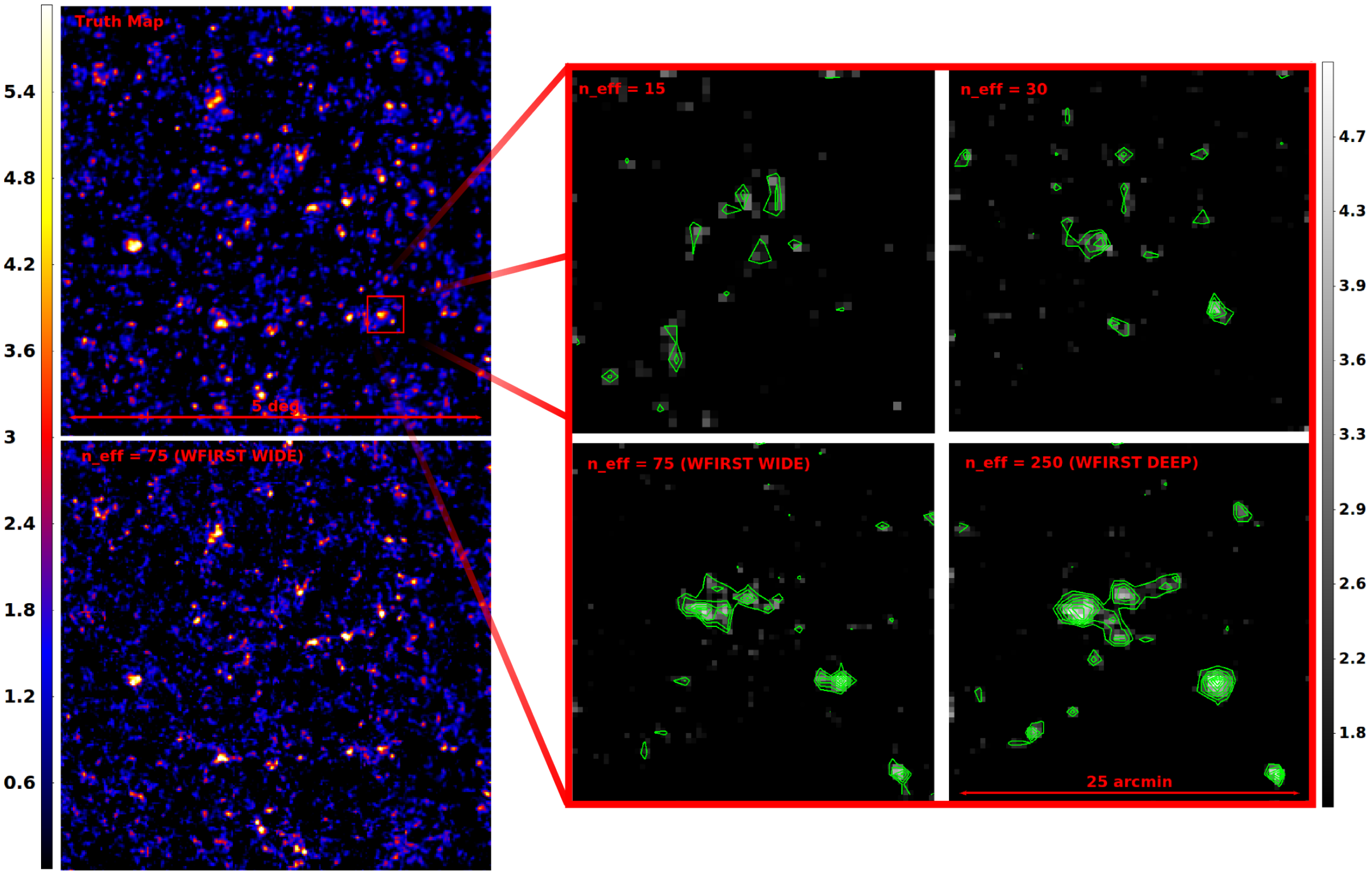

**Figure 2-3: Maps of the projected surface density of dark matter of a 5 deg² field of view (left panels) and an expanded region (25 arcmin²) on a side (right panels). The underlying mass distributions are taken from a large cosmological N-body simulation. The leftmost panel shows a map with the weak lensing shape noise expected from the WFIRST-2.4 HLS, at a source density n_eff = 75 arcmin$^{-2}$. The lowest levels shown (blue) are 1σ above the background. This map accurately recovers the large scale features of the true mass distribution (middle panel). The zoomed panels (with contours spaced at 0.5σ starting at 1.5σ) compare the fidelity of maps at the HLS depth to maps with the 15-30 arcmin$^{-2}$ source density characteristic of ground-based or Euclid observations, which miss much of the fine detail at small scale. Over selected areas of tens of deg², WFIRST-2.4 could carry out deeper imaging to achieve source densities of 200-300 arcmin$^{-2}$ (bottom right panel), yielding projected mass maps that are virtually indistinguishable from noiseless maps (not shown). Figure credit Julian Merten/Alina Kiessling (Caltech/JPL).**

The GRS allows measurement of the cosmic expansion history through the use of baryon acoustic oscillations. Sound waves that traveled in the photon-baryon fluid of the pre-recombination universe imprinted a characteristic scale on the clustering of matter, which was subsequently imprinted on the clustering of galaxies and intergalactic gas. The BAO method uses this scale as a standard ruler to measure the angular diameter distance $D_A(z)$ and the Hubble parameter H(z), from clustering in the transverse and line-of-sight directions, respectively. The BAO method complements the Type Ia supernova method in several respects: it measures distances against an absolute scale calibrated via CMB observations (instead of relative to the low-redshift Hubble flow), its precision increases towards *high* redshifts because of the larger comoving volume available for clustering measurements, and it can directly measure H(z), which is tied to the dark energy density through the Friedmann equation. Our current understanding of the BAO method suggests that it will be statistics-limited even for the largest foreseeable surveys; small corrections are required for the effects of non-linear gravitational evolution and non-linear galaxy bias, but these can be computed at the required level of accuracy using analytic and numerical techniques. The broad-band shape of the galaxy power spectrum P(k) provides a second "standard ruler" for geometrical measurements via the turnover scale imprinted by the transition from radiation to matter domination in the early universe, as well as a diagnostic for neutrino masses, extra radiation components, and the physics of inflation.

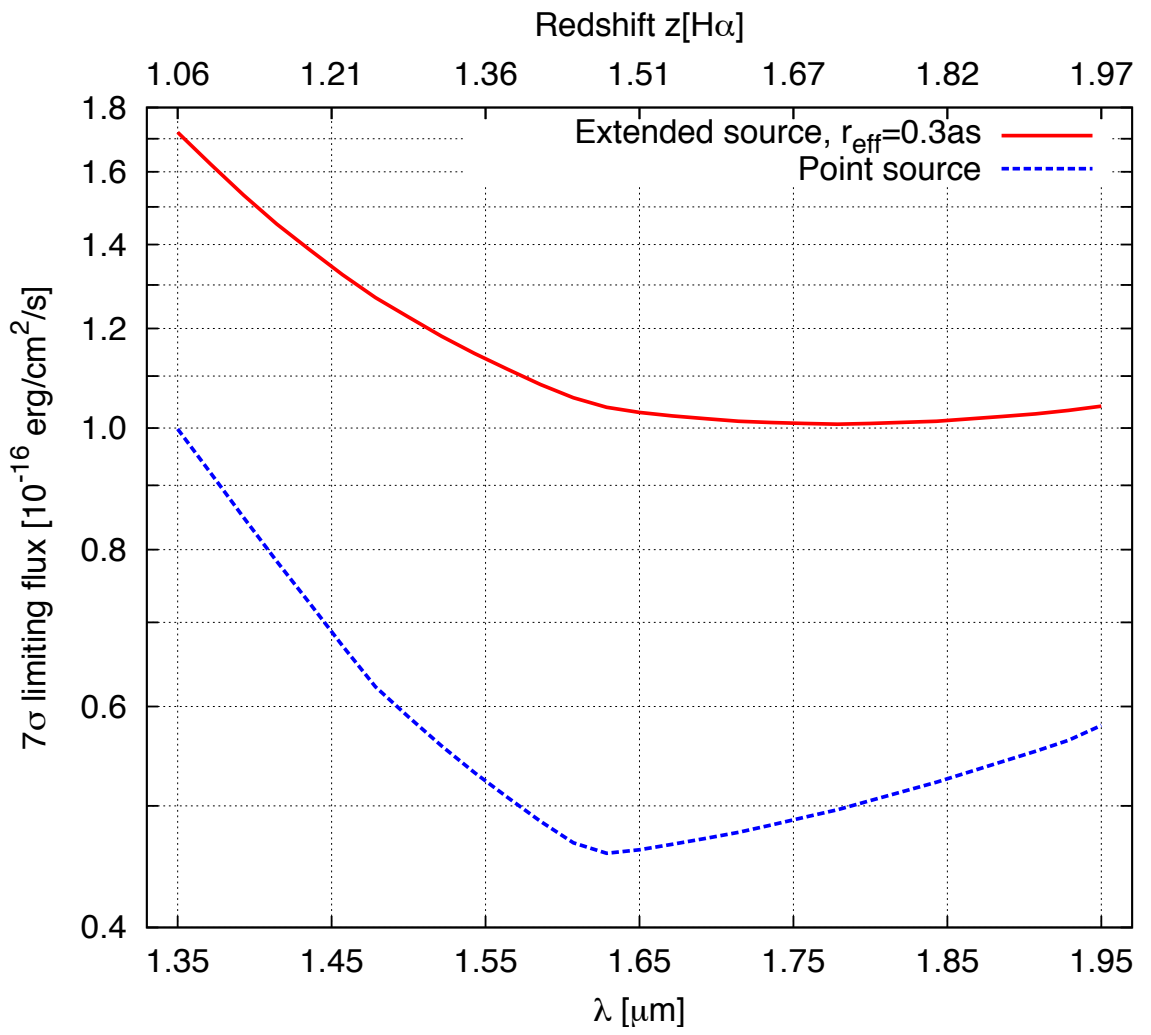

**Figure 2-4: The emission line sensitivity for the WFIRST-2.4 HLS spectroscopy survey. The blue curve shows 7σ point source sensitivities (for six exposures), and the red curve shows extended source ($r_{eff}$ = 0.3 arcsec, exponential profile) sensitivities. The depth is observed-frame (not corrected for Galactic extinction). This depth is for the reference zodiacal light level 1.3× the value at the ecliptic pole; 80% of the GRS has lower zodiacal background.**

The GRS conducted for BAO also offers a probe of the growth of structure via redshift-space distortions (RSD), the apparent anisotropy of structure induced by galaxy peculiar velocities. RSD measurements (in the large scale regime described by linear perturbation theory) constrain the product of the matter clustering amplitude and the clustering growth rate (specifically, the logarithmic derivative of clustering amplitude with expansion factor), complementing weak lensing observations that measure the matter clustering amplitude in isolation. Comparison of weak lensing and RSD growth measurements can also test modified gravity theories in which the potential that governs space curvature can depart from the potential that governs "Newtonian" acceleration of non-relativistic tracers.

| z | n ($Mpc^{-3}$) | dN/dz/dA ($deg^{-2}$) |
|---|---|---|
| 1.10 | 1.17E-03 | 10623 |
| 1.15 | 1.25E-03 | 11776 |
| 1.20 | 1.32E-03 | 12814 |
| 1.25 | 1.38E-03 | 13877 |
| 1.30 | 1.43E-03 | 14719 |
| 1.35 | 1.47E-03 | 15527 |
| 1.40 | 1.50E-03 | 16244 |
| 1.45 | 1.53E-03 | 16890 |
| 1.50 | 1.51E-03 | 16965 |
| 1.55 | 1.37E-03 | 15759 |
| 1.60 | 1.25E-03 | 14536 |
| 1.65 | 1.13E-03 | 13305 |
| 1.70 | 1.01E-03 | 12110 |
| 1.75 | 9.02E-04 | 10918 |
| 1.80 | 8.00E-04 | 9789 |
| 1.85 | 7.05E-04 | 8697 |
| 1.90 | 6.17E-04 | 7686 |
| 1.95 | 5.36E-04 | 6718 |
| 2.00 | 1.33E-04 | 1678 |
| 2.10 | 1.21E-04 | 1541 |
| 2.20 | 1.15E-04 | 1477 |
| 2.30 | 9.72E-05 | 1258 |
| 2.40 | 8.12E-05 | 1054 |
| 2.50 | 6.66E-05 | 866 |
| 2.60 | 5.35E-05 | 696 |
| 2.70 | 4.28E-05 | 556 |
| 2.80 | 3.27E-05 | 424 |

**Table 2-2: Comoving space density of galaxies, in comoving $Mpc^{-3}$ and number per unit z per $deg^2$, expected in the WFIRST-2.4 GRS. We include only Hα emitters at z < 2 and only [OIII] emitters at z ≥ 2.**

For measurements of BAO distances and expansion rates, the most important metric of a redshift survey is its total comoving volume and the product $nP_{BAO}$ of the mean galaxy space density n and the amplitude of the galaxy power spectrum P(k) at the BAO scale, approximately k = 0.2h $Mpc^{-1}$. For $nP_{BAO} > 2$, the BAO measurement error is dominated by sample variance of the structure within the finite survey volume and would not drop much with a higher tracer density. For $nP_{BAO} < 1$, the measurement error is dominated by shot noise in the galaxy distribution. For any given $nP_{BAO}$, the error on $D_A(z)$ and H(z) scales with comoving volume as $V^{-1/2}$.

To predict $nP_{BAO}$ for the space densities in Table 2-2, we adopt the prescription of Orsi et al.[13] for the bias factor between galaxy and matter clustering, b=1.5+0.4(z-1.5), which is based on a combination of semi-analytic model predictions and observational constraints. The clustering measurements of Geach et al.[14] suggest a somewhat higher bias for Hα emitters, which would lead to more optimistic forecasts.

Figure 2-5 plots $nP_{BAO}$ vs. z. The BAO scale is fully sampled ($nP_{BAO} > 1$) over the whole range $1.05 < z < 1.95$ probed by Hα emitters, with $nP_{BAO} > 2$ at $z < 1.8$. The strong decline at $z > 1.5$ arises because we assume that the Hα luminosity function does not evolve beyond the maximum redshift probed by the Colbert et al. data; this assumption could prove pessimistic, though extrapolation of a fixed luminosity function to z = 2.2 gives reasonable agreement with the measurements of Sobral et al.[125] at this redshift. While $nP_{BAO} \approx 4$ is "overkill" for BAO measurement at this scale, a high density survey allows better measurements of structure at smaller scales, better measurements of higher order clustering statistics, better characterization of galaxy environments, and more complete sampling of the population of star-forming galaxies, all beneficial for studies of galaxy formation and galaxy evolution. [OIII] emitters provide a sparse sampling of structure at z > 2; because of the large comoving volume, this sample of ~ 2 million galaxies yields useful cosmological constraints despite its relatively high shot noise.

Ground-based surveys like BOSS and the proposed MS-DESI experiment are likely to achieve $nP_{BAO} > 1$ out to z ~ 1.1, but reaching full sampling with ground-based observations becomes very difficult at higher z. Our forecasts for Euclid based on the same luminosity function assumptions and the exposure times in Laureijs et al.[15] imply space densities roughly 8, 16, and 30 times lower than those of WFIRST-2.4 at z = 1.1, 1.5, and 1.9, respectively, so Euclid BAO errors will be dominated by galaxy shot noise. Figure 2-6 presents a visual comparison of structure sampled at WFIRST-2.4 density and Euclid density, based on slices from the Millenium simulation[16] populated with semi-analytic galaxy formation modeling.[17] A survey at the WFIRST-2.4 depth recovers much of the fine detail present in the full dark matter distribution, which is lost at the much sparser sampling of the Euclid survey. We note that WFIRST-2.4 could carry out a shallow-wide GRS of 12,000 $deg^2$ in about one year of observing time, but this would be largely redundant with Euclid, while the deeper survey adopted for the DRM is complementary.[18]

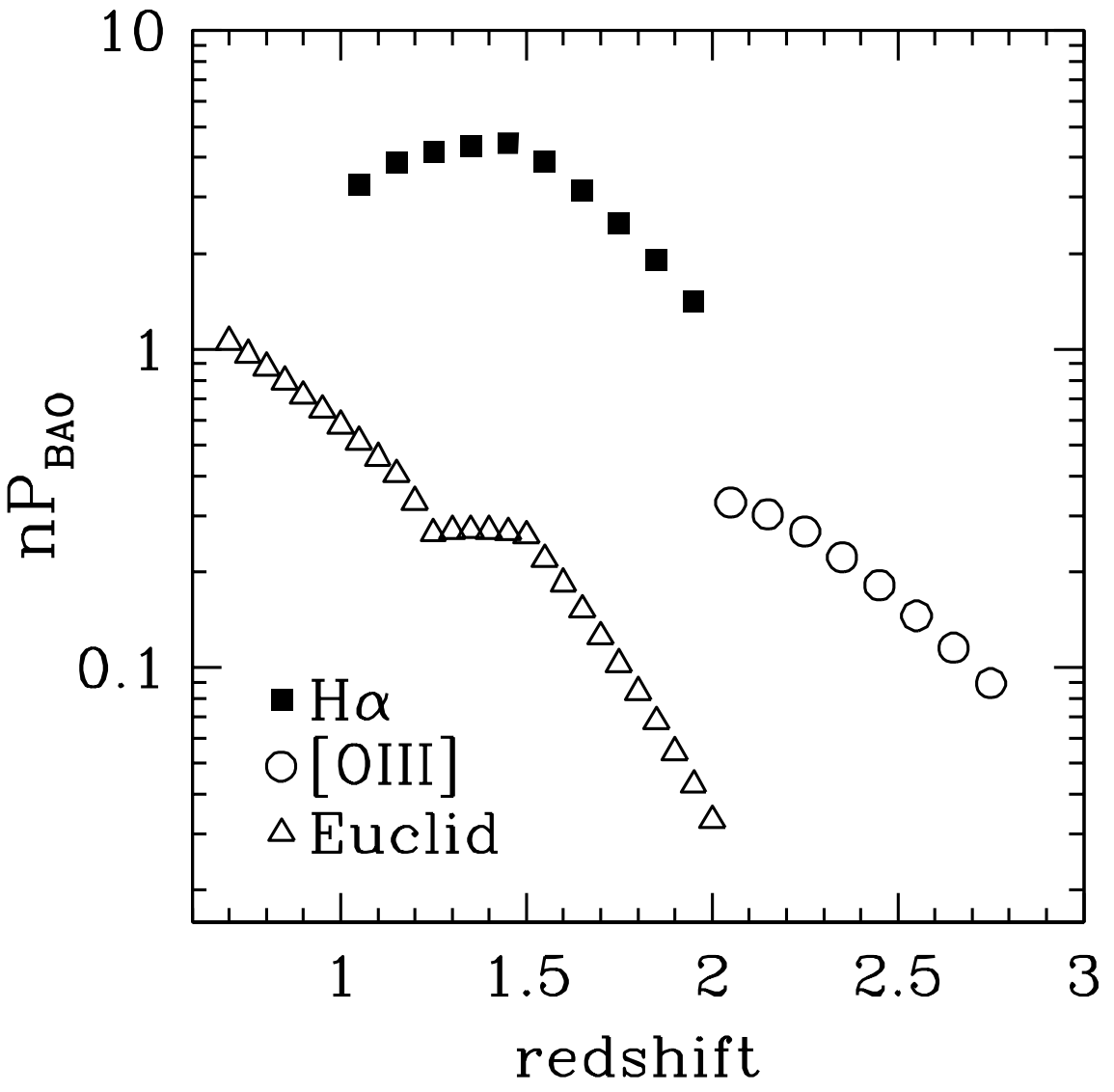

**Figure 2-5: Product $nP_{BAO}$ of the mean galaxy space density and the amplitude of the galaxy power spectrum at the BAO scale as a function of redshift for the WFIRST-2.4 GRS, based on the luminosity function of Hα emitters (squares) and [OIII] emitters (circles). Triangles show our estimate of the Euclid sampling density for the same luminosity function assumptions. For $nP_{BAO} > 2$, the statistical errors of BAO measurements are dominated by the sample variance of structure within the survey volume, while for $nP_{BAO} < 1$ they are dominated by shot noise in the galaxy distribution.**

### 2.2.3 *Tests of Cosmic Acceleration Models*

As shown in Appendix C, the WFIRST-2.4 supernova, imaging, and spectroscopic surveys will enable multiple independent measurements of cosmic expansion history and structure growth over the redshift range z = 0-3, each with aggregate precision at the ~ 0.1 – 0.5% level. This extremely high statistical precision demands that systematic biases be very tightly controlled to avoid compromising the measurements. The WFIRST-2.4 mission is designed with control of systematics foremost in mind, so that it can in fact realize the promise of its powerful statistics. For SNe, the use of a space-based observatory and near-IR observations already mitigates key systematics affecting ground-based surveys, and the use of an IFU on WFIRST-2.4 reduces systematics associated with photometric calibration and k-corrections and provides spectroscopic indicators that can be used to mitigate evolutionary effects. For WL, unique aspects of WFIRST-2.4 are the high surface density of sources and the control of systematics enabled by eliminating the atmosphere and having highly redundant multicolor data, with an observ-ing strategy that provides good sampling even when some of a galaxy's exposures are lost to cosmic ray hits or detector defects. For the GRS, the high space density of galaxies in the WFIRST-2.4 survey will make it possible to measure higher order clustering statistics and split the data into subsamples to constrain models of galaxy bias, which are the primary source of uncertainty in deriving cosmological constraints from RSD and intermediate-scale P(k) measurements. Cross-correlation with the WL shear maps (which have significant statistical power even out to z=2) will provide further tests of galaxy bias models and improved cosmological constraints. In all of these aspects of systematics control, the large aperture and sharp PSF of WFIRST-2.4 play a crucial role.

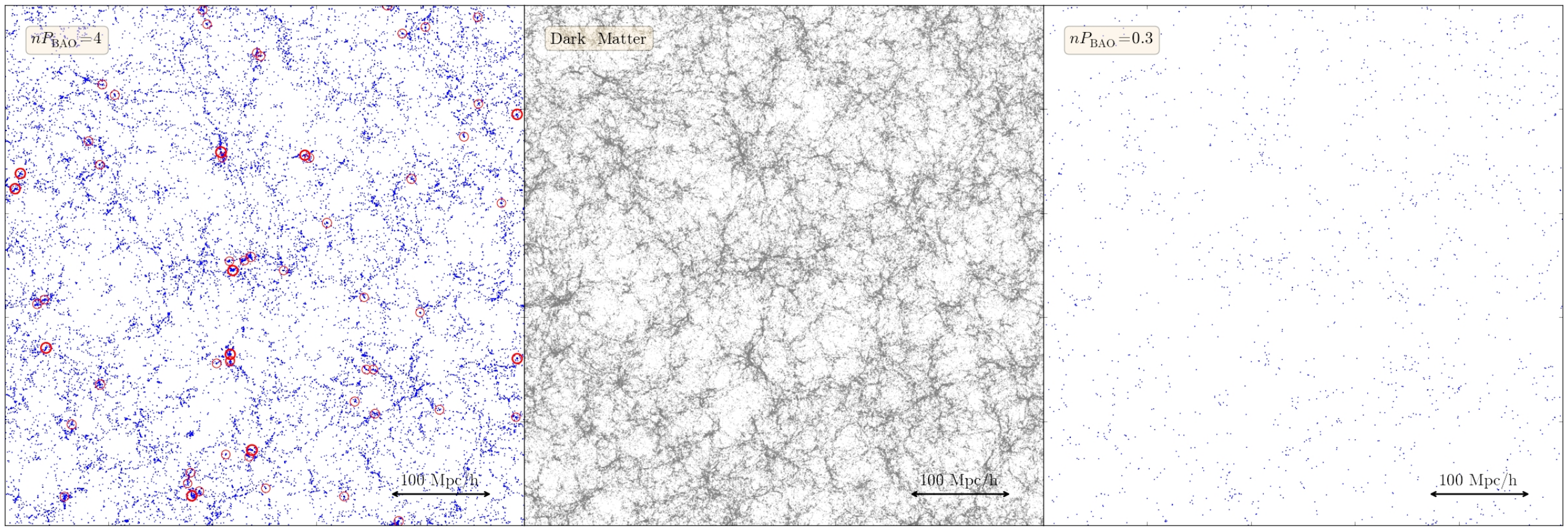

**Figure 2-6: Slices 500 $h^{-1}$ Mpc on a side and 30 $h^{-1}$ Mpc thick from the Millenium simulation at z = 1.5. Points in the left panel show semi-analytic galaxies[17] selected at a luminosity threshold that yields our predicted space density for WFIRST-2.4. Thin and thick red circles mark clusters with virial mass exceeding 5 × $10^{13}$ $M_{sun}$ and $10^{14}$ $M_{sun}$, respectively. The middle panel shows the dark matter density field, based on mass-weighting the full dark matter halo population. The right panel shows the galaxy distribution with a higher luminosity threshold that yields the space density predicted for the Euclid GRS at this redshift. At z = 1.5 this slice would subtend a solid angle (9.2 × 9.2) $deg^2$ with redshift depth $\Delta z$ = 0.022, so it represents a minuscule fraction (~ $10^{-3}$) of the GRS survey volume. Figure courtesy of Ying Zu.**

To characterize the ability of these measurements to test theories for the origin of cosmic acceleration --- i.e., to address the two questions posed at the start of §2.2, we adopt the framework put forward by the Dark Energy Task Force (DETF).[19] We assume a parameterized model in which the dark energy equation-of-state parameter evolves with redshift as $w(a) = w_0 + w_a(1\text{-}a) = w_p + w_a(a_p\text{-}a)$, where $a = (1+z)^{-1}$ is the expansion factor and $z_p = a_p^{-1}\text{-}1 \approx 0.5$ is the "pivot" redshift at which the errors on $w_p$ and $w_a$ are uncorrelated and the value of w(a) is best determined. Additional parameters that must be constrained by the measurements are the baryon and matter densities $\Omega_b h^2$ and $\Omega_m h^2$, the Hubble parameter h, the curvature $\Omega_k$, and the amplitude and spectral index of the inflationary fluctuation spectrum. Our analysis takes into account the degeneracies among these parameters and the ability of complementary cosmological observables to break these degeneracies.

With the assumptions described in Appendix C, we have computed Fisher matrices for the WFIRST-2.4 SN, GRS, and WL surveys, which we combine with the Fisher matrix for Planck CMB measurements. In addition to WFIRST-2.4 and Planck, our forecasts assume a local SN calibration sample with 800 SNe, and they include the anticipated BAO measurements at $z < 0.7$ from the SDSS-III BOSS survey.[20] Our WL Fisher matrix is based on cosmic shear only and does not include more uncertain contributions from higher order statistics or galaxy-galaxy lensing, which might significantly strengthen the anticipated constraints. We do include the expected constraints from clusters calibrated by weak lensing, using a Fisher matrix computed for us by Michael Mortonson and Eduardo Rozo following the methodology described by Weinberg et al.[7]

For the fiducial combination of measurements, we forecast 1$\sigma$ errors of 0.0088 on $w_p$ and 0.115 on $w_a$, after marginalizing over all other cosmological parameters in the model. The solid black ellipse in Figure 2-7 illustrates these constraints in the form of a $\Delta\chi^2$=1 error ellipse. This represents a dramatic improvement over the current state of the art, represented in Figure 2-7 by the green ellipse, with errors $\Delta w_p$ = 0.08 and $\Delta w_a$ = 0.96 taken from Anderson et al.'s[20] combination of the main current SN, BAO, $H_0$, and CMB data constraints. While current cosmological data are consistent with a cosmological constant, we obviously do not know what

WFIRST-2.4 will find. For the purpose of Figure 2-7 we have imagined that the true model (marked by the x) has $w_p$ = -1.022 and $w_a$ = -0.18. A value of w < -1 is exotic but possible in some theories, and it leads eventually to an accelerating rate of acceleration that causes the universe to end in a "Big Rip" at a finite time in the future. The baseline WFIRST-2.4 would discriminate this model from a cosmological constant at > $2\sigma$ significance, but the implications of such a result would be dramatic enough to demand stronger confirmation. The red ellipse, smaller in area by a factor of 2.5, represents the more stringent constraints that could be obtained with the "extended" dark energy program described below, which would unambiguously differentiate this model from a cosmological constant.

As a Figure of Merit (FoM) for dark energy experiments, the DETF proposed the inverse of the constrained area in the $w_p$-$w_a$ plane. We find FoM = $[\sigma(w_p)\sigma(w_a)]^{-1}$ = 990 assuming GR (and 942 when we allow deviations from GR parameterized by a free growth index $\gamma$). This value is represented by the blue shaded "Baseline" block in the upper panel of Figure 2-8. As a representation of likely pre-WFIRST constraints, the dashed blue block in this panel marks the forecast FoM = 131 for a combination of Planck and "Stage III" dark energy experiments such as BOSS and the Dark Energy Survey, taken from Table 8 of Weinberg et al.[21] The lower panel represents the growth index $\gamma$ --- as with the FoM, we plot $1/\sigma_\gamma$ so that higher values correspond to better constraints. Our forecast error for the baseline program is $\sigma_\Upsilon$ = 0.015, compared to the Stage III forecast of 0.15.

For Figure 2-7 and Figure 2-8, we have included only minimal external data that are strictly independent of WFIRST-2.4: Planck CMB, local SN calibrators, the BOSS BAO and RSD measurements at z < 0.7, and (implicitly) ground-based data for the photometric redshifts of the WFIRST-2.4 lensing source galaxies. These are important complementary data sets that are impossible (or do not make sense) to acquire from WFIRST itself. The inclusion of other pre-WFIRST data sets would, of course, increase the cumulative FoM. By the time of WFIRST-2.4, there is a good chance that ground-based redshift surveys such as eBOSS, the MS-DESI experiment, and the SuMIRe project with the Subaru PFS, will have measured BAO and RSD out to z = 1 over $10^4$ deg$^2$ or more. At z > 1 they may no longer be strictly independent of WFIRST-2.4, but because they will cover large sky areas at low sampling density ($nP_{BAO}$ < 1) instead of 2000 deg$^2$ at high sampling density, the measurements can probably be combined as though they are effectively independent. The ability to add pre-WFIRST SN and WL measurements to the cosmological constraints depends on whether they are limited by statistics or systematics and whether the systematics are correlated with any systematics in the WFIRST-2.4 analyses.

If WFIRST-2.4 development follows the schedule proposed in this report, then the LSST, Euclid, and WFIRST-2.4 dark energy experiments will be essentially contemporaneous, with all three of them reporting major results in the mid-to-late 2020s. As with ground-based surveys, it should be straightforward to combine WFIRST-2.4 GRS results with Euclid GRS results because of the complementarity between their narrow-deep and wide-shallow data sets. For SNe, we expect that the quality of IFU observations from WFIRST-2.4 will make its (spectro)photometry far better than anything achievable from the ground, but there may be advantageous ways to coordinate WFIRST-2.4 and LSST campaigns to get smaller statistical or systematic errors through additional light curve monitoring, better data on host galaxies, or the use of galaxy counts and shear maps to reduce the impact of lensing noise at high redshift (where it dominates the statistical error budget). The inclusion of SNe is a critical difference between WFIRST and Euclid, so it is worth noting that our forecast for *just* WFIRST-2.4 SNe + local SN calibrators + Planck CMB yields an FoM of 537, a fourfold improvement over the Stage III forecast.

For WL, the critical first task will be to cross-check LSST, Euclid, and WFIRST-2.4 measurements to test for systematics. Direct cross-correlation of shear maps from different observatories is a powerful test for shape measurement systematics, as these should be quite different among the three facilities. Photometric redshift uncertainties can be well constrained by combining WFIRST-2.4 and LSST data, to get deep 10-band photometry and good morphological information for all galaxies in the joint sample, to calibrate photo-z distributions via clustering cross-correlation with the imaging catalog,[22] and perhaps to obtain a direct spectroscopic calibration sample at faint magnitudes with the WFIRST-2.4 IFU. If these cross-checks and joint calibrations demonstrate that all three experiments are statistics-limited rather than systematics-limited, then their measurements can be combined to yield results much stronger than those from any one in isolation.

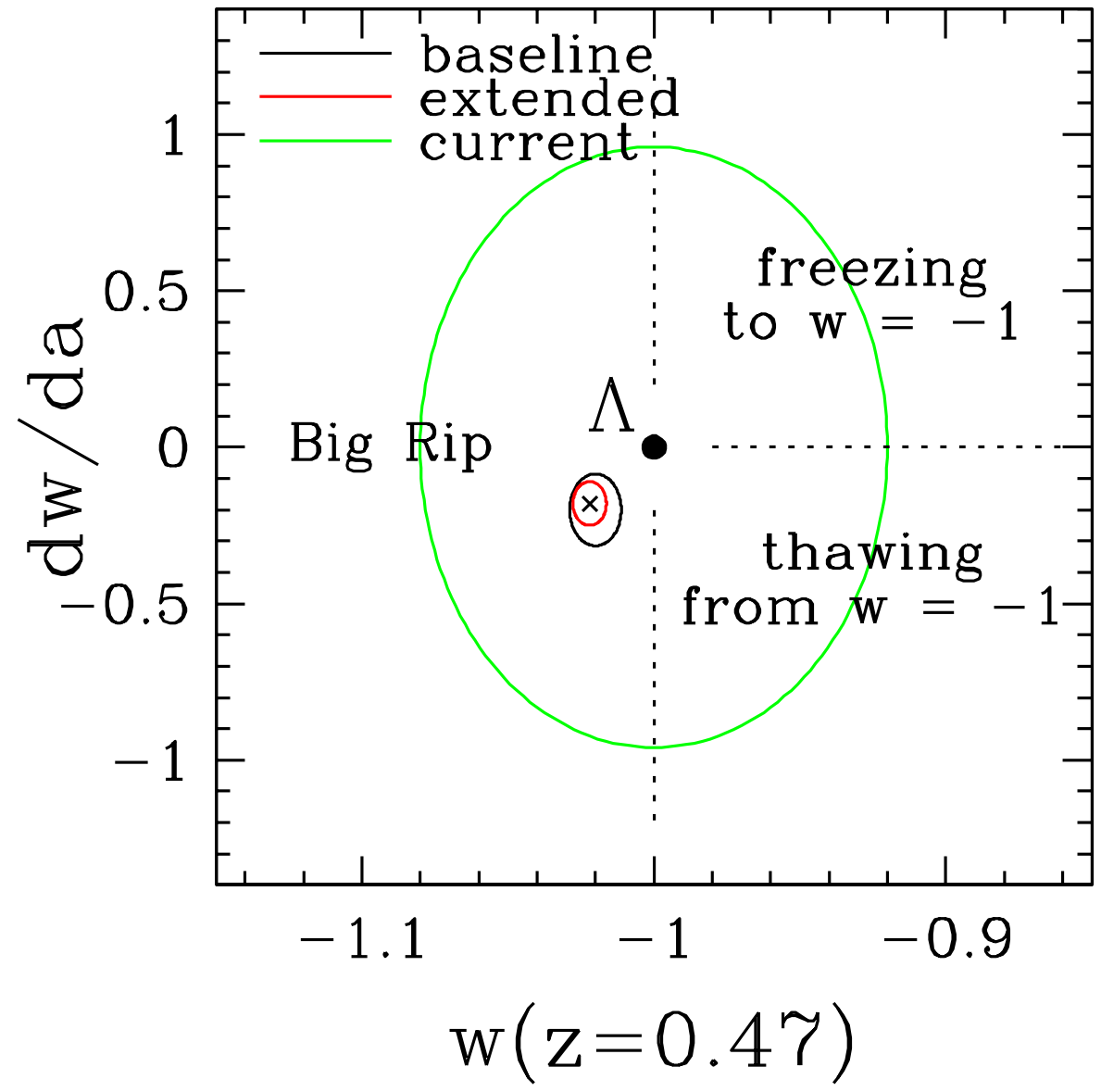

**Figure 2-7: $\Delta\chi^2 = 1$ error ellipses on the value of the dark energy equation-of-state parameter w at redshift z = 0.47 (the redshift at which it is best determined by WFIRST-2.4) and its derivative with respect to expansion factor dw/da. The green ellipse, centered here on the cosmological constant model (w = -1, dw/da = 0), represents current state-of-the-art constraints from a combination of CMB, SN, BAO, and $H_0$ data.[20] For this figure, we have imagined that the true cosmology is w(z=0.47) = -1.022 and dw/da = -0.18, well within current observational constraints. The black ellipse shows the error forecast for the baseline WFIRST-2.4 SN, GRS, and WL surveys, combined with CMB data from Planck, a local supernova calibrator sample, and BOSS BAO and RSD measurements at z < 0.7. The red ellipse shows the "extended" case in which the precision of the WFIRST-2.4 measurements (but not the Planck, local SN, or BOSS measurements) is increased by a factor of two, as a result of a longer observing program in an extended mission, better control of systematic uncertainties, or both. Legends indicate physically distinct regions of the parameter space: a cosmological constant (Λ), scalar field models that are "freezing" towards or "thawing" from w = -1, and models with w < -1 (often referred to as "phantom energy") in which increasing acceleration leads to a "big rip" at a finite time in the future.**

If the measurements from the WFIRST-2.4 prime mission are limited by statistics --- we have designed our requirements on systematic error control with this goal --- then the dark energy constraints could be improved considerably with additional observations in an extended mission. As an illustration, red, green, and cyan bars in Figure 2-8 show the impact of doubling the precision (i.e., multiplying all statistical+systematic errors by 0.5) of the WFIRST-2.4 GRS, WL, and SN measurements, respectively. If the multi-band imaging observations demonstrate good control of systematics, then it may be possible to carry out a wider area weak lensing survey in H-band only, in which case quadru-pling the area of the HLS (H-band imaging + spectroscopy) to 8000 deg$^2$ would require an additional 2.8 years of observation, yielding the "HLS×2" case illustrated by the blue bar. Doubling the precision of the SN survey would involve halving the systematic errors and quadrupling the sample size, which would take an additional 2.0 years of observing (including the overheads as estimated in §3.10). Doubling the precision of all three components, as shown by the purple bars in Figure 2-8, would increase the FoM from 990 to 2483 and $1/\sigma_Y$ from 66 to 109. These improvements, while substantial, are less than the naively expected factors of four[23] and two because the precision of the external data --- Planck, local SNe, and BOSS --- is held fixed.

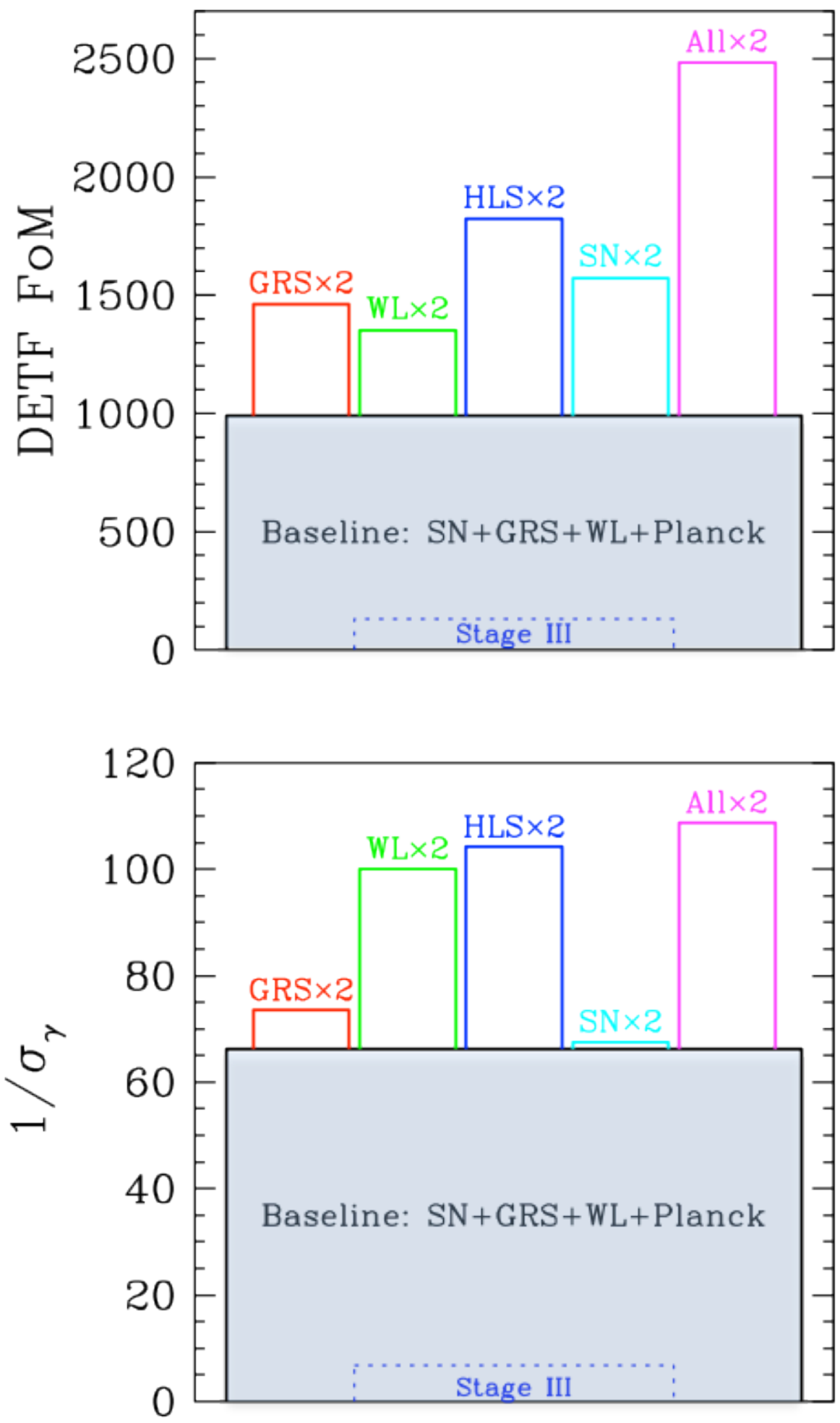

**Figure 2-8: Top: Figure of Merit FoM = $[\sigma(w_p)\sigma(w_a)]^{-1}$ for various assumptions. The blue shaded block shows the baseline case of FoM = 990 corresponding to the solid black contour of Figure 2-7. The blue dashed block shows the forecast FoM = 131 from Stage III experiments from Weinberg et al.[7] Red and green bars show the effect of increasing the measurement precision from the GRS or the WL survey by a factor of two, while the blue bar shows the effect of increasing both of them by a factor of two simultaneously. The cyan bar shows the impact of increasing the measurement precision from the SN survey by a factor of two. The purple bar shows the effect of increasing the precision of all three sets of measurement components by a factor of two, as described in the text. Errors for Planck, local SNe, and BOSS are held fixed throughout. Bottom: Same as top, but for the (inverse) $1\sigma$ error on the growth index $\gamma$.**

The gains illustrated by these alternative cases may be largely realized even within the prime mission, as our forecasts have generally been based on conservative assumptions. Our modeling of the SN survey has roughly equal systematic and statistical error contributions, so the performance would improve if we can keep systematics below our forecast level or if we can use spectroscopic diagnostics, host galaxy properties, and lensing corrections to reduce statistical errors. GRS precision will improve if theoretical modeling is robust enough to allow use of clustering information further into the non-linear regime. WL precision could be greatly improved by including constraints from galaxy-galaxy lensing and higher order statistics in addition to cosmic shear. In all cases we consider the potential gains to be large, but they are difficult to forecast because they rely on advances beyond the current state of the art. Of course, if these gains are realized and improve the return from the prime mission, then they would be amplified still further with the larger data set from an extended mission.

The motivation for and design of an extended WFIRST-2.4 dark energy mission will depend strongly on developments in the field over the next decade, and on the value of the extended data set (e.g., a larger area HLS) to other astronomical investigations. If existing data from WFIRST-2.4 and other facilities provide a solid but not undoubtable hint of deviation from a cosmological constant or GR-predicted growth, then the top priority may be to confirm the finding with higher precision or an independent measurement. If the data have already established a clear deviation, then the priority may be to investigate a new redshift range or different observable to better probe the underlying physical mechanism. Even if data remain compatible with a cosmological constant, advances on the theoretical side may lead to well motivated models that have testable predictions within reach of an expanded data set.[24] As a facility for mapping cosmic expansion and the growth of structure, WFIRST-2.4 is extraordinarily powerful along multiple dimensions, allowing considerable flexibility to adapt to a changing cosmological landscape.

### 2.3 High Latitude Surveys: General Astrophysics

WFIRST-2.4 will produce large data sets with homogenous observing conditions for each set. These data sets will be a treasure trove for Guest Investigators. The HLS will map ~2,000 deg$^2$ of sky in four broad NIR passbands down to a 5-sigma limiting AB magnitude of ~26.8 and will include a slitless spectroscopic survey component that will obtain R=600 spectra with a 7-sigma line flux sensitivity of $10^{-16}$ erg cm$^{-2}$ sec$^{-1}$ over the same region of sky (see Figure 2-4). While the survey is designed to obtain constraints on the dark energy equation of state from weak lensing and BAO measurements, a remarkable range of new astrophysical investigations will be enabled by the HLS data. We focus below on a few of the science cases that are especially well enabled by the access to a 2.4-meter wide-field survey telescope in space and that require no additional time other than that already allocated to the HLS.

#### 2.3.1 *The First Billion Years of Cosmic History*

Mapping the formation of cosmic structure in the first 1 billion years after the Big Bang is essential for achieving a comprehensive understanding of star and galaxy formation. Little is known about this critical epoch when the first galaxies formed. Determining the abundance and properties of such early objects is critical in order to understand (1) how the first galaxies were formed through a hierarchical merging process; (2) how the chemical elements were generated and redistributed through the galaxies; (3) how the central black holes exerted influence over the galaxy formation; and (4) how these objects contributed to the end of the "dark ages."

Finding candidate $z > 7$ galaxies has been pursued in two ways - from moderate area but very deep imaging surveys (e.g., UDF, Subaru Deep Field, CANDELS) and from images of strongly lensing clusters of galaxies.[25,26] Recent observations have identified at least 3 candidate objects at $z > 9$.[27,28,29] Two of these candidates were found in a survey of massive galaxy clusters. One of the lensed candidates is shown in Figure 2-9. While unlensed sources at these redshifts are expected to be extremely faint, with total AB magnitudes greater than 28, lensed sources can be 10 – 30 times brighter. This enhancement is particularly important be-cause it puts some of these sources within the reach of the spectrographs on the James Webb Space Telescope and from very large ground-based telescopes. Luminous z > 7 galaxies are extremely valuable as their spectra can be used to determine the epoch of the IGM reionization. Since only a tiny fraction of neutral hydrogen is needed to produce the high opacity of Ly$\alpha$ observed at z ~ 6, the damped Ly$\alpha$ absorption profile that results from even a partially neutral IGM[30] can be measured at low-spectral resolution. Furthermore, the early star formation rate (via Ly$\alpha$ and H$\alpha$ emission measurements[31]) can be estimated from the spectra of bright high-z galaxies.

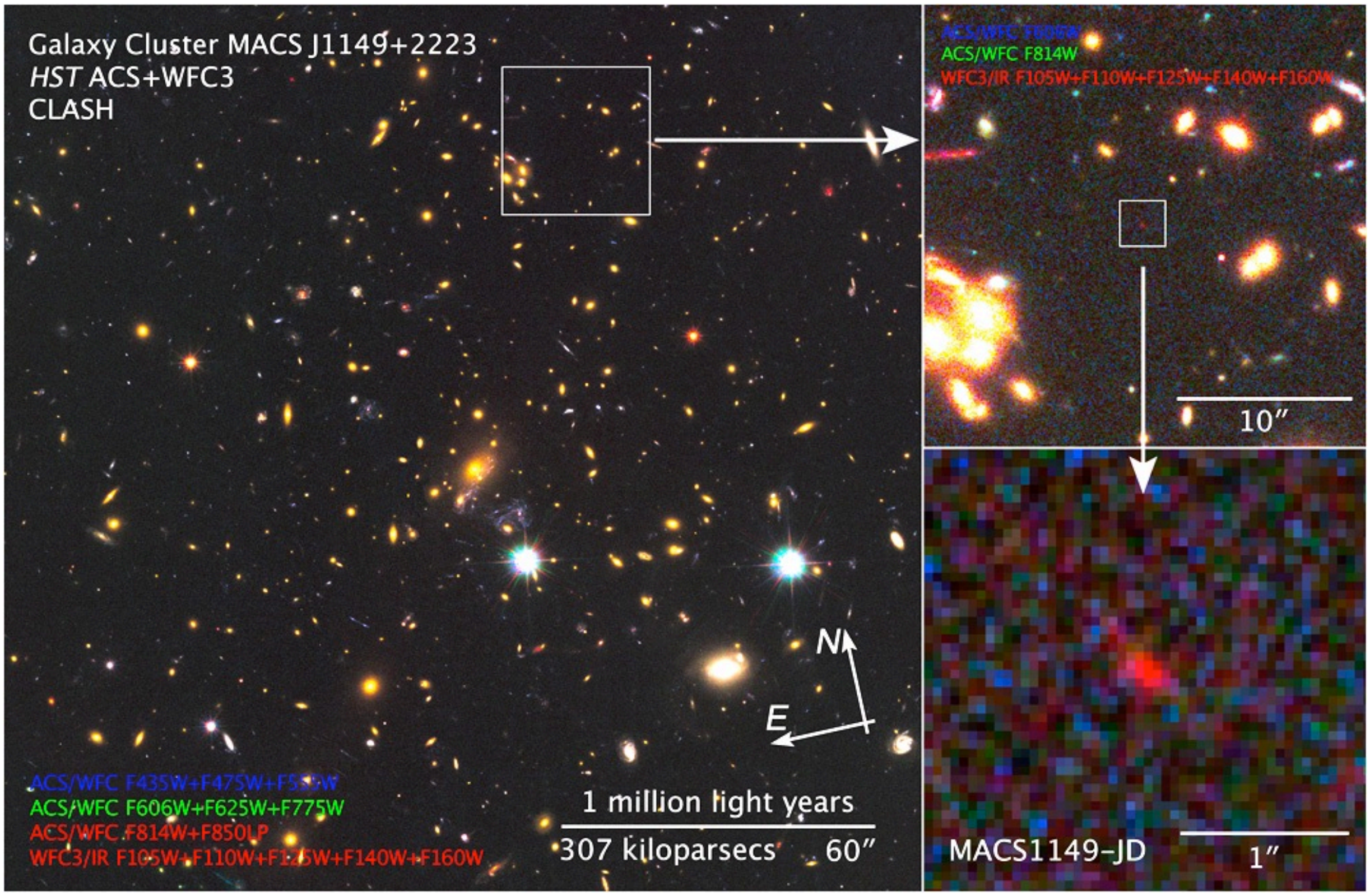

**Figure 2-9: A candidate for a galaxy at z=9.6, magnified by a factor of ~15 by the foreground cluster MACS J1149+2223 (z = 0.54). The object was found in an HST survey using the WFC3IR camera (Zheng et al. 2012). This young object is seen when the universe is only about 500 million years old.**

The HLS performed by the WFIRST-2.4 telescope as part of its weak lensing program will be superb for finding and studying objects in the early universe. A key advantage that the 2.4-meter aperture provides is that, in the same amount of time, the WFIRST-2.4 HLS will reach ~0.9 magnitudes fainter than the corresponding DRM1 HLS. The final 5-sigma limiting depth of the WFIRST-2.4 HLS is estimated to be J = 26.9 AB mag. Figure 2-10 shows the predicted cumulative number of objects that would be found at or higher than a given redshift for both gravitationally-lensed and unlensed regions in a 2,000 square degree survey. The predicted counts[32] for z~8 are extrapolated to higher redshifts using the few known z>9 candidates and bounded by assuming both a pessimistic (dM*/dz = 1.06) and optimistic (dM*/dz = 0.36) evolution of the characteristic magnitude of galaxies. The lensed source count predictions assume ~1 strongly lensing cluster per square degree. The mass models were adopted from the CLASH Multi-cycle treasury program.[26] The larger collecting area of WFIRST-2.4 relative to DRM1 will yield, in some redshift ranges, as much as 20 times as many z > 8 galaxies (Figure 2-11).

Having 10,000 or more luminous z > 8 galaxies will allow very stringent constraints to be placed on the early star formation rate density, on the amount of ionizing radiation per unit volume, and on the physical proper ties of early galactic structures. With its ~0.1 – 0.2 arcsecond resolution, combined with lens magnifications in the 10 – 30 range, the HLS images will allow us to measure structures on scales of 20 to 50 parsecs, thanks to the boost in spatial resolution provided by the cluster lenses. The HLS will certainly herald a remarka-ble era in probing the first 1 billion years of cosmic history.

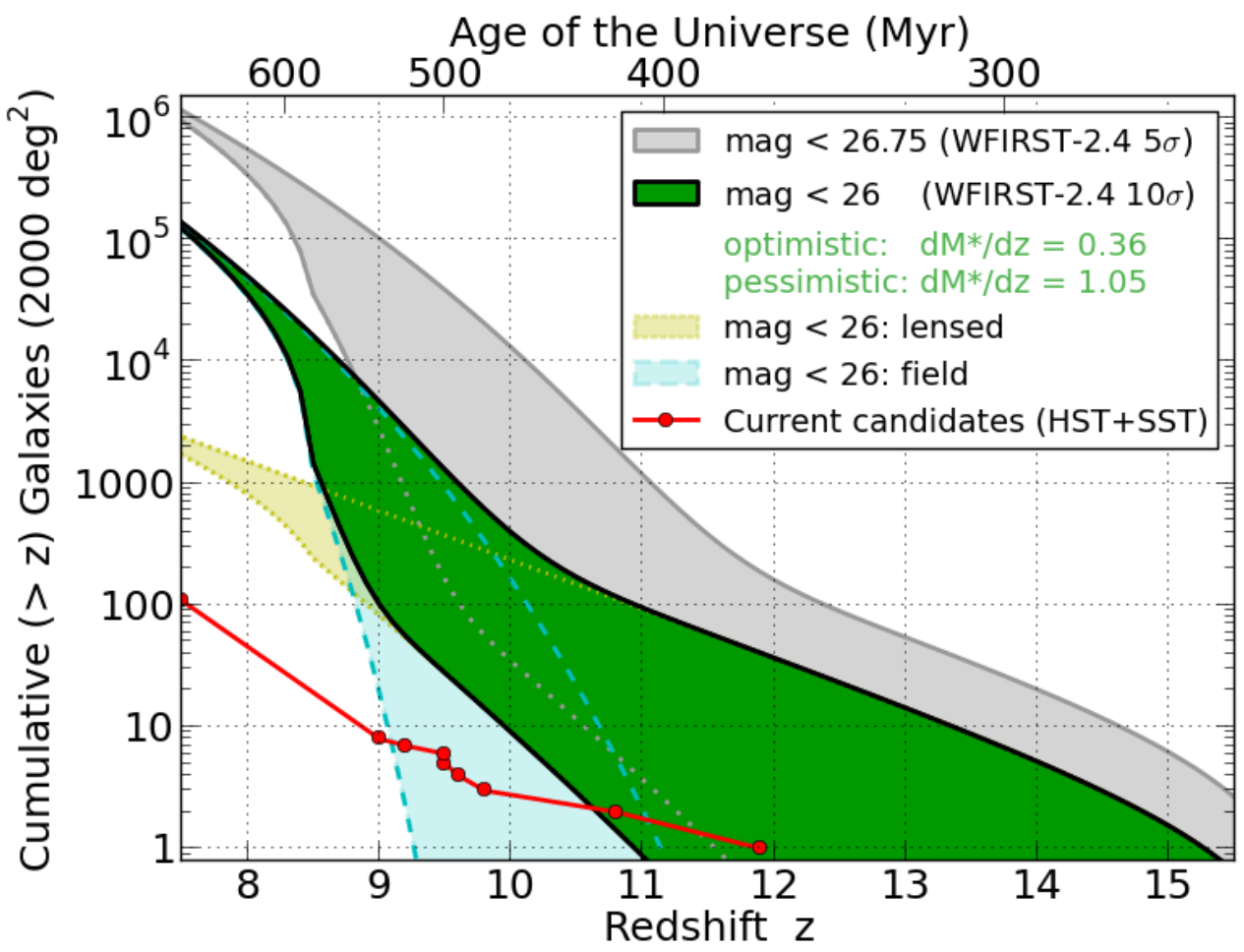

**Figure 2-10: Cumulative number of high-z galaxies expected in the HLS. JWST will be able to follow-up on these high z galaxies and make detailed observations of their properties. For understanding the earliest galaxies, the synergy of a wide-field telescope that can discover luminous or highly magnified systems and a large aperture telescope that can characterize them is essential; WFIRST-2.4 and JWST are much more powerful than either one alone.**

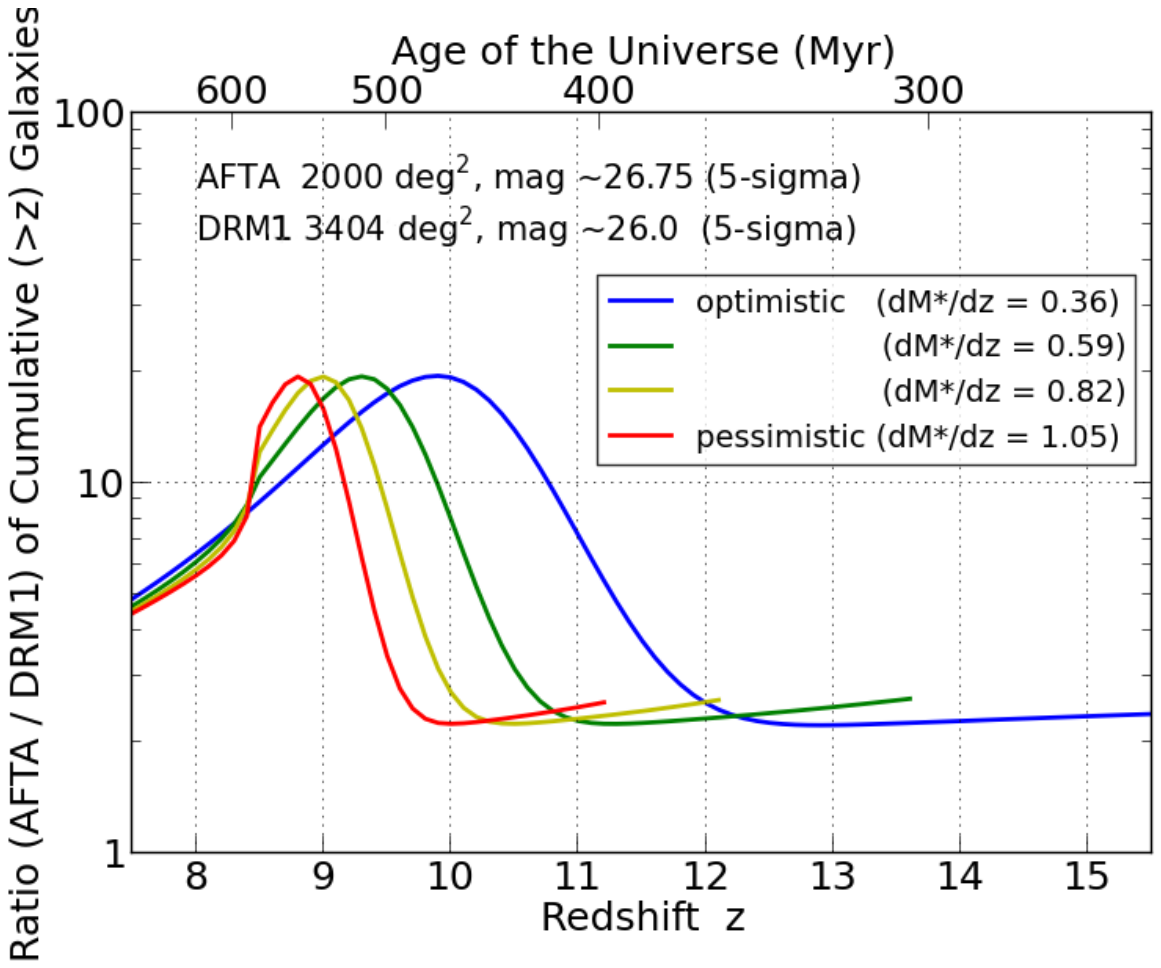

**Figure 2-11: The ratio of the cumulative number of high redshift galaxies detected with WFIRST-2.4 to the number detected with a smaller DRM1 version of WFIRST. The 2.4 m aperture yields up to 20 times more high-z galaxies.**

### 2.3.2 *Mapping Dark Matter on Intermediate and Large Scales*

Clusters of galaxies are important tracers of cosmic structure formation. All of their mass components including dark matter, ionized gas and stars are directly or indirectly observable. By its design, the HLS will be superbly suited to mapping weak lensing signatures. The signal-to-noise ratio of the shear signal is proportional to the square root of the surface density on the sky of galaxies that can be used to map the lensing. The WFIRST-2.4 HLS will produce catalogs with surface densities of 60 – 70 galaxies per square arcminute in a single passband and potentially as high as 80 – 100 galaxies per square arcminute in co-added multiband images. Achieving such densities for weak lensing measurements has already been demonstrated with the WFC3/IR camera on the Hubble Space Telescope. The WFIRST-2.4 HLS will thus enable WL maps that are a factor 3 – 5 higher number density of galaxies than any maps produced from the ground (even with 8 to 10 meter telescopes) as illustrated in Figure 2-12. The larger telescope and multiple band will produce much more robust and higher quality WL maps than Euclid (see Figure 2-3). On cluster scales (200 kpc – 2 Mpc), this higher galaxy density will allow the dark matter to be mapped to a spatial resolution of ~40 – 50 kpc. When combined with strong lensing interior to clustocentric radii of ~200 kpc, the central dark matter distribution can be mapped down to a resolution of 10 – 25 kpc. Comparing such maps around the hundreds of intermediate redshift clusters expected in the HLS to those from numerically simulated clusters would give us unprecedented insight into the main mechanisms of structure formation.

An especially interesting class of clusters are those in the process of merging / colliding (see Figure 2-13) where all mass components are interacting directly during the creation of cosmic structure. Multiple such mergers have been observed, with the Bullet Cluster being the most prominent example.[33] Paired with follow-up numerical simulations, such systems gave important insight into the behavior of the baryonic component[34] and set upper limits on the dark matter self-interacting cross section[35], which is of great importance in the search for the nature of dark matter. Recently, more complicated merging systems have been identified[36,37,38] offering a great opportunity to better characterize the nature of cosmic dark matter. The HLS will be well suited to producing highly accurate strong and weak lensing mass maps in the environs of these information-rich merging systems.

On the largest scales (>10 Mpc), the distribution of matter can best be measured via weak gravitational lensing, owing to the fact that the mass is predominant-ly dark matter and thus can only be detected gravitationally. The WFIRST-2.4 HLS's unique combination of spatial resolution, infrared observations and wide area will provide mass maps that have extremely high scientific impact. Mapping the large-scale distribution of dark matter can help identify and eliminate systematic effects: unexpected spatial variations in the mass distribution can hint at exciting new physics but could also indicate observational systematics not discoverable via a power spectrum measurement like the one that will be used to constrain dark energy via the HLS.

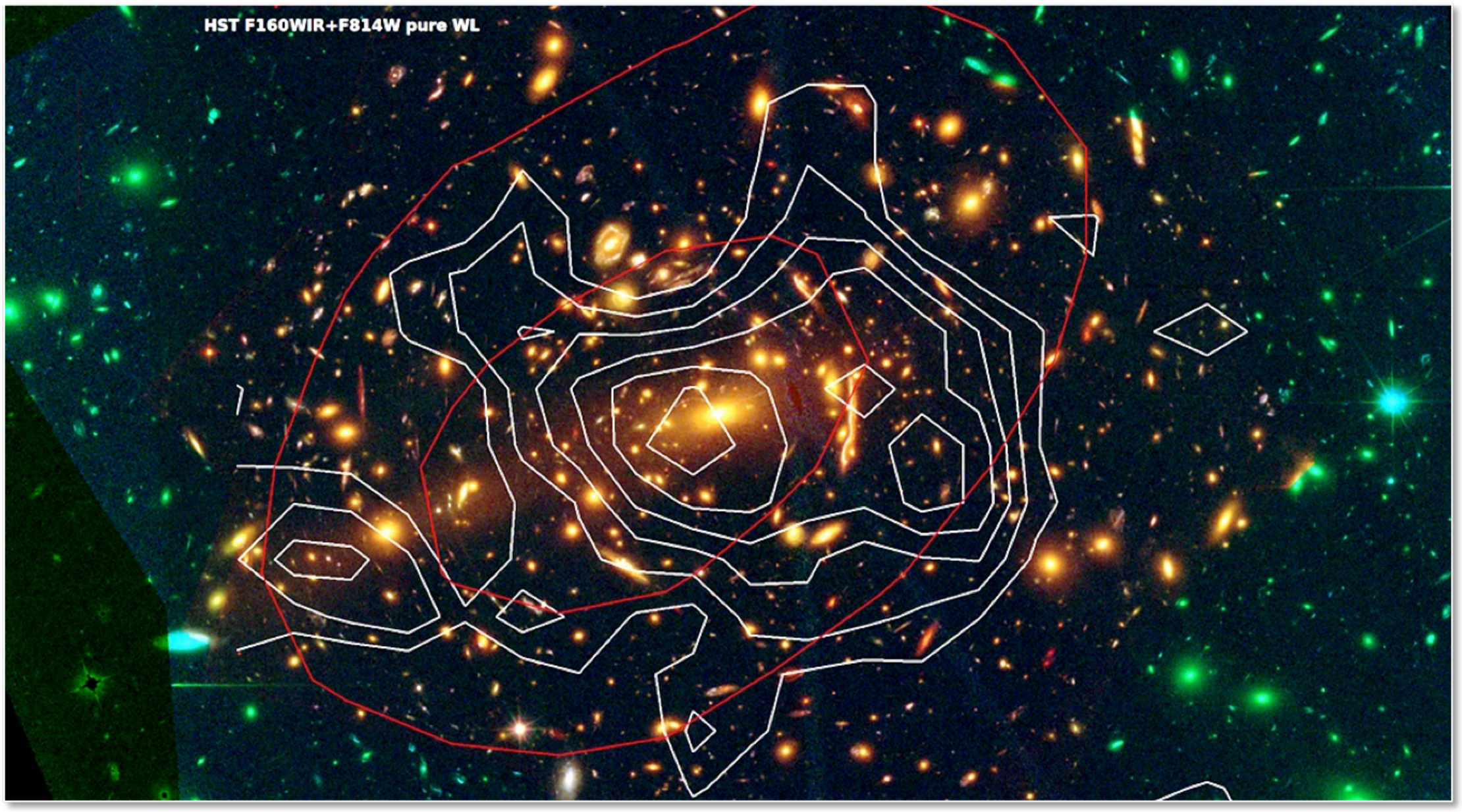

**Figure 2-12: Mass density contours around the cluster MACS J1206.2-0848 derived from a ground-based weak lensing survey with Subaru (red) vs. a weak lensing study with HST/ACS+WFC3 (white). The 10x higher surface density of lensed galaxies achieved from space yields ~3x higher spatial resolution maps. The HST data shown here is representative of the WFIRST-2.4 HLS. WFIRST-2.4 will make a map of this quality over 2,000 square degrees as part of its high latitude survey, a thousand-fold increase over the HST COSMOS mass map.**

### 2.3.3 *Kinematics of Stellar Streams in our Local Group of Galaxies*

One of the important auxiliary science products that will result from the resolution and sensitivity of the HLS is absolute proper motions for a very large number of field stars. From the ground, only bright galaxies and quasars can be used as high-precision absolute astrometric reference points. Unfortunately, their low density requires either very wide-field transformations or a bootstrapping approach that measures a target star relative to its local field population and larger field population relative to fixed reference points. WFIRST-2.4 will provide access to an enormous number of slightly resolved medium-brightness galaxies, and absolute motions can be measured directly within the field of each detector.

This local approach to measuring absolute motions has recently been used in the optical with HST to measure the absolute motions of the globular clusters, LMC, SMC, dwarf spheroidals and even M31, in addition to individual hyper-velocity and field stars. The strategy involves constructing a template for each galaxy so that a consistent position can be measured for it in exposures taken at different epochs. This template can be convolved with the PSF to account for any variations in focus or changes of the PSF with location on the detector. The same approach that worked with HST in the optical should work with WFIRST-2.4 in the IR. The WFIRST-2.4 detector pixel scale and sensitivity should be very similar to that of HST's WFC3/IR detector pixel scale. HST images of the Ultra Deep Field through the F160W (1.6 micron) passband show that there should be about 30 galaxies per square arcminute for which WFIRST-2.4 could measure a position to better than 5 mas in a 360 second exposure (see Figure 2-14). This gives us about 500 reference objects in each WFIRST-2.4 detector chip, enabling us to tie down the absolute frame to better than 0.5 mas in each exposure. The HLS 2-year baseline would allow absolute motions to be derived with systematic accuracies of about 125 μas/year. Folllow-on GO programs could extend this baseline to 5-years, enabling accuracies of about 50 μas/year. There are about 10 stars in the ~5 square arc-minute FOV of the UDF that can be measured with this accuracy, implying about 18 million halo stars in the high latitude imaging survey.

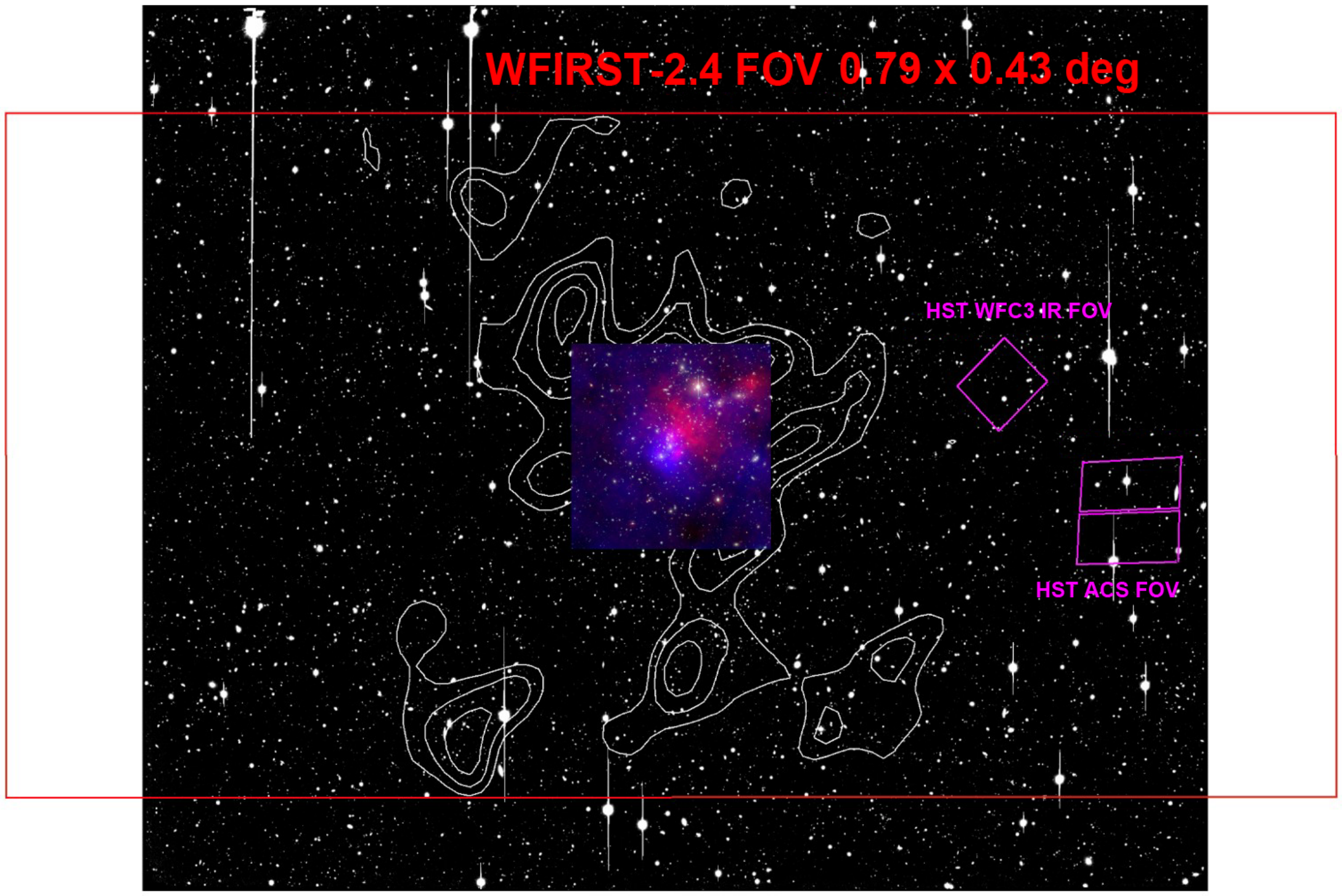

**Figure 2-13: Dark matter and cluster gas distribution around the merging cluster Abell 2744. Such maps require wide-field imaging. To map the entire area above with HST WFC3 or ACS would require 100 - 200 separate pointings, due to their small field of view. The same field is mapped with a single WFIRST-2.4 pointing (red rectangle.**

GAIA will overcome the sparse-galaxy issue by measuring positions across the entire sky in a single global solution, but it will not be nearly as sensitive as WFIRST-2.4. GAIA will achieve the above precision for G stars down to about V = 17, which allows the plentiful turnoff-star population to be probed out to about 2.5 kpc. WFIRST-2.4 will achieve the above precision for G stars down to V=20 and will allow turnoff stars to be probed out to about 10 kpc. There are about five times more stars within one magnitude of the turnoff than there are on the entire giant branch, and ten times more within two magnitudes of the turnoff, so WFIRST-2.4's depth will allow us to probe well beyond the thin disk with very good statistics.

The typical dispersion of stars in the halo is ~150 km/s, but since stream stars were gently stripped from objects with low dispersion, they typically have motions that are coherent to better than 10 km/s. Such 225:1 concentrations in phase space are easy to detect. Cur-rent ground-based stream studies are focused on HB/RGB/SGB stars, but access to the much more plentiful stars below the turnoff will increase by more than an order of magnitude the number of stars we have access to and the distance out to which we can study streams. By comparing the physical locations of streams and the motions within them, we can tease out the structure of the Galactic potential.

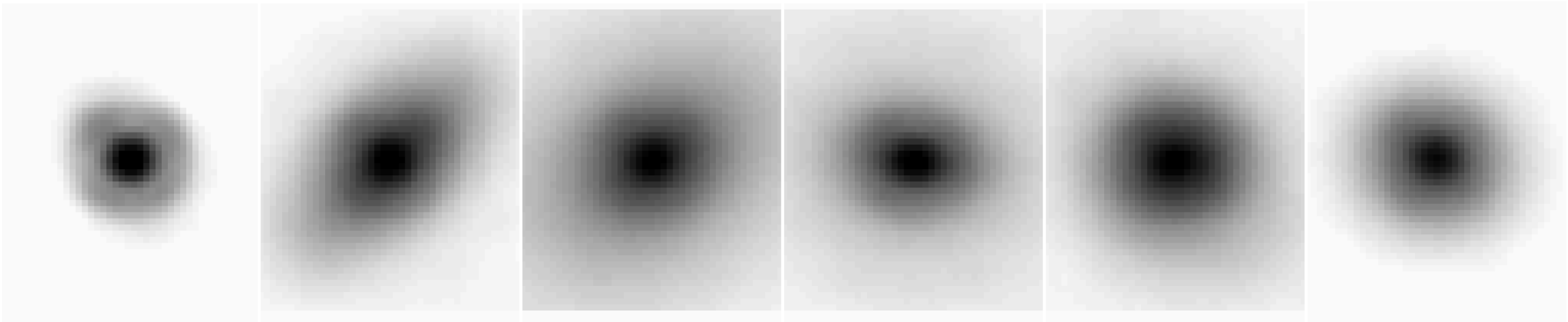

**Figure 2-14: A stellar PSF (left most image) compared to 5 galaxy images from the Ultra Deep Field as seen with HST's WFC3/IR detector - an imager with similar resolution to the WFIRST-2.4 imager. Each of these objects enables a position uncertainty good to 0.02 pixel (2 mas) in a *single* exposure.**

These stellar streams are powerful probes of the distribution of dark matter in our Galaxy and can determine whether our Galaxy has the large number of million solar mass subhalos predicted in the cold dark matter model. If the dark matter is not cold, but warm as suggested in some theories, these massive subhalos will not exist, and the stellar streams will be smooth and unperturbed[39].

### 2.3.4 *Discovering the Most Extreme Star Forming Galaxies and Quasars*

The supermassive black holes that sit in the centers of galaxies power active galactic nuclei (AGNs). There has been growing evidence that these supermassive black holes not only power these AGNs and quasars, but also play an essential role in the evolution of galaxies: (1) The stunning correlation between the masses of supermassive black holes, the AGN, with the masses of their host galaxies over a wide range of galaxy masses and galaxy types[40,41] and (2) numerical simulations show that star formation in the most massive galaxies must be suppressed by processes more powerful than supernovae and stellar winds in order to simulate galaxies resembling those we actually observe[42]. The most massive galaxies in the universe are the most affected by AGN feedback – they host the most massive black holes, and their star formation was truncated drastically about 10 billion years ago. The more massive the galaxy, the earlier this truncation happened, a trend called "downsizing".[43,44,45] With the HLS spectroscopic survey, we will be able to probe the epoch of downsizing of the most massive galaxies in the universe: z~2, which is the peak of the star formation density and near the peak of the quasar activity.

The HLS spectroscopic survey will enable the largest census yet performed of powerful emission-line galaxies and quasars up to (and possibly beyond) lookback times of ~90% of the current age of the universe. It will enable us to better understand the relationship between the accretion of matter by active galactic nuclei and the star formation occurring in the most massive galaxies in the universe. Several key star formation indicators in the spectral range of the grism will be usable to track star formation rates (SFR) in galaxies over the range 1 < z < 4.2. The H-alpha line will probe SFR that are at least 10 – 20 solar masses per year over the range 1 < z < 2, a cosmic epoch that is particularly challenging to study from the ground. The estimated surface density of such galaxies is about $10^4$ per square degree, implying the total survey will produce H-alpha measurements for over nearly 20 million galaxies. The OII[3727] line can be detected from star-forming galaxies lying between z = 2.6 and z = 4.2, but only for those systems with SFR in excess of ~200 solar masses per year. While such systems are less common, the WFIRST-2.4 HLS spectroscopic survey will allow us to accurately determine their space density over the key redshift range where the cosmic star formation rate has reached its highest value. At cosmic times from 200 – 600 million years since the Big Bang (8 < z < 15), Lyman-alpha emitting galaxies may be detectable if they have "attenuated" SFRs of at least 100 – 200 solar masses per year.

The HLS grism survey will discover ~2600 z > 7 quasars with H magnitude < 26.5, with an estimated 20% of those quasars being at z > 8. These are the lu-minous quasars whose existence tracks the assembly of billion solar mass black holes a mere few hundred million years after the Big Bang, and whose light can potentially illuminate the transition of intergalactic gas in the universe from neutral to an ionized phase (Gunn-Peterson effect[46]; Becker et al. 2001[47]; Djorgovski et al. 2001[48]; Fan et al 2003[49]). Such quasars must be found before their visible red light can be studied at high spectral resolution with JWST and ground-based 10-30-m class telescopes. The fainter quasars are particularly important to find because their region of local influence is smaller.[50] These are the same quasars that would be used to identify CIV in the IGM allowing the study of the chemical enrichment as a function of cosmic time. The HLS grism survey will also be a rich source of backlight targets for higher resolution spectroscopy with other facilities.

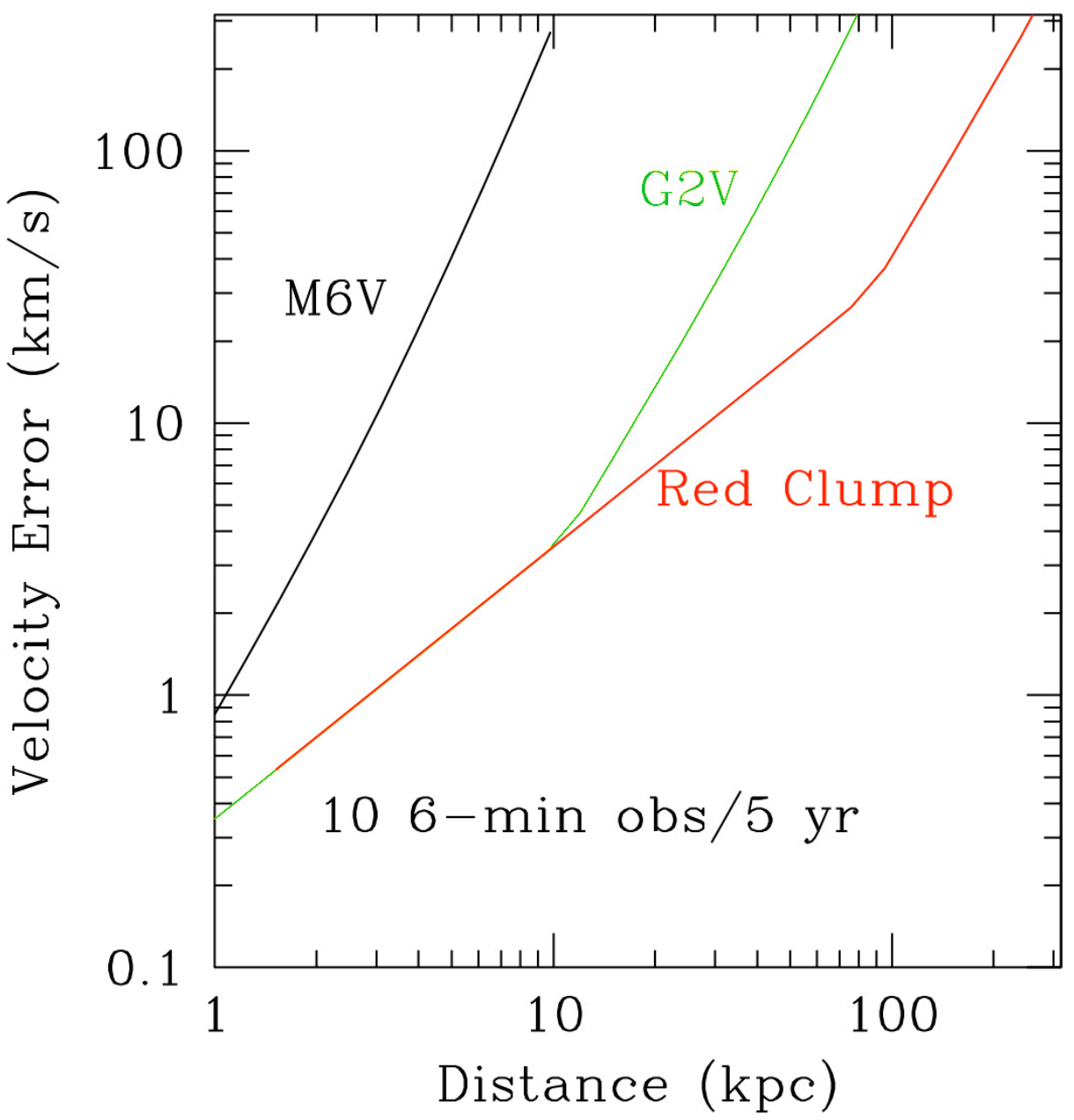

**Figure 2-15: Velocity uncertainty of different components of the Galaxy's stellar population as a function of distance from the Sun. These errors are typical of what WFIRST-2.4 can achieve over a 5 year period. As part of its high latitude survey, WFIRST-2.4 will be able to measure parallax distances to over 200 million stars and will be able to track the motions of stars much fainter than those observed by the soon-to-be launched GAIA satellite.**

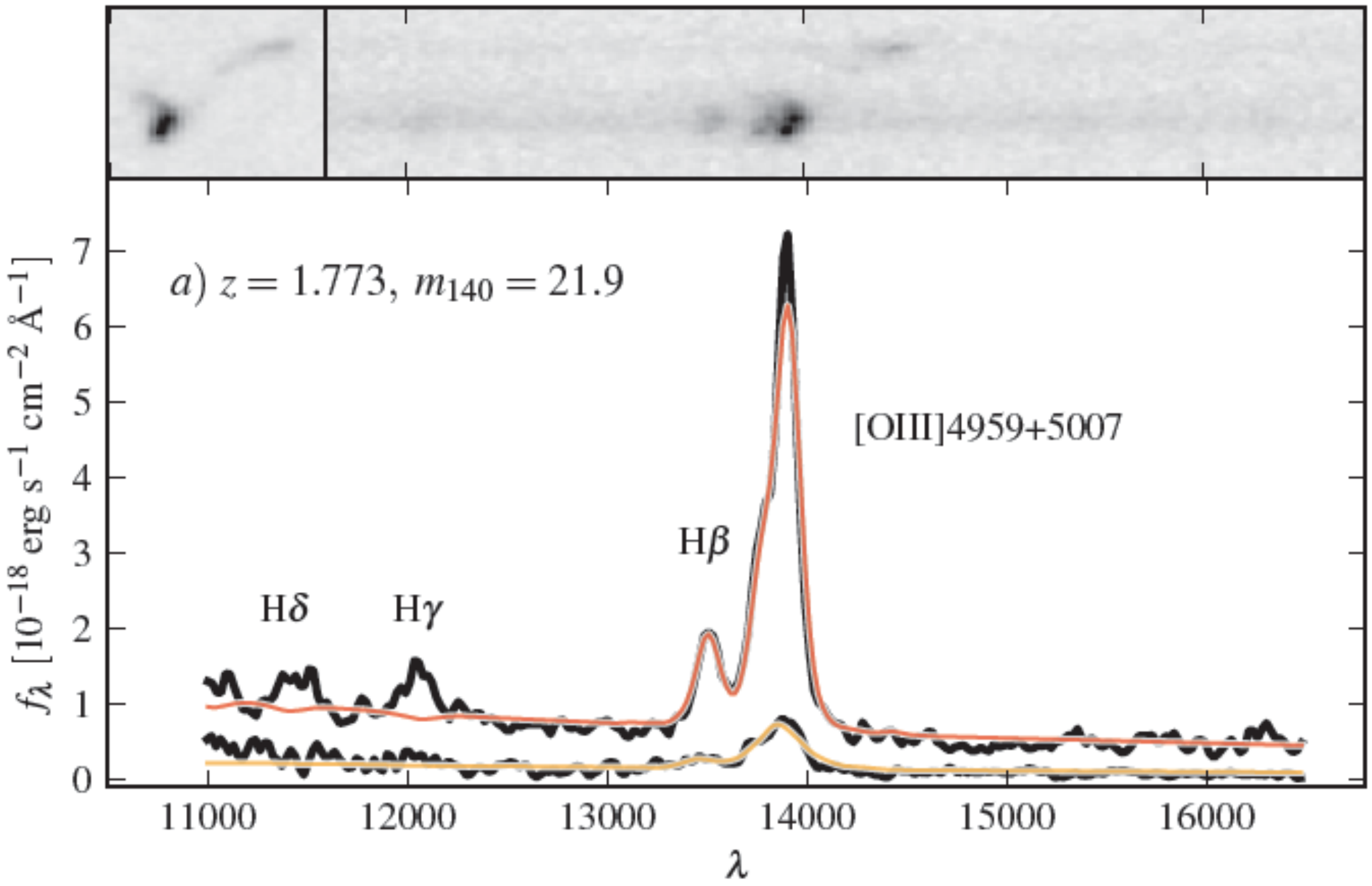

**Figure 2-16: The grism image (top) and the extracted spectrum of a z=1.77 star-forming galaxy. The WFIRST-2.4 grism will allow us to detect thousands of such galaxies in the HLS area.**

## 2.4 Guest Observer Program

The guest observer (GO) science program of WFIRST-2.4 will have high impact over a broad range of modern astrophysics. Taking Hubble as a guide, the GO program of WFIRST-2.4 will be at least as important to the astronomical community and the interested public as the dark energy and planetary microlensing components. The very wide field, large aperture, deep near-IR reach, excellent spatial resolution, and spectroscopic capabilities of the observatory will ensure that a multitude of GOs pursue numerous programs requiring degree-sized fields in the statistical realm for the first time. Examples include studies of stellar populations in the Milky Way and its neighboring galaxies and mapping large-scale structure through cosmic time. Many of these will be logical outgrowths of single object and small sample studies with HST, JWST, and other observatories, but they are likely to have outsized impacts just as SDSS, Kepler, and other wide field surveys already have.

Some capabilities of WFIRST-2.4 are unrivaled. For the foreseeable future, atmospheric seeing ensures that no ground-based facility will produce comparably sharp ultra-wide field images that compare to WFIRST-2.4. Likewise, the earth's atmosphere is opaque between the near-IR atmospheric J, H, and F184 bands, and even where it is not opaque, OH airglow ensures that the background will be large and highly time-variable compared to space. Compared to other space-based facilities, WFIRST-2.4 has 288x as many pixels as Hubble/WFC-3, and 210x JWST/NIRCam's FOV. Compared to Euclid, WFIRST-2.4 has approximately 4x the collecting area and about 2.5x the near-IR angular resolution (>6 improvement in PSF area). Euclid's dark energy surveys will certainly be useful for numerous other astrophysics investigations, but at this point Eu-clid is not planning to have a dedicated program for conducting GO pointed observations.[15]

WFIRST-2.4 will be complementary to these other observatories, and their combined data will be invaluable for attacking numerous problems in modern astrophysics. In particular, the wide field and moderately deep reach of WFIRST-2.4 will be highly complementary to the narrow field, higher resolution, and deeper reach of JWST in their overlapping near-IR wavelength regions. Likewise, WFIRST-2.4 will provide complementary wavelength coverage with much better spatial resolution and depth to LSST while still being able to sample wide areas of sky.

Compared to previous, smaller-aperture concepts of WFIRST, the significantly increased sensitivity, and higher spatial resolution over a similar field of WFIRST-2.4 will substantially increase the potential and value of its GO program. This is true for both dedicated GO observations and for guest investigator use of data sets produced for the dark energy and microlensing surveys. As for nearly all other missions, the astronomical community will be quite creative and will use WFIRST-2.4 for GO programs that the mission SDTs and SWGs will not conceive. If history is a guide, then these may prove to be among the highest impact programs executed with the telescope, rivaling or surpassing the results of the mission's primary dark energy and microlensing surveys. To illustrate this, the greater community has contributed 50 GO ideas on a multitude of science themes. These 1 page descriptions span a very broad range of topics including the solar system, exoplanets, stellar astrophysics, nearby galaxies, extragalactic astrophysics, and the complementarity / synergy of WFIRST-2.4 and other missions and surveys. These papers clearly illustrate that WFIRST-2.4 will truly be a 'Great Observatory' for the Twenty-first Century. These papers are included as an appendix to this report.

We now highlight some illustrative GO observations that would take advantage of WFIRST-2.4's strengths. These have been culled from the much more detailed GO examples presented in the earlier WFIRST SDT report[1] and the Princeton Workshop white paper[51], and many of these themes and programs are related to the community 1 page GO submissions. Because of the enhanced capability for a broad class of GO programs due to the larger aperture and the loss of capability for long wavelength (> 2.0 micron) studies, we have reallocate the galactic plane survey time to the GO program for the WFIRST-2.4 DRM.

Statistical studies of the stellar populations of the Milky Way and nearby galaxies may be one of the most obvious areas of WFIRST-2.4 GO science that will have outsized impacts. However, the higher resolution and speed of WFIRST-2.4 suggests a huge opportunity in deep studies of specific regions of the Milky Way disk. A prime example is the nature and origin of the initial mass function (IMF), particularly its extension to very high and very low masses. Surveys of the numerous low- to intermediate mass star forming regions within 1 kpc are still woefully incomplete. Ground-based near-IR surveys are either spatially complete with completely inadequate sensitivity (2MASS), or else they provide somewhat better but still inadequate sensitivity and resolution (e.g., UKIDSS now and VISTA in the future) over inadequate fields. WFIRST-2.4 surveys of the ~10 best studied young embedded clusters within 1 kpc to ~25 mag AB would reveal young brown dwarfs as well as some of the youngest Class 0 protostars in these nurseries. Combining WFIRST-2.4 and Spitzer mid-IR survey data will leverage these identifications and reveal much about the accretion properties of these stars early in their lives. Surveying more distant (D ~ 2–5 kpc) massive star forming regions will constrain high-mass star formation (numbers, luminosities) with only a single or a few deep WFIRST-2.4 fields in each near-IR broadband filter.

Closest to home, the wide field and excellent near-IR sensitivity will allow for sensitive surveys of both nearby field and open cluster brown dwarfs. A puzzling discovery from WISE surveys is a severe deficiency of brown dwarfs within the local volume, D < 8 pc: 33 brown dwarfs versus 211 normal stars is far less than typically predicted ratio of ~ 1:1.[52] Cataloging the substellar populations of nearby young clusters with WFIRST-2.4 will reveal the brown dwarf fraction with age and location, giving insight to their dispersal into the field. Broadband 1 – 2 micron photometry with WFIRST-2.4 will locate brown dwarfs in color – color and color-magnitude diagrams; these diagrams will also shed light on the mystery of why the near-IR colors and spectra of field brown dwarfs and extrasolar planets differ. Field brown dwarfs have colors and spectra consistent with a clearing of clouds as effective temperatures decrease to ~900 K (the L-T spectral type transition), but cool giant exoplanets, like those around HR 8799, seem to keep their clouds at these low temperatures, perhaps due to their lower surface gravities.[53] Surveying the brown dwarfs in a number of different young open clusters will revel the precise age-dependence of this effect. Progress in understanding the relations between brown dwarfs and giant planets will be crucial to interpreting WFIRST-2.4's own coronagraphic exoplanet observations.

Continuing out from the Galactic disk to its stellar halo, deep maps of the Milky Way halo will reveal the distribution and ages of the oldest stars in the Galaxy as well as its dwarf satellites.[54] WFIRST-2.4 will also be able to resolve individual stars in nearby galaxies and better separate stars from distant nearly point-like galaxies, providing accurate studies of their stellar populations and star formation histories for the first time. Near-IR performance is an important feature for these studies; the ability of near-IR light to penetrate dust lanes (Figure 2-17 of M31) and unique age-sensitive features in near-IR CMDs are key for these investigations. More details appear in §5.4.1 of the Dressler et al. Princeton whitepaper[51], Kalirai et al.,[55] and the Belsa & van der Marel, Geha, and van der Marel & Kalirai 1 page GO papers.

Many of the ~450 galaxies within 12 Mpc[56] have been studied with Spitzer, GALEX, and other missions / surveys, but few have been observed at high spatial resolution. WFIRST-2.4 images of these galaxies would provide color-magnitude diagrams (CMDs) that allow discerning their stellar populations and determining their ages. This would greatly leverage the existing data as well, providing a good look at the makeup of our local volume. Similarly, WFIRST-2.4 imaging data would probe the population and dynamical history of the core of the Virgo. These and related ideas appear in §5.4.2 of Dressler et al. Princeton whitepaper[51] as well as in the Dalcanton & Laine 1 page GO contribution.

The HLS will discover numerous galaxy clusters that can be followed up by WFIRST-2.4 itself with deeper fields. Likewise, medium – deep WFIRST-2.4 observations of LSST deep drilling fields will probe structure formation and lensing discoveries will test $\Lambda$CDM models. For example, WFIRST-2.4 could carry out a deep galaxy census covering 100 times the area of the famous Hubble Ultra Deep fields. Numerous AGN will be detected in these and other surveys, and these data will allow for some study of their co-evolution with their host galaxies. Coronagraphic WFIRST-2.4 observations will allow more detailed study of individual AGN host galaxy morphologies and halos. These studies are presented in more detail in §2.6.3 and 2.6.5 of the Green et al. SDT report[1], §5.5 of the Dressler et al. Princeton whitepaper[Error! Bookmark not defined.], and the Conselice, Merton & Rhodes, and Donahue 1 page GO papers.

WFIRST-2.4 data could also address the Cosmic Dawn theme of the NWNH Decadal Survey by characterizing the epoch, speed, and shape of cosmological reionization and perhaps characterizing reionization sources. This would be done with the luminosity function of high redshift quasars (mostly from the HLS), the high redshift galaxy luminosity function (Lyman-break galaxies), and Ly $\alpha$ emitters. The details of this exploration are presented in §5.6 of the Dressler et al. Princeton whitepaper[51], and related ideas appear in the Fan, Teplitz, and Whalen 1 page GO contributions.

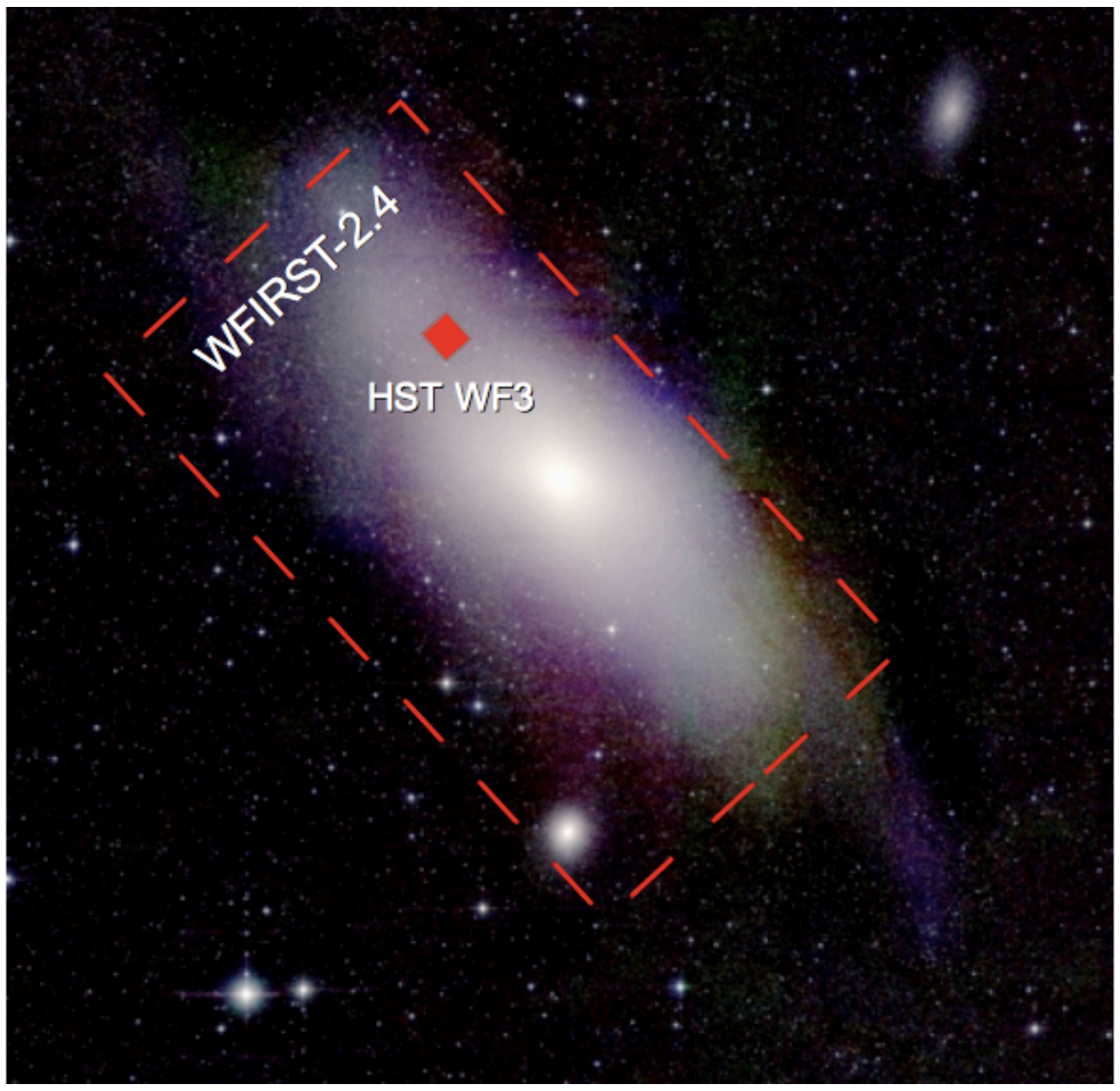

**Figure 2-17: 2MASS near-IR image of the Andromeda Galaxy M31. The dust lanes disappear at these wavelengths, revealing embedded and obscured stellar populations. The full field of the image is 1.4 x 1.4 degrees. The WFIRST-2.4 FOV is indicated by the dashed rectangle, and the field of the HST WF3 IR instrument is shown by the filled red rectangle. Both instruments have similar spatial resolutions and sensitivities, but WFIRST provides 260 times the field of view of HST WFC3/IR.**

WFIRST-2.4 will complement JWST in its quest for "first light" objects by searching for primordial stars too faint to be recognized as individuals by any telescope. The majority of the first light quasars and stars will not be seen individually, but their combined radiation may be detectable in the form of the cosmic infrared background (CIB). The source-subtracted CIB Spitzer/IRAC maps show no correlations with the visible HST/ACS sources down to m_AB>~28 and the signal has now been measured to ~ 1 degree scale.[57] The amplitude and shape of this rise is a direct measurement of the clustering properties of the sources responsible for the fluctuations, thus a primary key to understanding the nature of these sources. WFIRST-2.4 HLS data will re-veal the power spectrum of the CIB at sub-degree scales and help determine whether it is dominated by radiation from the first stars,[58] lower redshift stripped halo stars,[59] or other sources.

We list a number of these and other potential GO programs in Table 2-3. We intend this list to be a representative but by no means complete list of GO program areas that will likely take good advantage of the observatory's capabilities and have incredible scientific impact in the 2020 decade. WFIRST-2.4 will also be capable of conducting a reasonable number of these programs given that it will have ~1.5 years dedicated to its GO program.

## 2.5 Exoplanet Science with WFIRST-2.4

The first discovery of planetary companions to Sun-like stars was, along with the discovery of dark energy, one of the greatest breakthroughs in modern astronomy.[60,61,62] These discoveries have excited the astronomical community and the broader public as well. Since then, the pace of exoplanet discovery has increased each year. There are now nearly 900 confirmed exoplanets and Kepler has identified another 2700 candidates that await confirmation.[63]

Nature has surprised astronomers with the enormous and unexpected diversity of exoplanetary systems, containing planets with physical properties and orbital architectures that are radically different from our own Solar System. Since the very first discoveries, we have struggled to understand this diversity of exoplanets, and in particular how our solar system fits into this menagerie.

WFIRST-2.4 will advance our understanding of exoplanets along two complementary fronts: the statistical approach of determining the demographics of exoplanetary systems over broad regions of parameter space and the detailed approach of characterizing the properties of a handful of nearby exoplanets.

First, through its comprehensive statistical census of the outer regions of planetary systems using micro-lensing, including planets with separations spanning from the outer habitable zone to free floating planets, and analogs of all of the planets in our Solar System with the mass of Mars or greater, WFIRST-2.4 will complete the statistical census of planetary systems begun by Kepler. This science is described in §2.5.1.

| Program | Each pointing (sq deg) | # Targets | Comment |
|---|---|---|---|
| Embedded star formation < 1 kpc (Tau, Oph, Ser, Per, Ori) | 1 – 30 | ~10 | IMF: substellar to intermediate masses: calibrating models |
| 10-100 Myr open clusters (Pleiades, NGC clusters) | ~1 | ~10 | Mass Function to substellar masses, evolution, dynamics, brown dwarfs |
| Dense embedded star formation > 1 kpc (NGC 7538, W3, Gal Ctr) | ~ 0.25 (1 field) | ~15 | IMF: high mass end |
| Globular Clusters | 0.25 | 12 | Calibrate IR CMD with metallicity |
| Stellar populations of Galactic bulge, halo, satellites | HLS microlensing | HLS microlensing | Substructure, tidal streams, AGB ages from HLS and microlensing data |
| Ultra-faint MW dwarf galaxies including S/LMC | 0.25 – 4 | >10 | Star formation histories, metal-poor IMF, kinematics of LSST discoveries |
| Map local group galaxies (M31, M33, dwarf ellipticals) | ~2 - 20 | ~6 | Resolve disk structure, extinction maps, cluster dissolution to field |
| Mapping nearby galaxy thick disks and halos beyond local group | ~1 | > 24 | Halo substructure, population gradients, test dark matter, kinematics |
| Mapping the core of Virgo cluster | ~100 | 1 | Morphologies, luminosity function, superstar clusters, intracluster objects |
| AGN host galaxies (coronagraph) | 1E-6 | > 10 | Stellar populations, comparison to normal galaxies |
| Galaxy cluster (HLS) followup | 0.25 | > 24 | High z lensed objects, structure growth, interactions & dark matter |
| LSST Deep Drilling Fields | 0.25 | 1-4 | Galaxy Luminosity Function out to reionization epoch (z from dropouts) |
| QSOs as probes of cosmic dawn | HLS | HLS | Epoch, speed, patchiness of reionization using QSO spectra |

**Table 2-3: Examples of GO programs.**

Second, if WFIRST-2.4 is equipped with a coronagraph, it will be capable, for the first time in human history, of directly imaging planets similar to those in our Solar System. It will make detailed studies of the properties of giant planets and debris disks around nearby stars and will be the testbed for future coronagraphs capable of detecting signs of life in the atmospheres of Earth-like exoplanets. This option is explored in §2.5.2.

With these two complementary surveys, WFIRST-2.4 will provide the most comprehensive view of the formation, evolution, and physical properties of planetary systems. In addition, information and experience gained from both surveys will lay the foundation for, and take the first steps toward, the discovery and characterization of a "pale blue dot" — a habitable Earthlike planet orbiting a nearby star.

### 2.5.1 *Microlensing: Measuring the Demographics of Extrasolar Planets*

#### *2.5.1.1 Understanding the Origins and Evolution of Planetary Systems*

Canonical theories of planet formation and evolution originally developed to explain our Solar System[64] did not anticipate the incredible panoply of planetary systems that have been observed. They have since been expanded and altered to better describe the variety of planetary systems that we see. For example, the discovery of gas giant planets orbiting at periods of only a few days, as well as evidence for the migration of the giant planets in our own Solar System, have highlighted the fact that these theories must also account for the possibility of large-scale rearrangement of planet positions during and after the epoch of planet formation.[65,66] Many of these theories also predict a substantial population of "free-floating" planets that have been ejected from their planetary systems through interactions with other planets.[67,68]

In the most general terms, these formation theories should describe all of the relevant physical processes by which micron-sized grains grow through 13-14 orders of magnitude in size and 38-41 orders of magnitude in mass to become the terrestrial and gas-giant planets we see today. These physical processes are ultimately imprinted on the architectures of exoplanetary systems, specifically, the distributions of masses, compositions, and orbits.[69,70] Measuring these distributions, i.e., determining the demographics of large samples of exoplanets, is our best opportunity to gain insight into the physical processes that drive planet formation.

After nearly two decades of exoplanet discovery, our statistical census of exoplanets, as well as our understanding of planet formation and evolution, remains largely incomplete. This is because the currently confirmed exoplanets occupy a limited region of parameter space that is largely disjoint from both the planets in our solar system, and from the three largest reservoirs of planets that are predicted by many planet formation theories.[69,70]

Kepler has begun the process of creating a complete statistical census of exoplanetary systems by assaying the population of one of these reservoirs, namely small "hot" and "warm" planets.[71,63] Kepler has been enormously successful, discovering thousands of candidate exoplanets in a bewildering array of architectures. However, Kepler can only tell part of the story: the transit method used by Kepler is not sensitive to planets in orbits significantly larger than that of the Earth, including analogs to all of the outer giant planets in the Solar System. The crucial next step is to assay the population of planets in the cold, outer regions of planetary systems, and to determine the frequency of free-floating planets. These two populations of planets are invisible to Kepler, yet constitute the two other main reservoirs of planets predicted by theories. Fortunately, the gravitational microlensing method provides the perfect complement to the transit method, and so can be used to complete the census of exoplanetary systems begun by Kepler and so fully test these theories.

#### *2.5.1.2 Informing the Frequency and Habitability of Potentially Habitable Worlds*

Obtaining a census of planetary systems is also an important first step in the paramount goal of determining how common life is in the universe. Particularly essential is a measurement of frequency of potentially habitable worlds, commonly denoted $\eta_\oplus$. An accurate measurement of $\eta_\oplus$ provides a crucial piece of information that informs the design of direct imaging missions intended to search for biomarkers around nearby potentially habitable planet. Indeed, a primary goal of Kepler is to provide a robust measurement of $\eta_\oplus$. However, the importance of this number makes it highly desirable to confirm Kepler's estimate by a complementary method, working (in geometrical terms) from the outside in instead of the inside out. The WFIRST-2.4 microlensing survey is perfectly suited to provide this complement.

Furthermore, a measurement of $\eta_\oplus$ will only provide our first estimate of the frequency of *potential* habitable planets. The factors that determine what makes a potentially habitable planet actually habitable are not entirely understood, but it seems likely that the amount of water is a contributing factor. However, the origin of the water in habitable planets is uncertain. One possible source is delivery from beyond the snow line, the part of the protoplanetary disk beyond which it is cold enough for water ice to exist in a vacuum. As a result, the number, masses, and orbits of planets beyond the snow line likely have a dramatic effect on the water content of planets in the habitable zone.[72,73] Therefore, an understanding of the frequency of habitable planets requires the survey of planets beyond the snow line enabled by WFIRST-2.4.

#### *2.5.1.3 A Microlensing Survey with WFIRST-2.4*

A microlensing survey from space with WFIRST-2.4 is required to understand the origins, evolution, and demographics of planetary systems, improve Kepler's constraints on the frequency of potentially habitable planets, and understand the actual habitability of such planets. Together, the exoplanet surveys of Kepler and WFIRST-2.4 will yield a composite census of planets with the mass of Earth and greater on both sides of the habitable zone, overlapping at almost precisely that zone, allowing WFIRST-2.4 to *Complete the statistical census of planetary systems in the Galaxy, from the outer habitable zone to free floating planets, including analogs of all of the planets in our Solar System with the mass of Mars or greater.*

In particular, WFIRST-2.4 will be uniquely sensitive to four particularly important, broad classes of planets: (1) outer habitable zone planets, (2) cold planets, (3) free-floating planets, and (4) very low-mass planets down to the mass of Mars.

**Outer Habitable Zone Planets***:* WFIRST-2.4 will improve the robustness of our estimate of $\eta_\oplus$ by measuring the frequency of terrestrial planets just outside the habitable zone, as well as the frequency of Super-Earths in the habitable zone. With this information, WFIRST-2.4 will be able to estimate $\eta_\oplus$ with only modest extrapolation. The combination of WFIRST-2.4 and Kepler data will make it possible to robustly interpolate into the habitable zone from regions just outside of it, even if the frequency of habitable planets turns out to be small. Furthermore, Kepler's measurement of radius combined with WFIRST-2.4's measurement of mass will allow for statistical constraints on the densities and atmospheres of small planets in the habitable zone.

**Box 5**

**WFIRST-2.4 Exoplanet Discoveries**

Based on the forecast in Appendix D, WFIRST-2.4 will detect **2800** total bound exoplanets in the range of 0.1-10,000 Earth masses, including **1000** "Super-Earths" (roughly 10 times the mass of Earth), **300** Earth-mass planets, and **40** Mars-mass planets. This would enable the measurement of the mass function of cold exoplanets to better than ~10% per decade in mass for masses >0.3 $M_{Earth}$, and an estimate of the frequency of Mars-mass embryos accurate to ~15%.

In addition, WFIRST-2.4 will detect **40** free-floating Earth-mass planets, if there is one per star in the Galaxy.

These estimates are likely conservative. With better optimization of the target fields, and using preliminary data that suggest higher microlensing event rates in the target fields, WFIRST may detect up to 2 times more planets.

**Characterization of Systems Discovered by WFIRST-2.4**

Data from WFIRST-2.4 will likely allow estimates of the mass and distance to the host stars and planets to better than ~10% for the majority of the detected systems. In addition, spectroscopic follow-up with JWST will allow for the measurement of host star temperature and metallicities for a subset of the most massive hosts in the disk and bulge.

**Comparison to DRM1 Exoplanet Capabilities**

The estimated yields for WFIRST-2.4 are about 25% larger than the yields of the WFIRST-DRM design for the same total observing time. In addition, WFIRST-2.4 will allow significantly better characterization of the detected exoplanetary systems.

**Comparison to Euclid**

WFIRST will detect **5** times more planets than a 300 day survey with an extended Euclid mission.

**Cold planets*:*** Kepler's intrinsic ability to detect planets declines rapidly for planets with large separations. As a result, Kepler will be unable to find small planets with periods longer than one year. Simply put: Kepler is sensitive to "hot" and "warm" planets, but not to the "cold" planets in the outer regions of planetary systems, including analogs of all of the planets in our Solar System from Mars outward. In contrast, the exoplanet survey on WFIRST-2.4 is sensitive to planets from roughly the outer habitable zone outwards, including rocky planets with the mass of Earth up to the largest gas giant planets, and analogs of the ice giant planets in our Solar System.

**Free-Floating Planets:** WFIRST-2.4's microlensing survey can detect old, "free-floating" planets in numbers sufficient to test planet-formation theories. It will also extend the search for free-floating planets down to the mass of Earth and below, a task not possible with other techniques and not possible from the ground. This will allow it to address the question of whether ejection of planets from young systems is a phenomenon associated only with giant planet formation or also involves terrestrial planets.

**Mars-mass embryos:** WFIRST-2.4 is uniquely capable of detecting planets with mass as small as the mass of Mars or below in significant numbers. Since Mars-mass bodies are thought to be the upper limit to the rapid growth of planetary "embryos", determining the planetary mass function down to a tenth the mass of the Earth uniquely addresses a pressing problem in understanding the formation of terrestrial-type planets.

However, it is perhaps the unexpected and unpredictable returns of an exoplanet microlensing survey with WFIRST-2.4 that will prove to be the most enlightening for our understanding of the formation, evolution, and habitability of exoplanets. One of the most important lessons from the past two decades of exoplanet research is that there exists an enormous diversity of exoplanetary systems, generally far exceeding theoretical expectations. Indeed, one hallmark of the field is the fact that, whenever new regions of parameter space are explored, the subsequent discoveries necessitate revisions of our planet formation theories. Kepler is currently revolutionizing our understanding of the demographics of small, short-period planets. Because it will open up a similarly broad expanse of parameter space, and because the expected yields are similarly large, thus implying good sensitivity to rare systems, WFIRST-2.4 is essentially guaranteed to do the same. Furthermore, it will do so in a region of parameter space that is almost certainly critical for our understanding of planet formation and habitability.

In summary, WFIRST-2.4 will provide crucial empirical constraints on planetary systems that will allow us to address the three fundamental and interrelated questions: *How do planetary systems form and evolve?, What determines the habitability of Earthlike planets?, and What kinds of unexpected systems inhabit the cold, outer regions of planetary systems?* Without WFIRST-2.4's survey, we will have essentially no knowledge of the demographics of exoplanets in vast regions of parameter space that include analogs of Mars, Saturn, Neptune, and Uranus, and some basic questions of the formation, evolution, and habitability of exoplanets will remain unanswerable.

#### *2.5.1.4 Microlensing from Space is Required*

The microlensing technique for discovering exoplanets, illustrated in Figure 2-18, is well-developed and mature. Its capabilities have been amply demonstrated from the ground through several discoveries that have provided important new insights into the nature and diversity of planetary systems. These include the discovery of a substantial population of cold "Super-Earths",[74,75,76] the first discovery of a planetary system with an analog to our own Jupiter and Saturn,[77,78] and the detection of a new population of Jupiter-mass planets loosely bound or unbound to any host star.[79] Importantly, ground-based microlensing surveys have determined that cold and free-floating giant planets are ubiquitous: on average, every star in the Galaxy hosts a cold planet[80] and "free-floating" giant planets may outnumber the stars in our galaxy by two to one.

While exoplanet microlensing from the ground has had well-documented successes, realizing the true potential of the microlensing method, and achieving the primary science goals outlined above, is only possible from space with an instrument like WFIRST-2.4. Microlensing requires monitoring a very large number of stars in very crowded, heavily reddened fields toward the Galactic bulge continuously for months at a time. Therefore, reaching the needed sensitivity and number of planet detections requires steady viewing in the near-infrared without interruptions for weeks at a time, as well as high angular resolution, stable images over very wide-area fields. In other words, it requires a wide-field infrared survey telescope in space. Without such a mission, we will not be sensitive to Mars-mass embryos, we will not determine the frequency of low-mass free-floating planets, and we will not complete the census of planets begun by Kepler.

#### 2.5.1.5 *Expected Exoplanet Yields*

Using the methodology detailed in Appendix D, we find that ***WFIRST-2.4 will detect roughly 3000 bound planets, comparable to the expected final yield of Kepler.*** Figure 2-19 illustrates two example simulated microlensing events, including the detection of a bound Mercury-mass planet, as well as the detection of a free-floating Mars-mass planet. Table 2-4 shows our estimates for the total yield of bound planets, whereas Table 2-5 shows our estimates for the yield of free-floating Earth-mass planets. We estimate that WFIRST-2.4 will detect of order 2800 bound planets with masses in the range of 0.1-10,000 $M_\oplus$ and semi-major axes in the range of 0.3-30 AU. As illustrated in Figure 2-20, while massive planets can be detected over nearly this entire range of separations, lower-mass planets can only be detected over a narrower range of separations beyond the snow line. At least 10% of the detected bound plan-

**Time-series photometry is combined to uncover light curves of background source stars being lensed by foreground stars in the disk and bulge.**

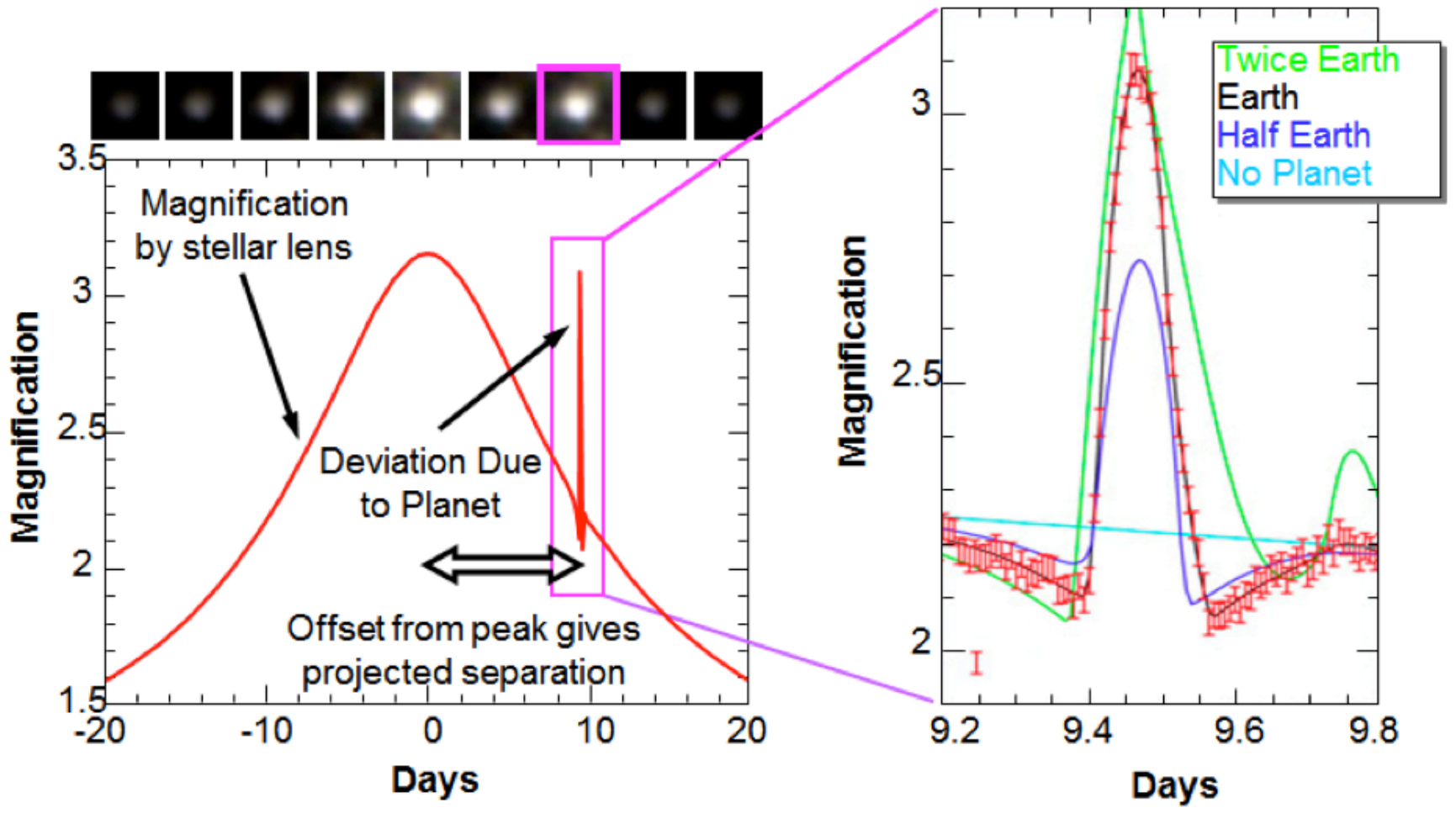

**Planets are revealed as short-duration deviations from the smooth, symmetric magnification of the source due to the primary star.**

**Detailed fitting to the photometry yields the parameters of the detected planets.**

**Figure 2-18: Schematic illustration of how WFIRST-2.4 discovers signals caused by planetary companions in primary microlensing events, and how planet parameters can be extracted from these signals. The left panel shows a simulated primary microlensing event, containing a planetary deviation from an Earth-mass companion to the primary lens. The offset of the deviation from the peak of the primary event, when combined with the primary event parameters, is related to the projected separation of the planet. The right panel shows an enlargement of the planetary perturbation. The width and precise shape of the planetary deviation yield the mass of the companion relative to that of the primary host lens.**

| $M/M_\oplus$ | Euclid-Extended (300 days) | DRM1 (432 days) | DRM2 (266 days) | WFIRST-2.4 (432 days) |
|---|---|---|---|---|
| 0.1 | 10 | 30 | 21 | 39 |
| 1 | 66 | 239 | 176 | 301 |
| 10 | 197 | 794 | 599 | 995 |
| 100 | 144 | 630 | 484 | 791 |
| 1000 | 88 | 367 | 272 | 460 |
| 10,000 | 41 | 160 | 121 | 201 |
| Total | 546 | 2221 | 1676 | 2787 |

**Table 2-4: Predicted "best estimate" yields for bound planets for various mission designs. The yields have been normalized using planet distribution function for cold exoplanets as measured from ground-based microlensing surveys by Cassan et al.[80] and using event rates extrapolated from observed rates for bright clump giant sources.**

| Euclid-Extended (300 days) | DRM1 (432 days) | DRM2 (266 days) | WFIRST-2.4 (432 days) |
|---|---|---|---|
| 5 | 33 | 27 | 41 |

**Table 2-5: Predicted yields for free-floating Earth-mass planets, assuming one free-floating Earth-mass planet per star in the Galaxy.**

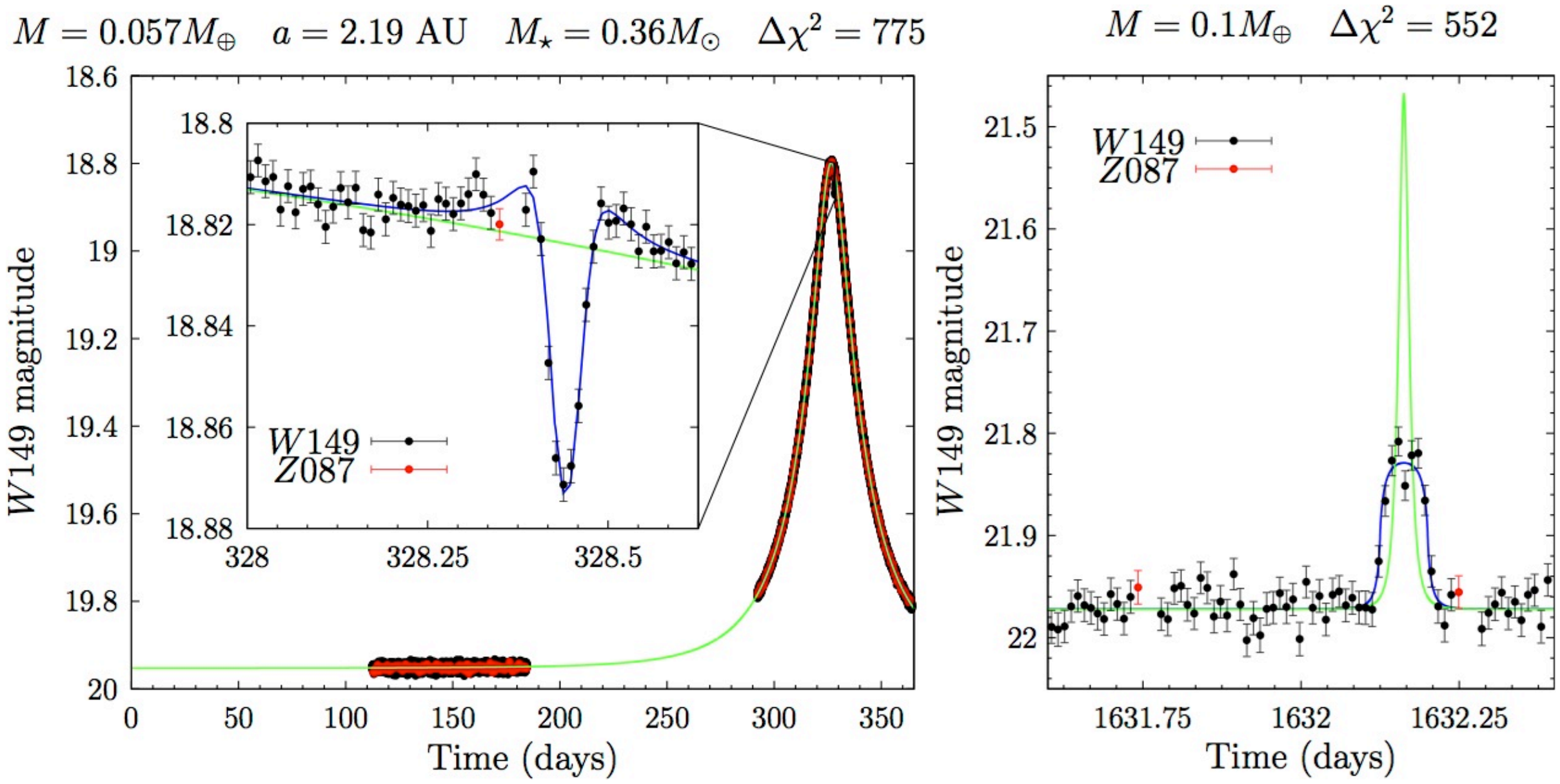

**Figure 2-19: Examples of simulated event light curves with detected planetary signals from simulations of a WFIRST-2.4 exoplanet survey. The left panel shows the detection of a Mercury-mass planet orbiting a 0.36 solar mass star with a semi-major axis of 2.19 AU. The right panel shows a simulated detection of a free-floating Mars-mass planet.**

ets will have mass less than three times the mass of the Earth, and WFIRST-2.4 will have significant sensitivity down to planets of roughly twice the mass of the Moon. WFIRST-2.4 will measure the mass function of cold planets to ~3%-15% in 1-dex bins of planet mass down to the mass of Mars. If free-floating planets are common, WFIRST-2.4 will detect a large number of them with masses down to that of Mars.

Table 2-4 and Table 2-5 also compare the WFIRST-2.4 yields to other WFIRST designs, as well as a 300-day microlensing mission with Euclid. WFIRST-2.4 will be substantially more capable than any of these missions, resulting in overall yields that are larger by factors of ~1.25 to 5. The WFIRST-2.4 yields for the smallest planets will be larger by factors of 1.3 to 4.

The primary microlensing survey will focus on fields located toward the Galactic center; however, a unique capability of microlensing is its sensitivity to very distant planetary systems, including planetary systems in other galaxies. Relatively modest surveys of serveral weeks towards M87 or the bulge of M31 would have significant sensitivity to extra-galactic Jupiter-mass planets[81], and be able to probe the frequency of planets in environments with stellar populations and star formation histories that differ substantially from those of the Milky Way disk and bulge.

#### *2.5.1.6 Physical Parameters of the Detected Systems*

In addition to a large yield of planets over a broad region of parameter space, the other dramatic improvement of a space-based microlensing mission over ground-based surveys is the ability to automatically estimate the masses of the majority of the planetary host stars and thus planetary companions. This qualitatively new capability is enabled by the small and stable point-spread functions (PSFs) afforded in space, which allow for many different types of measurements that are difficult or impossible from the ground.

- **Measurement of the host starlight:** PSFs of angular size <0.4" essentially resolve out most of the background stars, allowing one to isolate the time-variable flux of the microlensed source from any flux from the lens, or companions to the lens or source. Light from companions to the lens and source can be further distinguished from light from the lens alone based on observations taken over a very wide time baseline before or after the event, which will allow one to infer the relative lens-source proper motion when the source and lens are significantly displaced. This can be done either by measuring the elongation of the PSF, or by measuring the differential centroid shift of the PSF in two different passbands if they have substantially different colors.

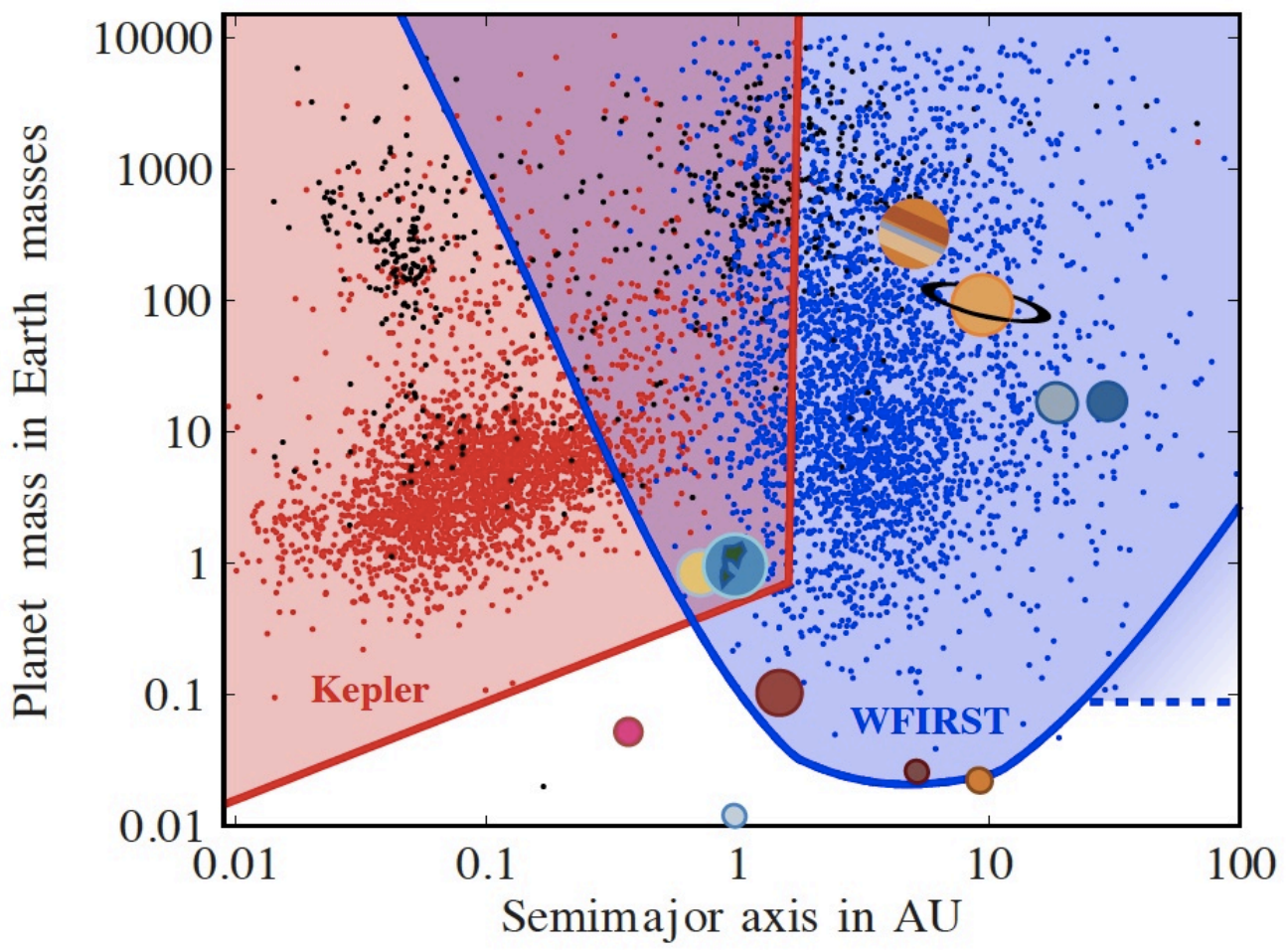

**Figure 2-20: The colored shaded regions show approximate regions of sensitivity for Kepler (red) and WFIRST-2.4 (blue). The solar system planets are also shown, as well as the Moon, Ganymede, and Titan. Kepler is sensitive to the abundant, hot and warm terrestrial planets with separations less than about 1.5 AU. On the other hand, WFIRST-2.4 is sensitive to Earth-mass planets with separations greater than 1 AU, as well as planets down to roughly twice the mass of the moon at slightly larger separations. WFIRST-2.4 is also sensitive to unbound planets with masses as low as Mars. The small red points show candidate planets from Kepler, whereas the small blue points show simulated detections by WFIRST-2.4; the number of such discoveries will be large with roughly 2800 bound and hundreds of free-floating planet discoveries.** ***Thus, WFIRST-2.4 and Kepler complement each other, and together they cover the entire planet discovery space in mass and orbital separation, providing the comprehensive understanding of exoplanet demographics necessary to fully understand the formation and evolution of planetary systems. Furthermore, the large area of WFIRST-2.4 discovery space combined with the large number of detections essentially guarantees a number of unexpected and surprising discoveries.***

- **Measurement of Higher Order Effects in the Micolens Lightcurves**: The small and stable PSFs delivered by WFIRST-2.4 will result in precise and accurate photometry during the microlensing events, with substantially higher precision and smaller systematics than is possible from the ground. This will enable the measurement of subtle but information-rich "parallax" distortions in the microlens light curves due to the orbit of WFIRST-2.4 about the Sun[82] and in the baseline case of a geosynchronous orbit, the orbit of WFIRST-2.4 about the Earth.[83] When combined with information routinely obtained in planetary microlens lightcurves, measurement of these effects will allow one to measure the mass and distance to the host for a subset of detections. Because these effects are generally larger for less massive hosts, this subset is complementary to the subset for which measurement of the host star mass via lens light is possible. Subtle distortions in the microlens light curves due to the orbital motion of the planet about the host will also enable constraints on the period, inclination, and eccentricity of the orbit in a subset of cases.
- **Astrometric Shifts due to Microlens Image Motions:** During a microlensing event, multiple images of the source are created whose positions and brightnesses vary as a function of time. While these images are typically separated by of order the Einstein ring radius $\theta_E$~1 mas and thus cannot be resolved, the changing fluxes and positions lead to a time-varying centroid shift of the unresolved source image, whose magnitude is proportional to $\theta_E$.[84,85] Astrometric precisions of a few 100 $\mu$as will allow the measurement of $\theta_E$ for more massive and/or nearby host stars. When combined with the measurement of the host light, this will allow for a measurement of the distance and mass of the host. When combined with microlensing parallax measurements, this effect will also allow for the detection and mass measurement of a significant sample of isolated neutron stars and black holes, thereby allowing for a determination of the mass function of isolated high-mass remnants, providing crucial constraints on supernova explosion mechanisms.[86] These measurements cannot be obtained in any other way.

A quantitative assessment of the potential of WFIRST-2.4 to measure the above effects and thus infer host star masses and distances requires detailed simulations, which have not yet been performed. However, using conservative assumptions, Bennett et al.

2007 demonstrated that a significantly less capable space-based microlensing mission would enable measurement of the host star and planet masses to an accuracy of ~10% for the majority of the detected systems. We expect WFIRST-2.4 to measure the masses to a similar accuracy for an even larger fraction of systems.

Finally, it is worth noting that there is likely to be substantial synergy between WFIRST-2.4 and JWST. For example, preliminary estimates indicate that it should be possible to estimate temperatures and metallicities for hosts brighter $K_{AB}$~20 using R~2700 spectra taken with NIRSpec. This would allow for the characterization of the turn-off hosts in the bulge, and well as even lower-mass hosts in the foreground disk. In addition, simultaneous observations of planetary signals from WFIRST-2.4 and JWST may allow for a measurement of microlens parallax, potentially enabling an estimate of the mass and distance to the host star.[87]

#### *2.5.1.7 Detection of Giant Exoplanets via Transits*

The WFIRST-2.4 microlensing Galactic bulge survey will also detect several thousand exoplanets via transits, with separations of less than a few tenths of an AU, and radii larger than Saturn.[88,89] Contamination by false positives is a significant concern, but the rate of this contamination can be quantified statistically using population synthesis analyses, and verified through follow-up observations in favorable cases.[90] Since the region of sensitivity to transits overlaps significantly with that of the microlensing survey, comparison of the results from the two techniques allows for a further check on the planet candidates, and ultimately the determination of the frequency of giant planets over all orbital separations (from 0 to infinity) from a single survey.

### 2.5.2 *Characterizing Nearby Worlds with WFIRST-2.4 with an Optional Coronagraph*

Our understanding of the internal structure, atmospheres, and evolution of planets was original developed through models that were tuned to explain the detailed properties of the planets in our own solar system. Our surveys of exoplanetary systems have led to the realization that there exists a diversity of worlds with very different properties and environments than those in our solar system, including gas giants under strong stellar irradiation, gas giants with massive heavy cores, water worlds, and "super-Earths" with masses intermediate to the rocky and the ice giants in our solar system. Subsequently, these models have had to be expanded and generalized to explain the properties of these new worlds, often including new and uncertain physics. Our understanding of these new worlds therefore remains primitive.

The best hope of understanding the physical properties of this diversity of worlds is through comparative planetology: detailed measurements of, and comparisons among, the properties of individual planets and their atmospheres. These measurements provide the primary empirical constraints on our models.

Understanding the structure, atmospheres, and evolution of a diverse set of exoplanets is also an important step in the larger goal of assessing the habitability of Earthlike planets discovered in the habitable zones of nearby stars. It is unlikely that any such planets will have exactly the same size, mass, or atmosphere as our own Earth. A large sample of characterized systems with a range of properties will be necessary to understand which properties permit habitability and to properly interpret these discoveries.

Detailed characterization is currently only possible for relatively rare transiting systems. Unfortunately, transiting planets are a relatively small subset of systems. Those that are bright enough for significant follow-up tend to be discovered from the ground. Ground-based transit discoveries tend to be a biased subset of systems, mostly including giant planets at short orbital periods that are subject to strong stellar irradiation. Of course, Kepler has detected many smaller and longer period transiting planets, but nearly all of these systems are far too faint for detailed characterization of the planets. Furthermore, atmospheric studies of transiting planets are only sensitive to very specific planet geometries and/or atmospheric pressures, therefore providing an incomplete view of the physical processes at work in the atmospheres of these systems.

Direct imaging surveys of planets orbiting nearby stellar systems offer a complementary and critical approach to studying the detailed properties of exoplanets. First, planets detected by direct imaging tend to be at longer orbital periods than those found by transits. Second, spectra of directly imaged planets provide powerful diagnostic information about the structure and composition of the atmospheres. Finally, these planets can be found around the closest stars, which tend to be the best characterized. While ground-based direct imaging surveys have made enormous strides in recent years, with new, much more capable surveys coming on line soon, such surveys are ultimately limited by the contrast achieved in the kinds of planets they can study, being sensitive only to warm/massive young planets. WFIRST-2.4 will detect analogs to our cold Jupiter for stars in the solar neighborhood.

Establishing a diverse sample of characterized systems, and thus advancing our knowledge of exoplanet composition, requires a high-contrast, coronagraphic space-based survey of the nearest stars. WFIRST-2.4 is the first step in doing that. The large aperture, excellent wavefront quality, high degree of stability, and available on-axis region of the image make the telescope an excellent platform for direct imaging of exoplanetary systems and debris disks and an ideal testbed for critical technology needed in future missions.

A high-contrast imaging survey of over a hundred bright nearby stars using a coronagraph on WFIRST-2.4 will discover and characterize a significant number of exoplanets from Neptune-like ice giants to large Jupiter size planets. Spectra will be obtained from more than a dozen of the currently known radial velocity planets, a number likely to grow as new RV planets are discovered. Combined with demographics from Kepler and microlensing, astrometric information from RV and Gaia, and size from a small number of transits, WFIRST-2.4 with a coronagraph will provide the most comprehensive comparative planetology yet obtained for planets outside the solar system. This new information will not only revolutionize exoplanet science, but it will also serve as an essential first practical demonstration of high-contrast imaging technology. This lays the groundwork for a mission that will achieve the ultimate goal of directly detecting a "pale blue dot" around a nearby star and searching for signatures of life on that planet. **The attendees at the January 2013 EXOPAG, representing a significant fraction, and broad cross-section, of the exoplanet community, unanimously supported the addition of a coronagraph to WFIRST-2.4, even at the possible cost of slowing progress in other technology development for exoplanet programs.**

In the remainder of this section we describe the science that can be expected from a coronagraph even with a conservative estimate of performance. The key direct imaging targets fall into two categories: planets (both rocky and giant), and disks (zodiacal disk and Kuiper belt analogs). We will show that with reasonable assumptions on coronagraph parameters, the advances in our knowledge of exoplanets atmospheres and disk structure as well as the increase in our understanding of high-contrast imaging technology would be great. The importance and value of such an instrument as a technology advancement cannot be overstated. The number one priority at medium scale for NASA in the decadal survey was technology advancement of high-contrast imaging in space. A coronagraph on WFIRST-2.4 directly responds to this recommendation, providing an unparalleled opportunity for space-flight experience with a very high-contrast coronagraph and wavefront control. Even should the combination of the complex pupil geometry, uncertain thermal variations, and geosynchronous orbit result in lower than expected contrast, significant disk science would still be achieved along with characterization of a small number of radial velocity planets while dramatically improving our understanding of how to use a coronagraph in space. On the other hand, should performance of the telescope meet or exceed requirements and should more advanced coronagraph designs prove feasible, then the scientific accomplishments would be profound, including, potentially, the first images of rocky planets.

#### *2.5.2.1 Coronagraphy for High-Contrast*

Exoplanets are orders of magnitude fainter than their parent stars. Imaging them thus requires a system to create *high contrast*, by which we mean the suppression of the diffraction pattern of the star, allowing the faint planet to be detected. A coronagraph is a set of optical elements that modifies the point spread function of the telescope to create a region around the stellar image where a dim companion can be extracted. Due to optical errors in any system, all coronagraphs must be designed together with wavefront control via one or more deformable mirrors (DMs). The combined coronagraph and wavefront control system is characterized by the contrast, inner working angle, and stability achieved. ***Contrast*** is the degree to which the instrument can suppress scattered and diffracted starlight in order to reveal a faint companion. ***Inner Working Angle*** (IWA) is the smallest angle on the sky at which it can reach its designed contrast. This angle is typically only a few times larger than the theoretical diffraction limit of the telescope. The resulting residual stellar halo must also be ***stable*** over the time scale of an observation, so that the halo can be subtracted to reveal an exoplanet or disk. In principle, with a well characterized and stable PSF, various subtraction methods that have been developed and used on both ground and space images can be employed to average the background photon noise and extract faint planets that are below the raw contrast level. The limit of an observable planet is then determined only by the photon noise and the available integration time. In practice, there is a systematic limit to the contrast of a recoverable planet given by the stability of the telescope (how rapidly the background speckle pattern changes) and the structure of the exozodiacal light (a confusion limit set by clumping in a disk).

Much of the current research in coronagraphy and wavefront control is directed at extending the limits of contrast, IWA, and detection limit. The capability of the coronagraph instrument on a future WFIRST-2.4 observatory will also be partially determined by the orbital environment and resulting stability of the telescope. Additionally, the recent history of planet imaging shows that recovering planets with up to factors of 10 fainter contrast than the background is regularly accomplished, both on large ground telescopes and from the Hubble Space Telescope[91,92,93]. In some cases, factors of 100 have been achieved. We thus characterize the coronagraph instrument by its detection limit, that is, the limiting magnitude of a recoverable planet relative to the star from the combination of the coronagraph and wavefront control and data processing.

In §3.4 of this report we describe an overall instrument architecture compatible with several specific types of coronagraphs, all employing wavefront control with one or two DMs. Each differs in the details of the implementation and range from conservative assumptions about the level of contrast and IWA achievable to aggressive approaches at small IWA and very high contrast. In the remainder of this section, we describe the science that can be achieved starting from a minimum contrast of 10 parts per billion (ppb) and an inner working angle of 0.2 arcsec down to our baseline detection contrast of 1 ppb. Ongoing technology developments may enable a more aggressive coronagraph than considered in this document, offering a combination of improved contrast and reduced inner working angle, and therefore potentially enabling direct imaging of rocky planets in habitable zones of nearby stars, provided that such planets exist around the few nearest stars. Recent work shows that conceptual designs exist for this aggressive option, even for WFIRST-2.4's centrally obscured aperture. Further study and laboratory demonstrations are however needed to validate such concepts for WFIRST-2.4, due to their lower technological maturity and their increased sensitivity to wavefront instabilities, chromaticity and stellar angular size.

#### *2.5.2.2 Exoplanet Discovery and Characterization*

Only a very small number of exoplanets have been imaged to date from the ground. We have obtained spectra on an even smaller number. All of them are at great distances from their parent star and are very large. Also, the limitations in contrast of a ground-based instrument means that all these planets are also extremely young – less than a hundred million years – and shining through their residual interior heat, rather than reflecting starlight. WFIRST-2.4 provides the first opportunity to observe and characterize planets like those in our solar system lying from 3 to 10 AU from their parent star. The coronagraph on WFIRST-2.4 operating at a contrast of 1 ppb will detect at least a dozen new planets in this range and characterize over a dozen of the known RV planets. Figure 2-21 is a scatter plot showing model planets of all types distributed around the brightest 200 or so stars within 30 pc along with the contrast and inner working angle limits of increasingly aggressive coronagraphs. Planets are distributed consistent with the radius distribution found by Kepler and extrapolated to larger semi-major axis and lower mass. There are roughly 4 planets per star with a mix of hydrogen-envelope gas giants, rocky planets at 1 to 1.4 Earth radii, and icy Neptune and Saturn like planets. The plot shows a snapshot of the random selection of planets in their orbit as viewed from Earth. At the 1 ppb contrast level and a 0.2 arcsec IWA, a substantial number of new gas and ice giants are accessible, thus increasing the number of known giant planets. Should further analysis and experiment show that $10^{-10}$ detection contrast or smaller IWA is achievable, then the number of possible planets to be discovered increases significantly, including a small number of water planets and Super-Earths.

Figure 2-21 also shows known planets already discovered by radial velocity (RV). We know that each of these RV planets exists, and we know the orbital size and approximate mass for each; we can thus estimate the expected contrast and angular separation from the parent star at the brightest position in their orbits, with estimates of their contrast based on albedo assumptions. From the figure, it is clear that a handful are brighter than a threshold contrast of $10^{-8}$, and more than a dozen are brighter than the expected contrast of $10^{-9}$, so there will be plenty of these targets available for study. In addition, the completeness of the radial velocity survey is expected to increase over the next decade as more planets at larger separations are confirmed, potentially increasing significantly the known planets for characterization reachable by WFIRST-2.4.

The science return from direct imaging of these RV systems will be large. For these planets themselves, we will learn how bright they are, which gives us an estimate of their reflectivity (albedo), and this in turn will be compared to planets that we do understand well, like Jupiter, a first step in comparative planetology. The brightness also depends on the diameter of the planet, but most of these have masses that are roughly the same as Jupiter, and at that mass level a planet tends to have a fixed diameter (at least theoretically); however, this is an area to be explored with real data from WFIRST-2.4. Direct imaging will allow the inclination angle of the orbit to be measured, presently not known from RV. Brightness (contrast) changes before and after the planet is at maximum elongation will tell us about clouds and gas in the atmosphere. If the coronagraph is equipped with polarizing filters, additional information on clouds and gas in the atmosphere becomes available. If we measure brightness at different wavelengths, say green and red, we will know the color of the planet, and this is directly tied to the absorbing gases in the atmosphere; for example, Jupiter and the other outer planets are brighter in the green than in the red because methane is a strong absorber of red light. Here too, polarization filters will play a role.

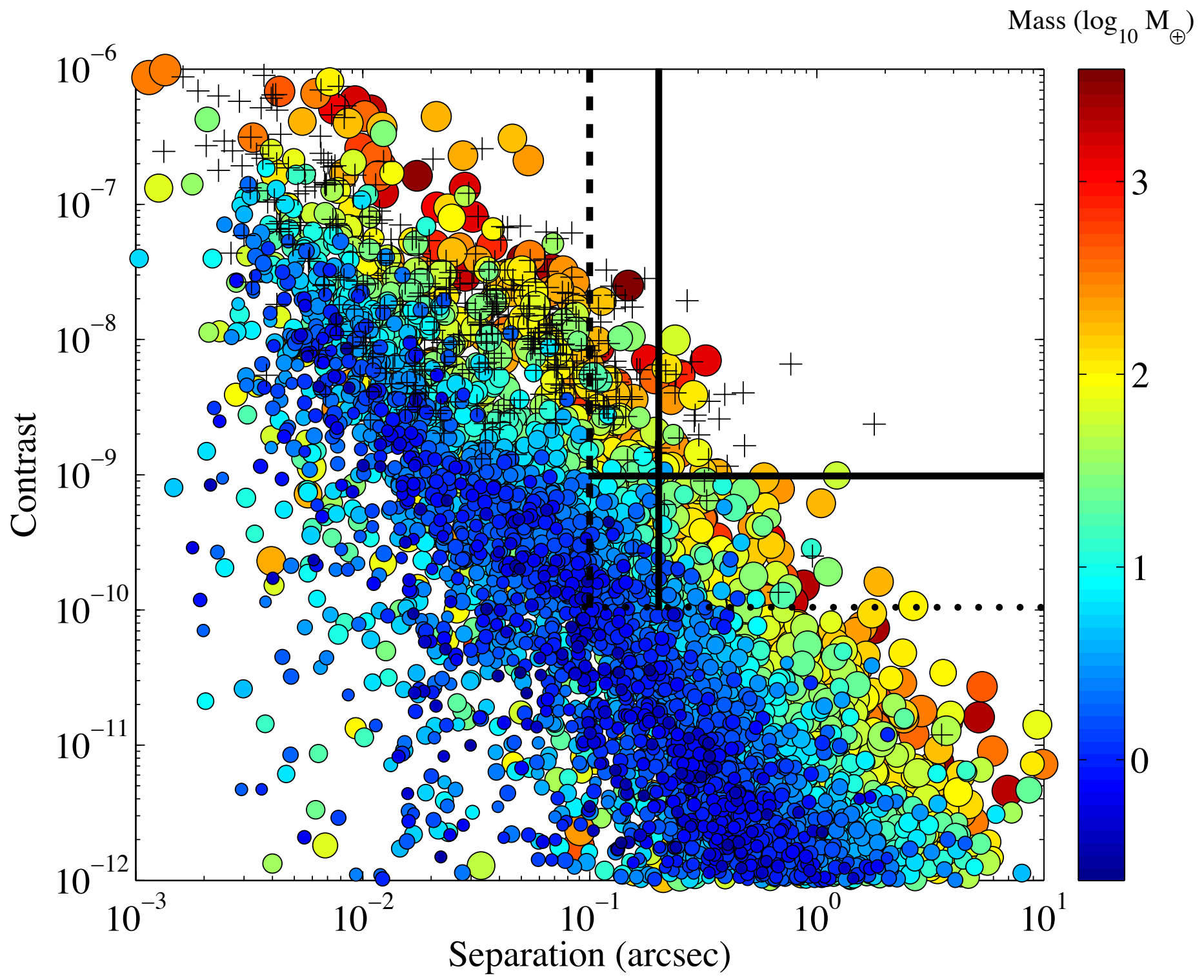

**Figure 2-21: This figure is a snapshot in time of contrast and separation for model planets, ranging in size from Mars-like to several times the radius of Jupiter, for about 200 of the nearest stars within 30 pc. Color indicates planet mass while size indicates planet radius. Crosses represent known radial velocity planets. Solid black lines mark the baseline technical goal of 1 ppb contrast and 0.2 arcsec IWA, while the dotted lines show the more aggressive goals of 0.1 ppb and 0.1 arcsec IWA.**

Experience from Kepler says that multi-planet systems are common, so we can expect to find other planets in these RV systems that are beyond the reach of the RV technique, for example with longer periods, which means potentially at larger angular separations. The WFIRST-2.4 search will be a major element in imaging campaigns for these targets.

Direct imaging of exoplanets provides a new and unique dimension to the pool of scientific information about exoplanets. Simple photometry can provide clues to planet composition, size and type. Spectra can distinguish clearly among planet types, provide an independent measure of planetary mass through gravity-sensitive spectral features, and potentially measure the metallicity of the planetary atmosphere. In our solar system, the giant planets are significantly enhanced in heavy elements compared to the sun or protosolar nebular. This metallicity is believed to be a signature of the process that formed them, gradually building up planetary cores from solid materials before accreting the outer gassy envelope. This same mechanism, in a different part of the solar system, produced rocky planets like Earth. If it operates the same way in most other solar systems, it would significantly strengthen the case that the known small planets include a large number of rocky examples. Measuring the metallicity of Doppler planets would be a powerful discriminator between different formation models.[94] For example, Figure 2-22 shows simulated spectra of two Jupiter-sized planets at 2 AU and a Neptune-sized planet with varying metallicity.[95] Atmospheric spectra of smaller planets will begin to reveal the diversity of exoplanet types and approach the question of trace biomarkers such as oxygen and methane.

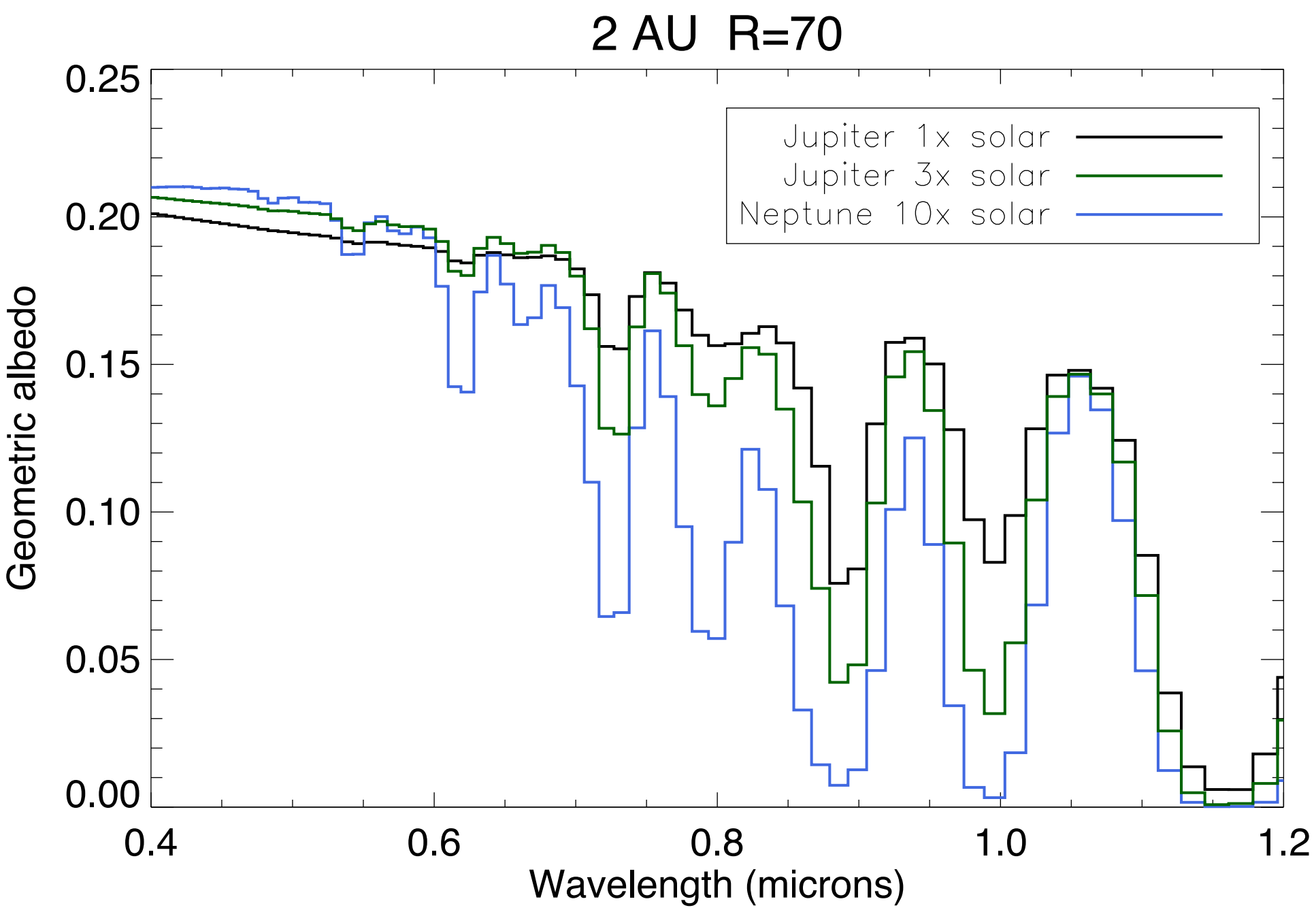

**Figure 2-22: Model exoplanet spectra (Cahoy et al 2010) for a Jupiter-mass planet with stellar metallicity (1x solar) and one enhanced in heavy elements by formation (3x), and a Neptune-like planet (10x). Spectra have been binned to the resolution of the WFIRST-2.4 coronagraph spectrometer, $\lambda/\Delta\lambda$ = 70. The three classes of planets are easily distinguishable.**

If equipped with a polarimeter, the WFIRST-2.4 coronagraph will provide an additional dimension of information. Since the reflected radiation that strikes planetary surfaces and atmospheres becomes partially polarized while the light from the central star is not, polarization measurements can both aid in detection and provide new science. Exoplanets are expected to show polarization signatures.[96] Several theoretical models have been developed to describe the observed polarization as a function of wavelength. The degree of polarization changes with wavelength across the UV, visible and near IR band-passes to reveal the structure of the exoplanet's atmosphere[97], climate[98], the nature of its surface[99], its orbital parameters[100,101] (inclination, position angle of the ascending node and eccentricity) and, possibly bio-signatures[102]. Analysis has shown that Jupiter-like gas giants may exhibit a degree of polarization as high as 60% at a planetary phase angle of 90-degrees. The flux and degree of polarization of starlight as a function of wavelength as reflected by three Jupiter-like exoplanets for phase angle 90 degrees is shown in Figure 2-23.[103]

#### 2.5.2.3 Exozodiacal Light and Disk Architecture

A star and its planetary system both form from the same circumstellar disk. As a cloud of gas and dust collapses, the embryonic system evolves a preferential axis of rotation. Material undergoes gravitational collapse along that axis but cannot collapse freely in the perpendicular direction; conservation of angular momentum prevents matter from falling inward. However eventually friction and energy loss slows the orbital motion and allows gravitation to take over, with matter collapsing inward to form a star, and locally in orbit, to form planets.

Sun-like stars accumulate their mass within a million years, and planets typically form within 10 million years. Small rocky planetesimals come together to become full-size planets. The smaller ones can become terrestrial planets, while the larger ones continue to gravitationally attract gas and become gas or ice giant planets.

Young solar systems are violent places, with many planetesimals, proto-planets, and planets crammed into a relatively tight space. Orbits are initially eccentric and these bodies interact dynamically with each other and their central stars, causing frequent collisions. These collisions produce debris, as evidenced by the material in the asteroid belt and Kuiper belt of our own solar system. We also see this debris in observations of nearby young stars; many stars in the solar neighborhood were observed by IRAS and Spitzer to have far-infrared excesses fluxes, indicative of reprocessing stellar radiation by small dust grains produced by collisions. Some of these debris disks have been resolved in reflected visible light with HST. WFIRST-2.4 will find many more such debris disks, and its data will provide significant clues to the natures of the planetary systems that produced them. In addition, polarization measurements, such as those of the AU Microscopii debris disk[104], shows the signature of primordial grain growth providing important clues to formation and evolution of exoplanetary systems. Some of the questions that may be answered include:

1. What is the amount and location about the star of circumstellar dust? The amount of debris seen in images will reveal the level of recent planetary activity. WFIRST-2.4 will be sensitive to several times the solar system's level of dust in the habitable zones and asteroid belts of nearby (~10 pc) sun-like stars.
2. What large-scale structures are present in disks? The high sensitivity and spatial resolution (0.05 arcsec is 0.5 AU at 10 pc) of WFIRST-2.4 images will reveal the presence of asteroid belts or gaps in disks due to unseen planets.
3. What are the sizes and compositions of the debris dust grains? Photometry at 2 well-separated wavelengths (e.g., 400 and 800 nm) provides constraints on dust grain size and composition, while polarimetry provides additional size information.

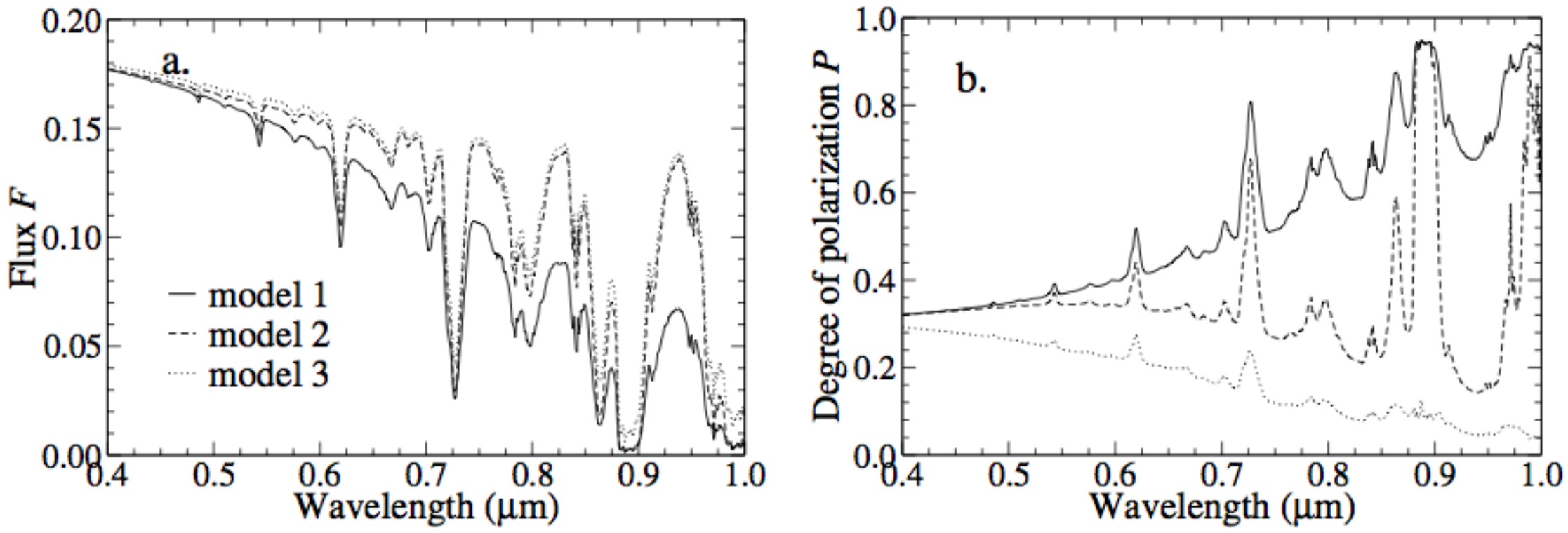

**Figure 2-23: Flux and the degree of polarization of starlight reflected by three Jupiter-like exoplanets for $\alpha$ = 90°. The planetary model atmosphere 1 (solid lines) contains only molecules, model 2 (dashed lines) is similar to model 1 except for a tropospheric cloud layer and model 3 (dotted) is similar to model 2, except for a stratospheric haze layer.**

WFIRST-2.4 will be the most sensitive observatory yet for studying warm dust near the habitable zones of nearby stars, providing several times the sensitivity and the resolution of the LBTI instrument. In our solar system, the zodiacal light is sunlight reflected from the zodiacal dust disk in the inner solar system, believed to come from asteroid collisions in the asteroid belt. In the outer solar system, the Kuiper belt is a zone of ice particles from collisions among comets. Each of these disks is continually replenished by collisions, and the location of the disks is controlled to a large extent by the planets. So our two disks are a sort of flashing red light around the sun, announcing that there are asteroids, comets, and planets in the system, and that these are all interacting by gravitation and collisions, without which the disks would disappear in a million years or so.

Likewise for other stars, their disks tell us that there are asteroids, comets, and planets active. Fortunately, many of these disks are expected to be bright enough to be seen with a coronagraph. We know that some are up to hundreds of times brighter than our own zodi and Kuiper belts. Figure 2-24 shows the expected signatures.

HST and other observatories have revealed the outer regions of several extreme / active disks, while WFIRST-2.4 will show us the inner disks of many nearby stars, more similar to our own solar system. Figure 2-25 shows an HST ACS image of the disk around the sun-like star HD 107146, but all material within 50 AU of the star is obscured by the coronagraph mask.[105] The WFIRST-2.4 coronagraph would allow us to see over 10 times closer to the star (130 mas IWA for 3 λ/D) and would be sensitive to detecting disks 1000 times fainter. This means that WFIRST-2.4 would be able to obtain comparable images of disks down to the habitable zones of many nearby stars, including ones within 10 pc of Earth.

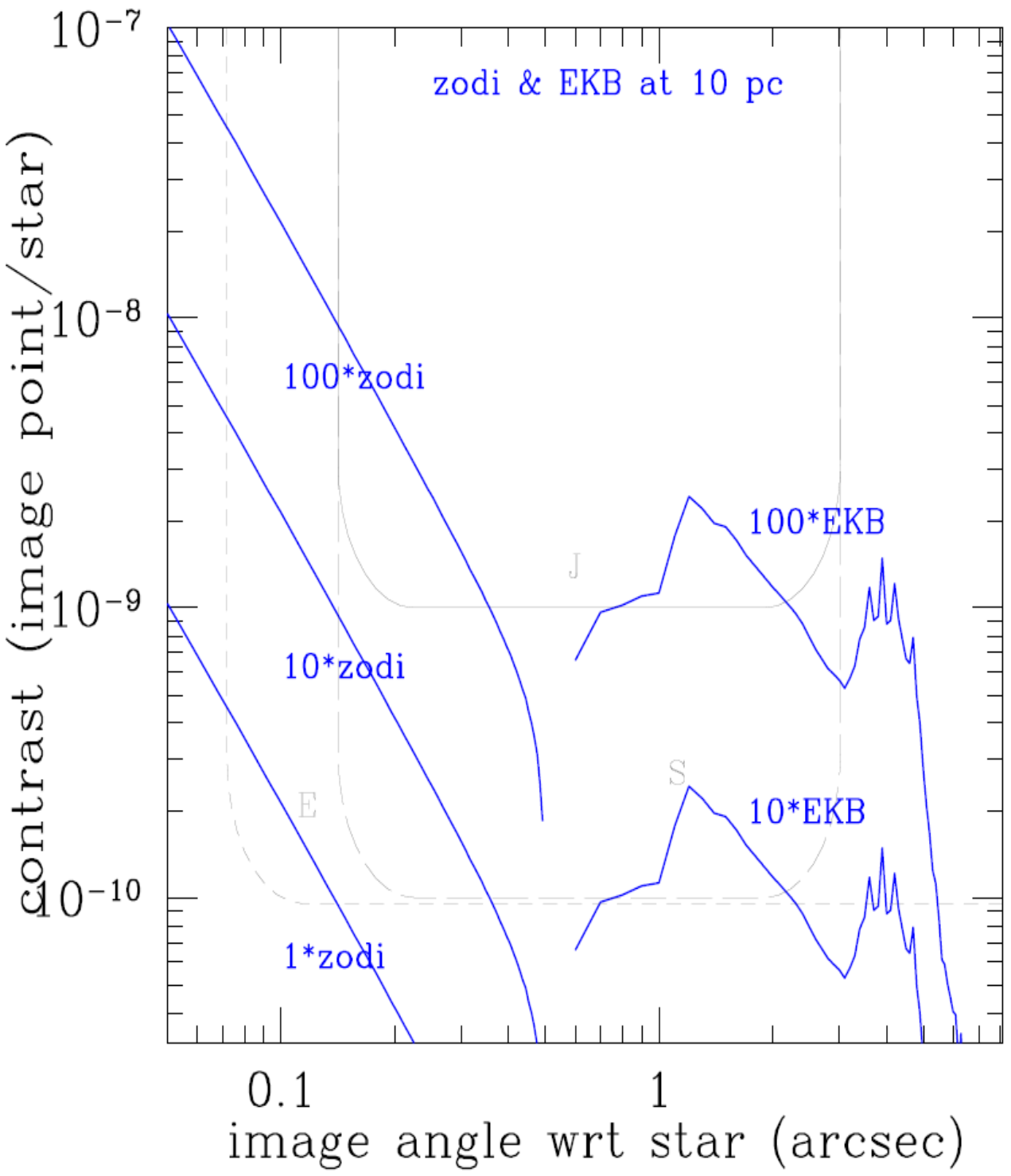

Figure 2-24: The brightness contrast of expected disks around nearby stars is illustrated here with the case of the solar system's zodiacal dust disk and Edgeworth-Kuiper Belt (EKB), with scaling factors of 1, 10, and 100, for the sun at a typical distance of 10 pc. Jupiter is located in the gap between these disks.

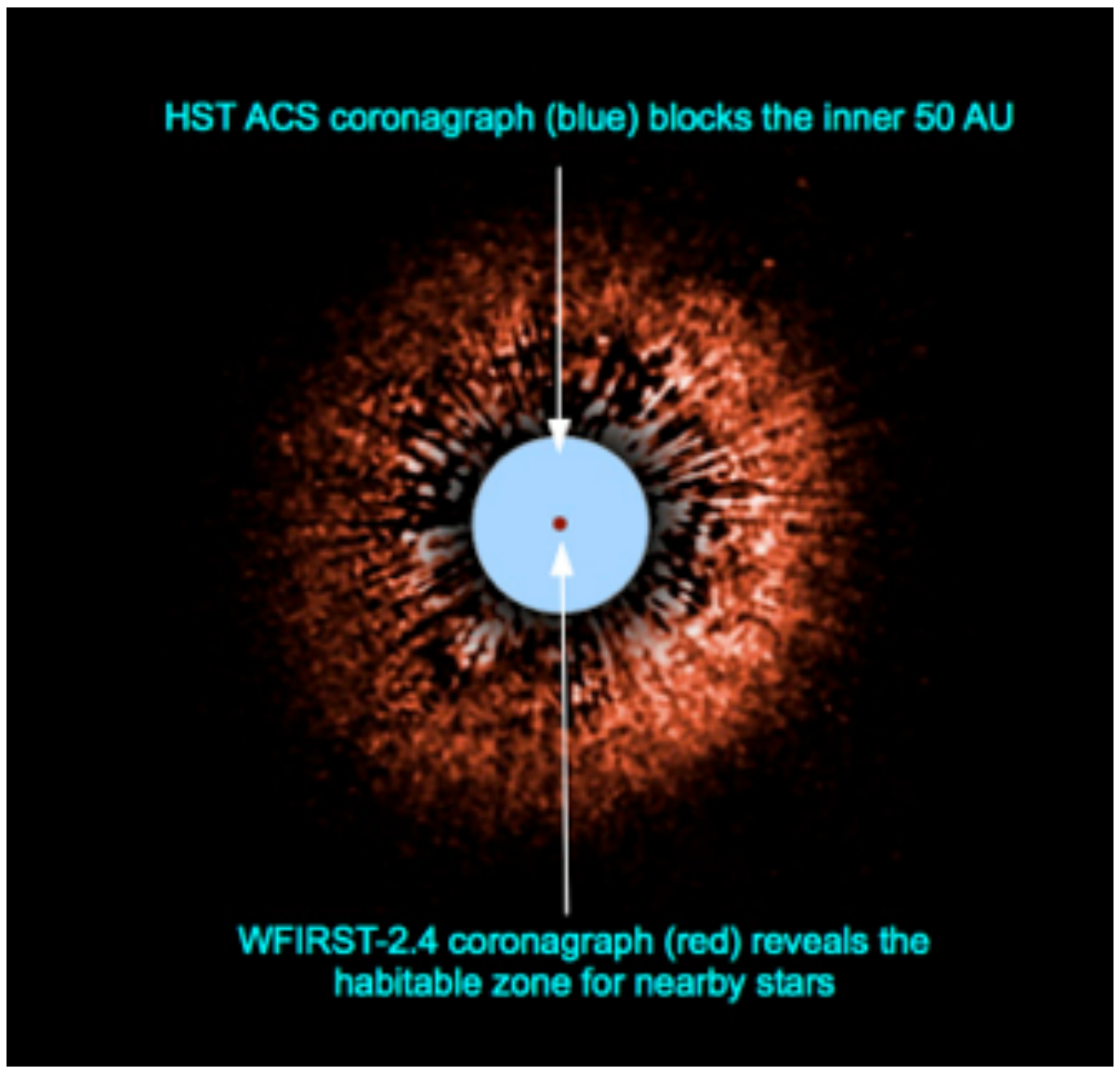

**Figure 2-25: HST ACS image of the debris disk around the young (~100 Myr), nearby (28 pc) sun-like (G2 V0 star HD 107146. This disk has 1000 - 10,000 times as much dust as our own solar system, and it is distributed out to a radius of approximately 180 AU (6.4"), while light within 50 AU (1.8") is obscured by the ACS coronograph occulting mask. This system is thought to have an inner hole in its disk at 31 AU[106], and the WFIRST-2.4 coronagraph could search for signs of planets within there.**

| The Scientific Objectives of the WFIRST-2.4 Mission |
|---|
| Determine the expansion history of the Universe and the growth history of its largest structures |
| Perform a Deep NIR Imaging and Spectroscopic Survey at High Galactic Latitudes |
| Complete the Statistical Census of Planetary Systems in the Galaxy |
| Discover and Characterize Dozens of Giant Planets and Circumstellar Disks about Nearby Stars* |
| Execute a Guest Observer program |
| **The Core Science Programs** |
| Dark Energy: A Three-Pronged Campaign using Type Ia Supernova, Baryon Acoustic Oscillations and Weak Lensing of Galaxies to Measure the Expansion and Growth History of the Cosmos |
| Exoplanet Microlensing: Detect Thousands of Exoplanets, Including Both Bound Cold Planets and Free-Floating Planets, with Masses Down to the Mass of Mars |
| High Contrast Coronagraphy*: Directly Image Planets about Nearby Stars from Earth-Orbit and Provide a Demonstration/Evaluation of the Rapidly Improving Technology |
| Guest Observer program |
| **Potential Guest Investigator Programs Using Data from the High Galactic Latitude NIR Survey** |
| Using Strong Lensing by Massive Clusters to Detect ~$10^4$ Galaxies at z>8 |
| Mapping Dark Matter on Intermediate and Large Scales |
| Kinematics of Stellar Streams in the Local Group of Galaxies |
| Discovering the Most Extreme Star Forming Galaxies and Quasars |
| **Potential Guest Observer Programs Requiring New Observations** |
| The Initial Mass Function Down to Brown Dwarfs in Taurus and Nearby Embedded Clusters |
| The Initial Mass Function Down to Brown Dwarfs of 10 – 100 Myr Clusters |
| The Brown Dwarf to Giant Planet Transition: Similarities and Differences Between the Two |
| Star Formation In the Milky Way Galaxy over Six Billion Years of Cosmic History |
| A Stellar Population Census of the Milky Way and Nearby Galaxies |
| Deep Color-Magnitude Diagrams of Globular Clusters to Determine Precision Ages and Metallicities |
| Wide Field Imaging of the Environments of One Hundred Nearby Galaxies |
| Linking Galaxy Evolution and Large Scale Structure |
| The Infrared Ultra Deep Field over 1 Square Degree |
| The Birth of Cosmic Structure During “Cosmic Dawn” ($z \approx 6$ to 10) |
| High Resolution Probing of LSST “Deep-Drilling” Fields to Test Models of $\Lambda$CDM |

**Table 2-6: Characteristics of the WFIRST-2.4 science program**

* Potential science if WFIRST-2.4 includes the optional coronagraph

## 2.6 Relation to Other Observatories

WFIRST-2.4 observations will be synergistic with the other major astronomical facilities planned for the next decade. The WFIRST-2.4 science program has many unique aspects that will enhance the science return from JWST, LSST, Euclid and the planned ground based facilities.

**JWST**. JWST stares deeply in the near-infrared and the mid-infrared at small portions of the sky. WFIRST-2.4 surveys large swaths of the sky at near infrared wavelengths. If operating concurrently, WFIRST-2.4 will significantly enhance the science return from JWST:

- WFIRST-2.4 will survey large areas of the sky and discover the most luminous galaxies at $10 < z < 20$. These rare objects will be superb targets for detailed study by JWST.
- WFIRST-2.4 will be capable of detecting powerful supernova explosions out to z~15. These first explosions will be exciting JWST targets.
- WFIRST-2.4's superb resolution should enable the HLS to discover 1000s of quad lensed QSOs. JWST/MIRI follow-up observations at mid-infrared wavelengths could then use the flux ratios of these lenses to probe the existence of the mini-halos predicted in cold dark matter models. If dark matter is warm, there would be a paucity of these halos.
- The combination of WFIRST-2.4 and JWST could study the progenitors of supernovae and their ejecta in nearby galaxies
- Simultaneous observations of microlensing events from WFIRST-2.4 at geosynchronous orbit and JWST in orbit around Sun-Earth L2 would provide a sufficient baseline to measure "microlens parallax" for many events, allowing for constraints on the planet and host star masses.
- The combination of optical observations of exoplanets with WFIRST-2.4's optical coronagraph and JWST's IR coronagraph would provide the broad spectral coverage that would probe the properties of the atmospheres of these planets.

These synergies are explored in more detail in Appendix B.

**LSST**. The Large Synoptic Survey Telescope (LSST) will survey a significant fraction of the sky at optical wavelengths at depths comparable to WFIRST (see Figure 2-2). The combination of the two telescopes will measure the photometry of stars, galaxies, and quasars in nine colors spanning nearly a decade in wavelength. The WFIRST-2.4 galaxy images will be much sharper with a PSF that is 12 times smaller in area; however, LSST optical data will provide an important complement that will enable astronomers to identify a host of interesting astronomical objects in this 9-color space. LSST will survey a larger area, but the 2000 deg$^2$ overlap with the WFIRST-2.4 HLS will allow better exploitation of the full LSST area, e.g., through characterizing correlations of optical colors with the near-IR spectral energy distributions (SED) and high resolution morphology, through providing robust photometric redshift calibrating effects on ground-based weak lensing measurements.

**Euclid**. While WFIRST-2.4 produces deep images of the sky at infrared wavelengths, Euclid will produce deep images of the sky at optical wavelengths (see Figure 2-2). Euclid depth in its single wide band is comparable to LSST in each of its six filters, but the angular resolution of Euclid is higher. Because optical photons are shorter wavelength, the two telescopes have comparable angular resolution so their combined galaxy images will provide a more complete image of galaxy properties. There are areas of overlapping science between the two projects: both aim to provide new insights into the nature of the dark matter. Euclid uses a combination of BAO measurements and gravitational lensing measurements in a single band to study dark energy. WFIRST-2.4 also uses these techniques; while it covers less area than Euclid, its redshift survey has an order of magnitude higher density of galaxies. Its weak lensing survey measures a 2.5x higher density of galaxies at higher redshifts, and will observe in multiple filters in multiples passes over the sky to maximize the robustness of its results. WFIRST-2.4 will also use supernovae as distance indicators a technique that complements BAO in multiple ways, including unequaled precision in the late-time universe when dark energy becomes dominant. While WFIRST-2.4 probes deeper into our past, Euclid will map a wider swath of the nearby universe. If launched early in the next decade, the 5-6 year prime mission of WFIRST-2.4 will be contemporaneous with the 6.25-year Euclid prime mission, so the two projects will carry out their major dark energy analyses largely independently and be able to cross-check findings in detail. Per unit time, WFIRST-2.4 is a substantially more powerful "dark energy machine" than Euclid, so if important discoveries about dark energy or modified gravity emerge from these investigations, they can be well characterized in extended-mission observations (see Sec 2.2.3).

## 2.7 Science Requirements

The SDT and Project have developed a requirements flowdown matrix for the mission. The four top-level scientific objectives for WFIRST-2.4 are:

- Produce a deep map of the sky at NIR wavelengths.
- Determine the expansion history of the Universe and the growth history of its largest structures in order to test possible explanations of its apparent accelerating expansion including Dark Energy and modifications to Einstein's gravity.
- Provide a guest observer program utilizing a minimum of 25% of the mission minimum lifetime.
- Complete the statistical census of planetary systems in the Galaxy, from the outer habitable zone to free floating planets, including analogs of all of the planets in our Solar System with the mass of Mars or greater.

These objects drive the requirements for the observatory capabilities and design.

With an optional coronagraph, WFIRST-2.4 acquires an additional top objective:

- Directly image giant planets and debris disks from habitable zones to beyond the ice lines, around nearby AFGK stars, at visible wavelengths, and characterize their physical properties by measuring brightness, color, spectra, and polarization while providing information to constrain their orbital elements with an optional coronagraph.

Table 2-6 summarizes the WFIRST-2.4 science program described in Sections 2.1 - 2.5. Figure 2-26 presents a top-level flow-down of the WFIRST-2.4 requirements. The Science Objectives above are the highest-level science requirements and appear at the top of the page. The derived scientific survey capability requirements of the observatory are listed in the left-hand boxes and data set requirements in the middle boxes. The top-level Observatory design/operations parameters are listed in the right-hand boxes. The detailed discussion of the basis for the requirements is given in the preceding subsections and the associated Appendices.

**WFIRST-2.4 Science Objectives:**

1) Produce a deep map of the sky at NIR wavelengths.
2) Determine the expansion history of the Universe and the growth history of its largest structures in order to test possible explanations of its apparent accelerating expansion including Dark Energy and modifications to Einstein's gravity.
3) Provide a guest observer program utilizing a minimum of 25% of the mission minimum lifetime
4) Complete the statistical census of planetary systems in the Galaxy, from the outer habitable zone to free floating planets, including analogs of all of the planets in our Solar System with the mass of Mars or greater.
5) Directly image giant planets and debris disks from habitable zones to beyond the ice lines, around nearby AFGK stars, at visible wavelengths, and characterize their physical properties by measuring brightness, color, spectra, and polarization while providing information to constrain their orbital elements with an optional coronagraph.

## WFIRST-2.4 Survey Capability Rqts

**NIR High Latitude Surveys**

**High Latitude Spectroscopic Survey**

- ≥1070 $deg^2$ per dedicated observing year (combined HLS imaging and spectroscopy)
- A comoving density of galaxy redshifts $n>6\times10^{-4}$ $Mpc^{-3}$ at z=1.9 according to Colbert + LF.
- Redshift range $1.10 \le z \le 1.95$ for Hα
- Redshift errors $\sigma_z \le 0.001(1+z)$, equivalent to 300 km/s rms
- Misidentified lines ≤5% per source type, ≤10% overall; contamination fractions known to $2\times10^{-3}$

**High Latitude Imaging Survey**

- ≥ 1070 $deg^2$ per dedicated observing year (combined HLS imaging and spectroscopy)
- Ability to measure a galaxy density of ≥60/$amin^2$, shapes resolved plus photo-*z*'s
- Additive shear error $\le 3\times10^{-4}$
- Multiplicative shear error = 0.2% per redshift slice (17 slices)
- Photo-z error distribution width ≤0.04(1+*z*),catastrophic error rate <2%
- Systematic error in photo-z offsets ≤ 0.002(1+z)

**Supernova SN-Ia Survey**

- >100 SNe-Ia per Δz=0.1 bin for all bins for 0.4 < z < 1.7, per dedicated 6 months
- Observational noise contribution to distance modulus error $\sigma_\mu \le 0.02$ per Δz=0.1 bin up to z = 1.7
- Redshift error σ ≤ 0.005 per supernova
- Relative instrumental bias ≤0.005 on photometric calibration across the wavelength range

**Exoplanet Surveys**

**Microlensing Survey**

- Planet detection capability to ~0.1 Earth masses ($M_\oplus$)
- Ability to detect ≥ 1500 bound cold planets in the mass range of 0.1-10,000 Earth masses, including 150 planets with mass <3 Earth masses
- Ability to detect ≥ 20 free floating Earth-mass planets

**Optional Coronagraphy Survey**

- Field of view from 0.1 – 1.5 arcsec at 400 nm, scaling linearly with wavelength up to 1000 nm (inner and outer radii of detection and characterization region)
- Allow for a survey of at least 150 stars with non-zero probability of detection
- Ability to image disks and map their structure with sub-AU angular resolution
- Single detection/characterization waveband of at least 10%

## WFIRST-2.4 Data Set Rqts

**NIR High Latitude Surveys Data Sets**

**High Latitude Spectroscopic Survey Data Set Rqts**

- Slitless grism, spectrometer, ramped resolution 550-800 over bandpass
- S/N ≥7 for $r_{eff}$ = 300 mas for Hα emission line flux at 1.8 μm ≥$1.0\times10^{-16}$ erg/$cm^2$-s
- Bandpass $1.35 \le \lambda \le 1.95$ μm
- Pixel scale ≤ 110 mas
- System PSF EE50% radius 240 mas (1.35 μm), 180 mas (1.65 μm), 260 mas (1.95 μm)
- ≥3 dispersion directions required, two nearly opposed
- Reach $J_{AB}$=24.0 AND ($H_{AB}$=23.5 OR F184$_{AB}$=23.1) for $r_{eff}$=0.3 arcsec source at 10 sigma to achieve a zero order detection in 2 filters

**High Latitude Imaging Survey Data Set Rqts**

- From Space: 3 shape/color filter bands (J,H,F184) and 1 color filter band (Y; for photo-z)
- S/N ≥18 (matched filter detection significance) per shape/color filter for galaxy $r_{eff}$ = 180 mas and mag AB = 24.7/24.6/24.1 (J/H/F184)
- PSF second moment ($I_{xx} + I_{yy}$) known to a relative error of ≤ $9.3\times10^{-4}$ rms (shape/color filters only)
- PSF ellipticity ($I_{xx}$-$I_{yy}$, 2*$I_{xy}$)/ ($I_{xx} + I_{yy}$) known to ≤ $4.7\times10^{-4}$ rms (shape/color filters only)
- System PSF EE50 radius ≤0.12 (Y band), 0.12 (J), 0.14 (H), or 0.13 (F184) arcsec
- At least 5 (H,F184) or 6 (J) random dithers required for shape/color bands, and 5 for Y at same dither exposure time
- From Ground: 5 filters spanning at least 0.35-0.92 μm
- From Ground + Space combined: Complete an unbiased spectroscopic PZCS training data set containing ≥ 100,000 galaxies ≤ mag AB = 24.6 (in JHF184 bands) and covering at least 4 uncorrelated fields; redshift accuracy required is $\sigma_z<0.01(1+z)$

**Supernova Survey Data Set Rqts**

- Minimum monitoring time-span for an individual field: ~2 years with a sampling cadence ≤5 days
- Cross filter color calibration ≤0.005
- Three filters, approximately Y, J, H for SN discovery
- IFU spectrometer, λ/Δλ ~100, 2-pixel (S/N ≥ 10 per pixel bin) for redshift/typing
- IFU S/N ≥15 per synthetic fiber band for points near lightcurve maximum in each band at each redshift
- Dither with 30 mas accuracy
- Low Galactic extinction, E(B-V) ≤0.02

**Exoplanet Surveys Data Sets**

**Microlensing Survey Data Set Rqts**

- Monitor ~3 square degrees in the Galactic Bulge for at least 250 total days
- S/N ≥100 per exposure for a J=20.5 star
- Sample light curves with wide filter W149 with λ =0.927-2.0 μm
- Photometric sampling cadence in W149 of ≤15 minutes
- ≤0.4" angular resolution to resolve the brightest main sequence stars
- Monitor microlensing events continuously with a duty cycle of ≥80% for at least 60 days
- Monitor fields with Z087 filter, 1 exposure every 12 hours
- Separation of >2 years between first and last observing seasons

**Optional Coronagraphy Survey Data Set Rqts**

- Detection limit after PSF calibration of 1 part per billion or better over entire field of view and wavelength band
- Spectral resolution of at least 70 from 400 to 1000 nm
- Multi-band polarimetric imaging with 5% accuracy and 2% precisions per spatial resolution element for disk characterization (grain size distribution, disk geometry) and identification of large polarimetric features in gas giants
- Critically sampled PSF at shortest wavelength

**WFIRST-2.4 Design/Operations Overview**

**Key WFIRST-2.4 Observatory Design Parameters**
- 2.4 m diameter telescope aperture
- 270 K telescope optical surfaces
- Bandpass 0.6 – 2.0 μm
- Pointing jitter ≤20 mas rms/axis (p/y), ≤2 as rms (roll)
- Coarse Pointing Accuracy <~3 arcsec rms/axis
- Fine (Relative/Revisit) Pointing Accuracy <10 mas rms/axis (p/y), ≤1 as rms (roll)
- ACS telemetry downlinked for pointing history reconstruction

**Imaging Mode:**
- Pupil Mask temperature: ~170 K
- Effective Area (avg over bandpass): 1.97 $m^2$ (Z), 2.06 $m^2$ (Y), 2.30 $m^2$ (J), 2.56 $m^2$ (H), 2.65 $m^2$ (F184), 2.41 $m^2$ (W),
- 6 band parfocal filter set on element wheel, driven by microlensing, SNe, WL, plus blank position
- FPA: 6x3 HgCdTe 4k x 4k SCAs, 2.1μm, ≤120K, 110 mas/pix
- FOV (active area) = ~0.281 $deg^2$; Bandpass 0.6 – 2.0 μm
- Fine guiding using guiding windows in each SCA in the FPA to control Pitch/Yaw/Roll
- WFE is diffraction limited at 1.2 μm
- 1 nm rms wavefront stability during the HLS imaging mode (requires 0.3 K thermal stability/obs.)

**Spectroscopy Mode:**
- IFU: R=100 (2-pix), 75 mas/pix
- GRS grism: R = 550-800, parfocal, zero deviation grism
- GRS grism Effective Area (avg over bandpass): 1.94 $m^2$ (1.35-1.95 μm)

| Name | Bandpasses (μm) |
|---|---|
| Z087 | 0.760 – 0.977 |
| Y106 | 0.927 – 1.192 |
| J129 | 1.131 – 1.454 |
| H158 | 1.380 – 1.774 |
| F184 | 1.683 – 2.000 |
| W149 | 0.927 – 2.000 |
| GRS Grism | 1.350 –1.950 |

**Key WFIRST-2.4 Operations Concept Parameters and Constraints**
- 5-year mission life (6 years with optional coronagraph), but consumables required for 10 yrs
- Science Field of Regard (FOR): 54˚ to 126˚ pitch off the Sun, 360˚ yaw
- Roll about line of sight ±15˚
- Gimbaled antenna allows observing during downlinks
- Slew/settle times: ~13 s for SCA gap steps, ~61 s for full field steps

**SNe-Ia Survey to *z* = 1.7, 6 months spread over 2 years**
- A sample 3-tiered survey with each tier optimized for a different redshift range
- Tier 1 for $z$<0.4: 27.44 $deg^2$ in Y and J bands
- Tier 2 for $z$<0.8: 8.96 $deg^2$ in J and H bands
- Tier 3 for $z$<1.7: 5.04 $deg^2$ in J and H bands
- The imager is used for SN discovery and the IFU spectrometer is used to type SN, measure redshifts, and obtain lightcurves
- IFU exposure times tailored for each individual supernova
- SNe dedicated time is distributed in a 5-day cadence over 2 years (30 hrs of SNe field monitoring would be done every 5 days for 2 years)
- Fields located in low dust regions ≤20˚ off an ecliptic pole

**Galaxy Shape + Galaxy Redshift Survey (~1,070 $deg^2$/yr, 2,000 $deg^2$ total)**
- HLS footprint to be within LSST footprint (or other deep optical survey) for photometric redshifts
- 2 imaging passes in each of J,H,F184 (shape) and Y (photo-z) filters
- Each imaging pass with the same filter is rotated from the other by ≥5˚, and includes 4 exposures (~184 sec each), with 5 exposures for each J band pass
- 90% of imaging field see ≥5 randomly dithered exposures (≥920 sec total) in Y,H,F184 bands, ≥6 exposures (≥1104 sec total) in J band
- 4 passes total, seasons chose to oppositely disperse 2 of the 4 on the sky
- Each spectrometry pass with the same prism is rotated from the other by ~5˚, and includes 2 exposures (362 sec each)
- 90% of spectroscopy field sees ≥6 randomly dithered exposures (≥2172 sec total)
- Zero order galaxy detection to GRS data set requirements provided in J,H or J,F184 bands by WL imaging passes

**Exoplanet Microlensing Survey (~2.62 $deg^2$ monitored every 15 min, 144 days/yr, 1.2 yrs total)**
- The Galactic Bulge is observable for two 72-day seasons each year
- The short revisit cadence impacts other observing modes while exoplanet data sets are being acquired. This, in combination with the field monitoring time span required for SN and the ≥60 days required for microlensing, limits the max number of Galactic Bulge seasons useable for exoplanet observations to six (over 5 yrs)
- In each season 9 fields are revisited on a 15 min cadence, viewing in filter W149, for light curve tracking except for one exposure every 12 hours that uses the Z087 filter for color
- Fields are revisited to an accuracy of 1 pixel rms; no precise dithers

**Optional Exoplanet Coronagraphy Survey**
- Availability of 3-6 hours in each observation for wavefront control and setup.
- Ability to spectrally characterize any planet detected in less than 24 hours of observing.
- Maximum of 10 days for any single spectral characterization.
- Capability of revisiting all stars with detected planets for orbit determination.
- Exoplanet imaging will use 1 year of allotted mission time, distributed among the 6 years.

**Figure 2-26: WFIRST-2.4 DRM requirements flowdown overview**

# 3 WFIRST-2.4 DESIGN REFERENCE MISSION

## 3.1 Overview

The WFIRST-2.4 Design Reference Mission (DRM) uses the existing 2.4-meter telescope hardware, along with heritage instrument, spacecraft, and ground system architectures and hardware to meet the WFIRST-2.4 science requirements. The WFIRST-2.4 DRM baseline payload (wide field instrument and telescope) provides the wide-field imaging and slitless spectroscopy capability required to perform the Dark Energy, Exoplanet Microlensing, and NIR surveys and an optional coronagraph instrument supports the Exoplanet Coronagraphy science (see Figure 3-1 for a fields of view layout).

The payload features a 2.4-meter aperture, obscured telescope, which feeds two different instrument volumes containing the wide-field instrument and the optional coronagraph (see Figure 3-2). The telescope hardware was built by ITT/Exelis under contract to another agency and is being made available to NASA. This existing hardware significantly reduces the development risk of the WFIRST-2.4 payload.

The wide-field instrument includes two channels, a wide-field channel and an integral field unit (IFU) spectrograph channel. The wide-field channel includes three mirrors (two folds and a tertiary) and a filter/grism wheel to provide an imaging mode covering 0.76 – 2.0 $\mu$m and a spectroscopy mode covering 1.35 – 1.95 $\mu$m. The wide-field focal plane uses 2.1 $\mu$m long-wavelength cutoff 4k x 4k HgCdTe detectors. The HgCdTe detectors are arranged in a 6x3 array, providing an active area of 0.281 deg$^2$. The IFU channel uses an image slicer and spectrograph to provide individual spectra of each 0.15" wide slice covering the 0.6 – 2.0 $\mu$m spectral range over a 3.00 x 3.15 arcsec field. The instrument provides a sharp PSF, precision photometry, and stable observations for implementing the WFIRST-2.4 science.

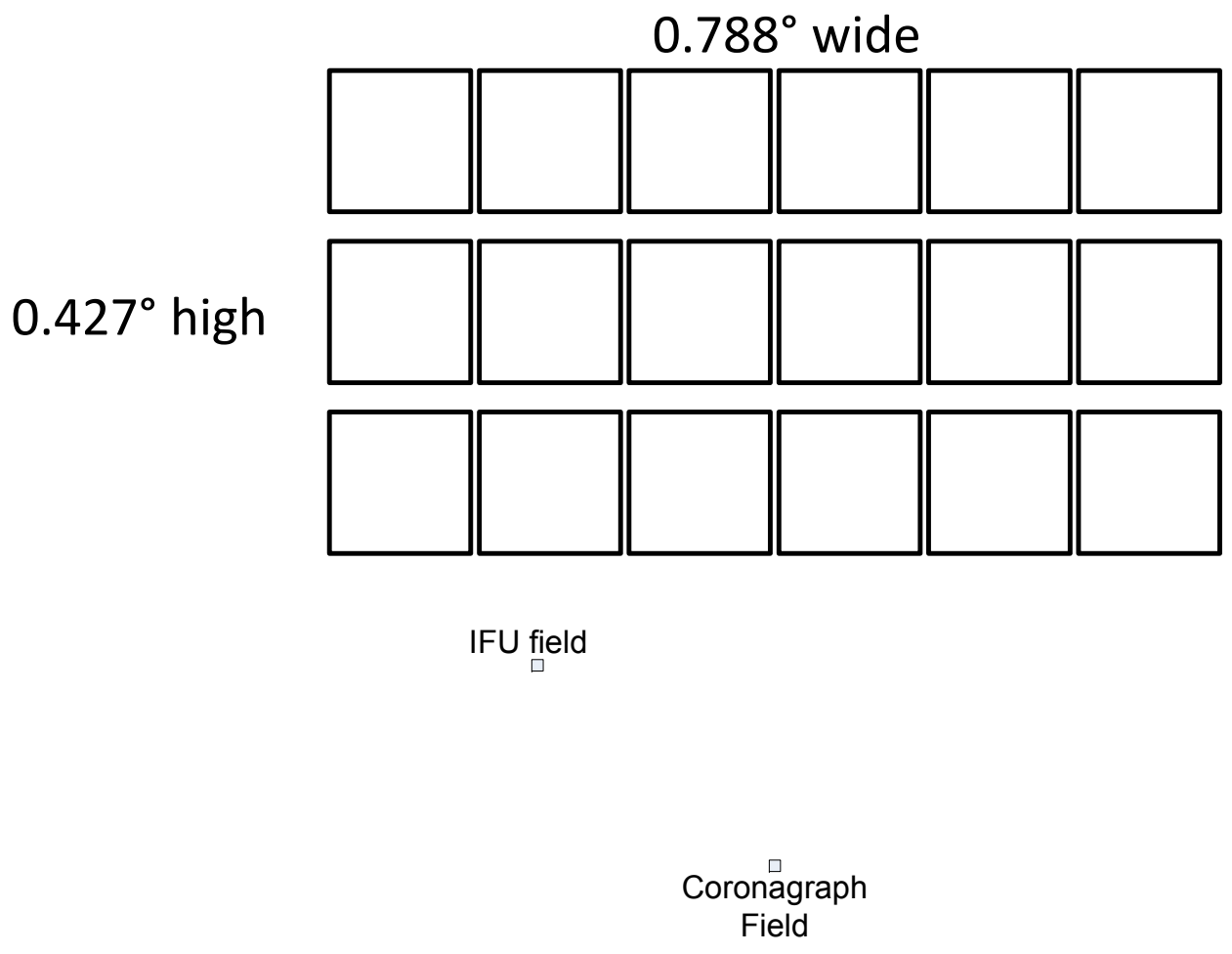

**Figure 3-1: The WFIRST-2.4 instrument field layout, as projected on the sky, showing the wide-field instrument (including the IFU) and the optional coronagraph.**

The coronagraph instrument includes an imaging mode, an integral field spectrograph, and a low order wavefront sensor to perform exoplanet detection and characterization. The coronagraph covers a spectral range of 0.4 – 1.0 $\mu$m, providing a contrast of $10^{-9}$ with an inner working angle of $3\lambda/D$ at 400 nm.

The SDT considered both geosynchronous and Sun-Earth L2 orbit options and selected a 28.5° inclined, geosynchronous orbit as the baseline for this study. The primary factor that drove the selection of this orbit is the ability to continuously downlink data to the ground and obtain a much higher science data rate. The SDT weighed these benefits against the higher radiation environment and slightly less stable thermal environment versus the Sun-Earth L2 orbit chosen by the previous WFIRST SDT. Preliminary radiation analysis was performed during this study to assess the impact of the electron flux environment on the HgCdTe detectors. The analysis shows that a three-layer sandwich of graphite epoxy and lead can reduce the event rate in the detectors caused by trapped electrons to a level below that of Galactic cosmic rays. A thick outer layer of graphite epoxy stops the electrons, the lead layer stops bremsstrahlung photons produced by the electrons as they are stopped, and the inner graphite epoxy layer stops tertiary electrons produced by the photons. A future analysis will optimize the shielding design by accounting for partial shielding provided by the other components of the payload. As is typical for all spacecraft in GEO, additional shielding is required on both the spacecraft and payload electronics for the higher total dose radiation environment. Preliminary thermal analysis shows that the HgCdTe detectors can be held passively at 120K while meeting the thermal stability requirements in the selected orbit.

An Atlas V 541 launch vehicle will inject the observatory into a geosynchronous transfer orbit. Using a bi-propellant propulsion system, the spacecraft will circularize at the 28.5° inclined, geosynchronous orbit. The mission life is 5 years (6 years with the optional coronagraph) with consumables sized to allow an extension for a total of 10 years. The spacecraft uses mature technology and redundant hardware to protect against any one failure prematurely ending the mission. It provides a fixed solar array/sunshield that allows operations over the full field of regard (see §3.10).

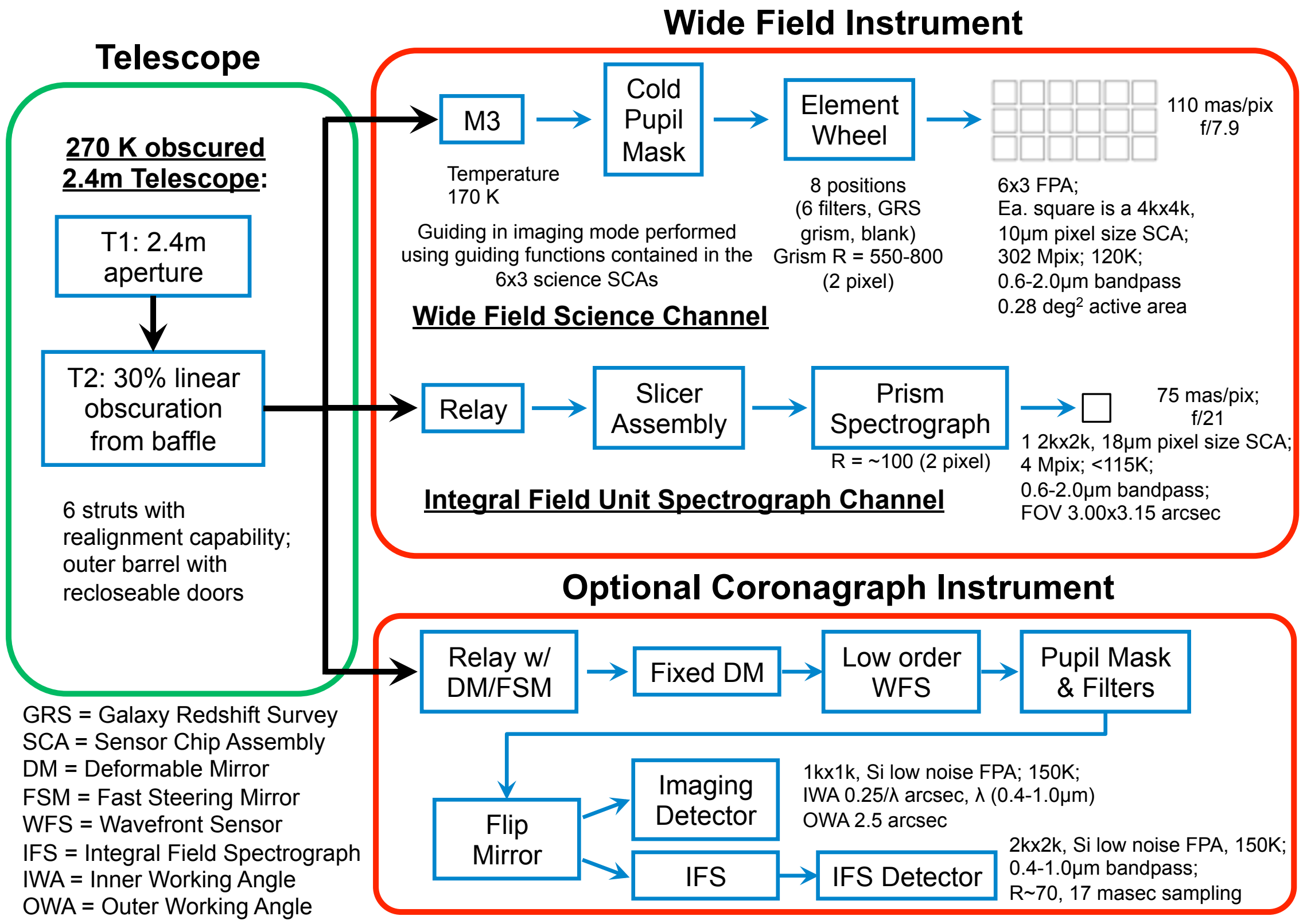

**Figure 3-2: WFIRST-2.4 payload optical block diagram.**

The SDT charter specified that the WFIRST-2.4 observatory be serviceable. For this study, it was decided to provide servicing at a module level, i.e. an entire instrument or a spacecraft module containing multiple electronics boxes. The modularity will also be a benefit during integration and test of the observatory (see Figure 3-3). On the payload side, an instrument carrier was designed to attach to the existing telescope metering structure interfaces and provide volumes for two instrument modules. The instrument carrier provides mechanical latches, with design heritage from HST, to interface the instrument modules to the carrier, contains harness to route power and data between the spacecraft and the instrument modules, and provides thermal isolation between the two instrument volumes.

High-accuracy pointing, knowledge, and stability are all required to resolve galaxy shapes and precisely revisit both the microlensing and SN fields and support coronagraphy. Pointing to between 54˚ and 126˚ off the Sun enables the observation of micro-lensing fields for up to 72 continuous days during each of the twice yearly Galactic Bulge viewing seasons.

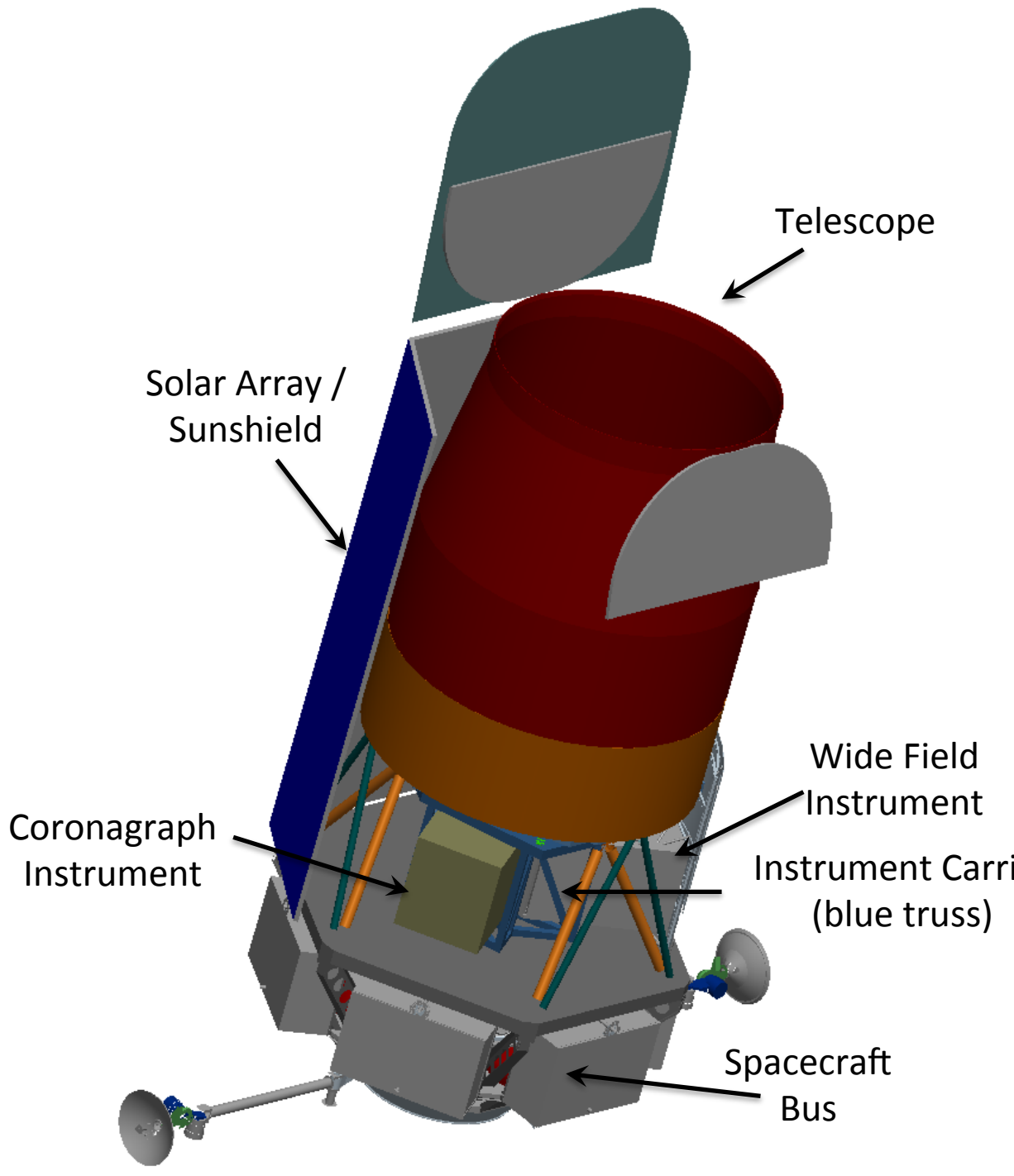

**Figure 3-3: WFIRST-2.4 Observatory configuration featuring the 2.4-m telescope, two modular instruments and a modular spacecraft. bus**

The microlensing survey requires large light gathering power (collecting effective area times field of view) for precise photometric observations of the Galactic Bulge to detect star + planet microlensing events. Multiple fields are observed repeatedly to monitor lightcurves of the relatively frequent stellar microlensing events and the much rarer events that involve lensing by both a star and a planet. In the latter case the planetary signal is briefly superposed on the stellar signal. Microlensing monitoring observations are performed in a wide filter spanning 0.93 – 2.0 μm, interspersed ~twice/day with brief observations in a narrower filter for stellar type identification.

The GRS measurement requires NIR spectroscopy to centroid Hα emission lines and NIR imaging to locate the position of the galaxy image relative to the dispersion window. Dispersion at R 550-800 (2 pixels) enables centroiding the Hα emission lines to a precision consistent with the redshift accuracy requirement. To address completeness and confusion issues, grism observations over at least 3 roll angles, two of which are approximately opposed, are performed over ~90% of the mapped sky. The bandpass range of 1.35 – 1.95 μm provides the required redshift range for Hα emitters.

The SN measurement also requires large light gathering power to perform the visible and NIR deep imaging and spectroscopy needed to identify, classify and determine the redshift of large numbers of Type Ia SNe. Precise sampling (S/N of 15) of the light curve every five days meets the photometric accuracy requirement, and the use of three NIR bands allows measurements of SNe in the range of $0.4 < z < 1.7$, providing better control of systematic errors at low z than can be achieved by the ground and extending the measurements beyond the z ~ 0.8 ground limit.

The WL measurement requires an imaging and photometric redshift (photo-z) survey of galaxies to mag AB ~24.6. A pixel scale of 0.11 arcseconds balances the need for a large field of view with the sampling needed to resolve galaxy shapes. Observations in three NIR filters, with ≥5 random dithers in each filter, are made to perform the required shape measurements to determine the shear due to lensing, while observations in an additional NIR filter are combined with color data from the shape bands and the ground to provide the required photo-z determinations. The GRS grism and the IFU, along with overlapping ground observations, are used to perform the photo-z calibration survey (PZCS) needed to meet the WL redshift accuracy requirement.

The coronagraphy measurements require imaging and spectroscopy to detect and characterize exoplanets. Coronagraphy requires high thermal stability of both the telescope and the coronagraph during observations. The relatively short initial observations focus on discovery of planets near the target star, while longer observations are required for planet spectroscopic characterization.

### 3.2 Telescope

The WFIRST-2.4 DRM is based on use of a repurposed, space flight qualified 2.4-meter, obscured two-mirror telescope (see Figure 3-4). Repurposing modifications will include conversion to a three mirror anastigmat (TMA) optical configuration to achieve a wide FOV capability, new internal stray light baffling optimized for the wide FOV, electronics replacements, new thermal blankets, and efforts to achieve a slightly lower operating temperature than originally specified. The operating temperature will be 270 K, which enhances performance for infrared wavelengths out to 2.0 μm while staying close to the original design specifications. The scientific advantage of this telescope is its much larger aperture relative to previous DRM concepts. The larger aperture increases both angular resolution and collecting area, allowing deeper and finer observations per unit observing time: The image is 1.9x sharper (the ratio of the PSF effective areas at H band, even after allowing for the obscured in the 2.4-m telescope). The fact that the majority of the hardware exists eliminates considerable cost, schedule and technical risk from the development program.

The telescope's pupil is obstructed by the secondary mirror baffle to a linear obscuration ratio of 30% (see Figure 3-5). The mounted secondary mirror has a 6 degree of freedom mechanism to adjust alignment on orbit, as well as a fine focus capability. The complete telescope includes an outer barrel assembly (OBA). The OBA provides a closed tube external light baffle, aperture doors, and telescope thermal isolation.

The telescope optics are made from ultra-low expansion material (Corning ULE®). The mirrors are significantly light-weighted and thermally stable. The telescope's structure is manufactured from mature low-moisture uptake composites to minimize mass and thermal distortions while providing superior stiffness and stability. The structure has active thermal control (heaters) and is isolated from the solar array by the OBA to minimize heat transfer into the telescope and instrument bay.

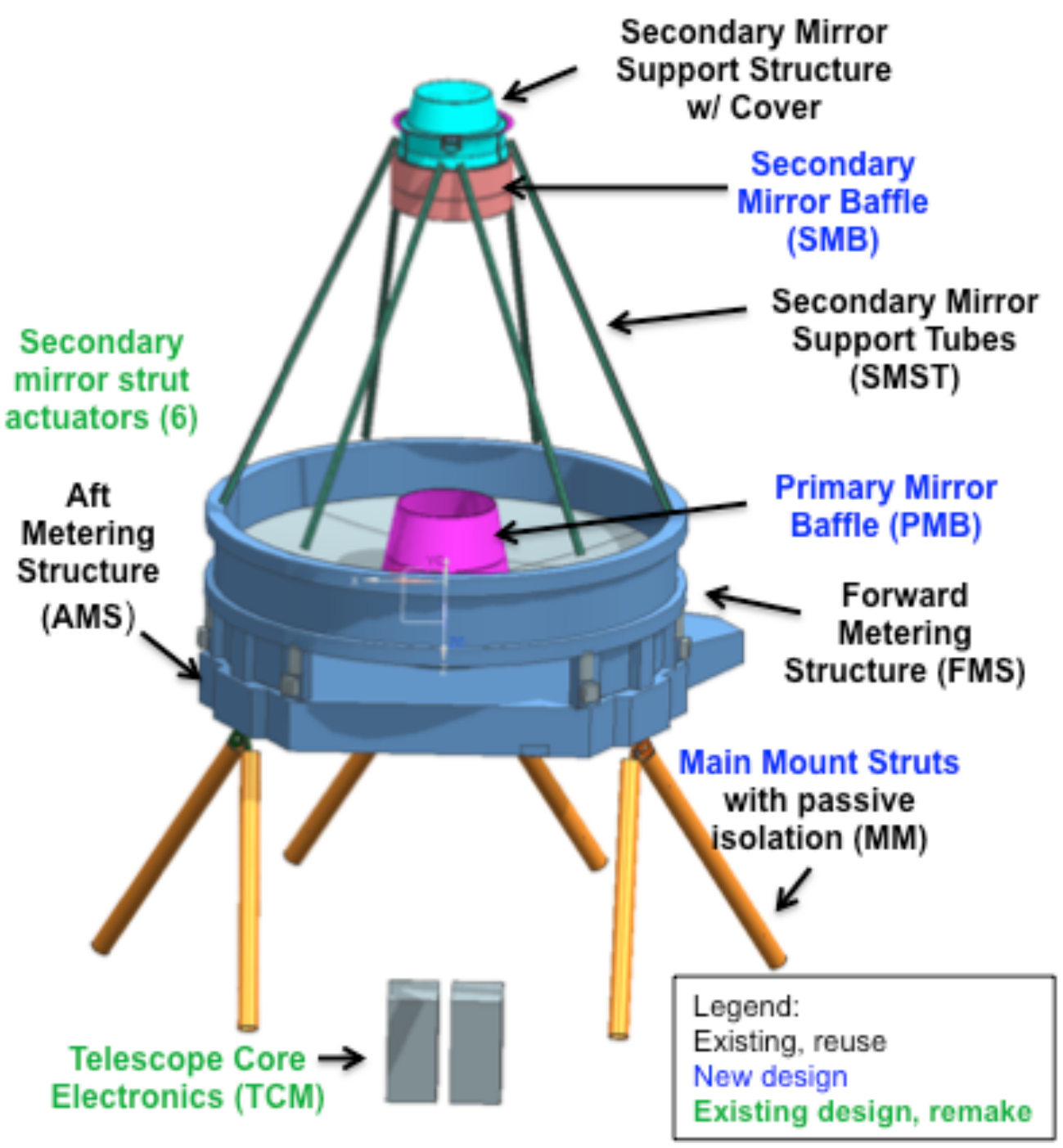

**Figure 3-4: The telescope components without the outer barrel assembly.**

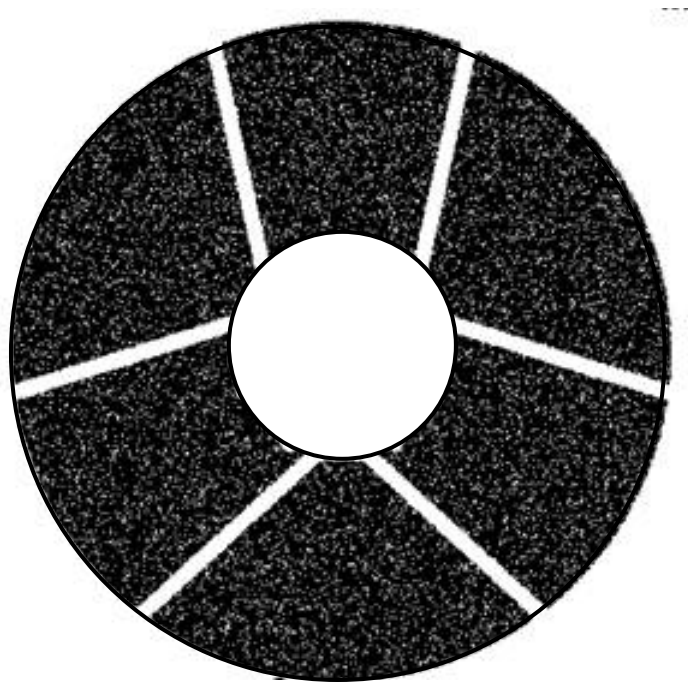

**Figure 3-5: The telescope entrance pupil**

The two telescope mirrors feed both a wide field instrument (see Figure 3-6), with a wide field of view channel and an integral field unit, and an optional nearly on-axis (in field) coronagraph instrument. The wide field channel includes a tertiary so that the system acts as a three mirror anastigmat (TMA); the other channels correct the residual aberrations in the first two mirrors. The overall three-mirror system is able to achieve excellent optical performance over a relatively wide FOV. The most important performance requirement for the telescope to enable the WFIRST-2.4 science is stability. Based on the low expansion materials, the thermal design, and the active heater control, the wavefront stability delivered to the wide field imager is modeled to be <0.15 nm rms over one 24-hour GEO orbit thermal cycle.

The telescope electronics were not included with the 2.4-m telescope. The current plan is to upgrade most of the electronics with the current commercial products. The electronics architecture is a traditional hierarchical design from spacecraft avionics through telescope control to local boxes to peripheral devices and heaters. The telescope control electronics are located in an electronics box in the spacecraft and perform higher control functions such as commanding actuators and managing thermal set-points. These higher-level commands are sent through cabling to local boxes on the telescope where they are converted to low-level signals and sent to the devices themselves. Each subsystem has a similar functional content with heaters, thermal sensors, and actuators. The exception to this overall architecture is control of telescope survival heaters. They are controlled directly by the spacecraft avionics to ensure operation in the event of an intermediate electronic failure.

The cost, schedule, and technical risk associated with the telescope development have already been significantly mitigated; the telescope hardware and supporting GSE largely exists and spare parts are available for most of the components. New stray light baffles, new thermal blankets, and new electronics derived from commercial products are low risk modifications.

## 3.3 Wide-Field Instrument

The wide-field instrument is divided into two modules, a cold instrument module (referred to as the instrument module below), containing the optics and the focal plane assembly, and the warm electronics module housed on the spacecraft. Optically, the instrument is divided into a wide-field channel, with both imaging and spectroscopy modes, and an integral field unit (IFU) channel. The key instrument parameters are shown in Table 3-1.

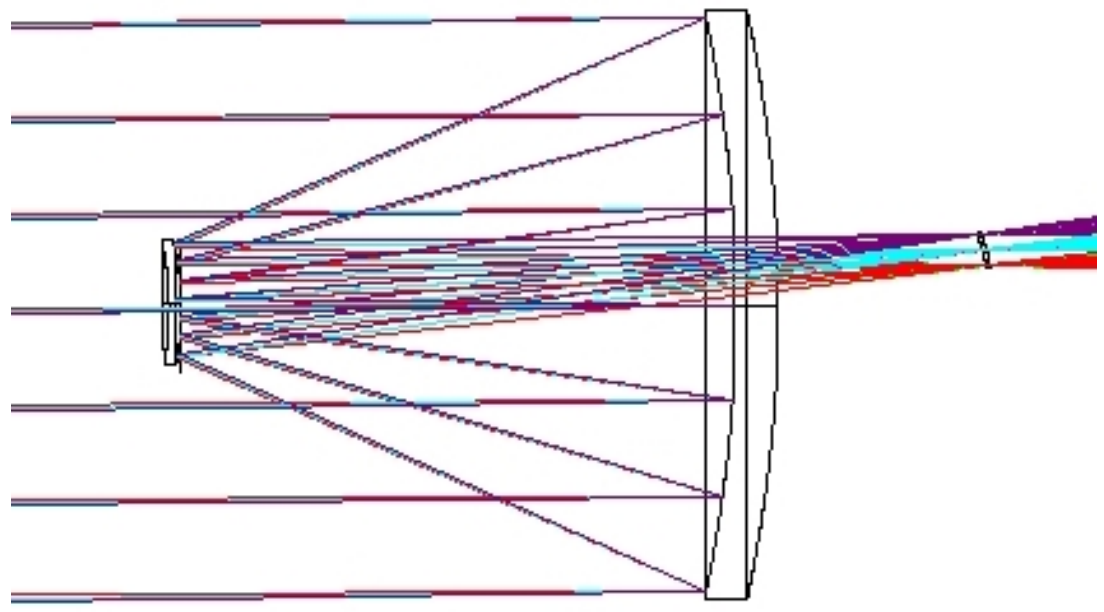

**Figure 3-6: Ray trace through the telescope to the wide field channel intermediate focus.**

The instrument module is kinematically mounted to and thermally isolated from the instrument carrier, which provides the load path between the telescope and the instrument module structure. The instrument module design is similar to the HST/WFC3 design (see Figure 3-7) and reuses the design of the WFC3 latches, which can be robotically engaged and disengaged, to mechanically interface to the instrument carrier. The instrument module also includes a grapple fixture to enable its removal by a robotic servicing vehicle. Guide rails on the instrument carrier align the instrument module latches with the instrument carrier latches. Connectors on the instrument module are designed to allow a blind mate to harness in the instrument carrier that runs to the warm instrument electronics module on the spacecraft.

The instrument module consists of an outer enclosure structure, which is flexured off of the latches and supports the instrument radiators and MLI blankets, and an inner structure, the optical bench, which supports the instrument cold electro-optical components (see Figure 3-8). The mechanical latches interface directly to the optical bench, via thermally isolating struts, providing a direct load path between the instrument carrier and the alignment-critical optical bench.

### 3.3.1 *Wide-Field Channel*

The wide-field channel optical train consists of a pair of fold flats, a tertiary mirror (M3), a cold pupil mask, an element wheel (EW), and the HgCdTe Focal Plane Array (FPA), see Figure 3-2. The FPA uses 18 4k x 4k 10 μm pixel (H4RG) HgCdTe Sensor Chip Assemblies (SCAs), with a detector wavelength cutoff of 2.1 μm (optical cutoff is limited to 2.0 μm via bandpass filters) and proximate SCA Control Electronics (SCE) boards. The SCAs are arranged in a 6x3 layout with a pixel scale of 0.11 arcseconds/pixel. An 8-position EW provides 6 filters, a dark position (for calibration) and a grism assembly for the GRS. The imaging mode is designed to a diffraction limit of 1.2 μm.

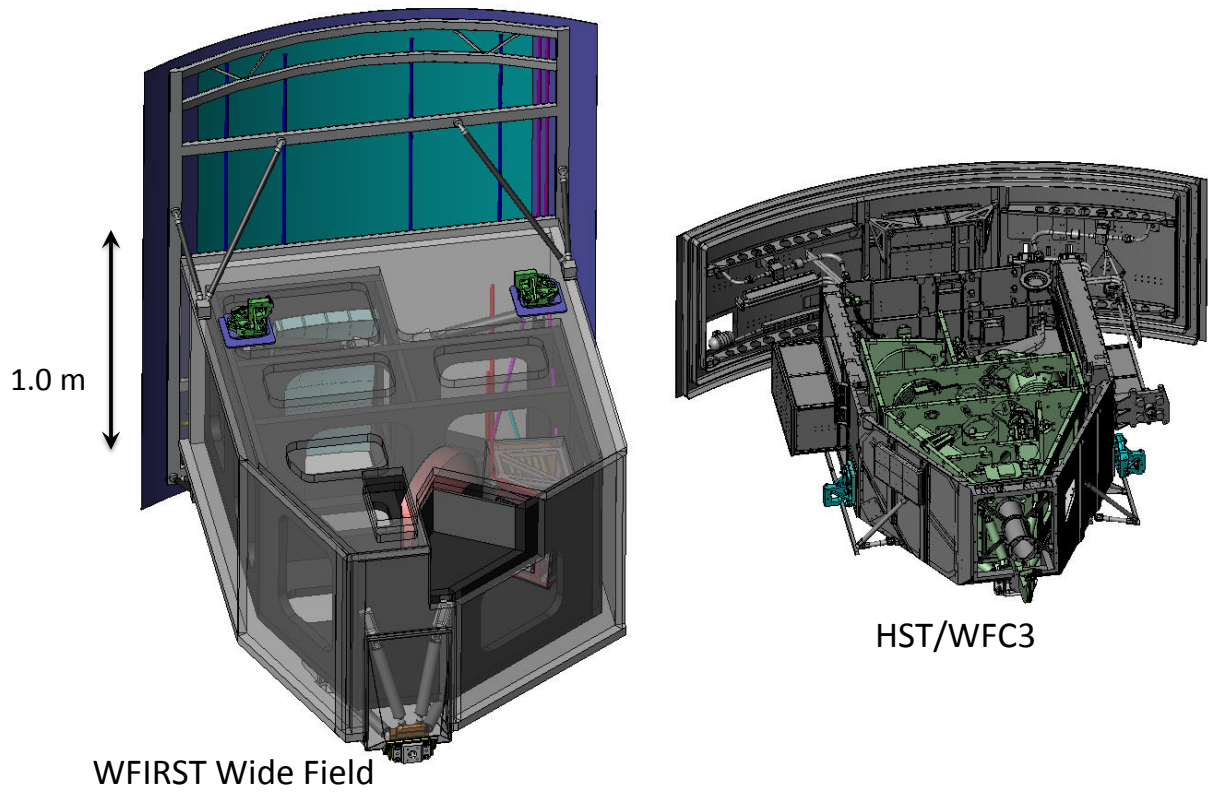

**Figure 3-7: Instrument comparison to HST/WFC3. WFIRST-2.4 wide field is designed to be similar to the WFC3. Both use composite optical benches, radial latches, passive radiators, and heat pipes for thermal control.**

The filters are fused silica substrates with ion-assisted, highly stable bandpass filter coatings with heritage to recent HST instruments

The 4-element GRS grism consists of 2 $CaF_2$ elements and $BaF_2$ and FK3 correctors. The grism has a spectral range of 1.35-1.95 μm. The equivalent dispersion $D_e=\lambda/(\delta\lambda/\delta\Theta)$, is obtained by multiplying by the angular size of a 2-pixel resolution element. Thus, for a value of R=675, the corresponding value $D_\theta$ is 675*2*0.11= 149 arcsec. A description of the EW complement is shown in Table 3-2. Figure 3-9 gives a detailed view of the grism assembly, its spectral dispersion, and its wavefront error over a 3x3 field point grid and its spectral range.

| Mode | Wavelength Range (μm) | Sky Coverage (active area) | Pixel Scale (arcsec/pix) | Dispersion | FPA Temperature (K) |
|---|---|---|---|---|---|
| Imaging | 0.76 – 2.0 | 0.281 deg$^2$ | 0.11 | N/A | ≤120 |
| GRS Spect. | 1.35 – 1.95 | 0.281 deg$^2$ | 0.11 | R=550-800 (2-pixel, grism in element wheel) | ≤120 |
| SN Ia Spect. | 0.6 – 2.0 | 3.00 x 3.15 arcsec | 0.075 | R=100 (2-pixel; IFU spectrograph, 1 slice maps to 2 pixels) | ≤115 |
| Fine Guiding | 0.76 – 2.0 | 0.281 deg$^2$ | 0.11 | Guide off wide-field focal plane using windowing function of H4RG | ≤120 |

**Table 3-1: Key Instrument Parameters**

**Mechanical**

The instrument structure, both the outer enclosure and the optical bench, will be made from aluminum honeycomb panels with composite facesheets of cyanate siloxane resin in a carbon fiber matrix. This mature, strong, light, low moisture absorption composite material is a good low thermal expansion match to the ULE® mirrors used in the instrument. The mirrors are stiff lightweight ULE® sandwich designs with heritage from Kepler, GeoEye and other programs. The optical bench structure has top and bottom panels with structural bulkheads between them, and is kinematically supported by and precisely aligned to the instrument carrier at three latch locations via

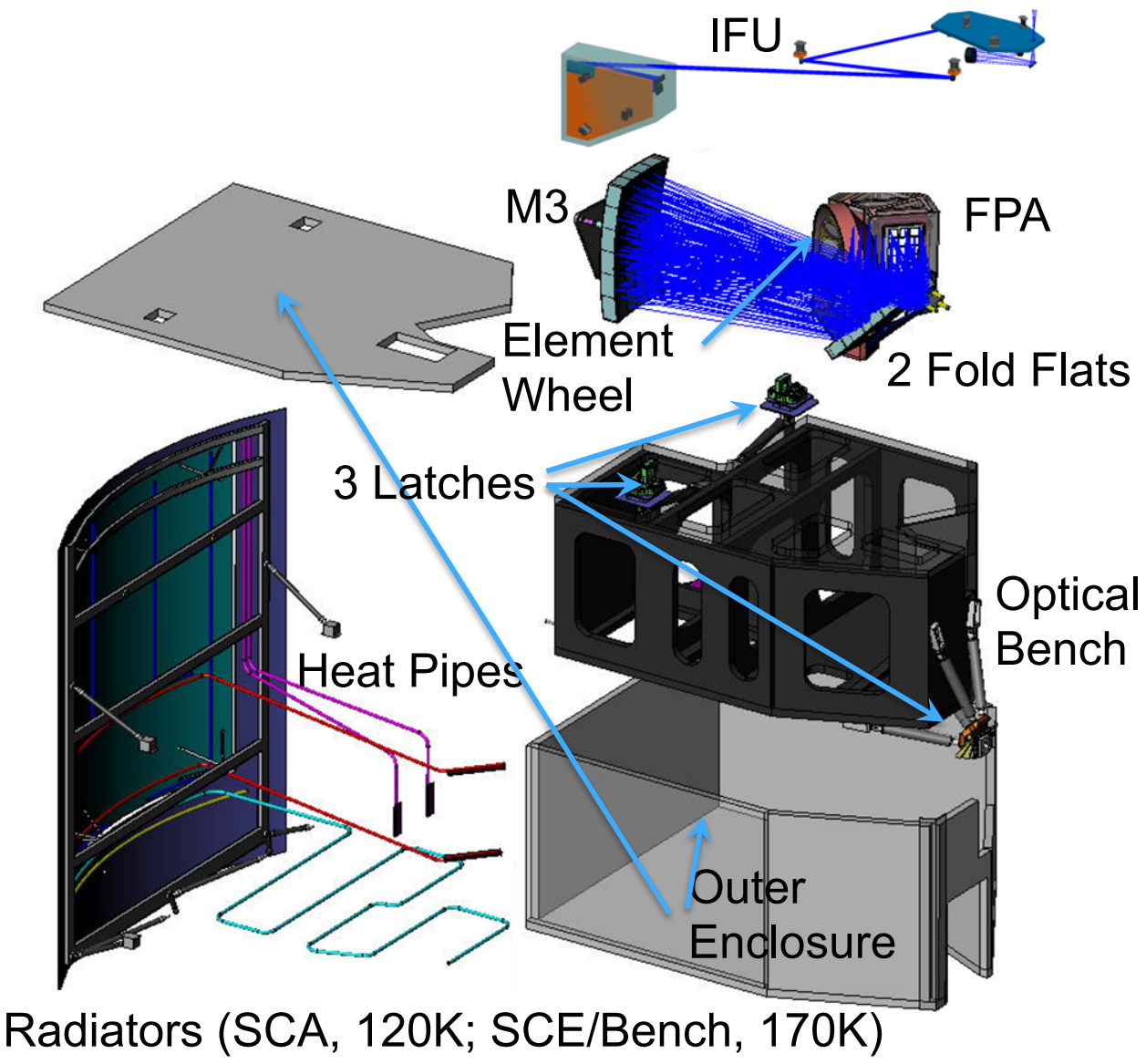

**Figure 3-8: Expanded view of the wide field instrument showing the major elements. The instrument electronics are located in a serviceable spacecraft module. The optical bench and outer enclosure are Al honeycomb panels with composite facesheets. Harnessing is not shown.**

| Band | Element Name | Min | Max | Center | Width | Dispersion |
|---|---|---|---|---|---|---|
| Element Wheel | | | | | | |
| Z | Z087 | 0.760 | 0.977 | 0.869 | 0.217 | 4 |
| Y | Y106 | 0.927 | 1.192 | 1.060 | 0.265 | 4 |
| J | J129 | 1.131 | 1.454 | 1.293 | 0.323 | 4 |
| H | H158 | 1.380 | 1.774 | 1.577 | 0.394 | 4 |
| | F184 | 1.683 | 2.000 | 1.842 | 0.317 | 5.81 |
| Wide | W149 | 0.927 | 2.000 | 1.485 | 1.030 | 1.44 |
| GRS | GRS Grism | 1.35 | 1.95 | 1.65 | 0.6 | $D_\theta$ = ~150 |

**Table 3-2: Filter and disperser descriptions in the wide-field channel spectral selection element wheel; Wavelengths are in μm. GRS grism dispersion is in units of arcsec with $D_\theta = \lambda/(\delta\lambda/\delta\Theta)$ where $\Theta$ is sky angle.**

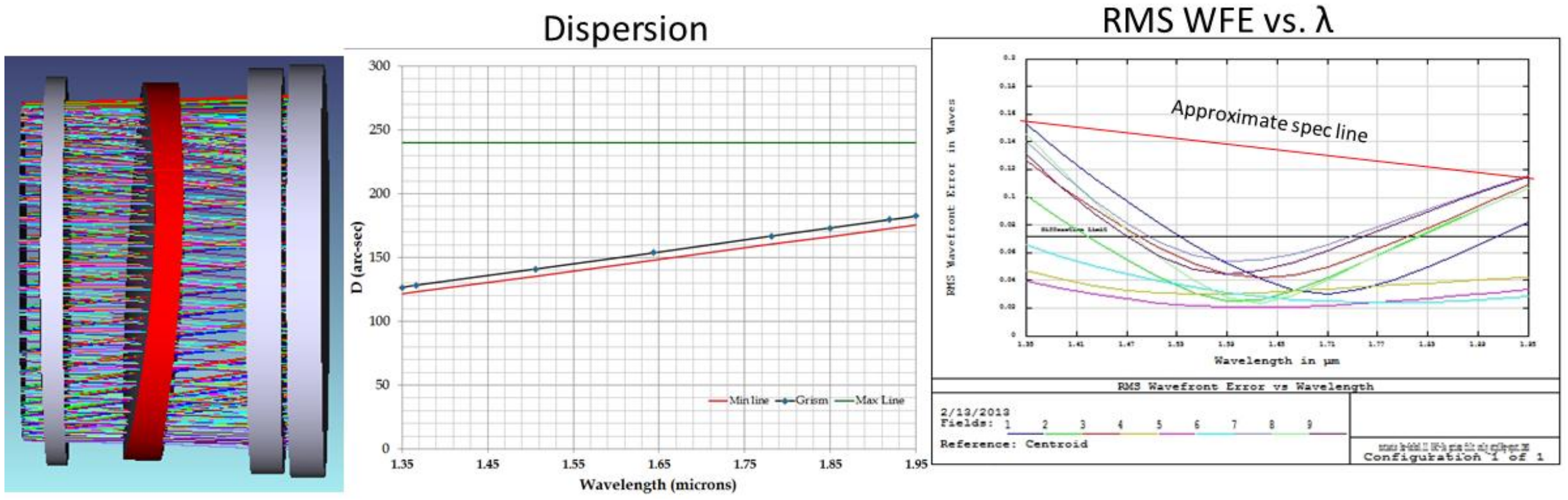

**Figure 3-9: The GRS grism assembly (left), its spectral dispersion (center), and its wavefront error over a 3x3 field point grid and spectral range (right).**

thermally isolating struts. Mirror and filter mounts are made from Ti and other materials, as needed, to appropriately athermalize the design. Thermally isolating flexures on the outer enclosure tie it to the instrument side of the latches in three locations without distorting or impacting the alignment of the optical bench. The outer enclosure supports inner and outer MLI blankets that thermally isolate the optical bench, two radiators that cool the optical bench and the FPA, the connector blind mate bulkhead, and the instrument servicing grapple

**Mechanisms**

The element wheel is the only mechanism routinely used in science operations. The EW's canted design allows for precise placement of any one of the 6 filters or the grism in a space-constrained volume of the instrument. All elements are transmissive, with modern ion-assisted, highly stable, thin film coating stacks. The wheel assembly includes a cold pupil mask (~100 mm diameter), which blocks the image of the highly emissive struts and obscurations in the telescope pupil so as to limit parasitic thermal input into the focal plane. The wheel mechanism includes a DC brushless motor with redundant windings that drive a spur and ring gear combination. Motor control is closed loop using a resolver for position feedback. Once positioned, the element is held in place by a detent arm that acts on the wheel via roller bearings. The detent arm also serves as a restraint during launch. No power is required to hold the wheel's position during science operations. All drive electronics for the wheel and detent arm are fully redundant with completely independent wiring. Figure 3-10 shows the EW including the detent arm.

A second mechanism is included to compensate for structural changes as the instrument ages and the cold alignment changes. Operation of this adjustment after commissioning is expected to be on an as needed basis to correct for misalignments due to long term changes such as moisture outgassing from the composite structural components in the telescope or instrument. This mechanism is attached to the F2 fold flat and allows for 2 rotational and one translational degrees of freedom. The mechanism uses piezo based actuators that do not require power to hold position. The resolution of the actuator is on the order of 5 $\mu$m with a total travel of +/- 2 millimeters. The mechanism is secured during launch using a launch lock. All drive electronics for the actuators are fully redundant with completely independent wiring.

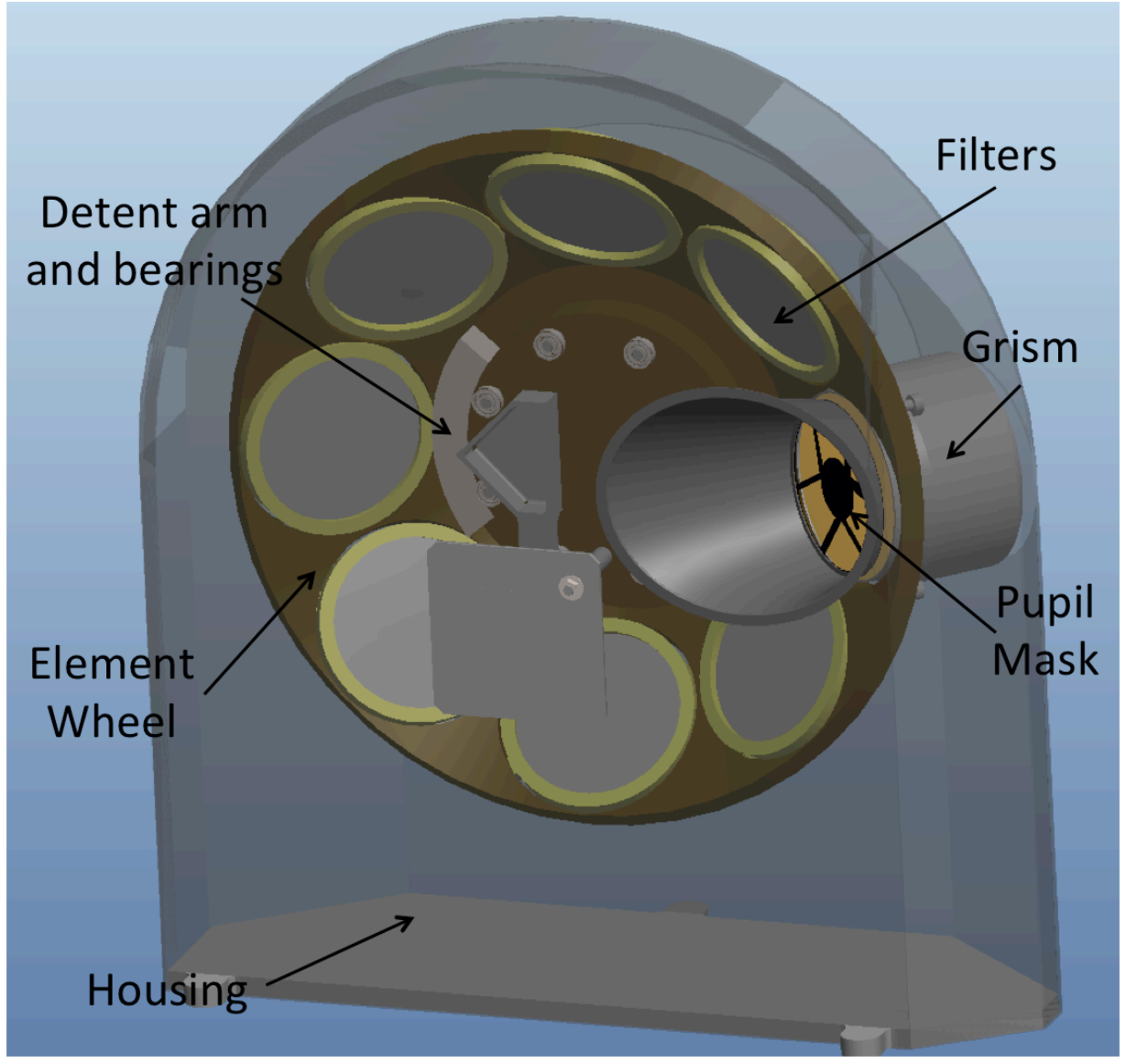

**Figure 3-10: Element wheel detail, including detent arm and bearings.**

**Thermal**

The wide field channel meets thermal requirements in the GEO orbit by combining a passive, cold-biased thermal design with precise heater control of the critical focal plane hardware. Constant conductance heat pipes are used to transport dissipated and parasitic heat loads to one of two passive radiators. The SCE/bench radiator includes integrated spreader heat pipes to cool the optical bench and SCE mounting plate to ≤170 K. The SCA radiator also includes spreader heat pipes and cools the SCA mosaic plate to ≤120 K. The bottom panel of the optical bench has an embedded ethane heat pipe that transports parasitic loads from the latches, optical aperture, and outer enclosure to the SCE/bench radiator. Two ethane heat pipes connect the SCE mounting plate to the SCE/bench and two methane heat pipes couple the SCA mosaic plate to the SCA radiator.

Proportional-Integral-Derivative (PID) heater control is required to meet SCA and SCE thermal stability requirements (±0.3K over a day, and ±10mK over any 150 sec observation) in the presence of environmental changes (primarily orbital variations in radiator Earth viewing. Three independent (and fully redundant) control zones control the top, middle, and bottom of the SCA mosaic plate, meeting the SCA stability and temperature requirements with margin and maintaining stable gradients across the focal plane (see Figure 3-11 and Figure 3-12). A separate, re-dundant PID controller is used to control one thermal zone on the SCE mounting plate, the SCE thermal variations to no more than ±1 K over any orbit.

**Focal plane assembly**

The FPA uses 18 state of the art H4RG SCAs mounted in a 6x3 pattern to a ≤120 K SiC mosaic plate. Each SCA has 4088 x 4088 active pixels, 10 μm in size, plus 4 extra reference pixels on each edge. The f/7.9 optical system maps each pixel to 0.11 arcsec square on the sky, providing an FPA active FOV of 0.281 deg$^2$. Readout wiring limits the extent to which the SCAs can be packed on the mosaic plate, with the minimum spacing in the x (6 SCA) direction being 2.5 mm and in the y (3) direction being 8.564 mm. The FPA (Figure 3-13) includes a light shield to limit direct illumination to active pixels to control stray light. Each SCA is constructed by hybridizing a HgCdTe sensor to a Si Read Out Integrated Circuit (ROIC) and then mounting that unit to its own SiC base. Flex cabling connects each SCA to its own SCA Control Electronics (SCE), with all 18 SCEs being mounted to a ≤170 K aluminum mounting plate located <12 inches from the SCA mosaic plate. Each SCA is (simultaneously, in sync with all others) read out non-destructively by its SCE in ~5.2s via 32 parallel outputs operating at 100 KHz. One 20 x 20 pixel guide window is also read out in sync at 20 Hz from each SCA (position in each SCA variable). All SCA-harness-SCE units are identical, simplifying production and sparing. A thermal shield with cable perforations separates the 120 K SCA and 170 K SCE plates to limit radiative coupling. Each plate is supported off the FPA's low thermal conductance composite truss structure using thermally isolating Ti flexures. The entire assembly is surrounded, except for the incoming optical beam, by a combined radiation and stray light shield enclosure. The shield extends around the second fold flat (F2) and is primarily designed to minimize the exposure of the SCAs and SCEs to the trapped electron environment in the geosynchronous orbit.

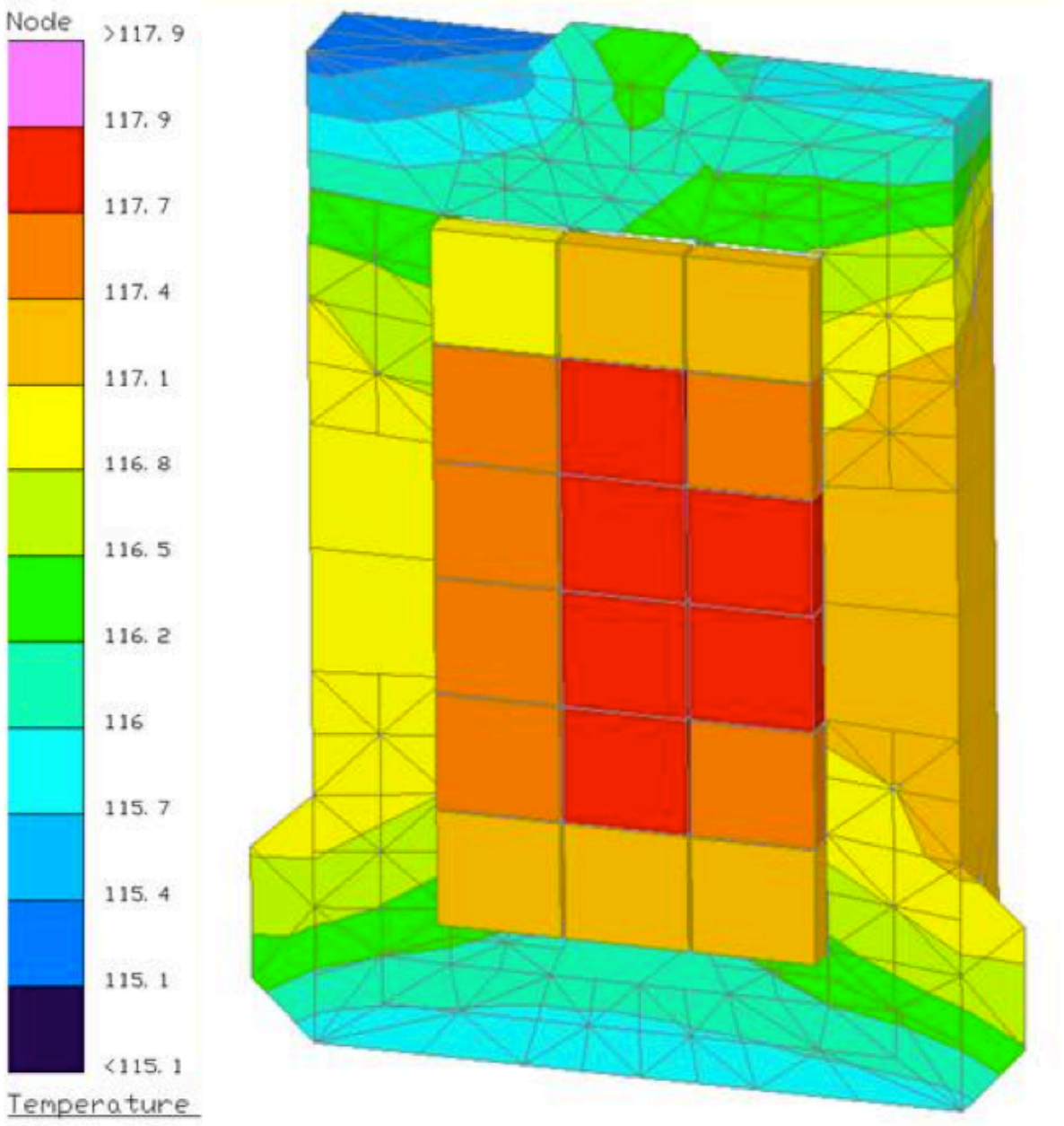

**Figure 3-11: Thermal map of the focal plane assembly, showing small (1K) gradients. This map is stable over several observing seasons and is insensitive to telescope pointing.**

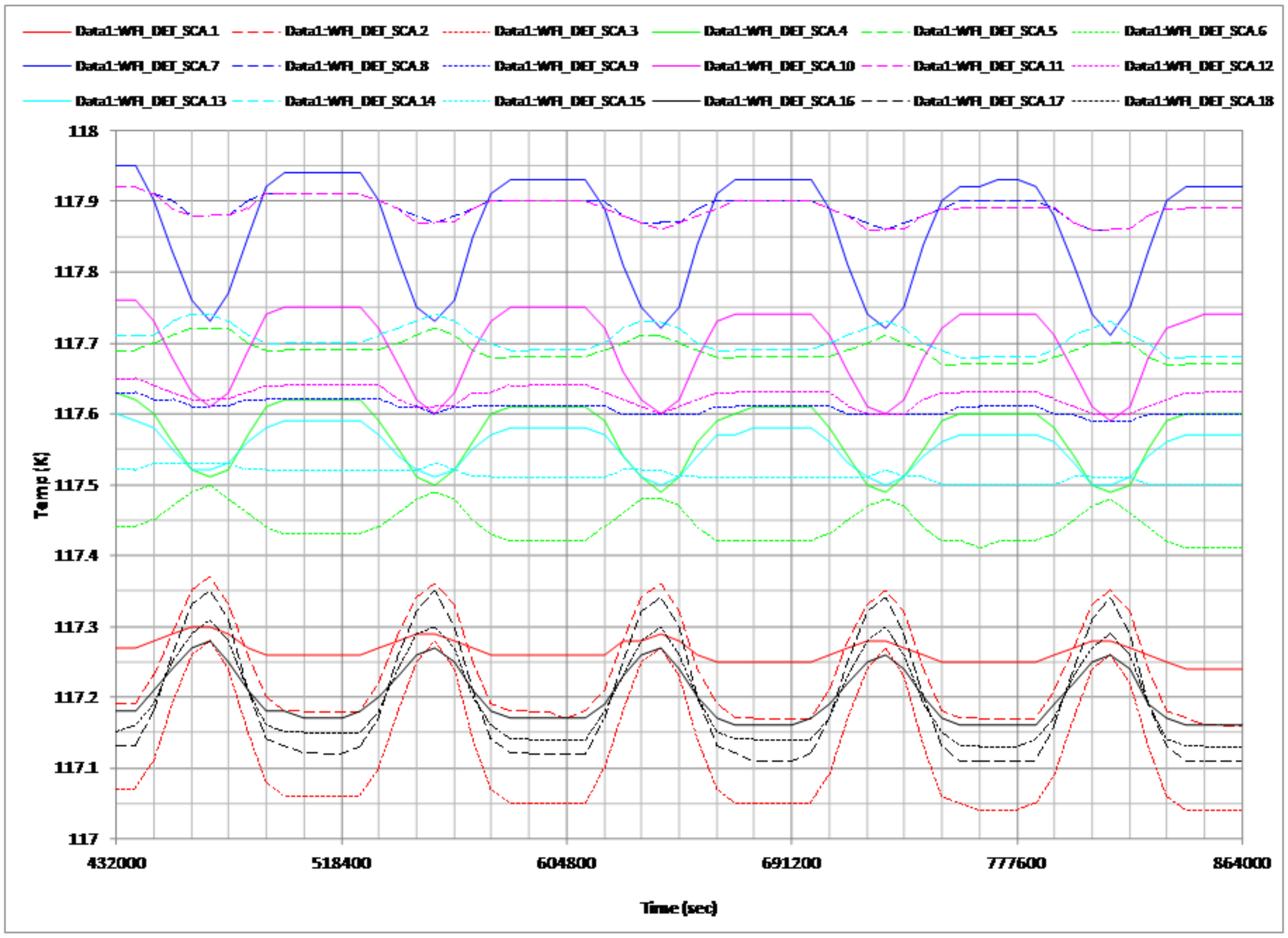

**Figure 3-12: Thermal stability of each of the SCAs over several days in an inertially fixed attitude. Short and long term stability requirements are met using a PID controller.**

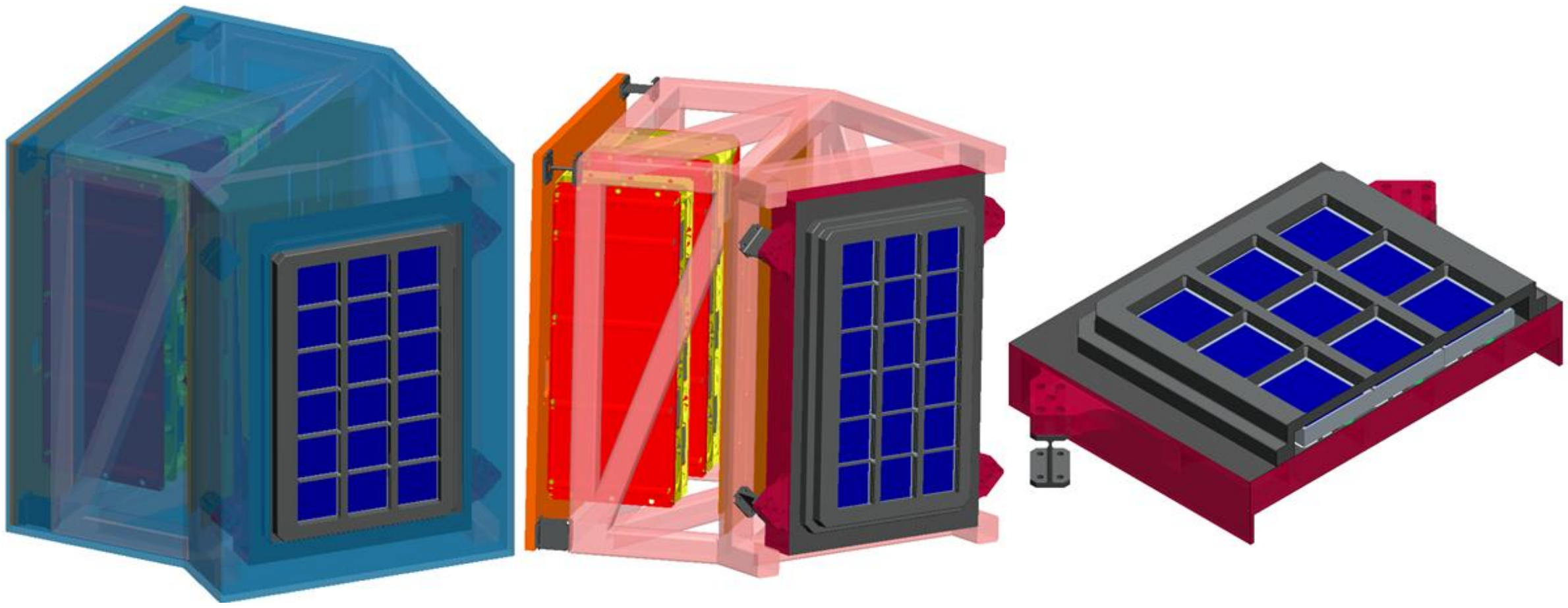

**Figure 3-13: Focal plane assembly; Left panel: entire assembly including SCAs (dark blue), light shield (grey), and radiation shield (light blue). Middle panel: view with radiation shield removed, showing composite structure (pink), SCE cards (red/yellow), and thermal standoffs (greeen). Right panel: cutaway view of SCA mosaic (SiC, maroon) and SCA carriers (grey).**

The H4RG 4k x 4k HgCdTe sensor itself is the only technology driver for the instrument. A discussion of the technology development requirements and plan is given below in §3.3.5.

**Wide Field Electronics and Wide Field Channel Signal Flow**

The wide field instrument electronics (handling both the wide field and IFU channels) are summarized (including redundancy) in the Figure 3-14 electronics block diagram. Four warm electronics boxes (Instrument Command and Data Handling box, or ICDH, Mechanisms Control Box, or MCB, and two Focal Plane Electronics, or FPE, boxes) are mounted to the serviceable instrument warm electronics module provided by the spacecraft, with interconnects to the cold instrument module being made via "blind mate" connections provided for both modules to harness permanently mounted to the spacecraft and the instrument carrier. ICDH functions are noted in detail in the block diagram, with key areas being the lossless data compression of the 19 science image data streams and their multiplexing for delivery to the spacecraft, the provision of the SCA and SCE thermal control loops, and the processing of the SCA guide windows to deliver 10 Hz quaternion updates to the spacecraft Attitude Control System. The MCB controls the EW to select filter/grism positions on a routine bases, and the F2 fold flat optical alignment mechanism is used during commissioning to align the optical pupil to the pupil mask. The signal flow off the focal plane is controlled by the cold SCEs mounted in the FPA and the warm SCE Control Unit (SCU) boards mounted in the FPE boxes on the spacecraft (one dedicated SCE/SCU chain for every SCA, all processing data in parallel). During an integration, image frames interleaved with guide window data are delivered to an SCU for processing every ~5.2 s. The guide window data is stripped out and sent to the ICDH for quaternion processing. The image frame data is handled two ways: averages of groups of four frames are passed to the ICDH for downlink, and Sample-Up-The-Ramp (SUTR)[107] processing is applied to all readouts. A table of cosmic-ray hits and pixel saturation data is passed to the ICDH for downlink at the end of each exposure. This table adds only a few percent to the total data volume, but the availability of corrections at the full 5.2 second readout ca-dence mitigates the more frequent incidence of radiation events in the geosynchronous orbit (compared to the Sun-Earth L2 orbit of prior WIRST studies) and maximizes the signal-to-noise that can be obtained in ground processing of the "multi-accum" style raw data. The ICDH applies lossless compression and multiplexes the 18 parallel data streams for transmission to the spacecraft. Preliminary studies indicate that compression factors greater than two can be achieved; a factor of 2 has been assumed in the data volume estimates.

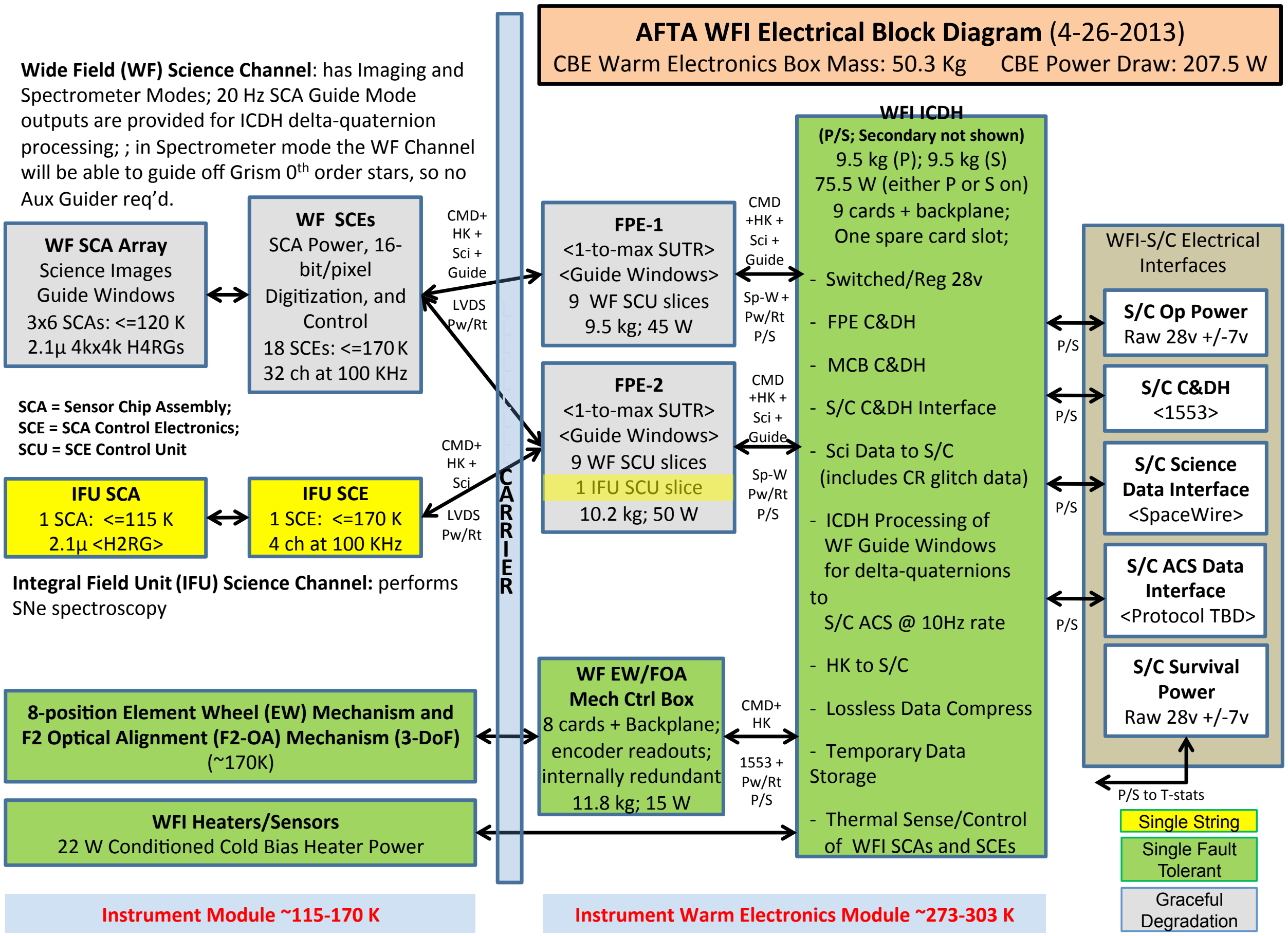

**Figure 3-14: Electronics block diagram for the wide-field instrument.**

### 3.3.2 *Integral Field Unit Channel*

The integral field unit is a separate instrument channel contained within the wide-field instrument. It uses a small field of view (3.00 x 3.15 arcsec) aperture to limit the sky background entering the instrument channel. An optical relay reimages this small field onto an image slicer and spectrograph covering the 0.6-2.0 µm spectral range (see Figure 3-15), resulting in a detector format wherein 21 slices, 3.0 x 0.15 arcsec each (see Figure 3-16), are imaged onto separate pixel sections of an 18 µm detector (H2RG, 2048 x 2048 x 18 µm). The slicer is based on the commercially available reflective image slicer from WinLight, a version of which has been through space environmental testing. While there are a substantial number of elements, they are small (a maximum 5 cm) and work in a slow optical beam with relaxed sta-bility tolerances.

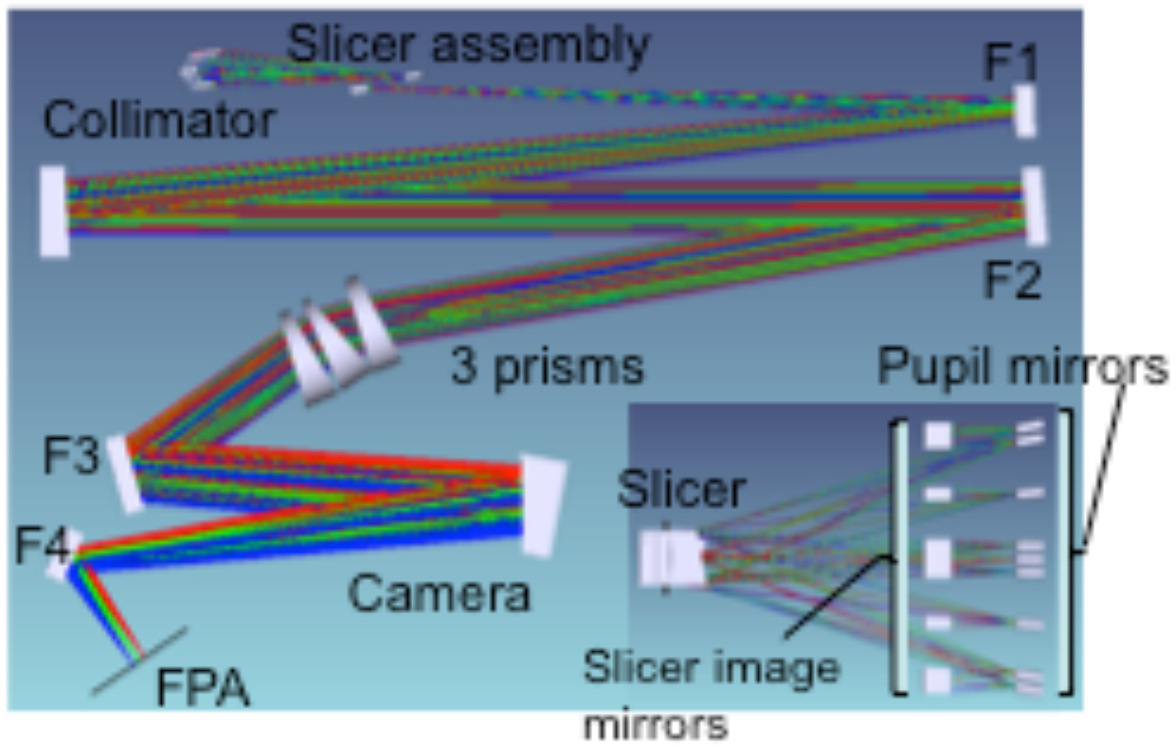

**Figure 3-15: Layout of the slicer assembly (inset) and spectrograph modules of the IFU. The relay is not shown.**

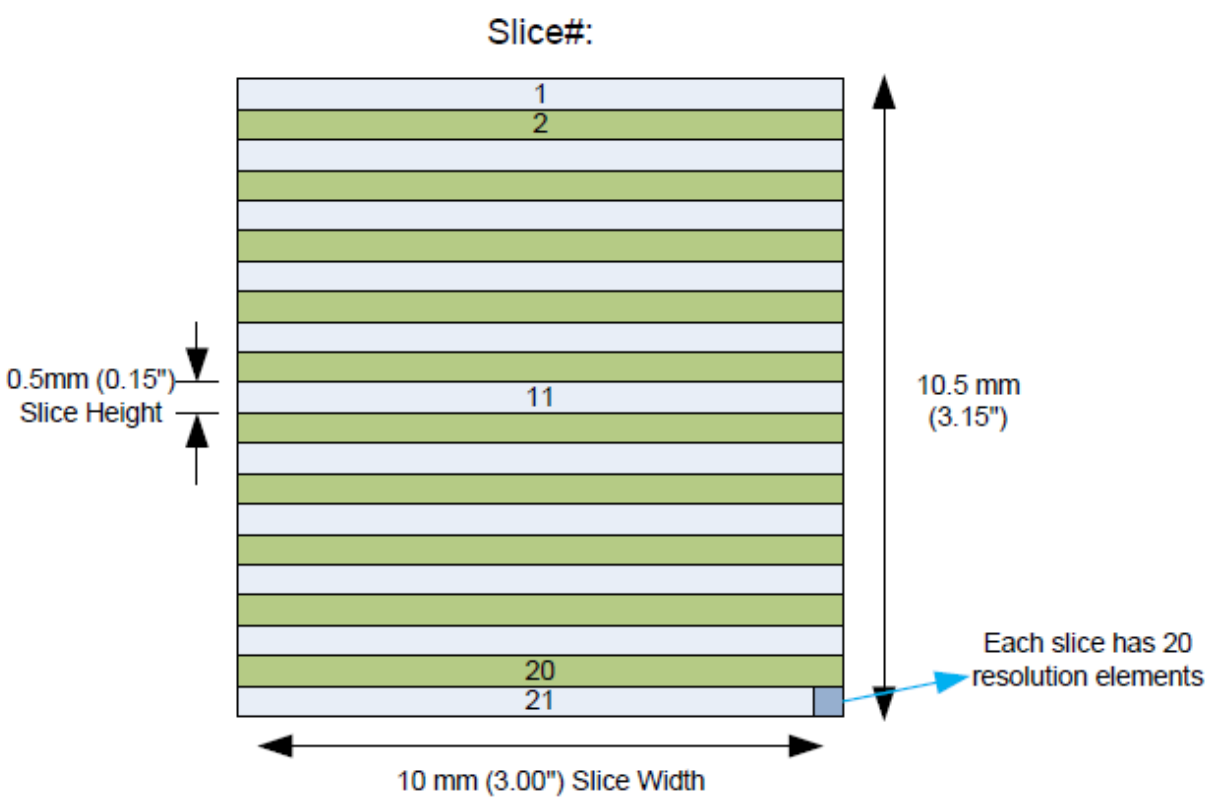

**Figure 3-16: The image slicer has 21 mirrors, 0.5 mm wide, each 0.15 arcsec field wide.**

The pixel scale at the focal plane is 0.075 arcsec. The resolving power is flattened by the use of a compound prism of materials Infrasil and SF11 to a typical value of 100 (2 pixels), see Figure 3-17a.

The slicer and spectrograph elements are packaged in a separate optical bench that is installed into the wide-field instrument housing. The opto-mechanical assembly is held at the same instrument temperature (170 K) as the rest of the wide-field instrument. The focal plane assembly includes a cryogenic heat pipe to maintain a temperature of 115 K at the focal plane and provide good thermal stability, so as to control the dark current and read noise that limits the signal to noise at each end of the spectral range.

While primarily designed around the need to observe high redshift type IA SNe, this observing mode would be of use in a GO program, to observe objects of interest, as the data product is a data cube (imagex, imagey, spectral position) with good stability and high S/N.

### 3.3.3 *Wide Field Instrument I&T*

The wide field instrument integration and test process provides early optical testing of small assemblies prior to full up instrument verification testing to minimize instrument I&T risk. The engineering development FPA, 2nd fold and EW will be verified as one unit with an elliptical mirror. The M3 and first fold will be verified separately. These two assemblies are then integrated into the full wide field instrument optical bench. Separately, the IFU will be integrated by building up the relay module, the slicer assembly, and the spectrograph. Again, these assemblies are verified separately prior to testing the full IFU channel and prior to its installation in the wide field optical bench. The wide field instrument integration effort is simplified by the availability of the telescope and instrument carrier along with a full aperture autocollimation flat mirror. This allows the actual telescope to be used for performance of the instrument in a Test As You Fly configuration. During payload I&T, the wide field instrument is tested with the telescope with its gravity offloader to provide the most realistic optical test. Once this testing is complete, the gravity offloader is removed from the telescope to place the telescope in flight configuration. At the same time, the engineering development FPA is removed and the flight FPA is installed. The use of the engineering FPA prior to final payload testing provides the maximum time to characterize the flight FPA. The payload is then retested prior to delivery to Observatory I&T (see Figure 3-18).

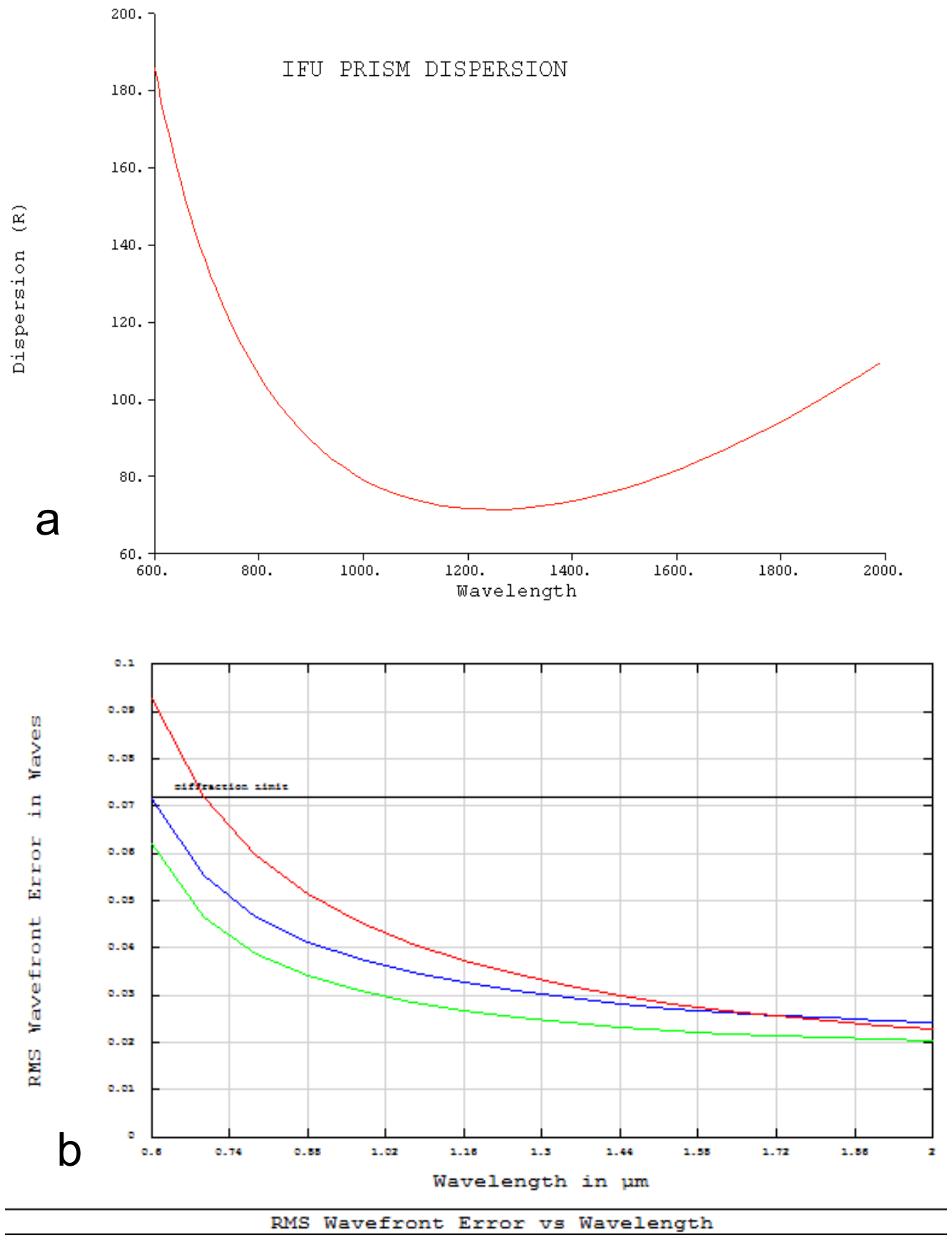

**Figure 3-17: The variation in resolving power of the IFU over the bandpass is shown in a. The diffraction limited imaging performance of an edge slice (center slices are better) at 3 field points (center, halfway out, and nearly at an edge of the slice, is shown in b.**

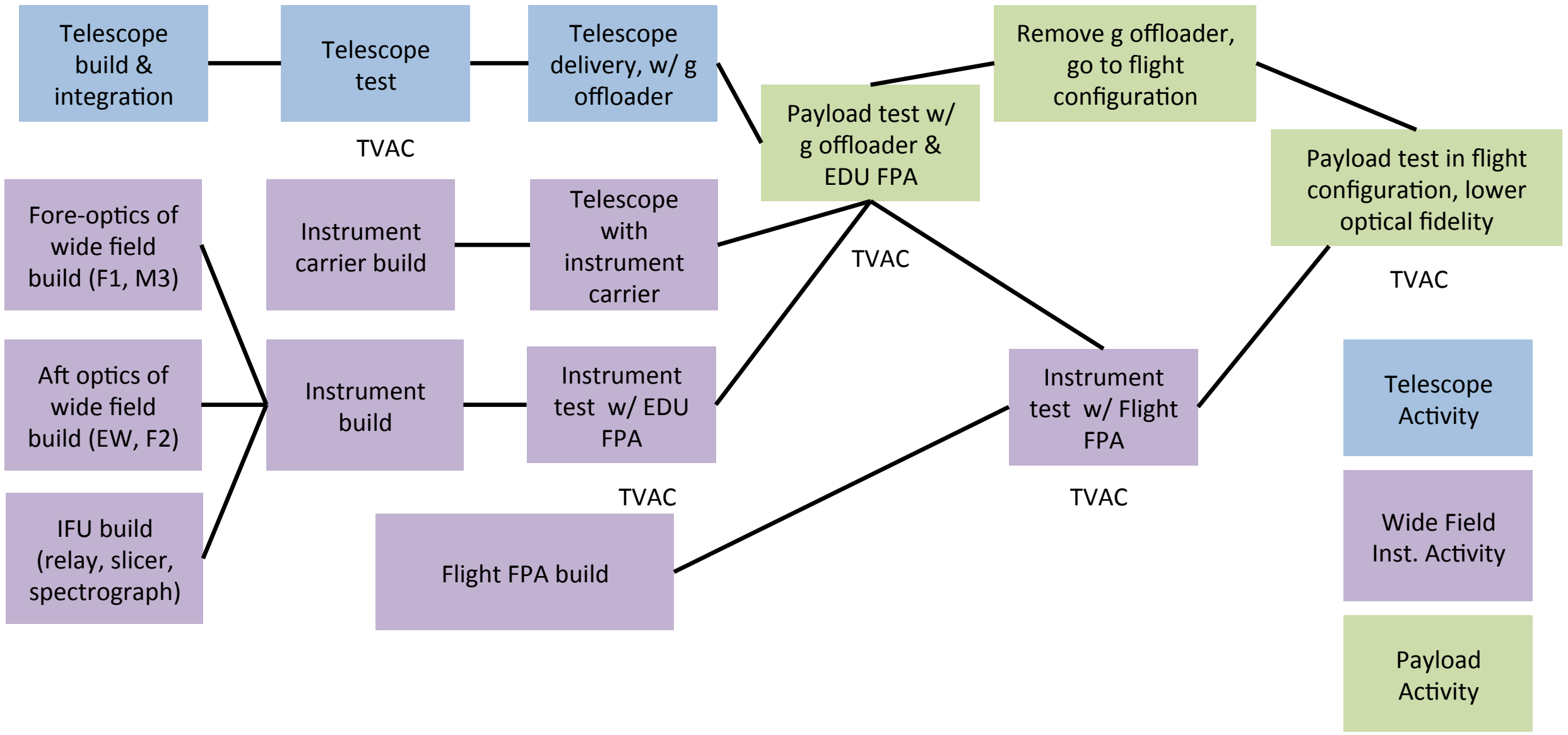

**Figure 3-18: I&T flow for the wide field instrument. The instrument is built up and tested in small assemblies to minimize the risk of finding instrument optical anomalies late in the instrument development.**

### 3.3.4 *Calibration*

The WFIRST-2.4 strategy is to use ground calibration methods to the maximum extent possible, reserving on-orbit calibration to verification of the ground results and extending the calibrations where ground calibration may not be effective. To maintain the calibration requirements over the entire mission, not only are the calibrations important, but so are measurements of calibration stability. The latter will determine the need for and frequency of on-orbit calibrations. The WFIRST-2.4 calibration program will place strong emphasis not only on the areas requiring calibration, but also on the verification of these calibrations, either on the ground or in orbit, using multiple techniques as cross-checks. The SN and Exoplanet fields are observed repeatedly over the lifetime of the mission, providing excellent opportunities to develop and use sky calibration standards.

All optical and detector components will be calibrated at the component, subsystem, and instrument levels. These data will be used to feed an integrated instrument calibration model that will be verified using an end-to-end payload-level thermal vacuum test. This test will involve a full-aperture (2.4 meter) diameter collimated beam that will test for optical wavefront error as well as photometry.

The exoplanet program imposes some constraints on photometric calibration, but also provides a unique opportunity to meet these calibration requirements. Proper measurements of the stellar and planetary light curves require stable relative calibration to 0.1% (relative to nearby stars in the same detector) over the course of the event. These observations use mainly a single filter. Over the course of the mission, the fields are sampled many tens of thousands of times with a random dither. Most of these observations will be of stars without microlensing events. If the star is not a variable star, then the relative calibration will be monitored during the extensive number of observations to establish stability. Slow variations across the field-of-view and over time can thus be monitored and corrected for. The absolute calibration and occasional color measurements using the bluer color filter have less stringent calibration requirements (1%) that will be met by the more stringent calibration requirements for Dark Energy that are described below.

The Dark Energy observational methods have different calibration demands on instrument parameters and their accuracy. The SN Survey places the most stringent demands on absolute and inter-band photometric calibration. White Dwarfs and other suitable sky calibration targets will be used to calibrate the absolute flux as well as linearity of the imager over several orders of magnitude. This linearity will be tested on the ground, and verified with an on-orbit relative flux calibration system (if necessary). Observations of astronomical flux standards will be extended across the detector by means of the "self-calibration' techniques described by Fixsen, Mosely, and Arendt[108] and employed for calibration of Spitzer/IRAC.[109] Similar techniques achieved ~1% relative calibration accuracy for the SDSS imaging data[110] despite the difficulties posed by variable atmospheric absorption. The intra-pixel response function (quantum efficiency variations within a pixel) will be fully characterized by ground testing.

For the WL Survey, the requirement for galaxy ellipticity accuracy places significant demands on both the optical and detector subsystems. The uniformity and stability of the point spread function (PSF) needs to be strictly controlled and monitored to ensure a successful mission. This drives the need to characterize the detector intra-pixel response and the inter-pixel response (capacitive cross-coupling with nearest neighbors) for both magnitude as well as spatial and temporal variations. It is likely that the combined PSF effects, including spacecraft jitter, will have some variability on time scales of a single exposure. These residual effects will be continuously monitored with the observatory attitude control system and field stars[111] and downlinked to provide ancillary information for the scientific data analysis pipeline.

The GRS relies primarily on the GRS grism, which does not drive the calibration requirements for the mission. Established calibration techniques used for other space missions should be adequate to meet the relatively loose photometric and wavelength calibration requirements. Since the GRS uses the same focal plane as the more demanding imaging surveys, the small-scale flat field, nonlinearity, dark current, IPC, and intra-pixel response calibration for these surveys should be adequate for the GRS.

Operated in parallel during the wide surveys, the IFU will obtain tens of thousands of SEDs for objects also observed with the imaging and grism surveys. We expect the network of these multiply-observed objects will be important in the final overall photometric calibration plan.

### 3.3.5 *Enabling Technology*

H4RG-10 near-IR array detectors are critical to enabling WFIRST-2.4. The H4RG-10 is a 4096 x 4096 pixel HgCdTe detector array that has 10 μm pixels. While building on NASA's heritage developing 1k x 1k pixel H1R detectors for Hubble/WFC3, and 2k x 2k pixel H2RG detectors for JWST, the H4RG-10 advances the state of the art by approximately quadrupling the number of pixels per unit focal plane area (see Figure 3-19). The higher pixel density simplifies mechanical and thermal system design, as well as integration and test. Moreover, because the cost of making of an astronomy-grade HgCdTe sensor chip assembly (SCA) scales roughly with the photosensitive area, the H4RG-10 has the potential to significantly reduce the cost per pixel.

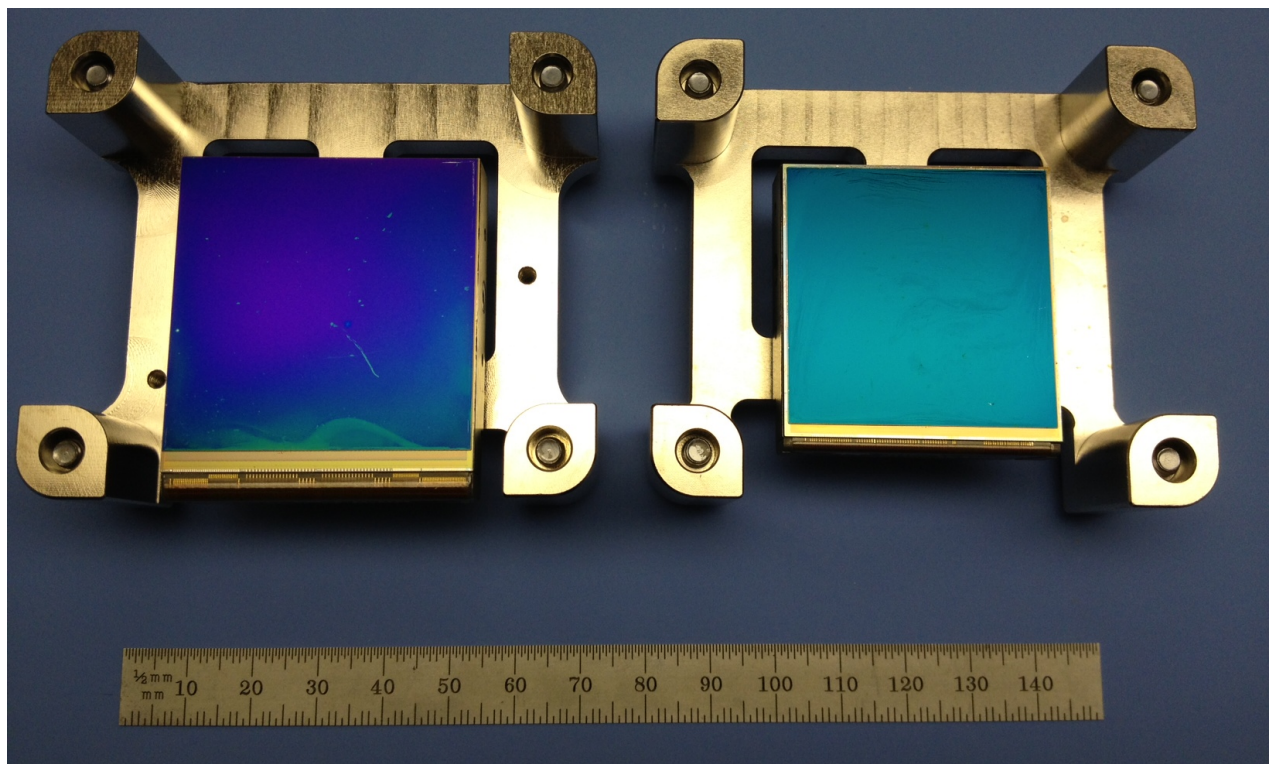

**Figure 3-19: Comparison of the size of an H4RG (4k x 4k pixels, 10 μm pixel size) on left to an H2RG (2k x 2k pixels, 18 μm pixel size) on the right.**

The H4RG-10 ROIC was originally developed for use with visible wavelength Si PIN detectors. In this context, NASA worked closely with the U.S. Naval Observatory to mature the H4RG-10 ROIC to TRL-6 for the Navy's Joint Milli-Arcsecond Pathfinder Survey (JMAPS) satellite. The only new element for WFIRST was hybridizing HgCdTe detectors to the H4RG-10 ROIC instead of Si PINs.

As a proof of concept, the WFIRST Project built and tested a small test lot of H4RG-10s. This was a success, both proving the concept and elevating the H4RG-10 to TRL-4. Out of a total of six devices constructed, five were functional, and four performed well enough to merit detailed characterization. The performance was very promising, with the exception of about 20% higher readout noise than is desired. These encouraging results formed the technical basis for a successful NASA ROSES/APRA Strategic Astrophysics Technology (SAT) proposal to continue H4RG-10 development. The SAT aims to achieve TRL-5, with TRL-6 as a goal.

After achieving TRL-6, the next step for WFIRST is to construct an engineering demonstration yield lot with a single design that validates the yield models (and therefore the flight development cost models). After this yield demonstration lot, the flight lot production can proceed.

### 3.4 Coronagraph Instrument

#### 3.4.1 *Coronagraph Design and Performance*

The coronagraph instrument will take advantage of the 2.4-meter telescope's large aperture to provide novel exoplanet imaging science at approximately the same cost as an Explorer mission, including the additional operations time. The coronagraph is also directly responsive to the goals of the Astro2010 by maturing direct imaging technologies to TRL 9 for an Exo-Earth Imager in the next decade. The coronagraph design is based on the highly successful High Contrast Imaging Testbed (HCIT), with modifications to accommodate the telescope design, serviceability, volume constraints, and the addition of an Integral Field Spectrograph (IFS). Key coronagraph characteristics are given in Table 3-3. Demonstrations in the HCIT have already exceeded the contrast performance in Table 3-3 for on-axis unobscured pupil coronagraphs. Figure 3-20 shows the coronagraph instrument layout, which can accommodate one of either a Complex Lyot image plane mask[112], a Shaped Pupil mask[113], or a Vector Vortex mask[114]. An instrument configuration has also been developed for Phase-Induced Amplitude Apodization with a complex mask (PIAA-CMC)[115]. The Exoplanet Exploration Program Technology Development for Exoplanet Missions (TDEM) has funded development for these as well as other coronagraph concepts. A trade study to be performed in the following months will identify

| | | |
|---|---|---|
| **Bandpass** | 400-1000 nm | Measured sequentially in five 18% bands |
| **Inner Working Angle** | 100 mas | at 400 nm, 3λ/D driven by challenging pupil |
| | 250 mas | at 1 μm |
| **Outer Working Angle** | 1 arcsec | at 400 nm, limited by 64x64 DM |
| | 2.5 arcsec | at 1 μm |
| **Detection Limit** | Contrast=$10^{-9}$ | Cold Jupiters. Deeper contrast looks unlikely due to pupil shape and extreme stability requirements. |
| **Spectral Resolution** | 70 | With IFS |
| **IFS Spatial Sampling** | 17 mas | This is Nyquist for λ = 400 nm |

**Table 3-3: Key coronagraph instrument characteristics**

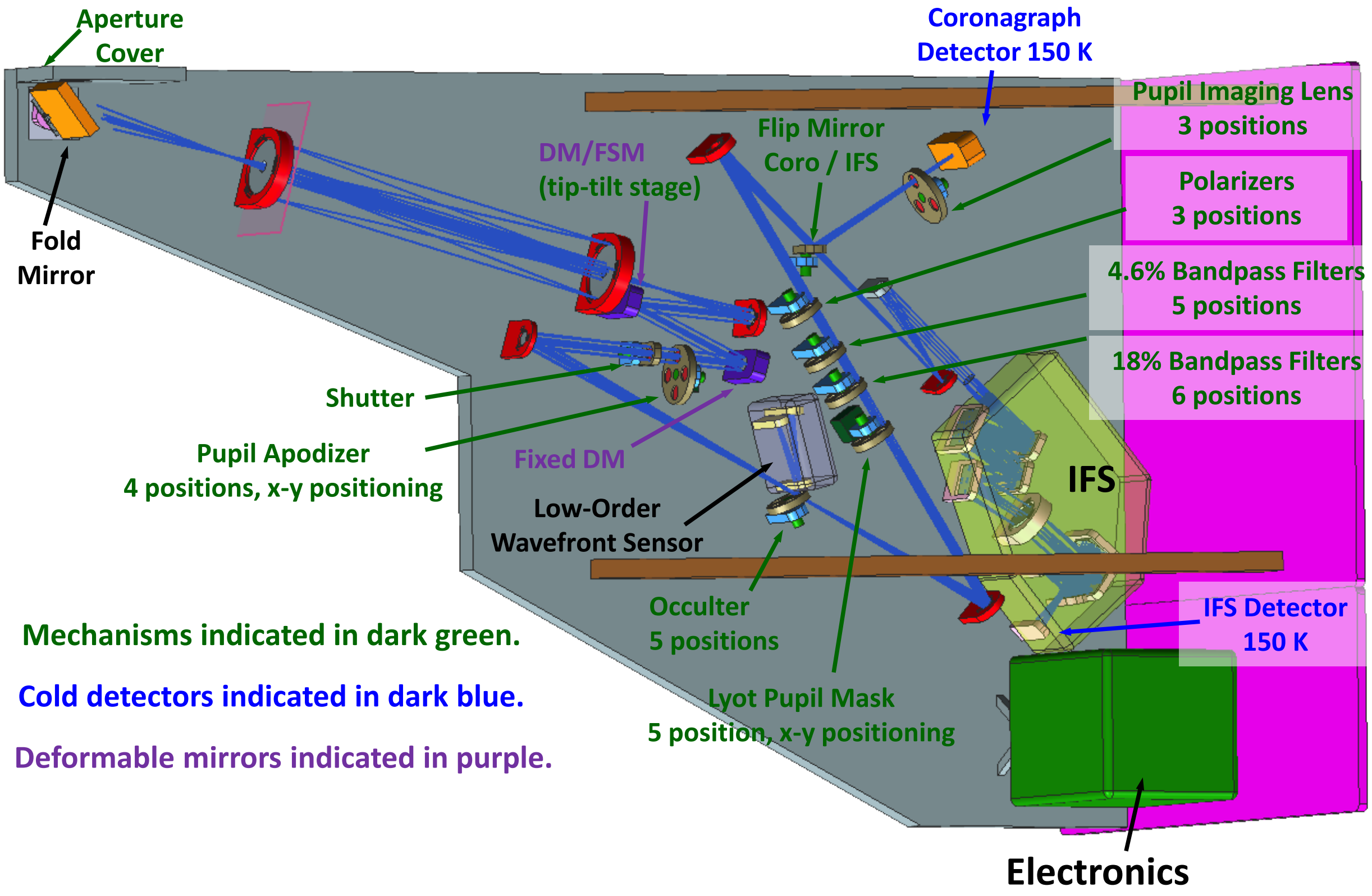

**Figure 3-20: Coronagraph layout within the robotically serviceable module.**

which coronagraph option is most appropriate for the WFIRST-2.4 coronagraph.

The snout on the left side of the volume serves to transfer the beam into the main body while avoiding interference with the surrounding wide-field instrument field of regard. All of the optics are mounted on a stiff, athermalized, planar optical bench that is attached near one side of the volume. A fixed pickoff mirror reflects the light to a pair of relay optics. The anamorphic first relay optic compensates for the aspheric primary and tilted secondary mirrors of the baseline telescope design. It is crucial that the beam remain stable to the micron level across the surface of this optic. Following the relay, an off-axis parabola (OAP) forms a pupil image where a microelectromechanical systems (MEMS) deformable mirror (DM) is placed. The mirror is on a tip-tilt stage and serves as the fine steering mirror. Along with a second DM, the two DMs form a sequential wavefront control system (WFCS) that compensates for both phase and amplitude errors in the telescope and coronagraph optics. Just downstream from the second DM, a filter wheel holds the apodizers that control diffraction before being reflected to the image plane, which is held in a second filter wheel. The light at the center of the image plane reflects off the mask to a low-order wavefront sensor (LOWFS). The LOWFS provides feedback for fine guiding as well as low-order aberration measurements for the wavefront control system. This is a crucial element of the system that relaxes the temporal-thermal requirements on the primary mirror and secondary mirror position. Following the image plane masks, an OAP collimates the beam and reimages the pupil onto a Lyot stop. This plane contains masks that strip off most of the remaining starlight while passing the planet light.

The portion of the coronagraph optical train between the pickoff mirror and the Lyot stop must be maintained at a cleanliness level of CL100 to ensure that instrument scatter is well below the exozodiacal background light level. Because the image plane mask removes most of the starlight, the downstream optics are not as critical.

After the Lyot plane, a pair of bandpass filter wheels selects either a 4.6% wide band for wavefront sensing, or an 18% band for planet detection and characterization. A polarizer wheel is included in this branch for polarimetry studies.

The design includes both a direct imaging detector and the lenslet-based IFS. The direct imager is a 1k x 1k format detector fed by a flip-in mirror. This branch also aids the WFCS with a pupil imaging lens for focus diversity estimation. The IFS arm has a 140x140 lenslet array that samples the image plane with a 17 milli-arcsec pitch. It has a nominally constant dispersion across the full 0.4 – 1 μm spectral range, with a free spectral range of 18%. The IFS detector has a 2k x 2k format with ultra-low noise.

### 3.4.2 *Coronagraph Operations*

Figure 3-21 shows the coronagraph observation timeline. After pointing at a chosen target, the telescope requires a roughly six hour thermal settling time before adequate stability is achieved. The initial observations will focus on discovery of planets near the target star, summing IFS channels into a single 18% spectral band to achieve adequate signal-to-noise ratios in relatively short time periods. When a planet is found, additional observations will be made for longer time periods with full spectral resolution for planet characterization. The required observation time depends strongly on the coronagraph performance parameters and the actual planet contrast. For this analysis a $10^{-8}$ instrument contrast and a $10^{-9}$ planet contrast are assumed. The design has 10% overall throughput to the image plane and 6% throughput to the IFS detector (including reflection losses) in the entire annulus around the target star. Other assumptions include a nearly noiseless detector (such as an EMCCD), a 4.8 magnitude star, and a signal-to-noise ratio of 5 for discovery and for 100% spectral line depth during spectroscopy. Figure 3-21 shows that the initial settling, wavefront control, and discovery observation take place within 24 hours. Spectroscopy covering 400-1000 nm in 5 sequential bands takes approximately 72 hours per band with an additional 6 hours per band allocated for wavefront control.

### 3.4.3 *Coronagraph Implementation*

High contrast coronagraphy requires extreme stability of the optical system over the observation time period. For any of the possible coronagraph concepts, WFIRST-2.4 coronagraph optics require ~0.25 μm relative stability during an observation. The coronagraph optical bench (shown in grey in Figure 3-20) must be positioned to within 1 mm with respect to the telescope, and this relative position must be stable to 0.5 μm during an observation period. The aluminum optical bench is thermally stabilized to <10 mK to achieve 0.25 μm internal stability. Three latches on the back of the optical bench will secure the coronagraph to the instrument carrier. The instrument carrier must be thermally stable enough to support the required dimensional stability, and a metrology and fine positioning system may be needed.

| Hour | Day 1 | Day 7 | Day 13 |
|---|---|---|---|
| 0-6 | Point telescope<br>Pointing control on<br>Telescope thermal settling time | IFS integration in 18% band #2 | IFS integration in 18% band #4 |
| 6-12 | Set contrast to $10^{-8}$ with multiple sets of probes<br>Wavefront estimation- IFS pixels summed to 4.6% bands- 4 probes | | |
| 12-18 | Science observation summing IFS pixels up to 18% band | | |
| 18-24 | IFS; 18% band #1 - Set contrast to $10^{-8}$ with multiple sets of probes<br>Wavefront estimation- IFS pixels summed to 4.6% bands- 4 probes | | |
| | Day 2 | Day 8 | Day 14 |
| 0-6 | IFS integration in 18% band #1 | IFS integration in 18% band #2 | IFS integration in 18% band #4 |
| 6-12 | | IFS; 18% band #3 - Set contrast to $10^{-8}$ with multiple sets of probes<br>Wavefront estimation- IFS pixels summed to 4.6% bands- 4 probes | |
| 12-18 | | IFS integration in 18% band #3 | |
| 18-24 | | | IFS; 18% band #5 - Set contrast to $10^{-8}$ with multiple sets of probes<br>Wavefront estimation- IFS pixels summed to 4.6% bands- 4 probes |
| | Day 3 | Day 9 | Day 15 |
| 0-24 | IFS integration in 18% band #1 | IFS integration in 18% band #3 | IFS integration in 18% band #5 |
| | Day 4 | Day 10 | Day 16 |
| 0-24 | IFS integration in 18% band #1 | IFS integration in 18% band #3 | IFS integration in 18% band #5 |
| | Day 5 | Day 11 | Day 17 |
| 0-6 | IFS; 18% band #2 - Set contrast to $10^{-8}$ with multiple sets of probes<br>Wavefront estimation- IFS pixels summed to 4.6% bands- 4 probes | IFS integration in 18% band #3 | IFS integration in 18% band #5 |
| 6-12 | IFS integration in 18% band #2 | | |
| 12-18 | | IFS; 18% band #4 - Set contrast to $10^{-8}$ with multiple sets of probes<br>Wavefront estimation- IFS pixels summed to 4.6% bands- 4 probes | |
| 18-24 | | IFS integration in 18% band #4 | |
| | Day 6 | Day 12 | |
| 0-24 | IFS integration in 18% band #2 | IFS integration in 18% band #4 | |

| | | |
|---|---|---|
| Telescope pointing and settling | Spectroscopy in 18% band #1 | Spectroscopy in 18% band #4 |
| Discovery using single 18% bandpass | Spectroscopy in 18% band #2 | Spectroscopy in 18% band #5 |
| | Spectroscopy in 18% band #3 | |

**Figure 3-21: Coronagraph observational timeline.**

The coronagraph optical bench assembly will be maintained at 290 K to ensure that the DMs are flat when unpowered. The LOWFS detector will be cooled to ≈250 K with a Peltier cooler. The coronagraph imaging detector and IFS detector will be cooled to ≈150 K by the radiators to minimize dark current. These radiators are located on the pink surface in Figure 3-20. Thermal isolation is provided at the three mechanical contacts to the instrument carrier (latch-es). Blanketing will isolate the coronagraph radiatively from the ~220 K instrument carrier. The coronagraph temperature will be maintained with its electronics power dissipation in addition to software-controlled heaters.

Figure 3-22 shows the coronagraph electronics block diagram. The spacecraft will supply redundant power at ≈29V to redundant DC-DC converters in the instrument. A redundant 1553 interface will handle commands and housekeeping, while redundant SpaceWire interfaces will be used for science data. Control and data handling will be performed by a RAD750 flight computer and a field-programmable gate array (FPGA). This capability will not be redundant, as its reliability has been shown to be very high. The two DMs will be driven by separate power supplies and electronics. If one DM or driver completely fails, the second DM will still allow much of the planned science. Individual DM elements or the drive electronics for individual DM elements can fail with minimal effect on DM performance. All actuators will have redundant windings, with redundant drive electronics (only one powered). Detector drive electronics will also be redundant. The spacecraft will read a limited number of internal coronagraph temperature sensors for thermal monitoring while the coronagraph is off.

The coronagraph image plane and IFS detectors must have extremely low noise. Dark current noise must be no more than 1 e- over a thousand to several thousand seconds, while read noise must be no more than 1 e- for the sum of 100-1000 reads. Silicon detectors can provide this low dark-current noise when cooled to ≈150 K. Very low read noise is more difficult to obtain. Two detector options are under consideration. The e2v electron-multiplying charge-coupled devices (EMCCDs) have demonstrated the required read noise. 1kx1k devices are already commercially available, and 4kx4k devices should be available by 2014 or 2015. However, radiation damage can increase the read noise considerably. Work is needed to determine whether proper operating conditions can mitigate the radiation-induced read noise. A second promising technology is Geiger-mode avalanche photodiodes, being developed by the University of Rochester and MIT Lincoln Laboratories. Further work will be needed to demonstrate arrays with adequate performance and array size.

Current best estimates of coronagraph mass, power and data rate are listed in Table 3-4. The data

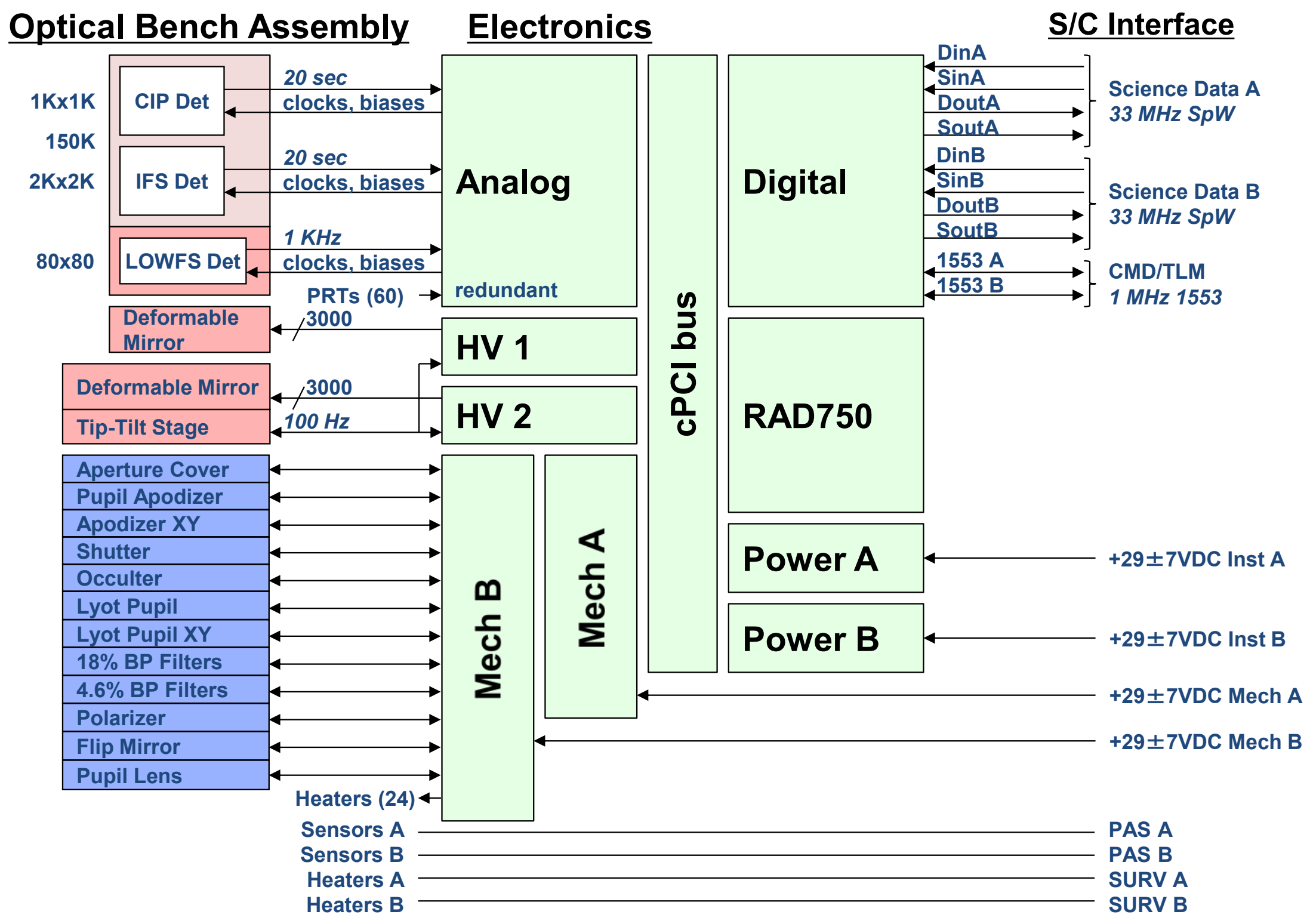

**Figure 3-22: Coronagraph electronics block diagram.**

rate depends strongly on how disruptive radiation hits are to the detector signals.

### 3.4.4 *Coronagraph Technology Development*

The WFIRST-2.4 coronagraph benefits from over 5 years of starlight suppression technology development funded by the NASA Exoplanet Exploration Program. The very successful HCIT at JPL with contributions from external investigators through NASA ROSES TDEMs has already exceeded contrast performance levels required by the WFIRST-2.4 coronagraph ($5 \times 10^{-10}$ at 10% bandwidth), thus validating that the physics and models of starlight suppression are understood to that level, and that wavefront sensing and control techniques are in hand. The HCIT results are based on coronagraph designs for off-axis, unobscured telescopes. The use of the 2.4-meter telescope does bring new challenges because of the obscuration and diffraction created by its on-axis design. A technology development approach is presented in Figure 3-23, which brings the coronagraph instrument system to TRL 5 by mission start in FY17 and to TRL 6 by instrument PDR at the end of FY18. An accelerated development approach is also feasible should additional funds be provided through the budget process for a mission start as early as FY15.

Deformable mirrors with capabilities similar to those required for the coronagraph are currently manufactured by Xinetics and Boston Micromachines. Xinetics DMs, which are used in the JPL coronagraph testbeds, utilize a bulk electrorestrictive material to deform the mirrors. This technology currently has a minimum mirror element size of 1 mm. Smaller mirror elements are needed to achieve the compact coronagraph design required for this mission. Boston Micromachines produces silicon micromachined DMs, which allow small element sizes as well as much thinner DMs. With SBIR and TDEM funding, Boston Micromachines is modifying the mirror design to improve reliability, reduce both large and small-scale deformations in the undeformed mirrors, and flight qualify their DMs. This work will be completed in 2014. If all current activities are successful, these DMs should meet the coronagraph requirements when these contracts are complete.

| | |
|---|---|
| Mass | 111 kg |
| Power | 80 W |
| Data Volume | 30 Gbits/day |

**Table 3-4: Current best estimates for coronagraph mass, power, and data volume.**

Detector development is also needed to meet the coronagraph requirements, which include very low noise and radiation tolerance. The e2v EMCCD detectors currently meet coronagraph performance requirements, but their performance may degrade to an unacceptable level after the expected radiation exposure. Significant test activities will be required to determine whether proper operating conditions will provide adequate performance through the entire mission. Geiger-mode avalanche photodiodes may theoretically provide the required performance and radiation tolerance. However, currently demonstrated performance is inadequate. Additional funding in this area may produce acceptable arrays in time for this mission.

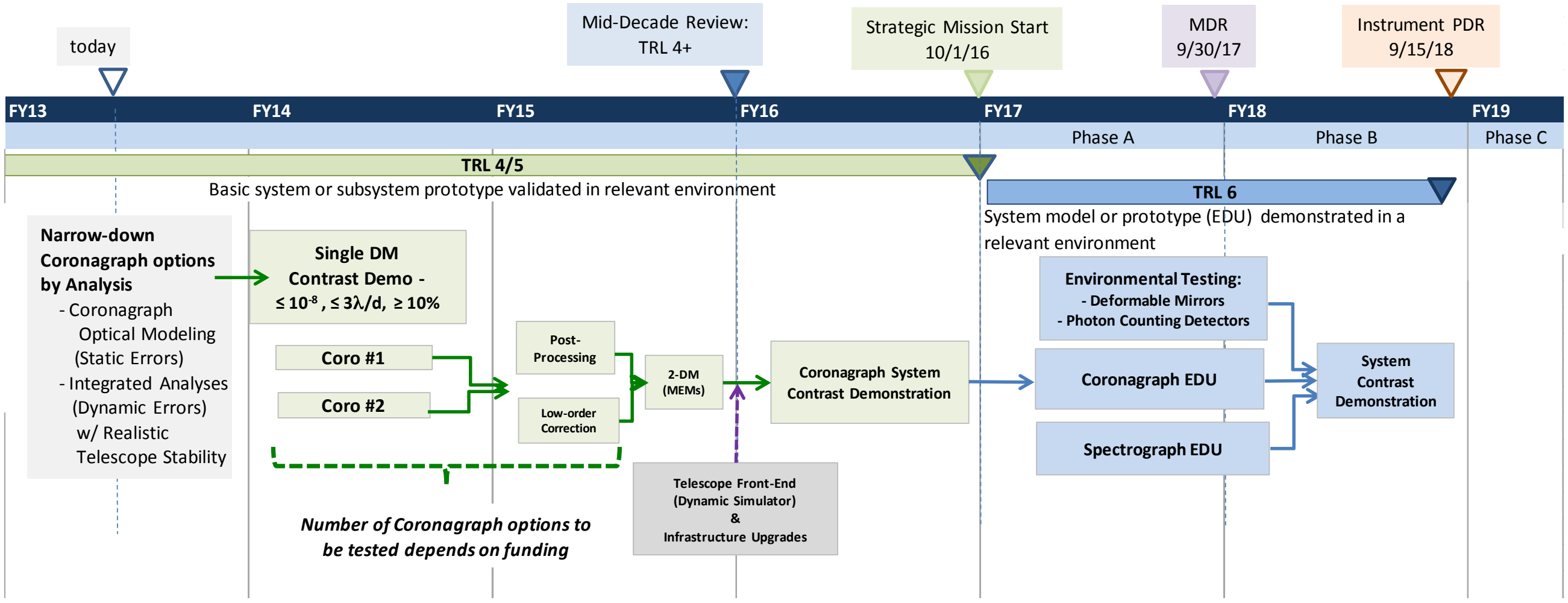

**Figure 3-23: Coronagraph technology development plan to TRL 6 by PDR (2018). An accelerated development approach is feasible in the event of an FY15 or FY16 mission start.**

The coronagraph elements (complex image plane mask, pupil apodizers, LOWFS) change significantly for obscured apertures and require further laboratory testing. Alternative designs such as PIAA-CMC and the vector vortex coronagraph may significantly improve the throughput and search space compared to the hybrid Lyot solution, but require further laboratory testing. A trade study will identify which of these mask options will be pursued for technology demonstration for the WFIRST-2.4 coronagraph. The LOWFS has been used to perform 0.01 λ/D rigid body tip-tilt estimation and control, more than adequate for WFIRST-2.4. But it has not yet been used at the $10^{-9}$ contrast level to estimate aberrations other than tip-tilt.

### 3.4.5 *Coronagraph Integration and Test*

The coronagraph integration and test (I&T) flow is shown in Figure 3-24. First the active components (DMs, detectors and actuators) are fully tested. Then the optical bench is assembled and aligned, with a possible seating shake and subsequent alignment check/adjustment. In parallel, the electronics are assembled and fully bench tested. Next the electronics are integrated with the optical bench, and full functional and performance testing is done in vacuum. It may be possible to do some testing in air, using short integration times to avoid detector saturation. Once functionality and performance have been verified, the full instrument assembly can be completed. EMC/EMI testing follows, with any necessary modifications to the electronics or shielding. Then the thermal blankets are installed, followed by mass properties and dynamics testing. After a functional test, the instrument is placed in a vacuum chamber for bake out, thermal cycling, and thermal balance tests. Finally, the full functional and performance testing is done in vacuum with cooled detectors.

## 3.5 Fine Guidance System (FGS)

The WFIRST-2.4 surveys each have unique science requirements that drive the observatory pointing requirements. The absolute pointing accuracy requirements of <10 milli-arcseconds for pitch and yaw and 1 arcsecond for roll are driven by SN requirements for returning precisely to previously observed objects. WL drives the pointing stability of <20 milli-arcseconds pitch and yaw and 2 arcseconds for roll over an exposure. The FGS uses the guide window function implemented on the H4RG SCAs to meet these precise pointing requirements. One bright star in each of the 18 SCAs in the wide-field focal plane is used for guiding. This guide mode is used in all wide-field channel observations, as well as IFU observations. During GRS spectroscopy, different guide software will be used, as the pitch and roll are controlled by looking at the stability in the cross dispersion position of the spectral continua of bright stars, and the yaw is controlled using the zero order images that al-so fall on the focal plane. The window data processing (centroid history of guide stars) is performed in the instrument electronics and is sent to the spacecraft for attitude control updates.

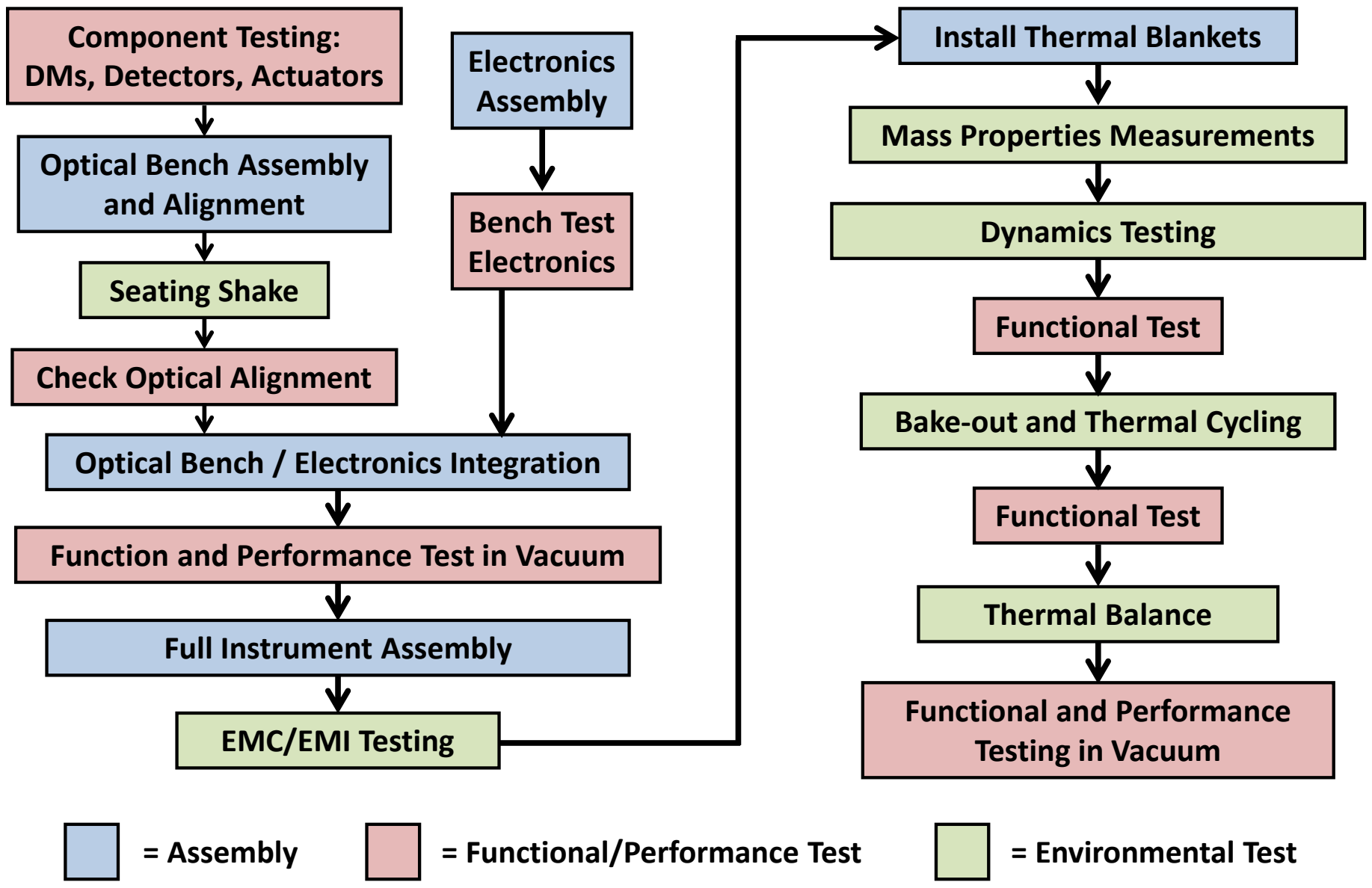

**Figure 3-24: Coronagraph I&T flow.**

## 3.6 Spacecraft

The WFIRST-2.4 spacecraft has been designed to provide all the resources necessary to support the payload in geosynchronous orbit using mature and proven technology. The design is based on the Solar Dynamics Observatory (SDO) spacecraft, which was designed, manufactured, integrated, tested, and qualified at GSFC and is currently operating in geosynchronous orbit. The spacecraft bus design provides cross strapping and/or redundancy for a single-fault tolerant design.

***Structures and Thermal***: The spacecraft bus design features three decks stiffened by six gussets and faced with module support plates all composed of aluminum honeycomb panels with M55J composite facesheets and sized to meet minimum launch vehicle frequency requirements while minimizing the spacecraft structure mass. The spacecraft bus provides the interfaces to the payload and the launch vehicle, transferring payload loads through the telescope struts, through the spacecraft gussets and into the payload attach fitting (PAF); see Table 3-5 for the Observatory mass breakdown.  The spacecraft top plate provides shear support of the bus, while also resisting bending. The propulsion subsystem is supported inside the bus on the prop support plate while the bottom deck provides the interface from the gussets to the PAF. The spacecraft design features 6 serviceable modules containing spacecraft, instrument, or telescope electronics with the electronics arranged in functional subsystems. The six modules are an electrical power module, a spacecraft C&DH module, a wide-field instrument electronics module, a telescope electronics module, an attitude control subsystem module, and a communications subsystem module. These modules are designed to be robotically serviceable, enabling refurbishment or upgrades to the observatory by replacing individual modules. Each of the spacecraft modules includes a kinematic Module Restraint System (MRS) based on the Multimission Modular Spacecraft (MMS) design, which was proven during the on-orbit module change out of the Solar Maximum Repair Mission. This MRS includes a beam interfacing to kinematic constraints on the top of each module and the accompanying floating nut/ACME threaded fastener on the bottom of each module.  These module interfaces can be robotically engaged and disengaged, providing a simple, repeatable mechanical interface to the spacecraft structure. The modules also include a grapple fixture to enable removal of the modules. The expected docking interface with the robotic servicer is a standard Marman band interface. Docking to the existing launch vehicle payload adapter fitting allows sufficient reach for the robotic servicer to remove and replace the spacecraft modules. Each module contains a set of connectors with alignment features to allow "blind mating" of the module harness connectors to connectors on the harness internal to the spacecraft bus connecting between modules and to the payload. This electrical interface has been demonstrated on the HST radial instruments. The electronics are mounted on the outboard panel of each of the modules with doublers used as needed for thermal conductivity. The spacecraft cold-biased thermal design maintains the spacecraft and payload electronics within their operational limits system using surface coatings, heaters and radiators. The bus also supports a multi-panel fixed solar array/sunshield to prevent the Sun from illuminating instrument radiators during science observations at roll angles up to 15° (see Figure 3-25).

| | Mass (kg) | Cont. (%) | Mass+Cont. (kg) |
|---|---|---|---|
| Wide-Field Instrument | 421 | 30 | 547 |
| Instrument Carrier | 208 | 30 | 270 |
| Telescope | 1595 | 11 | 1773 |
| Spacecraft | 1528 | 30 | 1987 |
| Observatory (dry) | 3752 | 22 | 4577 |
| Propellant (3σ) | 2544 | | 3095 |
| Observatory (wet) | 6296 | | 7672 |

**Table 3-5: Observatory mass breakdown**

***Electrical Power***: Three fixed, body-mounted solar array panels provide the observatory power. The solar array is sized to provide full observatory power at end of life with 2 strings failed at the worst case observing angles of 36° pitch and 15° roll with 30% power margin. Triple-junction, gallium arsenide solar cells operate with 28% efficiency and provide 4000 W of output power at end of life (at 0° roll and pitch) for an average orbit usage of ~2000 W (including 30% margin). The baselined geosynchronous orbit has two eclipse seasons of ~23 days per year with a maximum eclipse period of 72 minutes. A 160 A-hr battery is sized to accommodate all observatory loads during the maximum eclipse duration. An internally redundant power supply electronics box controls the distribution of power, providing unregulated 28 Vdc power to the spacecraft and payload, provides solar array power management, and controls battery charging.

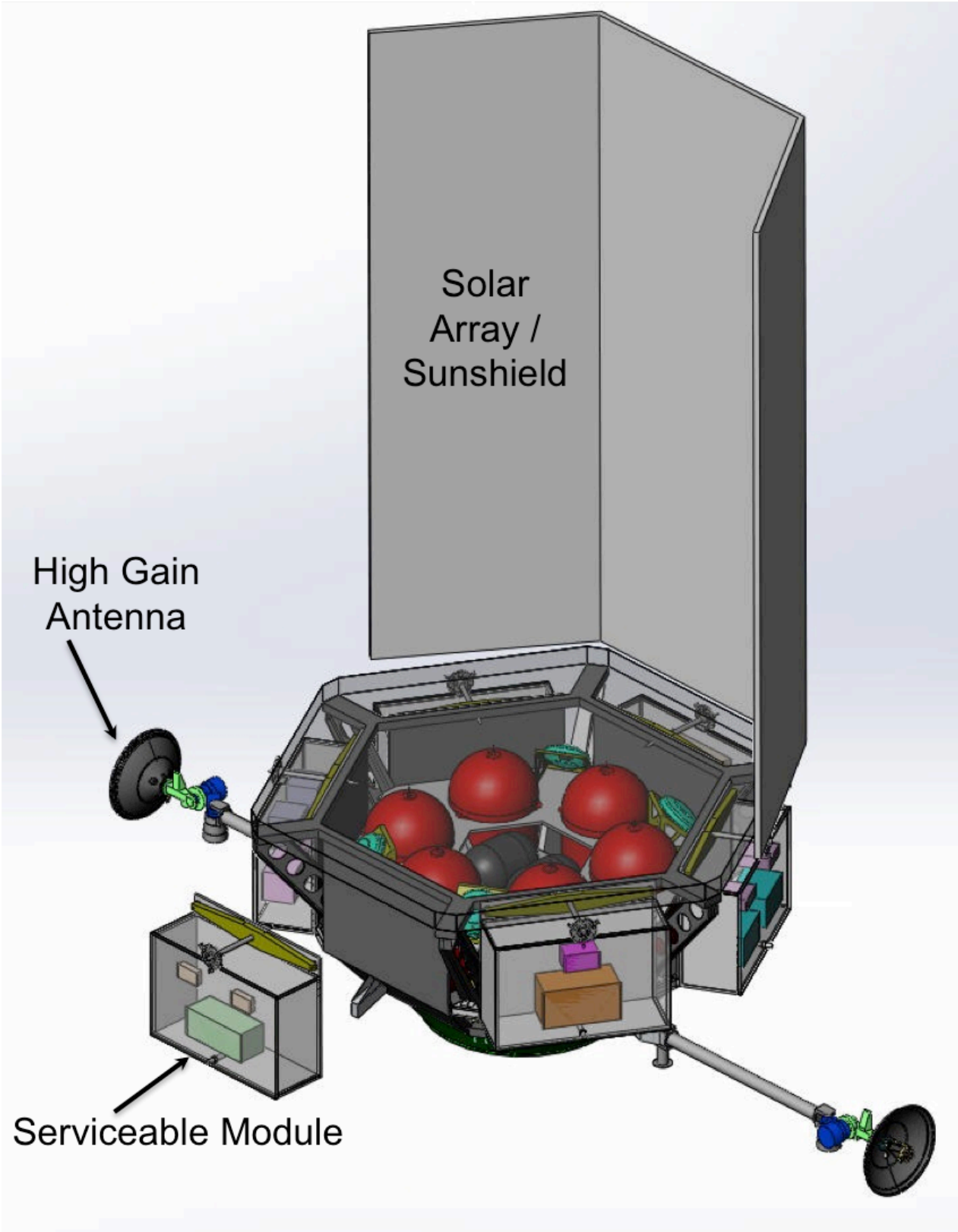

**Figure 3-25: Spacecraft bus with solar array/sunshield showing serviceable modules with spacecraft and payload electronics. The top deck of the spacecraft is removed to allow viewing of the propulsion subsystem.**

***Communications**:* The communication subsystem design leverages the SDO design to provide high reliability, continuous data downlink from the geosynchronous orbit. The communications subsystem uses S-band transponders to receive ground commands and to send real-time housekeeping telemetry to the ground via 2 omni-directional antennae. The high rate science downlink reuses the SDO design, using two dedicated Ka-band transmitters, each with a 0.75 m gimbaled antenna, to downlink science data at a rate of 150 Mbps without interrupting science operations. The SDO Ka-band transmitter uses a GSFC developed modulator with a 2.5 W SSPA.

***Command & Data Handling**:* The internally redundant command and data handling (C&DH) subsystem hosts the spacecraft flight software, provides onboard real time and stored commanding, receives payload and spacecraft housekeeping data, and provides fault management for the spacecraft health and safety as well as being able to safe the payload when necessary. The observatory uses a MIL-STD-1553B command/telemetry bus and uses SpaceWire for the high rate science data. The C&DH provides the interface between the instrument high rate science data and the Ka-band system and formats and encodes the instrument science data for downlink. With both the wide-field and coronagraph instruments operating, the C&DH will interleave this data onto the Ka-band downlink. Due to the direct, continuous science data downlink, no science data recorder is required for WFIRST-2.4. Control of the high gain antenna gimbal is also provided by the C&DH. The C&DH also provides the interface to the S-band transponders, performing command decoding and distribution as well as encoding for S-band downlink.

***Propulsion**:* The propulsion subsystem is a bi-propellant design using MMH/MON-3 as the fuel and oxidizer. The design is based on the successful SDO bi-propellant system developed by GSFC and currently operating in geosynchronous orbit. A 450 N (100 lbf) main engine provides the thrust required for orbit circularization from the geosynchronous transfer orbit provided by the launch vehicle. Eight 22 N (5 lbf) attitude control subsystem (ACS) thrusters are used for east-west stationkeeping and for momentum dumping from the reaction wheels throughout the duration of the mission. Three fuel and three oxidizer tanks (each 28" in diameter and 48" tall) are mounted on the prop deck internal to the spacecraft, which also supports two small helium pressurant tanks for this subsystem. The propellant load provides a total delta V of ~1580 m/s (~1550 m/s for orbit circularization and ~30 m/s for stationkeeping, momentum management and disposal) and covers 3$\sigma$ variations in launch vehicle performance, propulsion system performance, and flight dynamics errors. The propulsion subsystem control and safety inhibit electronics are contained in the Attitude Control Electronics box.

***Attitude Control**:* The spacecraft is three-axis stabilized and uses data from the inertial reference unit, star trackers, and the payload FGS to meet the coarse pointing control of 3 arcsec, the fine relative pointing control of 10 mas pitch/yaw and 1 arcsec roll, and stability of 20 mas pitch/yaw and 2 arcsec roll (all values RMS per axis). The internally redundant inertial reference unit provides precise rate measurements to support slew and settle operations. The star trackers (3 for 2 redundant) are mounted on the telescope aft metering structure so they are directly related to the telescope pointing. The star trackers are used for coarsely pointing to within 3 arcsec RMS per axis of a target. After that, the FGS takes over to meet the fine pointing requirements for revisits and relative offsets (see §3.5). A set of four 75 N-m-s reaction wheels is used for slewing as well as momentum storage. The wheels are mounted to and passively isolated from the prop support plate to allow stable pointing at frequencies higher than the FGS control band. The wheels can accumulate momentum for at least 7 days between dumps and the ACS thrusters are used to desaturate the wheels to manage momentum buildup. The internally redundant Attitude Control Electronics (ACE) control the attitude control subsystem and provide an independent safe hold capability using coarse sun sensors and reaction wheels, which keeps the observatory thermally-safe, power-positive and protects the instrument from direct sunlight. The ACE also contains the fine guidance electronics which control the fine pointing of the observatory based on the guide star centroids provided by the wide-field instrument C&DH.

### 3.7 Orbit

Several different orbit choices were considered for WFIRST-2.4 including geosynchronous Earth orbit, Sun-Earth L2, low Earth orbit, and highly elliptical Earth orbits. The best two options are Sun-Earth L2 and geosynchronous orbit, with a comparison given in Table 3-6. We have chosen to baseline a 28.5 deg inclined geosynchronous orbit with right ascension of the ascending node (RAAN) of 175 degrees and 105°W longitude, but have a mission architecture that works at Sun-Earth L2 as well. The geosynchronous orbit has the key advantage of continuous telemetry coverage with a single ground station. This enables a high data rate at low cost.

### 3.8 Optical Communications

The WFIRST-2.4 study team assessed the use of optical (laser) communications as a means of transmitting the science data to the ground. NASA is currently working on the Lunar Laser Communications Demonstration (LLCD), an optical communications demonstration on the LADEE mission, currently scheduled to launch in mid 2013. LLCD will demonstrate optical communications from lunar orbit at rates of 311 Mbps with pulse position modulation over ~16 hours in ~1 month of operations. NASA is also currently developing the Laser Communications Relay Demonstration (LCRD) as a hosted payload on a commercial communications satellite in geosynchronous orbit. The baselined data rate from geosynchronous orbit using DPSK modulation is 1.2 Gbps, but even higher data rates are expected.

The current WFIRST-2.4 Ka-band system downlinks data at a rate of 150 Mbps and is sufficient for the current DRM. The WFIRST-2.4 study team will continue to monitor the progress of the LCRD optical terminal as well as a 1.2 Gbps Ka-band transmitter that is currently in development at GSFC and is expected to reach TRL 6 in 2013. Should requirements change as the DRM matures necessitating a higher data rate, the WFIRST-2.4 study team would perform a trade study assessing the cost, risk and technical performance of the 1.2 Gbps LCRD optical terminal and the 1.2 Gpbs Ka-band transmitter.

| Parameter | Geosynchronous Orbit | Sun-Earth L2 | Comment / Impact |
|---|---|---|---|
| **Telescope operating temperature** | Limited by optomechanics | Limited by optomechanics | Long wavelength limit (2.x μm) is independent of Geo / L2 orbit |
| **Telemetry Downlink Rate** | >1 Gbps | Low | Geo allows more data down and continuous downlink |
| **Radiation** | p+ and e- | p+ only | Geo requires significant shielding of the focal plane |
| **Viewing constraints** | Moderate: Bulge, eclipses | Small | Geo constraints reduce efficiency but are tolerable |

**Table 3-6: WFIRST-2.4 orbit trade**

### 3.9 Ground System

The WFIRST-2.4 Mission Operations Ground System is comprised of three main elements: 1) the facilities used for space/ground communications and orbit determination, 2) the Mission Operations Center (MOC) and 3) the facilities for science and instrument operations and ground data processing, archiving, and science observation planning. For each element, existing facilities and infrastructure will be leveraged to provide the maximum possible cost savings and operational efficiencies. The functions to be performed by the ground system and the associated terminology are shown in Figure 3-26.

Two dedicated 18 m dual Ka and S-band antennae located in White Sands, NM are used for spacecraft tracking, commanding and data receipt. The antennae are separated by ~3 miles to minimize the chance of weather events interrupting the downlink. The two ground stations are within the beam width of the Ka-band antenna on the spacecraft so both receive the downlinked data simultaneously. Three 18 m antennae ground stations are currently in use at White Sands, one is dedicated to the Lunar Reconnaissance Orbiter (LRO) and two are dedicated to SDO. The two SDO antennae may be available for WFIRST-2.4, depending on how long beyond 2015, the end of the baseline 5-year mission, SDO operates. The antenna dedicated to LRO is expected to be available for WFIRST-2.4 as LRO is currently in its extended mission phase. The WFIRST-2.4 development cost assumes that one new ground station will be built with one of the three existing ground stations available as the WFIRST backup. The White Sands ground stations interface with the MOC for all commanding and telemetry. Tracking data is sent to the GSFC Flight Dynamics Facility.

The MOC performs spacecraft, telescope and instrument health & safety monitoring, real-time and stored command load generation, spacecraft subsystem trending & analysis, spacecraft anomaly resolution, safemode recovery, level 0 data processing, and transmission of science and engineering data to the science and instrument facilities. The MOC performs Mission-level Planning and Scheduling. With continuous downlink access, operations and communication

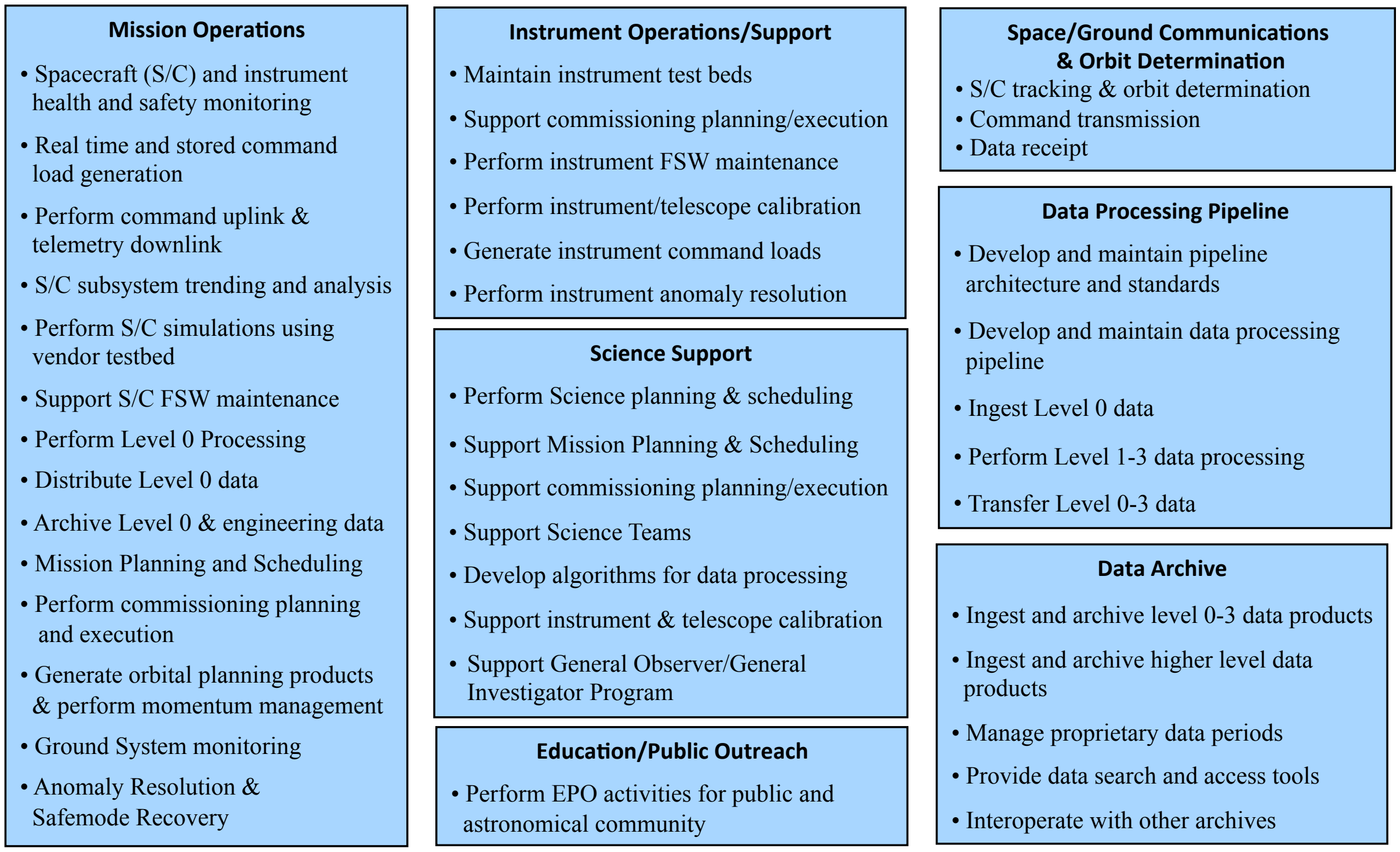

**Figure 3-26: WFIRST Ground System functions and associated terminology**

scheduling complexity is reduced including the elimination of DSN operations.

The science and instrument facilities maintain instrument test beds, perform instrument & telescope calibrations, assist in the resolution of instrument anomalies, perform instrument flight software maintenance, and generate instrument command loads.

These facilities are also responsible for science planning & scheduling, supporting mission planning activities carried out by the MOC, running the Guest Investigator (GI) and Guest Observer (GO) Programs, providing Science Team, GI, and GO support, and performing EPO activities for the public and the astronomical community. Data handling involves ingesting Level 0 science and engineering data from the MOC and performing Level 1-3 data processing for the Science Teams and GO community and transmitting these calibrated data to the data archive. All data will be archived. Data search and access tools are provided to the science community that enable efficient searches and delivery of archival data that ensure interoperability with NASA data archives and the Virtual Astronomical Observatory.

Several dedicated Science Teams will be funded over the prime phase of the mission to execute the primary dark energy, exoplanet, and NIR survey observing programs. In this period, the GO/GI program provides funding for analysis of public data and for pointed observations selected competitively. Operations costs and grants for the GO/GI program in the primary mission are fully included in the lifecycle costs.

### 3.10 Concept of Operations

WFIRST-2.4 will support a wide range of science programs during its primary mission. Each of these programs has unique constraints involving the field of regard, cadence, and S/C roll angles. Moreover, observations in GEO are subject to viewing constraints on daily (motion around the Earth), monthly (Moon avoidance), yearly (Sun aspect angle), and secular (orbital precession) timescales. The SDT therefore constructed an existence proof of a possible observing plan, which showed that the strategic science programs are all mutually compatible, while simultaneously enabling a robust GO program. **This is only an existence proof: the actual observing plan will be updated depending on the needs of the dark energy and exoplanet communities and the highest-ranked GO programs.**

The "existence proof" observing plan was built according to the following constraints:

- Mission duration: The science phase of the primary mission is taken to last 5 years (or 6 years if the optional coronagraph is included). A notional penalty of 3.3% is applied for times when science observations are not possible (e.g. safe holds).
- Orbit: The initial orbit is a circular GEO with initial inclination of 28.5°, RAAN 175°, and centered at longitude 105°W, and is integrated including perturbations from the Sun, Moon, and non-spherical Earth. The initial RAAN was chosen to allow microlensing observations without daily interruptions throughout the primary mission despite the significant (−7°/yr) orbital precession. A larger value of initial RAAN would provide adequate visibility of the Galactic bulge but would have increased thermal loading from the Earth on the WFI radiator.
- Viewing constraints: The angle ε between the line-of-sight and the Sun is constrained to 54—126°. The S/C roll angle relative to the Sun is constrained to ±15° (for $90<\varepsilon<110°$) or ±10° (otherwise). The LOS cannot point within 30° of the limb of the Earth or Moon. During normal survey operations the WFI radiator normal is required to be kept at least 47.5° away from the Earth. An exception is made for the microlensing and SN programs that use an inertially fixed attitude throughout the orbit, where the radiator angle is allowed to be as low as 27° but only with a specific thermal load vs. time profile. Overheads between any two observations are computed based on the RW torque, angular momentum capacity, and settling and detector reset time.
- High latitude survey: The HLS provides 2 passes over the survey footprint in each of the 4 imaging filters and 4 passes with the grism, all at different roll angles and with small steps to cover chip gaps. The footprint is in regions of high Galactic latitude (to suppress extinction and confusion) and is contained within the planned LSST footprint. The grism has 2 "leading" passes (looking forward in Earth's orbit) and 2 trailing passes to enable the single grism to rotate relative to the sky and provide counter-dispersion; the imaging mode has a leading and trailing pass in each filter.

- Microlensing: The microlensing program observes 10 fields in the Galactic bulge for continuous 72-day seasons, interrupted only by monthly lunar avoidance cutouts. The plan includes 6 seasons, including the first and last available season.
- Supernovae: The Type Ia supernova survey is carried out over 2 years; the duty cycle for actual observations (not including overheads) is 27%. The IFU observations, unlike the DRM1/2 slitless prism approach, do not have any roll angle constraints.
- Guest observer program: The GO program by definition cannot be "allocated" at this stage in the project. Moreover, in practice the planning of HLS observations will be re-organized based on the content of the GO program. For the purposes of the existence proof exercise, we have simply required that the time not used for other programs be ≥1.25 years, and that all portions of the sky are visible in multiple years during otherwise-unallocated time. In computing the unallocated time, we subtract the penalty for a typical 90° slew from each unallocated window. (This way, the slewing penalty between any two programs is charged against the time allocation of one of the programs but not both.)

### 3.10.1 *Observing plan without a coronagraph*

The sample observing plan is shown in Figure 3-27, and the footprint on the sky is shown in Figure 3-28. The time breakdown, including overheads, is as follows – 0.97 years for the microlensing; 0.62 years for the supernovae; and 1.89 years for the high latitude survey (imaging+spectroscopy). The unallocated time is 1.29 years.

The HLS imaging and spectroscopy depth forecasts assume a reference zodiacal sky brightness of 1.3 times the annual average at the pole. The example observing plan has 75% of the observations below the reference sky brightness, and 99% below 1.18 times the reference brightness. The large field of regard of WFIRST-2.4 will allow the GO program to target any point in the sky with multiple revisits even without moving the large survey programs, as shown in Figure 3-29. The ability to reschedule the non-time-critical HLS observations taking into account the distribution of GO programs will endow WFIRST-2.4 with even more flexibility.

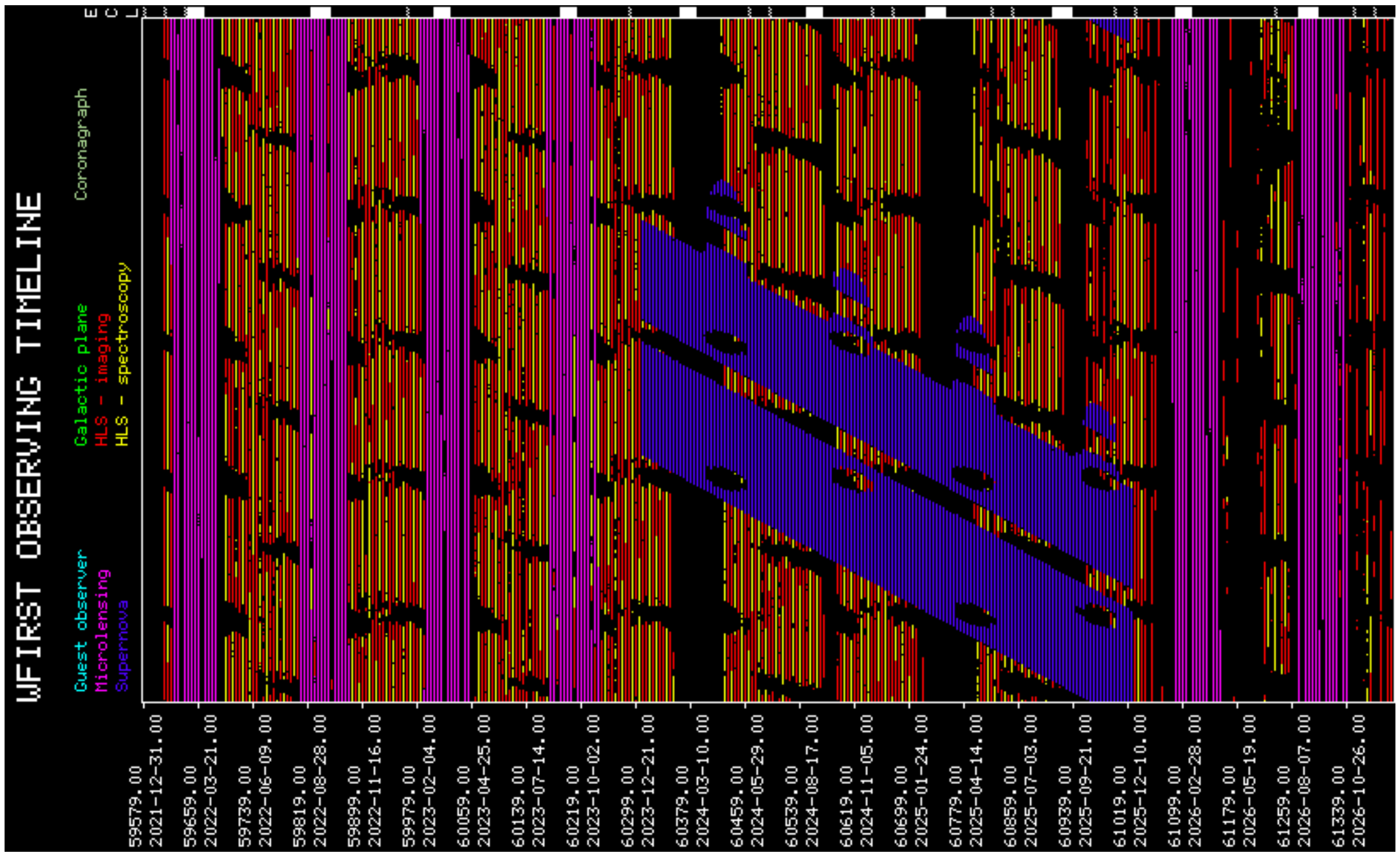

**Figure 3-27: The example WFIRST-2.4 observing sequence. Each column indicates a 5-day interval, with the programs color-coded. The row at top shows Earth eclipse seasons (solid rectangles) and passages through the lunar penumbra (vertical lines). The microlensing seasons (magenta) are visible with the lunar cutouts removed. The supernova program (blue) is spread over 2 years: observations are scheduled every 5 days, but are broken into chunks due to Earth viewing and radiator angle cutouts. The HLS fills in much of the remaining time. Note that Earth eclipse seasons occur during the microlensing campaigns. Unallocated time is shown in black: it is anticipated that this will be devoted to the GO program.**

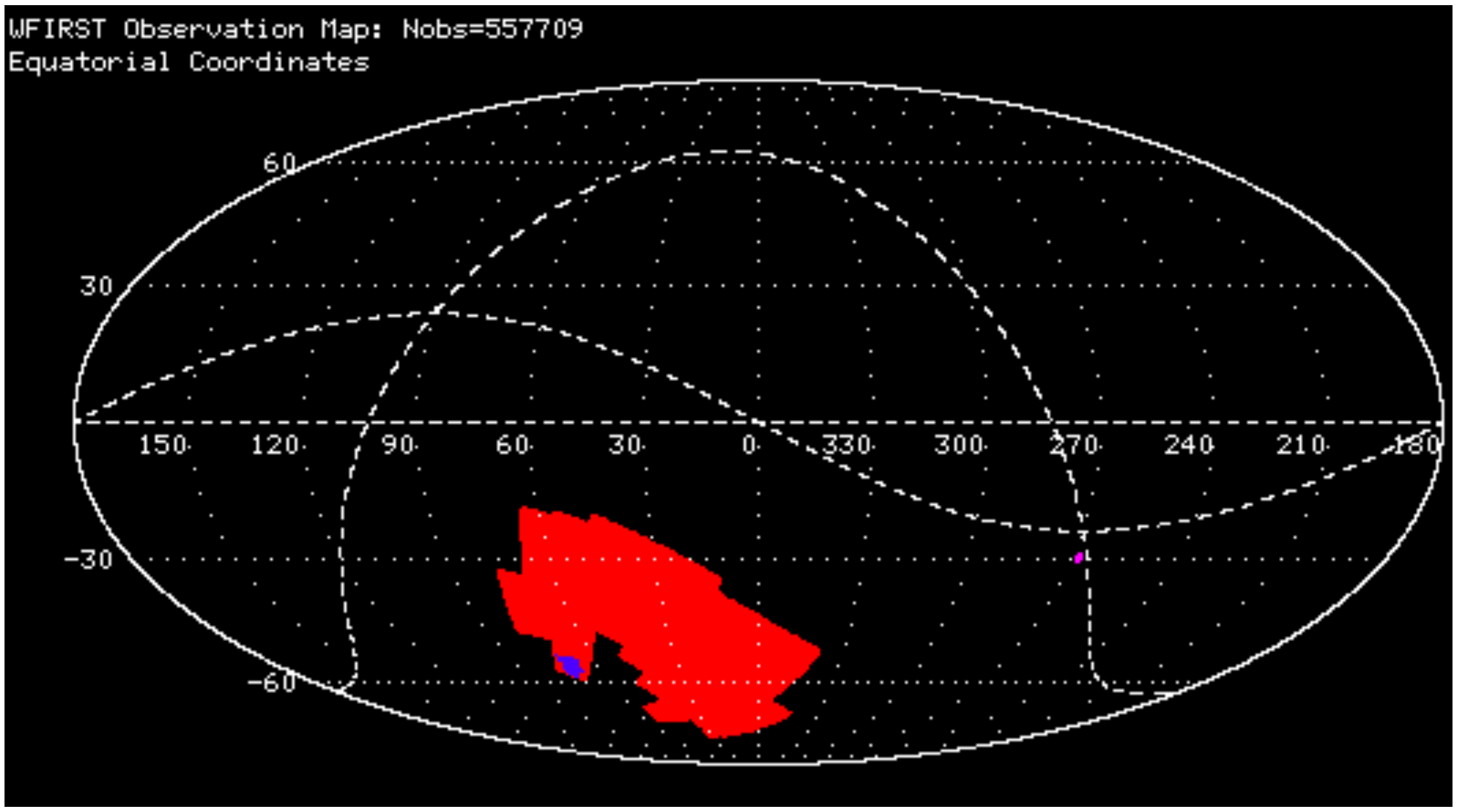

**Figure 3-28: The footprint of the WFIRST-2.4 observations. The red region shows the HLS, the blue shows the supernova survey, and the magenta spot shows the microlensing survey. The HLS footprint area is 2054 deg$^2$.**

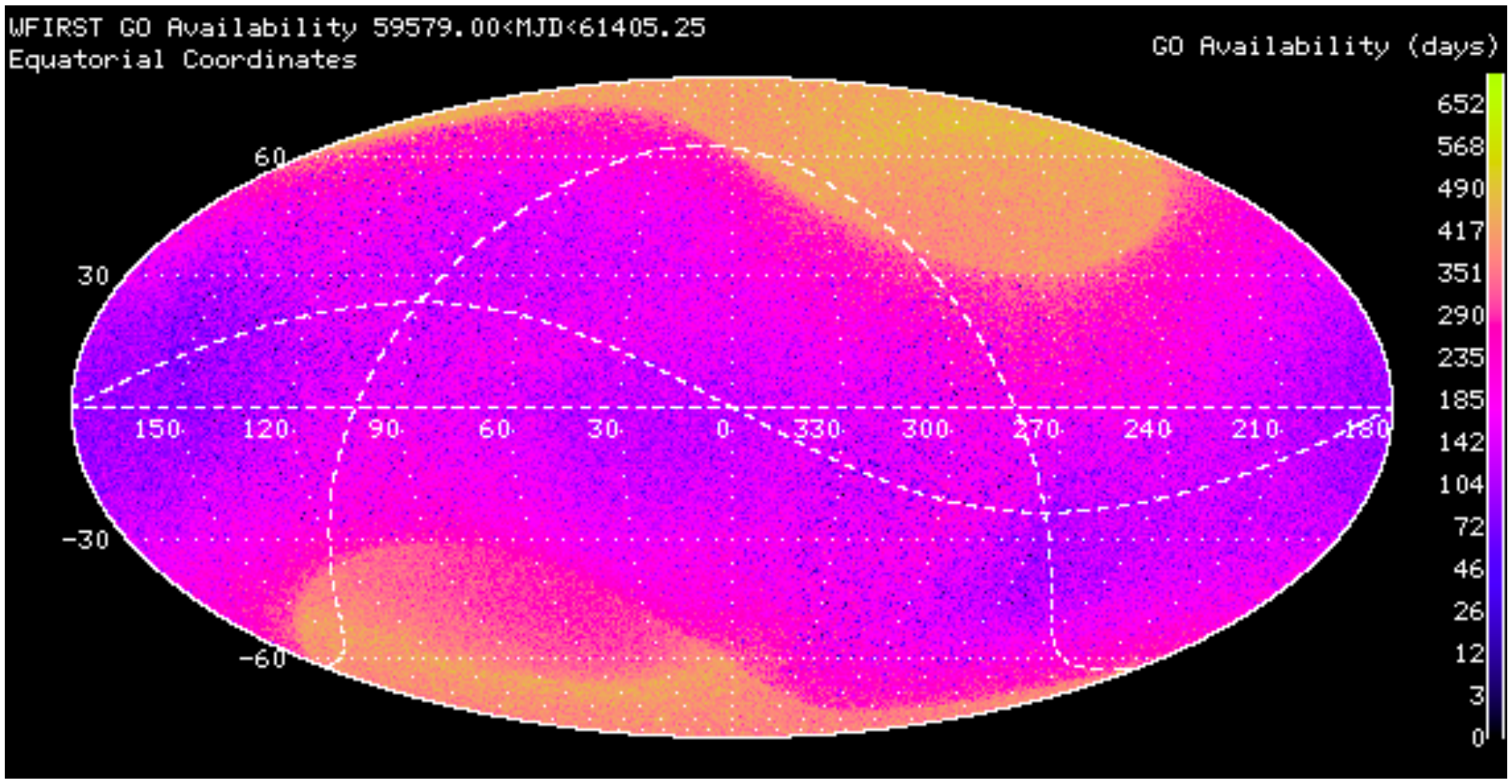

**Figure 3-29: The amount of time during the 5-year primary mission during which each field would be accessible by the GO program, with no re-scheduling of other observations, as determined by a Monte Carlo analysis. Fields within 36° of the ecliptic poles are available for the great majority of this time since they always satisfy the Sun angle constraint, with only occasional cutouts due to the Earth. Fields near the Ecliptic are available only part of the year. The Galactic bulge region near RA=270°, Dec=−30° has the lowest availability because most of the time when it is accessible it is used by the microlensing program. Nevertheless it still satisfies the requirement of being available in multiple years.**

The plan presented here is only one possible example of what WFIRST-2.4 might do, but it demonstrates that the strategic programs are all compatible, even given the complex constraints of a GEO mission.

### 3.10.2 *Observing plan including a coronagraph*

If the coronagraph option is exercised, the operations planner assumes the mission is extended to 6 years, with 1 year to be devoted to coronagraph observations. These observations are scheduled in 26 blocks of 2 weeks each, interspersed throughout the mission. The constraints on the WFI programs and the observing footprints are the same as in the case without the coronagraph: the only change is that the SN program is split into two 1-year chunks so as to provide more regular coronagraph observing periods. Since the supernovae are followed with the IFU and can be selected to have complete light curves within each supernova campaign, the penalty associated with this is minor.

The current operations planner assesses the availability of each potential target star during each block of coronagraph observing time. Out of a catalog of 239 potential target stars, in each of the 26 coronagraph observing blocks we find at least 24 to be continuously viewable over the full 2-week period with no violations of the Earth, Moon, or Sun pointing constraints. We find that 100 of the 239 targets meet the continuous viewing criterion during at least 1 of the coronagraph blocks throughout the mission (and 52 targets are visible in at least 4 blocks). The remaining targets would be viewable during some of the coronagraph blocks subject to daily Earth viewing cutouts (duration up to 5 hours) or the longer but less frequent Moon cutouts (up to 6 days): this would imply a loss of efficiency due to the need to re-acquire the target, allow for thermal settling, and perform the wavefront correction. The latter will be subject to optimization based on Monte Carlo simulations.

The overall observing sequence including the coronagraph is shown in Figure 3-30. Again all of the strategic programs are compatible, while allowing time for an ambitious GO program.

## 3.11 Cost & Schedule

The WFIRST-2.4 configuration builds on past WFIRST and JDEM mission design concepts. These encompass the JDEM-Omega configuration submitted to Astro2010, through the recent WFIRST design reference missions developed in collaboration with the WFIRST Science Definition Team (November 2010 thru July 2012). The unique difference of the WFIRST-2.4 design reference mission is that it incorporates the use of an existing 2.4-meter aperture telescope, along with a number of other payload and mission design simplifications. The development risk of the WFIRST-2.4 mission concept is lower than the previously studied 5 year WFIRST concept (IDRM) that was independently costed by the Aerospace Corporation due to the following characteristics:

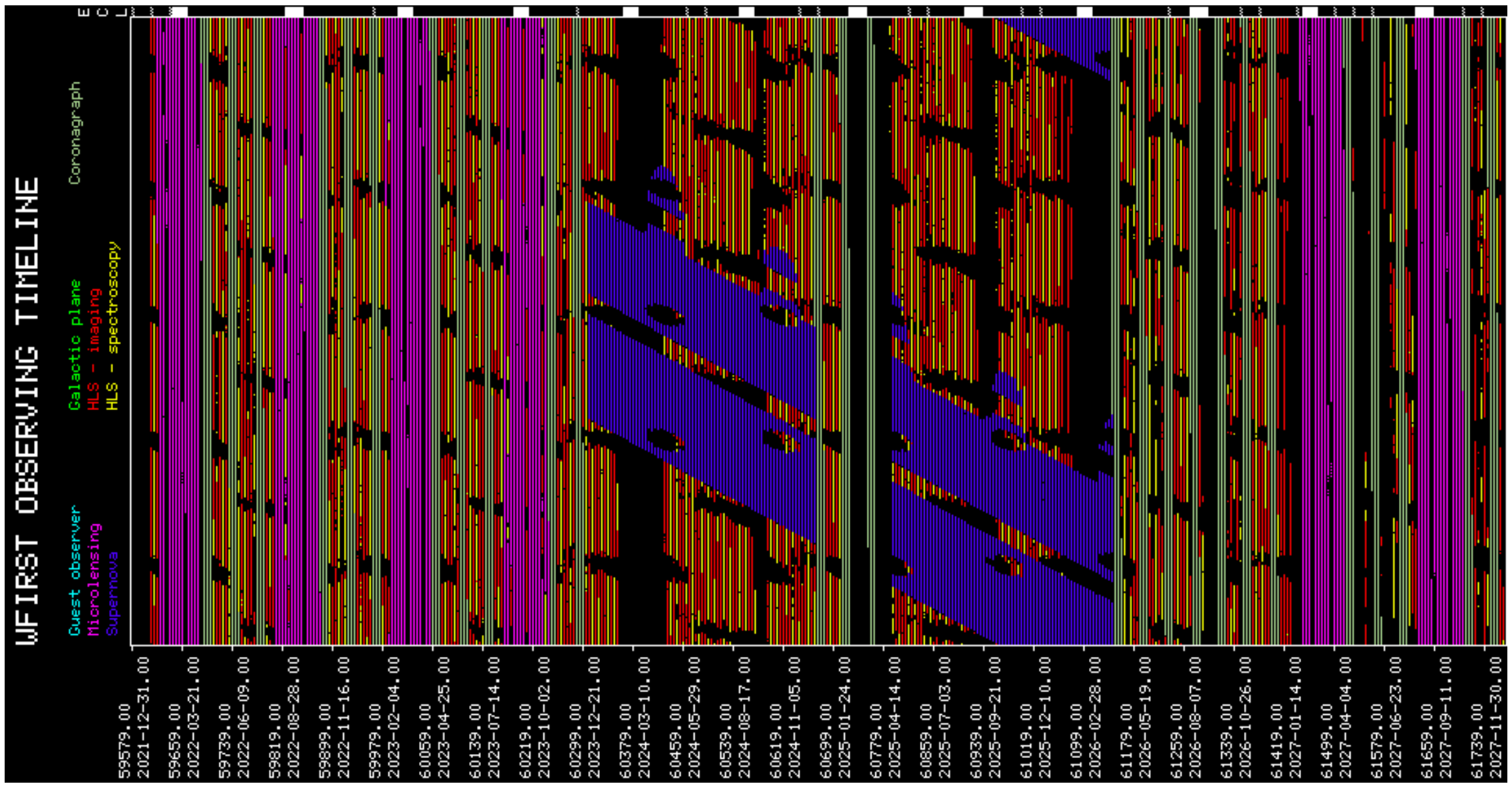

**Figure 3-30: The example observing plan (analogous to Figure 3-27, but with a coronagraph). The green blocks indicate the coronagraph observing periods.**

- Telescope already fabricated, large telescope optics well off of the critical path
- Simplified instrument optical train (2 channels vs. 4 channels)
- Half the number of detectors
- Simplified ground operations

The development phase for WFIRST-2.4 is 79 months, from preliminary design through launch (phase B/C/D). This development phase is preceded by 12 months to fully develop the baseline mission concept (Phase A) and several years of concept studies, many of which have already taken place. The observing phase (Phase E) of the mission is baselined and costed for five years. The development schedule is shown in Figure 3-31. The schedule estimate is at a 70% confidence level and includes seven months of funded schedule reserve.

The WFIRST-2.4 design reference configuration utilizes the 2.4-meter aperture telescope that was fabricated for another program, but subsequently not flown. The telescope optics and all of the supporting structure has been fabricated and integrated. A copy of the secondary mirror mechanisms and the telescope control electronics will be re-built to bring the telescope to flight configuration. The availability of this telescope hardware removes this critical piece of the payload from getting on the critical path and simplifies the number of trades that need to be performed by establishing interfaces early. The WFIRST-2.4 wide-field instrument is also a significant simplification compared to the IDRM. The wide-field instrument reduces the number of instrument channels from 4 to 2, with a very simple optical design in the wide-field channel (only one powered mirror). A single filter wheel is accommodated by the use of a grism for the GRS. The focal plane has one-half the number of detectors of the IDRM, simplifying integration and test of this critical assembly. The wide-field instrument does incorporate the addition of an IFU; however, electri-cally this capability is essentially a nineteenth detector and electronics channel. The optics for the IFU are small and high TRL.

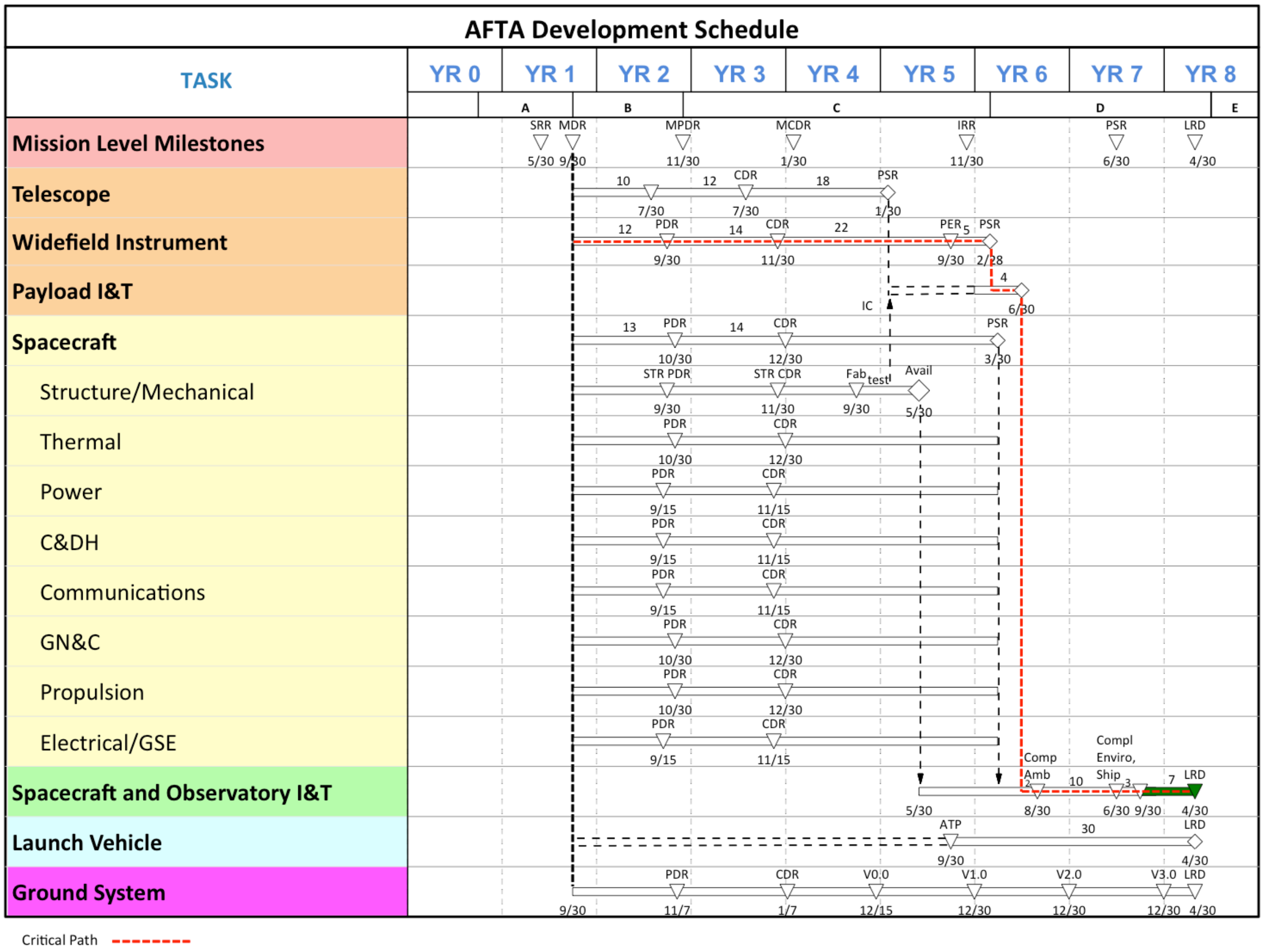

Figure 3-31: WFIRST-2.4 development schedule.

The WFIRST IDRM mission was estimated in 2011 at $1.63B (FY$12) by the WFIRST Project and the independent CATE estimate was 8% higher. The WFIRST-2.4 design reference mission is a simpler and lower risk implementation than the WFIRST IDRM reference configuration. A Lifecycle Cost Estimate (LCCE) for the WFIRST-2.4 reference mission is in process and is being developed using the same techniques (grassroots, parametric, and analogy) that have been applied to previous NIR survey missions. Project Management, Systems Engineering, Mission Assurance, and Integration & Test, are estimated using a grassroots approach and validated against analogous missions. Pre- and Post-launch Science will be projected based on the expected size of the WFIRST Science Announcement of Opportunity (AO). This estimate will include support for the AO-selected science teams in the mission development phase as well as the 5 years of operations. Additional funds for the Guest Observer and Guest Investigator Program are included in the Science estimate. The payload and spacecraft will be estimated primarily using parametric estimates, with the exception that the telescope estimates rely on historical estimates to cost the overall telescope effort. The parametric estimates will be constructed using master equipment lists (MELs) and historical cost databases and are adjusted for mass and complexity factors.

The build-up and integration philosophy of the WFIRST-2.4 observatory is based on the well-established practice of building small assemblies of hardware, thoroughly testing them under appropriate flight environments, and then moving on to the next higher level of integration with those assemblies. The WFIRST-2.4 telescope and instrument will be developed and individually qualified to meet mission environments. The critical path of the mission is through the development of the instrument. The instrument is qualified prior to integration with the telescope. Following ambient checkout, the entire payload is tested at temperature and vacuum at GSFC to verify the end-to-end optical performance. The spacecraft is then integrated to the payload, and checked at ambient. Following a successful ambient checkout, a complete observatory environmental test phase is performed. Upon successful completion of the observatory environmental test program, the observatory is readied for shipment to the KSC, where the launch campaign is conducted.

The parallel build-up of all of the mission elements, allows substantial integration activity to occur simultaneously, increasing the likelihood of schedule success. The early delivery of the telescope will allow the use of the telescope for early instrument optical checkout, allowing the retirement of integration risk even earlier in the program. Because all of the major elements of the observatory (telescope, instrument, and spacecraft) are located at GSFC approximately two years before the planned launch, there is considerable flexibility in optimizing the schedule to compensate for variation in flight element delivery dates, should workarounds be necessary. Over the two year observatory I&T period, the Project will have flexibility to reorder the I&T work flow to take advantage of earlier deliveries or to accommodate later ones. Should instrument or telescope schedule challenges arise there are options to mitigate the schedule impacts by reallocating the payload level environmental test period. Should instrument or spacecraft challenges arise, there are options to modify the workflow and pull other tasks forward to minimize risk and maintain schedule. The payload design includes access to the instrument volume when attached to the spacecraft, allowing late access to the instrument during observatory I&T. The WFIRST-2.4 observatory I&T flow is very achievable, given the planned schedule reserve and the opportunities available for workaround.

Fifty-three months are allocated to complete the WFIRST-2.4 instrument, from the start of preliminary design, through the delivery of the instrument, not including funded schedule reserve. The overall 79 month observatory development schedule includes seven months of funded reserve, further increasing the likelihood of executing the plan.

Early interface testing between the observatory and ground system is performed to verify performance and mitigate risks to schedule success. Prior to payload integration, interface testing between the spacecraft and the ground system is performed. Immediately following payload integration to the spacecraft, end-to-end tests are performed, including the payload elements. These tests are performed numerous times prior to launch to ensure compatibility of all interfaces and readiness of the complete WFIRST-2.4 mission team.

The WFIRST-2.4 reference configuration requires no new technologies, has an implementation strategy that is conservative, proven and amenable to workarounds, and has a schedule based on continuously retiring risk at the earliest possible opportunity. This WFIRST is executable within the cost and schedule constraints identified in this report, and is consistent with the New Worlds, New Horizons finding that WFIRST "…presents relatively low technical and cost risk making its completion feasible within the decade…".

# 4 PATH FORWARD

This report marks the completion of the first phase of study of the WFIRST-2.4 mission. NASA will now assess the study and decide by June 2013 if the agency will continue studying this version of WFIRST. Assuming that occurs, there are a number of activities that are worth studying, that were not possible in this initial time-constrained study, to fully exploit the scientific potential of what has already been identified as an extraordinary observing capability. These activities also further advance the WFIRST-2.4 DRM, which positions the mission for development following the schedule presented in §3.11. These activities will be undertaken by the SDT and Project office and reported to NASA/HQ at desired intervals, with reports in early 2014 and 2015. The following tasks will be undertaken over the coming year subject to the availability of funding:

1) Continue developing H4RG detectors for spaceflight application. Detailed characterization testing is planned following fabrication of the detector elements. H4RGs offer four times the pixels per device, over the H2RGs that are baselined for JWST, and are the detector of choice for ground and space applications of the future. Packaging these development detectors into a large focal plane array will also be performed to minimize potential risks that could threaten the hardware that is on the critical path of the WFIRST-2.4 mission. **This area remains the most critical for investment for minimizing overall risk of a WFIRST-2.4 mission.**

2) Assess the potential for pushing the wavelength cut-off for the wide field instrument further into the red. This involves determination of the feasibility of operating the wide field instrument and the telescope at colder temperatures. Design optimization, thermal analyses and some hardware testing in the case of the telescope will be performed to complete this task.

3) Determine requirements for on-orbit calibration of the wide field instrument (including the IFU) and investigate options for internal instrument calibration. Many aspects of the imaging and spectroscopic calibrations can be derived from the survey data and dedicated observations of standard objects. However, some aspects of the calibration may best be performed with internal light sources.

4) Optimize and refine the wide-field instrument and spacecraft designs and perform end-to-end structural/optical/thermal analyses of the observatory.

5) Continue assessing coronagraph options. The requirements for coronagraphy with WFIRST-2.4 need to be refined, and a choice made of technology baseline architecture. Development work is required to improve the TRL of the coronagraph. The September SDT meeting will be a suitable time for recommendations of a leading option and possible back-up technology.

6) Use the current operations planning tool to support ongoing trade studies (e.g. orbital inclination), provide input for engineering analysis (e.g. thermal stability, radiator loading from the Earth), and assess the WFIRST-2.4 science reach for candidate GO programs.

7) Continue refining the estimates of the expected performance of the microlensing survey. Detailed theoretical and observational studies are needed to provide more robust inputs to the survey yield and to assess the prospects for detection of habitable planets with WFIRST-2.4. Particularly important in this regard are robust empirical determinations of the microlensing event rates in the likely target fields. In addition, studies are needed to provide comprehensive estimates of the accuracy with which the physical properties of the detected planetary systems (including host star masses and distances) can be determined using all available methods. Particularly interesting are the prospects for geometrical measurement of host star masses using a combination of geosynchronous parallax and microlensing astrometry; the study of these two applications must carefully consider systematic errors in the achievable photometry and astrometry.

8) Model the performance of the grism survey. The modeling work will aid in refining the resolution requirement, and effects of source confusion and cosmic rays. The simulations will incorporate the image quality in non-first orders and ghost locations.

9) Study joint LSST & WFIRST science. There is a wealth of science that can be done by combining the LSST and WFIRST data sets. Explore coordinating the two observing programs to view the same regions at the same time. Work with the LSST project to understand the software interfaces required to enable merging the LSST and WFIRST data sets with minimal duplication of effort, and to generate consistent input simulated skies for the LSST and WFIRST instrument simulators.

10) Optimize the filter complement across the science areas. The 'placeholder' filters were chosen by the SDT as an example but were not subject to detailed optimization.

11) Optimize the IFU parameters. The interface to the telescope is defined, but the back-end parameters (camera focal length, dispersion curve) have not been subject to detailed optimization.

12) Develop briefing materials to inform the scientific community and government stakeholders of WFIRST-2.4 capabilities and discovery potential.

Progress on these tasks will be discussed on SDT / Project telecons over the coming year. There will be an SDT meeting in September 2013 where preliminary results will be presented.

# 5 CONCLUSION

The WFIRST-2.4 mission is highly compelling and is found in the current study to be feasible and affordable. It will revolutionize our understanding of the expansion of the universe, the birth and evolution of galaxies, and the formation of exoplanets. It has significantly enhanced performance compared to the previous DRM1 and DRM2 configurations, and compared to the Astro2010 requirements. The larger mirror gives finer imaging and improved sensitivity making the mission more powerful and more complementary to LSST and Euclid. In addition, JWST will greatly benefit from WFIRST's discovery of interesting objects in the wide-field survey. If equipped with a coronagraph, the larger 2.4-meter aperture enables rich scientific return at much lower cost than a dedicated smaller coronagraphic telescope mission, enabling the first images and spectra of exoplanets like those in our solar system and of dust disks around nearby stars. The conceptual design work described in §4 will further refine the observatory and its performance, enabling a faster and lower-risk development.

# Wide Field Infrared Survey Telescope (WFIRST-2.4)

## *A Collection of One Page Science Programs from the Astronomical Community*

Version 1.0
Mar 2013

http://wfirst.gsfc.nasa.gov/

This appendix contains a rich set of ~50 potential GO science programs that are uniquely enabled by WFIRST-2.4. The one pagers were put together by the broader astronomical community, and the SDT wishes to thank the community for this valuable contribution. The one pagers highlight the tremendous potential of WFIRST-2.4 to advance many of the key science questions formulated by the Decadal survey. These programs have not been vetted by the SDT, and we are not endorsing these specific studies. Additionally, the authors did not see the final version of the WFIRST-2.4 mission capabilities prior to their submissions, so inconsistencies may exist. As has proven extremely successful on other NASA Great Observatories, we expect that the selection of GO science programs will be made by a peer-reviewed Time Allocation Committee process.

# Table of Contents

# Table of Contents

## Extragalactic Astrophysics

## General

Hilke E. Schlichting (UCLA), hilke@ucla.edu

**A Full Portrait of the Kuiper Belt, including Size Distributions, Colors and Binarity**

Background

The Kuiper belt is a remnant of the primordial Solar system. It consists of a disk of icy bodies located at the outskirts of our planetary system, just beyond the orbit of Neptune, and is the likely source of short period comets. More than 1200 Kuiper Belt Objects (KBOs) have been detected since its discovery in 1992 (Jewitt & Luu 1993). In the Kuiper Belt, planet formation never reached completion and as a result it contains some of the least processed bodies in our Solar system. Its dynamical and physical properties illuminate the conditions during the early stages of planet formation and have already led to major advances in the understanding of the history of our planetary system. However, despite all these successes, many open questions remain: is the Kuiper Belt undergoing collisional evolution, grinding small KBOs to dust and therefore a true analogue to the dust producing debris disk around other stars? How did Neptune's outward migration proceed (Malhotra 1993, Tsiganis et al. 2005)? And what do the colors of KBOs imply about their formation and consequent evolution? Only a deep, uniform wide-field survey will be able to definitively answer these and other questions.

WFIRST

WFIRST will enable a uniform wide-field survey with unprecedented sensitivity of the Kuiper Belt. Since there are estimated to be more than 4 x $10^4$ KBOs with diameters greater than 100km (R ~ 24), WFIRST will increase the current number of known KBOs by two orders of magnitude. WFIRST has the potential to provide us with an almost complete census of KBOs with magnitudes of R< 26.7 ($m_{F087}$ ~ 27.4). This will yield the best measurement of the KBO size distribution below the observed break at R ~ 24, which will provide important constraints on the material properties of KBOs and their collisional evolution. In addition, WFIRST will enable us to make, for the first time, detailed comparisons between the size distributions of KBOs in different dynamical classes, shedding light onto the origin of the break in the KBO size distribution and the planet formation process itself. Furthermore, WFIRST will provide a detailed census of the resonant population in the Kuiper Belt and should discover 100s – 1000s of binaries, which together provide important constraints on Neptune's migration history. Finally, it will provide a uniform survey of KBO colors over a wide range of sizes. Comparison between the colors of small KBOs whose sizes are below the break radius with that of larger KBOs will, for example, show if some of the color diversity in the Kuiper belt can be attributed to collisional resurfacing.

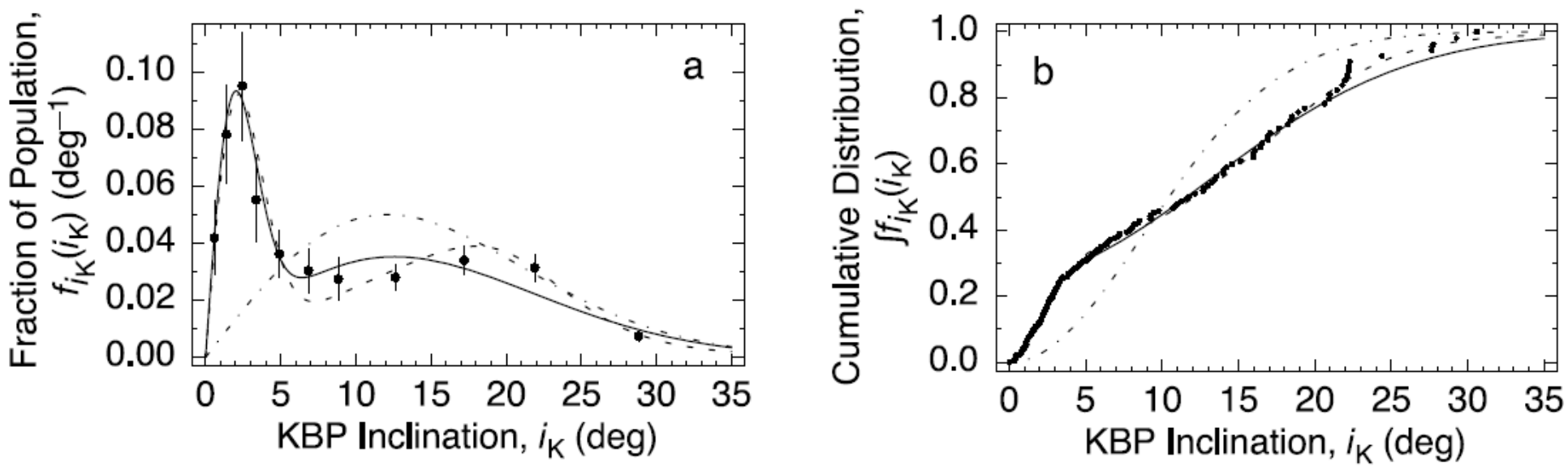

Caption: Unbiased inclination distribution of KBOs from Elliot et al. (2005).

Key Requirements

Coverage – Ideally ~14,000 deg$^2$ (+/- 20$^o$ of the ecliptic), but ~7,000 deg$^2$ suffices to discover the majority of KBOs (assuming distribution of fainter (smaller) KBOs is similar to brighter counterparts)
Cadence – 15 minute exposures, revisit each field three times separated by ~30-60 minutes
Wavelength Coverage – Need two NIR filters to get colors

Charles Alcock (Harvard-Smithsonian Center for Astrophysics), calcock@cfa.harvard.edu

# The Outer Solar System from Neptune to the Oort Cloud

Background

The end stage of planet formation in our solar system, when there was still a significant disk of proto-planetary material from which the planets were growing, is believed to have concluded with very significant migration of the giant planets. The outward drift of Neptune modified the outer disk, which led to the capture of many Kuiper Belt objects into mean motion resonances with Neptune. Large numbers of small bodies were ejected from the disk, many of which now inhabit the Oort Cloud. These populations of small bodies now inhabit a region approaching the dynamical outer boundary of the solar system (>50,000 AU).

Although the greatest distance at which a solar system object has been directly observed is <100 AU, new long period comets, first observed when they enter the planetary region, provide clear evidence for a population with semi-major axes >10,000 AU (the Oort Cloud). The intermediate volume between 100 AU and 10,000 AU is unexplored because objects at these distances are much too faint to detect in reflected sunlight (Sedna is presently near perihelion at 87 AU).

These objects may, however, be detected when they briefly occult bright stars (e.g., Nihei et al 2007, AJ, 134, 1596; Schlichting et al 2012, ApJ, 761, 150). Two events attributed to small Kuiper Belt Objects have been reported in HST/FGS time series (Schlichting et al 2012; one of these events shown in the insert). The event rate is ~$10^{-4}$/star-hour for events of this kind, and is expected to be smaller for more distant populations, but there is no observational basis for any such estimate.

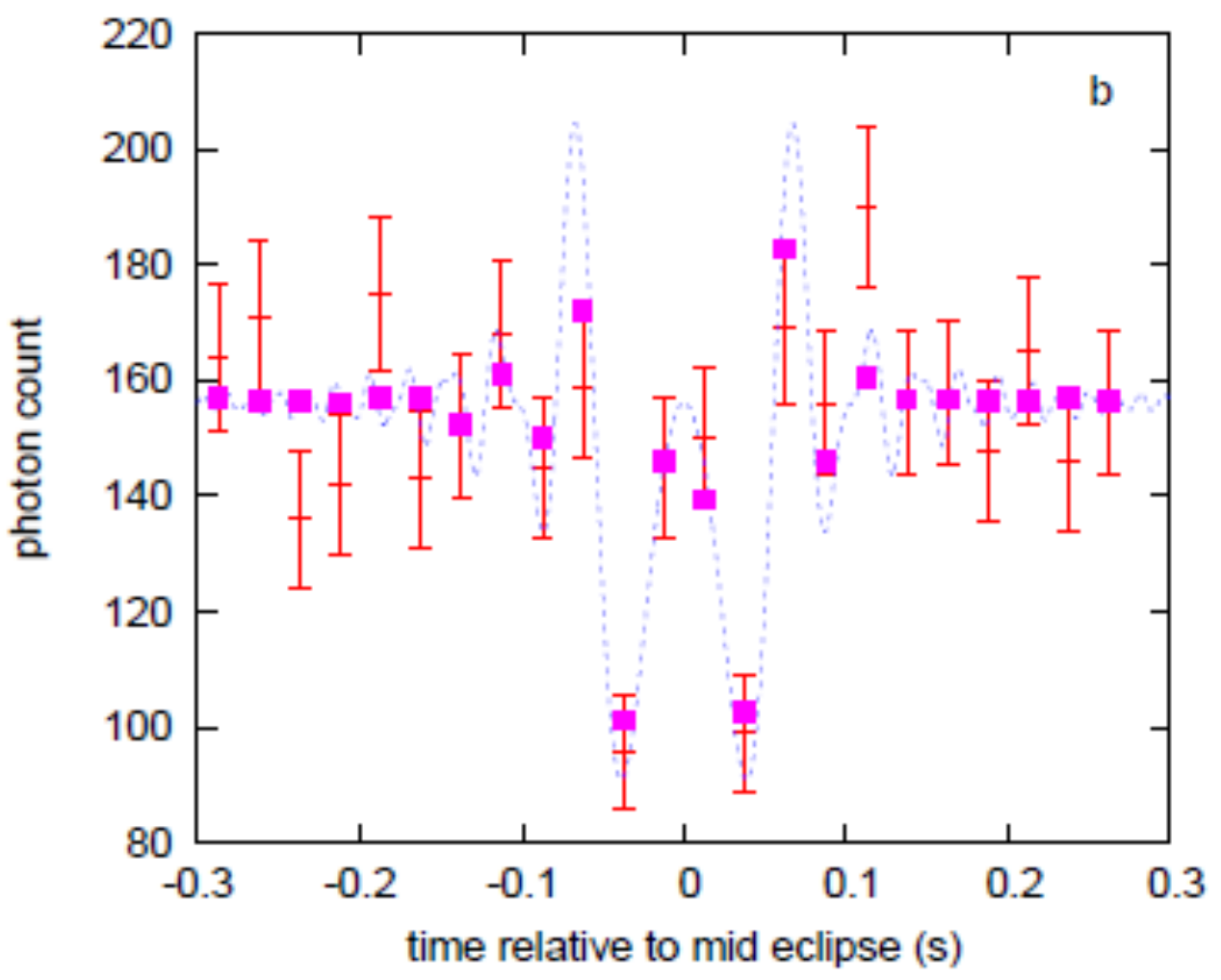

WFIRST

WFIRST could perform a highly significant survey for occultations of bright stars with no interference to its main programs by exploiting the capability of the Teledyne HxRG devices to selectively read pixels in small windows while not impacting the accumulation of photons over the majority of the focal plane. The subregions enable high cadence (~40 Hz) photometry of these stars followed by an on-board search for events.

Key Requirements

- *This is a program that runs in parallel with the core WFIRST surveys. We believe that the proposed read out mode should have no impact on the core WFIRST survey programs.* (If the WFIRST extragalactic surveys avoid the ecliptic; there may be added value by including survey fields in the ecliptic for the purposes outlined here.) We are looking forward to exploring these questions with the WFIRST team.
- Selective high cadence (10 – 40 Hz) reads of small windows on the H2RG detectors (demonstrated 32 windows per device in lab demo in Cambridge)
- On-board photometry of bright stars at 10-40 Hz to search for candidate occultation events (this has been implemented in a lab demo in Cambridge using an FPGA).
- Identified candidate events to be telemetered to ground for detailed analysis.

David R. Ardila (NHSC/IPAC), ardila@ipac.caltech.edu

# Free-floating Planets in the Solar Neighborhood

Background

As part of its primary mission, WFIRST is likely to refine the estimates for the population of very faint isolated objects, all the way down to planetary masses. Indeed, using microlensing observations, Sumi et al. (2011) conclude that isolated, free-floating Jupiter-mass planets are twice as common as main-sequence stars, and at least as common as planets around stars. Delorme et al. (2012), using data from the 335 $deg^2$ Canada-France Brown Dwarf Survey InfraRed project, report a free-floating 4-7 $M_{Jup}$ candidate which may belong to the AB Doradus moving group (50-100 Myrs).

The possibility of free-floating planets in the solar neighborhood is an exciting one: as with the study of low-mass brown dwarfs, isolated planets provide excellent laboratories to understand the evolutionary processes at play in planets around other stars. They also provide constraints on the efficiency of molecular cloud collapse mechanisms of planet formation versus expulsion from protoplanetay disks due to scattering with other planets (e.g. Veras et al. 2009). Ultimately, the discovery of a nearby free-floating planet will have a strong and lasting public impact. It is not impossible that the closest body to the solar system could be a Jupiter mass planet expelled from a young planetary system!

The Investigation

We propose to search for young (<100 Myrs) Jupiter-mass planets within 5 pc of the Sun. According to Spiegel and Burrows (2012) a 100 Myr, 1 $M_{Jup}$ has $M_J$=24.7 mag ($m_J$ AB=24.1 mag at 5 pc), $M_H$=25.9 mag ($m_H$ AB = 25.7 mag at 5 pc), and $M_K$=27.4 ($m_K$ AB=27.7 mag at 5 pc). In J and H these magnitudes are well within the capabilities of the Galactic plane WFIRST survey, if performed with a 2.4 m telescope. We limit the estimate to t<100 Myrs, d<5 pc as beyond those limits the luminosity function falls down very quickly, detection efficiency becomes poor, and ground-based follow-up is difficult. The survey area required to find a significant number of objects is uncertain, but this investigation could be carried out as ancillary to the 1240 $deg^2$ Galactic Plane survey. Multi-epoch observations are crucial to determine proper motions and rule out distant objects. Ground-based spectroscopic follow-up is required to confirm the surface gravities.

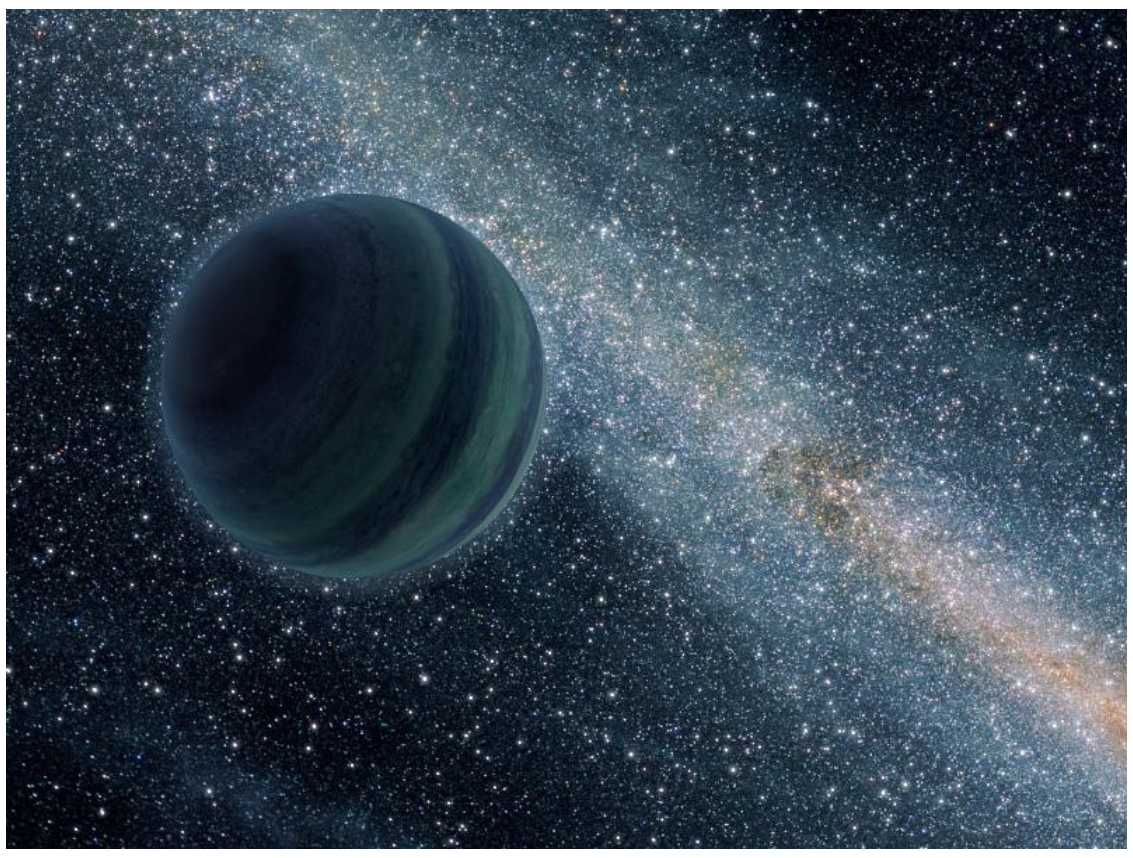

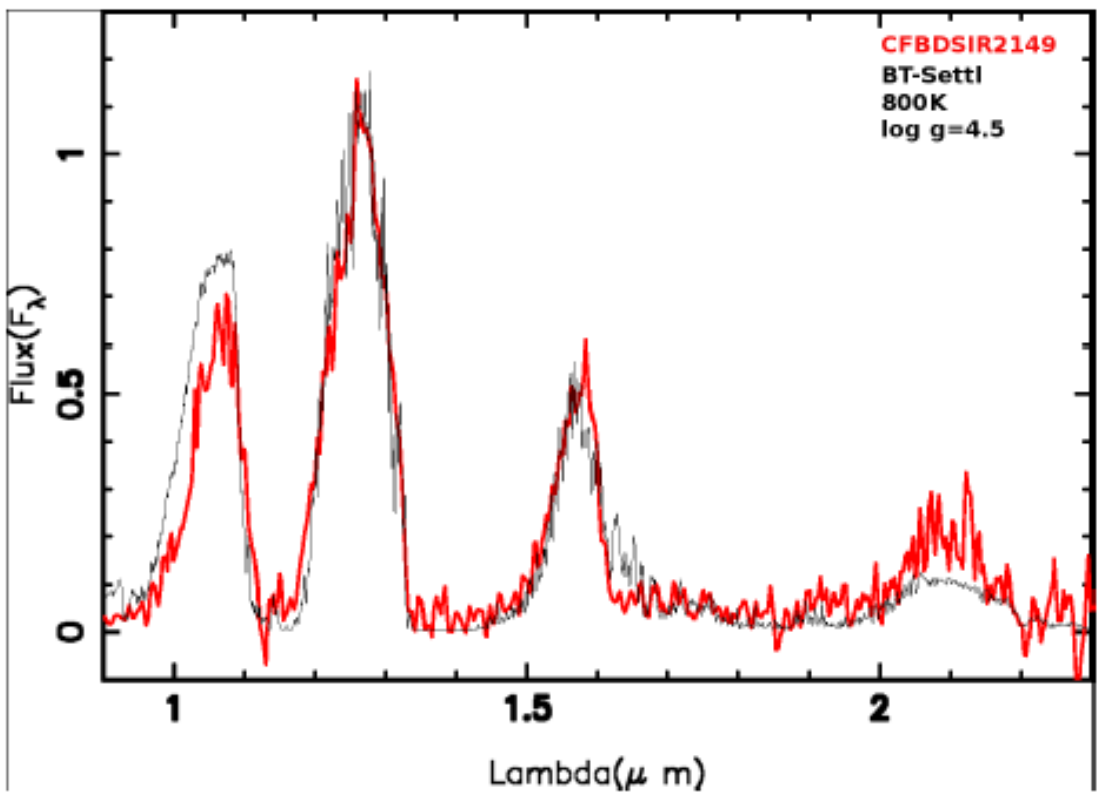

Fig. 1. Left: Artist conception of a free-floating planet in the solar neighborhood. Right: Spectrum of CFBDSIR2149-0403, a candidate free-floating Jupiter-mass object (Delorme et al. 2012).

Key Requirements

Depth – 24 AB mags in J, 26 AB mags in H; Field of View – Galactic plane survey; Cadence – Three epochs to obtain proper motion; Wavelength Coverage – YJHK

Matthew J. Holman, Charles Alcock, Joshua A. Carter (Harvard-Smithsonian Center for Astrophysics), Eric Agol (University of Washington), Michael Werner (JPL)
mholman@cfa.harvard.edu

# Measuring Planet Masses with Transit Timing Variations

## Background

The *Kepler* mission has demonstrated that systems with multiple Earth-size to Neptune-size planets are common. However, these planets' masses are inaccessible to current and future radial velocity (RV) or astrometric measurements; hence, we are blind to their bulk densities, which have important implications for planet composition, habitability, and formation.

Gravitational interactions among transiting planets in such systems can lead to observable deviations of the transit times from strict periodicity (Agol et al. 2005, Holman et al. 2005). In fact, roughly sixty of these systems show transit timing variations (TTVs). Analysis of these TTV signals has become a key tool in confirming *Kepler* planets and measuring their masses (Holman et al. 2010; Lissauer et al. 2011; Ford et al. 2012; Fabrycky et al. 2012; Steffen et al. 2010; Carter et al 2012).

The mass sensitivity of the TTV method is governed by the orbital elements of the planets, the precision with which individual transit times can be measured, the total number of observations, and the total time span of those observations. Transit times measured by WFIRST would be ~2.4 times more precise than those measured by *Kepler*, the ratio of aperture diameters. By obtaining more precise transit light curves of selected *Kepler* target stars, and by significantly extending the time baseline of *Kepler*, WFIRST could measure the masses of Earth-size planets near the habitable zones of their stars. As mentioned, this is well beyond the limits of RV observations and will not be feasible for *Kepler* stars even with the TMT.

## WFIRST

The idea would be to use either WFIRST's imager or an auxiliary camera optimized for highly precise photometry to observe transits of selected *Kepler* target stars. These stars would host multiple transiting planets, one (or more) of which would be near its habitable zone. Furthermore, we would select for systems that already show TTVs in the *Kepler* data. The frequency of transits in any given system would range from monthly to yearly, depending upon the stellar properties. The duration of the observations would typically be tens of hours. Only a small number of the most promising targets would be selected to minimize the impact on WFIRST's surveys. Future targets might come from missions such as TESS. (If, as has been proposed by Perlmutter, Werner, et al., the WFIRST IFU is used for transit spectroscopy of exoplanets, the same data might be used for transit timing, depending upon the performance of the instrument.)

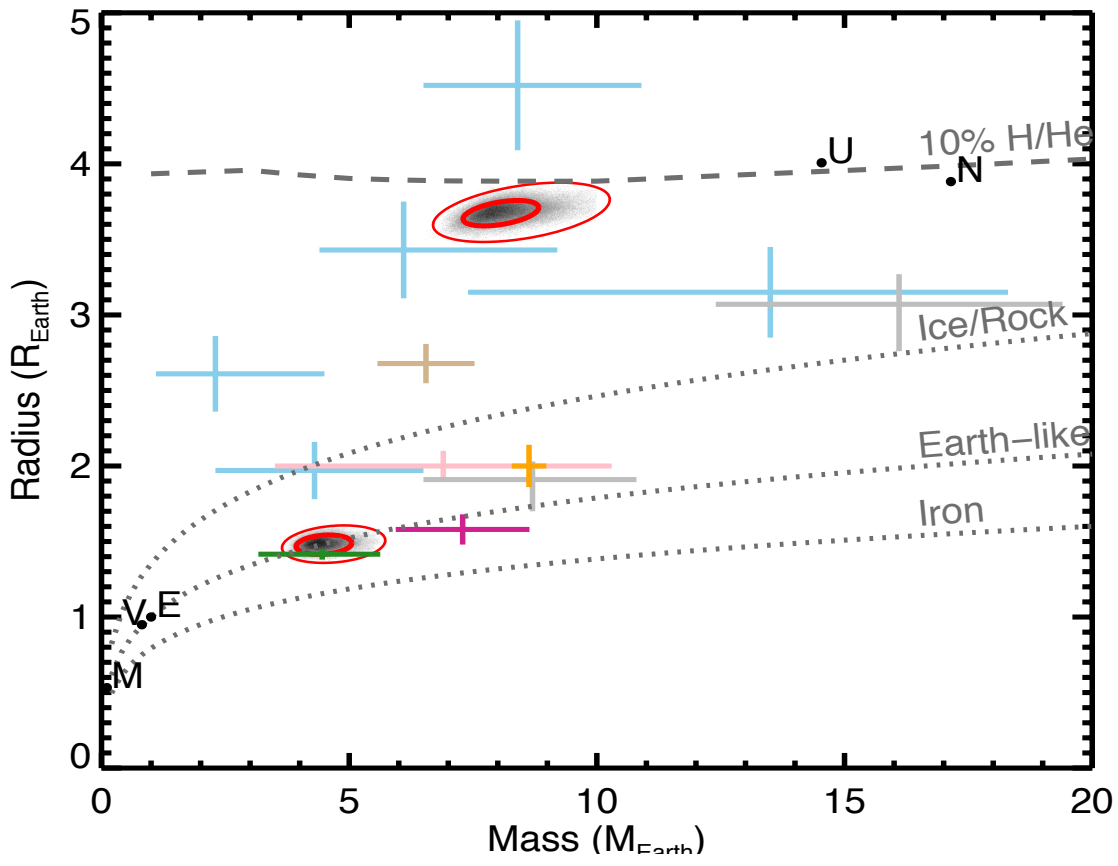

Estimates of masses and radii for a number of exo-planets, mostly from Kepler. The error ellipses refer to the well-studied Kepler 36 system. WFIRST will provide mass estimates for upward of 100 Kepler exoplanets with known radii, determining their densities, placing them on the theoretical curves, and revealing the frequency of planets of different type and composition (Carter et al., 2012).

## Key Requirements

Photometric precision – The precision with which an individual transit is measured is directly related to the depth of the transit, the S/N of the observations, and the duration of the ingress and egress of the transit. Ideally, as is the case with *Kepler*, these observations would be photon-limited.
Saturation – We would be observing bright stars. It would be essential to know that all of the photons are captured, even for heavily saturated targets.
Observing cadence – To resolve the ingress and egress of the transit of a small planet, exposure times less than a few minutes are required. This is easily satisfied by WFIRST.
Scheduling flexibility – The idea would be to use WFIRST to observe selected transits of *Kepler* targets. The approximate times of these transits would be known in advance, but it would be necessary to insert these periods of transit observation into the overall schedule.

Carl J. Grillmair (Caltech), carl@ipac.caltech.edu

## Exoplanet Spectroscopy with WFIRST

Provided a low-resolution spectroscopic capability is incorporated into the final design, NRO WFIRST could be used without serious modification for spectroscopy of transiting exoplanets. Exoplanet atmospheres could be studied both in transmission and in emission, and the effective wavelength range (.7 to 2 microns) of NRO WFIRST is well-matched to the most interesting molecular signatures we expect to see in hot-Jupiter-like atmospheres. Key diagnostic molecules would include water, methane, carbon dioxide, and carbon monoxide, which are particularly important for investigating atmospheric radiation balance, temperature structure, equilibrium/nonequilibrium chemistry, and photochemical processes.

Exoplanet spectroscopy can be carried out with either a slit or a slitless system. A slitless system necessarily entails some risk of spectral contamination by other sources in the field, but in most cases this could be overcome if the spacecraft has some degree of rotational freedom. For a traditional slit, to avoid jitter-induced, time-variable slit losses, exoplanet spectroscopy would benefit from using a slit many times wider than the FWHM of the spatial PSF. While the critical information is contained in the time domain rather than the spatial domain, an integral field unit would also be suited to exoplanet spectroscopy, provided that pointing-induced “slit losses” can be minimized. The optimal system would again use fibers, lenslets, or image slicers whose elements subtend an area several times larger than the spatial PSF.

Since the primary molecular bandheads are quite broad, high spectral resolution is not required, particularly for an initial survey. The low resolution, R = 75 grism considered in DRM1 for would be more than adequate. By obtaining spectra over multiple eclipses, WFIRST can build up signal-to-noise ratio for the faintest or most interesting sources, or search for variations in composition, temperature structure, or other signs of global climate change.

Systematics will be dominated by pointing performance and detector stability. With the large field of view available, pointing drifts can be very accurately determined in post-processing using the many thousands of other sources in the field. Without any design optimizations specific to exoplanet observations, HST and Spitzer have both demonstrated that, with suitable calibration and decorrelation techniques, systematics can be brought down to a level of 30-70 ppm (Demory et al., 2011, A&A, 544, 113, Todory et al. 2012, ApJ, 746, 111).

Some of the brightest exoplanet spectroscopy candidates will saturate NIRSpec and thus be essentially unobservable with JWST. If NRO WFIRST includes the higher resolution, R=600 prism for the galaxy redshift survey in DRM1, then WFIRST will in principle be capable of obtaining exoplanet spectra for targets ~1.6 magnitudes brighter than JWST. In the post-HST era, WFIRST may become our only means for obtaining high-precision, space-based spectra for such objects in this particularly interesting wavelength range.

Angelle Tanner (MSU), at876@msstate.edu
David Bennett (U Notre Dame)

# WFIRST: Additional Planet Finding Capabilities - Transits

## Background

Since the detection of the first transiting planet, HD209458, in 2000 over 200 planets have been detected with this method. The recently extended Kepler mission will eventually open the flood gates of such systems with over 2000 planetary candidates. The goal of the Kepler mission is to detect habitable earth-mass planets around solar-type stars and it is close to that goal. An advantage to the detection of a transiting planet is that we are able to place stringent limits on the mass and radius of the planet, thus determining its density. Finding transiting planets also allows us to study the atmospheric composition of hot Jupiters and super-Earths through transit spectroscopy and photometry.

## WFIRST

The primary objective of the WFIRST exoplanet science program is to complete the statistical census of planetary systems in the Galaxy with a microlensing survey. This program will cover 500 days of observation time over five years and is expected to be sensitive to habitable Earth-mass planets, free floating planets and all solar system analog planets except Mercury. Using estimates of the anticipated signal-to-noise of the data as well as the observing cadence and number of stars in the microlensing study, It is expected that WFIRST will obtain light curves with this cadence for ~$3X10^{8}$ stars and a photometric precision of 1%. With this sensitivity and observing cadence the WFIRST microlensing light curves can also be used to detect up to 50,000 Jupiter transits around main sequence stars and about 20 super-Earth transits around the brightest M dwarfs in the field of view. The statistics on the populations of hot Jupiters from the jovian transits will shed light on the mechanisms responsible for planetary migration when combined with information on the properties of the host star.

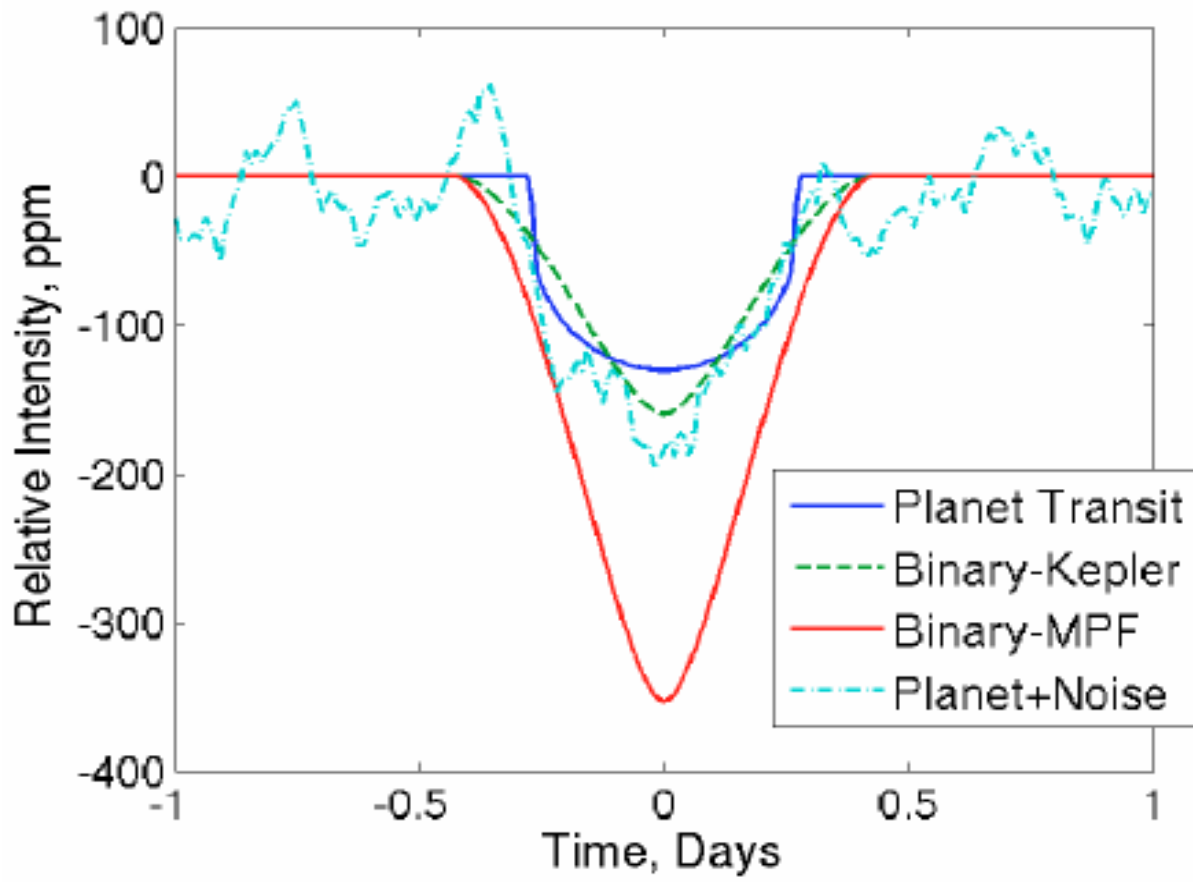

Caption: Simulated planetary transit light curve from the Microlensing Planet Finder mission (light blue, noise added) similar to WFIRST (Bennett et al. 2010) compared to Kepler (Borucki et al. 2010).

## Key Requirements

Duration – 500 days (μL survey)
Precision – 1% photometry
Cadence – Every 15 minutes
Sensitivity – J < 19 at 5 sigma
S/N – ~100 for all 300 transits of a Jupiter mass planet in a 3 day orbit

Angelle Tanner (MSU), at876@msstate.edu

# WFIRST: Additional Planet Finding Capabilities - Astrometry

Background

While the first exoplanet discovered solely through astrometric observations remains elusive, this method of exoplanet detection has been used to confirm their existence and determine the mass of the planet. While not all planets transit their star, they do all make them wobble. It is just a matter of getting a large number of data points with good astrometric precision. For instance, the 0.64$M_J$sini Jupiter mass planet discovered around the M dwarf GJ 832 at an orbital separation of 3.4 AU, period of 9.35 years and distance of 5 parsecs should produce an astrometric wobble with about a 1 milli-arcsecond amplitude.

WFIRST

The primary objective of the WFIRST exoplanet science program is to complete the statistical census of planetary systems in the Galaxy with a microlensing survey that will cover 500 days of observation time over five years. In order to maximize the number of microlensing events, WFIRST will observe a set of adjacent fields in the Galactic bulge. Observing in the near infrared significantly reduces the effects of extinction relative to visible wavelengths, increasing the number and apparent brightness of the background stars. A total of seven fields will be observed. The exposure time will be 88 seconds per field, with a slew and settle time of 38 seconds between fields. With this observing cadence and WFIRST's 0.15" spatial resolution, the positions of the ~$3X10^8$ stars collected for the microlensing survey can also be used to detect hundreds of > 10 $M_J$ planets in 100 day orbits around 0.05 – 0.3 M¤ M dwarfs. While the astrometric precision of a single observation may be on the order of 1 milli-arcseconds, WFIRST will collect a few thousand observations over the course of the microlensing program.

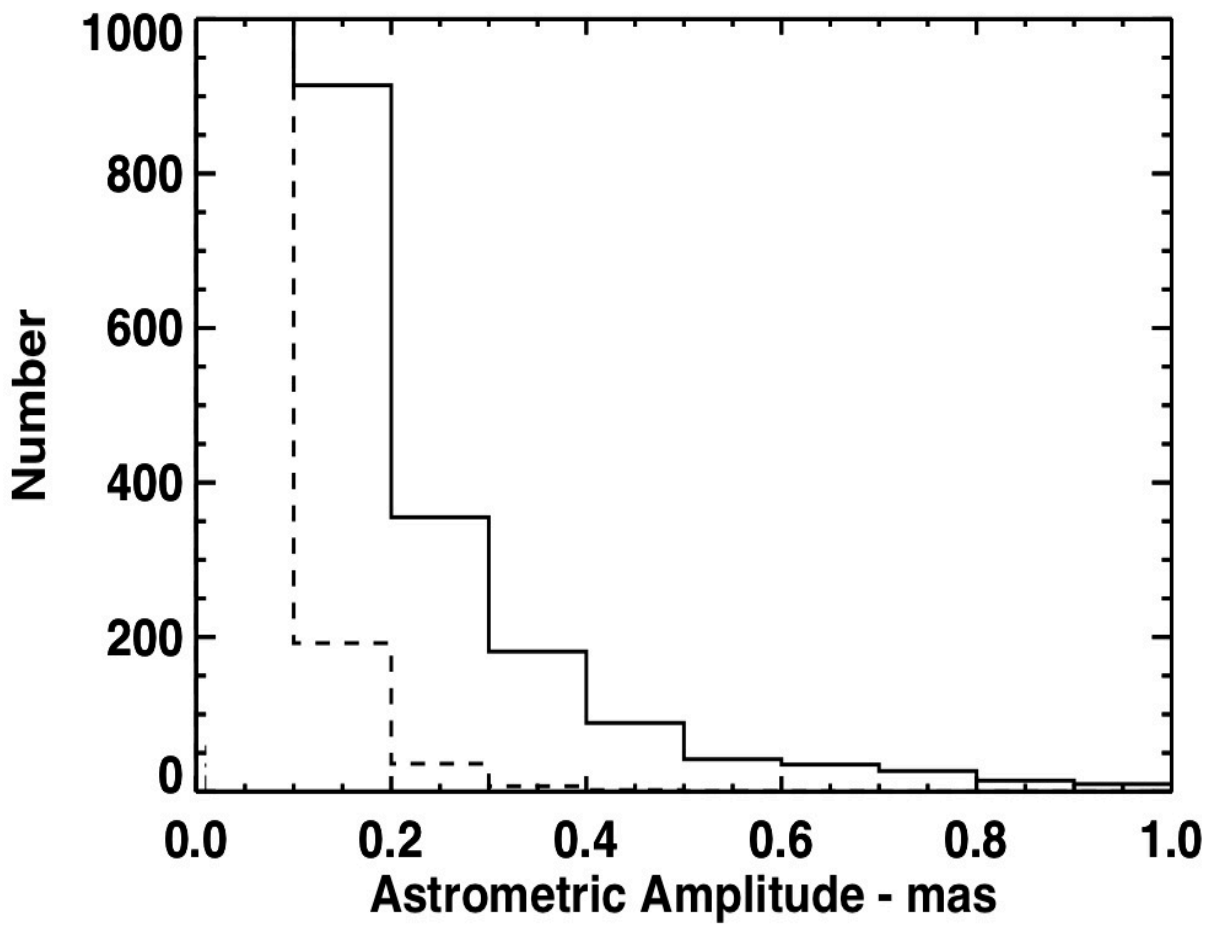

Caption: The astrometric signals of 10 $M_J$ (dashed) and 40 $M_J$ (solid) companions in a 100 day orbit around 10% of the 28000 M dwarfs in the FOV. With > 20000 observations, final astrometric precisions will be < 0.05 mas

Key Requirements

Duration – 500 days (µL survey)
Resolution – 0.15"
S/N – 100 for a J < 19 mag star
Precision – 1 mas single measurement

Lynne Hillenbrand (Caltech), lah@astro.caltech.edu

# Stellar and Substellar Populations in Galactic Star Forming Regions

## Background

Studies of low mass populations (encompassing stars, brown dwarfs, and even free-floating planetary mass objects) are hindered in star forming regions due to observational and astrophysical effects. These include: the large angular sizes of the nearby molecular clouds harboring newborn and young stellar objects, the significant extinction and reddening, and the excess emission due to accretion processes and circumstellar dust. Nevertheless, it is possible to disentangle the complexities of these effects with a combination of multi-color photometry and spectroscopy. For young stellar objects, Y-J is the best wavelength range for sampling the photospheric temperature and J-H and J-K colors are surface-gravity sensitive. Extinction, however, is degenerate with intrinsic near-infrared colors and therefore *spectroscopy is needed* in order to accurately de-redden observed colors and enable comparisons to predicted colors and magnitudes (or temperatures and luminosities). Science goals include extension of our knowledge of the initial mass function in young regions -- where the low mass objects are brighter by many orders of magnitude than they are on the main sequence -- down to and below the opacity limit for fragmentation within the molecular cloud. We see deeper into the mass function in these regions than anywhere else. What is the lowest mass object that can form like a star? How does the age distribution at 3 and 30 $M_{JUP}$ compare to that for 0.3, and 3.0 $M_{SUN}$ stars in the same cluster?

## WFIRST

WFIRST will enable sensitive, systematic large-scale surveys of nearby star forming regions. These areas are typically hundreds of sq. deg. and have been completely surveyed before only by the shallow 2MASS and WISE. UKIDSS has provided increased depth but with limited (albeit wide-area) coverage. The next steps with WFIRST will enable probes of the substellar mass function to near its bottom, over required wide areas. (JWST, by contrast, will have more depth but restricted spatial coverage). A unique capability of WFIRST will be its R=200 spectroscopy, the potential of which is illustrated in the Figure.

## Key Requirements

- Large area surveys covering known star forming regions, with possible extension in to the galactic plane depending on confusion.

- Wavelength coverage beyond the H- opacity minimum, i.e. out to K-band. Photometry in all available filters.

- Grism spectroscopy. R = 200 is sufficient to spectral type M, L, and T-type objects probing temperatures below 1000K.

- Depth to $M_H$ = 18 (15) to reach 1 (3) $M_{JUP}$ at 1 Myr (Vega System).

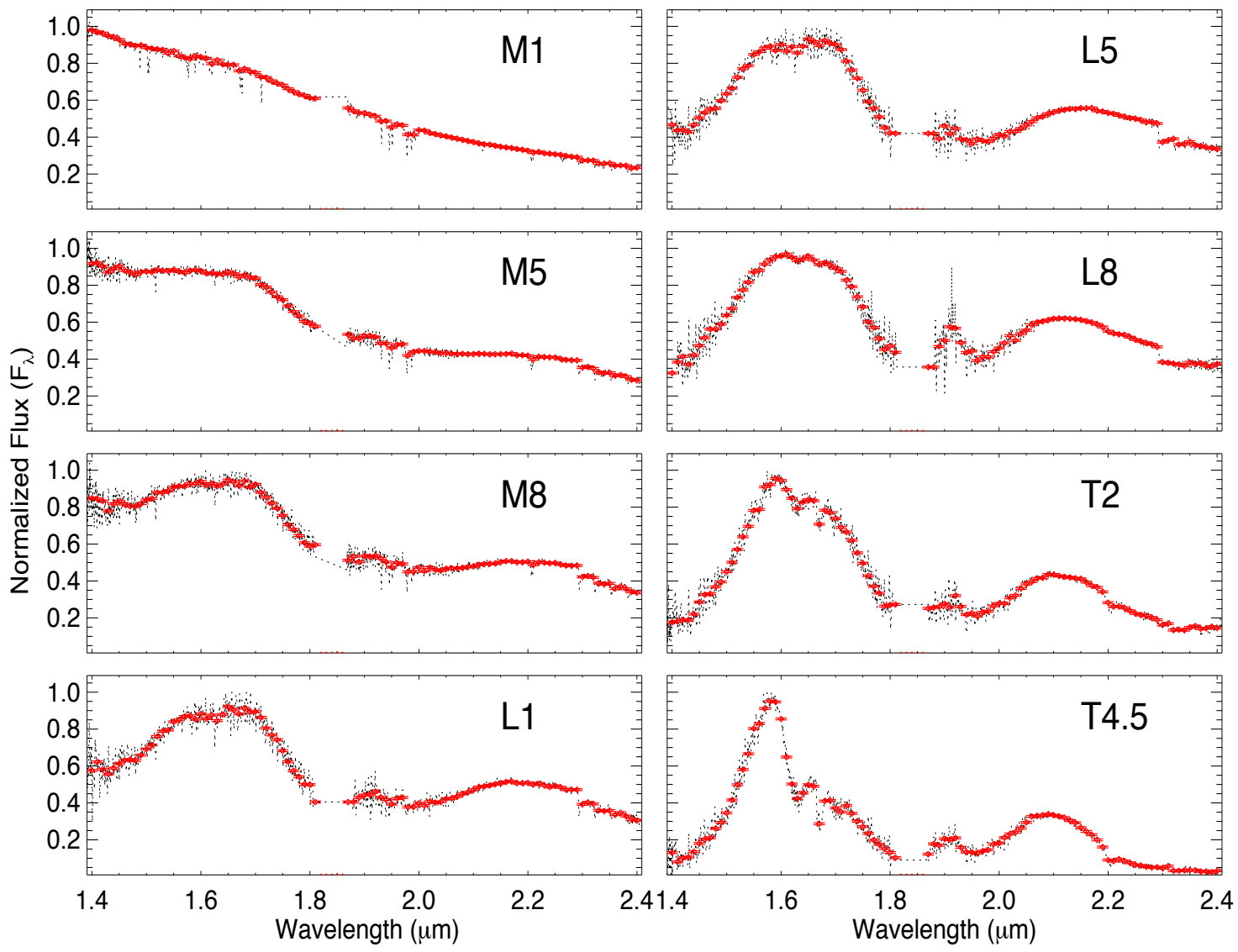

Figure credit: Emily Rice

*Angelle Tanner (MSU), at876@msstate.edu,*
*Adam Burgasser (UC San Diego), aburgasser@ucsd.edu*

## ***Identifying the Coldest Brown Dwarfs***

Background

One of the key scientific contributions of large infrared surveys such as 2MASS, SDSS, UKIDSS and WISE has been the discovery of brown dwarfs in the immediate vicinity of the Sun, sources with insufficient mass to sustain core hydrogen fusion. These include members of the newly-defined L, T and Y spectral classes whose atmospheres are distinctly planetary-like in composition ($H_2O$, $CH_4$ and $NH_3$ gases, mineral and ice clouds). While roughly 1000 brown dwarfs have been uncovered to date, some as cool as ≈300 K, the bulk of the Galactic brown dwarf population has likely cooled to temperatures that have only become recently accessible due to WISE. Because to brown dwarf thermal evolution, both the shape of the substellar mass function below 0.1 $M_\odot$ and the minimum "stellar" mass remain too poorly constrained (uncertain by factors of 3-5) to robustly constrain formation models. Current surveys of cool brown dwarfs are also highly incomplete to modest distances (e.g., up of 100 pc), inhibiting studies of Galactic spatial and velocitiy distributions, key statistics for dynamical formation models. A large-scale, volume-complete assessment of the space density of brown dwarfs over a broad temperature range (300—3000 K) would allow us to probe stellar formation processes as a function of both mass and time (given the inherent time-dependent cooling of brown dwarfs), providing simultaneous tests to brown dwarf formation and evolutionary models.

WFIRST

The combination of wide-field coverage, depth, time resolution and filter selection makes WFIRST a highly efficient survey machine for the coldest brown dwarfs. A focused infrared GO survey using at least three WFIRST filters has the potential to detect a few thousand L dwarfs and a few hundreds T dwarfs based on LSST expected performance and a smaller WFIRST field of view. The combination of LSST visual rizy and WFIRST NIR photometry will allow for a thorough census of brown dwarfs over the MLT spectral range, as LSST will be less sensitive to the cooler L dwarfs. A WFIRST brown dwarf photometric survey could be completed as a separate GO program as it would only require a couple deep (mag AB ~ 25) fields separated over a couple years for additional proper motion information independent from LSST. Some of the stars found with a LSST+WFIRST combined data set could have their parallaxes determined with the LSST data or with a focused infrared parallax programs. The brown dwarfs discovered from this survey could also be followed up with high-resolution infrared spectra from >10-m class telescopes to further constrain spectral types of the object and investigate multiplicity.

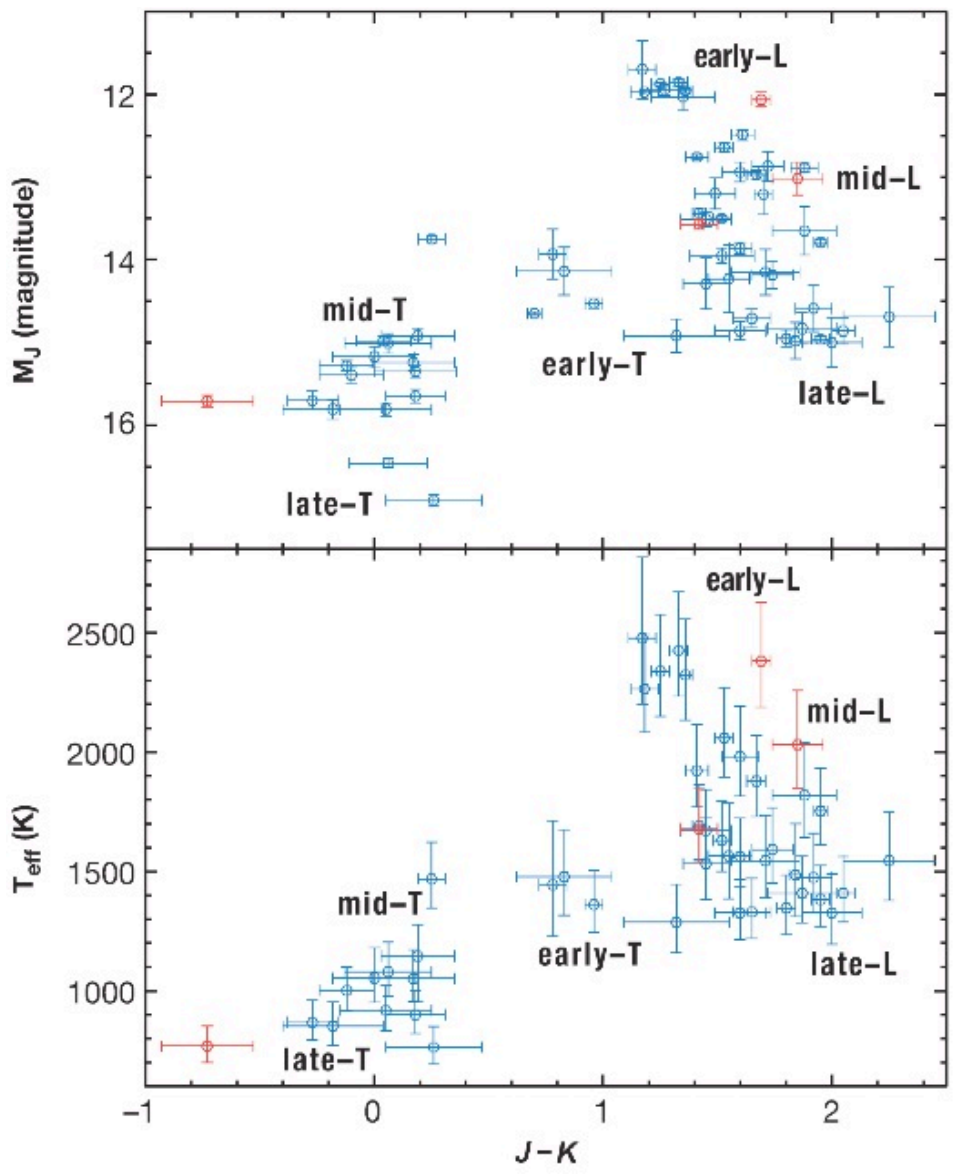

Caption: Near-IR CMD of nearby brown dwarfs utilizing data from the 2MASS and other near-infrared surveys that have wavelength coverage (JHKs) similar to WFIRST (Kirkpatrick et al. 2005). A targeted WFIRST GO brown dwarf survey will go much deeper than 2MASS and, thus, will be able to extend the bounds of a complete a census of nearby MLT brown dwarfs.

Key Requirements

Coverage – > 2500 $deg^2$ high galactic latitude, > 1500 $deg^2$ galactic plane (HL survey)
Bands – Three of the four bands - F087, F111, F141, or F178
Sensitivity – Mag AB ~ 25

Jason Kalirai (STScI), jkalirai@stsci.edu

## Stellar Fossils in the Milky Way

Background

98% of all stars will end their lives quiescently and form white dwarfs. These remnants are remarkably simple as they contain no nuclear energy sources. Over time, white dwarfs will radiate away thermal energy and cool, thereby becoming dimmer and redder. In old stellar populations, such as the Galactic halo, a significant fraction of the mass is now tied up in white dwarfs and the properties of these stars hold clues to infer the nature (i.e., the age and mass function) of their progenitors.

WFIRST

The largest sample of white dwarfs studied to date comes from the SDSS, which has increased the known population to over 20,000 remnants (Eisenstein et al. 2006). The brightest white dwarfs have $M_V$ = 10, therefore SDSS is mostly sensitive to the luminosity function out to less than a 1 kpc. WFIRST will discover and characterize the luminosity function of white dwarfs down to a much larger volume, across different sightlines. In the Galactic disk, the spatially-dependent structure of these luminosity functions correlates to peaks in the star formation history (see Figure below). The faintest stars, also in the Galactic halo, provide a robust estimate of the formation time of the first populations (Harris et al. 2006). Follow up spectroscopy of these stars can yield their fundamental properties (temperature, gravity, and mass), which can be connected to the progenitor masses through the well-measured initial-final mass relation (Kalirai et al. 2008). A wide field survey of the Milky Way disk and halo will provide a complete characterization of this remnant population in the Milky Way.

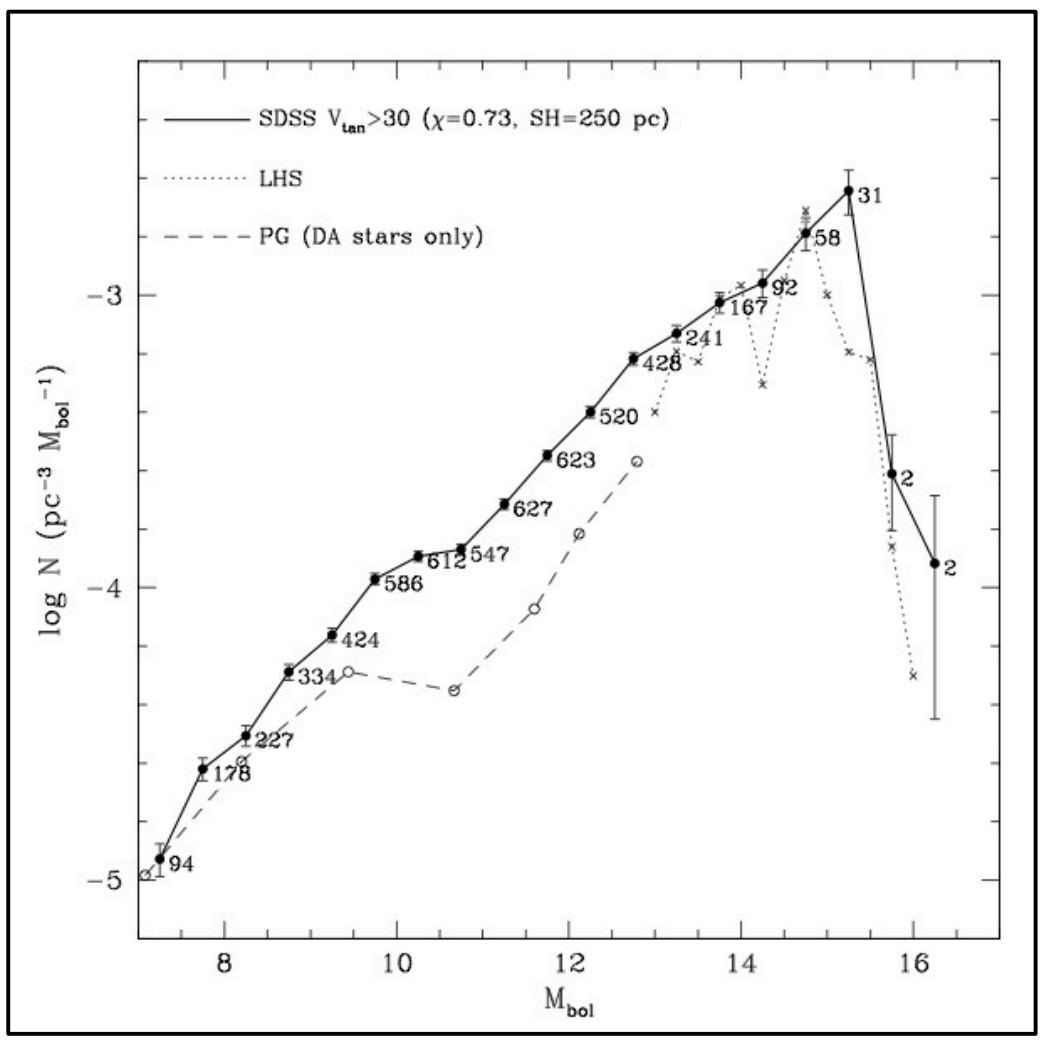

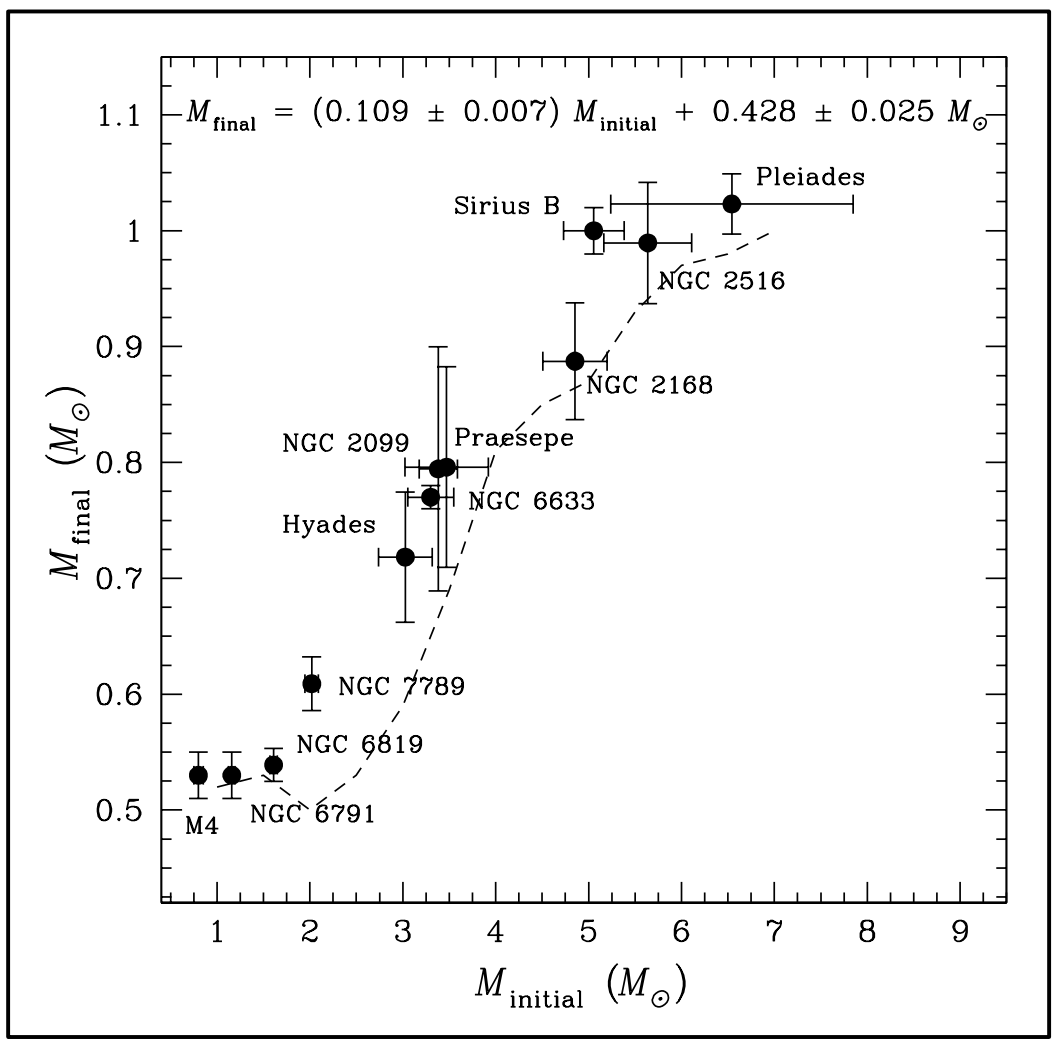

Caption: The luminosity function of white dwarfs in the Galactic disk (Harris et al. 2006) shows a turnover at the faint end (left), corresponding to the age of the oldest stars in the Galactic disk. The initial-final mass relation of stars is shown in the right panel (Kalirai et al. 2008).

Key Requirements

Depth – Measure faintest white dwarfs in the local halo ($M_V$ = 17).

Field of View – Sample range of stellar environments in Galactic disk and different probes in halo.

Cadence – Second epoch for proper motions would enable kinematic separation of components and a much cleaner selection from contaminants.

Wavelength Coverage – Two NIR filters for color-magnitude diagram analysis

Jason Kalirai (STScI), jkalirai@stsci.edu

# The Infrared Color-Magnitude Relation

Background

Star clusters in the Milky Way (MW) have served as the primary tools to measure the color-magnitude relation of stars, and to calibrate its dependency on stellar properties such as age and metallicity. This relation is a key input to test stellar evolution models, and in turn to carry out population synthesis studies that aim to interpret the integrated light of astrophysical sources across the Universe (e.g., Bruzual & Charlot 2003). For decades, this work has primarily focused on the interpretation of visible-light color-magnitude diagrams (CMDs).

WFIRST

WFIRST will enable high-precision IR CMDs of stellar populations. The figure below illustrates the morphology of the IR CMD of the globular cluster 47 Tuc, from a 3 orbit (depth) exposure with the WFC3/IR camera on HST (Kalirai et al. 2012). The sharp "kink" on the lower main sequence is caused by collisionally induced absorption of $H_2$. Unlike the visible CMD, the inversion of the sequence below the kink is orthogonal to the effects of distance and reddening, and therefore degeneracies in fitting fundamental properties for the population are largely lifted. The location of the kink on the CMD is also not age-sensitive, and therefore can be used to efficiently flag 0.5 Msun dwarfs along any Galactic sightline with low extinction. A WFIRST two-stage survey will first establish the IR color-magnitude relation and the dependency of the "kink" on metallicity through high-resolution, deep imaging of Galactic star clusters. Second, this relation can be applied to field studies to characterize the stellar mass function along different sightlines, the dependency of the mass function on environment, and to push to near the hydrogen burning limit in stellar populations out to 10's of kpc.

Key Requirements

Depth – Well dithered exposures extending down to the H burning limit in clusters with [Fe/H] = -2.2 to 0.0 (i.e., 10 kpc)

Field of View #1 – Single pointings for globular clusters covering appreciable spatial extent

Field of View #2 – Wide field survey of Galactic plane sampling over star forming regions and spiral arms

Cadence – One image per galaxy

Wavelength Coverage – Two NIR filters for color-magnitude diagram analysis

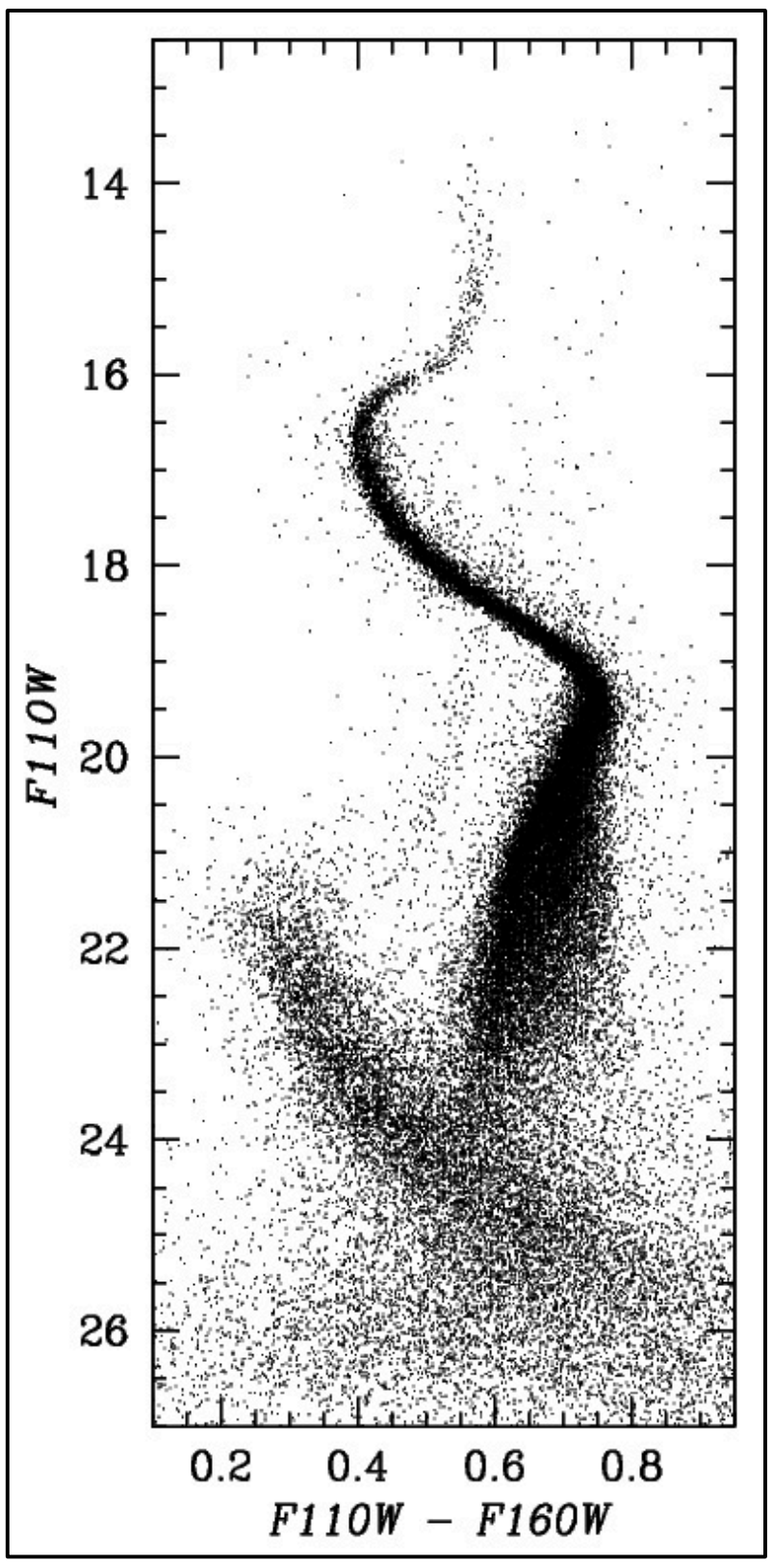

Caption: IR color-magnitude diagram for the nearby globular cluster 47 Tuc, constructed from a 3 orbit (depth) observation with HST/WFC3/IR (Kalirai et al. 2012). The kink in the lower main-sequence of the cluster is caused by $H_2$ opacity. The fainter main-sequence represents stars from the background SMC galaxy.

David R. Ardila (NHSC/IPAC), ardila@ipac.caltech.edu

## Finding the Closest Young Stars

Background

Young, nearby stars provide us with unique laboratories to study stellar evolution and planet formation. Suspected planets in their surroundings will be hot and inflated and therefore easier to image that in older systems. Since the identification of TW Hya as a nearby (50 pc), young (10 Myrs) T Tauri star (Rucinski & Krautter 1983), about 100 stars within 100 pc have been classified as young (<100 Myrs). The co-moving, co-eval associations to which they belong are extended and sparse, which makes them difficult to identify. While determining the age of a single star is subject to large systematic uncertainties, placement of the group members on a color-magnitude diagram provides a robust way to determine their age.

The Investigation

There are two components to this investigation: completing the low-mass population of young nearby associations and finding new associations. The first part involves imaging known nearby associations: the β Pictoris moving group, the ε and η Chamaeleontis clusters, the TW Hya association itself, the Tuc/Hor, Columba, AB Doradus, and Argus Associations. These all range in ages from 5 to 50 Myrs. For the second component, we will select known isolated young stars to search for low-mass stars with their know space motions. Determining membership to the association requires determination of the UVW velocity components and medium resolution spectroscopy to search for youth indicators. The 2.4m WFIRST will be able to detect all the Y0 ($M_H$=20 mag, Kirkpatrick et al. 2012) dwarfs within 100 pc ($m_H$(AB)≈26.3 mag). The role of WFIRST is to determine proper motions of very low-mass objects (best done in the NIR, where they are brightest) in an efficient way, given its large field of view. Repeated visits (at least three and preferably five) and ground-based spectroscopic follow-up will be necessary.

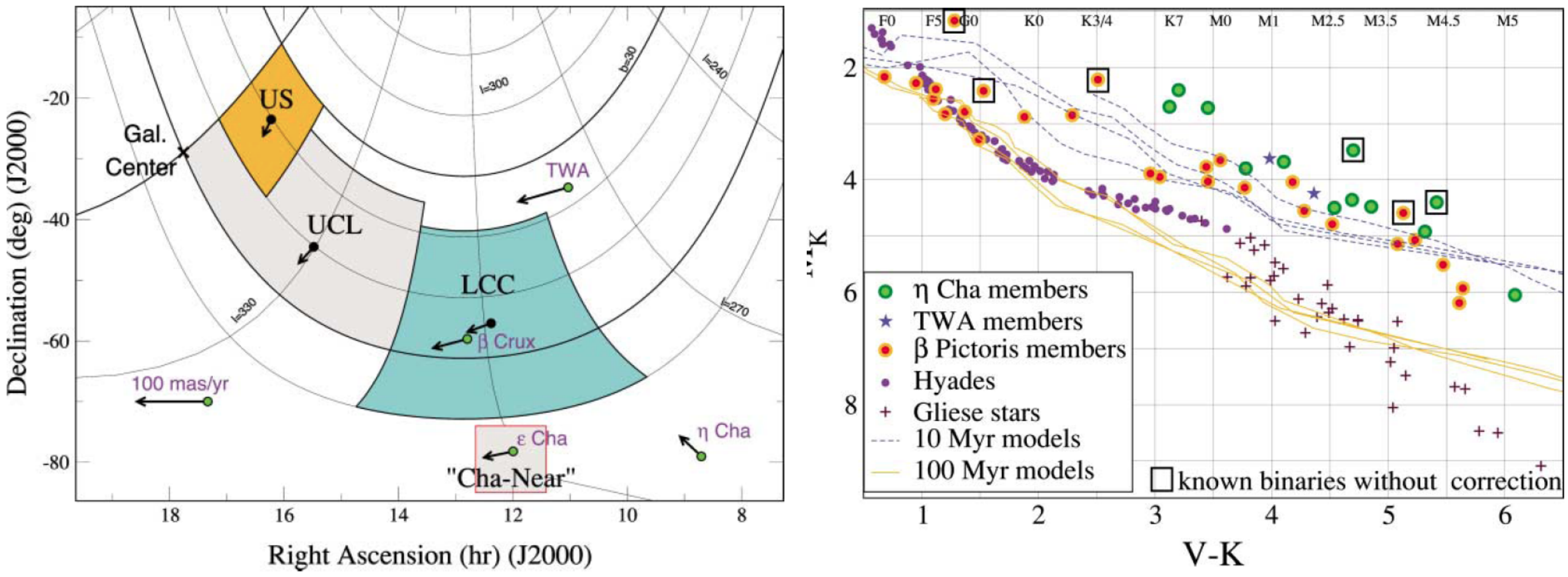

Fig. 1: Left: Nearby southern young associations. Right: CMD diagram for associations nearby the sun. From Zuckerman & Song (2004)

Key Requirements

Depth – 26.3 AB mags in H to reach all Y0s within 100 pcs

Field of View – 0.375 square degrees (DRM1)

Cadence – At least three epochs, separated by one year

Wavelength Coverage – YJHK

John Stauffer (IPAC), stauffer@ipac.caltech.edu

## The Most Distant Star-Forming Regions in the Milky Way

Background

Wide field imaging surveys of nearby spiral galaxies allow us to study how the star-formation process depends on global properties (Fe/H, mean gas surface density, disk kinematics). By their nature, those studies are limited in their abilities due to the distance of the target galaxies from the Earth. Similar studies in the Milky Way could do much better, but are limited by interstellar extinction and confusion. A space-based near-IR survey of the outer galaxy can overcome many of these problems and allow use of the Milky Way as a laboratory for how the star-formation process depends on metallicity, gas surface density and triggering mechanisms. The most distant star-forming regions in the galaxy are expected to have [Fe/H] ~ -1.0 (Brand et al 2003) – comparison of the IMF of such clusters to those in the inner galaxy could help interpret colors of high redshift galaxies where low metallicities are also expected.

WFIRST

The expected edge of the star-forming disk of the Milky Way is at of order R(G) ~ 20 kpc, or about 12 kpc distant from Earth. Towards the outer-galaxy where confusion is much reduced, a WFIRST JHK survey is capable of reaching stars down to 0.1 Msun at 1 Myr at 12 kpc. Two-color J-H, H-K diagrams can then be used to identify YSOs with warm, dusty circumstellar disks. Multi-epoch imaging of candidate clusters can further identify additional members by their variability. A single epoch survey of the entirety of quadrants 2 and 3 of the disk over +/- one degree in latitude could be accomplished in less than a month of observing time with WFIRST. Alternatively, the same amount of time could be used to sample a smaller longitude range but wider latitude range (to account for larger gas scale height and disk warp at large galactocentric distance).

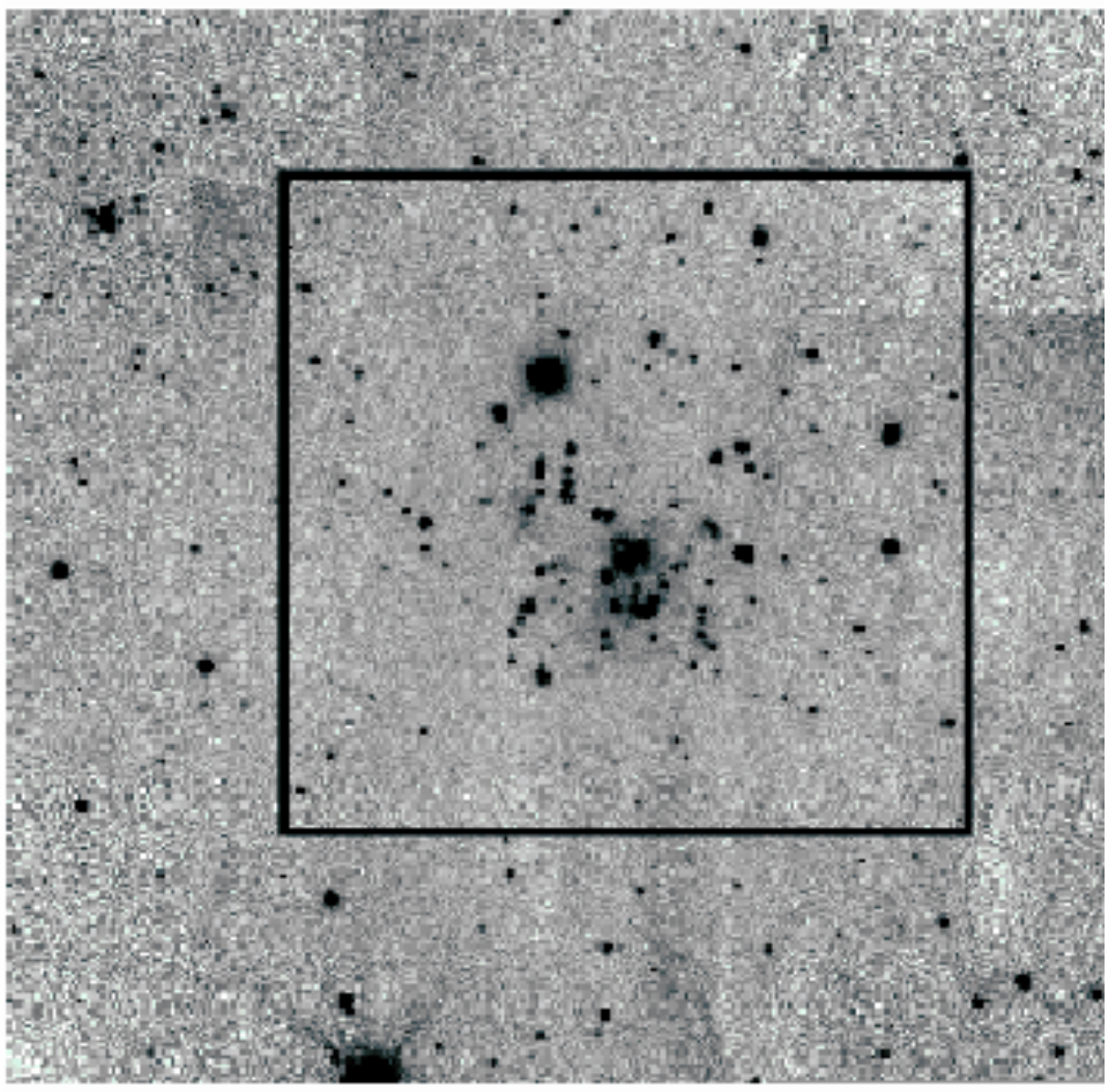

Caption: Near-IR image of the star-forming region WB89-719, aka IRAS 06145+1455, the most distant known star-forming region at R(G) ~ 20.2 kpc (Brand and Westerloot 2007, AA 464, 909).

Key Requirements

Depth – To below 0.1 M(sun) at R(G) = 20 kpc, or to K ~ 19 (5 sigma)
Field of View – The outer MW plane – 90 < l < 270, -1 < b < 1
Cadence – One epoch all bands; multiple epochs on specific regions
Wavelength Coverage – JHK filters for color-color and CMD analysis

Frantz Martinache (Subaru Telescope) frantz@naoj.org

# Super-resolution imaging of low-mass stars with Kernel-phase & precision wavefront calibration with Eigen-phase

Background

In the high contrast regime or at very small angular separations, traditional image analysis techniques fail to interpret some of the content, as faint companions and structures are buried under diffraction features. In such situations, it is advantageous to adopt an interferometric point of view instead, and examine the Fourier properties of images. With an instrument providing Nyquist-sampled non-saturated images, and when wavefront errors become small (Strehl ~ 50 %), it is possible to extract self-calibrating observable quantities called kernel-phases (a generalization of the notion of closure-phase), which exhibit a compelling property: they do not depend on residual wavefront errors, therefore enabling the detection of features in the super-resolution regime. The technique was already successfully applied on HST/NICMOS archival data, and uncovered previously undetected companions around nearby M- and L-dwarfs in an unambiguous fashion (Martinache, 2010, ApJ, 724, 464; Pope et al, 2013, PASP, submitted). The same linear model that leads to kernel-phases also provides Eigen-phases, which in combination with an asymmetric pupil mask enable high precision wavefront characterization (Martinache, 2013, PASP, submitted), important for coronagraphs and general telescope performance.

Kernel-phase on NRO telescope

A space-borne telescope is the ideal observatory for the study of cool dwarfs and their companions in the near-to-mid IR. Ground based observations of such objects indeed typically require Laser Guide Star Adaptive Optics (LGSAO) observations, which do not exhibit satisfactory performance yet; and the detection limits are strongly limited by the background, beyond the K-band.

The self-calibration properties of Kernel-phase enable detection of companions and high-precision relative astrometry in a regime of angular separation (< 100 mas in the near-IR) that currently eludes any other imaging technique. Repeated observations allow to fully characterize orbital parameters, which combined with RV or wide field astrometry, provide high precision dynamical masses of these otherwise elusive objects (e.g. Martinache et al, 2007, ApJ, 661, 496; Martinache et al, 2009, ApJ, 695, 1183).

The application of the technique is however not restricted to nearby low-mass stars. Any observation usually performed with high-performance AO, for instance, resolved observations of the Galactic center could benefit from Kernel-phase.

Key requirements

Sampling: PSF should be Nyquist sampled at the shortest wavelength of interest

Wavefront quality: Strehl better than 50 % at the shortest wavelength of interest

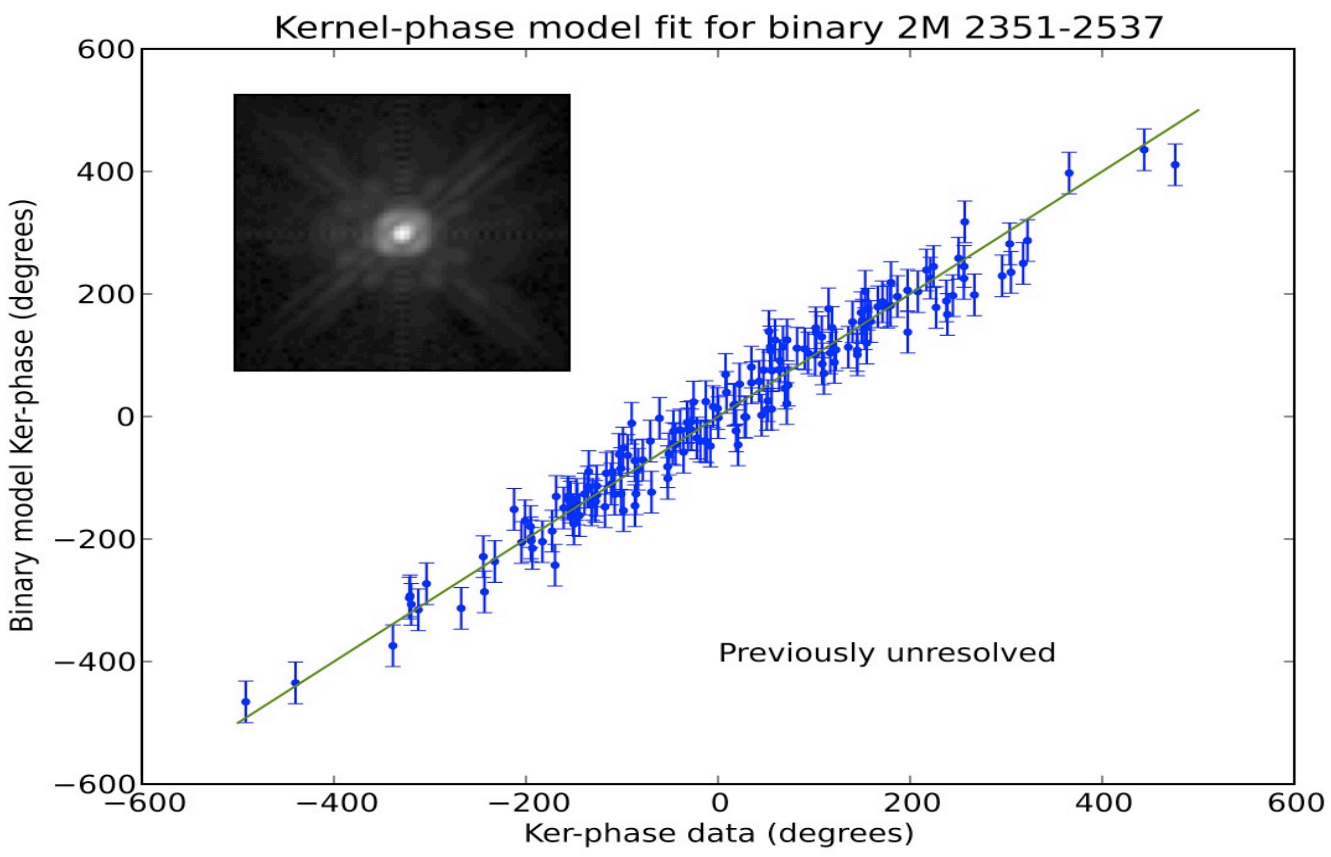

Example of NICMOS Ker-phase super-resolution (64 milli-arc second separation) discovery around the nearby L-Dwarf 2M 2351-2537 (Pope et al, 2013).

Kailash Sahu (STScI), ksahu@stsci.edu

**Detecting Neutron Stars and Black Holes through Astrometric Microlensing**

Stars with initial masses of ~8–20 M⊙ are expected to end their lives as NSs, and those with greater masses as BHs (Heger et al. 2003). As much as 10% of the total mass of our Galaxy should be in the form of NSs and BHs (Oslowski et al. 2008). However, a vast majority of stellar remnants are expected to be single, either primordially or due to disruption of binaries by supernova (SN) explosions (Agol & Kamionkowski 2002). Such isolated massive remnants are extremely difficult to detect directly, and in fact no isolated BH has ever been unambiguously found within our Galaxy. Mass measurements of NSs and BHs mostly come from observations of binary systems which show NS masses to be concentrated around 1.4M⊙, and BH masses to be a narrow Gaussian at 7.8 ± 1.2 M⊙ (Ozel et al. 2010). However, theoretical models (Fryer & Kalogera 2001) predict that the compact-remnant distribution should be a continuous distribution from 1.2 to 15M⊙. The discrepancy between the observed and predicted mass distribution of BHs is generally attributed to the fact that, BHs in binaries are a biased and minority sample. What is missing is an unbiased mass distribution for isolated stellar remnants.

**Astrometric Microlensing:** Isolated stellar remnants can be detected through microlensing. The microlensing survey programs such as OGLE and MOA have so far detected more than 8000 microlensing events. If stellar remnants constitute a few percent of the total mass, many of these observed microlensing events must be due to stellar remnants. However, microlensing light curves are degenerate with respect to the mass, velocity and distance of the lens. A route to resolving these degeneracies arises from the fact that microlensing, in addition to amplifying the brightness of the source, produces a small shift in its position (Dominik and Sahu, 2000). Thus if high-precision astrometry is added to the photometry, the deflection of the source image can be measured, and thus the mass of the lens determined unambiguously. The sizes of the astrometric shifts, however, are such that they require a 2m-class space based telescope.

**WFIRST/NRO:** Astrometric precisions of order 300 microarcsec have been achieved with HST/WFC3, so we expect that a similar astrometric precision is achievable with NRO. The maximum astrometric deflection caused by a 0.5 M⊙ lens at 2 kpc is 400 μas (see Figure 1). This implies that NRO can measure the astrometric deflection caused by almost all NSs and BHs.

In order to unambiguously measure the mass of the lens, the remaining quantity to be determined is the distance to the lens. Fortunately, the timescales of microlensing events caused by NSs and BHs are long ( >30 days), so photometric observations of these events with NRO will provide clear measurements of the parallax distances.

From our HST observations towards the galactic bulge, we estimate that NRO will enable monitoring of about 20 million stars towards the Galactic bulge in a single pointing, covering about 0.25 square degrees. By monitoring 10 fields in the Galactic bulge, 200 million stars can be easily monitored continuously. Since the optical depth towards the Galactic bulge is about 3 * 10(-6), this will lead to the detection of 600 microlensing events at any given time, and 5000 events per year. Taking the current statistics of microlensing events, about 10% of them (500 events) are expected to have $T_E$ > 30 days, and at least three dozen of them are expected to be due to NSs and BHs. Observations taken over 5 years will lead to the detection and mass measurement of well over 100 NSs and BHs. These measurements will provide (i) constraints on SN/GRB explosion mechanisms that produce NSs and BHs, and (ii) a quantitative estimate of the mass content in the form of stellar remnants.

Scott Gaudi (OSU), gaudi@astronomy.ohio-state.edu, Matthew Penny (OSU)

# Proper Motions and Parallaxes of Disk and Bulge Stars

## Background

Measurements of the kinematics and three-dimensional structure of stars in the Galactic Bulge and inner disk allow for the determination of the dynamical mass in these populations, and provide important clues to their formation and evolutionary history. The dominant formation mechanism of bulges in the universe (i.e., secularly-grown pseudobulges versus merger-driven classical bulges) remains poorly understood (i.e., Kormendy & Kennicutt 2004), and our bulge provides the nearest and thus most accessible example with which to test the predictions of various formation models. Nevertheless, there are a number of challenges to obtaining such measurements for the bulge, in particular the small proper motions and parallaxes, and the large and variable extinction towards the bulge. To date, information on the kinematics of bulge stars has been limited to primarily radial velocities, luminous stars or stars with large proper motions, or a few deep but narrow pencil-beam surveys with HST. Direct geometrical distances to individual stars in the bulge have been essentially unavailable.

## AFTA-WFIRST

AFTA-WFIRST will enable high-precision proper motion measurements and rough (~10%) parallaxes to essentially all the ~$10^8$ bulge and foreground disk stars in the ~2 square degree microlensing survey field-of-view with magnitudes of J<22. For the exoplanet survey, AFTA-WFIRST will achieve a SNR~100 per 85s observation for stars with J<22. Assuming a resulting astrometric precision of a $\sigma_{AST}$ ~ mas per observation (i.e., $\sigma_{AST}$ ~ FWHM/SNR~0.15"/100~mas), and N~10^4 observations, the final mission uncertainty on the measured proper motions over a T~4 year baseline will be $\sigma_{\mu\mu}$~ sqrt(12/N)*( $\sigma_{AST}$/T) ~ 0.01 mas/yr. The typical proper motion of a star in the bulge is $\mu$~100 km/s/(8000 pc) ~ 3 mas/year, and thus individual stellar proper motions will be measured to ~0.3%. Similarly, the fractional uncertainty on the parallax of a star in the bulge will be $\sigma$_pi/Pi ~ $\sigma$ $_{AST}$/Pi/sqrt(2N) ~ 10%, where Pi ~ 1/8 mas is the typical parallax of a star in the bulge. AFTA-WFIRST will provide proper motions and parallaxes that are more accurate by a factor of ~5 than WFIRST DRM1 at J~22, and will provide proper motions and parallaxes for roughly twice as many stars to a given precision. Note that these estimates assume that systematic uncertainties can be controlled to better than ~0.01 mas. Because these observations will be taken in the NIR, they will reach below the bulge MS turnoff and will be relatively unaffected by extinction. The occasional observations in bluer filters will help distinguish between bulge and disk populations. With these measurements, AFTA-WFIRST will provide unprecedented measurements of the kinematics and structure of the bulge.

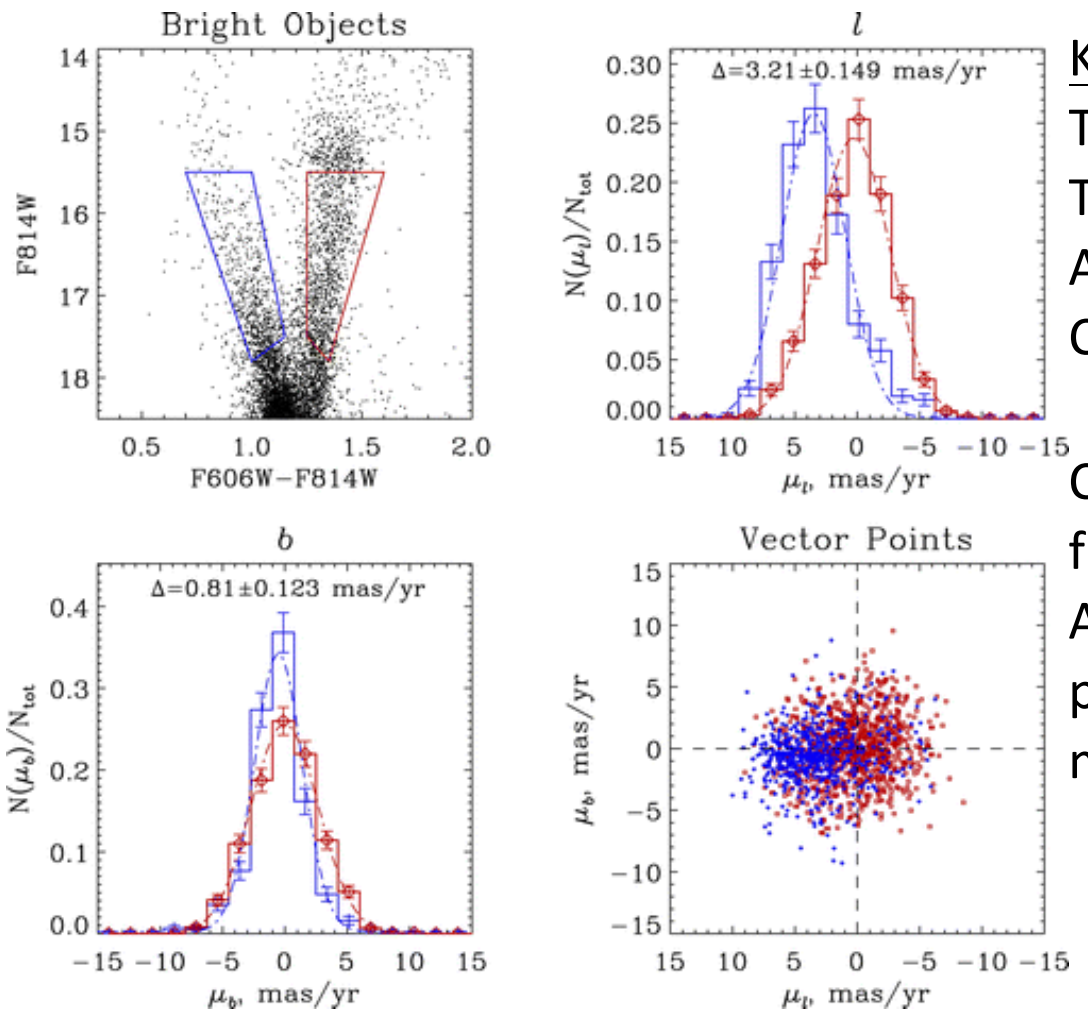

Key Requirements

Total Span of Observations –T~ 4 years

Total number of epochs - ~10^4

Astrometric precision per epoch for J<22 of ~ mas

Control of systematic errors to better than ~0.01 mas

Caption: CMD and bulge, disk proper motion distributions from an HST study using ASC/WFC (Clarkson et al 2008, ApJ, 684, 1110). AFTA-WFIRST will provide individual proper motions that are roughly an order of magnitude more accurate than HST.

Daniel Stern (JPL), daniel.k.stern@jpl.nasa.gov

# Quasars as a Reference Frame for Proper Motion Studies

## Background

Luminous quasars behind the Magellanic Clouds or the Galactic bulge provide several unique and important scientific applications. First, they provide the best reference frame for proper motion studies (e.g., Kallivayalil et al. 2006a,b; Piatek et al. 2008). For example, recent improvements measuring the proper motion of the Magellanic Clouds all relied on Hubble observations of fields centered on quasars. The results were surprising, as the Clouds were found to be moving significantly faster than previous estimates (e.g., van der Marel et al. 2002). The tangential motion of the SMC was also found to differ from that of the LMC. This implies that the Clouds may not be bound to each other or the Galaxy, and may instead be on their first pericentric passage. Precisely measuring the proper motion of the Magellanic Clouds will also improve modeling of the Magellanic Stream, which is a sensitive probe of the Galactic potential. Finally, bright background quasars are useful background probes for absorption studies of the interstellar medium.

## WFIRST

The deep, wide-field, high-resolution imaging capabilities of WFIRST will provide a significant improvement for this science, vastly improving the statistics relative to previous Hubble studies. The precise astrometry afforded by space-based observations, combined with the large number of background AGN identified from their broad-band spectral energy distributions (SEDs), including mid-IR data (e.g., Kozlowski et al. 2011, 2012), will allow large numbers of both AGN and Galactic/Magellanic Cloud stars to be identified in each field.

## Key Requirements

Depth – To provide high S/N detections of large numbers of AGN
Morphology – To precisely measure quasar positions
Grism – For spectroscopic confirmation of quasar candidates
Field of View – Wide-area to improve statistics
Wavelength Coverage – single band sufficient

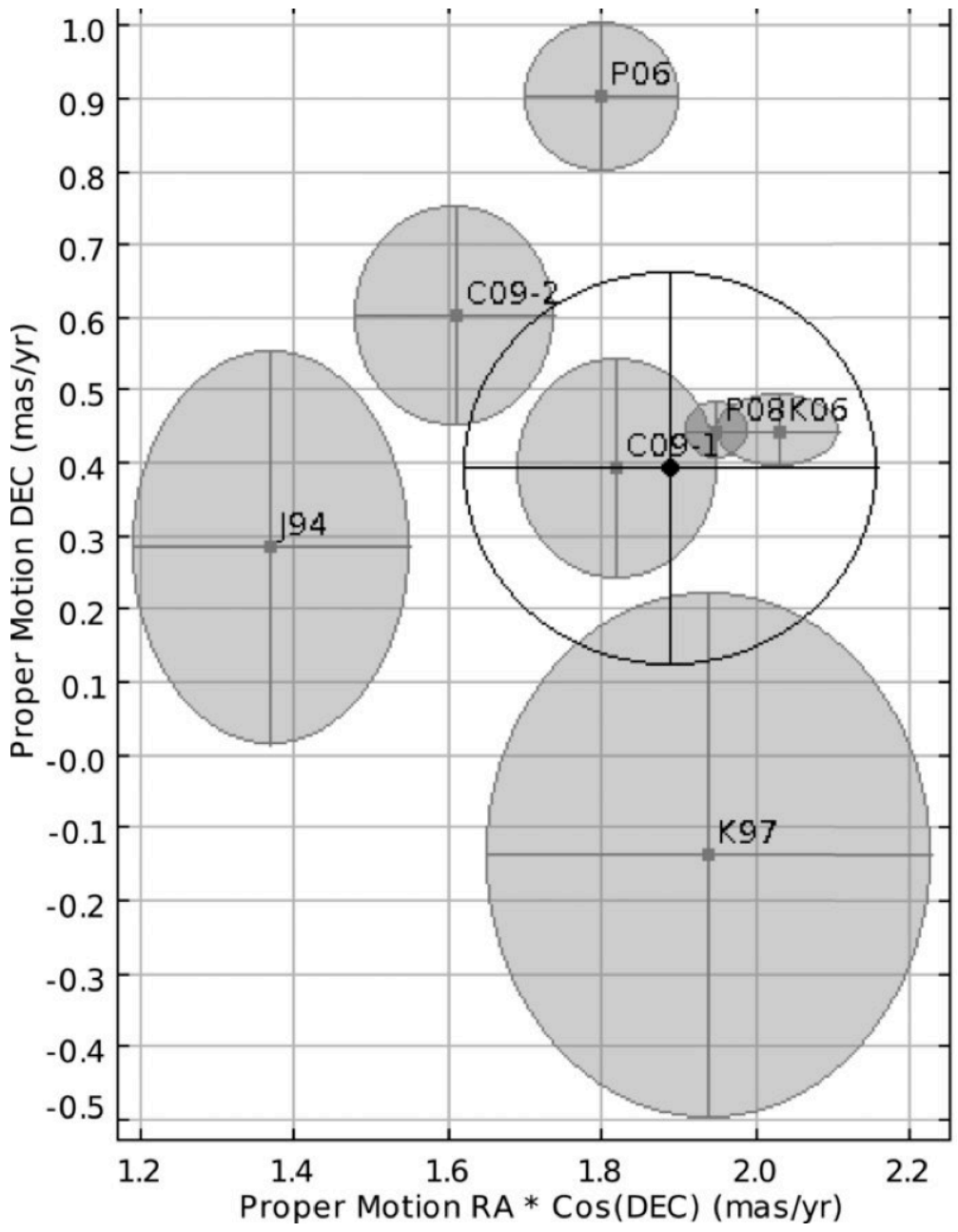

Caption: Recent measurements of the proper motion of the Large Magellanic Clouds, from Vieira et al. (2010).

Gurtina Besla (Columbia) gbesla@astro.columbia.edu; Roeland P. van der Marel (STScI) marel@stsci.edu

# The Detection of the Elusive Stellar Counterpart of the Magellanic Stream

## Background

As the most massive satellites of the Milky Way (MW), the Large and Small Magellanic Clouds (LMC and SMC) play an important role in our still developing picture of the buildup and evolution of the Local Group. In the $\Lambda$CDM paradigm, MW-type halos are expected to have built up the majority of their mass by the accretion of LMC-type subhalos; *the evolution and disruption of the Magellanic System is thus directly relevant to our understanding of how baryons are supplied to the MW.*

The Magellanic Clouds (MCs) are undergoing substantial gas loss, as is evident by the stream of H I trailing (the Magellanic Stream) and leading the MCs (the Leading Arm), in total stretching 200$^\circ$ across the sky (Mathewson et al. 1974; Nidever et al. 2010). Despite extensive modeling and multi-wavelength studies of the system, the dominant mechanism for the formation of this extended gas distribution is unknown. Leading theories are: tidal stripping of the SMC (by the MW, Gardiner & Noguchi 1996; or by the LMC alone, Besla, Hernquist & Loeb 2012; see Figure) or hydrodynamic processes (ram pressure stripping, Mastropietro et al. 2005; stellar outflows, Nidever et al. 2008). *Distinguishing between these formation scenarios is critical to the development of an accurate model for the orbital and interaction histories of the MCs with each other and with the MW.*

In the tidal models, stars are removed in addition to gas. In contrast, stars are not removed in any of the hydrodynamic models. *The detection of stellar debris in the Magellanic Stream would prove conclusively that the Stream is in fact a tidal feature, ruling out models based on hydrodynamic processes.* Although previous searches for stars in sections of the Stream have yielded null results (e.g., Guhathakurta & Reitzel 1998, Bruck & Hawkins 1983), the predicted optical emission from the stellar debris in e.g., the Besla, Hernquist & Loeb 2012 model ( >32 mag/arcsec$^2$; V-band ), is well below the observational limits of these prior surveys.

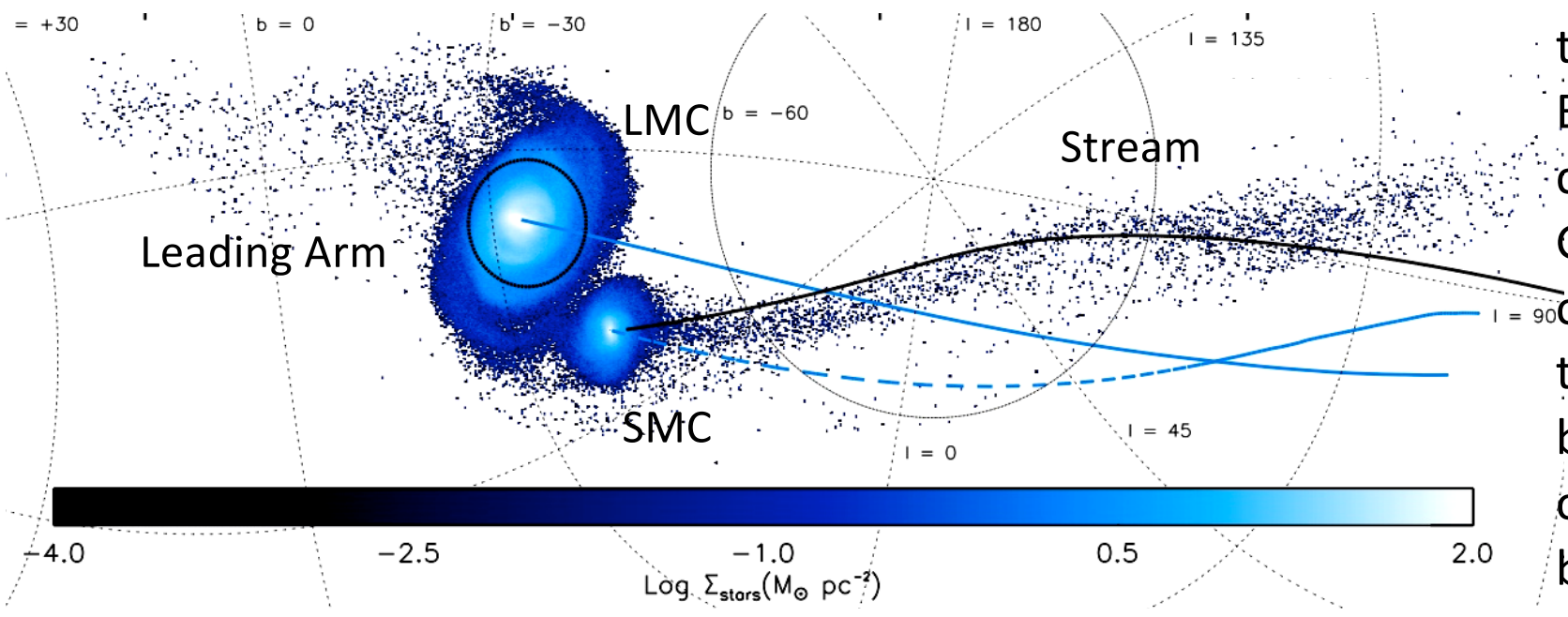

Caption: The stellar counterpart to the Magellanic Stream from Besla, Hernquist & Loeb (2012); color indicates stellar density. Galactic coordinates are overplotted. The past orbits of the MCs are indicated by the blue lines. The current location of the gaseous Stream is marked by the solid black line. Densities < 0.01 M⊙/pc$^2$ are typical.

## WFIRST

WFIRST will enable a uniform, wide-field, deep survey across sections of the gaseous Magellanic Stream to ascertain the existence of a faint stellar counterpart. Using MC main sequence stars as tracers of stellar debris has been proven to reach equivalent population surface brightness of >33 mag arcsec$^{-2}$ based on CTIO Mosaic-2 imaging (Saha et al. 2010); with WFIRST's wider field of view, larger surface brightness limits can be reached. Furthermore, a wide area survey will uncover any spatial offsets between the gaseous and stellar stream, which may occur owing to hydrodynamic gas drag.

## Key Requirements

Depth – NIR mag limit of >23.5 (to reach I-band surface brightness >33 mag/arcsec; cf. Saha et al. 2010)
Field of View – contiguous mapping across the Stream (~ 20 degrees)
Wavelength Coverage – Two NIR filters for color-magnitude diagram analysis

Marla Geha (Yale University), marla.geha@yale.edu

# Near-field Cosmology: Finding the Faintest Milky Way Satellites

Background

Since 2005, fourteen dwarf galaxy satellites have been discovered around the Milky Way, more than doubling the satellite population. These 'ultra-faint' galaxies are the least luminous and most dark matter-dominated galaxies in the known Universe. However, the census of Milky Way satellites is far from complete. Accurately estimating the total number of luminous satellites is key to understanding the 'Missing Satellite Problem', and constraining galaxy formation models on the smallest scales.

All of the ultra-faint galaxies have so far been discovered in the SDSS as slightly statistical over-densities of resolve stars. Due to the SDSS magnitude limits, less than 1% of the total Milky Way volume has been searched for these faint galaxies. Various assumptions for incompleteness suggest that the Milky Way may contain anywhere from 60 to 600 luminous satellites out to 300 kpc (Tollerud et al. 2008). Given even optimistic predictions, we expect one Milky Way satellite per 100 square degrees.

Detecting ultra-faint galaxies with WFIRST

Detecting ultra-faint galaxies throughout the full Milky Way volume requires imaging down to the main sequence turnoff at 300 kpc (Walsh et al. 2008), corresponding to roughly r = 27 mag. High spatial resolution is critical, as background galaxies which are otherwise unresolved from the ground, vastly overwhelm the number of stars at these magnitudes. Star-galaxy separation will severely limit the number of distant Milky Way satellites found in ground-based surveys. A deep wide-field WFIRST survey would allow an unprecedented search for the faintest dwarfs to the outer limit of the Milky Way halo.

Characterizing ultra-faint galaxies with WFIRST

In addition to finding ultra-faint galaxies, WFIRST is well positioned to characterize these galaxies. High-precision photometry of the stellar main-sequence turnoff will yield accurate ages for these systems which appear to have exclusively old stars (e.g., Brown et al 2012). Furthermore 80 microarc/year proper motions will allow *internal* measurements of motion in these systems, allowing estimates of dark matter mass and density (i.e., <2 km/s for 50 stars @ 100 kpc, in 3 years).

Depth – Deep point sources imaging to the main sequence turnoff at 300 kpc ($m_r \sim 27$)

FOV – Wide field imaging (1000+ sq deg) to cover large volume of Milky Way halo

Cadence – Repeat observations for proper motions

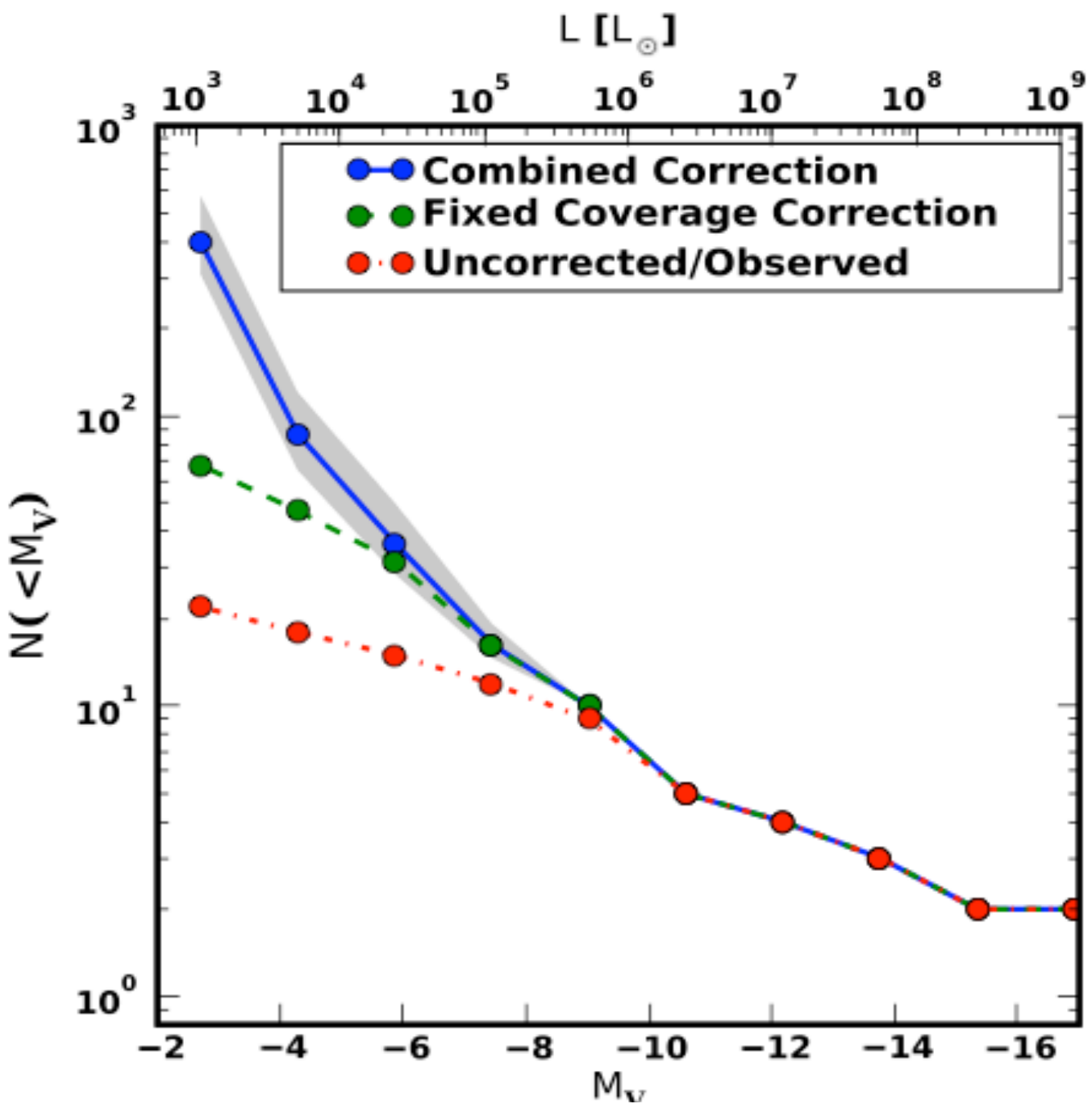

Caption: The luminosity function of known Milky Way satellites (red) as a function of luminosity. Green and blue curves are the predicted satellite numbers based on different assumptions for incompleteness (taken from Tollerud et al. 2008).

Alis J. Deason (UCSC) alis@ucolick.org; Kathryn V. Johnston (Columbia);
David N. Spergel (Princeton University); Jay Anderson (STScI)

# The Mass of the Milky Way

## Background

The mass of our Galaxy is a fundamental --- yet poorly constrained --- astrophysical quantity. Several attempts have been made to measure the total mass of the Milky Way using kinematic tracers, the orbits of the Magellanic Clouds, the local escape speed and the timing argument. The results of this extensive list of work are distressingly inconclusive with total masses in the range 0.5-3 x $10^{12}$ Msol. This somewhat confused picture is partly caused by the difficulties of the task. Full kinematic analyses of tracer populations are hampered by small sample sizes and lack of complete information on the phase-space coordinates. However, the dearth of distant halo tracers is now beginning to be rectified: Deason et al. (2012) have compiled a sample of stellar halo stars, with measured line-of-sight velocities, out to r ~ 150 kpc. This project is ongoing, and it is anticipated that the number of distant halo stars tracers will substantially increase in the next few years. However, regardless of the number of halo tracers, the well-known *mass-anisotropy* degeneracy still plagues any mass estimate. At large distances in the halo, our measured line-of-sight velocity closely approximates the radial velocity component, so we have very little handle on the tangential motion of the halo stars.

## WFIRST

The WFIRST high latitude survey (2500 $deg^2$) will achieve an absolute proper motion accuracy of 80 mu as/yr for a carbon giant at r ~ 100 kpc (J ~ 16). This corresponds to an uncertainty of ~ 40 km/s in tangential velocity, which is below the expected velocity dispersion at these distances. This will allow us, for the first time, to break the mass-anisotropy degeneracy in the distant halo. Furthermore, with full 3D velocity information we will be able to confirm or rule out any associations between random halo tracers, which is a key assumption in any dynamical analysis.

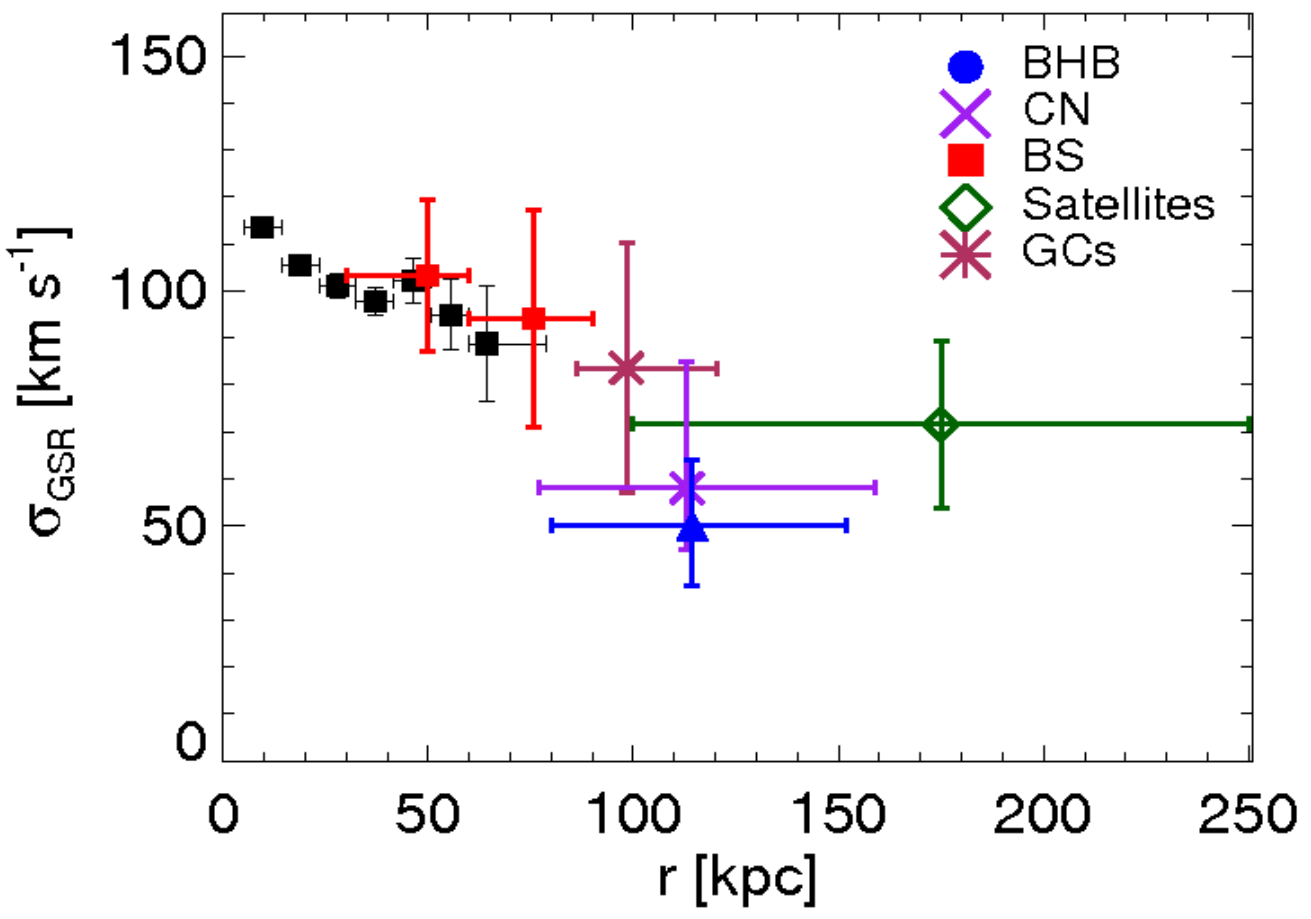

## Key Requirements

High latitude survey (2500 $deg^2$) needed for good statistics of distant tracers. The anticipated sample size (approx. several tens) will be comparable to current samples of distant halo tracers (r > 50 kpc), with radial velocity measurements.

Caption: Radial velocity dispersion of halo stars (adapted from Deason et al. 2012). At large distances the dispersion drops remarkably: the implications for the MW mass depend strongly on the (as yet unknown) velocity anisotropy.

Louis E. Strigari (Stanford University), strigari@stanford.edu
Kathryn V. Johnston (Columbia University), David N. Spergel (Princeton University),
Jay Anderson (STScI) jayander@stsci.edu

# Distinguishing between cold and warm dark matter with WFIRST

Background

The theory of cold dark matter predicts that the central density profiles of dark matter halos are cusped, rising like 1/r towards the centers of galaxies. Warm dark matter theories, on the other hand, predict that halos are much less dense in their centers. In order to distinguish between these theories, and place strong constraints on the nature of particle dark matter, it is essential to measure the density profiles of the most dark matter-dominated galaxies in the universe.

The dwarf spheroidal (dSph) satellite galaxies of the Milky Way are particularly interesting systems in which it is possible to probe the nature of dark matter. They are near enough that it is possible to measure the velocities of individual stars, and from these measurements extract the dark matter mass profiles. In recent years, through high-resolution multi-object spectroscopy the line-of-sight velocities of hundreds of stars have been measured, providing important measurements of their integrated dark matter mass profiles within their half-light radius. Though there are some hints that their dark matter distributions may be less cuspy than is predicted by cold dark matter theory, precise measurements of their central dark matter mass profiles have remained elusive. Because line-of-sight velocities provide only one component of the motion of a star, there is a well-known degeneracy between the central slope of the dark matter mass profile and the velocity anisotropy of the stars. The only way to break this degeneracy is to measure stellar proper motions in dSphs.

WFIRST

Because of its pointed nature and large field-of-view, WFIRST will play a key role in studying the dark matter distributions of dSphs. For example, Draco and Sculptor are at a Galactocentric distance of 80 kpc. At this distance, a tangential velocity of 10 km/s corresponds to a proper motion of 27 micro-arcseconds per year. The figure illustrates the projected uncertainty on the log-slope of the dark matter density profile of a dSph at a distance of 80 kpc, as a function of the number of stars with measured proper motions for different assumed errors for the transverse velocities. These simulations assume that the dSph has 1000 measured line-of-sight velocities, in addition to the indicated number of proper motions. The addition of proper motions clearly improves the constraint on the central dark matter density profile, and will ultimately provide a valuable method to rigorously distinguish between cold and warm dark matter candidates.

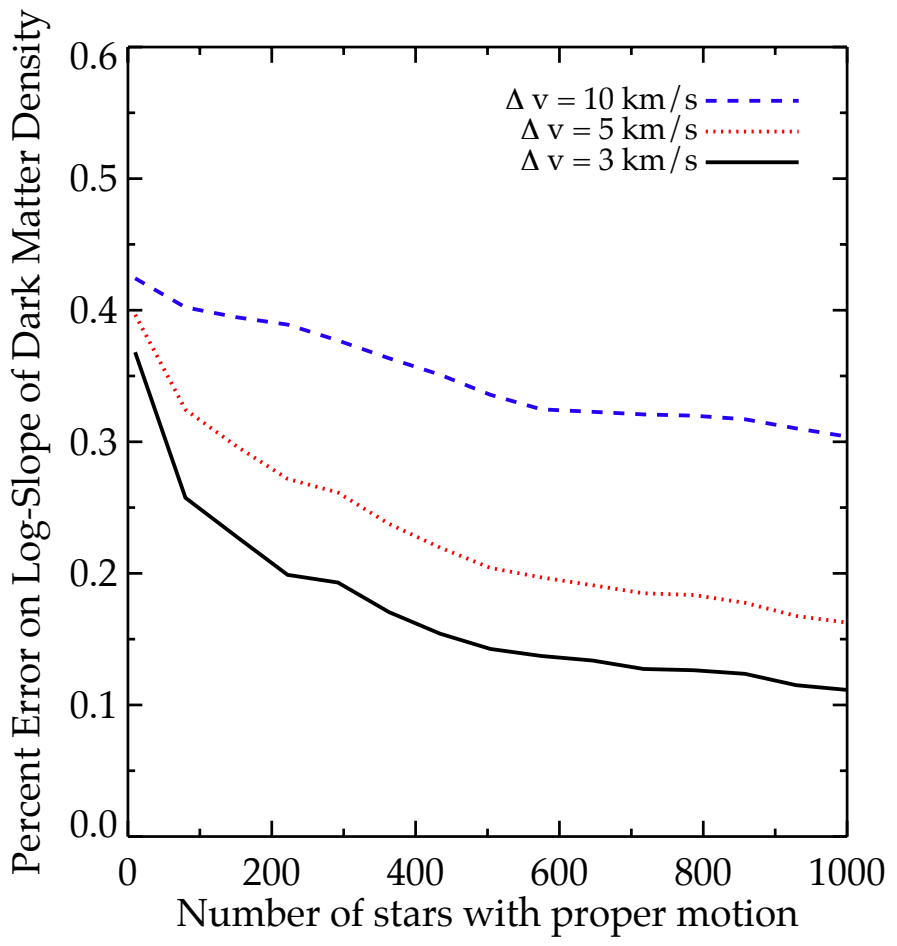

Key Requirements

Astrometry figures suggest that 10 6-minute integrations over 5 years would achieve 90 micro-arcsecond/year accuracy for positions of the brightest stars in these objects. This survey would require additional visits to increase the accuracy to levels where the slope could be measured.

Kathryn V. Johnston (Columbia University) kvj@astro.columbia.edu
David N. Spergel (Princeton University); Jay Anderson (STScI)

# Finding (or Losing) those Missing Satellites

Background

Within the standard cosmological context, the Milky Way's dark matter halo is thought to be embedded with thousands of dark matter subhalos with masses greater than $10^6 M_\odot$. If such a population exists, its influence is expected to be detectable in the form of gaps in tidal globular cluster tidal streams such as those associated with GD-1 (d=10kpc) and Pal 5 (d=20kpc). Indeed, current photometric studies suggest there may be gaps in Pal 5's streams. An alternative explanation for these features in streams is that they are a natural consequence of the clumping of debris along the orbit. These scenarios may be distinguished – and hence the presence of "missing satellites" either proved or ruled out – by mapping the debris in phase-space. Stars on either side of gaps caused by encounters with subhalos should exhibit sharp discontinuities in energy and angular momenta, and can potentially deviate beyond the spread expected for tidal stripping alone. In contrast, if the gaps are simply a natural clumping, the phase space structure should be more continuous and stay within the range expected for tidal stripping.

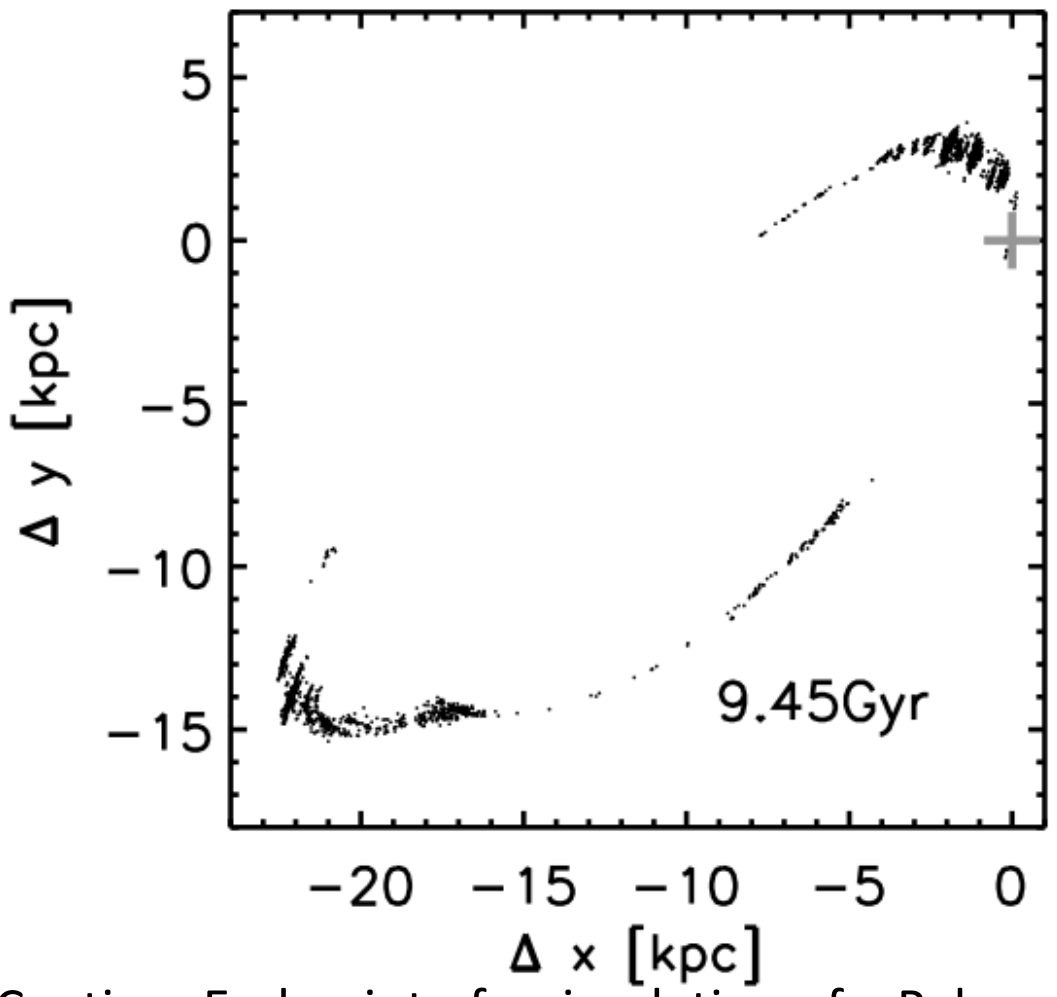

Caption: End point of a simulation of a Pal 5-like stream bombarded by the expected subhalo population. Cross indicates where the main body (not shown) would be.

WFIRST

WFIRST can contribute uniquely to this goal by virtue of its pointed nature and large field of view, which will allow proper motions of stars in selected fields in streams to be probed. With 10 6-minute integrations spread over 5 years, proper motions for dwarfs in globular cluster streams (such as those associated with GD1 and Pal 5) could be assessed with few km/s accuracies. This is only a little larger than the expected dispersion of stream stars, and comparable to the velocity offsets expected from the satellites original orbit. Hence any differences in orbital properties for stars around a gap beyond this range could be assessed by averaging properties on either side.

Kathryn V. Johnston (Columbia University) kvj@astro.columbia.edu
David N. Spergel (Princeton University); Jay Anderson (STScI)

# Mapping the Potential of the Milky Way with Tidal Debris

<u>Background</u>

The disruption of satellite stellar systems around the Milky Way has left tidal debris in the form of streams of stars, spanning tens of degrees (in the case of globular clusters) or even entirely encircling the Galaxy (in the case of the Sagittarius dwarf galaxy). By virtue of their origin, we know more about samples from these structures than we do about random samples – we know that each sample consists of stars that were once apart of the same object. We can exploit this knowledge to use tidal debris as much more sensitive probes of the Galactic potential than possible with a purely random sample. Moreover, while debris structures have been found at small (~5kpc) and large (~100kpc) distances from the Sun, they are more apparent and longer-lived in the outer halo of the Galaxy – precisely the region where we know least about the potential. Using stellar streams, the Milky Way is the one galaxy in the Universe where we can hope to map the 3-D shape, orientation and mass of a dark matter halo at all radii within the virial radius.

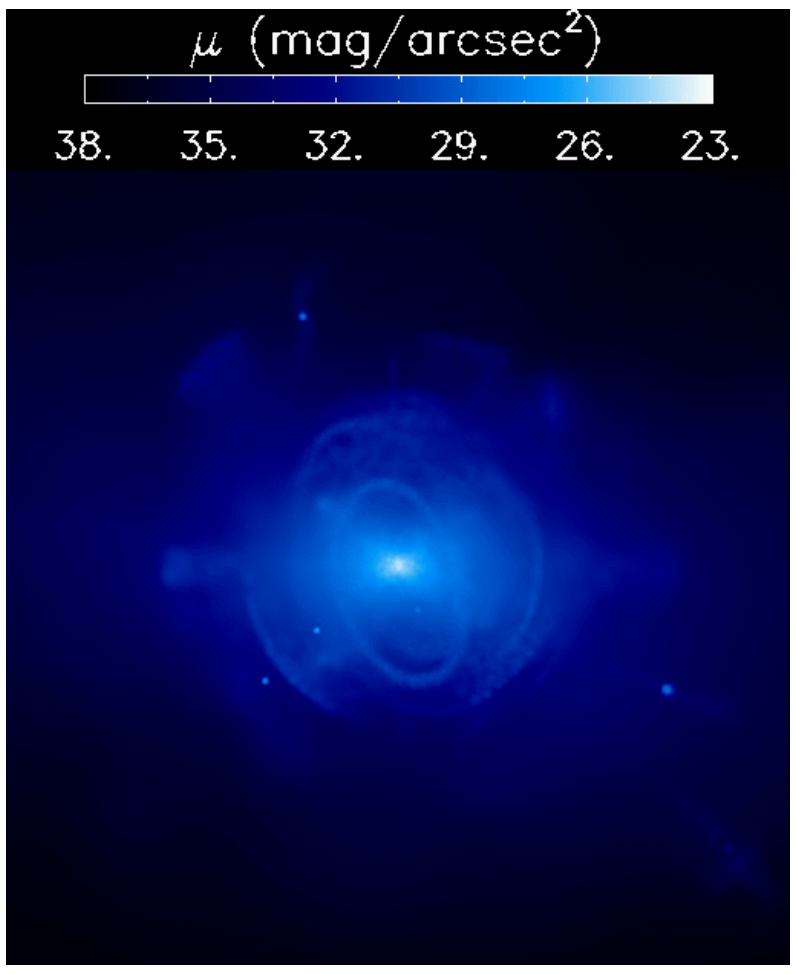

Caption: External view of simulated debris around a galaxy. Box is 300kpc on each side.

<u>WFIRST</u>

WFIRST can contribute uniquely to this goal by virtue of its pointed nature and large field of view, which will allow proper motions of stars in selected fields in streams to be probed even at large Galactocentric radii. With 10 6-minute integrations spread over 5 years, proper motions for dwarfs in the globular cluster streams (such as those associated with GD1 and Pal 5) in the inner halo (< 20kpc) and for giants in satellite debris (e.g. Sagittarius and the Orphan Stream) in the outer halo (> 50 kpc) could be estimated with few km/s and tens of km/s accuracies respectively. These numbers are only a little larger than the expected dispersion of stream stars, and comparable to the velocity offsets expected from the satellites original orbit. Hence, averaging over the debris stars in several fields along each stream would produce an accurate map of stream centroids in phase-space and at different orbital phases – something beyond the reach of current surveys and a key step forward for mapping the Galactic dark matter halo.

Roeland van der Marel and Jason Kalirai (STScI), marel@stsci.edu; jkalirai@stsci.edu

## Dissecting Nearby Galaxies

Background

Recent wide-field imaging surveys of the Milky Way and M31 (Juric et al. 2008; McConnachie et al. 2009) have enabled a new landscape to test cosmological models of galaxy formation on small scales (Bullock & Johnson 2005). Direct imaging of stellar halos can provide detailed insights on the surface brightness profile, chemical abundance gradients, stellar ages, level of substructure, and quantity of dwarf satellites and star clusters. This input represents a key constraint to unravel the formation and assembly history of galaxies within the hierarchical paradigm (Font et al. 2011), but is limited to the detailed study of just two fully resolved spiral galaxy halos. Current surveys of galaxies outside the Local Group (Spitzer/SINGS, HST/ANGST) have primarily involved either deep pencil-beam probes of narrow fields of view, or wide-field coverage at shallow depth. Similarly, optical ground-based surveys such as SDSS and Pan-Starrs lack the depth to study the tracer population, red giant branch stars, outside the Local Group.

WFIRST

WFIRST will enable a uniform 1.) wide-field, 2.) deep, and 3.) high-resolution study of the full extent of over a hundred nearby galaxies in the Local Volume. Accurate photometric characterization of the top three magnitudes of the red giant branch in each galaxy can map the halo structure, surface brightness profile, substructure content, and metallicity gradient. Variations in these properties, and their connection with the host environment and galaxy luminosity will provide tight constraints to cold dark matter models of galaxy formation.

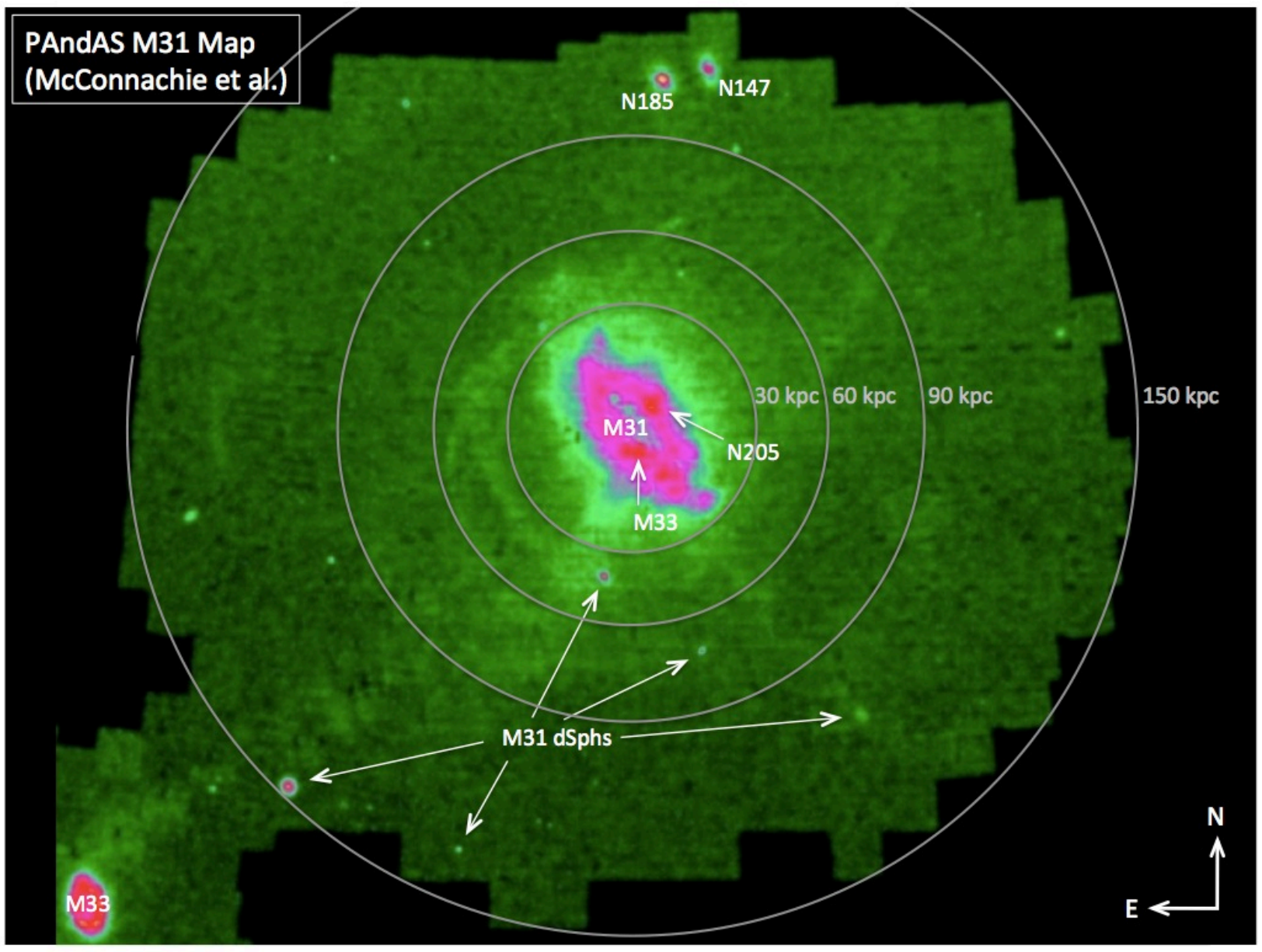

Caption: A (very) wide-field map of M31 from the PAndAS survey (McConnachie et al. 2009) reveals the clearest picture of substructure in a spiral galaxy's halo.

Key Requirements

Depth – Several magnitudes of the red giant branch in ~100 galaxies (i.e., out to d = 5 Mpc)
Field of View – Full halo extent out to 150 kpc (tiling in nearby galaxies)
Cadence – One image per galaxy
Wavelength Coverage – Two NIR filters for color-magnitude diagram analysis

P. Guhathakurta (University of California, Santa Cruz), raja@ucolick.org
R. Beaton (University of Virginia), rlb9n@virginia.edu

# Galaxy Evolution from Resolved Stellar Populations: Halo Age Distributions of the Local Volume

Background: Large galaxies like the Milky Way are formed hierarchically through dozens of satellite accretion events that not only build up the stellar halo, but also contribute to the evolution of the galaxy as a whole. Statistical studies comparing satellite and field dwarfs indicate that star formation is quenched early in the accretion process (e.g., Geha et al. 2012). Thus, the youngest stars in a satellite can be used to date the initial infall of that satellite into its parent halo. Asymptotic giant branch (AGB) stars are prime tracers of these accreted populations as they are present in stellar populations over a wide range of ages (4 Myr to 10 Gyr). As shown in Figure 1 for the LMC, the NIR magnitude and color of an AGB star is strongly dependent on its age. The NIR CMD is therefore an ideal tool for quantifying the stellar age distribution of nearby galaxy halos.

WFIRST: Observational programs of this nature are difficult to carry out from the ground and to date this technique has only been applied to the SMC, LMC and, as a pencil-beam survey in M31. With WFIRST, however, it is feasible to explore the halo age distribution in a representative sample of nearby galaxies to complement Hubble and Spitzer Legacy programs (LVLS, ANGST), that have observed recent and old star formation tracers in all galaxies within 3.5 Mpc and all S/Irr within 11 Mpc. Further, WFIRST will allow an expansion beyond existing HST NIR imaging to obtain full spatial coverage of the M31 disk, bulge and halo. Key questions addressed by these observations will be: (1) what is a 'typical' galaxy's accretion history? (2) over what physical scales does satellite accretion vs *in-situ* star formation dominate? and (3) is the age distribution smooth or discontinuous, and what does this imply about the typical progenitor over time?

Key Requirements
Depth – Reliable well below the TRGB for galaxies over 1-11 Mpc (Ks~22-30)
Spatial Resolution – Morphological distinction between stars and galaxies
Field of View – Wide-field for efficient mapping of nearby galaxy halos
Wavelength Coverage – J,H,Ks to separate AGB stars as in Figure 1

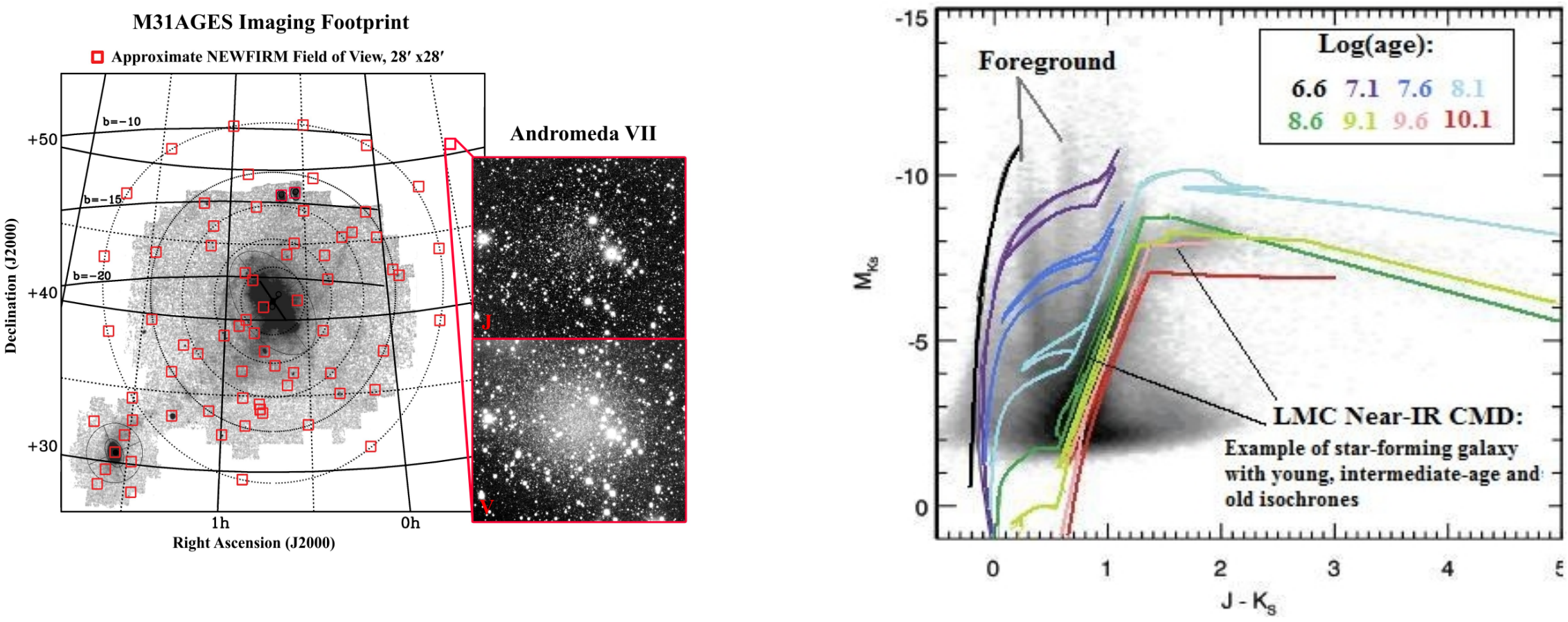

Figure 1. (Left) An example CMD for the LMC emphasizing the clean separation of AGB stars by age (Martha Boyer, private communication). (Right) The M31 Asymptotic Giant Exploration Survey footprint is the first large scale age mapping project in a large galaxy. With WFIRST it is feasible to complete filled area surveys of this nature for a representative sample of galaxies.

Seppo Laine (SSC, Caltech), seppo@ipac.caltech.edu
David Martinez-Delgado (MPIA), Carl Grillmair (SSC, Caltech), Steven R. Majewski (UVa)

## Substructure around Galaxies within 50 Mpc

Background

A lot of the attention on galaxy evolution has been focused on interactions between nearby roughly equal mass galaxies (e.g., Toomre & Toomre 1972). However, minor mergers and dwarf galaxy accretion events are far more common, and are critical in the evolution of larger galaxies by building their halos, bulges, bar and spiral and even globular cluster systems. These events may leave relics in the circumgalactic environment in the form of discrete tidal stellar streams and intracluster light. The investigation of such streams and their significance in hierarchical galaxy formation has just begun (e.g., Martinez–Delgado et al. 2010). The prerequisites include deep red visual and near-IR imaging of the peak of the stellar energy distributions of galaxies over large areas.

WFIRST

WFIRST can probe the circumgalactic environments with sufficient depth and spatial resolution around galaxies up d = 50 Mpc, and with its large field of view it can perform these observations very efficiently. Combined with ground-based surveys in the visual wavelengths, the ages, metallicities and masses of the stellar populations of disrupted companions can be studied for a large sample of galaxies, providing much tighter constraints on the minor merging history in the CDM models of galaxy formation.

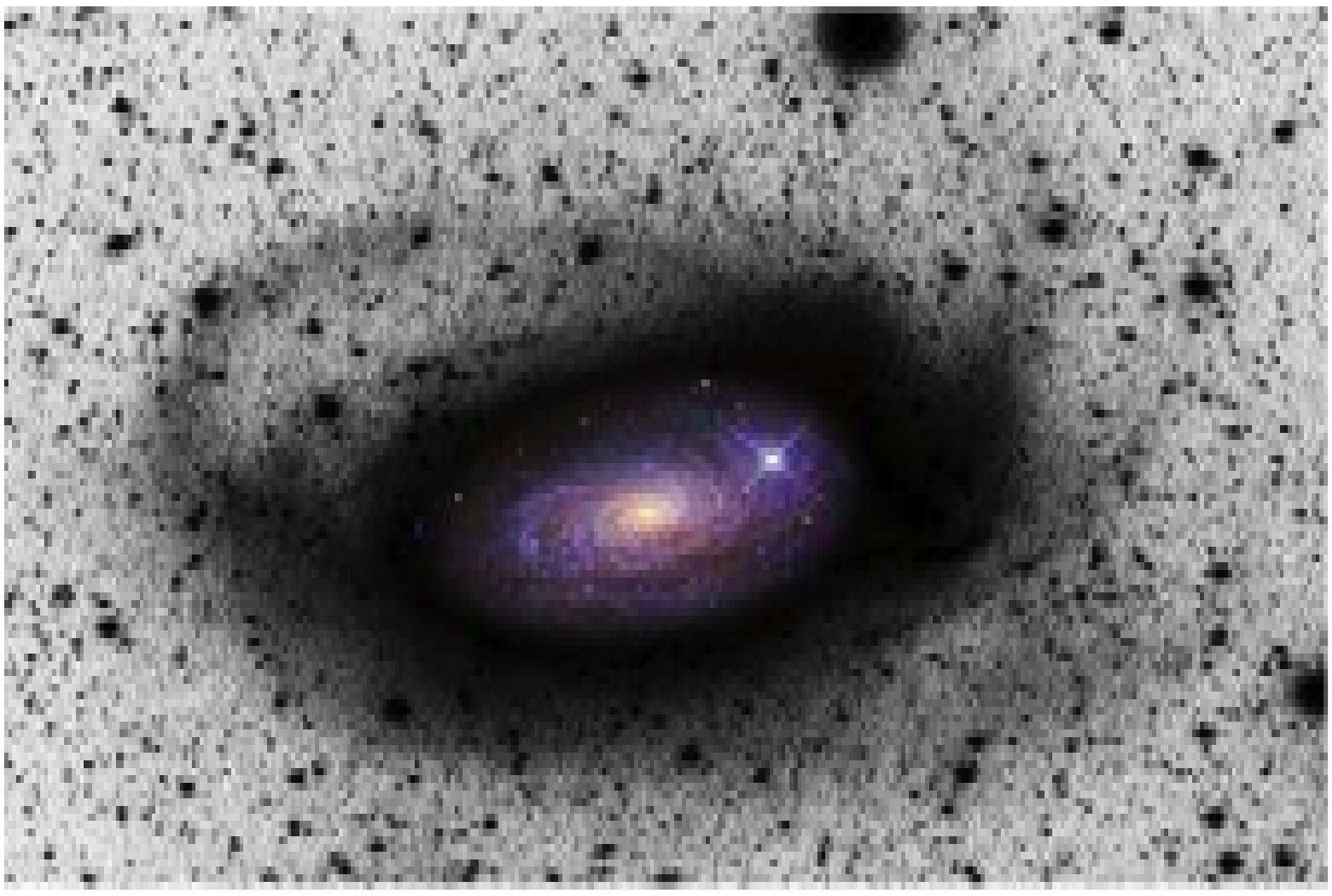

Key Requirements

Field of View – Mapping the halos out to 100 kpc (35′–7′ at 10–50 Mpc)
Depth – 27 mag arcsec$^{-2}$ (AB mag) at 1.2 μm
Resolution – Resolving the width of features down to 100 pc (2″ – 0.4″ at 10-50 Mpc)
Cadence – Only one epoch of imaging is required (assuming sufficient depth achieved)
Wavelength Coverage – ~0.7 μm to (>1.5 μm) is optimal for C–M diagnostics of stellar populations.

Julianne Dalcanton (University of Washington), jd@astro.washington.edu

## Resolved Stellar Populations in Nearby Galaxies

Background

From the UV through the near-IR, galaxies' SEDs are dominated by stars. These stars encode a rich record of key astrophysical processes. From the distribution of stars in a CMD, one can map out the history of star formation, which in turn can be used to anchor models of the physics that controlsgalaxy evolution (i.e., what physical conditions lead to star formation? how does energy from evolving stars affect the ISM? how do evolving stars affect the global SED?), and models for interpreting the observations made of galaxies across cosmic time (i.e., what is the IMF? what timescales and star formation rates are associated with traditional star formation rate indicators? etc). CMDs also contain rich, redundant information about the evolution of the stars themselves, allowing one to constrain the physics of some of the most rapidly evolving, rare stars. Many of these classes of stars, such as AGB and red core Helium burning stars, are incredibly luminous, and can have a significant impact on the overall SED in the near-IR. Unfortunately, the internal evolution of these stars is sufficiently complex that it cannot be derived theoretically from first principles, and instead must be extensively calibrated for a wide range of metallicities and stellar masses. The net result is that stars in nearby galaxies lay the foundation for building an accurate interpretation of extragalactic observations.

WFIRST:

The main factor limiting stellar population studies is angular resolution. With ground-based telescopes, light from individual stars is blended together, prohibiting the study of stars beyond the Local Group. While adaptive optics can alleviate this problem, stellar population studies require a level of photometric accuracy and stability that cannot yet be achieved routinely for long integrations over large fields of view. In contrast, the superb resolution and stability available with a 2.4m space-based telescope allows individual red giant stars to be resolved at distances out to ~5 Mpc in the main body of galaxies, or out to even larger distances in low surface brightness regions (halos, intercluster light, dwarf galaxies, etc). High luminosity stars, which more easily rise above the crowding limit, can also be detected out to greater than ~10 Mpc, allowing the detection of massive main sequence stars, blue and red supergiants, and AGB stars. WFIRST's wavelength coverage is optimized for the cooler of these stars, whose bolometric luminosity peaks longward of 7000 Angstroms; however, many hotter stars can be sufficiently luminous to still be detectable at WFIRST's wavelengths. The near-IR coverage of WFIRST also minimizes the impact of dust on interpretations of CMDs. Existing studies in the Magellanic Clouds (Nikolaev & Weinberg 2000, Boyer et al 2011) have demonstrated that there would be significant added value if WFIRST detectors were sensitive out beyond 1.7 microns, even with the high background expected for a non-cryogenic mission.

Key Requirements

Depth – Less necessary than resolution, given that most stars are in regions that are limited by crowding/confusion rather than photon counting. In uncrowded regions, depth allows detection of fainter, more numerous stars, giving greater sensitivity to stellar halo substructure

Morphology – Resolution absolutely critical to maximizing depth and separating stars/galaxies

Grism – Spectral characterization of AGB subtypes and separation from core He-burning stars

Field of View – Wide-area surveys required to identify rare populations

Wavelength Coverage – minimum of 2 filters. Including an optical filter as far to the blue as possible maximizes sensitivity to stellar temperature, separating features on the CMD. Including a filter longward of H-band increases ability to separate AGB subtypes (O-rich, C-rich, etc).

Chris Mihos (mihos@case.edu) and Paul Harding (paul.harding@case.edu)

# Deep Surface Photometry of Galaxies and Galaxy Clusters

Background

The low surface brightness outskirts of galaxies hold a wealth of information about processes driving their evolution. Interactions and accretion events in galaxies leave behind faint, long-lived tidal tails and stellar streams which can be used to trace their interaction history (e.g., Martinez-Delgado et al 2010). In galaxy clusters, tidally stripped material mixes to form the diffuse intracluster light (ICL; e.g., Mihos et al 2005), whose structure and luminosity provides constraints on the dynamical history of the cluster (Rudick et al 2009). The photometric properties of the extended disks of spiral galaxies probe star formation in low-density environments (e.g., Bigiel et al 2010), as well as dynamical models for stellar migration in galaxy disks (Sellwood & Binney 2002; Roskar et al 2008). However, accessing this information is quite difficult, as optical surface photometry must reach down to $\mu_V > 27$, or < 1% of the ground-based night sky. Working to an equivalent depth in the near IR is impossible from the ground, since the IR sky is much brighter and significantly more variable than in the optical. Furthermore, this level of accuracy must be achieved over a wide field of view (>~0.5 degree to study nearby galaxies and clusters.

WFIRST

WFIRST can take advantage of the low IR background levels from space to deliver deep, wide-field surface photometry in the near IR. Compared to optical surface photometry, working in the IR has the advantage of maximizing the signal from the old stellar populations that comprise the ICL, providing a more direct tracer of stellar mass at low surface brightness, and minimizing the effects of scattered light from Milky Way galactic cirrus and internal extinction from the target galaxies. Proper baffling and PSF stability will be critical to reduce scattered light and allow for the subtraction of extended stellar wings in the imaging; high spatial resolution will help resolve and reject faint background sources from contaminating the deep surface photometry.

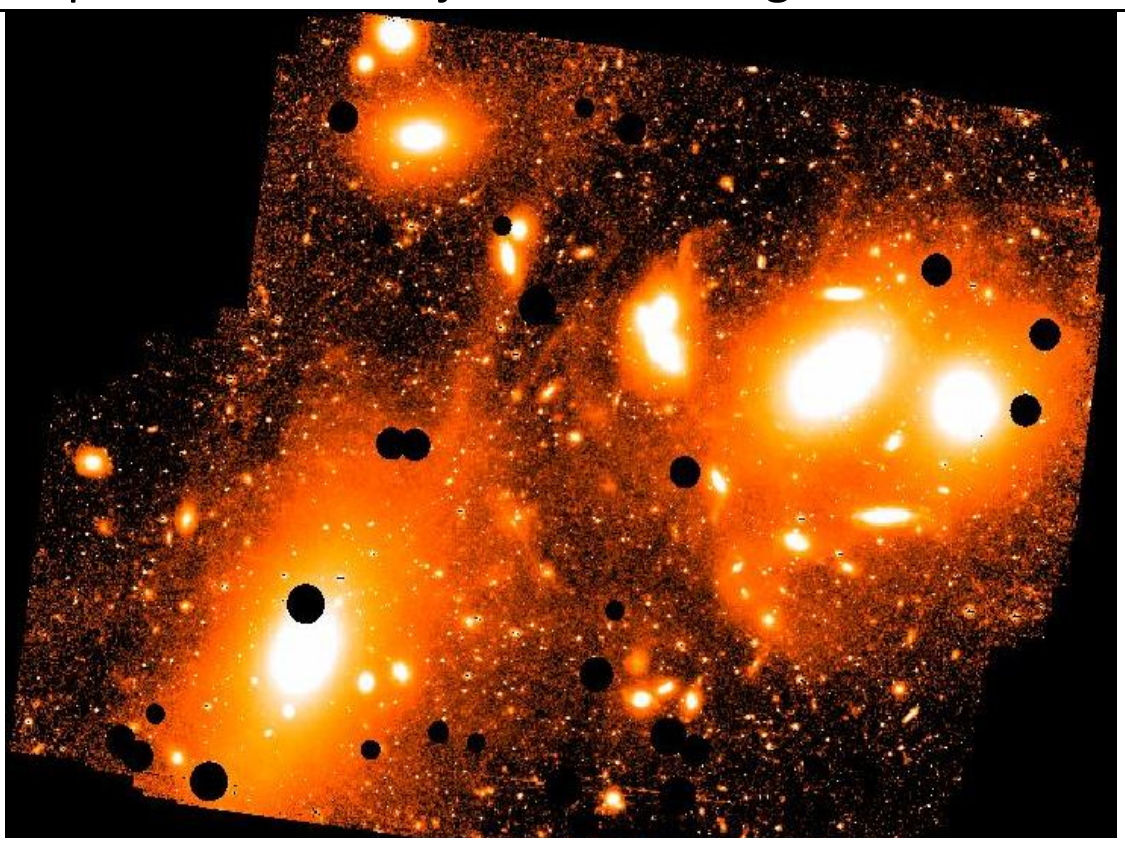

Deep imaging of diffuse ICL in the Virgo Cluster (Mihos et al 200 showing tidal streams and extended galaxy halos. The image cov deg² to a limiting depth of $\mu_V = 28.5$.

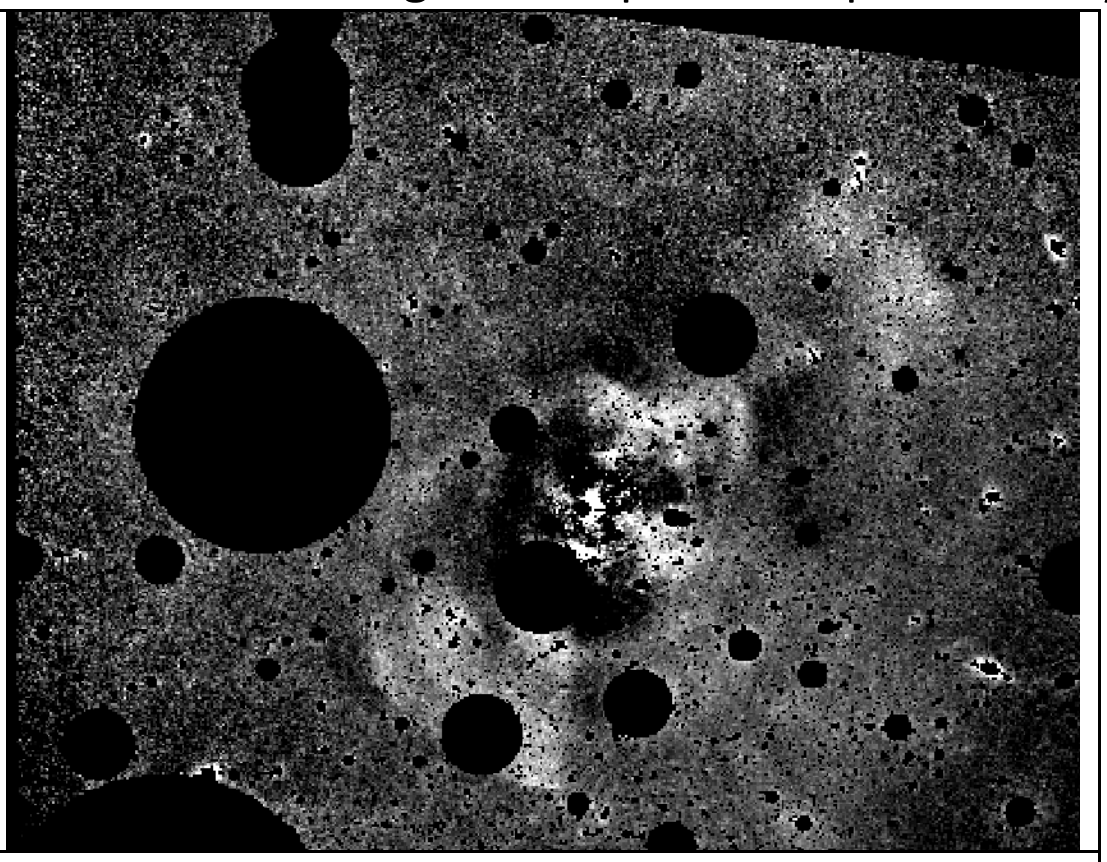

The Virgo elliptical galaxy M49, after subtraction of smooth lig profile, showing a complex system of diffuse tidal shells from a accretion event (Janowiecki et al 2010).

Key Requirements

*Depth* – Limiting J-band surface brightness of ~ 27 mag/arcsec$^2$ (~ 0.001 MJy/sr)
*FOV* – Wide field necessary to cover nearby galaxies and galaxy clusters
*High spatial resolution* – To reduce photometric contamination from faint background objects.
*Baffling and PSF stability* – To limit scattered light and allow accurate subtraction of extended stellar wings
*Wavelength* – Infrared bands needed to sample peak of old population SED, better trace stellar mass distribution, and minimize contamination due to scattering and absorption from Galactic and target galaxy dust

Christopher J. Conselice (Nottingham), conselice@nottingham.ac.uk

## Galaxy Structure and Morphology

Introduction

There are several major issues within galaxy formation and evolution where WFIRST/NRO telescope will make a big impact. One of these critical aspects of a WFIRST/NRO mission will be to study galaxy structure - i.e., sizes and morphologies of galaxies on a larger area than is possible with Hubble Space Telescope, or from the ground using adaptive optics. Galaxy structure is critical for deciphering the processes which drive galaxy assembly and allow us to move beyond simply counting properties of galaxies to study directly their assembly processes through e.g., mergers, gas accretion, and star formation. Yet this has proved difficult to do for galaxies at z > 1. Properties such as the asymmetry of a galaxy's light, its concentration, and the clumpy nature of this light all reveal important clues to the galaxy formation process.

WFIRST/NRO

The Hubble Space Telescope has shown how powerful this resolved structural approach is for understanding galaxies. Only very small fields have been observed with Hubble, and most of those within the observed optical. To study the formation of galaxies will require high resolution imaging in the near-infrared, which WFIRST/NRO will provide, especially if a large 2.4m mirror is utilized. This will bring the pixel size down to 0.13-0.11 arcsec, allowing us to perform WFC3 type surveys over at least a few degrees, which will be much larger than any near infrared survey of this type for distant galaxies yet performed. This will allow us to directly measure in an empirical way how galaxies, and the stars within them, assembled over most of cosmic history to within a Billion years of the big bang.

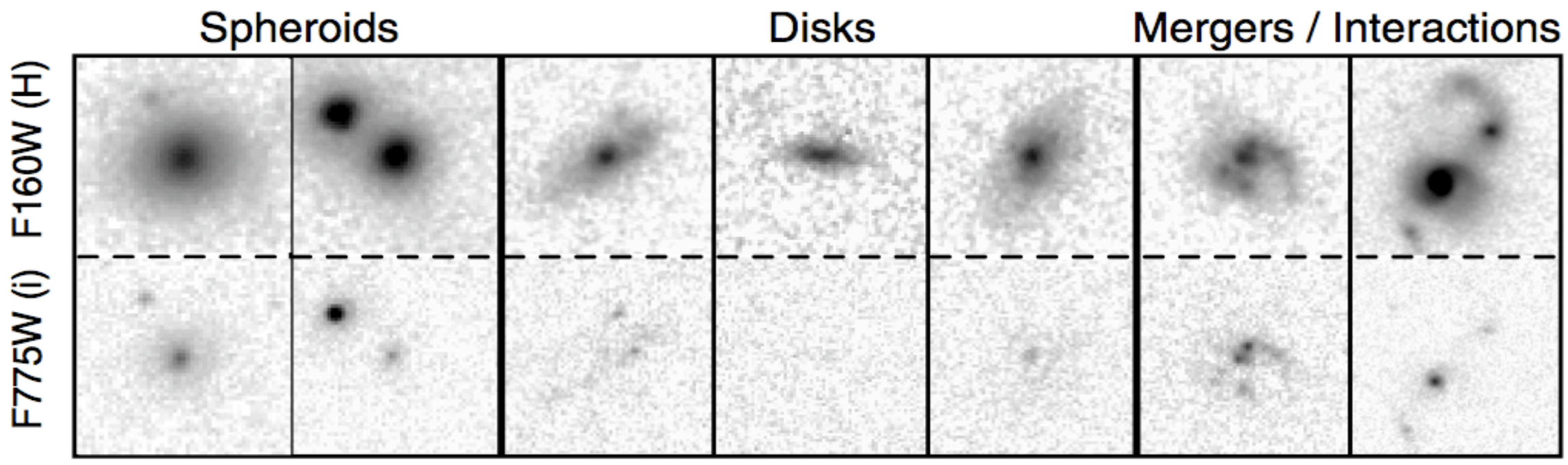

Caption: How infrared light (F160W) shows features of distant galaxies not seen in optical light (F775W) (from Kocevski et al. 2011)

Key Requirements

Depth – Deep enough to examine galaxy structure down to ⅕ L* up to z = 4
FoV – Large survey areas to obtain statistically representative galaxy populations
Wavelength – Need > 1.6 microns to examine the rest-frame optical structures of galaxies

Daniel Stern (JPL), daniel.k.stern@jpl.nasa.gov

## Strong Lensing

Background

Strong gravitational lensing is a remarkable physical phenomenon. Massive objects distort space-time to the extent that light sources lying directly along the line of sight behind them can appear multiply-imaged (see Schneider 2006 for a review). When this rare alignment occurs, we are given: (1) the opportunity to accurately infer the mass and mass distribution of the lensing object; and (2) a magnified view of the lensed source, often probing luminosities and size scales that would not be accessible with current technology. While the approximately 200 galaxy-scale strong lenses we know of currently generally come from low- resolution, ground-based surveys, the bulk of the scientific potential of these rare systems requires (and has relied upon) high-resolution follow-up studies with the Hubble Space Telescope (Fig. 1; e.g., Browne et al. 2003, Bolton et al. 2006). The same has been true for the similar number of cluster-scale strong lens systems (e.g., Smith et al. 2005).

Simply making a precise measurement of lens statistics provides cosmographic information from the total lens counts, the so-called "lens redshift test" (e.g., Capelo & Natarajan 2007). If the lens mass distribution is well-constrained, as in the case of "compound lenses", where multiple sources line up behind the lensing galaxy or cluster, the lens geometry can be used to measure ratios of distances — compound lenses are standard rods for probing the Universe's expansion kinematics with high precision (Golse et al. 2002, Gavazzi et al. 2008). However galaxy-scale compound lenses are rare, typically just 1% of galaxy-scale lenses, and only by imaging a substantial fraction of the sky can we find such golden lenses.

WFIRST

Based on Hubble surveys of a few square degrees, we expect strong lenses to have an abundance of about 10 per sq. deg. (Faure et al. 2008), suggesting that the Hubble-era sample of lenses observed at high resolution will be expanded by several orders of magnitude by the WFIRST weak-lensing survey (e.g., Marshall et al. 2005), the majority of these will be galaxy-scale lenses. The number of observable galaxy-scale lenses is a very strong function of angular resolution: the factor of six degradation in resolution from a diffraction limited 1.5-meter space-based telescope relative to typical ground-based conditions incurs a two order of magnitude decrease in the number of lenses identifiable from the ground. WFIRST, with its high-resolution and wide-field capabilities, will revolutionize the field of strong gravitational lensing, enabling fundamental, new cosmological and astrophysical probes.

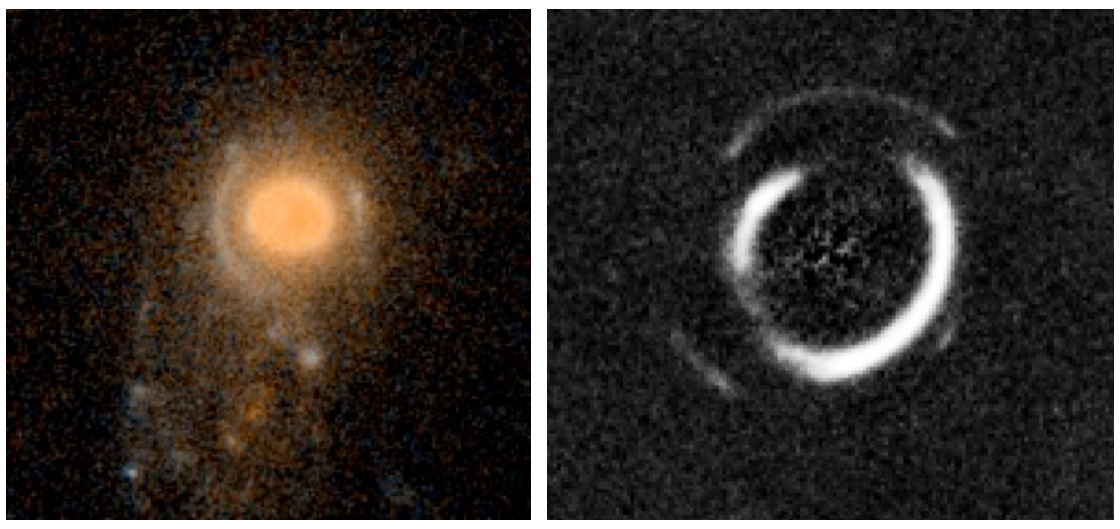

Caption: Strong gravitational lenses imaged with Hubble. WFIRST will increase the number of strong lenses known by a factor of ~100, thus also identifying rare configurations — such as lensed SNe, lenses with higher order catastrophes in their caustics, and compound (e.g., double lens-plane) lenses — which have even richer scientific potential. [From Moustakas et al. 2007 and Gavazzi et al. 2008.]

Key Requirements

Depth – To provide high S/N detections
Morphology – To precisely measure positions
Field of View – Wide-area to detect rare events
Wavelength Coverage – Single band sufficient

P. Appleton, J. Rich , K. Alatalo, P. Ogle (Caltech), S. Cales (U. Concepcion) & L. Kewley (ANU)
apple@ipac.caltech.edu

## Searching for Extreme Shock-dominated Galaxy Systems from 1 < z < 2

Background: The importance of turbulence and shocks at heating gas in galaxies is only just beginning to emerge through both visible IFU imaging and mid-IR molecular hydrogen observations of collisional galaxies, and galaxies with powerful AGN winds. Recent models of fast shocks predict optical emission-line ratios, which overlap the Low Ionization Nuclear Emission Line (LINER) region of the traditional BPT diagnostic diagram (Fig1a). Amazingly (Fig1b & c) the distinct nature of shocked gas can be diagnosed even when H$\alpha$ and [NII] are blended—as in the case of a low-res GRISM spectrometer. Although many LINERs are likely to be heated by sub-Eddington accretion onto a black hole, there are clear examples of LINER emission where galaxy-wide shocks strongly dominate the optical spectrum, e. g. NGC 1266 (Alatalo et al. 2012), or the giant shocked filament in Stephan's Quintet (Xu et al. 2003; Appleton et al 2006). The shocks may suppress star formation, as in the case of some radio galaxies, where the jet may drive the heating (Ogle et al 2010; Nesvadba et al. 2011). Even old radio galaxies may still show remnants of shocked gas (Buttiglione 2013). Discovering more "pure-shock" objects at higher z, will allow us to investigate places in the universe where strong turbulence is operating to quench star formation. Finding "nascent" pre-starforming disks in a highly turbulent phase will help us understand how galaxy disks form and evolve (e. g. in cold flows), or how AGN feedback may influence galaxy hosts.

WFIRST: Large galaxy surveys such as SDSS and BOSS have already begun to turn up rare objects in the correct area of phase-space in the local (z < 0.2) universe (see comparison of Fig1a with b). WFIRST's near-IR GRISM capability will allow for a similar exploration at much higher z, where turbulent galaxy disks may be more common. Assuming a GRISM spectrograph with 160 < R < 260 in the wavelength range 1.1 to 2 microns, WFIRST will detect galaxies over a range 1 < z < 2 in the required lines. Because the [OI]/H$\alpha$ ratio is so extreme in pure shocked systems, it may not even be necessary to detect the [OIII] and H$\beta$ lines to form a sample of extreme shock-dominated systems. Based on simulations and studies using a similar GRISM on HST/WFC3 (WISP), 2000-3000 emission line galaxies/sq. degree would be detectable (> 1 x $10^{-16}$ ergs/s/cm$^2$) over the required range. With large planned deep redshift BAO surveys envisaged in the WFIRST Interim Report covering ~2000-3500 sq. degree of sky/yr, even if only 0.5% of all emission-line galaxies are shocked galaxies (probably conservative given that 7% of SDSS galaxies are LINERs), then we expect to find tens of thousands of shocked galaxies over a few years of operation. These BAO samples will allow us to investigate how these galaxies change with redshift and local environment—for example are they correlated with galaxy over-density or filaments? Significant new ancillary data sets (e. g. deep SKA-precursor radio galaxy surveys) will also be available then, allowing for smaller, deeper, targeted GO searches of specific dense proto-clusters, or quasar environs.

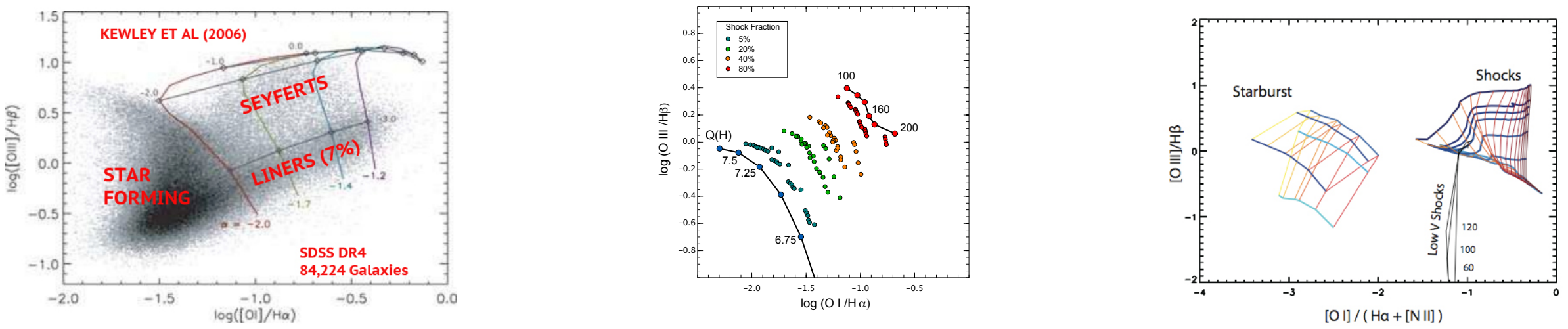

*Fig.1a Classical BPT diagram showing separation of LINERs (7%) from Seyferts and SF galaxies (Kewley et al. 2006). Fig1b. Some LINERS may be created in high-speed shocks (Rich et al. 2012). Models show range of shock velocities and mixing fractions with SF. Fig1c. Even when H$\alpha$+[NII] is blended, the shock models separate cleanly from the starburst sequence on the $x$-axis.*

Key Requirements for GRISM Survey

Depth :– 1-2 x$10^{-16}$ ergs s$^{-1}$cm$^{-2}$ detection threshold; Areal coverage:- 2000-3000 deg$^2$ (BAO type planned coverage)
Spectral Resolution (R > 150) ; Wavelength Coverage – 1.1-2 $\mu$m

Julian Merten, Jason Rhodes (JPL / Caltech), jmerten@caltech.edu

# Mapping the Distribution of Matter in Galaxy Clusters

## Background

Clusters of galaxies are important tracers of cosmic structure formation. All of their mass components including dark matter, ionized gas and stars are directly or indirectly observable. Furthermore, the complicated effects of baryons are less dominant, although not negligible (Duffy+10). This allows us to directly compare numerically simulated galaxy clusters with real observations. This comparison is usually done via parametrized 1D density profiles, like the NFW profile (Navarro+96); where the distribution of parameters in NFW fits to simulated halos (Bhattacharya+11) and observed halos (Fedeli+12) are compared. However, these comparisons do not make use of the full 2D density distribution as can be inferred from gravitational lensing, X-ray or SZ observations of clusters and which show that individual clusters are usually non-spherical and highly sub-structured. We therefore propose a full 2D, morphological characterization of galaxy clusters by means of e.g. mathematical morphology (Serra+65) or Minkowski functionals (Kratochvil +12). These techniques will be applied to simulated clusters and a large sample of clusters observed with WFIRST. A comparison of the two will give unprecedented insight into the main mechanisms of structure formation.

## WFIRST

Key to such analysis is a map of the mass distribution in galaxy clusters created with the technique of gravitational lensing (Bartelmann11). Methods which combine weak and strong gravitational lensing are particular successful (Bradac+06, Merten+10, Merten+11, Meneghetti+11) and only space-based observations deliver the depth and resolution needed for a detailed reconstruction of a galaxy cluster. In the regime of weak gravitational lensing, space-based observations result in a four or more times higher density of background galaxies, which can be used to map the distribution of matter in the full cluster field. This high density of background objects is key for a reconstruction on high spatial resolution (Massey+12). In the regime of strong gravitational lensing, only the crisp images from space are able to resolve fine structures in strongly lensed galaxies within the full cluster core. These structures are key to trace substructure in gravitational lens systems (Postman+11). In the figure above, we highlight all these requirements with the example of the complex cluster Abell 2744 (Merten+11). The reconstruction in panel (1) is based on ground-based weak lensing and shows only low S/N encoded by the color scale. When adding weak lensing from space in panel (2), the S/N increases significantly in the areas where these observations are available (indicated by the white frames). Adding space-based strong lensing in the panels (3) and (4) provides the resolution needed to compare to a simulation on the the basis of 2D morphology. WFIRST matches all the requirements above and provides the tremendous advantage of a wide FOV, which is similar to the best ground-based telescopes, but combines it with the advantages of a space-based observatory.

## Key Requirements

FOV – to cover the full area of the merger scenario
Depth – to measure shapes of a large number of background galaxies (60-100 arcmin^-2)
Multi-band – to reliably distinguish foreground from background galaxies through photo-z

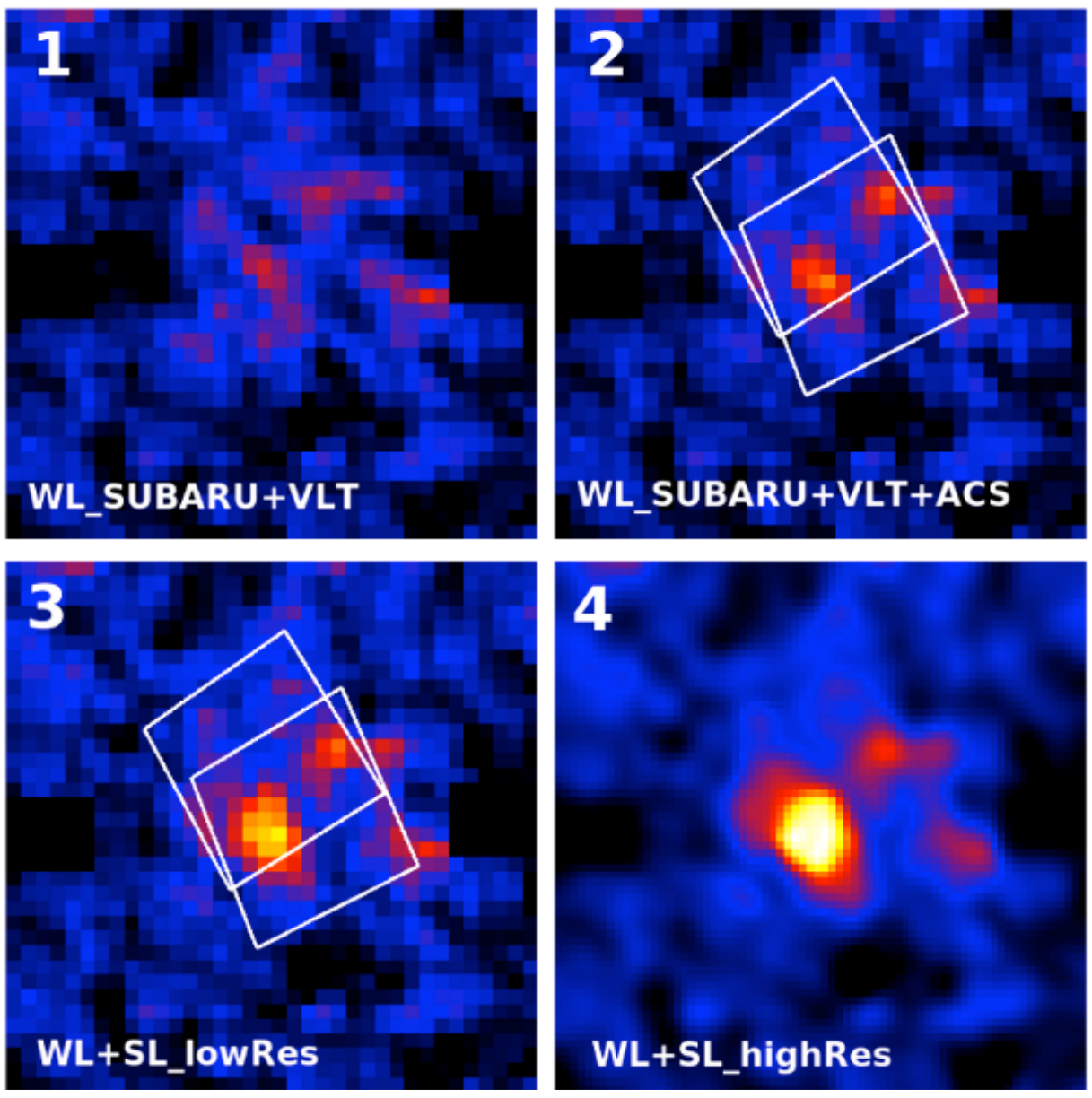

Julian Merten, Jason Rhodes (JPL / Caltech), jmerten@caltech.edu

# Merging Clusters of Galaxies

Background

Clusters of galaxies are the ideal tracers of cosmic structure formation. They are made of dark matter (~75%), hot ionized gas (~23%) and stars (~2%), and all these fundamental mass components are directly or indirectly observable in the optical, X-ray or radio. Particularly interesting are merging clusters of galaxies, where all mass components are interacting directly during the creation of cosmic structure. Multiple such mergers have been observed, with the Bullet Cluster being the most prominent example (Clowe+06). Paired with follow-up numerical simulations, such systems gave important insight into the behavior of the baryonic component (Springel+07) and set upper limits on the dark-matter self-interacting cross-section (Randall+08), which is of great importance in the search for the nature of dark matter. Recently, more complicated merging systems have been identified (Merten+11, Clowe+12, Dawson+12) offering a great opportunity to boost our understanding of dark matter.

WFIRST

Key in the analysis of a cluster merger is the precise mapping of the matter distribution and finding the exact positions of peaks in the distribution of the different mass components. In the case of dark matter, weak and strong gravitational lensing are the main tools of interest. As has been shown for many merger cases (Bradac+09, Merten+11, Jauzac+12, Clowe+12), space-based observations are required to achieve the necessary spatial resolution. This is true for strong gravitational lensing in order to resolve fine features in the shape of e.g. gravitational arcs (Postman+11) and for weak gravitational lensing where a high effective density of weakly lensed background galaxies is required (Massey+12). Currently, only the HST allows for such studies, but they are limited by its small FOV. The figure below shows an example. The complex merger Abell 2744 is shown on a Subaru/Suprimecam FOV, together with the WFC3/UVIS and ACS footprints (pink). As indicated by the white mass contours, the full merger scenario exceeds the area which was mapped to high resolution shown by the colorful overlay (Merten+11). Mapping the full field of interest with HST would need many separate pointings. But as is indicated by the red frame in the figure above, the full merger field can easily be mapped with a single WFIRST pointing with all advantages of a space-based observation mentioned above.

Key Requirements

FOV – to cover the full area of the merger scenario
Depth – to measure shapes of a large number of background galaxies (60-100 arcmin^-2)
Multi-band – to reliably distinguish foreground from background galaxies through photo-z

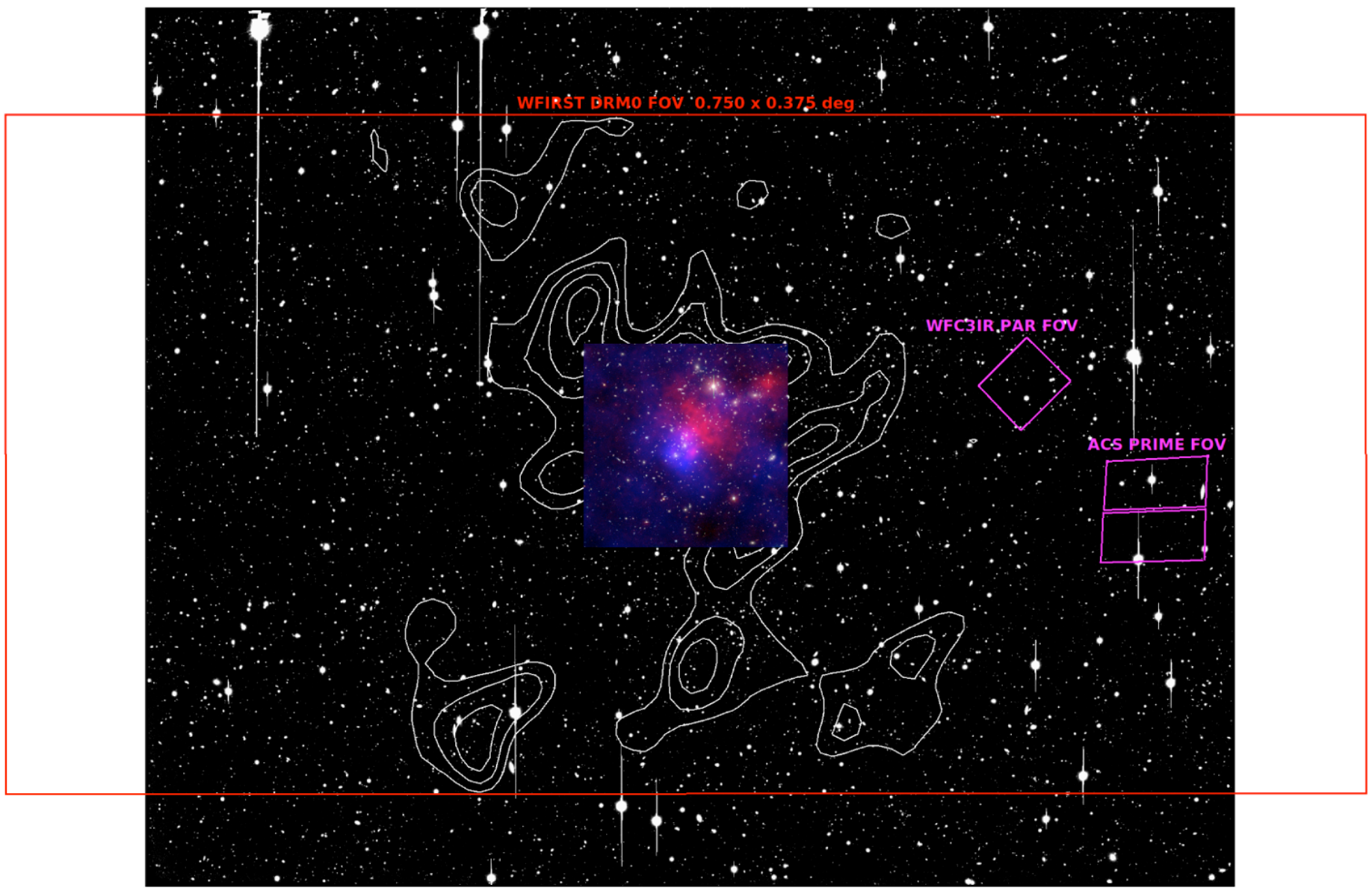

Appendix A: One

Priyamvada Natarajan (Yale University); priyamvada.natarajan@yale.edu

## Group-Scale Lenses: Unexplored Territory

**Background**

With the precision cosmology enabled by instruments like WMAP, we now have extremely robust constraints on the cold dark matter model and the structure formation paradigm. These results suggest that the bulk of the mass in the universe is locked on group scales, with typical values of ~ $10^{13}$ solar masses. The full census of properties of groups is currently incomplete. Some key questions that remain to be settled are whether groups have a large-scale dark matter halo or whether the individual galaxies that comprise them have independent dark matter halos.

The group environment also enables probing the nature of interaction, if any, between dark matter and the baryonic component – whether and how significant the adiabatic contraction in the center of the halo is can be addressed with a large sample of group scale lenses. Finding groups using observed strong lensing features in large surveys has been demonstrated to be a powerful technique as recent results from the CFHTLS SARCS survey suggest (More et al. 2012 and references therein). Multiple image separations ranging from 10 – 20 arcsec corresponds to group scale masses and can be unambiguously detected.

**WFIRST**

A proposed multi-band survey by WFIRST will enable the discovery of several hundred hitherto undetected groups. CFHTLS found 30-45 arcs from group scale lenses in a survey of 170 sq. degrees. A 1000 – 2000 sq. degree WFIRST survey will provide between 200 – 400 groups. Expected image separation distributions are expected to range from 10 – 20 arcsec.

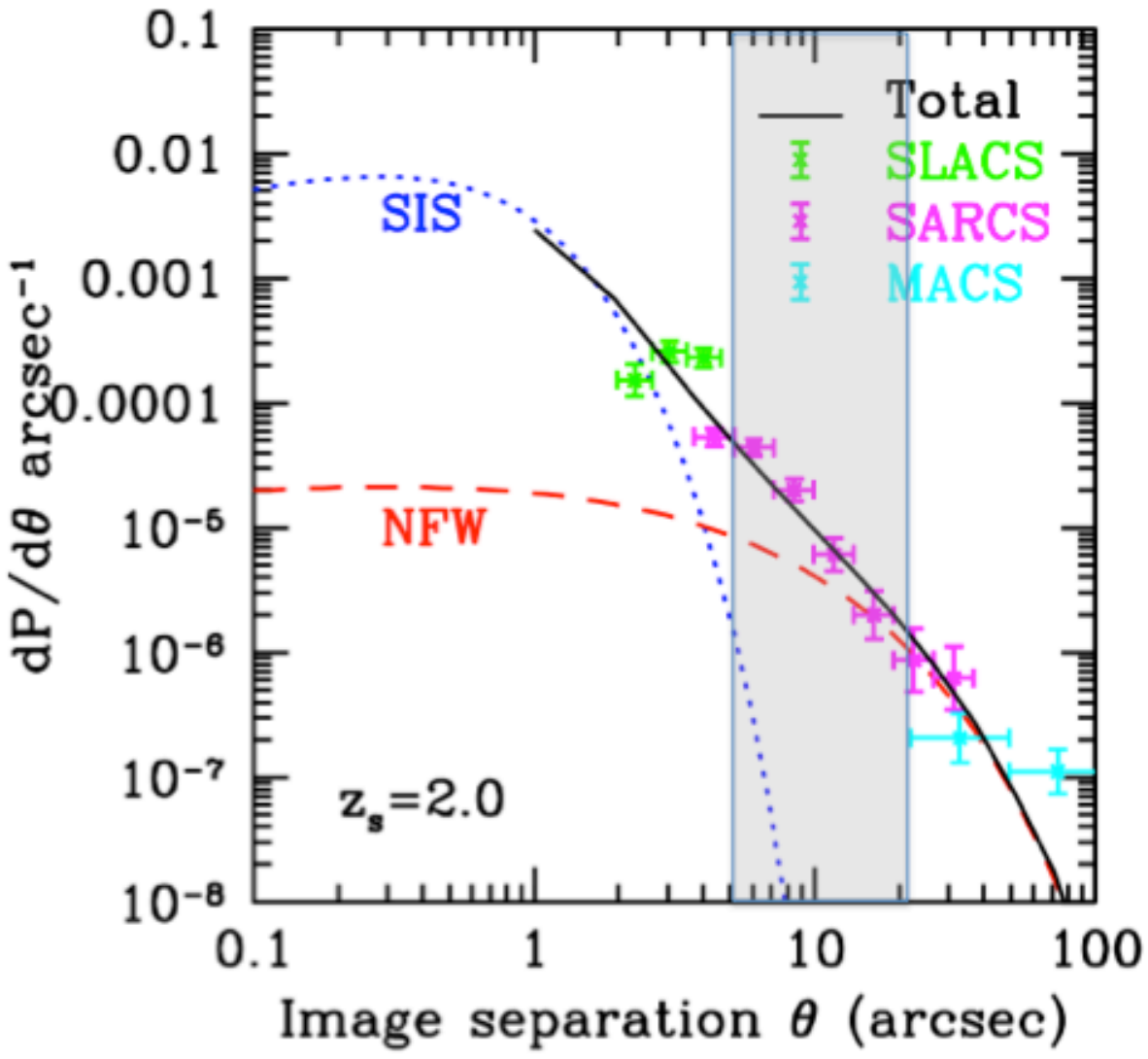

Caption: The targeted image separation range for group scale lenses is marked in gray.

**Key Requirements**

Wavelength coverage: I, Y, J, H, (K) – I band for identifying the arcs and as many bands as possible for optimal photo-z determination.

Megan Donahue and Mark Voit (Michigan State University), donahue@pa.msu.edu

# Finding and Weighing Distant, High Mass Clusters of Galaxies

Background

The mean density of matter and dark energy and the initial fluctuation spectrum determine the abundance and growth of clusters of galaxies (e.g Voit 2005). The most dramatic effects are experienced by the abundance of the most massive clusters. Furthermore, the abundance of the most extremely massive clusters tests the assumption of Gaussianity of the power spectrum at Mpc scales. The key to cluster cosmology lies in accurate estimates of both the number counts (as a function of redshift) and the gravitating masses of clusters of galaxies. Finding clusters by their projected lensing mass, by the overdensity of red sequence galaxies, and by their Sunyaev-Zeldovich decrement on the CMB would provide mutual verification of cluster existence, hot gas (baryonic) content, presence of a prominent red sequence. Spitzer IRAC surveys have discovered over one hundred z>1 candidates in 7.25 sq degrees (Eisenhardt et al. 2008), yielding at least one massive cluster at z=1.75 (Brodwin et al. 2012; Stanford et al. 2012).

WFIRST

Survey fields obtained for WFIRST weak lensing shear and/or baryon acoustic oscillation studies will also allow a census of massive clusters. Photometric redshift observations, either ground-based or from WFIRST, would be required to estimate the redshift of newly discovered clusters.

Key Requirements

Depth – Well-dithered exposures sufficient to obtain shape measurements for galaxies at z>z_cluster; shape measurements benefit from 3-4 repeats

Field of View – A wide-field survey of 8300 sq degrees include 100 10^14 h^-1 and 2800 10^13.7 h^-1 z=2-2.5 clusters; for z=1.5-2.0, the numbers rise to 1660 and 20,000 clusters for the same masses, respectively; the most massive clusters are the most rare, the larger the survey, the better the probe of the high-sigma tail of the mass distribution

Wavelength Coverage – NIR filters

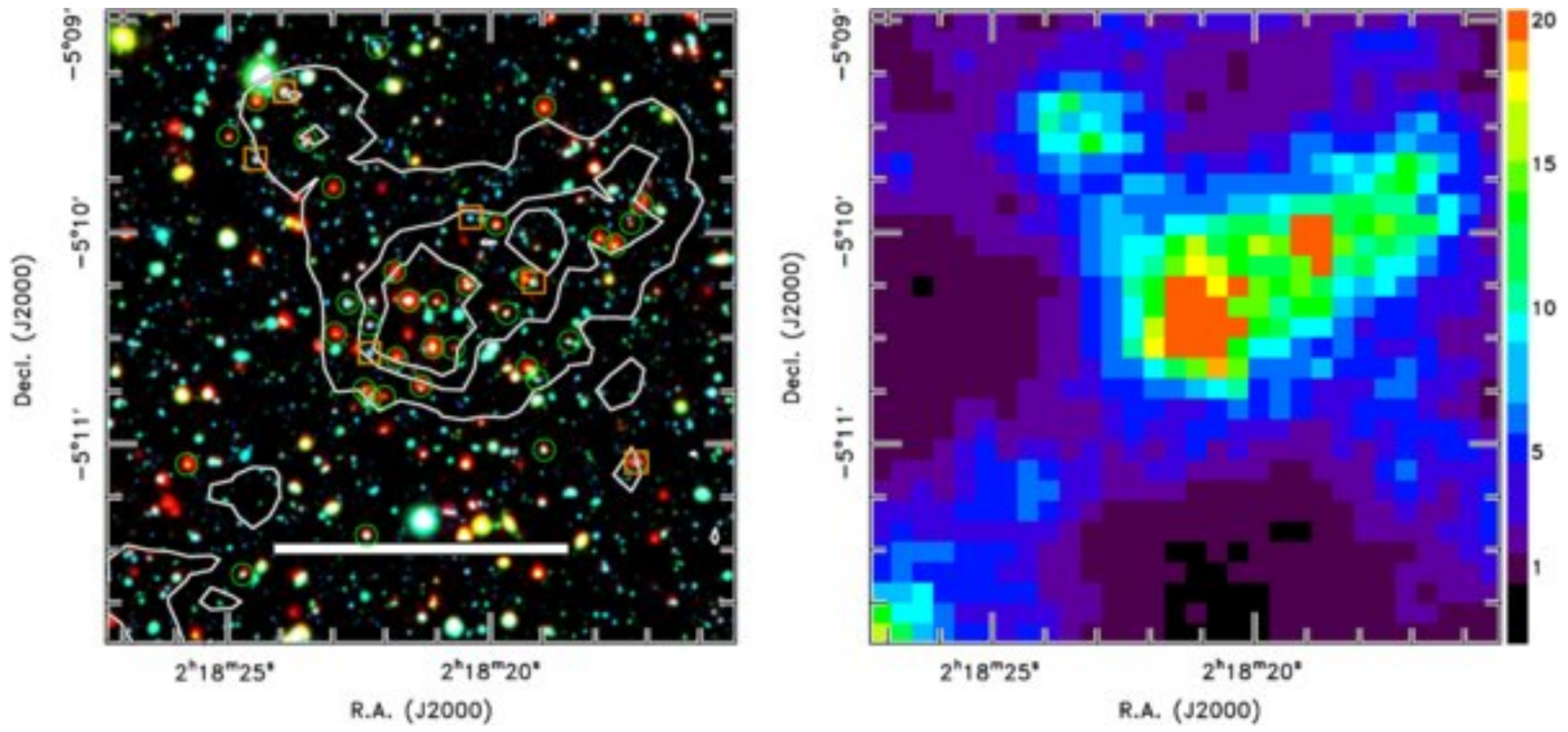

Caption: Left panel shows a false-color image of a Spitzer-selected cluster at z=1.6 (Papovich et al. 2010). Blue corresponds to the Suprime-Cam B band, green to the Suprime-Cam i band, and red to the Spitzer 4.5 μm band. The right panel shows the surface density of galaxies, color coded in units of standard deviations above the mean.

Megan Donahue (Michigan State University), donahue@pa.msu.edu

## The Evolution of Massive Galaxies: the Formation and Morphologies of Red Sequence Galaxies

Background

The most massive galaxies in the present day universe are red, usually dormant, elliptical galaxies. They appear in the color-magnitude diagrams of galaxies in clusters of galaxies as the red sequence (RS). The red sequence represents a distinct population of galaxies whose stars formed at high redshift (z>3) and have passively evolved since then. These galaxies also experience significant interactions and mergers over their lifetime. The red sequence feature is so prominent that it can be used to discover distant clusters of galaxies and to estimate their redshifts. Current studies suggest that while the stars in these galaxies formed early, the assembly of the galaxies we see locally occurred relatively late. Studies of the RS galaxies as a function of redshift, cluster mass, and local environment are critical tests of our understanding of galaxy evolution.

WFIRST

WFIRST will enable sensitive IR color-magnitude diagrams of RS galaxies in clusters and in the field at redshifts representing the epoch at which the red sequence is beginning to assemble, z~1.5-2.5. The spatial resolution will allow galaxy sizes to be estimated; as mergers progress, the stellar orbits get progressively more puffed out. The best photometric redshifts will be obtained from filters which span the 4000 Angstrom break and which are medium-width (not too broad, see FourStar results from Spitler et al. 2012.)

Key Requirements

Depth – Well dithered exposures sensitive to K~25-26 (AB)
Field of View – Targeted follow-up of high-z candidates (from Sunyaev-Zeldovich surveys or other techniques). Cluster size is ~ 1-2 arcmin, ~ independent of z.
Field of View – Survey for z=1.5-2.5 clusters: ~ 100 sq degrees would include ~20 clusters at M~10^14 h^-1 solar masses if you had to find the clusters first.
Wavelength Coverage – Medium-band NIR filters.

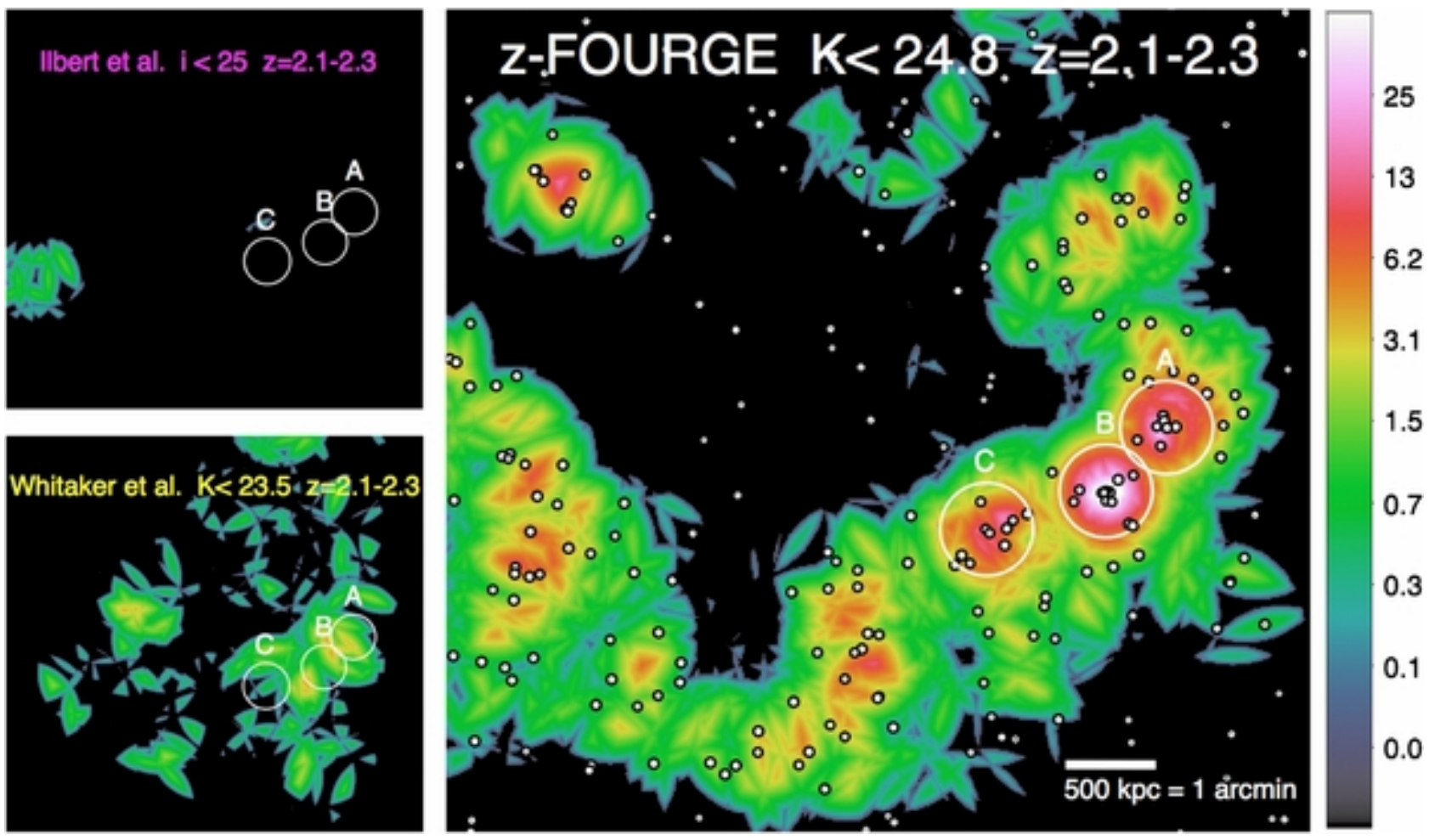

Caption: Nearest-neighbor surface density maps for z = 2.1-2.3 in a 9' × 9' region in the COSMOS field (Spitzer et al. 2012).

Sangeeta Malhotra and James Rhoads (ASU), Daniel Stern (JPL), daniel.k.stern@jpl.nasa.gov

# Probing the Epoch of Reionization with Lyman-Alpha Emitters

Background

Observations of the Gunn-Peterson effect in the most distant quasars tell us that the reionization of the Universe concluded at redshift z~6 (e.g., Fan et al. 2006). Since that time, the gas between the galaxies has remained largely ionized, while at earlier cosmic epochs this intergalactic medium (IGM) was largely neutral. The exact time, or times, that this cosmic phase change occurred is a fundamental question for galaxy formation, telling us both when the first generation of stars formed in the Universe and how much effect they had on their neighboring systems. Various observations paint potentially conflicting pictures of this "epoch of reionization", with quasars suggesting a late reionization, and both the cosmic microwave background and Lyman-alpha emitting galaxies (LAEs) suggesting an earlier reionization. Theory suggests that multiple epochs of reionization are possible. WFIRST will be a premiere facility for studying the late stages of the cosmic reionization epoch (see Figure).

NRO/WFIRST

At the resolution of the WFIRST grism, LAEs will be easily identified from their strong, narrow Lyα emission and their diminished flux blue-ward of this emission. Simple arguments dictate that LAEs provide a powerful probe of reionization: Lyα photons injected into a neutral medium are strongly scattered, thus strongly suppressing, or even eliminating, detectable Lyα emission. The transmitted fraction depends upon the size of the local cosmological HII region surrounding the source, and therefore on the ionizing luminosity and age of the source (e.g., Santos 2004) as well as contributions from associated, clustered sources (e.g., Wyithe & Loeb 2004). Nevertheless, we expect a rapid decline in the observed space density of LAEs as the reionization epoch is approached: a statistical sample of LAEs spanning the reionization epoch, as the WFIRST grism would uniquely provide, will present an extremely robust probe and two-dimensional map of reionization (e.g., Malhotra & Rhoads 2004; Stern et al. 2005; Ouchi et al. 2010). In particular, such work will require much larger fields than are accessible with Hubble or expected to be observed with JWST.

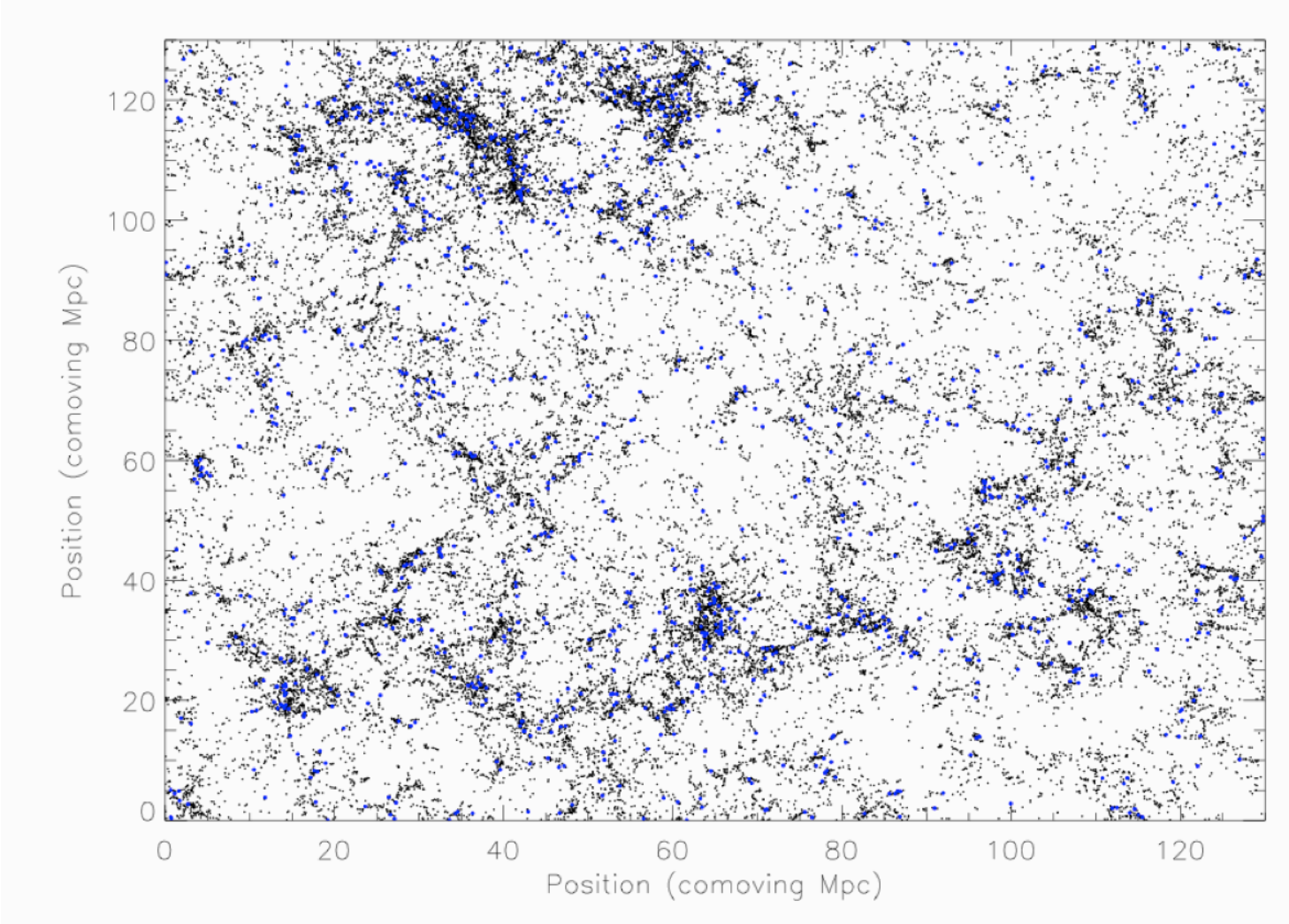

Caption: Simulated large scale structure at z=6.6. Black dots are collapsed haloes, blue dots are LAEs. This corresponds to a FOV of 0.73 degrees on a side. The ACS FOV is roughly 8% of this, and is therefore much too small to study the spatial distribution of LAEs or reionization bubles efficientyly (based on Tilvi et al. 2009).

NRO/WFIRST could measure the redshift of reionization using an observational effort equivalent to a typical Hubble Treasury program. 200-300 hours of slitless spectroscopic integration, intensively covering a one square degree field, would yield an expected sample of ~ 2000 Lyman alpha galaxies beyond redshift 8 in the absence of neutral intergalactic gas. The redshift where this sample is cut off by Lyman alpha scattering in a neutral IGM would be measured with high statistical confidence at any redshift up to z ~ 12. More detailed information is provided in the Dressler et al. white paper.

Daniel Stern (JPL), daniel.k.stern@jpl.nasa.gov

## Obscured Quasars

Background

Obscured, or type-2 quasars are expected to outnumber unobscured, type-1 quasars by factors of 2-3. They are predicted by models of active galactic nuclei (AGN), and are required to explain the hard spectrum of the cosmic X-ray background. Until recently, however, our census of this dominant population of AGN has been lacking since such systems are difficult to identify in optical and low-energy X-ray surveys. High-energy missions to date have had limited sensitivity, while the recently launched NuSTAR mission has a very limited field-of-view. Mid-infrared surveys, initially with Spitzer (e.g., Lacy et al. 2004; Stern et al. 2005) and more recently with WISE (Stern et al. 2012; Assef et al. 2013) have dramatically changed the situation. In its shallowest fields, WISE identifies 60 AGN candidates per deg2 with a reliability in excess of 95%. In higher latitude, deeper parts of the WISE survey, this surface density rises to >100 AGN per deg2. Comparably powerful AGN identified in optical surveys have surface densities of only ~20 per deg2 (e.g., Richards et al. 2006), implying that WISE has finally realized the efficient identification of the dominant luminous AGN population across the full sky.

WFIRST

While mid-infrared observations with WISE have identified this dominant population of obscured, luminous quasars, WFIRST will be required to characterize its properties. Working with deep ground-based optical data, WFIRST will provide photometric redshifts for this population, something not possible from the mid-infrared data alone since the identification relies on the power-law mid-infrared spectra of luminous AGN. Photometric redshifts will allow us to probe the cosmic history of obscured black hole growth, and relate it to galaxy formation and evolution. AGN feedback is expected to play an important role shaping the present-day appearances of galaxies. Measuring the clustering amplitude of obscured and unobscured quasars will probe AGN unification scenarios. For the classic orientation-driven torus model, clustering should be the same for both populations. On the other hand, if obscured AGN are more common in merging systems, with the obscuration caused by galactic-scale material, then the clustering amplitudes are expected to differ.

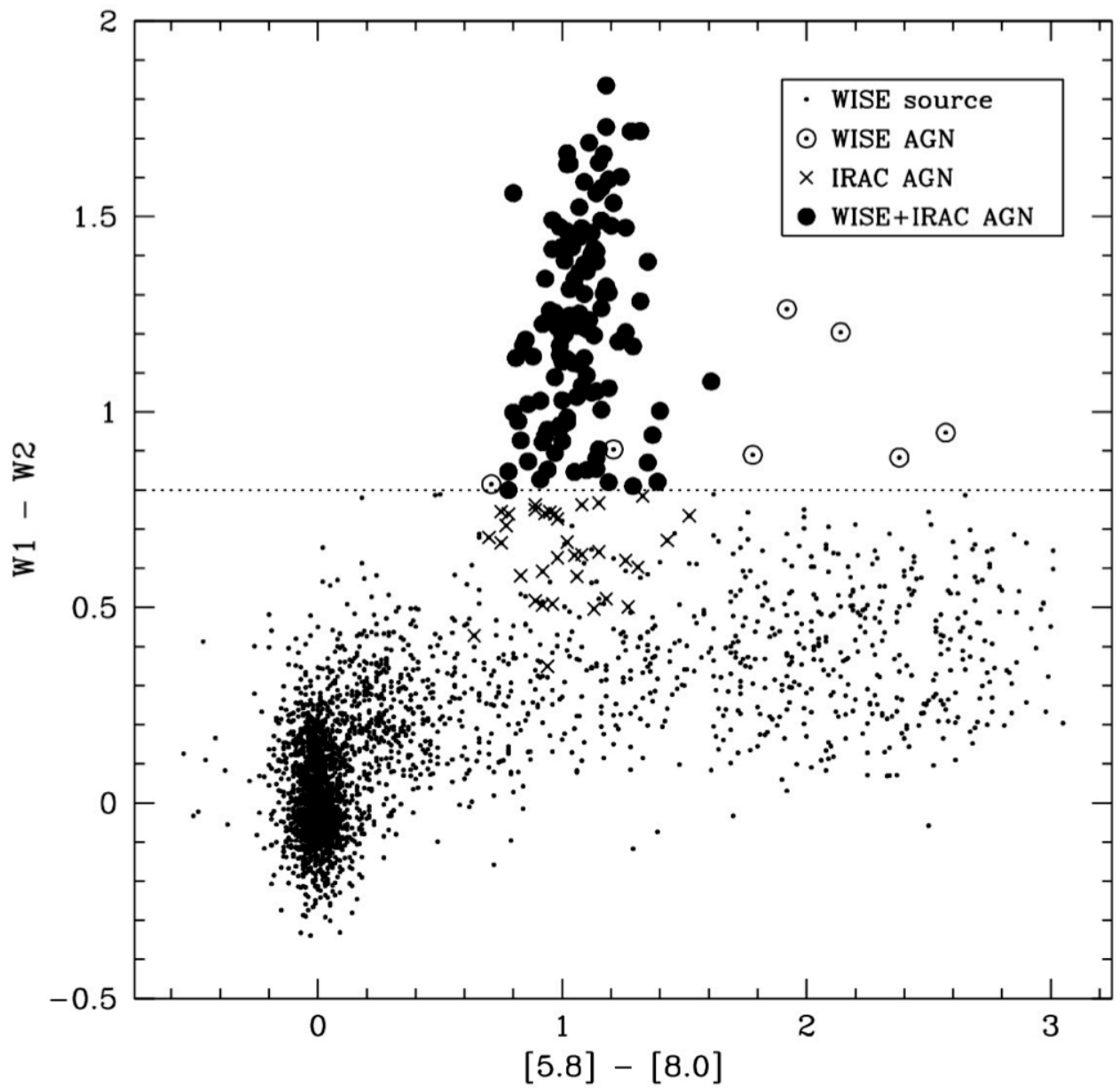

Caption: Mid-infrared color-color diagram, illustrating that a simple WISE color cut of W1-W2 ≥0.8 identifies 62 AGN per deg2 with 95% reliability assuming the complete reliability of the Stern et al. (2005) Spitzer AGN selection criteria. From Stern et al. (2012).

Key Requirements

Depth – Provide robust photometric redshifts for sub-L* populations to z~2
Field of View – Wide-area surveys required for clustering analysis
Wavelength Coverage – >2 NIR filters for robust photometric redshifts

Daniel Stern (JPL), daniel.k.stern@jpl.nasa.gov

# The Faint End of the Quasar Luminosity Function

## Background

The primary observable that traces the evolution of quasar (QSO) populations is the QSO luminosity function (QLF) as a function of redshift. The QLF can be well represented by a broken power-law: $\Phi(L) = \phi_*/[(L/L_*)\alpha + (L/L_*)\beta]$, where $L_*$ is the break luminosity. Current measurements poorly constrain the corresponding break absolute magnitude to be $M_* \simeq -25$ to $-26$ at λ=1450 Å. The bright-end slope, α, appears to evolve and flatten toward high redshift, beyond z ~ 2.5 (Richards et al. 2006). The faint-end slope, β, is typically measured to be around −1.7 at z~2.1; it is poorly constrained at higher redshift, but appears to flatten at z ~ 3 (Siana et al. 2008). At yet higher redshift, however, the situation is much less clear due to the relatively shallow flux limits of most surveys to date. The true shape of the QLF at z > 4 is still not well measured, and the evolution of $L_*$ and the faint-end slope remain poorly constrained (Glikman et al. 2010, 2011; Ikeda et al. 2010). Studying the faint end of the high-redshift QLF is important for understanding the sources responsible for re-ionizing the Universe. While Vanzella et al. (2010), studying faint z~4 galaxies in the GOODS fields, finds that Lyman-break galaxies (LBGs) account for <20% of the photons necessary to ionize the intergalactic medium (IGM) at that redshift, Glikman et al. (2011) find that QSOs can account for 60±40% of the ionizing photons. Furthermore, new quasar populations appear when one studies faint, high-redshift quasars. Glikman et al. (2007), in their initial work on the faint end of the high-redshift QLF, found strong NIV] 1486 emission in ~10% of the QSOs surveyed, likely associated with early starbursts.

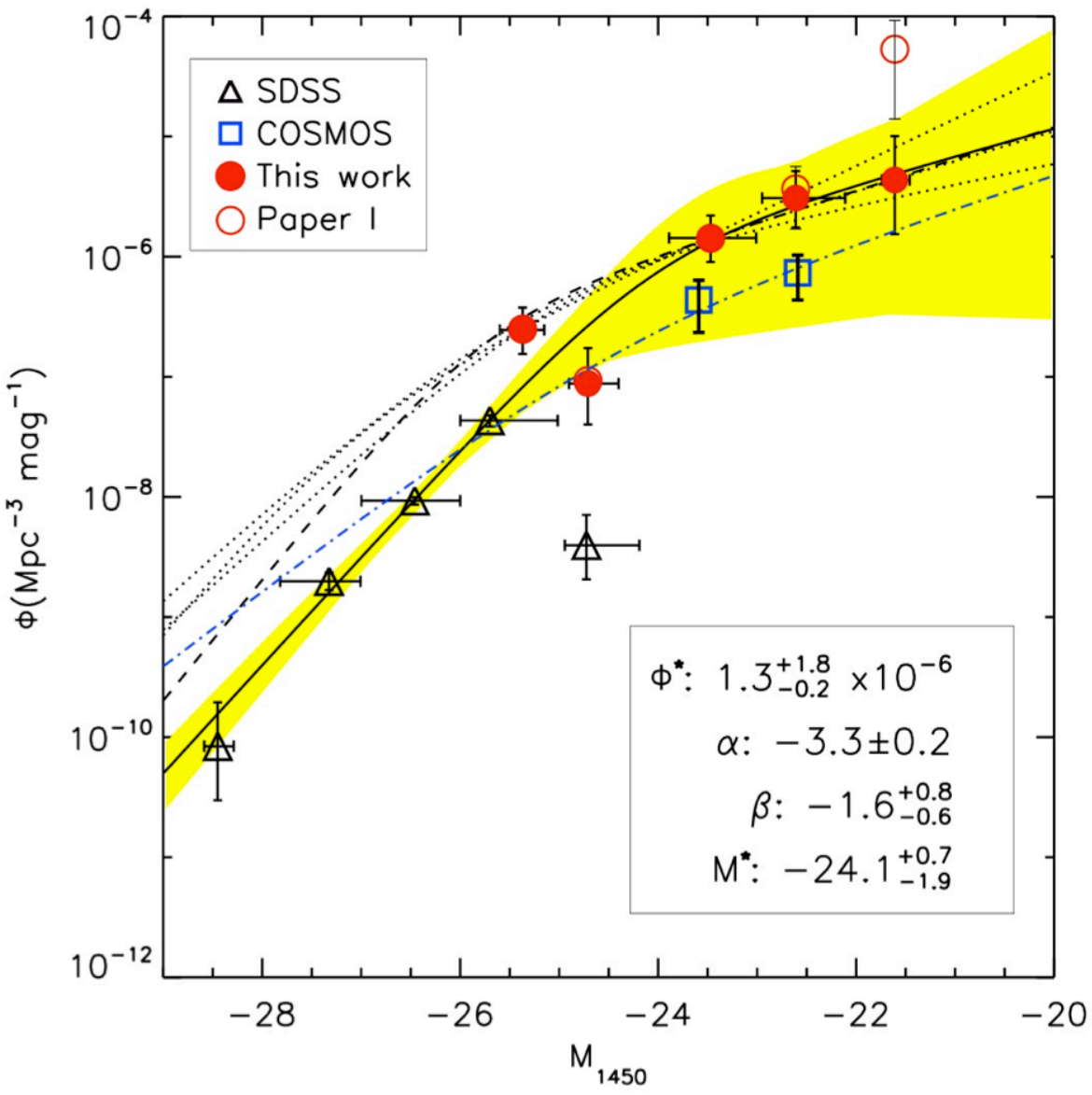

Caption: QLF at z~4 from Glikman et al. (2011). Note the substantial uncertainties below the knee in the QLF.

## WFIRST

The deep, wide-field imaging data from WFIRST DE surveys will provide critical information for constraining the faint end of the high-redshift QLF. While at z~4, deep ground-based data from surveys such as LSST will be essential, WFIRST grism spectroscopy will play an essential role in confirming the high-redshift AGN nature of the photometric candidates. At yet higher redshifts, z>7, deep surveys from WFIRST alone will probe the earliest epochs of nuclear activity.

## Key Requirements

Depth – To study sub-L* QSO populations at high redshift
Morphology – To separate stars from galaxies (though note that faint QSOs at high redshift are likely to show significant host galaxy light, and thus morphology should be used to characterize the populations, but not as a selection criterion)
Grism – For spectroscopic confirmation and characterization
Field of View – Wide-area surveys required to identify rare populations
Wavelength Coverage – >2 NIR filters to robustly identify QSOs at the highest redshifts

Xiaohui Fan (University of Arizona), fan@as.arizona.edu

## Strongly Lensed Quasars

Background

Strong gravitational lensing, when multiple images of distant objects are produced by massive objects in the foreground, is a powerful and unique tool for both cosmology and galaxy/AGN physics (e.g., Oguri et al. 2012). Applications of strongly lensed quasars include: using time delay among different lens components to measure $H_0$ and to constrain the expansion history of the Universe (Fassnacht et al. 2002); using lens models to measure galaxy mass and structure (Bolton et al. 2006); using flux ratio anomalies to probe dark matter halo properties and the existence of halo subtracture (Keeton et al. 2006); using AGN microlensing to probe accretion disk structure (Kochanek et al. 2006); using the quasar lensing fraction to study magnification bias and the quasar luminosity function (Richards et al. 2006); using high spatial resolution imaging of lensed systems to study properties of quasar host galaxies and environments (Peng et al. 2006); using rare examples of quasars acting as gravitational lenses to study the cosmic history of the black hole-bulge relation (Courbin et al. 2012). However, strong lensing is a rare event, requiring wide-field, high resolution imaging to establish a large sample for statistical studies, and to uncover those with ideal lensing configurations for cosmological tests (so-called "Golden Lenses").

WFIRST

The deep, wide-field imaging data from WFIRST surveys will provide a treasure trove for lensing studies. With a pixel size of ~0.1", WFIRST will resolve any multiply imaged system with separation >0.2". This is expected to account for the vast majority of all strong quasar lenses, and will allow accurate photometry of lensed image components as well as lensing galaxies. Multicolor photometry will provide accurate photometric redshifts for sources and lenses. Figure 1 presents the expected number/redshift distributions of lensing galaxies (lenses) and lensed quasars (sources) in a total survey area of 20,000 deg, with a limiting magnitude of ~24 AB and image separation of >0.5". This represents a nearly two orders of magnitude increase in size from current samples (Inada et al. 2012). Strong lensing is also one of the key areas for LSST science. WFIRST will have strong synergy with LSST, providing deeper imaging with higher spatial resolution and more accurate photometry and photometric redshifts.

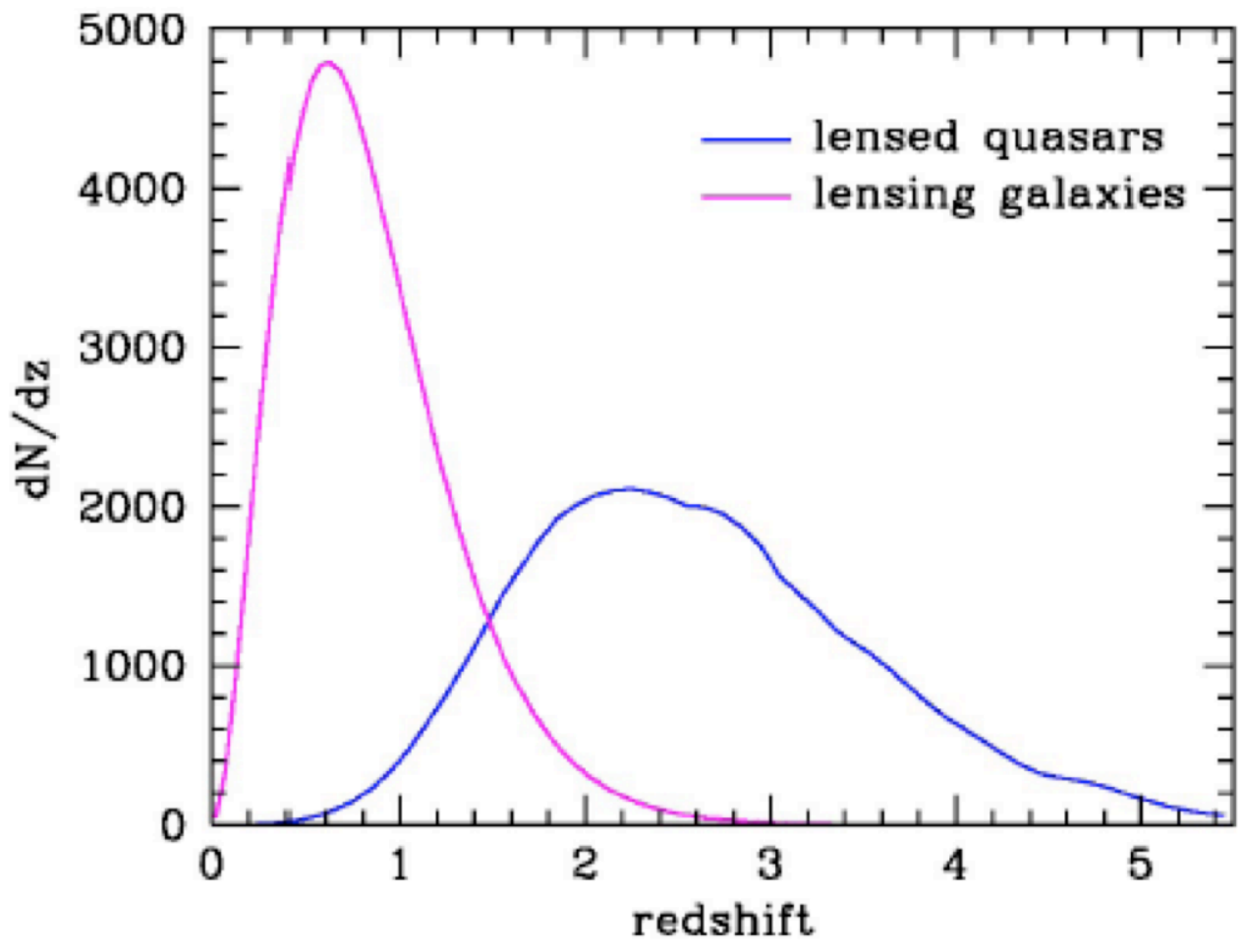

Key Requirements

Depth – To sample a wide range of quasar lum. and to detect faint lensed image components

Spatial Resolution – This is the key: a pixel size and resolution <0.3" is needed to uncover the majority of lenses and to allow accurate photometry of lensed components

Field of View – Wide-area surveys required to establish large sample and to find the rare unique systems

Wavelength Coverage – >2 NIR filters for photo-z of both lenses and sources

Xiaohui Fan (University of Arizona), fan@as.arizona.edu

## High-Redshift Quasars and Reionization

Background

Luminous quasars at high redshift provide direct probes of the evolution of supermassive black holes (BHs) and the intergalactic medium (IGM) at early cosmic time. The detection of z>7 quasars (e.g., Mortlock et al. 2011) indicates the existence of billion solar mass BHs merely a few hundred million years after the Big Bang, and provides the strongest constraints on the early growth of supermassive BHs and their environments. Spectroscopy of the highest redshift quasars reveals complete Gunn-Peterson (1965) absorption, indicating a rapid increase in the IGM neutral fraction and an end of the reionization epoch at z=6-7 (Fan et al. 2006). Current observations suggest a peak of reionization activity and emergence of the earliest galaxies and AGNs at 7<z<15, highlighting the need to expand quasar research to higher redshift.

WFIRST

While ground-based surveys such as LSST and VISTA will make progress in the z=7-8 regime in the coming decade, strong near-IR background from the ground will limit observations to the most luminous objects and to z<8. Wide-field, deep near-IR survey data offered by WFIRST will fundamentally change the landscape of early Universe investigations. Figure 1 shows the predicted number of high-redshift quasars in a WFIRST survey based on current measurements at z~6 (Jiang et al. 2008; Willott et al. 2010) and extrapolation to higher redshift with a declining number density following the trend seen at z=3-6. WFIRST should allow robust identifications of a large sample of reionization-epoch quasars up to z>10, if they exist at those epochs. WFIRST grism will provide direct spectroscopic confirmation and characterization of these high-redshift quasars, while JWST and next-generation extremely large telescope high-resolution spectroscopic observations will measure IGM and BH properties. Key questions to be addressed by these observations are: (1) when did the first generation of supermassive BHs emerge in the Universe; (2) how and when did the IGM become mostly neutral; (3) did quasars and AGNs play a significant role in the reionization process?

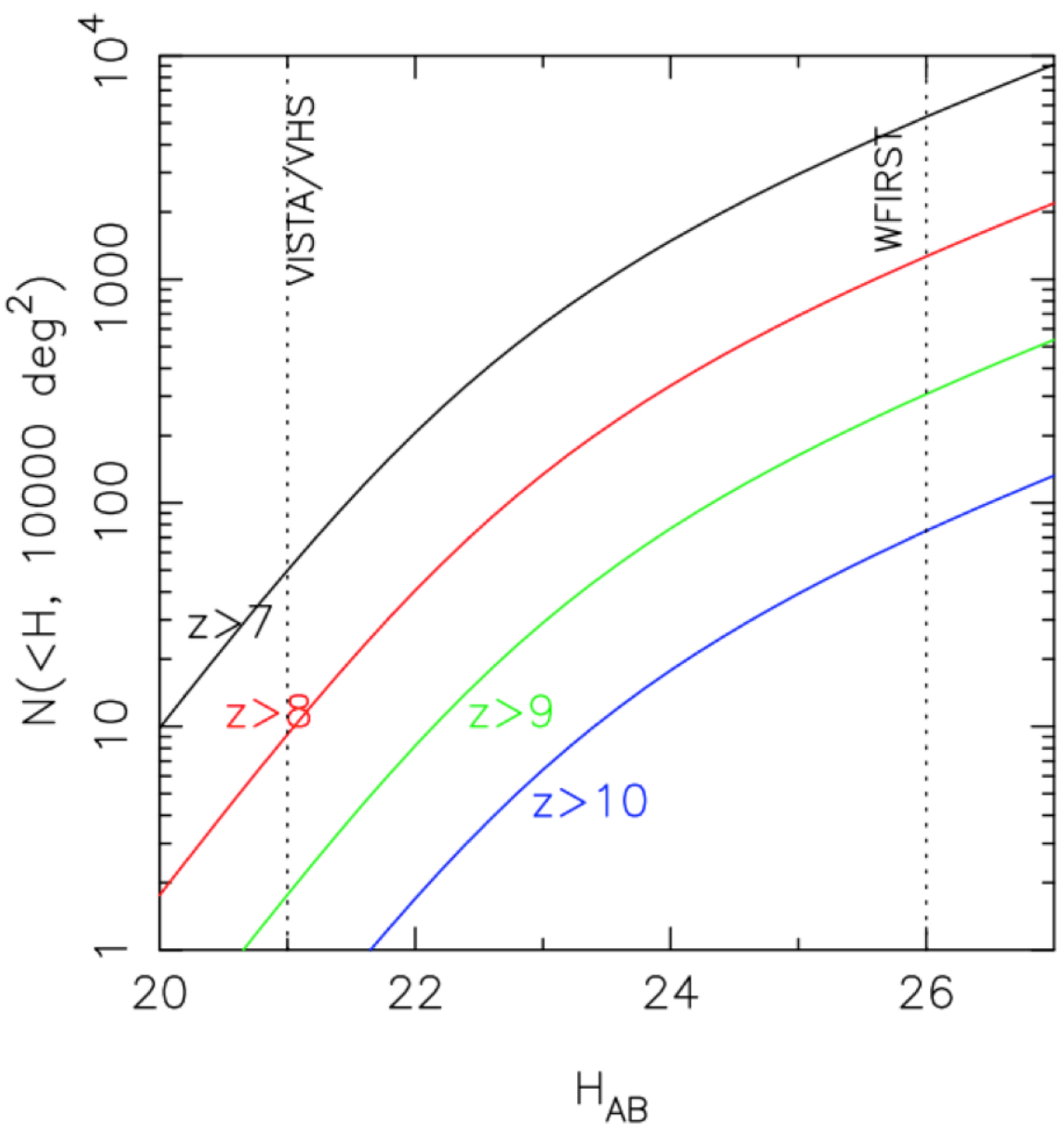

Key Requirements

Depth – To study sub-L* QSO populations at high redshift

Morphology – To separate stars from galaxies; one of the main contaminations is expected to be low-redshift red galaxies

Grism – For spectroscopic confirmation and characterization

Field of View – Wide-area surveys required to identify rare populations

Wavelength Coverage – >2 NIR filters to robustly identify QSOs at the highest redshifts

Harry Teplitz (IPAC), hit@ipac.caltech.edu

## Characterizing the sources responsible for Reionization

Star–forming galaxies are likely responsible for reionizing the Universe by z~6, implying that a high fraction of HI–ionizing (Lyman continuum; LyC) photons escape into the IGM. Measurements of galaxies at z<3.5 show the average escape fraction, $f_{esc}$, to be very low or undetected at all redshifts. At redshift z~3, high $f_{esc}$ (≥ 50%) have been reported in only about 10% of LBGs and Lyα emitters though significant mysteries remain for even these few detections ($f_{esc}$ > unity, and unknown numbers of interlopers).

Current studies suggest that $f_{esc}$ may evolve with redshift (Figure 1) and/or be higher in low mass galaxies. A recent study of the radiation transport of LyC at z= 3-6 in galaxies drawn from cosmological SPH simulations predicted substantial LyC ($f_{esc}$ =8–20%) emission from galaxies with halo masses $M_{halo}$ <~ $10^{10}$ M*, but little or nothing from more massive systems. In addition, $f_{esc}$ is found to increase with decreasing metallicity and SFR – both of which are thought to positively correlate with $M_{halo}$.

The LyC is best studied at z<3 (in the UV) due to the increasing opacity of the IGM to LyC photons higher z. However, current studies are severely limited by the need to find appropriate analogs to the low mass/metallicity objects that are the likely sources of LyC photons during Reionization.

Figure 1: $L_{yC}/f_{1500}$ measurements (versus UV luminosity) at z ~ 1 and z ~ 3 corrected for the average IGM attenuation at the relevant wavelengths (Siana et al. 2010).

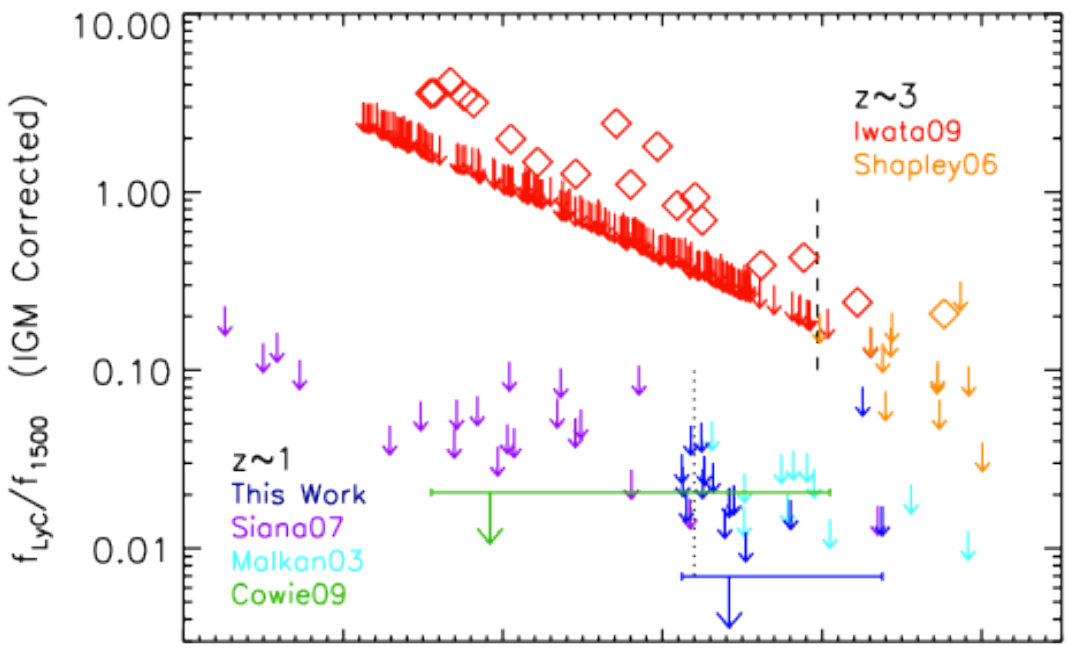

WFIRST

A wide-area NIR grism survey with WFIRST presents an important opportunity to find LyC-emitting sources through their rest-frame optical emisison lines. HST grism surveys discovered a population of young, low-mass starbursts, selected based on their very strong emission-lines ($EW[H\alpha]_{Rest}$ > 200Å). These galaxies are ideal for testing the hypothesis of a mass dependent $f_{esc}$: they are among the most metal poor objects known at these redshifts,and their average emission–line corrected stellar mass is $10^7$—$10^8$M*, i.e., ~30 times less massive that typical M* galaxies at the same redshifts. HST surveys are severely limited by lack of spatial coverage in finding such rare sources. A WFIRST survey would enable UV follow-up -- either archival (GALEX, HST, SWIFT-UVOT) or with new observations from ground-based U-band (at z~3) and possibly future UV telescopes. Ground-based observations would be particularly helped by space-based source identification that would spatially resolve possible foreground interlopers.

In addition, WFIRST would allow us to identify obscured Type-2 AGNs. The possibility that faint AGNs play a substantial role in Reionization has not yet been ruled out observationally.

Key Requirements

Grism – To find low mass/metallicity line-emitters to enable (possibly archival) UV follow-up
Depth – These sources are faint
Field of View – Wide-area surveys required to identify rare populations

Daniel Whalen (Los Alamos National Laboratory), dwhalen@lanl.gov

# Finding the First Cosmic Explosions With WFIRST

## Background

Population III stars ended the cosmic Dark Ages at z ~ 25 and began the reionization and chemical enrichment of the early universe (Whalen et al 2004; Joggerst et al. 2010). They also determined the luminosities and spectra of primitive galaxies at 10 < z < 15 and may be the origin of z ~ 7 SMBHs (Mortlock et al 2011). Unfortunately, despite their extreme luminosities individual Pop III stars and many primeval galaxies lie beyond the detection limits of even next-generation observatories such as JWST and 30m class telescopes. However, their masses can be directly inferred from their supernova explosions, which can be 100,000 times more luminous than the stars or the protogalaxies in which they reside.

## WFIRST

Recent numerical simulations indicate that Pop III pair-instability supernovae (PI SNe) at z ~ 15 – 20 will reach AB magnitudes of ~ 26 at 2.2 microns (Whalen et al 2012a,b), as shown at right for 150, 175, 200, 225 and 250 solar mass progenitors. These magnitudes are within the proposed NIR detection limits for WFIRST. This epoch may be optimal for detecting PI SNe because of the rise of strong Lyman-Werner UV backgrounds from the first generation of Pop III stars, which delayed subsequent star formation in cosmological halos until they grew to larger masses by accretion and mergers. The larger virial temperatures of more massive halos in this era likely led to the formation of more massive stars (O'Shea & Norman 2008) that died as PI SNe. Numerical simulations of Pop III star formation rates through cosmic time suggest SN event rates of ~ 1 sq. deg.$^{-1}$ yr$^{-1}$ at 15 < z < 20, so WFIRST could detect hundreds of these transients over its lifetime. Detections of this volume will enable WFIRST to build up the the reliable estimates of the Pop III IMF. The discovery of SNe by WFIRST at slightly lower redshifts, z ~ 10 – 15, will probe star formation in primeval galaxies and reveal their positions on the sky for more detailed spectroscopic followup by JWST.

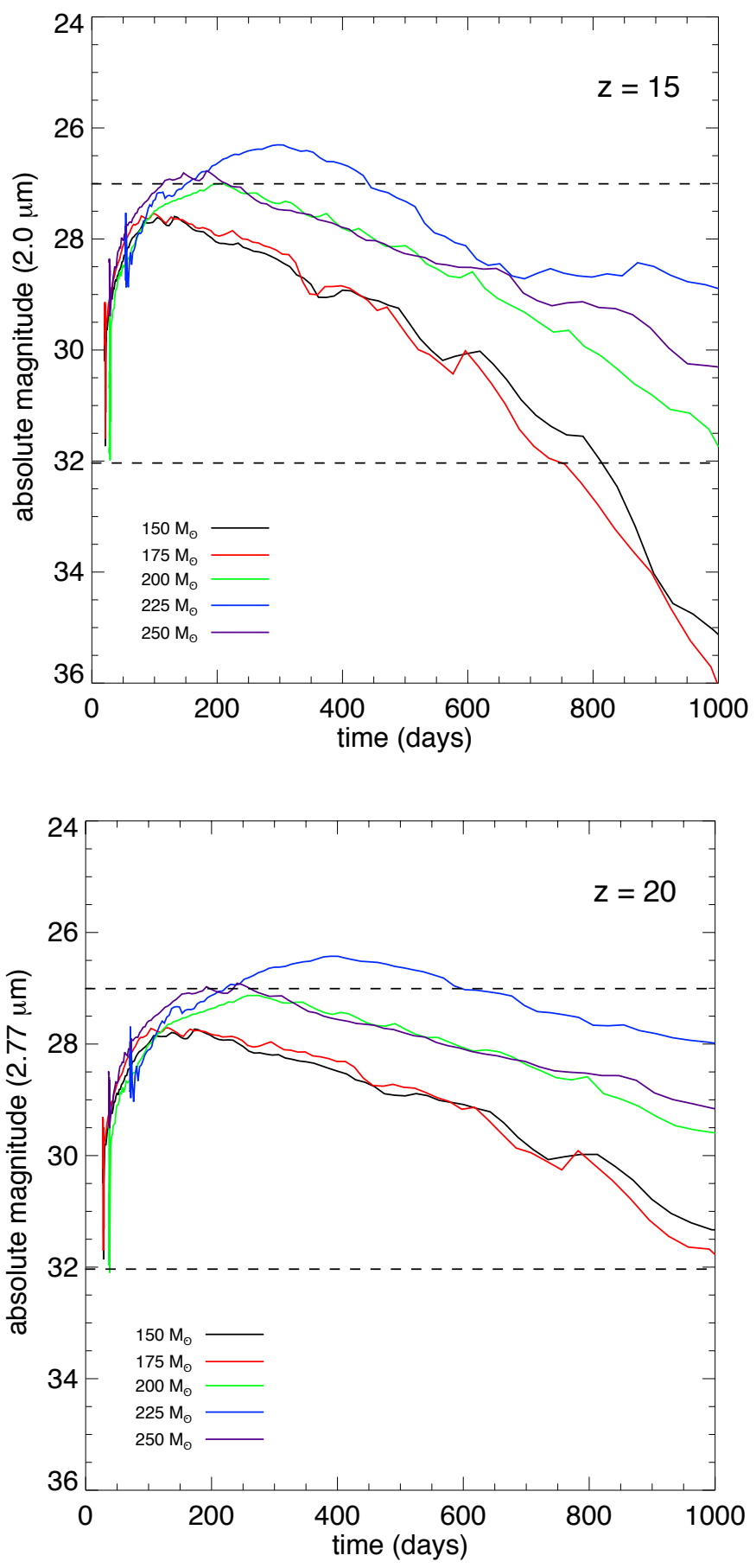

## Key Requirements

Sensitivity --> AB mag 27

Field of View --> Wide-area surveys, which are required for collecting large numbers of transients

Wavelength Coverage --> 2 filters to cover NIR peaks of Pop III PI SNe from z = 10 – 20 (F200, F277)

Claudia Scarlata (UMN, scarlata@astro.umn.edu) and Harry Teplitz (IPAC, hit@ipac.caltech.edu)

## Resolved stellar population studies in z~2 SF galaxies

Observations show that "normal" SF galaxies were in place at z~0.5, with stellar population and scaling relations consistent with passive evolution into the stable population observed locally. Looking back to z > 2, dramatic changes appear. Massive star-forming galaxies along the so-called main sequence at these epochs tend to be thick, clumpy disks, forming stars at rates (100 $M_\odot$/yr) much higher than is observed in the thin, quiescent disks observed at z < 0.5.

Gravitational instabilities in gas–rich turbulent disks can produce dense "clumps" supported by cold streams of gas. These structures could eventually migrate into a bulge component over time scales < 10 dynamical times, or < 0.5 Gyrs. As the clumps merge, the growing spheroid/bulge will have a stabilizing effect on the disk, and eventually it will be massive enough (~20% of the galaxy stellar mass) to prevent further fragmentation. Constraining this self-regulating process requires accurate measurements of the ages of both the clumps and the diffuse stellar component.

*Figure 1: z=2.5 galaxy from Elmegreen et al. (2005). WFIRST+IFU will provide low resolution spectra over scales < 2Kpc, accurately allowing age-dating of the stellar population.*

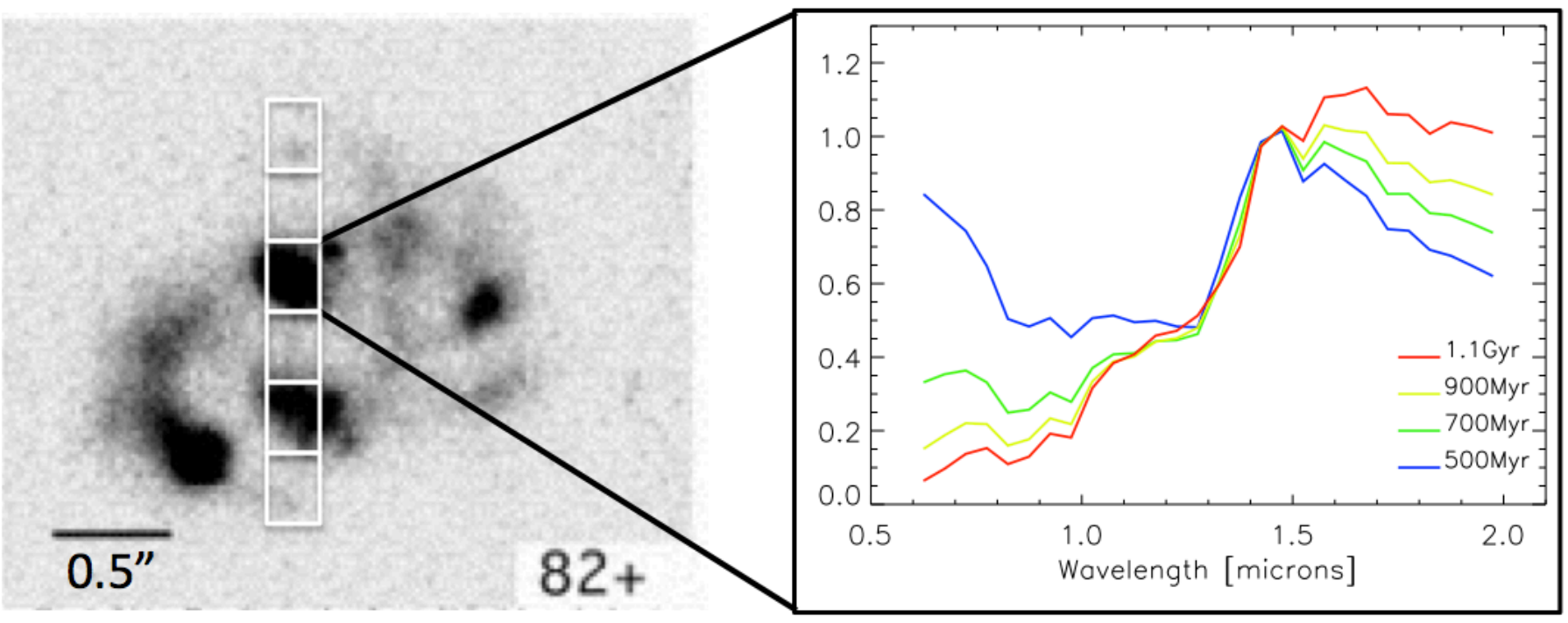

WFIRST

A wide-area NIR grism slitless survey with WFIRST will discover a large number of z~2 star-forming galaxies via the detection of their Hα and [OIII] emission lines. These objects are ideal targets for followup with the WFIRST IFU spectrometeter, which will provide rest-frame optical spectra on spatial scales <2 Kpc. The low-resolution IFU spectra will allow us to constrain the shape of the age sensitive part of the spectral energy distribution of the z~2 galaxies in a much greater detail than with broad band imaging. This, in turn, will allow for an accurate measurement of the spatially resolved stellar population properties, crucial to constrain galaxy formation models. Moreover, the stellar population maps resulting from the IFU data will be used as input to interpret the higher spectral resolution slitless grism data in an iterative process to obtain maps of emission-line intensities, which will be used to derive SFR and dust extinction surface density.

Key Requirements

Grism – To find z~2 emission line galaxies

IFU – depth & resolution needed to derive spatially-resolved stellar population properties

Michael Strauss (Princeton University), strauss@astro.princeton.edu

# Synergy Between LSST and WFIRST

Background

WFIRST and LSST were the two highest priority space and ground-based missions in the 2010 Decadal Survey. The primary mission of both facilities is to carry out wide-field imaging surveys of the sky in multiple bands; these surveys will enable a wide range of science, including weak lensing, large-scale structure in the distribution of galaxies, supernovae, the evolution of galaxies, and stellar populations. LSST is an optical telescope, and will do deep repeated broad-band photometry; the near-infrared and spectroscopic capabilities of WFIRST will allow much more accurate photometric and spectroscopic galaxy redshifts in regions of survey overlap.

WFIRST

The spectroscopic survey of WFIRST will allow redshifts to be measured for the brighter galaxies in the survey area. To go deeper will require photometric redshifts, which photometry in the near-infrared alone will not allow. LSST will survey 18,000 square degrees in the Southern Hemisphere in six filters (*u,g,r,i,z,y*) to a point-source depth of *r*~27.5. The overlap with WFIRST should be several thousand square degrees, in which we will have photometry from the *u*-band to H or K, allowing for excellent photometric redshifts. WFIRST image quality will allow galaxy shapes to be measured to fainter magnitudes than LSST can go; combining with the photometric and spectroscopic redshifts will allow much enhanced cosmic shear measurements as a function of cosmic time. The broad wavelength coverage will also allow accurate estimates of star formation rates, stellar masses, and dust content for galaxies at a wide range of redshifts, as well as for stellar populations in the Milky Way through SED fitting. The multi-band photometry will reveal and characterize rare populations of stars and galaxies, including "dropouts" at high redshift. LSST will visit any given point on the sky roughly 1000 times over ten years, giving detailed information on variability and proper motions. If the operations of WFIRST and LSST overlap in time, there will be an opportunity for combining the data streams to look for variability in both. LSST discoveries can also be followed up by WFIRST in its GO program.

Key Requirements

Depth – To match the photometric S/N in J/H/K for galaxies seen in LSST
Sky Coverage – To overlap the LSST footprint over at least 1000 square degrees
Resolution – Approach the diffraction limit with well-sampled images to measure galaxy shapes
Grism – For spectroscopic redshifts of large populations of galaxies.
Wavelength Coverage – >2 NIR filters to extend the galaxy SEDs to beyond the observed optical
Data Processing – Allow joint processing with LSST data to put photometric measurements on a common scale to optimize matching and cross-comparison of the two datasets.

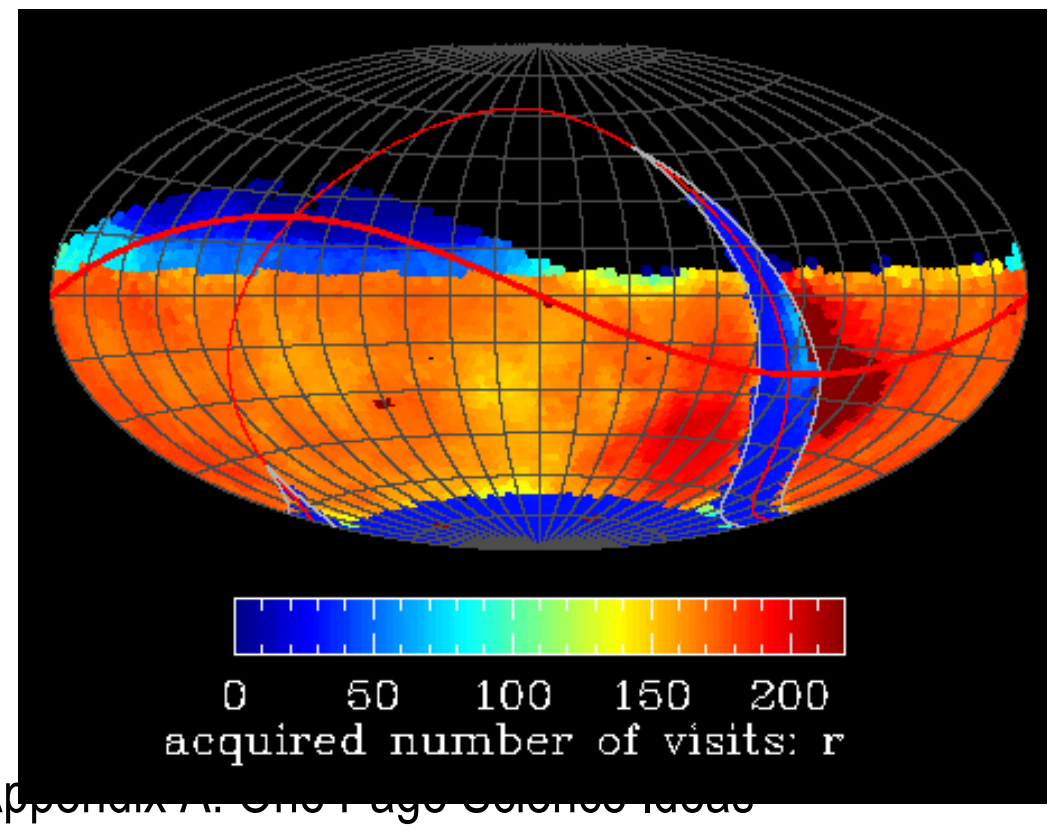

A realization of the sky coverage of LSST in the r-band, represented in Equatorial coordinates. The color represents the number of visits across the sky in a simulation of LSST operations: low Galactic latitudes, the Southern equatorial cap, and regions in the North along the ecliptic receive fewer visits. There is similar coverage in the u, g, i, z, and y filters. From the LSST Science Book (2009).

P. Capak (Caltech), capak@astro.caltech.edu

## Synergies between Euclid and WFIRST

The Euclid and WFIRST missions are complementary and have a number of synergies for both cosmological measurements and for extragalactic science. These are summarized below:

1) A large uncertainty for both Euclid and WFIRST are systematics and assumptions in the Weak Lensing (WL) and Baryon Acoustic Oscillation (BAO) measurements (e.g. Refregier et al. 2004 ApJ, 127, 310, Hirata & Eisenstein 2009, astro2010,127). Euclid and WFIRST probe different optimizations for these measurements, mitigating many of these systematic effects. In particular, the weak lensing measurements will use different detector technology and probe different wavelengths providing a truly independent measurement of the shear signal. In addition both the Lensing and BAO surveys will probe different galaxy populations with different biases, yielding different systematic effects in the measurements.

2) WFIRST will add 1-2um images of similar depth and resolution as those provided by Euclid in the optical (0.6-1um) enabling a range of galaxy evolution science. Star formation is known to drop precipitously at $z<2$ (Hopkins et al. 2006, ApJ, 651, 142), and this is believed to be a largely secular (internal) process best probed with large statistics (e.g. Peng et al. 2010, ApJ, 721, 193, Wuyts et al. 2011, ApJ, 742, 96). Studying what is causing this global decline in star formation requires resolved high-resolution rest frame UV measurements that probe the instantaneous star formation rate and resolved rest frame optical measurements to probe the accumulated stellar mass. WFIRST will not independently probe the rest frame Ultra-Violet (UV) at $z<\sim 1.8$ and Euclid has poor sensitivity and resolution to the rest-frame optical light at $z>0.7$, so both instruments will be required to conduct these studies.

At $z>4$ the global star formation rate density is rapidly increasing, peaking at $z\sim 2$ (Hopkins et al. 2006, ApJ, 651, 142). However, WFIRST alone will only be able to select and study these first galaxies at $z>8$ due to its lack of an optical channel, while Euclid will not have the near-IR sensitivity to probe "typical" systems at $z>5$. So, again both instruments are required to study these first epochs of galaxy formation.

3) If WFIRST includes longer wavelength (1.5-2.5um) coverage the spectra will complement the 0.9-2um spectra from Euclid. This increased wavelength coverage will expand the redshift range over which standard spectral line diagnostics (e.g. Kewley & Ellison 2008, ApJ, 681, 1183) can be used from $1<z<2$ with Euclid and $2<z<2.8$ with WFIRST to $1<z<2.8$ with the combined instruments. This redshift range is crucial because star formation is known to peak and start declining in the $1<z<3$ redshift range and the combination of Euclid and WFIRST will enable studies of ionization parameters, AGN content, and metalicity during this key epoch.

Paul Schechter (MIT), schech@mit.edu

# The Shapes of Galaxy Haloes from Gravitational Flexion

Background

Gravitational "flexion" is the next higher order gravitational lensing effect after shear (e.g. Bacon et al. 2006, MNRAS, 365, 414). Shear is proportional to second derivatives of the gravitational potential; flexion is proportional to third derivatives. Flexion causes otherwise symmetric galaxies to appear lopsided and curved. Flexions are greatest at relatively small distances from gravitational lenses – within roughly 10 Einstein radii. But while shear signals are degenerate with the much larger intrinsic ellipticities of lensed galaxies, the intrinsic lopsidedness and curvature of galaxies is quite small. A single flexion measurement can therefore give significant signal-to-noise. These properties make flexion a far better tool than shear for the study of the shapes of the dark matter haloes of galaxies (Hawken & Bridle 2009, MNRAS, 400, 1132; Er & Schneider 2011, A&A, 528, 52). But, this comes at the cost of deeper exposures. In order to take full advantage of the small intrinsic scatter in galaxy lopsidedness and curvature, one must obtain higher signal-to-noise.

WFIRST

The wider-versus-deeper question for shear measurements is quite subtle (e.g. Bernstein 2002, PASP, 114, 98), and it is even trickier for flexion measurements, but we suspect that the optimum depth will be somewhat closer to that achieved by WFIRST's supernova program than by its weak lensing program. Given the extraordinary ultimate power of flexion measurements – with uncertainties in the mean ellipticity of galaxy dark matter haloes of 0.00003 per steradian, according to Hawken & Bridle (2009) – flexion measurements in the supernova survey fields will, for the purpose of measuring galaxy halo shapes, be competitive with shear measurements over the much larger weak lensing survey.

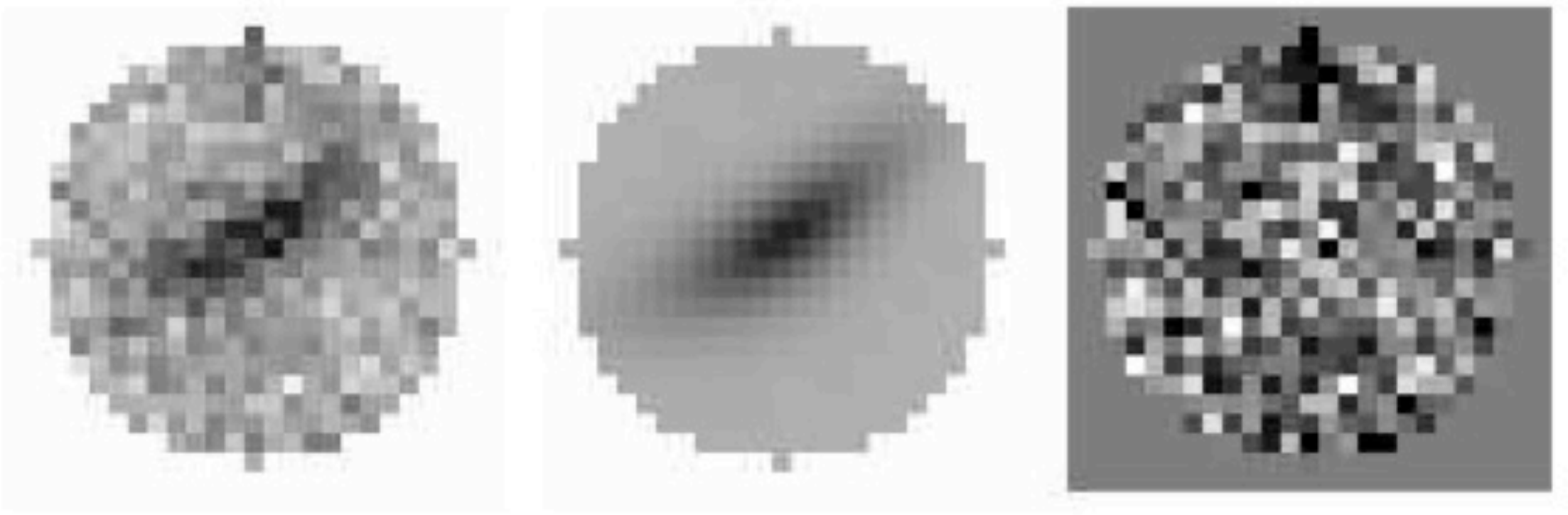

Caption – A 30 minute Magellan exposure of a flexed galaxy observed in the Sloan r' filter (left), an intrinsically elliptical galaxy model distorted by flexion (center), and the difference between the two (right).

Vandana Desai and Harry Teplitz, desai@ipac.caltech.edu, hit@ipac.caltech.edu

# WFIRST and IRSA: Synergy between All-Sky IR Surveys

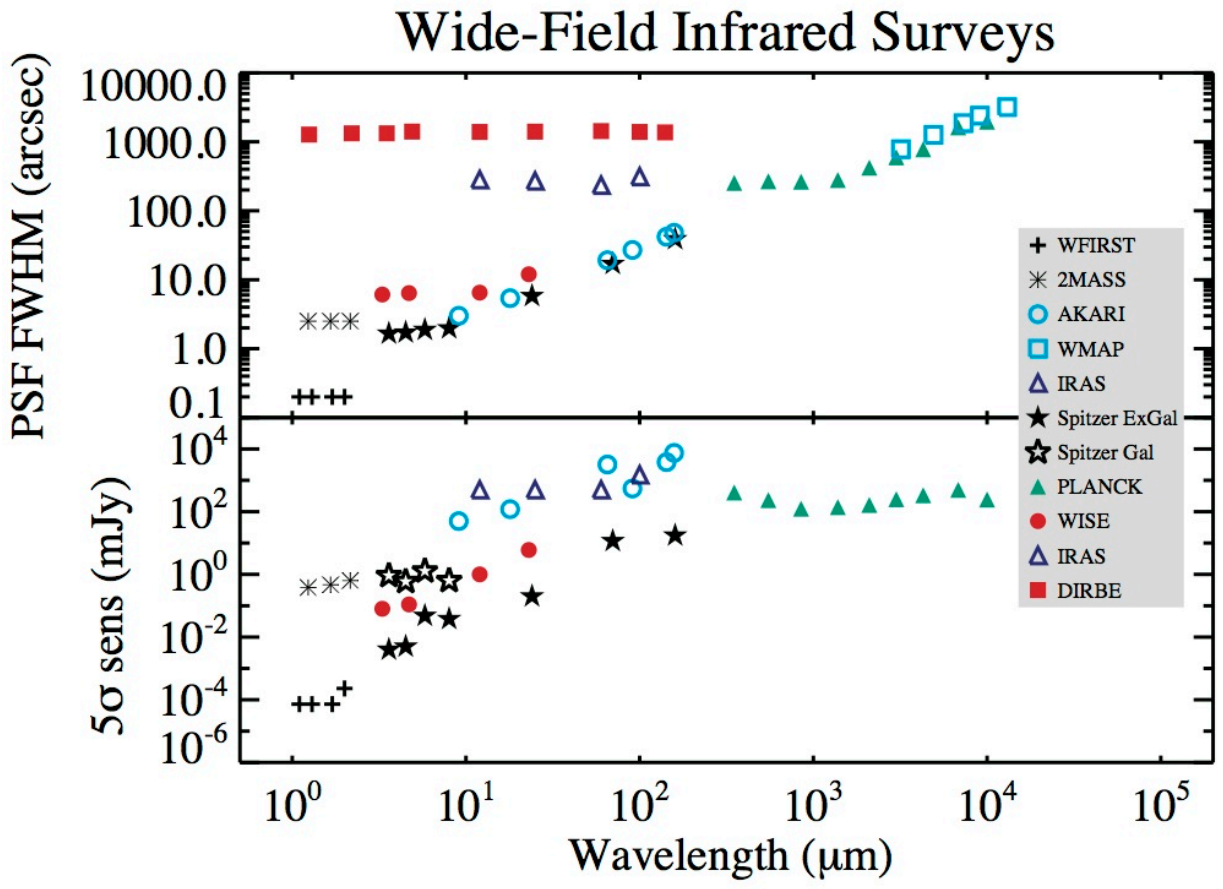

IRSA curates and serves calibrated data products from NASA's infrared and submillimeter missions, including Spitzer, WISE, Planck, 2MASS, and IRAS. In total, IRSA provides access to more than 20 billion astronomical measurements, including all-sky coverage in 20 bands. These holdings enable a wide range of astronomical research; here we describe some areas that particularly benefit from the synergy between existing IR all-sky surveys and WFIRST.

**Discovery Space:** IR surveys have revealed populations of astronomical sources that were previously undetected or not recognized at shorter wavelengths (e.g. ULIRGs, SMGs). All-sky surveys also allow the discovery of extremely rare sources. Mining existing long-wavelength IR surveys for interesting objects will be greatly facilitated by the WFIRST contributions of (1) deep near-IR imaging for improved color selection, and (2) high-quality photometric and spectroscopic redshifts to remove interlopers. The result will be complete, pure samples that can be followed up to support full multiwavelength analyses.

**Multiwavelength Analyses:** The wide-area and unprecedented sensitivity of WFIRST (and LSST) will in many cases entirely eliminate the need for dedicated follow-up imaging of sources identified at longer wavelengths. In cases where follow-up is needed via the WFIRST GO program, existing IR surveys will be an important resource not only for selecting objects, but also for planning observations.

Together with wide-field imaging at shorter wavelengths, all-sky IR surveys support multiwavelength analyses, mitigating selection biases and providing tighter constraints on physical models. Multiwavelength analyses rely on robust object associations across the electromagnetic spectrum. However, matching the LSST and WISE catalogs (for instance) will be complicated by their very different resolutions. WFIRST, matched in sensitivity to LSST, will be an important wavelength bridge for determining the optical properties of far-IR objects, or the far-IR properties of optically-selected sources.

**Morphologies:** Existing IR surveys have fairly low resolution and are unsuitable for determining the morphology of the detected infrared emission. WISE and Spitzer data have revealed many stars with infrared excesses, presumably from structures of gas and dust that may eventually evolve into planetary disks. WFIRST will determine the spatial distribution of a statistically significant fraction of these structures. WISE and Spitzer, along with submillimeter surveys, have successfully uncovered rare classes of high redshift (z=2) galaxies. WFIRST will determine morphologies for many of these galaxies, elucidating their role in the formation of quasars and today's most massive galaxies. In both of these examples, WFIRST will yield larger samples than would be practical with JWST.

**Time Domain:** In addition to the multiple epochs planned within the WFIRST mission, WFIRST can be used in conjunction with WISE and UKIDDS to (1) discover and study moving objects; and (2) measure proper motion distances to the closest stars and brown dwarfs.

## Near Infrared Counterparts of Elusive Binary Neutron Star Mergers

Mansi Kasliwal (Carnegie), Samaya Nissanke (Caltech), Christopher Hirata (Caltech)

Background

Within the next decade, ground-based gravitational wave (GW) detectors are expected to routinely detect the mergers of binary compact objects (neutron stars and black holes). The NS-NS and NS-BH mergers contain ultra-dense ($>10^{14}$ g/cm$^3$) matter and hence provide a laboratory for some of the most extreme astrophysical processes in the Universe. Electromagnetic observations, sensitive to the composition and thermodynamic state of the matter, will be required to complement the GW signal, which constrain the NS and BH masses. Except in the lucky case of a gamma ray burst that happens to be beamed toward us, we will have to rely on isotropic electromagnetic counterparts. Tidal debris or accretion disk winds are expected to expel neutron-rich material from the merger resulting in an isotropic radioactivity-powered remnant. Dubbed "kilonova", the counterparts should be detectable in the optical/NIR with durations of order days and luminosities of order $10^{40}$—$10^{42}$ erg/s. Follow-up of binary mergers will be challenging: GW detectors lack intrinsic angular resolution, and simultaneous fitting of an event in several detectors can only localize a source to an error ellipse of a few to hundreds of deg$^2$. This ellipse will contain not just the binary merger counterpart, but also many unrelated "false positive" transients. The eventual goal is to obtain not just a detection, but also a light curve and spectroscopy to constrain the velocity, opacity, and composition of the debris.

WFIRST

Theoretical predictions for the counterpart suggest that the emission peaks in the NIR due to opacity considerations (Barnes and Kasen 2013). Hence, the exquisite wide-field mapping speed, red sensitivity and rapid response capability of WFIRST in geosynchronous orbit are well-suited to this search. Specifically, we suggest following up binaries localized within 6 deg$^2$ and detection S/N>12 by a network of 5 GW detectors. During a 5-year mission, we expect ~30 such binaries within the WFIRST field of regard at a median distance of 230 Mpc (Nissanke, Kasliwal and Georgieva 2012). At this distance, a depth of $H_{AB}$=24.6 mag corresponds to $M_{H,AB}$=−12.2 or $\lambda L_\lambda = 6\times10^{39}$ erg/s. The rate of NS-NS mergers is currently uncertain by 1—2 orders of magnitude: the follow-up strategy and criteria for a ToO trigger will be adjusted based on the actual event rate. One possible follow-up plan, requiring a total of 27 hours per binary (including overheads), would involve 2 filter (J+H to mag 24.6 AB at S/N=10) and grism (S/N=5 per synthetic R=30 element at $H_{AB}$=22) observations of the entire error ellipse as quickly as possible. After this, we envision imaging mode (J+H) revisits at 4 epochs from t ~ 1—30 days, and a later reference epoch. Once the counterpart candidate list is reduced to ≤10 candidates based on the grism spectrum and the photometric evolution, we would use the IFU to acquire a high S/N spectrum (S/N=5 per synthetic R=30 element at mag 24.9 AB).

Key Requirements

Wide-field, high sensitivity NIR camera for counterpart search and detection
Field of regard – only events in an accessible region of sky (59% for WFIRST) can be followed up
Target of opportunity capability, rapid response time (several hours)
Spectroscopy – grism (for prompt, blind spectra) + IFU (deep spectra of candidate counterparts)

# Wide Field Infrared Survey Telescope

## *Synergistic Science Programs between WFIRST-2.4 and JWST*

Version 1.0
Mar 2013

http://wfirst.gsfc.nasa.gov/

This appendix contains a rich set of 10 potential synergistic science programs that are uniquely enabled by WFIRST-2.4 observations prior to or coincident with JWST. The papers were put together mostly by STScI staff, with some contributions from other astronomers, and the SDT wishes to thank the community for this valuable contribution. The one pagers highlight the tremendous potential of WFIRST to advance many of the key science questions formulated by the Decadal survey. These programs have not been vetted by the SDT, and we are not endorsing these specific studies. Additionally, the authors did not see the final version of the WFIRST mission capabilities prior to their submissions, so inconsistencies may exist.

Title: Scientific synergies obtained by the simultaneous operation of WFIRST-NRO and JWST

Author: John W. MacKenty, Space Telescope Science Institute

Contributors: Martha Boyer[1], Larry Bradley[2], Dan Coe[2], Harry Ferguson[2], Karl Gordon[2], Paul Goudfrooij[2], Jason Kalirai[2], Knox Long[2], Peter McCullough[2], Marshall Perrin[2], Massimo Stiavelli[2], and Tommaso Treu[3]
(1=GSFC, 2=STScI, 3=UCSB).

Summary: This White Paper considers a set of scientific investigations which require the availability of observations from WFIRST-NRO prior to or coincident with JWST observations. These scientific synergies offer a substantial increase in the scientific return from both missions and will explore a range of pressing and compelling questions ranging from the origins of our Universe, the nature of dark matter, the assembly of galaxies, the fates of stars, and the properties of exoplanets.

Background: This White Paper assumes a WFIRST mission built upon a NRO 2.4 meter telescope with a core scientific capability of near-infrared imaging comparable in sensitivity and resolution to HST/WFC3-IR over a field of view of order 0.25 square degrees. These investigations assumes wavelength coverage from 0.8 to ~2 microns and some investigations require or would benefit from extending coverage (but not spatial sampling) to 0.4 microns, from the inclusion of a richer set of visible and medium band filters, or, in one instance, a visible light high resolution camera/coronagraph.

Science Investigations:

1. Studies of the earliest galaxies (Dan Coe and Larry Bradley)
   - Large solid angle survey with WFIRST to discover the most luminous sources at $10<z<20$ using the Lyman break drop-out technique that is used by HST to discover $z>11$ galaxies.
   - JWST NIRSPEC and MIRI detailed spectroscopy enabled by the discovery with WFIRST-NRO of sufficiently bright sources yields clues to ionization, metal growth, and possibly masses of these infant galaxies.
2. Finding the first stars via high-z pair instability supernovae (Massimo Stiavelli)
   - Discovery of PISN from zero metallicity first generation stars at $z>10$ to explore the first generation of stars, their initial mass function, and the initial enrichment of the ISM.
   - PISN have 1-2 year timescales and would be observable with WFIRST to z~15-20. Their expected source density of 1-2 per square degree requires WFIRST for discovery and JWST to determine the properties of their host galaxies. Working in the infrared, WFIRST will (post-SWIFT), have an unparalleled capability to discover rare $z>6$ sources (e.g. GRBs and supernovae) due to the ultraviolet opacity of the IGM prior to reionization.
3. Probing dark matter via lensed QSOs (Tommaso Treu)
   - Measurement of dark matter sub-halos down to $10^6$ Msun will test CDM versus self-interacting or warm dark matter models.
   - WFIRST has the unique capability to discovery 1000s of quad lensed QSOs in a large solid angle sky survey which requires HST resolution. JWST/MIRI is then required to measure flux ratios of the lensed sources at mid infrared wavelengths to avoid stellar microlensing.
4. Progenitors of supernovae and other highly variable objects (Knox Long)
   - Within 50 Mpc, where significant resolution is possible to identify individual stars or their environments, we expect 50 core collapse SN and 15 SN1a per year.

- JWST follow-up of these nearby supernovae will be significantly enhanced if pre-detonation images permit the identification of the type of progenitor star. WFIRST surveys at HST resolution over a broad range of wavelengths provide pre-detonation images of nearby galaxies.

5. Globular clusters in nearby galaxies (Paul Goudfrooij)
   - Massive star clusters trace early star formation history and galaxy assembly.
   - Wide Field, high resolution and sensitive NIR photometry is required to select samples outside the local group. JWST/NIRSPEC MOS spectroscopy is required to obtain metallicities to derive star formation timescales.

6. AGB stars and dust production in the local volume (Martha Boyer & Karl Gordon)
   - AGB stars are major producers of dust, especially carbon, in the ISM. Understanding both the mechanisms of dust production and its implications for the infrared properties of galaxies requires improved understanding of this short lived phase of stellar evolution.
   - Wide field near infrared observations with WFIRST will establish a sample of AGB stars suitable for detailed follow-up with spectroscopy and longer wavelength observations using JWST.

7. Galactic streams in nearby galaxies (Harry Ferguson)
   - Tidal streams from the merging of dwarf galaxies into nearby large galaxies provide clues to the assembly history of these galaxies and a key test of CDM models.
   - WFIRST has the unique capability to discover and map streams surrounding ~100 massive galaxies within 5 Mpc. JWST will then provide the deepest imaging and infrared spectroscopy to yield constraints on the ages, kinematics, and chemical abundances of the streams via observations of individual stars.

8. Selecting planets for exoplanet atmospheric characterization (Peter McCullough)
   - TESS should discover 100s of exoplanet systems which WFIRST could monitor to obtain planetary masses via the transit timing variation method. The most suitable candidates could then be observed during transits with JWST's broad suite of instruments to obtain detailed information on their atmospheres.

9. Simultaneous observations of exoplanets with WFIRST-NRO and JWST (Marshall Perrin)
   - The inclusion of a white light coronagraphic instrument with an integral field spectrometer on WFIRST-NRO opens the possibility of simultaneous observations with the JWST coronagraphs to achieve broader spectral coverage.
   - Studies of exoplanet atmospheres, indications of weather and seasonal variations, and (with the inclusion of a polarimeter capability within the WFIRST coronagraph), detailed measurements of scattered light from circumstellar dust in exoplanetary systems will be possible.

10. Resolved Studies of Galaxy Formation (Jason Kalirai)
    - Measuring the surface brightness, abundance gradient, velocity dispersion, star formation history, and substructure in the halos of nearby Milky Way type galaxies provides a view into their formation history, and an input to test models of galaxy formation.
    - Observations with WFIRST-NRO will reveal starcount maps for nearby galaxies, and can be used to locate a pristine population of halo stars. Follow up observations with JWST can be used to directly measure turnoff ages and ground based 30-meter telescopes to measure stellar velocities.

1. <u>Studies of the earliest galaxies</u> (Dan Coe and Larry Bradley)

The first galaxies likely formed ~ 100 – 400 Myr after the Big Bang (z ~ 11 – 30) contributing to reionization and the end of the "dark ages", a process likely complete by ~ 900 Myr (z ~ 6). The nature of these early galaxies and the degree to which they contributed to reionization are the current frontier outstanding questions of extragalactic astronomy. The Hubble and Spitzer Space Telescopes have yielded candidates with robust photometric redshifts as distant as z ~ 10.8 or 420 Myr (Coe13) and perhaps z ~ 11.9 (Ellis13, though the authors urge caution due to the single-band detection). The former is gravitationally lensed into multiple images, the brightest magnified by a factor of ~8 to AB mag 25.9, enabling more detailed follow-up study. Significant samples of high-redshift candidates have been discovered back to ~600 Myr after the Big Bang (~100 candidates at z ~ 8: Bradley12, Oesch12, Ellis13). Deep Hubble imaging has yielded the fainter, more numerous population, while wider, shallower Hubble surveys as well as very wide field (~square degree) ground-based surveys have discovered the rarer relatively brighter galaxies. No survey has yet achieved the area, depth, and wavelength coverage required to yield significant numbers of z > 9 galaxies within the universe's first 500 Myr.

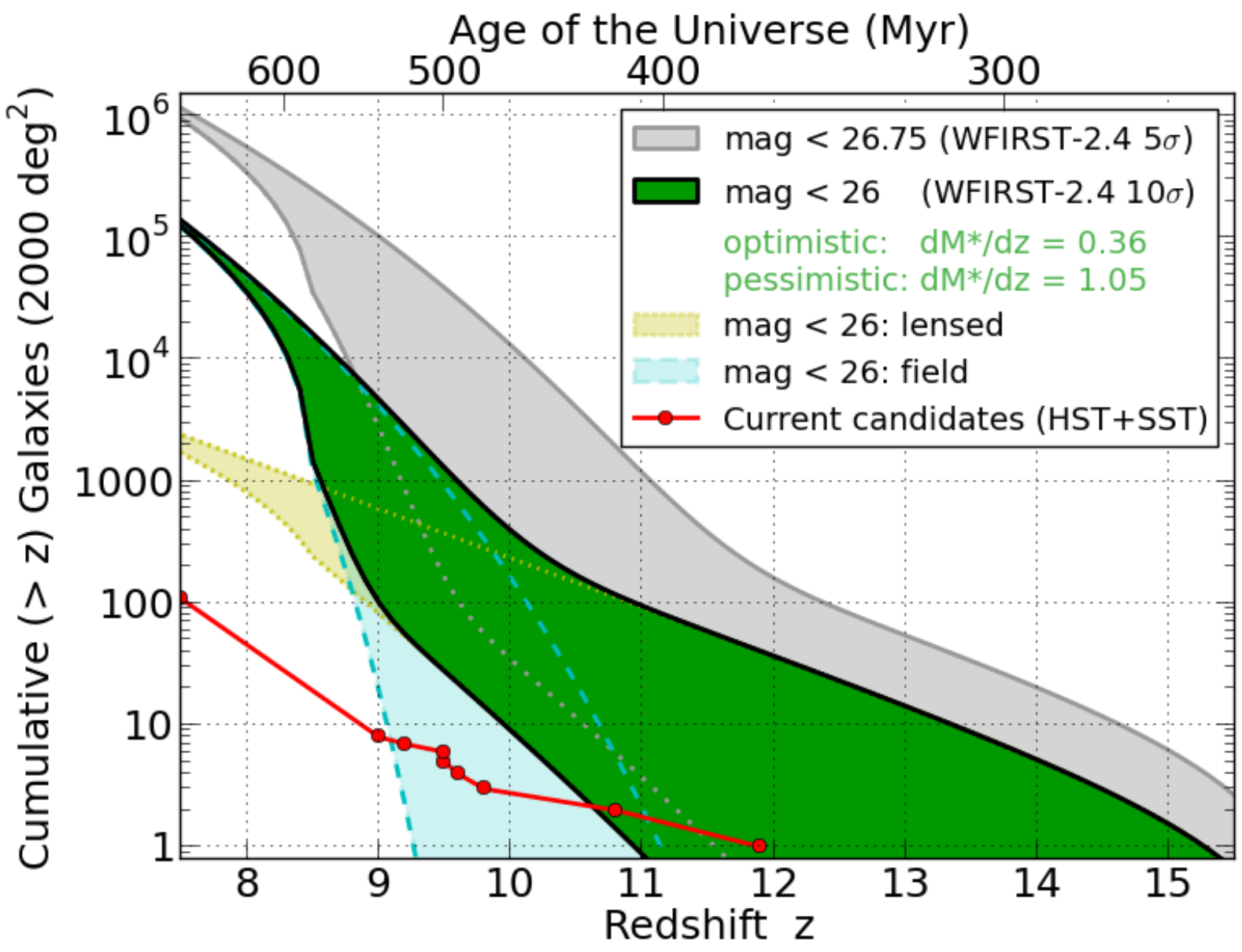

The wide and deep near-IR survey made possible by NRO WFIRST will transform our understanding of the early universe by yielding ~100,000 high-redshift (7.5 < z < 15) candidates bright enough (AB mag < 26) for follow-up study, that could allow for detailed stellar population studies, metallicities, and kinematics with the James Webb Space Telescope and large ground-based telescopes. The figure at right shows a range of expectations for cumulative (in redshift) numbers of high-redshift galaxies in a 2,000 square degree survey. This is based on current measurements at z ~ 8 (Bradley12) extrapolated to higher redshifts both optimistically and pessimistically based on constraints from the handful of ~8 current z > 9 candidates (Bouwens11, Zheng12, Coe13, Bouwens13, Ellis13, Oesch13). Contributions from gravitationally lensed galaxies are estimated by roughly assuming one strongly lensing cluster per square degree and adopting an average lens model from the CLASH multi-cycle treasury program (Postman12). Most ten-sigma detections (green) will be properly identified, but pushing down to five-sigma detections will result in >50% incompleteness relative to the numbers shown in gray. Key questions to be addressed by these observations are: (1) how rapidly did early galaxies build up; (2) what was their contribution to reionization; (3) were early galaxies primarily composed of pristine Population III stars?

2. Finding the first stars via high-z pair instability supernovae (PISN) (Massimo Stiavelli)

A central goal of the JWST mission is the observation of the first stars and galaxies. Actually observing a single population III star may actually be possible when it dies as a pair-instability supernova. Such supernovae are very luminous but also very rare. WFIRST-NRO provides the most viable means of discovering such objects as this requires deep large area near infrared surveys with high angular resolution and good photometric precision over long timescales. JWST would then provide detail studies of both the SN and its host galaxy.

Present day star formation mechanisms rely upon gas cooling via metals and dust to condense gas into stars. As these heavy elements originated in the first generation of stars (Population III) stars, those stars are expected to have formed via different and very uncertain mechanisms which likely resulted in a relatively higher proportion of very high mass stars. Such stars can end their lives as Pair Instability SuperNovae (PISN) with extremely high luminosities that visibly brighten their host galaxies. At $z>10$, time dilation will make these SN visible for periods of several years requiring long term (but infrequent) monitoring with deep images to discover PISN.

PISN have been discussed by Weinmann and Lilly 2005 (ApJ, 624, 526). They find SN rates in the range 0.2-4 sq deg per year at redshift 15-25. These rates are in some sense even an upper limit for the following reasons: (1) recent studies have found even lower masses for Pop III stars largely due to some evidence for fragmentation. If the mass function extends down to few tens of solar masses and at the same time the upper mass cutoff goes down to 300 Msun or even much lower, the range of masses giving rise to PISN shrinks and they become even rarer. (2) The full impact of negative feedback (especially radiative but also chemical, i.e. If there is enrichment you don't form Pop IIIs) have been underestimated until the last 5 years or so (e.g. Trenti and Stiavelli 2009, ApJ, 694, 879). There is a lot of uncertainty on Pop III from $H_2$ and even higher uncertainty on those from atomic H. While 2 PISN per sq deg per year is a moderately optimistic but not unreasonable rate if fragmentation is rare, it would be hard to sustain that rate with the amount of fragmentation that some are finding now.

3. Probing dark matter via lensed QSOs (Tommaso Treu)

Measuring the mass function of subhalos down to small masses (e.g. 1e6 Msun) is a unique test of cold dark matter and of the nature of dark matter in particular. The CDM mass function is supposed to be a powerlaw rising as $M^{-1.9}$ until Earth-like masses. A generic feature of self interacting dark matter and warm dark matter is to introduce a lower mass cutoff. Current limits set that cutoff at somewhere below 1e9 Msun. We know from MW problems that there is a shortage of luminous satellites at those masses, but that's never going to be conclusive because subhalos can only be detected with traditional methods if they host stars.

One powerful way to detect subhalos independent of their stellar content is via the study of large samples of strong gravitational lenses, in particular the so-called flux ratio anomalies (see Treu 2010, ARA&A, 48, 87 and references therein). Small masses located in projection near the four images of quadruply-lensed quasars cause a strong distortion of the magnification and therefore alter the flux ratio. However, anomalous flux ratio can also be caused by stellar microlensing if the source is too small. The only way to avoid microlensing is to go to wavelengths such as mid-IR where the lensed quasar emission is sufficiently extended. Large sample of thousands of lenses are needed to achieve sensitivity down to 1e6-1e7 solar masses required to probe the nature of dark matter. Unfortunately, at the moment only 3 dozen quads are known, and only a handful of those are bright enough at mid-IR to be observable from the ground (e.g., Chiba et al. 2005, ApJ, 627, 53). JWST-MIRI is so powerful that it will be able to observe any quad discovered in the next decade with integration times of order seconds to minutes. However, JWST will be limited for this application by the number of known quads. Ground based surveys before JWST launches will probably discover of order 100 quads, still short of the number require to fully realize its potential as a dark matter experiment.

Discovering thousands of quads will required a close to full sky survey at HST-like resolution (Oguri & Marshall, 2010, MNRAS, 405, 2709), i.e. WFIRST. The ability to resolve and therefore discover quads is a very strong function of resolution and therefore WFIRST will be unmatched in its ability to find them. Of course if WFIRST comes after JWST, those will not be available for mid-IR follow-up at JWST resolution.

4. The progenitors of SNe and other highly variable objects (Knox Long)

JWST will carry out detailed studies of SNe in nearby galaxies, particularly since its IR capability will allow it to probe the late phases of a SN explosion to determine the nature and distribution of the ejecta and to understand the conditions in SNe that lead to dust creation (see, e.g. Barlow 2009). However, a crucial aspect of understanding SNe is to connect a SN explosion to its progenitor star. The combination of WFIRST and JWST are crucial to rapid progress in this area.

The importance of high quality imagery for the advancement of this subject has been made clear with HST. Indeed, some HST imagery exists for about 25% of the SNe (and luminous blue variable outburst) exploding in galaxies within 28 Mpc (Smart et al 2009). (Over about a decade, 50% of the SNe actually occurred in galaxies that were observed with HST, half occurred outside HST's small FOV) HST images have resulted in the identification of at least 4 progenitors, including discovery of the binary progenitor for SN1993J (Maund et al 2004) and enabled studies of the local stellar populations in the remainder. As a result, we know that the most common type of core collapse SNe (type II-p) arise from red supergiants and that the minimum and maximum masses for these SNe is about 8 Msol and 17 Msol (Smart et al. 2009).

But WFIRST, with its large FOV, will do much better. For example, it be quite straightforward to create a uniform treasure archive of nearby galaxies at distances within, say 50 Mpc, that extended out well beyond the D25 contours. About 50 core collapse SNe and about 15 Ia SNe are expected in this volume every year. It would also be possible with a little effort to survey larger galaxies, up to and including M31 and M33. Not only would this enable some of the science associated with the PHAT multi-cycle treasury program of M31 (which an allocation of 830 orbits with HST covers only a portion of the galaxy), but it would also ensure that when the next SN goes off in one of these galaxies, that pre-exposure images would be able to identify the progenitor down to (depending on crowding) a mass of about 3.5 Msun (Dalcanton 2012). One could also imagine shallower surveys to map the entirety of the SMC or LMC down to considerably lower mass limits.

Furthermore, the needs for pre-explosion images are not just limited to the progenitors of SNe, but includes a wider range of other objects: the eta-Car like explosions of luminous blue variables, supernova imposters, such as SN2008S and optical transients such as those seen in M85 and NGC300. All of these objects are likely candidates for JWST observations. JWST would be used not only for spectroscopic observations during the explosion that would characterize the SNR, but, if the WFIRST mission were complete, would carry out the NIR imagery after the SN had faded to find out which object had been disrupted as a result of the SN explosion.

5. Near-field cosmology and galaxy evolution using globular clusters in nearby galaxies (Paul Goudfrooij)

The increasing realization that the study of star clusters has direct relevance for the basic processes involved in how galaxies assemble and evolve over time has placed this field at the forefront of extragalactic research in recent years. A fundamental, yet unresolved, question in the formation of early-type galaxies is the assembly and chemical enrichment histories of their 'halo' and 'bulge' components which are represented by the metal-poor and metal-rich GC subpopulations, respectively. Working together, WFIRST-NRO and JWST can make unprecedented progress in solving this question.

a) Globular Clusters as Fossil Records of the Formation History of Galaxies

Star formation plays a central role in the evolution of galaxies and of the Universe as a whole. Studies of star-forming regions in the local universe have shown that star formation typically occurs in a clustered fashion. Building a coherent picture of how star clusters form and evolve is therefore critical to our overall understanding of the star formation process. Most star clusters disrupt within a few Gyr after they form, building up the field star population in the process. However, the most massive and dense star clusters remain bound and survive for a Hubble time. These globular clusters (GCs) constitute luminous compact sources that can be observed out to distances of several tens of megaparsecs. Furthermore, star clusters represent the best known approximations of a "simple stellar population", i.e., a coeval population of stars with a single metallicity, whereas the field stars in galaxies typically constitute a mixture of populations. Thus, studies of GC systems can constrain the *distribution* of stellar ages and metallicities whereas measurements of the integrated light of galaxies can only provide luminosity-weighted averages of these key quantities. Consequently, GCs represent invaluable probes of the star formation rate and chemical enrichment occurring during the main star formation epochs within their host galaxy's assembly history (see, e.g., review of Brodie & Strader 2006).

The study of extragalactic GCs was revolutionized by the Hubble Space Telescope (HST). The main reason for this is that the size of GCs is well-matched to diffraction-limited optical imaging with a 2-m class telescope: a typical GC half-light radius of ~3 pc at a distance of 15 Mpc corresponds to 0.05 arcsec on the sky, which is roughly the diffraction limit (and detector pixel size) for HST at 600 nm. This yields very high quality photometry of GCs relative to ground-based imaging by beating down the high galaxy surface brightness in the central regions of galaxies. Furthermore, it also allows robust measurements of globular cluster radii, and hence of their dynamical status.

Notwithstanding the important progress that HST imaging has facilitated in this field, there is one critical property of GC systems that HST imaging *cannot* address well. GC systems around massive nearby early-type galaxies extend far into the galaxy halos, covering several tens of arcminutes on the sky (e.g., Goudfrooij et al. 2001; Zepf 2005), while the FOV of HST images is only 3.3 × 3.3 arcmin$^2$. This is illustrated in Figure 1. Obviously, wide fields of view (> 20' per axis) are required to accurately determine total properties of GC systems (e.g., total numbers of clusters per unit galaxy luminosity, color or metallicity distributions, trends with galactocentric distance). Furthermore, the faint outer halos of galaxies are thought to hold unique clues regarding the early assembly history of galaxies, and bright GCs constitute one of the very few probes that can be studied at high S/N in these environments.

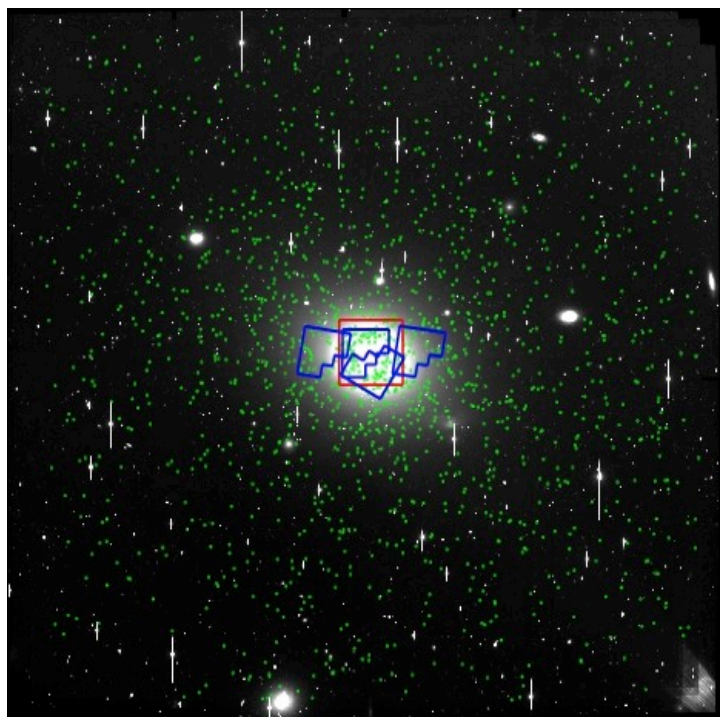

**Figure 1: R-band KPNO 4-m/MOSAIC image of the giant elliptical galaxy NGC 4472 in the Virgo cluster of galaxies, covering a 36' × 36' FOV. Footprints of available HST/ACS and HST/WFPC2 images are drawn in red and blue, respectively. Globular cluster candidates from Rhode & Zepf (2001) are indicated as green dots. Note the small fraction of globular cluster candidates covered by HST images, implying the need for large and uncertain extrapolations when trying to extend conclusions from the HST studies to the full systems of globular clusters. Figure taken from Zepf (2005).**

b) New Constraints on the History of Star Formation and Chemical Enrichment of Early-Type Galaxies

A key discovery of HST studies of GC systems of luminous galaxies was that their optical color distributions are typically bimodal (e.g., Kundu & Whitmore 2001; Peng et al. 2006). Figure 2 shows an example. Follow-up spectroscopy of bright GCs using 10-m-class telescopes indicated that both ``blue'' and ``red'' populations are typically old (age > 8 Gyr), implying that the color bimodality is mainly due to differences in metallicity (e.g., Puzia et al. 2005). In broad terms, the metal-rich GC population features colors, metallicities, radial distributions, and kinematics that are consistent with those of the spheroidal (`bulge') component of early-type galaxies. In contrast, the metal-poor GC population has a much more radially extended distribution, and is likely physically associated with metal-poor stellar halos such as those found around the Milky Way and M31 (e.g., Bassino et al. 2006; Goudfrooij et al. 2007; Peng et al. 2008).

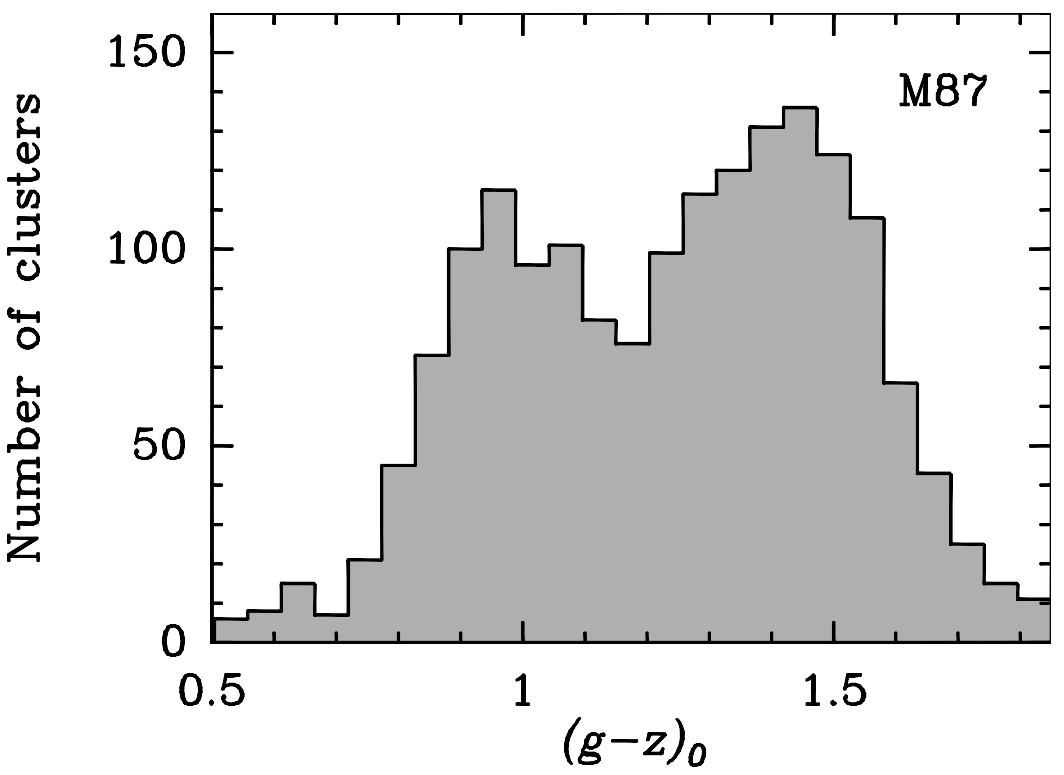

**Figure 2: *g−z* color distribution of globular clusters in the massive elliptical galaxy M87 from Peng et al. (2006). Note the obvious color bimodality, which has been confirmed to be mainly due to differences in metallicity, and which is common among massive early-type galaxies in the local universe.**

The bimodality in optical colors of GCs constitutes one of the clearest signs that star formation in luminous early-type galaxies must have been episodic. However, we emphasize that the optical color distributions do not significantly constrain *when* these events occurred, or in what order. This is because optical colors alone cannot generally distinguish between different combinations of age and metallicity (the ``age-metallicity degeneracy''). A general understanding of the age and metallicity distributions of GC systems requires braking this degeneracy. There are two primary and complementary ways to do this.

**(1) The combination of optical and near-IR photometry**. The main power of this method (using color-color diagrams) is the ability to identify age differences (of order 25% for high-quality data), due to the fact that near-IR colors are primarily sensitive to metallicity while optical colors (e.g., *r-z*) are sensitive to both age and metallicity. This approach resulted in the identification of substantial populations of intermediate-age metal-rich GCs in the inner regions of several early-type galaxies (Goudfrooij et al. 2001; Puzia et al. 2002; Hempel et al. 2007; Georgiev et al. 2012). The current limitation of this method is twofold. While HST has a powerful near-IR channel in its WFC3 instrument, its use is limited to the *innermost regions* of nearby galaxies due to its relatively small footprint of 2' × 2' (cf. Fig. 1 above). The NIRCam instrument to be installed on the 6.5-m James Webb Space Telescope (JWST) will reach 2 mag fainter than HST in a given integration time, but its footprint is similarly small. Conversely, while near-IR imaging instruments with reasonably large fields of view are starting to become available on large ground-based telescopes (e.g., 7.5' × 7.5' for HAWK-I on the VLT), contamination of GC candidate samples by compact background galaxies is a major concern for ground-based spatial resolution (see, e.g., Rhode & Zepf 2001). As demonstrated by HST, imaging at 0.1 arcsec resolution effectively eliminates this concern due to the marginally resolved nature of GCs (cf. Section 1). Thus, the study of galaxy formation and evolution by means of accurate GC photometry will benefit tremendously from space-based wide-field imaging in the 0.6 – 2 μm range. A relatively simple multi-chip Optical-NIR camera installed on one of the two 2.4-m NRO telescopes would be ideal for this (and many other) purpose(s). Their fast (f/1.2) primary mirror could easily yield a useful field of view of hundreds of square arcminutes per exposure at a resolution of 0.1 arcsec, providing accurate photometry of virtually *all* GCs associated with nearby galaxies with very little contamination. Such an instrument would not only place crucial constraints on the assembly history of massive early-type galaxies, it would also allow the selection of the best targets for follow-up multi-object spectroscopy to infer their chemical enrichment history (see below).

**(2) Follow-up Space-Based Multi-Object Spectroscopy**. The main strength of this technique lies in the presence of intrinsically strong absorption lines of several key elements in the 0.7 – 2.5 μm region (e.g., O, Mg, Al, Si, Ca, Fe), which facilitates accurate determinations of overall metallicities and element abundance ratios that can be used to infer typical timescales of star formation (e.g., Valenti et al. 2011). Currently, this technique is only available from the ground and is therefore significantly hampered by the high surface brightness of the diffuse light of the inner regions of the host galaxies. In practice, this limits the application of this technique currently to the outer regions of galaxies. This has caused a general lack of crucial spectroscopic information for the metal-rich GCs, which are located mainly in the inner regions. While future developments in the area of adaptive optics systems on large telescopes will enable high spatial resolution imaging and spectroscopy from the ground, they will do so only over a small (< 1') FOV which is not useful for spectroscopy of extragalactic GCs. This science can however be expected to advance dramatically with the advent of NIRSpec on the JWST with its multi-slit array and high-efficiency medium-resolution gratings.

6. AGB stars and dust production in the local volume (Martha Boyer & Karl Gordon)

When low- to intermediate-mass stars (0.8 < M < 8 Msun) begin to ascend the asymptotic giant branch (AGB), pulsations levitate material from the stellar surface and provide density enhancements and shocks, encouraging dust formation and re-processing (e.g., Bowen 1998; van Loon et al. 2008). This dust is subsequently released to the interstellar medium via a strong stellar wind driven mainly by radiation pressure on the grains. The composition of the dust depends on the atmospheric chemistry (the abundance of carbon relative to oxygen) which is altered by dredging up newly formed carbon to the surface of the star (the 3rd dredge-up; Iben & Renzini 1983). The efficiency of the 3rd dredge up (which depends on the stellar mass) and the initial metallicity of the star determine whether or not a star will ultimately become a carbon star, with metal-poor stars becoming carbon-rich more easily owing to a lack of oxygen available to tie up the newly dredged-up carbon into CO molecules.

While all stars between 0.8 and 8 solar masses will pass through the AGB phase, the phase itself is short, making AGB stars rare. Despite this, they may be among the most important contributors of dust in galaxies. The most significant known dust producers are AGB stars and supernovae (SNe) and it is unclear which dominate the dust production in the Universe since the dust-production rate from SNe is highly uncertain. However, even if assuming the most generous estimates for SNe dust production, the AGB stars come out as at least equally important over the lifetime of galaxies like the Magellanic Clouds (e.g., Matsuura et al. 2009; Boyer et al. 2012). AGB stars are certainly the most important contributors of carbon dust, as their total dust input is dominated by the carbon stars and SNe are thought to produce mainly silicates.

The infrared flux of galaxies is often used to infer the total masses and star formation rates of galaxies. Since AGB stars are among the brightest sources and radiate strongly from 1-20 microns, they can significantly affect these measurements. In the Magellanic Clouds, AGB stars account for up to 30% of the 3-4 micron flux, despite being < 5% of stellar population (Melbourne & Boyer 2013). Without accounting for the AGB stars, stellar masses measured in the near-IR could be biased too high, especially in galaxies with large intermediate-aged populations.
Despite their importance, very little is known about the details of the evolution and dust production in AGB stars. AGB stars in galaxies more distant than the Magellanic Clouds have been difficult to observe at IR wavelengths owing to limits in sensitivity and resolution, leaving us with observations of a handful of stars in a random smattering of nearby galaxies. In the Magellanic Clouds, the Spitzer SAGE programs (Gordon et al. 2011; Meixner et al. 2006) have provided an incredible wealth of new information about AGB stars with their fully complete photometric catalogs in the IR, but IRS spectra was limited to only the brightest AGB stars even in these nearby galaxies. AGB stars in the Milky Way have been difficult to identify, leaving us with samples biased towards only the most extreme examples. Optical surveys do not suffice for AGB observations, as up to 60% of a galaxy's AGB population can be obscured by circumstellar dust (Boyer et al. 2009).

The James Webb Space Telescope will go a long way towards understanding these important stars owing to its incredible sensitivity in the IR. For the first time, whole populations of AGB stars can be observed both photometrically and spectroscopically in the IR within the complete suite of galaxy environments available in the Local Volume, potentially shifting our understanding of AGB stars and AGB dust production in profound ways. Since the lifetime of JWST is limited, it is crucial that we identify AGB stars in the Local Volume before, or at least in tandem with the JWST mission so that IR spectra can be obtained for an unbiased sample of stars. The most efficient way to identify AGB stars is to use near-IR photometry since the spectral energy distributions of most AGB stars peak from 1-3 microns, and since circumstellar dust does not obscure the stars at these wavelengths in the vast majority of cases (95% of the AGB stars in the LMC are detected in the K-band with 2MASS data).

Near-IR photometry also has the potential to separate carbon-rich from oxygen-rich AGB stars, depending on the exact filters chosen. This is due to molecular absorption in AGB atmospheres (e.g., TiO, CN, $C_2$) in the 0.5-2 micron region. The J-K color successfully separates C from O stars in the Magellanic Clouds with 85% confidence (Riebel et al. 2012), allowing for the construction of a representative and un-biased sample of sources for follow-up spectrosco-py. The near-IR wide filters on HST/WFC3, on the other hand, do not separate C-rich from O-rich stars, making it difficult to calibrate stellar evolution models.

With filters similar to 2MASS J, H, and K, and with resolution and sensitivity similar to HST, WFIRST could identify all AGB stars of interest within the Local Volume, allowing for the most efficient usage of JWST.

7. Galactic streams in nearby galaxies (Harry Ferguson)

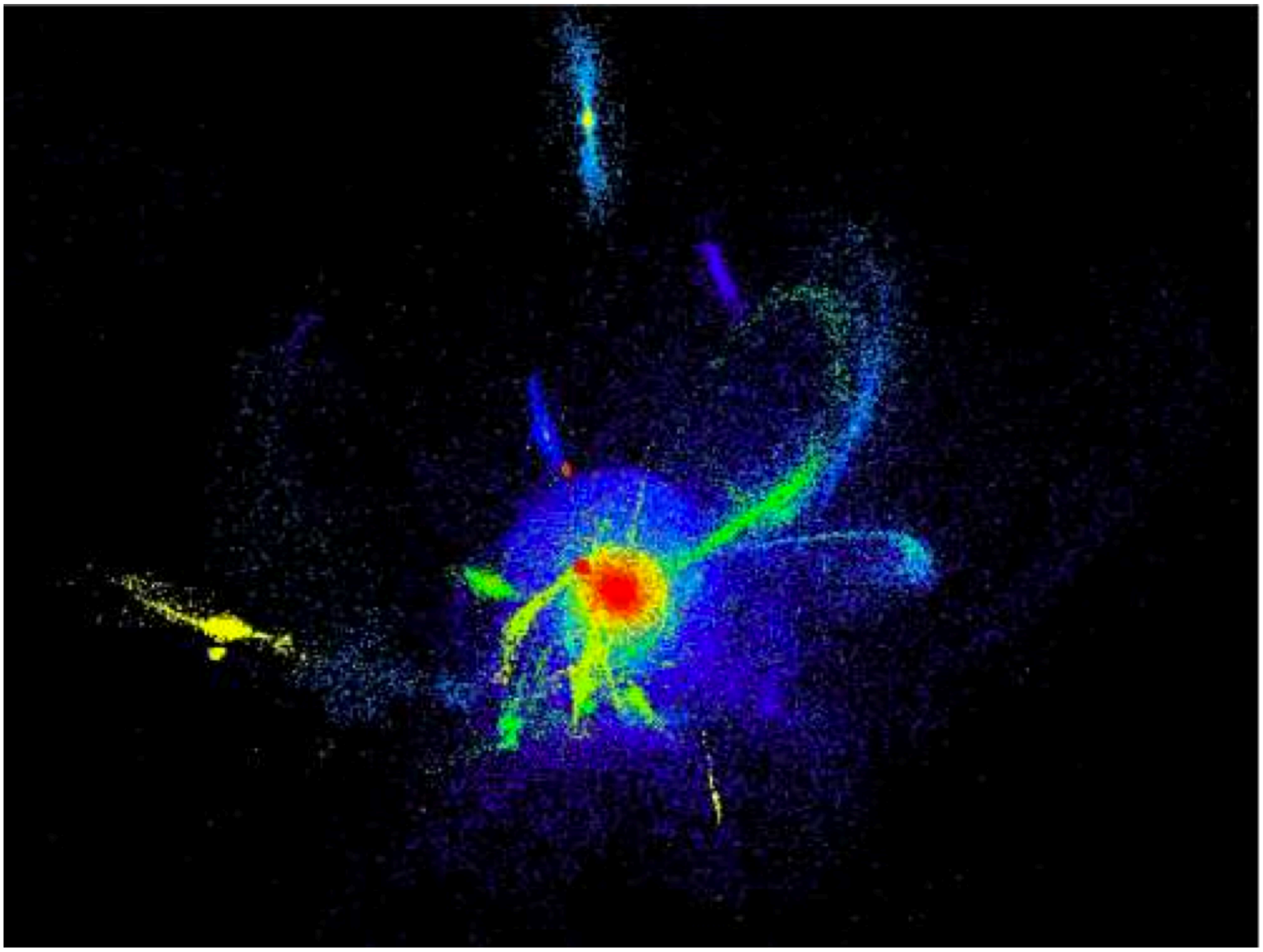

**Figure 1: Simulations of tidal streams around a z=0 elliptical galaxy; colors code stellar age (from Bournaud 2010 ASPC, 423, 177).**

WFIRST's wide-field capabilities, low background, and high spatial resolution will revolutionize the study of streams of stars around nearby galaxies, formed by the tidal destruction of infalling dwarf galaxies. Outside of the Local Group, these streams are typically below the surface-brightness limits for detection from ground-based telescopes, but can be revealed with high-resolution observations, where individual RGB stars can be distinguished from background galaxies. Numerical simulations of galaxy formation suggest that there are streams around every L* galaxy.

The TRGB absolute magnitude is $M_J$ = -5.67-0.31[Fe/H] (Valenti 2004, MNRAS, 354, 815). With observing times of an hour or two per galaxy, WFIRST can detect streams down to TRGB + 3 mag around all the 100 most massive galaxies within 5 Mpc. With observing times per galaxy of 10-100 hours, WFIRST can detect streams in galaxies to the distance of the Virgo cluster. WFIRST will provide breakthrough statistics on the number and morphology of the streams and the colors of their stars. The dissolution of dwarf galaxies in the halos of massive galaxies depends sensitively on (1) their dark-matter contents, (2) star-formation and feedback in the dwarf galaxies before they enter the massive halo and (3) interaction of the gas in the dwarf galaxies with the gas in the more massive halo. Item (1) is a fundamental assumption of LCDM models, but is currently hard to reconcile with all of the observed properties of satellite dwarf galaxies. Items (2) and (3) are very poorly constrained in galaxy-formation models; they provide the wiggle room for LCDM to survive the current discrepancies. So observations of streams will be very important. The same observations, of course, will pick up low-surface-brightness dwarf galaxies, which will also be important.

JWST does not have the field of view to carryout out such imaging studies, but has the spectroscopic sensitivity to follow up these observations. A rough rule of thumb is that JWST NIRSpec can obtain an R=100 spectrum of any star detected by WFIRST, with the same exposure time that WFIRST used to detect the star. This will improve the constraints on the metallicities of the streams, and, by doing so, help to constrain the ages of the streams. At higher resolution, R=2700, JWST, with observing times of 10-100 hours for galaxies within 5 Mpc, can obtain kinematics and more precise chemical abundance measures of individual elements. Phase-space densities of the streams will provide strong constraints on the mass and shape of the parent dark-matter halo (e.g. Penarrubia et al. 2006, ApJ, 245, 240), and may also allow tests of alternative models of gravity (e.g. Kesden 2009, Phys Rev. D. 80, 83530.)

The combination of JWST and WFIRST greatly exceeds the individual contribution of each mission. WFIRST is the finder scope. It obtains broad-brush statistics on the prevalence of streams. JWST probes the details that are necessary to really constrain the physics.

8. Selecting planets for exoplanet atmospheric characterization (Peter McCullough)

This discussion assumes that the WFIRST-NRO telescope is agile, i.e. can point all over a large part of the sky without much overhead. So, for example, to take a 1 minute picture in each of half of the 88 constellations on the sky, the total wall-clock time to be not hugely larger than 44 minutes, say 2 hours, or an overhead of ~50%. It's not important for these programs what the overhead fraction is, so long as 1-min exposures don't cost the program ~1 hour each, as they would for HST or JWST.

Generically speaking, the WFIRST-NRO will be an excellent time-series photometer for monitoring transits (or eclipses) of exoplanets. Compared to HST, the WFIRST-NRO will not have the problem of Earth occultation for ~45 minutes every 96 minutes that chop up transit observations, and the WFIRST-NRO could be more flexible in its ability to schedule observations of specific timed events like transits, again because of HST's LEO. However, for many of the programs described below, the WFIRST-NRO is similar in capability to HST, and the greatest potential advantage of WFIRST-NRO could be that it is operational, if HST is not, e.g. due to HST's old age.

Planetary masses and radii are the most basic characterizations we can perform on an exoplanet. Traditionally, radii come from transits, and masses come from stellar radial velocities, RVs. For small (low-mass) planets and long period orbits, the RVs can be much less than 1 m/s and are difficult, if not impossible to measure, even with a perfect spectrograph because of intrinsic stellar RV jitter.

a) Single-planet systems

The comparison of the IR transit light curves from the WFIRST-NRO with the visible (CCD) light curves of Kepler will also refine and validate the radii of those planets. The comparison of WFIRST-NRO (IR) observations of secondary eclipses of exoplanets with Kepler (Visible) observations will disentangle the reflected - and thermally- emitted light from each planet. This implies measurements of the temperature of the exoplanet's "surface" although for the technique to be practical, the exoplanet typically must be much hotter than the ~300K habitable zone and its surface area must be relatively large compared to its host star for the planet/star contrast to be measurably large.

b) Multi-planet systems

Kepler has found that a good fraction of small transiting planets come in pairs or multiples. The gravitational perturbations of one planet on another can enable their respective planetary masses to be determined without stellar radial velocities RVs (although those sometimes help, if available) so long as precise transit light curves are obtained, such that the ingress and egress times can be measured precisely. Currently, the best photometry comes from Kepler, a broadband 1-m telescope in space; however, Kepler's 4-year mission extension will last only until 2016 (TBR). The WFIRST-NRO could be pointed at Kepler-discovered multi-planet systems around the times of predicted transits, to monitor transit timing variations, TTVs, in order to refine the masses of those planets. Importantly, the precision of mass measurement (via TTVs) improves quadratically with the baseline of time of the TTVs: i.e. twice as long a baseline implies four times greater precision.

TESS will discover a similar number of transiting planets as Kepler (thousands), but TESS's planets, by design of the TESS mission, will orbit much brighter stars, enabling characterizations of the TESS planets, especially their atmospheres, that is impractical with Kepler-discovered exoplanets. The WFIRST-NRO could provide validation of TESS exoplanets much like SST has done for Kepler exoplanets, via IR light curves of transits and eclipses. However, a penalty TESS pays for its all-sky coverage is that its longest dwell time (at the ecliptic poles) is ~1 year, and for most of the sky, TESS dwells only for ~1 month. The latter implies that of exoplanets that TESS discovers, and those with periods of a few months, a large fraction will have only a single transit observed by TESS. So could WFIRST-NRO help? The following is a quick estimate. The numbers could be inaccurate. Also, while the following paragraph addresses how WFIRST-NRO could assist in finding a 2nd transit, there will be a similar number of TESS – discovered multi-planet systems in which the outer planet already has two transits discovered, and WFIRST-NRO could jump straight to the much-less costly (in WFIRST-NRO time) subsequent monitoring at specific times to measure TTVs.

TESS will discover a few dozen transiting planets of special interest: small, rocky-sized multi-planet systems around very bright stars, with the larger orbital periods from 3 months to 6 months, but which will have only 1 transit observed by TESS of the outer planet (and too small an RV signature to pursue it that way). If the WFIRST-NRO could bounce around the sky from one TESS-selected star to another, snap a picture in order to detect another transit when it occurs, and return to each star quickly enough to be sure not to miss the transit, that would be very useful. An alternative approach would be to fly an armada of small (~10-cm) purpose-built telescopes, each one monitoring one TESS selected star at a time. Or TESS could re-observe the sky in an extended mission beyond its 2-year baseline mission. Would the WFIRST-NRO do this program even if it could? Not sure yet, but for reference, Spitzer stared at a single star GJ 1214 continuously for 20 days, and HST observed HD 17156 for 10 days continuously, so perhaps the WFIRST-NRO might observe ~50 stars continuously for ~180 days in order to firmly establish their orbital periods and pave the way for subsequent monitoring at specific times in order to measure TTVs and hence planetary masses (and radii). Then JWST could observe the scientifically "best" ones from that set for atmospheric characterization.

In the last scenario, the telescopes must work in series in this order: 1) TESS discovers hundreds of exoplanets, 2) for the multi-planet systems, WFIRST-NRO measures some of their masses by TTVs, and hence helps identify the "scientifically best" exoplanets for continued follow up, and 3) JWST measures those "best" planets' atmospheric properties.

9. Simultaneous observations of exoplanets with WFIRST-NRO and JWST (Marshall Perrin)

Motivated by recent dramatic growth in studies of exoplanets, there is significant community interest in support of including a very high contrast coronagraphic camera as a second science instrument alongside the wide-field instrument on WFIRST-NRO. Such a coronagraph would operate at contrasts up to 1e9 in visible light (likely at least 0.6-1 micron total range with instantaneous spectral coverage ~ 10-20%, observed with an integral spectrograph with resolution R~50-100). This would enable low-resolution imaging spectrophotometry of Jupiter-mass planets around nearby stars, as well as extremely sensitive observations of scattered light from circumstellar dust. The inner working angle of this coronagraph would be well matched to JWST's coronagraphs, of order 0.2", with the smaller telescope diameter being offset by the shorter operating wavelengths.

We highlight here three specific areas of synergy between an WFIRST-NRO coronagraph and the coronagraphs on JWST: (a) broad spectral coverage studies of atmospheric properties and chemistry in exoplanets, (b) the exciting possibility of time domain observations of weather or seasonal changes in exoplanetary atmospheres, which could benefit substantially from simultaneous operation of the two observatories, and (c) characterization of the complete range of bodies in nearby systems, including dusty debris belts and analogs to the still-poorly-understood Fomalhaut b.

a) Understanding Exoplanet Atmospheres.

Brown dwarfs and giant planets have atmospheric (photospheric) properties that change dramatically as they cool over time, due to complex processes such as cloud formation and on-equilibrium atmospheric chemistries that are not yet well understood (e.g. Fortney & Nettelmann 2010). While sample sizes remain small, initial indications from e.g. the HR 8799 planets and 2M1207 suggest that planetary-mass companions have colors distinctly different from brown dwarfs, hinting at significantly different physics. The complexity of atmospheric chemistries imprints itself in atmospheric spectra in a variety of ways, including for instance absorption and scattering features from clouds of H2O and NH3, and also the effects of high altitude photochemical hazes (e.g. Cahoy et al. 2010). In the case of Jupiter, such photochemical hazes are one of the dominant contributors to atmospheric opacities, causing its overall reddish color (cf. Marley et al. 1999). Unlike transiting planets, directly imaged planets are (at least potentially) accessible to spectral characterization across a broad range of wavelengths spanning the bulk of their spectral energy distributions, offering a path to interrogate atmospheric properties through observations in both scattered light and thermal emission. Current spectral models for exoplanet atmospheres are subject to degeneracies when fitting only near-infrared data; optical spectrophotometry (0.6-1.0 microns) can break these degeneracies and aid in unambiguous retrieval of atmospheric parameters. For instance, a Cycle 19 HST program (PI: Barman) has recently obtained WFC3 0.9 μm observations of HR 8799 to allow robust tests of cloud models that are not possible on the basis of near-IR data alone. The combination of a high contrast optical coronagraph on WFIRST-NRO with the near- to mid-infrared capabilities of JWST would enable extending such studies to the large sample sizes needed to robustly characterize extrasolar planet populations.

Advanced adaptive optics optics systems on ground-based 8-10 m (and potentially larger) telescopes will likely be competitive with JWST's contrast performance at near-IR wavelengths (1-2 microns), but atmospheric effects will in general prevent very high contrast (1e9 or better) observations from the ground, and will certainly preclude such observations at wavelengths substantially less than a micron even on timescales of 10-20 years from now. A very high contrast coronagraph in space is the only route to such data.

b) Weather and Seasons on Exoplanets

In addition to characterizing the bulk properties of exoplanets, we seek to understand them as dynamical systems, worlds in their own right. In several cases for close-in transiting planets, time-resolved observations have enabled a rough mapping of variations in brightness and temperature across their surfaces, providing insights into global circulation and weather patterns (e.g. Knutson et al. 2009, 2011; Cowan et al. 2009). Likewise, observations of brown dwarfs in the field indicate that substantial fraction (perhaps 10-20%) show variability on short time scales indicative of weather (Radigan et al. 2011); in some case the variability is as high as 25% in the J band. Initial models suggest that this variability can be explained by patchy clear regions in an otherwise cloudy atmosphere, possibly indicating holes or breaks in the cloud layer. (Radigan et al. 2012)

Recent modeling has shown that appears feasible to extend this technique of time-resolved mapping to direct imaging observations, and use the rotational modulation of exoplanets to recover their properties, including potentially surface properties, cloud cover variations and weather. See Kostov & Apai 2012 for a detailed discussion, and also Cowan and Strait 2013. These will be technically challenging observations but the potential payoff is large. This technique would strongly benefit from simultaneous studies with both JWST and WFIRST-NRO to allow multi-wavelength characterization of any variations and better wavelength coverage to discern spectral signatures of spectral components. While additional study of optimal observing techniques is needed, initial simulations suggest greater modulation at shorter wavelengths (Kostov & Apai 2012), indicating that WFIRST-NRO can be a key contributor here. Because of the time variability of these effects, simultaneous observations are required with both telescopes if broad spectral models are to be fit.

Depending on mission lifetimes, it may also be possible to observe seasonal changes. For instance, over the years of the Cassini mission to Saturn, we have witnessed the seasonal transformation of the northern hemisphere from pale blue to straw yellow (and the inverse in the southern hemisphere), believed to be driven by haze production due to increased ultraviolet illumination with the changing angle of incidence of sunlight. The potential extended mission lifetimes of JWST or WFIRST-NRO are comparable to a significant fraction of an orbital period for Jovian planets on 3-5 AU orbits. Maximizing the temporal overlap between the two missions might improve our ability to detect such seasonal changes.

c) Studies of Planetary System Architectures as traced by Small Bodies

A coronagraphic instrument on WFIRST-NRO could also potentially provide polarimetric measurements of light scattering from both planets and circumstellar dust - crucial because JWST lacks any polarimetry capabilities whatsoever. (In fact, several of the leading coronagraph concepts require polarization beam splitting for optimal operation, for instance the vector vortex coronagraph, so polarimetric capabilities may arise naturally as an inherent part of the starlight suppression methods.) Studies of solar system planets and theoretical modeling of exoplanets have demonstrated that polarimetry is a key diagnostic for assessing the presence of clouds or hazes in planetary atmospheres, so polarimetry could help with the above mentioned studies. But moreover, measurements of polarization can provide key constraints on dust grain properties in circumstellar disks (e.g. Graham et al. 2007, Perrin et al. 2009). When combined with observations of dust thermal emission in the mid-infrared with JWST and submillimeter with ALMA, scattered light imaging and polarimetry will enable comprehensive characterization of grain distributions, and therefore constraints on parent planetesimal populations in nearby debris disks.

Consider also the case of Fomalhaut b (Kalas et al. 2008), now confirmed by recent STIS coronagraphy to definitely be a physical companion on a disk-crossing-orbit, but whose physical nature remains very unclear. It has been suggested to possibly be a dust cloud from a recent collision of two KBO-type objects, or perhaps a giant planet surrounded by a swarm of moonlets in a collisional grinding process (See Currie et al. 2012 and Galicher et al. 2012). Having found one such object already with HST, it seems likely that further examples exist and may be found by JWST. Combined optical and infrared measurements, especially with optical polarimetry, would be a compelling tool to determine the physical nature of Fomalhaut b and similar objects. (Indeed, if the ACS HRC coronagraph was still operational, it would today be near mandatory to obtain ACS polarimetry of Fomalhaut b). Because JWST lacks polarimetric capabilities, a WFIRST-NRO coronagraph including polarimetry would provide an extremely valuable adjunct capability that would enhance our ability to characterize small bodies and dust grains. When combined with infrared observations of giant planets and their atmospheres, JWST and WFIRST-NRO together hold the potential to achieve comprehensive understanding of nearby planetary systems spanning from giant planets to asteroid and debris belts.

10. Resolved studies of galaxy formation (Jason Kalirai)

The detailed study of nearby galaxies can impact our fundamental knowledge of the processes that shape the formation and assembly of galaxies. The signatures of early accretion events can be directly seen in the halos of galaxies, where recent star formation is non-existent and stellar orbital time scales are of order the Hubble time. Perhaps the best example of a fully resolved spiral galaxy's halo is that of M31, as shown in the Figure below (McConnachie et al. 2009). From this star count map, we can measure the surface brightness profile, the level of substructure, and the abundance gradient of the stellar halo, and directly compare these signatures to those predicted in modern N-body simulations of galaxy formation (e.g., Bullock & Johnston 2005).

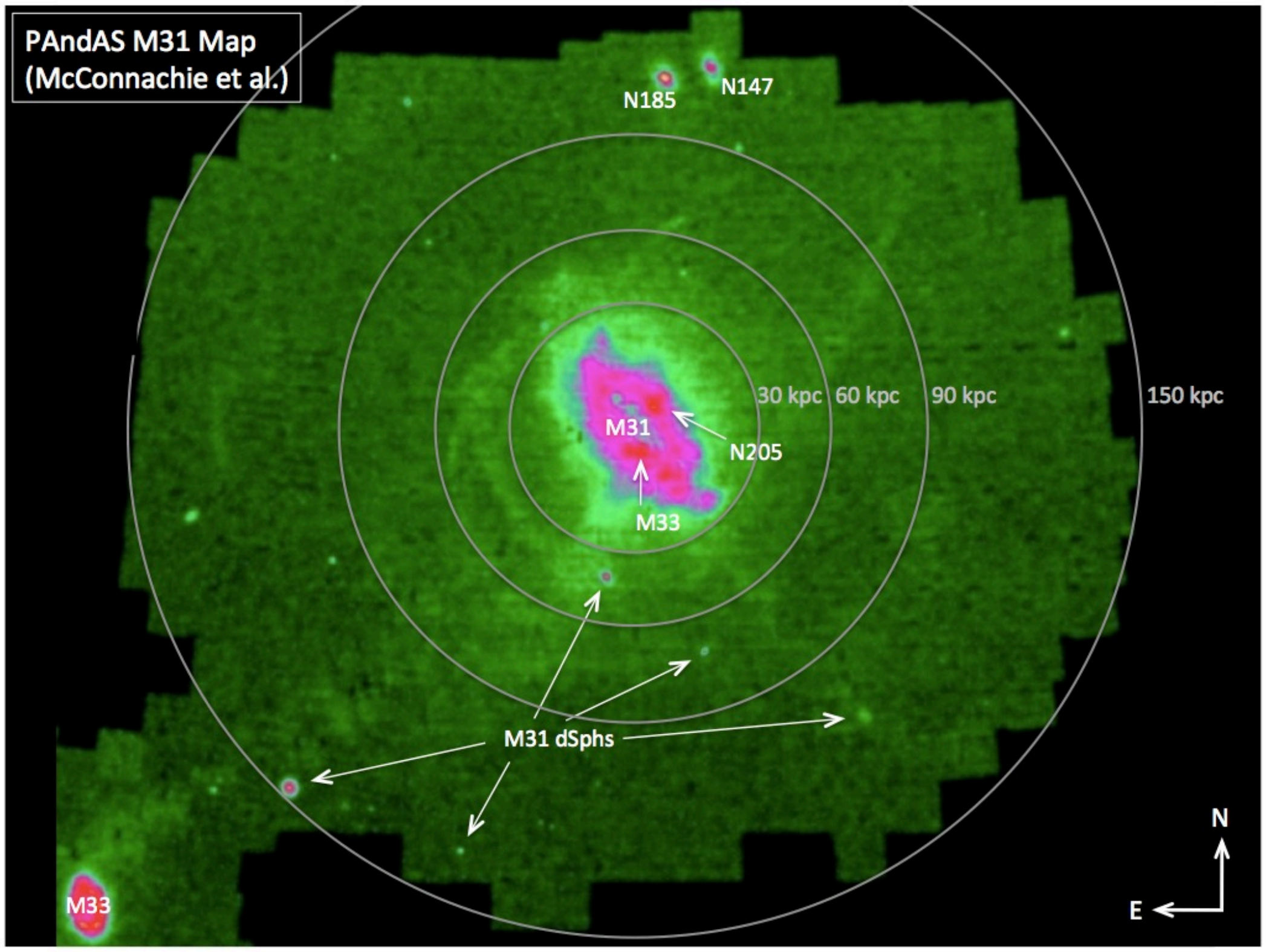

Unlike our map of M31, most nearby galaxies contain a patchwork of high precision data, often focused on the visible bulge and disk. Improving this situation is difficult since it requires a telescope with

1.) high throughput to measure the stellar tracers in old stellar population, red giants, out to several Mpc,
2.) high resolution to provide robust morphological separation between faint extragalactic contaminants and stars in the sparse halos,
3.) wide field of view to probe stellar distributions to out beyond 100 kpc in a limited number of pointings

WFIRST satisfies all of these requirements. With a modest investment of telescope time of about 1 month, WFIRST can survey several magnitudes of the red giant branch in 100 of the closest galaxies over their complete spatial extents. Such a survey would provide the means to study the diversity of stellar halos in detail, and to correlate their properties with galaxy size, luminosity, environment, and other factors.

Despite this wealth of new information, WFIRST, like Hubble, will be limited to measuring direct ages for old stellar populations to galaxies within the Local Group. The main-sequence turnoff in a 13 Gyr population is too faint to measure with a 2.4 meter telescope beyond the distance of Andromeda. JWST's NIRCam and NIRISS instruments can easily do this, out to several Mpc. However, blind pencil beams will run the risk of encountering recent accretion events and therefore not revealing the age of the oldest component of the galaxy, and so the JWST observations need WFIRST to establish the spatial maps beforehand. The JWST observations can be then fine tuned to explore quiescent areas of the stellar halos, with the appropriate density to reveal robust star formation histories.

Taken together, the combination of star formation histories (JWST), level of substructure (WFIRST), abundance gradients (WFIRST and JWST), surface brightness profiles (WFIRST) and velocity dispersions (30-meter) could be established for a dozen galaxies, and would provide an unprecedented data set to test N-body models of galaxy formation and evolution.

## Appendix C WFIRST-2.4 Dark Energy Measurement and Forecast Methods

### C.1. The Supernova Survey

Figure C-1 shows the number of SNe expected in each $\Delta z = 0.1$ bin of redshift. The drops at the redshift boundaries of the shallow and medium tiers are evident. In the deep tier we choose a fixed number N=136 in each redshift bin for IFU follow-up, from the larger number discovered in the imaging survey, to keep the time for IFU spectra within the time allocated to the supernova survey. The total number of predicted SNe is 2725, with a median redshift z = 0.7.

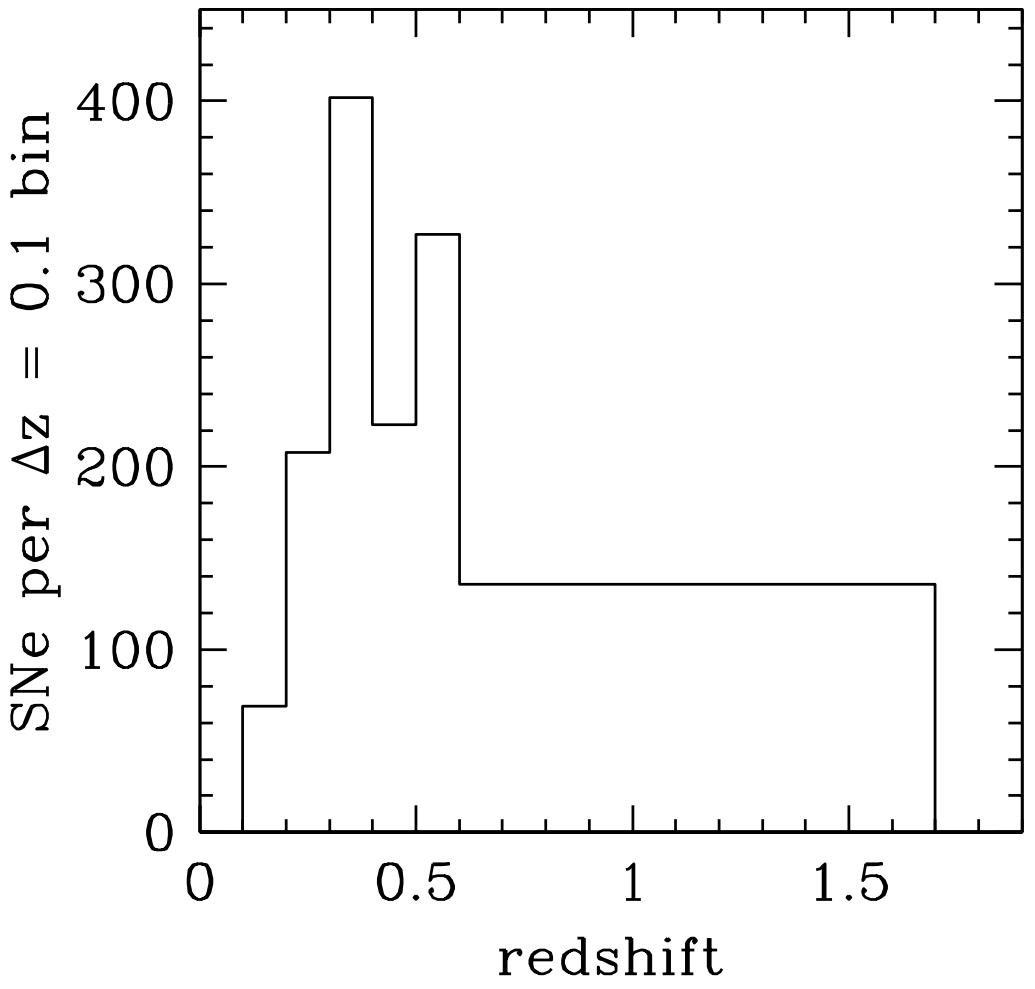

**Figure C-1: Expected number of Type Ia SNe in each Δz = 0.1 redshift bin. For z > 0.6 there are, by design, 136 SNe followed up with spectroscopic observations in each bin. The total number of SNe is 2725.**

To guide the design and forecast the performance of the supernova survey, we have adopted the following error model. The photometric measurement error per supernova is $\sigma_{meas} = 0.08$ magnitudes based on seven IFU spectra with S/N=15 per filter band and one deep IFU spectrum with S/N=47 per filter band. We assume an intrinsic dispersion in Type Ia luminosities of $\sigma_{int} = 0.08$ mag (after correction for light curve shape and spectral properties); this is at the low end of current estimates, but studies of nearby supernovae that use full spectral time series (which the IFU spectroscopy will provide) achieve values in this range. The other contribution to statistical errors is gravitational lensing magnification, which we model as $\sigma_{lens} = 0.07 \times z$ mags. The overall statistical error in a $\Delta z = 0.1$ redshift bin is then

$$\sigma_{stat} = [(\sigma_{meas})^2 + (\sigma_{int})^2 + (\sigma_{lens})^2]^{1/2} / \sqrt{N_{SN}},$$

where $N_{SN}$ is the number of SNe in the bin. We assume a systematic error per bin of

$$\sigma_{sys} = 0.01\,(1+z) / 1.8 \text{ mag},$$

with no correlation of errors between redshift bins. This corresponds to the "optimistic" systematics case from the Green et al. report[1] because we expect the IFU spectrophotometry to minimize systematics associated with photometric calibration and K-corrections and to reduce evolutionary systematics.

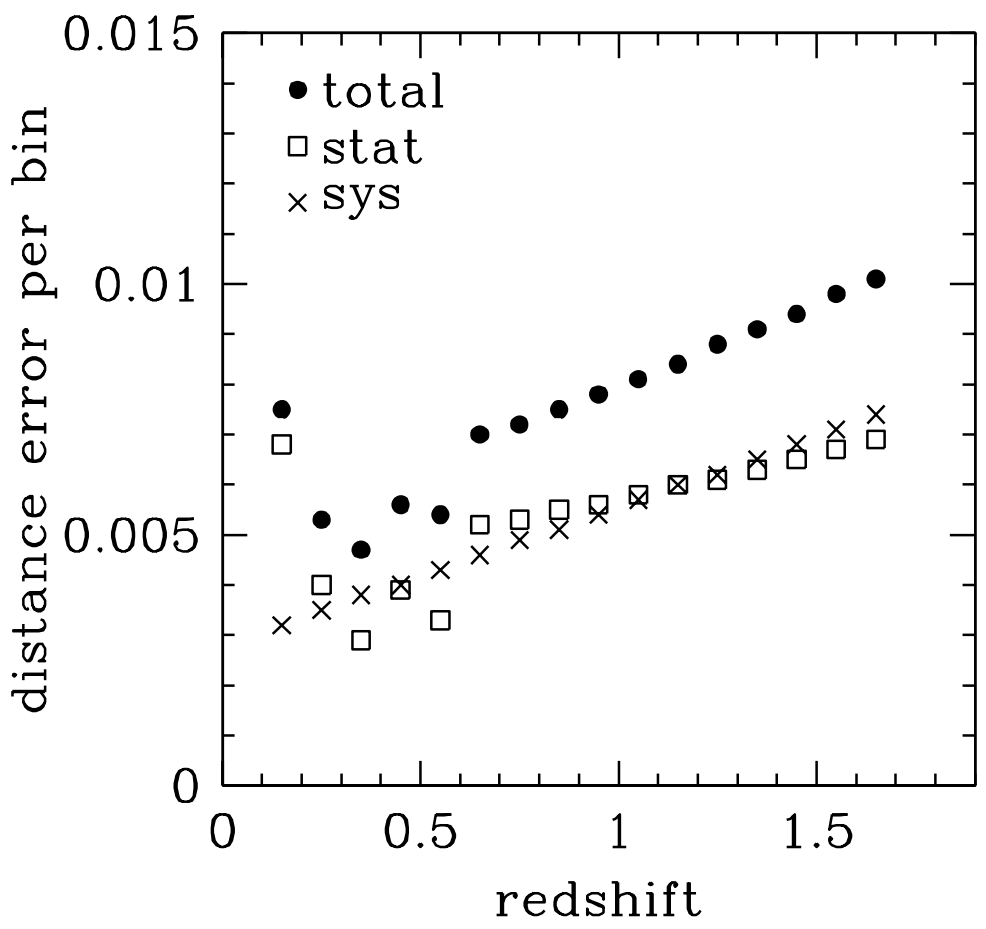

**Figure C-2: Fractional errors in distance per Δz = 0.1 bin. Squares and crosses show the statistical and systematic contributions, respectively, and filled circles show the total error.**

The total error for a bin is just the quadrature sum of the statistical and systematic errors. Filled circles in Figure C-2 show the total distance error (which is half the flux error) based on the WFIRST-2.4 design and the error model above; these are the same errors shown for luminosity distance in Figure C-5. By design, the statistical and systematic errors are comparable, except in the lowest redshift bin where the volume is small. In general, one gains by improving statistics in low redshift bins (where the exposure times are shorter) until one hits the systematics limit, and then one gains by going to higher redshifts, which continue to provide new information. (Our assumption that systematic errors are uncorrelated across redshift bins is important in this regard.) The total error is about 0.5% per bin near z = 0.5, then climbs to 1% at the highest redshifts because of the increased contribution of lensing to the statistical error and the assumed redshift behavior of the systematic error.

Our error model is based on current empirical understanding of the supernova population, which is improving rapidly thanks to extensive surveys of nearby and moderate redshift SNe. The design and analysis of the WFIRST-2.4 SN survey, including the optimal divi-sion of observing time among redshift tiers, will ultimately depend on lessons learned from ground-based surveys and HST/JWST observations between now and the launch of WFIRST. It is clear, however, that the near-IR coverage, sharp imaging, and large collecting area of WFIRST-2.4 will allow a supernova cosmology survey of extraordinary quality, far surpassing that of any other planned facility on the ground or in space.

The scientific goals of the WFIRST-2.4 SN survey lead to the following requirements:

**Survey Capability Requirements**

- >100 SNe-Ia per $\Delta z$=0.1 bin for all bins for 0.2 < z < 1.7, per dedicated 6 months
- Observational noise contribution to distance modulus error $\sigma_\mu \leq 0.02$ per $\Delta z$=0.1 bin up to z = 1.7.
- Redshift error $\sigma \leq 0.005$ per supernova
- Relative instrumental bias ≤ 0.005 on photometric calibration across the wavelength range.

**Data Set Requirements**

- Minimum monitoring time-span for an individual field: ~2 years with a sampling cadence ≤ 5 days
- Cross filter color calibration ≤ 0.005
- Three filters, approximately Y, J, H for SN discovery
- IFU spectrometer, $\lambda/\Delta\lambda$ ~100, 2-pixel (S/N ≥ 10 per pixel bin) for redshift/typing
- IFU S/N ≥ 15 per synthetic filter band for points near lightcurve maximum in each band at each redshift
- Dither with 30 mas accuracy
- Low Galactic extinction, E(B-V) ≤ 0.02.

## C.2. The HLS Imaging Survey

Figure C-3 illustrates the HLS tiling strategy, which consists of 4 exposures, each offset diagonally by slightly more than a chip gap. The sky is then tessellated with this pattern (i.e. it is repeated in both the X and Y directions spaced by the field size). This is considered to be 1 "pass" and achieves a depth of 2—4 exposures. A second pass at a different roll angle (but the same filter) achieves a depth of 4—8 exposures (90% fill at ≥5 exposures). The J band has 5 exposures instead of 4 in one of the passes, since we are attempting WL shape measurement in this band and it has the tightest sampling requirements.

With this strategy and the image combination algorithms of Rowe et al.[116], the J, H, and F184 imaging fully samples the telescope PSF over 90% of the survey area, even in the presence of defects such as cosmic ray hits. The ≥5 exposures per filter are separated into two passes several months apart at different roll angles, facilitating photometric calibration embedded in the survey. Y-band images are not fully sampled and will therefore provide lower fidelity shape measurements, so we have not included them in our calculations of weak lensing constraints on dark energy. They still make an important contribution to determining photometric redshifts of the source galaxies, in combination with the other bands and with ground-based optical imaging from LSST. The use of three filters for shape measurements allows multiple internal cross-checks: the WL signal is achromatic but many systematic errors may not be, so correlating shapes measured in different bands is a powerful test. Multiple filters are also important for addressing wavelength-dependent PSF issues in galaxy shape measurements;[117,118] with a single shape measurement filter, the difference in PSF between galaxies and calibration stars can easily become a limiting systematic that is both difficult to detect and problematic to correct, especially in galaxies that have a radial color gradient.

The statistical power of a WL survey depends mainly on the total number of galaxies with reliable shape measurements, which is the product of the survey area and the effective number density of usable source galaxies $n_{eff}$ (*i.e.*, the number density down-weighted by the measurement noise in the ellipticity). For the WFIRST-2.4 HLS, after removing sources lost to CR hits, the total effective densities are 61, 55, and 43 galaxies arcmin$^{-2}$ in the J, H, and F184 bands, respectively. The total number of unique objects in all bands is 66 arcmin$^{-2}$, while co-adding the J+H bands leads to $n_{eff}$ = 75 arcmin$^{-2}$. Galaxies generally have higher S/N in the redder bands, which leads to the increase of $n_{eff}$ between J and H, but the larger PSF size and narrower filter width in F184 leads to a drop in $n_{eff}$. The area of the HLS imaging survey is 2000 deg$^2$, so with ~70 galaxies arcmin$^{-2}$ WFIRST-2.4 provides a total of 500 million WL shape measurements. The use of three shape measurement bands allows the construction of 6 auto- and cross-correlation shear power spectra, providing the redundancy needed to ensure that dark energy constraints are not biased by systematic errors. Comparison of these spectra will enable end-to-end consistency tests of the long chain of corrections that are necessary in any weak lensing analysis. Since they are based on (almost) the same set of galaxies, the differences of these spectra can diagnose even systematic errors that are well below the statistical error of the survey; the lensing signal itself should be achromatic, but many of the measurement systematics would not be. For some purposes (e.g., cluster-galaxy lensing, which is likely to have lower systematics than cosmic shear and therefore require less redundancy) it will be possible to measure shapes from the sum of the J and H-band images, for which we estimate $n_{eff}$=75 arcmin$^{-2}$.

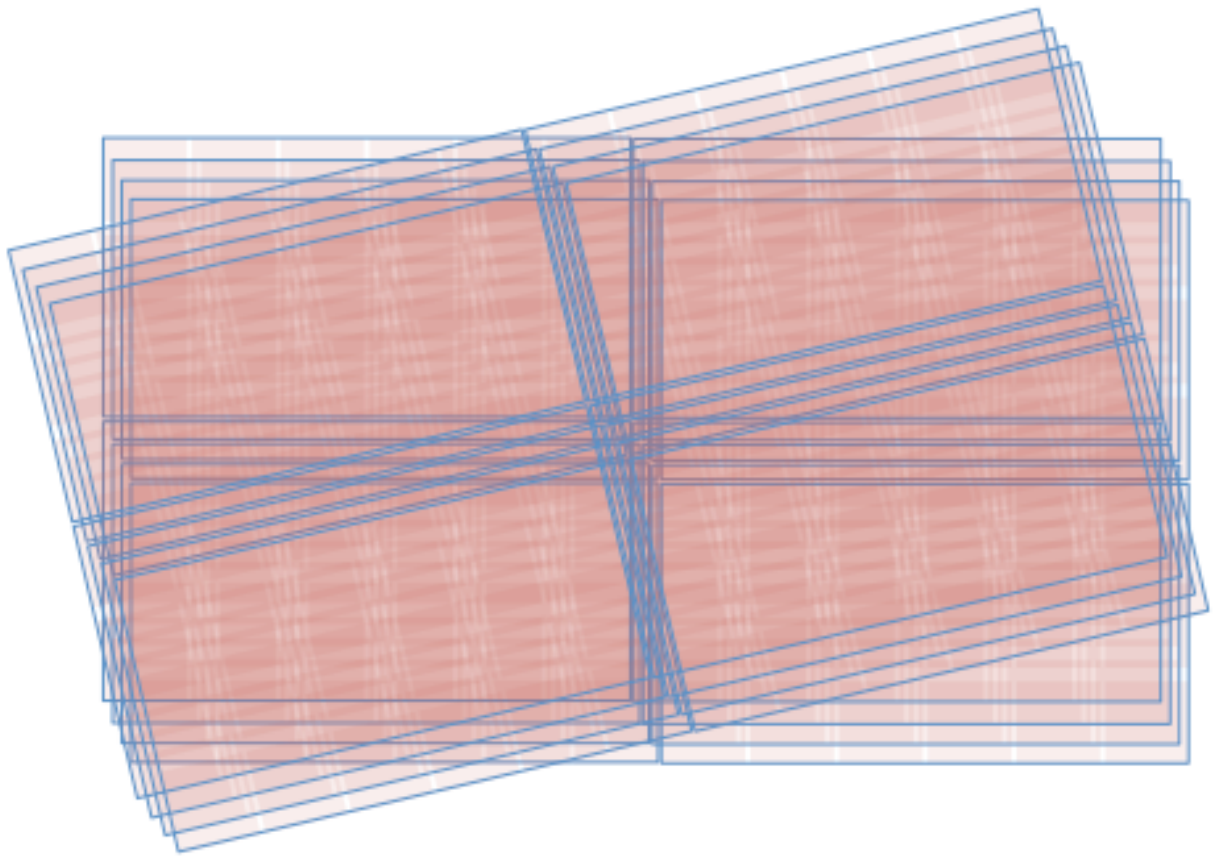

**Figure C-3: Illustration of the rolled tiling strategy in the Y, H, and F184 band filters. The J-band observations have one additional exposure in one of the bandpasses.**

These effective surface densities are about a factor of two higher than the values for DRM1, thanks to the greater depth and higher angular resolution of the WFIRST-2.4 HLS. They are also about a factor of two above the values typically quoted for Euclid and LSST, which are based on cuts that include galaxies more poorly resolved or with lower S/N than the ones we use here; the actual gain with WFIRST-2.4 may wll be closer to a factor of 3-4. Longer exposures with WFIRST-2.4 can achieve much higher $n_{eff}$, comparable to the 200-300 galaxies arcmin$^{-2}$ in the deepest HST weak lensing surveys. Greater depth comes at the expense of survey area, but for interesting objects (e.g., clusters and superclusters) one can carry out long-exposure programs to obtain detailed dark matter maps.

Figure C-4 shows the predicted cosmic shear angular power spectrum for 17 tomographic bins of source galaxy photometric redshift together with the projected statistical errors for the WFIRST-2.4 WL survey. For multipole L less than a few hundred, the statistical errors are dominated by sample variance in the lensing mass distribution within the survey volume, while for high L or the extreme redshift bins the errors are dominated by shape noise, *i.e.*, the random orientations of the source galaxies that are available to measure the shear. For a given photo-z bin the statistical errors at

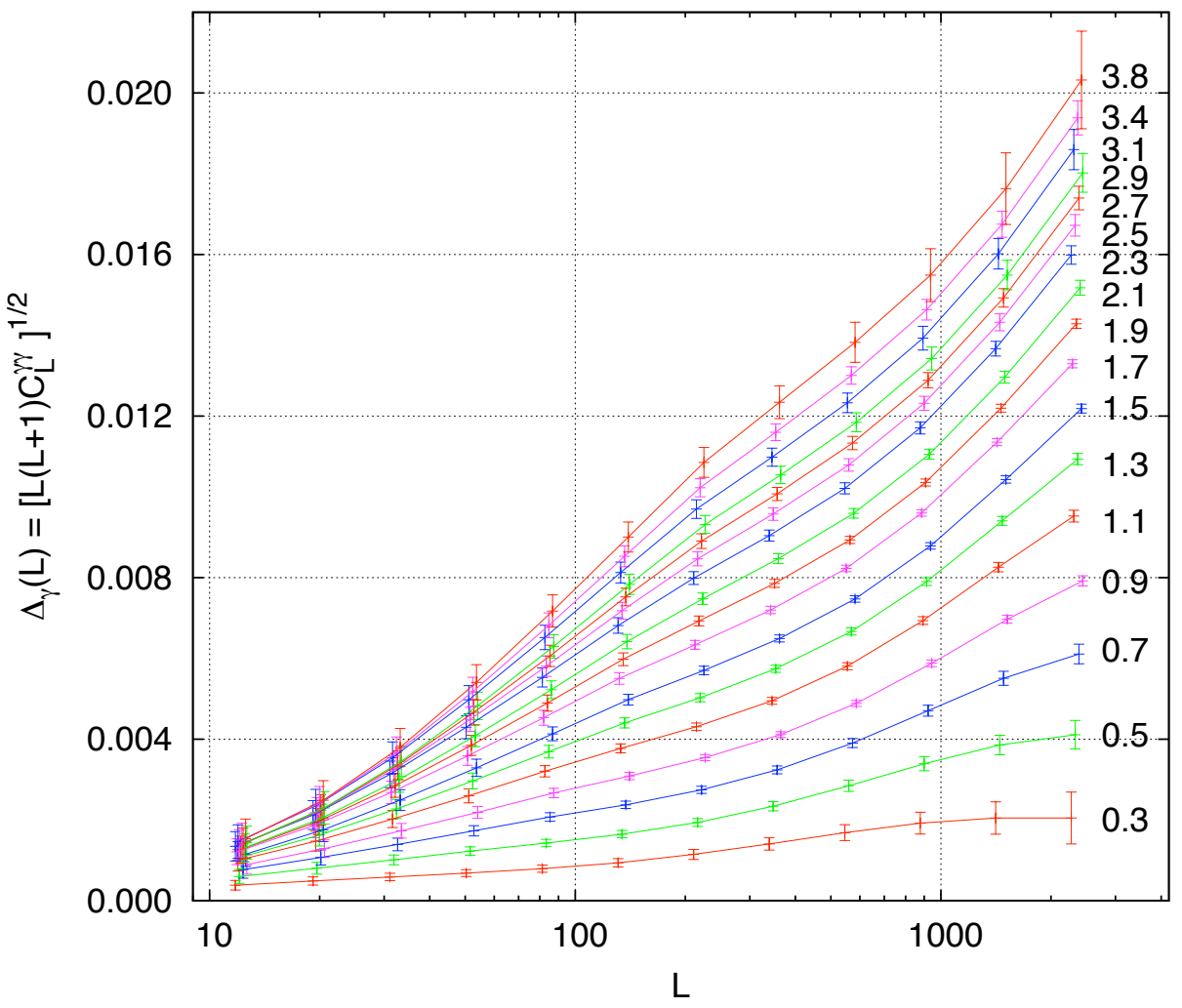

**Figure C-4: The predicted cosmic shear angular power spectrum and associated statistical errors for the WL survey, for 17 tomographic bins of source photometric redshift as labeled. L is the angular multipole, and $L(L+1)C_L^{\gamma\gamma}$ is the contribution to the variance of cosmic shear per logarithmic interval $\Delta \ln L$.**

different L are uncorrelated. Across photo-z bins, the errors at lower L are partly correlated because the same foreground structure contributes to the lensing of all galaxies at higher redshift; the errors decorrelate at high L when they become dominated by shape noise, which is independent for each set of sources. While Figure C-4 displays the auto-spectra for the 17 photo-z bins, there are also 17×16/2 = 136 cross-spectra that provide additional information.

When forecasting constraints from the WL survey, we use the Fisher matrix calculation described in the JDEM Figure of Merit Science Working Group (FoMSWG) report.[119] This calculation makes relatively conservative assumptions about intrinsic alignments, which are expected to be the most important source of astrophysical systematics. We marginalize over a multiplicative shear calibration error of 0.2% in each of our 17 redshift slices, assumed independent so that the overall shear calibration error is smaller by $\sqrt{17}$, and an additive shear calibration error of $3\times10^{-4}$, which affects all bins equally. We also marginalize over systematic (but uncorrelated) uncertainties in the *mean* photo-z of each tomographic bin of $\Delta z = 0.002(1+z)$. With this level of systematic uncertainties, cosmic shear WL measurements would remain statistics-limited even in an extended survey of ~ $10^4$ deg$^2$, and we have therefore adopted them as capability requirements for the HLS imaging survey. They in turn lead to demanding constraints on knowledge of the PSF second moment and ellipticity and on the photo-z calibration data, as listed in the data set requirements.

Cosmic shear measures the correlation of galaxy ellipticities produced by the correlated weak lensing shear from a common foreground matter distribution, predominantly dark matter. An alternative use of WL data is to cross-correlate the galaxy shear map with a foreground cluster or galaxy distribution --- cluster-galaxy or galaxy-galaxy lensing. This is (nearly) equivalent to measuring the average tangential shear profile around the foreground objects, i.e., the average elongation of source galaxies perpendicular to the projected lens-source separation. These forms of lensing analysis bring in the additional information provided by the cluster or galaxy maps, which mark the expected locations (including photometric and in some cases spectroscopic redshifts) of dark matter overdensities. The measurement systematics of cluster-galaxy and galaxy-galaxy lensing are also likely to be smaller than those affecting cosmic shear analysis.

As emphasized by Oguri & Takada[120] and Weinberg et al.[7], the combination of cluster abundances and cluster-galaxy lensing (which effectively calibrates the cluster mass-observable relation) can provide cosmological constraints that are competitive with and distinct from the cosmic shear constraints from the same WL survey. Clusters can be readily identified from the HLS galaxy maps, and the eROSITA mission will also provide a deep all-sky survey of X-ray selected clusters. In Figure C-6 below, we use the approach of Weinberg et al.[7] to estimate the constraints from cluster-galaxy lensing in the WFIRST-2.4 HLS, assuming that clusters can be reliably identified down to $10^{14}M_{sun}$. This estimate likely understates the power of this approach because it does not include the degeneracy breaking from measurements at multiple mass scales, nor does it utilize the independent information in the large scale cluster-galaxy signal (far beyond the virial radius). A Guest Observer program could target high mass clusters from eROSITA or other surveys outside of the HLS area, both to sharpen dark energy tests and to characterize the detailed distribution of dark matter within clusters.

Galaxy-galaxy lensing provides insight into the dark matter halo masses and extended environments of different classes of galaxies (e.g., Mandelbaum et al.[121]). In combination with galaxy clustering measurements, it can also be used to derive constraints on dark matter clustering. We have not included cosmological forecasts for galaxy-galaxy lensing because they depend on uncertain assumptions about how well one will be able to model the non-linear bias of galaxies with respect to dark matter. However, plausible forecasts suggest that constraints on structure growth from this approach will be competitive with and perhaps even more powerful than those from cosmic shear alone.[122] We have also not included constraints from higher order lensing statistics (e.g., the shear bispectrum), which could have substantial power at the high $n_{eff}$ of the WFIRST-2.4 WL maps.

One of the most challenging tasks for the WFIRST-2.4 weak lensing analysis is highly accurate calibration of photometric redshifts, i.e., the mean and distribution of true redshifts for galaxies in bins of photometric redshift. Achieving this calibration requires a sample of ~ $10^5$ galaxy redshifts, with high completeness, to the depth of the weak lensing imaging survey.[123] While large ground-based telescopes will play a crucial role in creating this calibration data set, which is also needed for interpreting LSST and Euclid weak lensing, the IFU on WFIRST-2.4 can also make a vital contribution, especially for the faintest and highest redshift galaxies. Parallel observations with the IFU during wide-field imaging for the HLS and supernova surveys should yield ~10,000 redshifts of faint galaxies that fall serendipitously within the 3.00"x3.15" IFU field of view, rising to ~80,000 if it proves practical to incorporate few-arcsec offsets in dithers to get galaxies into the IFU. A targeted IFU spectroscopic observing program could yield high completeness for galaxies whose redshifts would be difficult to obtain from the ground, perhaps in coordination with deep imaging programs. In addition to direct spectroscopic calibration, cross-correlation between WL sources and large scale structure[124] in the GRS will play an important role in photo-z calibration.

The Green et al. report[1] includes a more extensive discussion of WL systematics and the ways that the design of WFIRST is intended to mitigate them. We will not repeat all of this discussion here, but we note that the systematics considerations are essentially the same for WFIRST-2.4 and that our mitigation strategies are similar.

**Survey Capability Requirements**

- ≥ 1070 $deg^2$ per dedicated observing year (combined HLS imaging and spectroscopy)
- Effective galaxy density ≥ 60 per $arcmin^2$, shapes resolved plus photo-z's
- Additive shear error ≤ $3x10^{-4}$
- Multiplicative shear error = 0.2% per redshift slice (17 slices)

- Photo-z error distribution width ≤ 0.04(1+z), catastrophic error rate <2%
- Systematic error in photo-z offsets ≤ 0.002(1+z)

**Data Set Requirements**

- From Space: 3 shape/color filters (J, H, and F184), and one color filter (Y; only for photo-z)
- S/N ≥ 18 (matched filter detection significance) per shape/color filter for galaxy $r_{eff}$ = 180 mas and mag AB = 24.7/24.6/24.1 (J/H/F184)
- PSF second moment ($I_{xx}$ + $I_{yy}$) known to a relative error of ≤ $9.3\times10^{-4}$ rms (shape/color filters only)
- PSF ellipticity ($I_{xx}$-$I_{yy}$, $2I_{xy}$)/($I_{xx}$ + $I_{yy}$) known to ≤ $4.7\times10^{-4}$ rms (shape/color filters only)
- System PSF EE50 radius ≤ 0.12 (Y band), 0.12 (J), 0.14 (H), or 0.13 (F184) arcsec
- At least 5 (H,F184) or 6 (J) random dithers required for shape/color bands and 5 for Y at same dither exposure time
- From Ground: 5 filters spanning at least 0.35-0.92 μm, i.e. from blueward of z=0 Balmer break out to the beginning of WFIRST Y (LSST ugriz is sufficient)
- From Ground + Space combined: Complete an unbiased spectroscopic PZCS training data set containing ≥ 100,000 galaxies with ≤ mag AB = 24.6 (in J,H,F184 bands) and covering at least 4 uncorrelated fields; redshift accuracy required is $\sigma_z$<0.01(1+z)

## C.3. The HLS Spectroscopic Survey

The spectroscopy observing strategy is similar in concept to the HLS imaging strategy described in §2.2.2. The spectroscopic mode has 4 passes, each with 2 exposures offset diagonally by slightly more than a chip gap, and achieving 90% fill at ≥6 exposures. Successive passes are offset by a roll angle of ~180° to provide counter-dispersion, suppressing systematic errors in wavelength determination associated with the astrometric offset between the continuum galaxy image and the Hα emitting region. The time per exposure is 362 seconds, and the total time to survey 2000 deg$^2$ is 0.65 years, compared to 1.26 years for the HLS imaging. Figure 2-4 shows the limiting flux of the HLS spectroscopic survey for a point source and for an extended source with $r_{eff}$ = 0.3 arcsec and an exponential profile.

To predict galaxy yields, we use the Hα and [OIII] luminosity functions from J. Colbert et al. (in prep.) based on blind grism spectroscopic surveys with HST's WFC3. Our Hα computations assume a threshold of 7σ (matched filter significance) for detection of the Hα emission line and 70% completeness. To be conservative we have not assumed any enhancement in S/N from the [N II] doublet, although at the WFIRST dispersion this will always be a partial blend with Hα. The galaxy yields in the table are averaged over the depth histogram of the survey. Over 90% of the survey bounding box is observed at ≥6 exposures and hence reaches ≥80% of the mean galaxy density. The depth variations are typically on scales of order the chip gaps, which are 1.4—2.2 Mpc comoving. Hα moves out of the WFIRST-2.4 bandpass at z > 1.95, but [OIII] can be detected over 1.7 < z < 2.9. We again adopt a 7σ significance threshold, in this case for a matched filter optimally combining the λ5007Å and λ4959Å features, with 70% completeness. We caution that the statistics of high-redshift [OIII] emitters remain quite uncertain, and even estimates of the Hα luminosity function at these redshifts show significant variation from one study to another. With these caveats, Table 2-2 presents our predicted galaxy space density for the HLS as a function of redshift, using Hα at z < 2 and [OIII] at z ≥ 2. Space densities are in comoving Mpc$^{-3}$. If we used the Hα luminosity function of Sobral et al.[125] then our predicted space densities would be lower by a factor of ~2 at z ~ 1.5 but similar at z ~ 2.

To forecast errors in cosmological measurements from the GRS, shown in Figures C-5 and C-6, we use the formalism of Wang et al.[5] For the $D_A(z)$ and H(z) errors we use BAO information only, modeling the power spectrum up to a maximum comoving wavenumber $k_{max}$ = 0.5 h Mpc$^{-1}$ and assuming that reconstruction will reduce non-linear suppression of the BAO signal by correcting 50% of the Lagrangian displacement.[126,127]

For the $f(z)\sigma_m(z)$ constraints we use the full redshift-space power spectrum up to $k_{max}$ = 0.2 h Mpc$^{-1}$. BAO analysis can work to higher $k_{max}$ because non-linear gravitational evolution and galaxy bias produce only broad-band distortions that do not significantly shift the location of the localized BAO feature. WFIRST-2.4 still provides good sampling on these smaller scales, and if we used full P(k) analysis with $k_{max}$ = 0.2 h Mpc$^{-1}$ our $D_A(z)$ errors would in fact be slightly worse, because the additional information in the broad-band P(k) does not fully compensate for the loss of BAO information at k = 0.2-0.5 h Mpc$^{-1}$. However, full P(k) analysis (or correlation function analysis) to scales $k_{max}$ ~ 0.2 h Mpc$^{-1}$ is already the state-of-the-art for current redshift surveys,[4,5,128,129] and with improvements in theoretical modeling it should be possible to work further into the non-linear regime, especially as the clustering data themselves (on smaller scales and from higher order statistics) provide strong constraints on the models. Full P(k) analysis brings in additional information from the broad-band shape, from the magnitude and scale dependence of redshift-space distortions, and from the constraint on the product $H(z)D_A(z)$ that comes from demanding statistical isotropy in the real-space clustering.[130] On the timescale of WFIRST-2.4, it may well be possible to derive constraints from GRS data much tighter than we have forecast by exploiting smaller scales and using more sophisticated "reconstruction" modeling to reduce the degradation of cosmological information from non-linear evolution. The number of Fourier modes measured in the survey grows as $(k_{max})^3$, so the potential gains are large.

The requirements for the galaxy redshift survey component of the dark energy program are:

**Survey Capability Requirements**

- ≥ 1070 $deg^2$ per dedicated observing year (combined HLS imaging and spectroscopy)
- A comoving density of galaxy redshifts $n>6x10^{-4}$ $Mpc^{-3}$ at z=1.9 according to Colbert + LF.
- Redshift range $1.10 \le z \le 1.95$ for Hα
- Redshift errors $\sigma_z \le 0.001(1+z)$, equivalent to 300 km/s rms
- Misidentified lines ≤ 5% per source type, ≤ 10% overall; contamination fractions known to $2\times10^{-3}$

**Data Set Requirements**

- Slitless grism, spectrometer, ramped resolution 550-800 over bandpass
- S/N ≥ 7 for $r_{eff}$ = 300 mas for Hα emission line flux at 1.8 μm ≥ $1.0x10^{-16}$ $erg/cm^2/s$
- Bandpass $1.35 \le \lambda \le 1.95$ μm
- Pixel scale ≤ 110 mas
- System PSF EE50% radius 240 mas (1.35 μm), 180 mas (1.65 μm), 260 mas (1.95 μm)
- ≥ 3 dispersion directions required, two nearly opposed
- Reach $J_{AB}$=24.0 AND ($H_{AB}$=23.5 OR $F184_{AB}$=23.1) for $r_{eff}$=0.3 arcsec source at 10 sigma to achieve a zero order detection in 2 filters.

## C.4. Dark Energy Observables

We have used the survey strategies and systematic error models described in the preceding subsections to forecast measurement precision on cosmological observables, and to compute the expected constraints on dark energy and growth parameters reported in §2.2.3. The latter forecasts account for the ability of multiple observational probes to break degeneracies in a high-dimensional model space. However, it is also useful to look directly at the anticipated precision on observables, which provide a more model-independent characterization of the power of the measurements. Aggregate precision values over broad ranges of redshift (as reported in Figure 2-1, for example) capture the power of the data while also highlighting the necessary level of systematics control.

Figure C-5 examines the anticipated errors on the distance and H(z) measurements. For the SN survey, the median error is ~ 0.8% *per Δz = 0.1 redshift bin.* For independent measurements in $N_{bin}$ bins, the aggregate precision on a constant multiplicative offset of the the luminosity distance scale is $(\Sigma_i\ [\Delta \ln D_L(z_i)]^{-2})^{-1/2}$. Blue and red points show the aggregate precision for the z < 1 and z > 1 portions of the SN survey, 0.20% and 0.34%, respectively, plotted at the error-weighted mean redshifts. The aggregate precision over the full 0 < z < 1.7 range is 0.18%. For BAO, typical measurement errors in the 1 < z < 2 redshift range probed by Hα emitters are ~ 1.2% per Δz = 0.1 bin in $D_A(z)$ and ~2.1% in H(z). Errors on H(z) are larger because there is one line-of-sight direction vs. two transverse directions, but H(z) measurements are more sensitive to the dark energy equation of state because they directly probe the total energy density, which is the sum of the matter, radiation, and dark energy densities. The aggregate precision over 1 < z < 2 (blue squares) is 0.40% for $D_A(z)$ and 0.72% for H(z). Statistical errors are much larger for the [OIII] emitters at z > 2, but they still afford aggregate precision of 1.3% on $D_A(z)$ and 1.8% on H(z), allowing diagnostics of models with unusual early time evolution of dark energy.

As context for assessing these projected errors, curves in Figure C-5 show the fractional changes in the predicted $D_L(z)$, $D_A(z)$, and H(z) for several models relative to a model with a flat universe, a cosmological constant, and $\Omega_m$ = 0.267. Dotted and short-dashed curves show the effect of changing the dark energy equation-of-state parameter from w = -1 to w = -0.96 or w = -1.04, while maintaining a flat universe. Long-dashed and short-dashed curves show the effect of changing the space curvature to $\Omega_k = \pm 0.002$ while maintaining w = -1. Isolated changes to w or $\Omega_k$ with all other parameters held fixed tend to produce models that are easily ruled out by CMB data. Here we have followed the strategy of Weinberg et al.,[7] where for each change to w or $\Omega_k$ we adjust the values of $\Omega_m$, $\Omega_b$, and *h* in a way that keeps the CMB power spectrum almost perfectly fixed. A simple comparison of the model differences to the aggregate precision error bars in Figure C-5 suggests that WFIRST-2.4 could readily discriminate these variants from a ΛCDM model. The Fisher matrix forecasts reported in §2.2.3, which account for parameter degeneracies and for the complementary information content of different probes, confirm this impression, with an expected 1σ error on the equation-of-state parameter $\Delta w = 0.0088$.

Figure C-6 presents corresponding precision forecasts for observables related to structure growth. The top panel shows forecast errors from RSD on $\sigma_m(z)f(z)$, the product of the matter fluctuation amplitude and the logarithmic growth rate $f(z) = d\ln\sigma_m(z)/d\ln a$. These errors are computed by the methodology of Wang et al.[4], assuming that the redshift-space power spectrum can be modeled up to $k_{max} = 0.2\ h\ \mathrm{Mpc}^{-1}$. To reduce the degeneracy between RSD and the anisotropy induced by the Alcock-Paczynski effect,[130] for this panel (but not other plots in this section) we include a prior for Planck CMB measurements when computing the $\sigma_m(z)f(z)$ errors. The predicted aggregate error is 1.2% at z = 1-2 and 3.2% at z=2-3.

Cluster abundances calibrated by weak lensing constrain a combination of parameters that is approximately $\Omega_m^{0.4}\sigma_m(z)$, where $\Omega_m$ is the z=0 matter density in units of the critical density. For the middle panel of Figure C-6, we have used the approach of Weinberg et al.[7] to compute expected errors on this quantity for the surface density and redshift distribution of weak lensing galaxies anticipated in the HLS imaging survey, assum-ing that clusters can be reliably identified down to a mass threshold of $10^{14}$ $M_{sun}$.[131] The per-bin error bar initially shrinks with increasing redshift, as the volume element grows yielding more clusters, then grows at high redshift as the space density of clusters declines and the surface density of *background* galaxies for weak lensing calibration drops. The predicted aggregate precision is 0.14% for $z < 1$ and 0.28% for $z > 1$.

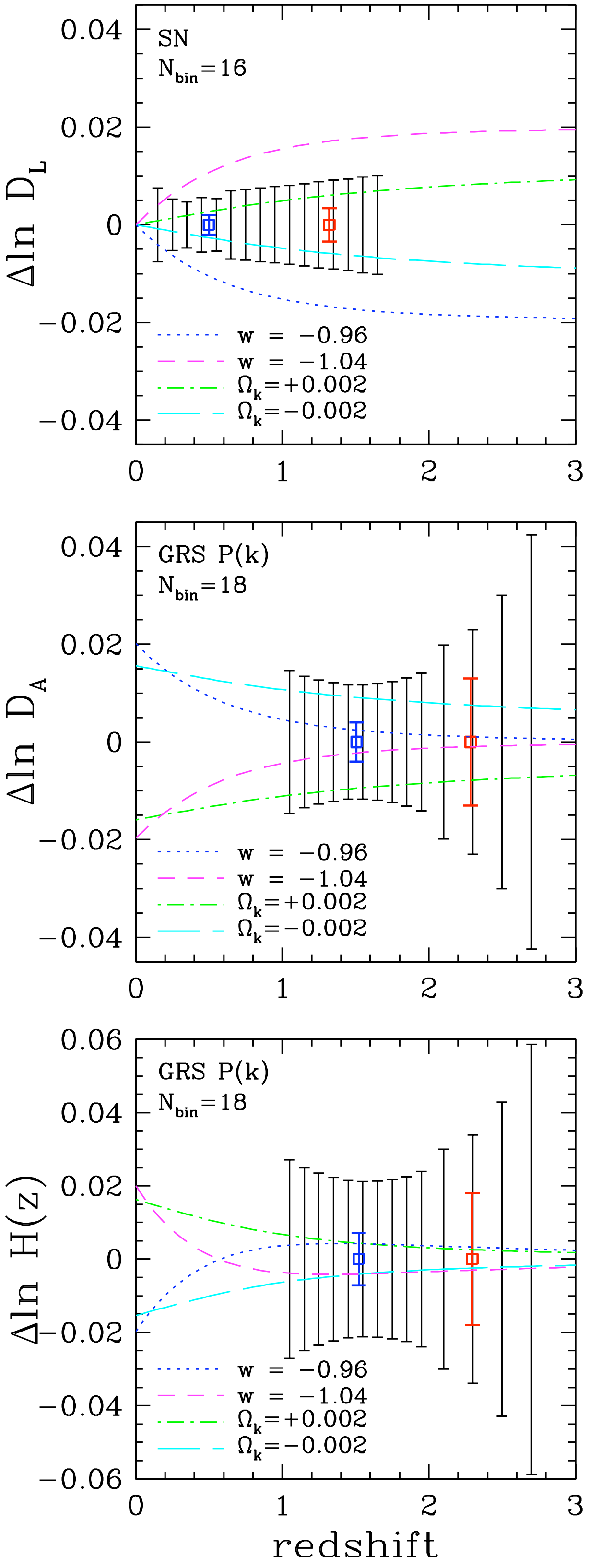

**Figure C-5: Measurements of expansion history from the SN and galaxy redshift surveys. Black error bars show forecast 1σ fractional uncertainties in narrow redshift bins for luminosity distance (top) from the SN survey and for angular diameter distance (middle) and Hubble parameter (bottom) from BAO measurements in the GRS. In the lower panels, errors for z > 2 bins are based on [OIII] emitters. Blue and red points with error bars show the aggregate precision in the low and high redshift ranges, respectively (z<1 vs. z>1 for SNe, z<2 vs. z>2 for BAO). Curves in each panel show the impact of changing the equation-of-state parameter w or the space curvature $\Omega_k$ relative to a fiducial flat ΛCDM model. Any of these single-parameter changes would be easily ruled out by the combined data set.**

As context for interpreting these predicted errors, we again show in each panel the predicted impact of changing dark energy parameters away from those of a fiducial ΛCDM model. Dotted and short-dashed curves show the impact of maintaining GR but changing w to -0.96 or -1.04, which alters structure growth by changing the history of H(z). Long-dashed and dot-dashed curves show the effect of maintaining w = -1 but changing the index of the growth rate $f(z) \approx [\Omega_m(z)]^\gamma$ from its GR-predicted value of $\gamma \approx 0.55$, by $\Delta\gamma = \pm 0.05$. This is a simple way of parameterizing the possible impact of modified gravity on structure growth, but other forms of GR modification could produce stronger growth deviations at high redshift, as could GR-based models with early dark energy. Our Fisher matrix calculation, which includes the cosmic shear measurements, forecasts a $1\sigma$ error on $\Delta\gamma$ of 0.015.

It is difficult to abstract WL constraints into a form similar to those of Figures C-5 and C-6, with errors on an "observable" as a function of redshift. First, the WL signal depends on both the amplitude of structure and the distance-redshift relation: for sources at $z_s$ and lenses at $z_l$, the quantities $\sigma_m(z_l)$, $D_A(z_l)$, and $D_A(z_s)$ all affect the shear. Second, the errors on the WL signal for two different bins of source redshift can be strongly correlated because matter in the foreground of both redshifts can lens the galaxies in both bins. For WL one obtains errors that are more nearly uncorrelated by considering different multipoles $C_L$ of the angular power spectrum, which is the basic quantity measured in a cosmic shear analysis. Figure C-4 above illustrated the predicted errors on the power spectrum in tomographic redshift bins. However, this plot shows only the $N_{bin}$ = 17 auto-spectra, while much of the cosmological information resides in the $N_{bin} \times (N_{bin}-1)/2$ = 136 cross-spectra. As a representative aggregate precision, we have computed the error on $\sigma_m$ considered as a single parameter multiplying the amplitude of linear matter fluctuations at all redshifts. The predicted precision is 0.13%, accounting for all error correlations and marginalizing over systematic measurement and modeling uncertainties as described in §2.2.2 below.

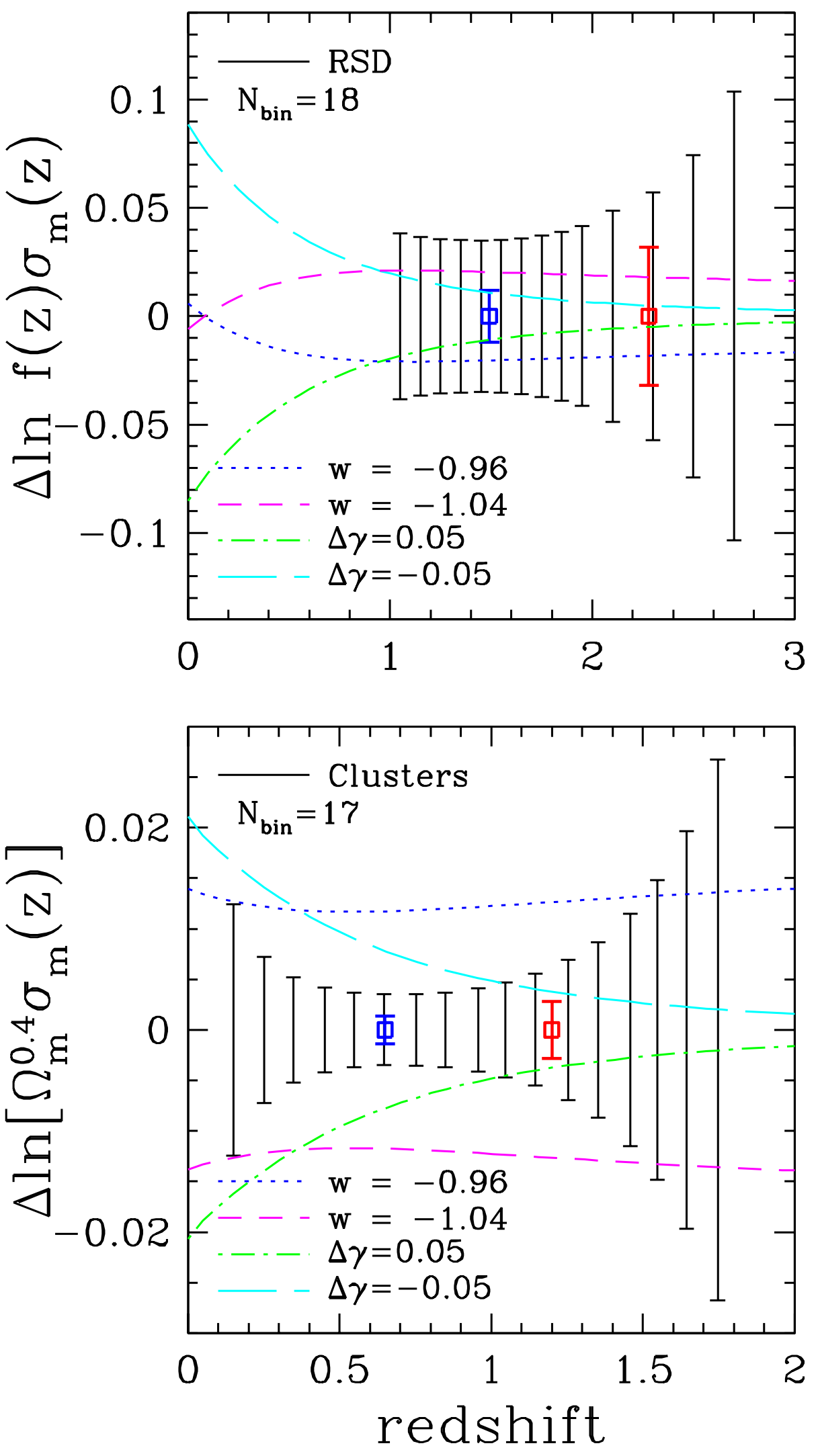

**Figure C-6: Measurements of structure growth from redshift-space distortions (RSD) in the GRS and the cluster mass function with the mass-observable relation calibrated by stacked weak lensing of the cluster mass-observable relation in each redshift bin. Black error bars show forecast 1σ fractional uncertainties in narrow redshift bins on the parameter combinations $f(z)\sigma_m(z)$ and $\Omega_m^{0.4}\sigma_m(z)$ best constrained by RSD and clusters, respectively. Blue and red points with error bars show the aggregate precision in the low and high redshift ranges ($z < 2$ vs. $z > 2$ for RSD, $z < 1$ vs. $z > 1$ for clusters). Curves in each panel show the impact of changing the equation-of-state parameter w or the growth index γ relative to a fiducial flat ΛCDM+GR model.**

In sum, WFIRST-2.4 will make multiple independent measurements of the cosmic expansion history and the history of structure growth over the range $0 < z < 3$, each with aggregate precision well below 1%, using methods with complementary information content and completely different sources of systematic uncertainty. This represents an enormous advance over the current state-of-the-art in precision, redshift leverage, and robustness. These measurements will allow far more stringent tests of the ΛCDM cosmological model and have the potential to reveal subtle signatures of evolving dark energy or departures from GR.

# Appendix D WFIRST-2.4 Microlensing Requirements, Simulation Methodology, and Future Work

## D.1. WFIRST-2.4 Microlensing Survey and Data Requirements

The WFIRST-2.4 microlensing survey detects planetary companions to stars via short-duration, relatively low-amplitude, planetary deviations to the longer timescale microlensing events caused by their host stars (see Figure 2-18). The survey and data requirements for the microlensing exoplanet survey can be understood by the following qualitative considerations. Microlensing events require extremely precise alignments between a foreground lens star and a background source star, and are both rare and unpredictable. Furthermore, the probability that a planet orbiting the lens star in any given microlensing event will give rise to a detectable perturbation is generally much smaller than unity, ranging from a few tens of percent for a Jupiter-mass planet and a typical low-magnification event, to less than a percent for planets with mass less than that of the Earth. These planetary perturbations have amplitudes ranging from a few percent for the lowest-mass planets to many tens of percent for the largest perturbations, but are brief, ranging from a few days for Jupiter-mass planets to a few hours for Earth-mass planets (see Figure 2-18). Also, the time of the perturbations with respect to the peak of the primary event is also unpredictable. For these reasons, detecting a large number of exoplanets with microlensing requires monitoring of a very large number of stars continuously with relatively short cadences and good photometric precisions of a few percent. Practically, a sufficiently high density of source and lens stars, and thus a sufficiently high microlensing event rate, is only achieved in lines of sight towards the Galactic bulge. However, these fields are also crowded, and this high star density means that high resolution is needed to resolve out the individual stars in order to achieve the required photometric precisions and to identify the light from the lens stars.

Quantitative estimates of the set of exoplanet survey and data requirements necessary to meet a given survey yield are listed below. These estimates were based on the results of sophisticated simulations of a microlensing exoplanet survey, which we describe in the next section. However, we can understand the order of magnitude of these requirements using simple back-of-the-envelope estimates, as follows. Consider, as a specific example, the survey requirement of detecting at least 150 Earth-mass planets. The typical detection probability for an Earth-mass planet at *2AU* is ~1.5%, and thus ~150/0.015 ~ $10^4$ microlensing events must be monitored to detect ~150 such planets, assuming every star hosts such a planet. The average microlensing event rate in the WFIRST-2.4 microlensing survey fields is ~$5\times10^{-5}$ events/year/star, and thus $10^4/5\times10^{-5}$ ~200 million star-years must be monitored. The typical stellar density down to J=23 is ~$10^8$ stars per square degree, and thus we have the first data requirement below that at least ~2 square degrees must be monitored. As illustrated in Figure 2-18, in order to detect and accurately characterize the perturbations due to such planets, which typically last a few hours and have amplitudes of several percent, photometric precisions of a few %, continuous monitoring, and cadences of less than 15 minutes are needed. Finally, given the areal density of ~$10^8$ stars per square degree, an angular resolution of $10^{-4}$ degrees (0.4 arcseconds) is needed to resolve the faintest stars.

The remaining three data requirements below ensure our ability to accurately measure the parameters of the primary events, which typically last ~40 days, and allow one to infer the angular size of the source star from its color and magnitude, separate the light from the lens and source, and to measure the relative lens-source proper motion. All three are required to measure the mass and distance to the primary lens (needed for determining the mass and separation of the detected planets) for the majority of events.[132] In particular, the spacecraft must be able to point continuously or nearly continuously at the Galactic bulge for longer than the duration of the typical primary microlensing events – greater than 40 days. A broad near-IR filter and a narrower bluer filter are also required, the first for the primary science data to achieve the requisite signal-to-noise ratio on the target stars, and the second to obtain color information in order to characterize of the source stars.

### Survey Capability Requirements

- Planet detection capability down to ~0.1 Earth masses
- Ability to detect ≥ 1500 bound cold planets in the

mass range of 0.1-10,000 Earth masses, including 150 planets with mass <3 Earth masses

- Ability to detect ≥ 20 free floating Earth-mass planets

***Exoplanet Data Requirements:***

- Monitor ~3 square degrees in the Galactic bulge for at least 250 total days.
- S/N of ≥100 per exposure for a J = 20.5 star.
- Sample light curves with wide filter W149 with $\lambda$ =0.927-2.0 $\mu$m.
- Photometric sampling cadence in W149 of ≤15 minutes.
- Better than 0.4" angular resolution to resolve the brightest main sequence stars.
- Monitor microlensing events continuously with a duty cycle of ≥80% for at least 60 days to measure basic light curve parameters.
- Monitor fields with Z087 filter, 1 exposure every 12 hours in order to measure the color of the microlensing source stars.
- Separation of >2 years between first and last observing seasons to measure lens-source relative proper motion.

There are additional, secondary requirements that may bear on the WFIRST-2.4 design and its ability to carry out a microlensing mission. First, the storage capacity and downlink rate must be sufficient to reserve and transmit the ~800 full-frame image per day during the bulge pointings. Second, there are requirements on the pointing accuracy and slew-and-settle times. Third, observations in the second filter must be obtained every ~12 hours. This, in turn, leads to a requirement on the total number of filter changes during the microlensing experiment of >1000.

## D.2. WFIRST-2.4 Microlensing Survey Yields: Methodology

We estimate the exoplanet yields for WFIRST-2.4 using the simulation methodology described in Penny et al.[133] This simulation was also used to estimate the yields of the WFIRST-DRM1 and WFIRST-DRM2 designs, as well as Euclid. We expect the relative yields between different survey designs (given our assumptions) to be more robust than the absolute rates, for reasons described in D.5.

We assume observations are carried out in 6 seasons of 72 days each, for a total of observation time of 432 days. We have assumed observations are continuous during each season. In reality, observations will need to be interrupted when the moon passes through the Galactic bulge. However, the minimum moon avoidance angle is not known, and therefore it is not clear how much time will be lost during these moon cutouts. In §3.10, we adopted a conservative moon avoidance angle of 30 degrees for the notional observing plan, which leads to a nearly 80% reduction in the total observing time. Should such large avoidance angles be required, the microlensing program will need to be augmented by another season in order to realize the yields described here.

To convert sensitivities to expected yields, we have adopted our best estimate of the planet distribution function from Cassan et al.[80], which has the form

$$dN/d\log M_p d\log a = f\,(M_p/95M_\oplus)^{\alpha}$$

where f = 0.24/dex$^2$ and $\alpha$ = -0.73. We assume that this power-law distribution saturates at a constant value of 2/dex$^2$ for $M_p$<5.2 $M_\oplus$. This is likely conservative, as Kepler has found several multi-planet systems that have more than 2 planets per decade in semi-major axis.[134,135] Note that the exponent of this power-law distribution with planet mass is slightly steeper but of opposite sign to the exponent of the approximately power-law dependence of the detection efficiency on planet mass, and thus the number of expected planets is a relatively slowly rising function of decreasing mass in this model in the regime where we assume the exponent $\alpha$ holds. Note also that the cold planet frequency below ~10$M_{Earth}$ is very poorly constrained. This ignorance is, of course, one of the primary justifications for an exoplanet survey with WFIRST-2.4.

We chose 10 target fields, with locations indicated in Figure D.1. We assume a cadence of 15 minutes, which leads to exposure times of 52 seconds per field assuming a 38 second slew time. Observations in the Z087 filter are taken of each field every 12 hours with an exposure time of 290 seconds. We simulated 8 of these fields, and assumed that the yields of the two fields at positive Galactic latitude had yields equal to the average of the other fields. The total survey area is 2.81 square degrees. Note that we did not fully optimize the survey relative to the number, location, and cadence of the target fields. We therefore expect our results to be conservative.

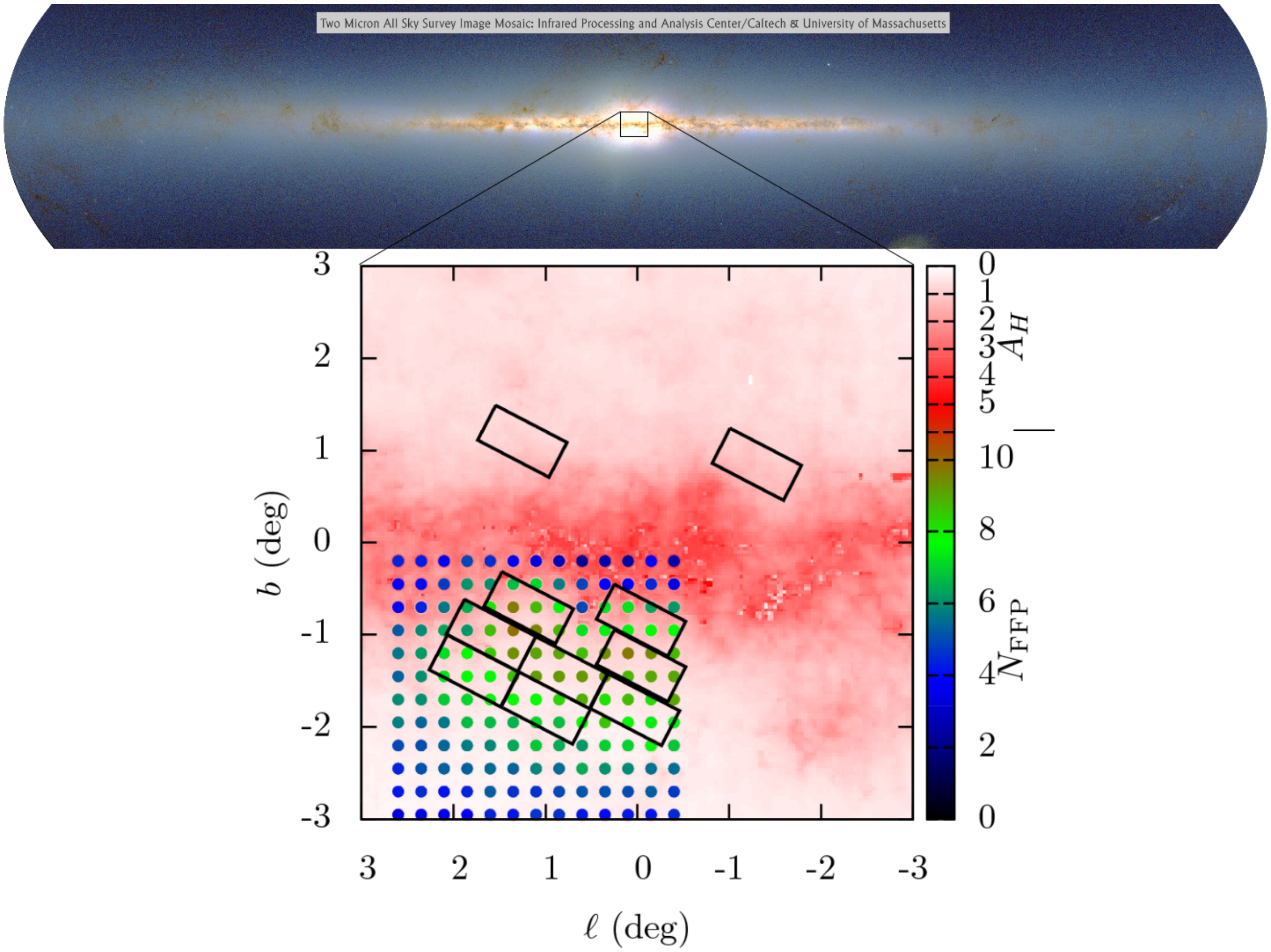

**Figure D.1: The upper figure shows an image of the Galactic plane from 2MASS, whereas the bottom panel zooms in on the ~36 square degrees around the Galactic center. In the bottom panel, the solid black rectangles show the locations of the 10 baseline WFIRST-2.4 fields in Galactic coordinates, with a total active area of 2.81 deg$^2$. These are compared to rates of free-floating Earth-mass planet detections as a function of location shown as the solid circles, and a map of the extinction in H-band from Gonzalez et al.[113] The WFIRST-2.4 fields were centered on locations of relatively low extinction and relatively high planet detection rates.**

## D.3. WFIRST-2.4 Microlensing Survey Yields: Baseline Results

The basic results are summarized in Box 5, Table 2-4, Table 2-5, and illustrated in Figure 2-20. In order to highlight the intrinsic sensitivity of the survey, Figure 2-20 shows the region of parameter space where WFIRST-2.4 is expected to detect at least 3 planets, assuming one planet per star at each mass and semi-major axis. Table 2-4 then reports the predicted yields for planets of a range of masses assuming the Cassan et al. mass function, where we have averaged the yields for a given mass over a range of semi-major axes between 0.03-30 AU. The majority of these planets will be detected in regions near the peak sensitivity of WFIRST-2.4 near at roughly 2 AU.

For comparison, we show the results for the WFIRST-2.4, DRM1, and DRM2, as well as Euclid in Table 2-4 and Table 2-5. All three assume the same Cassan et al.[80] mass function. The yields for WFIRST DRM1 assume a total of 432 days observations, whereas the yields for DRM2 assume a total 266 days. For Euclid, we have assumed a 300 day survey, which may be possible with an extended mission.

We also quote the expected yields for free-floating planets, assuming that there is one free-floating planet with a given mass per star in the Galaxy. The yields for Earth-mass free-floating planets are compared for different designs in Table 2-5. For other masses and WFIRST-2.4, we find yields of 4 (Mars-mass), 204 (10 Earth masses), 746 (100 Earth masses or Saturn mass), and 2472 (1000 Earth masses or ~3 Jupiter masses) planets.

## D.4. Comparison Between WFIRST-DRM1 and WFIRST-2.4 Yields

WFIRST-2.4 is intrinsically more capable than the previous WFIRST designs. This improvement is due to the net result of two competing effects. First, the larger aperture but somewhat smaller bandpass results in a net increase in the photon collection rate of ~ 2.28 times that of DRM1. Second, because of the smaller field of view of the WFIRST-2.4 instrument, a larger number of pointings are required to monitor the same total area of the bulge. Therefore, in order to maintain the same cadence, the exposure time per field must be smaller. Including overheads, the net photon count per sample taking into account all these effects is ~1.42 larger for WFIRST-2.4 than DRM1. The effect of this increased photon count on the yields depends primarily on the form of the cumulative number of detections as a function of the minimum threshold for detection, which roughly follows a power law of the form $N(>\Delta\chi^2) \propto (\Delta\chi^2)^a$, with the exponent a depending on the mass and separation of the planets of interest. Averaged over separations between 0.03-30 AU, the exponent varies from a =-0.8 for low-mass planets of ~$0.1 M_{Earth}$, to a =-0.6 for high-mass planets of mass >100 $M_{Earth}$. Since $\Delta\chi^2 \propto N_{phot}$, this number of detections scales with the photon rate in the same way.

The net result of these two effects is a yield for WFIRST-2.4 that is a factor of ~1.25 times larger than that predicted for WFIRST DRM1 averaged over planet mass, ranging from a 24% increase for giant planets to a 31% increase for Mars-mass planets. The larger relative increase in the yields of the lowest-mass planets simply reflects the fact that these planets are near the edge of the mission sensitivity for DRM1. In fact, our choice for the number of fields in the baseline survey design (10) is better optimized for high-mass planets than low-mass planets. Therefore, the yields of low-mass planets could be further increased by choosing a smaller number of fields, although of course this would come at the expense of the number of high-mass planet detections. This further suggests that WFIRST-2.4 may have significant sensitivity to other classes of planets that were just beyond the region of peak sensitivity for DRM1, e.g., very small separation planets and massive habitable planets.

## D.5. Uncertainties and Future Work

The basic input ingredients that are required for estimating the yield of a microlensing exoplanet survey are a (1) model for the spatial, kinematic, and luminosity distribution of source stars, (2) model for the spatial, mass, and kinematic distributions of the host lenses, which are then used to estimate microlensing event rates and event parameter distributions, and (3) a model for the probability distribution of planets as a function of planet mass and semi-major axis, and host mass and distance.

Unfortunately, for the regimes of interest for WFIRST-2.4, the properties of the populations of sources and host lenses are poorly constrained by empirical data, leading directly to relatively large uncertainties in the final yields. In particular, the magnitude distribution of source stars in the fields of interest has not been measured in the passbands and to the faint magnitudes that will be probed by WFIRST-2.4. Similarly, the microlensing event rates for some of the fields of interest have not been measured, as these fields typically suffer high optical extinction, and to date all microlensing surveys have been performed in the far optical.

The event rates and source stars densities adopted in this report were scaled to match published microlensing optical depth estimates[136,137,138,139,140] and source star luminosity function[141] from fields further from the plane than the preferred WFIRST-2.4 fields. The MOA team has recently performed a preliminary analysis of the 2006-2007 MOA-II data[79] and used this to determine the optical depth and event rate toward each of the 22 MOA fields, each covering 2.2 $deg^2$. Four of these fields contain 45% of all the analyzed microlensing events and overlap with (or are very close to) some of the proposed WFIRST-2.4 fields. These results indicate a significantly higher optical depth than assumed in this report. Because they are based on preliminary, unpublished MOA-II data and analysis, we have chosen to be conservative and not use these for our baseline yields. However, we note that they have much higher statistical weight than the published results, and if they are correct, our planet yields are significantly underestimated.

Another, somewhat subtler, source of uncertainty in the planet yields that may significantly impact the choice of target fields is the relative frequency of planets in the Galactic bulge versus disk. The relative contribution of the bulge and disk lenses to the event rate varies as a function of Galactic latitude *b*, with bulge lenses expected to dominate at low $|b|$.[142] If the Galactic bulge happens to be devoid of planets, e.g., because of an extreme radiation environment during a starburst-like bulge formation event,[143] then it would likely be desirable to avoid such low latitude fields.

These uncertainties suggest the necessity for several avenues of future work to prepare for the WFIRST-

2.4 mission. First, measurements of the source luminosity function to faint magnitudes with HST should be performed in the potential target fields, preferably in the near infrared (1-2 microns). It is possible that archival data exists that would be suitable for this purpose. Second, a near-IR microlensing survey of the potential WFIRST fields should be performed from an excellent site, both to measure the microlensing event rate in these fields, and to estimate the relative bulge and disk contributions to the planet detection rates. The latter will also require follow-up observations with AO facilities and/or HST in order to measure or constrain the distances to the detected systems.

Additional work is also needed in order to provide detailed and quantitative estimates of the ability to measure detailed properties of the detected planetary systems, including masses, distances, and orbital parameters from the WFIRST-2.4 data alone, and host star temperatures and metallicities from follow-up observations with JWST and/or future ELTs

Finally, the yields for planets near the edge of the sensitivity of WFIRST-2.4 suffer from additional uncertainties beyond those arising the sources discussed above. This is due to the relatively strong dependence of number of detections on $\Delta\chi^2$ in these regimes, which includes planets with very small or very large separations, and very low mass planets. As a result of the strong scaling with $\Delta\chi^2$, small differences in the assumptions and approximations needed to make these predictions result in large changes in the estimates of the number of detected planets.

Predictions for the yield of habitable planets suffer from all of the uncertainties above, but are also sensitive to additional assumptions, such as the mass-bolometric luminosity relationship for stars in the bulge and disk, the age and metallicity of the stars in the bulge and disk, and the precise definitions for the mass and semi-major axes boundaries of the habitable zone. Therefore, the yields of habitable planets are particularly uncertain. Initial estimates indicate that WFIRST-2.4 will be sensitive to habitable planets, if the intrinsic frequency is large. However, substantially more work needs to be done to provide robust estimates of the habitable planet yield.

## Appendix E WFIRST-2.4 Coronagraphy

As described in §2.5.2, the optional exoplanet direct imaging survey using a coronagraph will survey the nearest stars in visible light, detecting and characterizing planets ranging in size from Neptune-like to larger than Jupiter. This will provide a wealth of new science, complementing microlensing and ground-based surveys, and will provide the first spectral data of planets similar to those in our Solar System. Additionally, disk imaging will provide new insights into the formation mechanisms of planetary systems as well as revel signposts to hidden planets about those target stars. The survey, data-set, and operations requirements listed in Figure 2-26 for this science program arise from practical considerations of what is possible with current coronagraph and wavefront control technology and from detailed Monte-Carlo simulations of possible mission scenarios. These simulations show that in 1 year of mission time it is possible to survey well over 100 nearby stars for planets and disks, down to the habitable zone around the closest (less than 10 pc). The baseline capability described in §3.4, meets these requirements and enables the rich science described in §2.5.2. In this brief appendix, we summarize the analyses and simulations that were used to determine these requirements.

### E.1. Exoplanet Population Modeling

In order to derive the coronagraph instrument performance required to achieve the mission science goals, it is necessary to model distributions of orbital and planetary physical parameters (mass, radius and albedo) of the target planet population. To do so, we use distributions of parameters derived from Doppler spectroscopy and transit photometry surveys and corrected for the selection biases of the particular instruments used to collect the original data sets.[144,145,146] There are two main difficulties in this task. First, observing baselines for these surveys are relatively short, with complete data available only for orbits of less than 100 days for all planetary types, and less than 2000 days for giant planets. Second, there are inconsistencies between the radius distribution provided by transit photometry (and specifically the Kepler data set) and the minimum mass distributions derived from Doppler spectroscopy due either to incomplete data or unmodeled physical effects in planetary density distributions.

Despite these difficulties, the use of existing population statistics is our best approach towards modeling a realistic planetary population, and so we pursue this approach while noting that one of the main science objectives of the direct imaging and microlensing science programs is to resolve inconsistencies in the current statistics and expand the population of known planets to much longer periods. Because the detectability of exoplanets via imaging is dependent on radius we begin by sampling the known radius distribution from the Kepler data set. We define logarithmic mass bins as in Fressin et al.[146] and use the selection bias-corrected occurrence rates from that paper to populate each bin. In order to extrapolate the statistics for short period planets (85 days for Earth-radius planets up to 418 days for 'large Neptunes' and Giants), we assume that occurrence statistics are constant for all periods and extrapolate using the power-law period distribution (exponent = 0.26) derived from the 8 year Keck planet search Doppler spectroscopy survey.[144] Because we expect a drop-off in higher period planets, we introduce a damping term to this distribution equivalent to an exponential drop-off after 60 AU.[147]

Having sampled the radius distribution (using uniform logarithmic functions for each radius bin) we derive the mass distribution by applying density models from Fortney et al.[148] These are comprised of continuous density functions for planets modeled as mixtures of ice/rock and rock/iron and 4D lookup tables for giant planets with the density parameterized by planet core mass and planet-star separation. The semi-major axes for the planets are sampled from the same period distribution described above, biased such that the largest planets (> 15 $R_{Earth}$) are placed within 0.05 AU to account for their inflation. This procedure produces a logarithmic mass distribution that is broadly consistent with the minimum mass distribution from Howard et al.[145] although failing to reproduce all of the specific features. However, the resultant mass distribution matches the mean occurrence rates in all mass bins (i.e., ~10% solar-type star hosting gas giant planets with periods between 2 and 2000 days).[149]

The remaining orbital parameters are sampled so as to produce isotropically oriented orbits (uniformly distributed argument of periapsis and longitude of the ascending node, and sinusoidally distributed inclination). Finally, the geometric albedo is interpolated from models in Cahoy et al.[150] based on the mass and semi-major axis, with heavy metal abundances approximately matching those observed for equivalently sized solar system planets. The phase function used to calculate instantaneous planet-star flux ratios at the time of an observation is either taken to be isotropic (a Lambertian sphere) or interpolated from models in Sudarsky et al.[151] based on the semi-major axis.

The total number of planets sampled is determined by generating a Poisson random variable for each radius bin with distribution parameter given by the product of the occurrence rate for that bin and the total number of target stars. This produces an overall occurrence rate of approximately 3 planets per sun-like star, which is marginally higher than the extrapolated occurrence rate from Doppler spectroscopy surveys, but is broadly consistent with the results of transit and microlensing surveys. When the input target list is the 1926 main sequence, non-binary stars within 30 pc of the Earth available in the SIMBAD database, the resulting distribution of planets (at random points in their orbit given by a uniform mean anomaly distribution) is the result presented in Figure 2-21.

## E.2. Coronagraph Performance Requirements

As described in §2.5.2.1, a coronagraph is characterized by the depth of its detection contrast, the field of view (its inner and outer working angles), and the overall stability of the telescope and wavefront control system. The inner and outer working angles define the region of search space in which the high detection contrast is achieved. Smaller inner working angle allows visibility closer to the target star, thus imaging smaller planets (with lower albedo) and planets close to the habitable zone. Larger outer working angle allows imaging a larger fraction of the exosolar system, capturing the large, outer gas giant planets and outer regions of the debris disks.

The inner working angle and contrast achievable is a key characteristic of a specific coronagraph implementation and it is typically a tradeoff with contrast and throughput. Additionally, smaller inner working angle results in a higher sensitivity to low order aberrations induced by telescope instability. Small inner working angle also becomes challenging as the pupil geometry of the telescope becomes more complex, as is the case with the 2.4-m telescope. Figure 2-21 shows that at a detection limit of 1 ppb and an inner working angle between 0.1 and 0.2 arcsec, a population of giant planets is likely to exist that is detectable by WFIRST-2.4. Our current requirement of 0.1 arcsec at 400 nm (corresponding to a 3 $\lambda/D$ inner working angle) and a contrast of $10^{-9}$ thus arises from a balance between what has been achieved in the laboratory with current coronagraph approaches, estimates of stability of the 2.4-m telescope, and the potential science yield. Larger inner working angles or less contrast virtually eliminates any potential exoplanet science, though a significant number of disks would still be visible.

The outer working angle of the discovery space is determined by the "pitch" of the deformable mirror, that is, by the number of actuators across each linear dimension. The maximum spatial frequency that a wavefront control system can correct is roughly determined by half the number of actuators. The current requirement on the outer limit of the field of view assumes a 96 x 96 actuator deformable mirror. Again from Figure 2-21, this angle captures almost all of the possible planets (given current models of planet albedo) and the outer regions of most disks.

## E.3. Detection and Observation Requirements

While a number of techniques are in use and continue to be developed for extracting planets from noisy and cluttered backgrounds in ground images,[93,152] the analysis here used a simple matched filtering approach described in Kasdin and Braems[153]. In all cases, a critical sampled PSF is required to extract the planet from the background of speckles from residual wavefront error and exozodiacal light. The detection waveband requirement arises from a compromise between throughput and the ability of the wavefront control system to correct in broadband. Laboratory results show that our current approaches are robustly able to correct in 10% bands while simulations show that the resulting throughput allows for detection in reasonable integration times.

The spectroscopic resolution of the integral field spectrograph was set by optimizing the signal-to-noise on strong atomic and molecular bands expected to be seen in the atmospheres of gas and ice giant planets around nearby stars. Multiple methane ($CH_4$) bands spanning 460 - 990 nm will appear in cool giant atmospheres, and they have widths of 15 - 20% of their wavelengths (e.g., Karkoschka[154]). Water ($H_2O$) features span 650 - 950 nm, and these are relatively narrow, a few percent of the wavelength. If a giant planet has a warmer atmosphere than Jupiter (i.e., located within ~1 AU of the Sun in our Solar System), then its clouds clear and its visible spectrum is dominated by a strong blue color due to Rayleigh scattering. Strong absorptions from sodium (Na) and potassium (K) alkali metals are also expected near 400 and 590 nm (both about 2% wide) in that case, and their detection would help confirm the atmospheric temperature.[150] A spectroscopic resolution of R=70 (1.4% of the wavelength) is sufficient to resolve the $CH_4$ features well while also delivering maximum signal-to-noise on and confirming the narrow natures of the $H_2O$ and atomic features.

Requirements for observations of exozodiacal disks are driven by the desire to observe down to the habitable zones of the nearest stars, to be sensitive to nearby disks with less than 10 times of our own zodiacal disk contrast (10 zodi), and to have spatial resolution of ~0.5 AU, sufficient to resolve structures and gaps in disks that are generated by seen or unseen planets. Fortunately these requirements are satisfied by the same performance parameters that satisfy the coronagraphic exoplanet imaging requirements. A final contrast of $1 \times 10^{-9}$ is sufficient to detect a 2 zodi disk in a solar system twin at 10 pc in a single WFIRST-2.4 resolution element in about 15 hours. A 3 λ/D IWA allows probing this system to 1.6 AU from its star, at the edge of its habitable zone. The WFIRST-2.4 spatial resolution is 50 mas at 550 nm, corresponding to 0.5 AU at 10 pc.

The WFIRST-2.4 system uses an imaging polarimeter with its coronagraph to form a very powerful system for exoplanet science. Exoplanetary systems form within dust and gas clouds surrounding stars. The polarimeter will image details within the observable dust cloud. A polarization precision of 0.5% along with a retardance precision of 5 degrees per resolution element will enable us to distinguish between dust grains and planets and thus directly observe the evolution of planetary systems as well as identify the compositional properties of the dust. This polarimeter will also be used to measure the composition of the atmosphere, the climate, and the nature of the exo-planetary surface features such as solid, liquid, rock & soil. The measured standard deviation of the nightly averaged Stokes Q/I and U/I parameters of less than $5 \times 10^{-6}$ enables the identification of aerosol particles[98] in the exoplanet's atmosphere, as well as an indication of the nature of the surface of the exoplanet.

## E.4. Mission Simulation and Operations Requirements

The generation of ensembles of mission simulations has been extensively described in Savransky et al.[155] The primary goal is to combine a description of the instrument, the constraints of the observatory, modeling of an observation of an exoplanetary system, and an automated survey scheduler to produce multiple realizations of an exoplanet survey. The resulting ensemble of survey simulations is then analyzed to produce distributions of science metrics (i.e., number of planetary detections and spectral characterizations) as well as engineering and programmatic metrics (i.e., spacecraft fuel use and average time per target). We employ both the framework and specific codebase described in the paper above.

The instrument is parameterized by its point spread function, angular-separation dependent coronagraph throughput curve, total throughput of all remaining optics, telescope aperture, pixel angular extent, operating wavelengths, and detector efficiency. The primary instrument constraints derived from its design are the minimum and maximum angular separations of detectable planets (inner and outer working angles) and the contrast of residual, static speckle noise in the search region. The (space) observatory constraints are given by its orbit and allowable sun angles during observations, and the settling time required between repointings of the spacecraft. Integration times for detection and characterization are calculated based on a desired false positive and false negative rate, as in Kasdin and Braems[153].

The scheduling algorithm is derived as a locally minimum cost path through a directed graph whose edge lengths are determined by a linear cost function representing the transition from one target to another. The cost function incorporates terms serving as heuristics for the probability of planetary detection (as parameterized in Figure E.1), and attempts to maximize the number of unique planetary detections, the number of complete spectral characterizations, and the number of planets observed up to 4 times for confirmation and preliminary orbit fitting. From these simulations comes the operations requirements listed in Figure 2-26. Figures E.2 and E.3 show sample outputs of the simulation for four cases considered at different inner working angles and detection contrast.

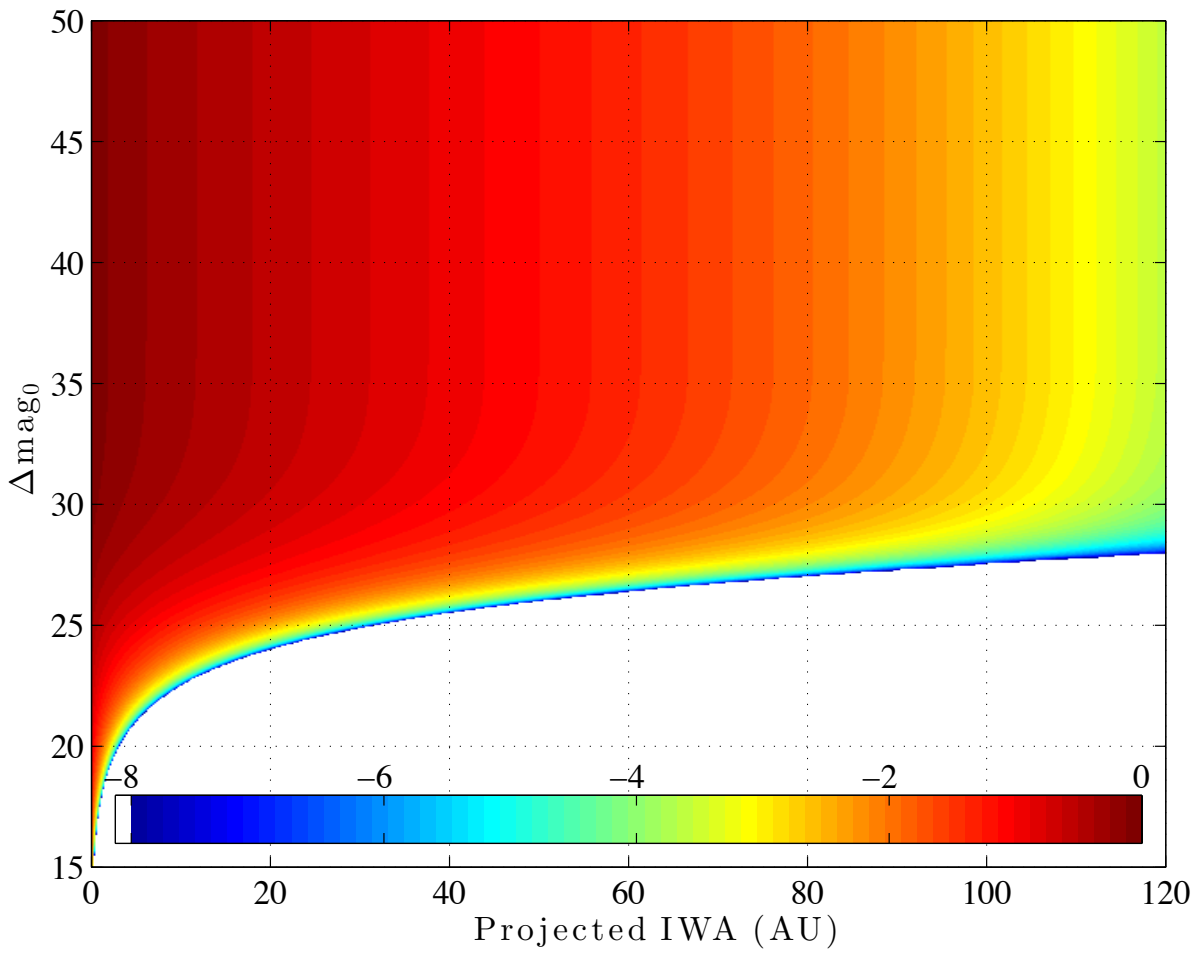

**Figure E.1: Cumulative density function of direct imaging observables, parameterized by the instrument constraints of projected inner working angle and limiting difference in magnitude between star and planet. The value of each bin represents the portion of the planetary population that could be observed by an instrument capable of achieving the corresponding projected IWA and maximum delta magnitude for a particular target system. The planetary population is the one described in §E.1 and the colorscale is logarithmic in powers of 10.**

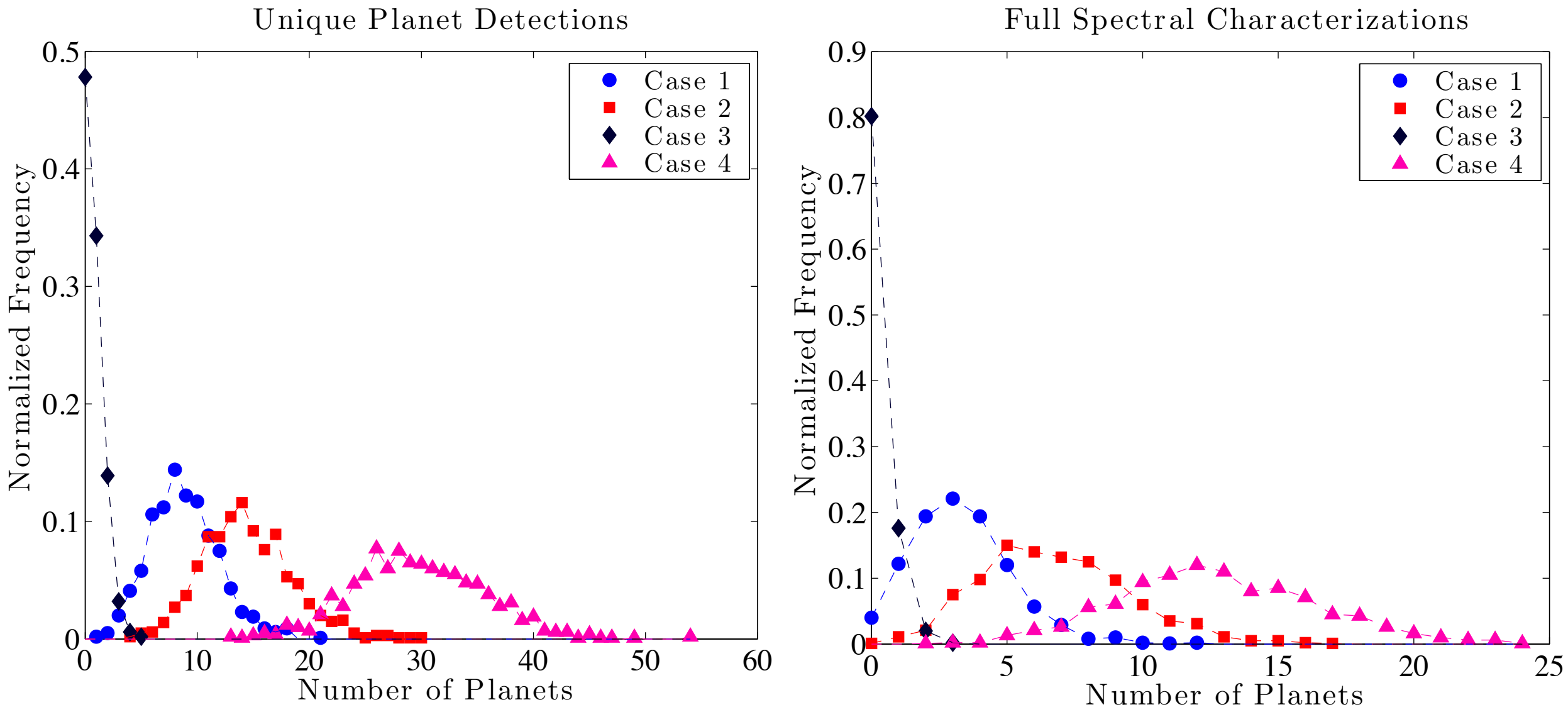

**Figure E.2: Probability distributions from Monte-Carlo mission simulation of number of unique planets found and spectrally characterized for four cases: (1) 0.2 arcsec IWA and 1x10$^{-9}$ detection contrast, (2) 0.1 arcsec IWA and 1x10$^{-9}$ detection contrast, (3) 0.2 arcsec IWA and 1x10$^{-8}$ detection contrast, and (4) 0.2 arcsec IWA and 1x10$^{-10}$ detection contrast.**

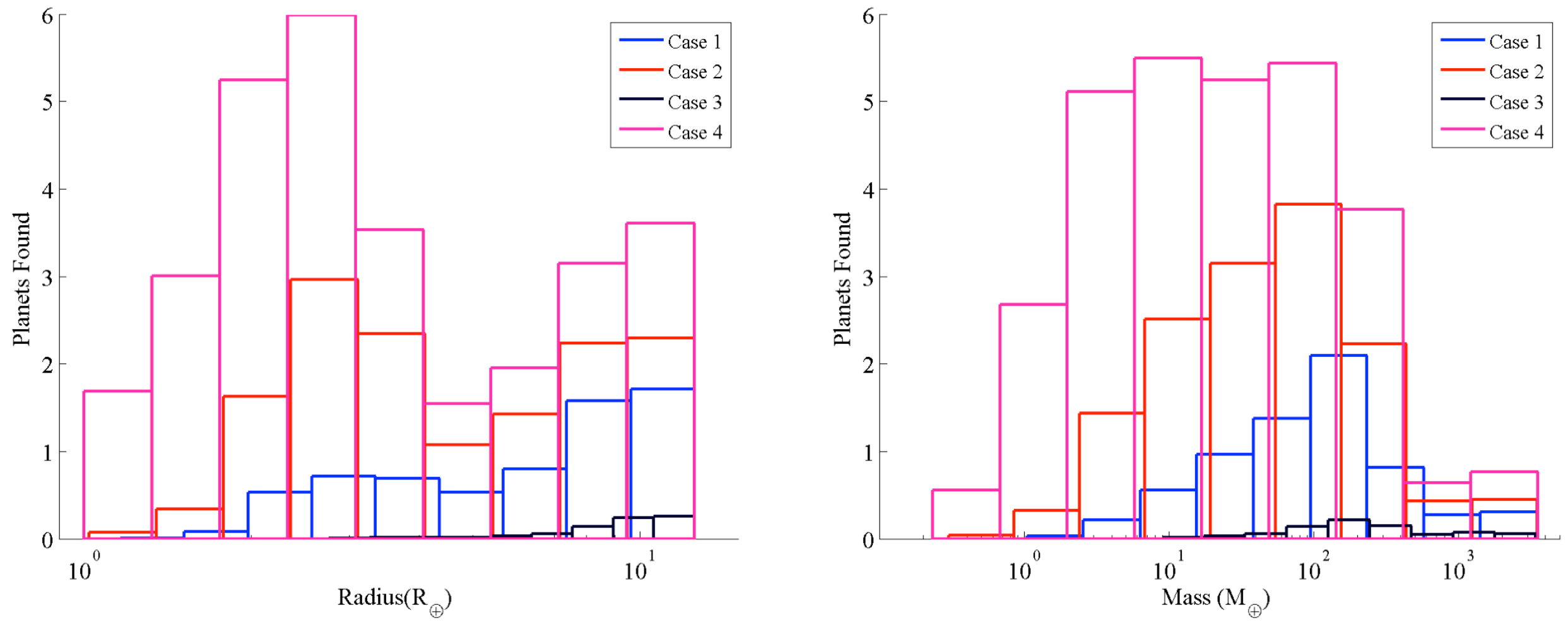

**Figure E.3: Distributions of unique planet detections from Monte Carlo mission simulation as a function of planet radius discovered and planet mass.**

## Appendix F Acronym List

| | |
|---|---|
| 2MASS | 2 Micron All Sky Survey |
| $\Lambda$CDM | Lambda Cold Dark Matter |
| A&A | Astronomy and Astrophysics |
| ACE | Attitude Control Electronics |
| ACS | Advanced Camera for Surveys |
| ACS | Attitude Control System |
| AFTA | Astrophyiscs Focused Telescope Assets |
| AGN | Active Galactic Nuclei |
| AJ | Astronomical Journal |
| AO | Announcement of Opportunity |
| ApJ | Astrophysical Journal |
| ApJL | Astrophysical Journal Letters |
| ApJS | Astrophysical Journal Supplement Series |
| APRA | Astrophysics Research and Analysis |
| AU | Astronomical Unit |
| BAO | Baryon Acoustic Oscillations |
| BOSS | Baryon Oscillation Spectroscopic Survey |
| C&DH | Command and Data Handling |
| CCD | Charged Coupled Device |
| CDR | Critical Design Review |
| CIB | Cosmic Infrared Background |
| CMB | Cosmic Microwave Background |
| CMD | Color-Magnitude Diagram |
| COSMOS | Cosmic Evolution Survey |
| DE | Dark Energy |
| DETF | Dark Energy Task Force |
| DM | Deformable Mirror |
| DOF | Degree of Freedom |
| DRM | Design Reference Mission |
| DRM1 | Design Reference Mission #1 |
| DRM2 | Design Reference Mission #2 |
| DSN | Deep Space Network |
| E(B-V) | Extinction (B-V) |
| eBOSS | Extended Baryon Oscillation Spectroscopic Survey |
| EE | Encircled Energy |
| EMCCD | Electron Multiplying Charge Coupled Device |
| EOL | End of Life |
| EOS | Electromagnetic Observations from Space |
| EPO | Education and Public Outreach |
| ESA | European Space Agency |
| EW | Element Wheel |
| F2 | Fold Flat #2 |
| FGS | Fine Guidance Sensor |
| FoM | Figure of Merit |
| FoMSWG | Figure of Merit Science Working Group |
| FOR | Field-of-Regard |
| FOV | Field-of-View |
| FPA | Focal Plane Array |
| FPE | Focal Plane Electronics |
| FPGA | Field-Programmable Gate Array |
| FSW | Flight Software |
| FY | Fiscal Year |
| GALEX | Galaxy Evolution Explorer |
| GHz | Gigahertz |
| GI | Guest Investigator |
| GO | Guest Observer |
| GR | General Relativity |
| GRS | Galaxy Redshift Survey (includes BAO & RSD) |
| GSFC | Goddard Space Flight Center |
| HCIT | High Contrast Imaging Testbed |
| HgCdTe | Mercury Cadmium Telluride |
| HLS | High Latitude Survey |
| HST | Hubble Space Telescope |
| I&T | Integration and Test |
| ICDH | Instrument Command and Data Handling |
| ICE | Independent Cost Estimate |
| IDRM | Interim Design Reference Mission |
| IFS | Integral Field Spectrograph |
| IFU | Integral Field Unit |
| IMF | Initial Mass Function |
| IR | Infrared |
| IRAC | Infrared Array Camera |
| IRAS | Infrared Astronomical Satellite |
| IWA | Inner Working Angle |
| JCL | Joint Confidence Level |
| JDEM | Joint Dark Energy Mission |
| JMAPS | Joint Milli-Arcsecond Pathfinder Survey |
| JWST | James Webb Space Telescope |
| KDP | Key Decision Point |
| KSC | Kennedy Space Center |
| L2 | Sun-Earth 2$^{nd}$ Lagrangian Point |
| LADEE | Lunar Atmosphere and Dust Environment Explorer |
| LCCE | Lifecycle Cost Estimate |
| LCRD | Laser Communications Relay Demonstration |
| LLCD | Lunar Laser Communications Demonstration |
| LOWFS | Low-Order Wavefront Sensor |
| LRD | Launch Readiness Date |
| LRG | Luminous Red Galaxy |
| LRO | Lunar Reconnaissance Orbiter |
| LSST | Large Synoptic Survey Telescope |
| M3 | Tertiary Mirror |
| mas | Milli-Arc-Seconds |
| Mbps | Megabits per Second |
| MCB | Mechanism Control Box |
| MCR | Mission Concept Review |
| MEL | Master Equipment List |
| MEMS | Microelectromechanical System |
| MMS | Multimission Modular Spacecraft |
| MNRAS | Monthly Notices of the Royal Astronomical Society |
| MOC | Mission Operations Center |
| Mpc | Megaparsec |
| MRS | Module Restraint System |
| MS-DESI | Mid-Scale Dark Energy Spectroscopic Instrument |
| NASA | National Aeronautics and Space Administration |
| NIR | Near Infrared |
| NIRCam | Near Infrared Camera |
| NRC | National Research Council |
| NWNH | New Worlds, New Horizons in Astronomy and Astrophysics |
| OAP | Off-Axis Parabola |
| OBA | Outer Barrel Assembly |
| OTA | Optical Telescope Assembly |
| OWA | Outer Working Angle |
| PAF | Payload Attach Fitting |
| PASP | Publication of the Astronomical Society of the Pacific |
| PDR | Preliminary Design Review |
| Photo-z | Photometric Redshift |
| PIAA-CMC | Phase-Induced Amplitude Apodization Complex Mask Coronagraph |
| PID | Proportional-Integral-Derivative |
| PPB | Parts Per Billion |
| PSF | Point Spread Function |
| PSR | Pre-Ship Review |
| PZCS | Photo-Z Calibration Survey |
| QE | Quantum Efficiency |
| QSO | Quasi-Stellar Object (Quasar) |
| RAAN | Right Ascension of the Ascending Node |
| RMS | Root Mean Square |
| ROIC | Readout Integrated Circuit |
| ROSES | Research Opportunities in Space and Earth Sciences |
| RSD | Redshift Space Distortion |
| RV | Radial Velocity |
| S/C | Spacecraft |
| S/N | Signal/Noise |
| SAT | Strategic Astrophysics Technology |
| SCA | Sensor Chip Assembly |
| SCE | Sensor Cold Electronics |
| SCG | Science Coordination Group |
| SCU | SCE Control Unit |
| SDO | Solar Dynamics Observatory |
| SDT | Science Definition Team |
| SDSS | Sloan Digital Sky Survey |
| SFR | Star Formation Rates |
| SED | Spectral Energy Distribution |
| SiC | Silicon Carbide |
| SIR | Systems Integration Review |
| SN | Supernova |
| SNe | Supernovae |
| SNR | Signal to Noise Ratio |
| SRR | System Requirements Review |
| SuMIRe | Subaru Measurement of Images and Redshifts |
| SUTR | Sample Up The Ramp |
| SZ | Sunyaev-Zel'dovich |
| TBD | To Be Determined |
| Tbits | Terabits |
| TBR | To Be Resolved |
| TDEM | Technology Development for Exoplanet Missions |
| TMA | Three Mirror Anastigmat |
| TRL | Technology Readiness Level |
| TRMD | Transition from Radiation to Matter Domination |
| UKIDSS | UKIRT Infrared Deep Sky Survey |
| ULE | Ultra Low Expansion Fused Silica |
| US | United States |
| Vdc | Volts Direct Current |
| VISTA | Visible and Infrared Survey Telescope for Astronomy |
| WFC3/IR | Wide Field Camera 3/Infrared channel |
| WFCS | Wavefront Control System |
| WFIRST | Wide-Field Infrared Survey Telescope |
| WFIRST-2.4 | WFIRST mission using a 2.4-m telescope |
| WISE | Wide-field Infrared Survey Explorer |
| WL | Weak Lensing |
| WMAP | Wilkinson Microwave Anisotropy Probe |

## Appendix G References